\documentclass[11pt,a4paper,twoside,openright]{book}
{}
{}

\usepackage{graphicx} 
\usepackage{float}
\usepackage{color}
\usepackage[toc,page]{appendix}
\usepackage{chngcntr} 
\usepackage{enumerate}
\usepackage{latexsym,amsthm,amssymb,amsfonts,amsbsy,amsmath}
\usepackage{amsmath}
\usepackage{amscd} 
\usepackage{amsfonts}
\usepackage{amssymb}
\usepackage{mathrsfs}
\usepackage{amsthm}
\usepackage{frcursive}
\usepackage{bm} 
\usepackage{bbm} 
\usepackage{slashed} 
\usepackage{feynmp} 
\usepackage{cancel}
\usepackage{xcolor}
\newcommand\Ccancel[2][black]{\renewcommand\CancelColor{\color{#1}}\cancel{#2}}
\usepackage{tikz}

\usepackage[latin1]{inputenc} 
\usepackage[T1]{fontenc} 
\usepackage[francais, english]{babel}

\definecolor{lightred}{RGB}{255,127,127}
\definecolor{lightgreen}{RGB}{127,255,127}
\definecolor{lightblue}{RGB}{127,127,255}
\definecolor{linkcolor}{rgb}{0,0,0.6}
\usepackage[ pdftex,colorlinks=true,
pdfstartview=FitV,
linkcolor= linkcolor,
citecolor= linkcolor,
urlcolor= linkcolor,
hyperindex=true,
hyperfigures=false]
{hyperref}

\makeatletter

\@addtoreset{equation}{section}
\makeatother

\usepackage{tabularx} 

\usepackage{fancyhdr} 

\theoremstyle{plain}
\newtheorem{theorem}{Theorem}[section]
\newtheorem{definition}[theorem]{Definition}
\newtheorem{lemma}[theorem]{Lemma}
\newtheorem{proposition}[theorem]{Proposition}
\newtheorem{corollary}[theorem]{Corollary}
\newtheorem{conjecture}[theorem]{Conjecture}

\definecolor{myGreen}{rgb}{0.0,0.7,0.0}

\usepackage[nosort]{cite}
\usepackage{pdfpages}

\DeclareFontEncoding{LS1}{}{}
\DeclareFontSubstitution{LS1}{stix}{m}{n}
\DeclareSymbolFont{stixsymbols}{LS1}{stixscr}{m}{n}
\SetSymbolFont{stixsymbols}{bold}{LS1}{stixscr}{b}{n}
\DeclareMathSymbol{\kay}{\mathalpha}{stixsymbols}{"6B}

\title{Integrable models with twist function and affine Gaudin models}
\author{Sylvain Lacroix}

\newcommand{\mathfrc}[1]{\mbox{\small\cursive\slshape#1}}

\newcommand{\noi}{\noindent}

\newcommand{\dd}{\text{d}}
\newcommand{\p}{\partial}
\newcommand{\g}{\mathfrak{g}}
\newcommand{\f}{\mathfrak{f}}
\newcommand{\n}{\mathfrak{n}}
\newcommand{\h}{\mathfrak{h}}
\newcommand{\bo}{\mathfrak{b}}

\newcommand{\so}{\mathfrak{so}}
\newcommand{\su}{\mathfrak{su}}
\newcommand{\spc}{\mathfrak{sp}}
\renewcommand{\sl}{\mathfrak{sl}}

\newcommand{\C}{\mathbb{C}}
\newcommand{\Kb}{\mathbb{K}}
\newcommand{\s}{\sigma}

\newcommand{\Ad}{\text{Ad}}
\newcommand{\ad}{\text{ad}}

\newcommand{\M}{\mathcal{M}}
\newcommand{\E}{\mathcal{E}}
\newcommand{\R}{\mathbb{R}}
\newcommand{\N}{\mathbb{N}}
\newcommand{\Nc}{\mathcal{N}}
\newcommand{\W}{\mathcal{W}}
\newcommand{\Z}{\mathbb{Z}}
\newcommand{\CP}{\mathbb{P}^1}
\newcommand{\Sc}{\mathbb{S}^1}
\newcommand{\Zc}{\mathcal{Z}}
\newcommand{\Tc}{\mathscr{T}}
\newcommand{\J}{\mathscr{J}}
\newcommand{\K}{\mathscr{K}}
\newcommand{\Hc}{\mathcal{H}}
\newcommand{\Yc}{\mathcal{Y}}
\newcommand{\Y}{\mathcal{Y}}
\newcommand{\Cc}{\mathcal{C}}
\newcommand{\Pc}{\mathcal{P}}
\newcommand{\Q}{\mathcal{Q}}
\newcommand{\Oc}{\mathcal{O}}
\newcommand{\Id}{\text{Id}}
\newcommand{\Lc}{\mathcal{L}}
\newcommand{\Ac}{\mathcal{A}}
\newcommand{\Lct}{\widetilde{\mathcal{L}}}
\newcommand{\Rc}{\mathcal{R}}
\newcommand{\Rct}{\widetilde{\mathcal{R}}}
\newcommand{\ti}[1]{_{\bm{\underline{#1}}}}
\newcommand{\Tr}{\text{Tr}}
\newcommand{\Pexp}{\text{P}\overleftarrow{\text{exp}}}
\newcommand{\lsb}{\left[}
\newcommand{\rsb}{\right]}
\newcommand{\lrb}{\left(}
\newcommand{\rrb}{\right)}
\newcommand{\lwb}{\left\lbrace}
\newcommand{\rwb}{\right\rbrace}
\newcommand{\F}{\mathcal{F}}
\newcommand{\YB}[3]{\mathcal{Y}\ti{#1#2#3}(\lambda_{#1},\lambda_{#2},\lambda_{#3})}
\newcommand{\vp}{\varphi}
\def\res{\mathop{\text{res}\,}}

\newcommand{\etat}{\widetilde{\eta}}
\newcommand{\Rt}{\widetilde{R}}
\newcommand{\gt}{\widetilde{g}}
\newcommand{\jt}{\widetilde{j}}
\newcommand{\hti}{\widetilde{h}}

\newcommand{\Gdz}{G^{\text{diag}}_0}
\newcommand{\Dz}{G_0 \times G_0}

\newcommand{\Xt}{\widetilde{X}}
\newcommand{\RgXt}{\tilde{R}_{\tilde{g}}\tilde{X}}
\newcommand{\lt}{\widetilde{\lambda}}
\newcommand{\Lg}{\mathcal{L}(\mathfrak{g})}
\newcommand{\diag}{\text{diag}}
\newcommand{\fs}[2]{f^{#1}_{{\color{white} #1}\, #2}\;}
\newcommand{\ft}[2]{f^{#1}_{{\color{white} #1}\, #2}}

\newcommand{\Span}{\text{Span}}
\newcommand{\Aut}{\text{Aut}}
\newcommand{\Inn}{\text{Inn}}
\newcommand{\ach}{\check{\alpha}}
\newcommand{\och}{\check{\omega}}
\newcommand{\rch}{\check{\rho}}
\newcommand{\fd}{(\cdot,\cdot)}
\newcommand{\PB}{\lwb \cdot, \cdot \rwb}
\newcommand{\X}{\mathcal{X}}
\newcommand{\ev}{\text{ev}}
\newcommand{\LB}{[\cdot,\cdot]}
\newcommand{\End}{\text{End}}
\newcommand{\Ker}{\text{Ker}}
\newcommand{\pl}{(\!(\lambda)\!)}
\newcommand{\Ct}{\widetilde{C}}
\newcommand{\Tp}{^{\mathsf{T}}}
\newcommand{\U}{\mathscr{U}}
\newcommand{\Diff}[1]{\text{Diff}(#1)}
\newcommand{\gd}{\mathfrak{g}_{0,\rm diag}}
\newcommand{\Ls}{\mathscr{L}}
\newcommand{\Hs}{\mathscr{H}}
\newcommand{\Ms}{\mathscr{M}}
\newcommand{\Qs}{\mathscr{Q}}
\newcommand{\rc}{\mathfrc{r}}
\newcommand{\Ta}{\mathsf{T}}
\newcommand{\D}{\mathcal{D}}
\newcommand{\Co}{\text{Conn}_\g(\Sc)}
\newcommand{\Jc}{\mathcal{J}}
\newcommand{\Loop}{\mathfrak{\g}[t,t^{-1}]}
\newcommand{\Tt}{\mathcal{T}}
\newcommand{\Kd}{\mathsf{K}}
\newcommand{\Dd}{\mathsf{D}}
\newcommand{\gft}{\widetilde{\mathfrak{g}}}
\newcommand{\Nt}{\widetilde{\nabla}}
\newcommand{\Obs}{\text{Obs}}
\newcommand{\pk}{\pi_{\bm k}}
\newcommand{\Scc}{\mathcal{S}}
\newcommand{\rp}[3]{#1^{(#2)}_{[#3]}}

\newcommand{\zb}{\bm{z}}
\newcommand{\lb}{\bm{\lambda}}
\newcommand{\vl}{v_{\bm{\lambda}}}
\newcommand{\Di}[1]{#1_{(\infty)}}
\newcommand{\Hl}{H_{\bm{\lambda}}}
\newcommand{\Ll}{L_{\bm{\lambda}}}
\newcommand{\Il}{I_{\bm{\lambda}}}
\newcommand{\Ev}{\mathcal{E}_{\text{vac}}}
\newcommand{\Evk}[1]{\mathcal{E}_{\text{vac},#1}}
\newcommand{\Ewk}[1]{\mathcal{E}_{i,#1}}
\newcommand{\Exo}{\mathcal{E}_{i}^{\,\text{on}}}

\newcommand{\Po}{\Psi^{\text{on}}}

\newcommand{\sld}{\mathfrak{sl}(2,\mathbb{C})}
\newcommand{\sldb}{\bm{\mathfrak{sl}(2,}\pmb{\mathbb{C}}\bm{)}}
\newcommand{\Opz}{\text{Op}^{\text{RS}}_{\null^L \g,\zb}\bigl(\mathbb{P}^1\bigr)}
\newcommand{\Zg}{\mathcal{Z}_{\bm z}(\g)}
\newcommand{\Zgc}{\mathcal{Z}^\sigma_{\bm z}(\g)}
\newcommand{\Fun}[1]{\text{Fun}\left(#1\right)}
\newcommand{\op}[1]{\text{op}_{#1}\bigl(\mathbb{P}^1\bigr)}
\newcommand{\Op}[1]{\text{Op}_{#1}\bigl(\mathbb{P}^1\bigr)}
\newcommand{\mop}[1]{\text{mOp}_{#1}\bigl(\mathbb{P}^1\bigr)}
\newcommand{\Opsld}{\text{Op}^{\text{RS}}_{\mathfrak{sl}(2,\C),\zb}\bigl(\mathbb{P}^1\bigr)}
\newcommand{\htg}[1]{\text{ht}(#1)}
\newcommand{\pn}{p_{-1}}
\renewcommand{\a}{\mathfrak{a}}
\newcommand{\Cog}{\text{Conn}_\g \bigl(\CP\bigr)}
\newcommand{\Lang}{\null^L \mathfrak{g}}
\newcommand{\pch}{\check{p}}
\newcommand{\nc}{[\nabla]_{\text{can}}}
\newcommand{\Pcw}{\Psi_{\bm c}(\bm w)}
\newcommand{\Pcwo}{\Psi^{\text{on}}_{\bm c}(\bm w)}
\newcommand{\Pcwt}{\Psi_{\bm{\widetilde{c}}}(\bm{\widetilde{w}})}
\newcommand{\Pcwto}{\Psi^{\text{on}}_{\bm{\widetilde{c}}}(\bm{\widetilde{w}})}
\newcommand{\Opzc}{\text{Op}^{\text{RS},\,\sigma}_{\null^L \g,\zb}\bigl(\mathbb{P}^1\bigr)}
\newcommand{\bl}{\null^L\mathfrak{b}_+}
\newcommand{\Nl}{\null^L N_+ (\mathcal{M})}
\newcommand{\Opp}[1]{\text{Op}^{\,\varphi}_{#1}\bigl(\mathbb{P}^1\bigr)}
\newcommand{\Ps}{\mathscr{P}}
\newcommand{\Oppz}{\text{Op}^{\,\vp}_{\null^L \g,\zb}\bigl(\mathbb{P}^1\bigr)}

\makeatletter
\def\cleardoublepage{\clearpage\if@twoside \ifodd\c@page\else
\hbox{}
\thispagestyle{empty}
\newpage
\if@twocolumn\hbox{}\newpage\fi\fi\fi}
\makeatother

\begin{document}

\thispagestyle{empty}

\includepdf[offset=8 -30]{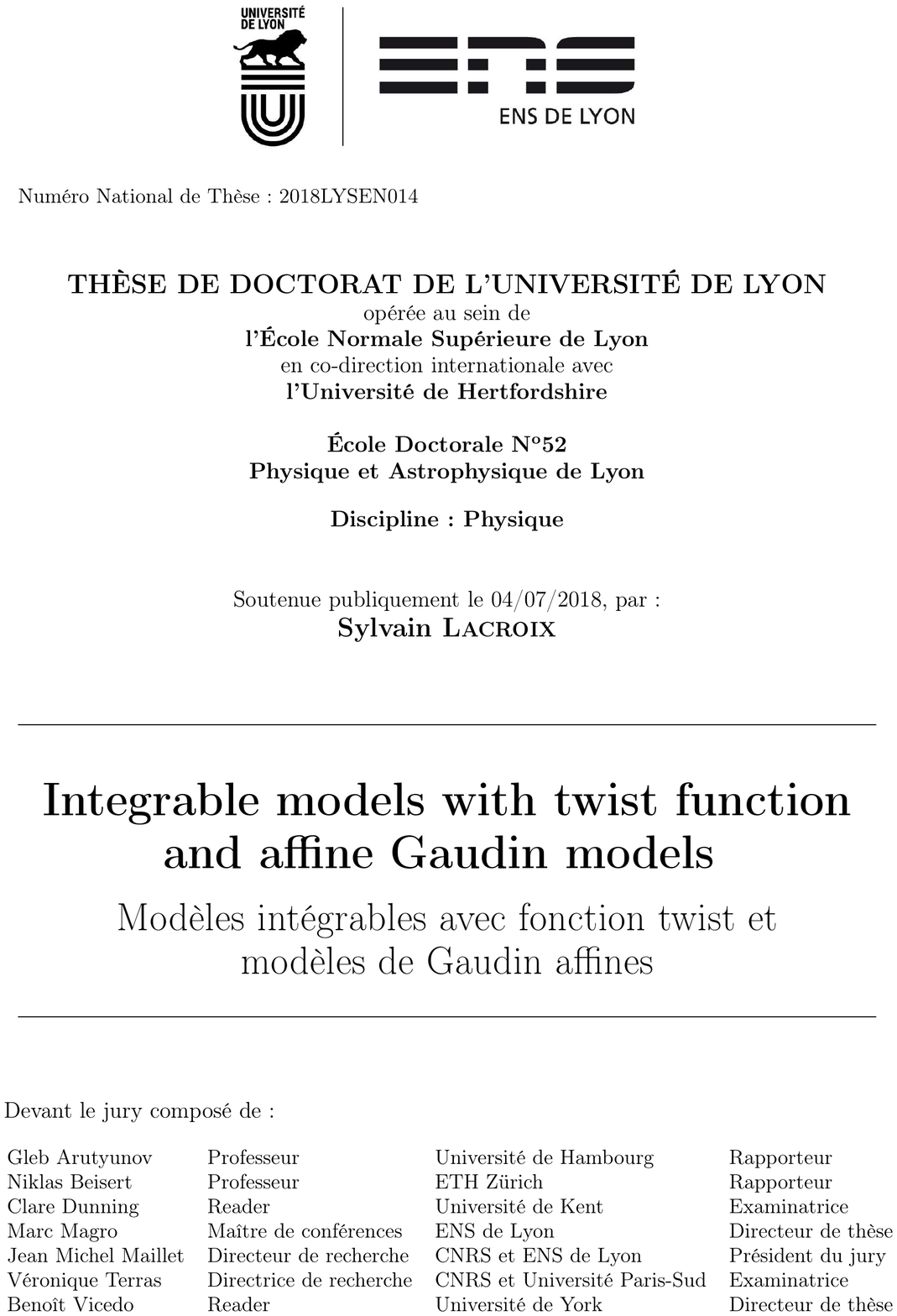}

\newpage \thispagestyle{empty}~

\newpage

\thispagestyle{empty}

~\vspace{70pt}~

\begin{center}
\Huge \textbf{Integrable models with twist function \\ and affine Gaudin models \\~\\~\\}

\LARGE Sylvain Lacroix \\ ~\\

\LARGE PhD thesis
\end{center}\vspace{45pt}

\begin{center}
\begin{minipage}[L]{0.48\textwidth}
\hspace{-30pt}\includegraphics[scale=0.45]{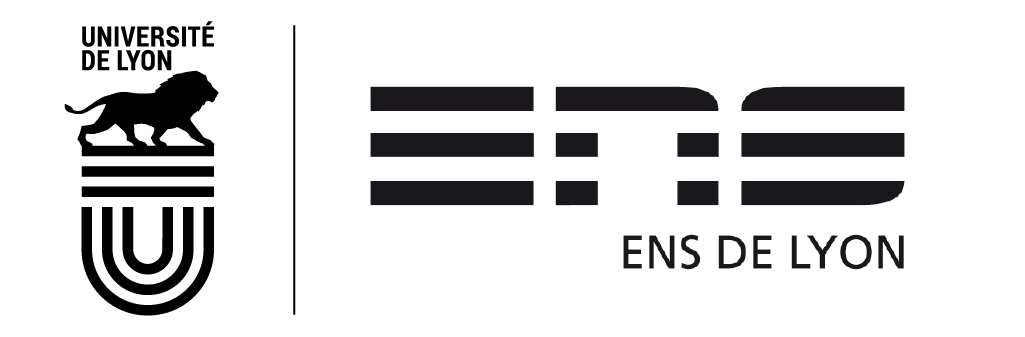}
\end{minipage}
\begin{minipage}[R]{0.48\textwidth}
\hspace{60pt}
\includegraphics[scale=0.55]{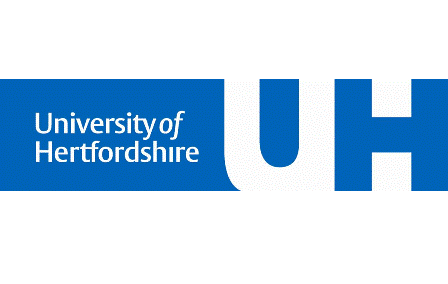}
\end{minipage}
\end{center}\vspace{45pt}
\begin{center}
\begin{minipage}[L]{0.20\textwidth}
\hspace{-35pt}
\includegraphics[scale=0.75]{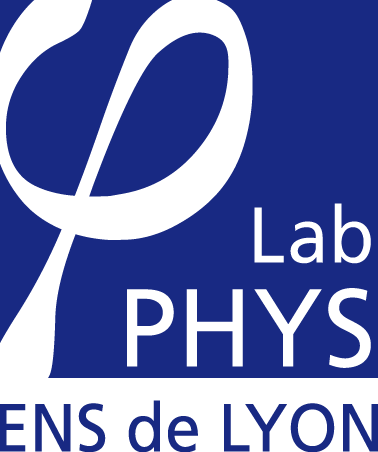}
\end{minipage}
\begin{minipage}[L]{0.30\textwidth}
\hspace{-20pt}
\includegraphics[scale=0.39]{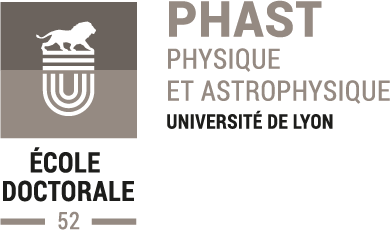}
\end{minipage}
\begin{minipage}[L]{0.25\textwidth}
\hspace{15pt}
\includegraphics[scale=0.4]{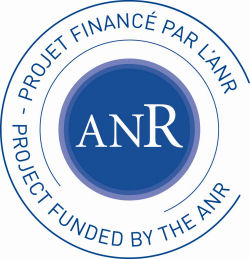}
\end{minipage}
\begin{minipage}[L]{0.20\textwidth}
\hspace{25pt}\includegraphics[scale=0.5]{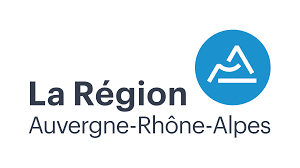}
\end{minipage}
\end{center}

\newpage

\cleardoublepage
\chapter*{Foreword}

 \setcounter{page}{3}

\noi This thesis is the final dissertation of my doctorate, that I prepared under the cosupervision of Marc Magro (at the ENS de Lyon) and Benoit Vicedo (at the University of Hertfordshire). It is mostly based on the articles or preprints I published, together with collaborators, during my PhD. These publications and preprints are listed below:\\~\\
\begin{tabular}{cl}
\cite{Delduc:2015xdm} & \textbf{On the Hamiltonian integrability of the bi-Yang-Baxter $\bm\sigma$-model} \\ 
 & Fran\c{c}ois~Delduc, Sylvain~Lacroix, Marc~Magro and Beno\^it~Vicedo, \\
 & \href{http://dx.doi.org/10.1007/JHEP03(2016)104}{\emph{JHEP} {\bf 1603} (2016) 104}, [\href{https://arxiv.org/abs/1512.02462}{{\tt 1512.02462}}].\\[15pt]
 
\cite{Delduc:2016ihq} & \textbf{On q-deformed symmetries as Poisson-Lie symmetries and application to} \\
 & \textbf{Yang-Baxter type models}\\
 & Fran\c{c}ois~Delduc, Sylvain~Lacroix, Marc~Magro and Beno\^it~Vicedo \\
 & \href{http://dx.doi.org/10.1088/1751-8113/49/41/415402}{\emph{J. Phys.} {\bf
  A49} (2016) 415402}, [\href{https://arxiv.org/abs/1606.01712}{{\tt   1606.01712}}].\\[15pt]
  
\cite{Lacroix:2017isl} & \textbf{Local charges in involution and hierarchies in integrable sigma-models} \\
 & Sylvain~Lacroix, Marc~Magro and Beno\^it~Vicedo \\
 & \href{http://dx.doi.org/10.1007/JHEP09(2017)117}{\emph{JHEP} {\bf 1709} (2017) 117}
  [\href{https://arxiv.org/abs/1703.01951}{{\tt 1703.01951}}].\\[15pt]
  
\cite{Lacroix:2016mpg} & \textbf{Cyclotomic Gaudin models, Miura opers and flag varieties} \\
 & Sylvain~Lacroix and Beno\^it~Vicedo \\
 & \href{http://dx.doi.org/10.1007/s00023-017-0616-8}{\emph{Ann. Henri Poincar\'e} {\bf 19} (2018) 71-139} 
  [\href{https://arxiv.org/abs/1607.07397}{{\tt 1607.07397}}].\\[15pt]
  
\cite{Lacroix:2018fhf} & \textbf{Affine Gaudin models and hypergeometric functions on affine opers} \\
 & Sylvain~Lacroix, Beno\^it~Vicedo and Charles A.~S.~Young \\
 & Preprint arXiv (april 2018): \href{https://arxiv.org/abs/1804.01480}{[{\tt 1804.01480}]}\\[15pt]
  
\cite{Lacroix:2018cag} & \textbf{Cubic hypergeometric integrals of motion in affine Gaudin models} \\
 & Sylvain~Lacroix, Beno\^it~Vicedo and Charles A.~S.~Young \\
 & Preprint arXiv (april 2018): \href{https://arxiv.org/abs/1804.06751}{[{\tt 1804.06751}]}\\[15pt]
\end{tabular}

The main ideas of the articles and preprints \cite{Delduc:2015xdm,Lacroix:2016mpg,Lacroix:2018fhf,Lacroix:2018cag} are summarised in this thesis, with less technical details. These summaries are meant to be read independently of the articles. The entirety of the articles~\cite{Delduc:2016ihq,Lacroix:2017isl} is embodied in the thesis.\\

This work is partially supported by the French Agence Nationale de la Recherche (ANR) under grant ANR-15-CE31-0006 DefIS. I also acknowledge a grant from the region Auvergne-Rhone-Alpes for supporting my mobility in the United-Kingdom.

\cleardoublepage
\chapter*{Acknowledgements / Remerciements}

\thispagestyle{empty}

First and foremost, I would like to express my profound gratitude to my PhD supervisors, Marc and Beno\^it. Their enthusiasm, patience and encouragements have made my PhD a wonderful adventure and my experience with them has more than strengthened my will to pursue in research. I have learned a lot at their side, both about science and about the everyday life of an academic. I cannot thank them enough for their presence, their guidance and their patient answers to my (numerous) questions.\\

I am grateful to Gleb Arutyunov and Niklas Beisert for being the referees for my PhD thesis. I also want to thank Clare Dunning, Jean Michel Maillet and V\'eronique Terras for being part of my defense committee.\\

I would also like to warmly thank Charles and Fran\c{c}ois, with whom I had the opportunity to collaborate on different projects and to discuss about many topics. I have really enjoyed and benefited from these interactions. Similarly, I thank  Ben and Takashi for our collaborations and discussions. My acknowledgments also go to Jean Michel for his advices and helps, both scientific and professional. More generally, I would like to thank all the people with whom I had the chance to discuss and interact and who welcomed me in the scientific community. I also want to mention those who made summer schools and workshops both interesting and exciting, in particular Alejandro, Alessandra, Rouven and Istvan.\\

I wish to thank the Laboratoire de Physique and the D\'epartement de Physique and in particular their directors Thierry and Francesca, for making those three years possible and such a great experience of research and teaching. I am also grateful to the \'Equipe 4 for welcoming me and in particular to its director Jean Michel and to Henning, Karol, Krzysztof and Giuliano for various discussions and advices.  For the same reason, I would also like to thank all other members of the Laboratoire and D\'epartement, especially Eric and Sylvain. I also wish to acknowledge the secretary's offices of the Laboratoire and the D\'epartement, in particular Laurence for her help with all my missions and travels. Similarly, I would like to thank the members of the School of Physics, Astronomy and Mathematics, at the University of Hetfordshire, for welcoming me there, in particular Yann.\\

I was very happy to share these three years at the Laboratoire with other PhD and internship students. I thank Alexandre, Camille, Louis and Fabio for our nice and quite numerous discussions about physics and mathematics. I thank Baptiste, Salvish, Lavi-Kumar, Paul and Takashi, with whom I've shared an office. More generally, I would like to thank Jeremy, Lucile, Jason, Yannick, Benjamin, Arnaud, the different Cl\'ement, J\'er\^ome and the other PhD students, as they contributed quite a lot to my nice stay in the Laboratoire. I also had the chance during my PhD to co-supervise an internship student, Guilhem, whom I thank for his interest and his investment in the project.\\

I thank the ENS de Lyon, the Ecole Doctorale PHAST and the University of Hertfordshire for giving me the opportunity and the means to prepare this doctorate.\\

Finally, I would like to thank my family and friends for their great support, during my PhD and before. I could have never done it without you.

\renewcommand{\baselinestretch}{0.985}\normalsize
\tableofcontents
\renewcommand{\baselinestretch}{1.0}\normalsize

\cleardoublepage
\chapter{Introduction}

\section{Integrability in field theory}

\paragraph{Field theories.} Field theory has many applications in various domains of physics. For example, electromagnetism, general relativity and hydrodynamics are described by classical field theories. The Standard Model of particle physics, which describes the fundamental components of matter and their interactions, is based on particular models of quantum field theory, the gauge theories. Similarly, field theory models are also used in condensed matter theory and statistical physics. Most methods of field theory are based on perturbative expansions. They apply to models describing fields whose interactions are controlled by a small parameter. The physical observables of these models are then expressed as power series expansions in this parameter.

In quantum field theory, such methods led to the development of Feynman diagrams, which proved to be extremely successful for many applications. For instance, the use of perturbative methods in the Standard Model of particle physics has allowed the computation of many physical observables, such as the probabilities of creation and annihilation of particles. These theoretical predictions agree with the experimental observations with a tremendous accuracy. A particular recent example of the theoretical and experimental success of the Standard Model is the observation of the Higgs boson at the Large Hadron Collider, in 2012.

Despite this success, perturbative methods turn out to be insufficient in other situations. For example, the theory of Quantum Chromodynamics (which is the gauge theory describing the strong nuclear interaction in the Standard Model) possesses a strongly-coupled regime which cannot be described by perturbative expansions (as there is no small perturbation parameter in this case). This motivates the development of non-perturbative methods.

\paragraph{Integrable field theories.} Some of these methods belong to the theory of integrability. This theory, which will be described in more details in the next sections of this introduction, allows to compute exactly some physical observables, for certain models called integrable systems. Integrable field theories are scarce, as the requirements for being integrable are quite constraining. In particular, almost all techniques for finding and studying integrable field theories are restricted to two-dimensional field theories. However, integrable field theories are still a domain of growing interest in theoretical research. For example, the existence of exactly solvable models allows one to test and develop further the more usual techniques of field theory, using integrable theories as toy models for a deeper understanding of general field theory. Moreover, the theory of integrability has proven to be mathematically quite rich and has allowed many deep developments at the boundary between physics and mathematics.

A very important development in the domain of integrable field theories came recently with the AdS/CFT correspondence~\cite{Maldacena:1997re,Gubser:1998bc,Witten:1998qj}. This correspondence conjectures a dictionary between some physical observables of quantum gravity models defined on Anti de-Sitter (AdS) manifolds and observables of dual Conformal Field Theories (CFT). The most studied example of the AdS/CFT correspondence concerns, on the CFT side, the $\mathcal{N}=4$ super-symmetric Yang-Mills theory in four dimensions. This is a gauge theory, such as Quantum Chromodynamics, which possesses an additional property of invariance under super-symmetry. Although this (extended) super-symmetry prevents the theory to be a good phenomenological description of particle physics, it shares several similarities with non super-symmetric Yang-Mills theories and is thus a very good toy model for theoretical high-energy physics. The AdS dual of this theory is the Green-Schwarz superstring on the background $AdS_5 \times S^5$~\cite{Metsaev:1998it}, \textit{i.e.} the cartesian product of the five dimensional Anti de-Sitter manifold with the five dimensional sphere.

This superstring theory has been shown to be integrable at the classical level~\cite{Bena:2003wd,Magro:2008dv,Vicedo:2009sn}. This is the reason why this particular example of AdS/CFT correspondence is the most studied. Indeed, one can then apply integrability techniques to obtain results on both the string model and the four dimensional Super-Yang-Mills theory (see for instance~\cite{Beisert:2010jr}).

\section{Classical integrable models with twist function}

Part of this PhD thesis is devoted to the study of a certain family of classical integrable field theories, called models with twist function. In this section of the introduction, we introduce and motivate this class, particularly through its relation with integrable $\s$-models, and present the main questions concerning classical integrable models with twist function addressed in this thesis.

\subsection{Classical integrable field theories}
\label{SubSec:IntroIntegrability}

\paragraph{Liouville integrability.} The theory of integrability started within the Hamiltonian formalism of classical mechanics, which deals with models with a finite number of degrees of freedom. Such a model is said to be integrable (in the sense of Liouville) if it possesses  sufficiently large number of independent conserved quantities in involution, \textit{i.e.} whose Poisson brackets vanish (more precisely, this number must be half the dimension of the phase space of the model). By a theorem of Liouville, such a model can be solved exactly by quadratures. Concretely, this means that the solutions of this model can be formally expressed and can be explicitly computed after performing some algebraic manipulations and integrations.

\paragraph{Integrable field theories.} The notion of integrability for a classical field theory is more subtle. As it possesses an infinite number of degrees of freedom, a natural necessary condition for a field theory to be integrable is to admit an infinite number of conserved charges in involution. Yet, one cannot give a precise definition of such a condition as there is not a unique notion of infinity: how can one ensure that the infinity of charges that one has is ``enough'' to solve the model ? Thus, there is no universally accepted definition of integrability for field theories.

\paragraph{Lax pairs.} In this thesis, we will be interested in two-dimensional field theories. We make the choice of defining integrability in this context through the notion of Lax pairs. A model is said to admit a Lax pair~\cite{Lax:1968fm} if its equations of motion can be recast as a zero curvature equation (or Lax equation) on a two-dimensional connection. The spatial and temporal components of this connection are described by two matrices $\Lc$ and $\M$, valued in some Lie algebra $\g$, depending on the fields of the model and on an auxiliary complex parameter $\lambda$, called the spectral parameter. The Lax equation then reads
\begin{equation*}
\p_t \Lc(\lambda) - \p_x \M(\lambda) + \bigl[ \M(\lambda), \Lc(\lambda) \bigr] = 0, \;\;\;\;\; \forall \, \lambda\in\C,
\end{equation*}
where $\LB$ denotes the Lie bracket of $\g$. This ensures the existence of an infinite number of conserved charges of the model extracted from the monodromy of the Lax matrix $\Lc$.

This zero curvature equation can be seen as the compatibility condition of an auxiliary system of linear partial differential equations. This auxiliary system is at the basis of the so-called Inverse Scattering Method~\cite{Gelfand:1951,Marchenko:1955}. This led to rich developments in the theory of solitons, for example for the Korteweg-de Vries~\cite{Gardner:1967}, non-linear Schroedinger~\cite{Zakharov:1970tma} and sine-Gordon~\cite{Ablowitz:1973} equations (see also the books~\cite{Babebook,fadtakbook87} and the lectures~\cite{Torrielli:2016ufi}).

\paragraph{Maillet bracket.} The existence of a Lax pair allows the construction of an infinite number of conserved charges. However, it does not ensure that these charges have vanishing Poisson brackets, as required for the theory to be integrable. As these charges are constructed from the monodromy of the Lax matrix $\Lc$, their Poisson brackets are related to the Poisson bracket between the different components of $\Lc$.

A first sufficient condition for the charges extracted from the monodromy to be in involution is given by the Sklyanin bracket~\cite{Sklyanin:1982tf}. This condition requires the Poisson bracket of $\Lc$ to be expressed in terms of commutators of $\Lc$ with another matrix called the $\Rc$-matrix, which is skew-symmetric. Moreover, it requires this Poisson bracket to be proportional to the Dirac $\delta$-distribution in the space coordinate (and thus not to contain derivatives of this distribution). Such a bracket is called ultralocal. It describes for example the sine-Gordon model and the non-linear Schroedinger equation.

A generalisation of this condition, which allows the Poisson bracket of $\Lc$ to contain a term proportional to the first derivative of the $\delta$-distribution, has been found by Maillet~\cite{Maillet:1985fn,Maillet:1985ek}. Using the standard tensorial notations, it is given by
\begin{align*}
\hspace{-60pt}\lwb \Lc\ti{1}(\lambda,x), \Lc\ti{2}(\mu,y) \rwb
&= \lsb \Rc\ti{12}(\lambda,\mu), \Lc\ti{1}(\lambda,x) \rsb \delta(x-y) - \lsb \Rc\ti{21}(\mu,\lambda), \Lc\ti{2}(\mu,x) \rsb \delta(x-y)\\
& \hspace{60pt} - \left( \Rc\ti{12}(\lambda,\mu) + \Rc\ti{21}(\mu,\lambda) \right) \p_x\delta(x-y). \notag
\end{align*}
It also involves a $\Rc$-matrix $\Rc\ti{12}(\lambda,\mu)$, which contrarily to the case of the Sklyanin bracket is not skew-symmetric.  As it contains a derivative of $\delta$, the Maillet bracket is said to be non-ultralocal.

\subsection[Integrable $\s$-models]{Integrable $\bm{\s}$-models}
\label{SubSec:IntroSigma}

\paragraph{A new class of integrable theories.} Important examples of integrable two-dimensional field theories are given by the integrable (non-linear) $\s$-models, such as the Principal Chiral Model (PCM) on a Lie group. They have applications in various domains of physics and mathematics. For instance, one particular integrable $\s$-model, called the non-linear $O(3)$ model, plays a role in condensed matter as the continuum limit of certain spin chains, describing the magnetic properties of materials~\cite{HALDANE1983464}.

Integrable $\s$-models also play an important role in high-energy physics and more precisely in string theory. The recent and archetypal example is the $AdS_5 \times S^5$ Green-Schwarz superstring mentioned above in the context of the AdS/CFT correspondence. The classical limit of this theory is described as a two-dimensional $\s$-model with the Metsaev-Tseytlin action~\cite{Metsaev:1998it}. It has been shown by Bena, Polchinski and Roiban in~\cite{Bena:2003wd} that this model admits a Lax pair representation and by Magro that the Hamiltonian Lax matrix satisfies a Maillet bracket~\cite{Magro:2008dv}.

Integrable $\s$-models regained interest in the mathematical physics community even more recently with the discovery of integrable deformations of these models. These are models depending on continuous deformation parameters, such that the integrability property is preserved for any value of these parameters. These deformed models thus form a whole new class of integrable field theories, which is quite interesting regarding the rarity of integrable models.

The first examples of such integrable deformations of $\s$-models were found for particular target spaces of low dimension. For instance, integrable deformations of the PCM on the group $SU(2)$ were discovered by Cherednik in~\cite{Cherednik:1981df}, by Balog, Forg\'acs, Horv\'ath and Palla in~\cite{Balog:1993es} (see also~\cite{Rajeev:1988hq}) and by Fateev in~\cite{Fateev:1996ea}. Similarly, an integrable deformation of the $O(3)$-model was proposed by Fateev, Onofri and Zamolodchikov in~\cite{Fateev:1992tk}. We will now discuss more general integrable deformations.

\paragraph{Yang-Baxter type deformations.} A general deformation of the PCM on an arbitrary Lie group, called the Yang-Baxter model, was discovered by Klim\v{c}ik in~\cite{Klimcik:2002zj,Klimcik:2008eq} (for the group $SU(2)$, this deformation coincides with a particular limit of~\cite{Cherednik:1981df}). The extension of this deformation for symmetric-space $\s$-models was discovered by Delduc, Magro and Vicedo in~\cite{Delduc:2013fga} (recovering~\cite{Hoare:2014pna} in the case of the $O(3)$-model, the deformation proposed in~\cite{Fateev:1992tk}). This deformation was also extended by the same authors to the superstring model on $AdS_5 \times S^5$ mentioned above~\cite{Delduc:2013qra,Delduc:2014kha} (we will come back on this deformation later). In~\cite{Klimcik:2008eq}, Klim\v{c}ik proposed a two-parameter deformation of the PCM, the Bi-Yang-Baxter model, which can be seen as a further deformation of the Yang-Baxter model and which was proven to admit a Lax pair in~\cite{Klimcik:2014bta} (the Bi-Yang-Baxter model for $SU(2)$ has been identified in~\cite{Hoare:2014pna} with the deformed model proposed in~\cite{Fateev:1996ea}).

One of the features of these deformations is to break a global symmetry of the undeformed model. It was shown in~\cite{Delduc:2013fga,Delduc:2014kha} (and before that in~\cite{Kawaguchi:2011pf,Kawaguchi:2012gp} for the PCM on $SU(2)$) that the conserved charges associated with this symmetry in the undeformed model are deformed in a set of non-local conserved charges satisfying a $q$-deformed algebra. More precisely, the Poisson brackets of these charges form a Poisson-Hopf algebra, which can be thought of as a classical limit of a quantum group. In the case of the Yang-Baxter model, it was also shown recently~\cite{Delduc:2017brb} that this deformed algebra is part of a bigger, infinite, one: a $q$-deformed affine Poisson-Hopf algebra (here also, the case for the group $SU(2)$ was already treated in~\cite{Kawaguchi:2012gp,Kawaguchi:2012ve}).

We shall refer to all of these deformations as Yang-Baxter type deformations (they are also called $\eta$-deformations or $\varkappa$-deformations in the literature), as they are constructed from a solution of the modified Classical Yang-Baxter Equation (mCYBE). One can also construct~\cite{Kawaguchi:2014qwa} deformations associated with solutions of the (homogeneous) Classical Yang Baxter Equation. Contrarily to inhomogenous deformations, these models do not possess $q$-deformed conserved charges. As these charges will be one of the subjects of interest of this thesis, we will use the denomination Yang-Baxter type deformation to refer to deformations associated with solutions of the (inhomogeneous) mCYBE.

\paragraph{Other deformations of $\bm\s$-models.} Another type of integrable deformation of $\s$-models, called $\lambda$-deformation, was also constructed by Sfetsos in~\cite{Sfetsos:2013wia} for the PCM and Hollowood, Miramontes and Schmidtt in~\cite{Hollowood:2014rla} for symmetric space $\s$-models. This deformation was also extended to the Green-Schwarz superstring model on $AdS_5 \times S^5$ in~\cite{Hollowood:2014qma}.

The PCM also possesses another type of integrable deformation, which is obtained by adding a Wess-Zumino term (see~\cite{Wess:1971yu,Novikov:1982ei} for the construction of this term) to the action of the PCM, multiplied by an arbitrary deformation parameter. This model was first considered in~\cite{Witten:1983ar} for a particular value of this parameter which makes the model conformal. It was later shown in~\cite{Abdalla:1982yd} that this model is integrable for any value of the deformation parameter.

The PCM can be further deformed by combining several types of deformations. For example, one can combine the addition of a Wess-Zumino term with a Yang-Baxter deformation (as shown by Delduc, Magro and Vicedo in~\cite{Delduc:2014uaa}) and with a Bi-Yang-Baxter deformation and so called TsT transformations (as shown quite recently by Delduc, Hoare, Magro and Kameyama in~\cite{Delduc:2017fib}, recovering a 4-parameter deformation introduced by Lukyanov in~\cite{Lukyanov:2012zt} for the group $SU(2)$).

\paragraph{Deformations of the $\bm{AdS_5\times S^5}$ superstring.} As explained above, both the Yang-Baxter type deformation and the $\lambda$-deformation have been extended to the Green-Schwarz superstring on background $AdS_5\times S^5$. These deformations have been the subjects of many recent works in the mathematical physics community.

For the Yang-Baxter deformation of the $AdS_5 \times S^5$ superstring, the existence of $q$-deformed symmetries mentioned above can also be seen on the tree-level light-cone S-matrix. Indeed, it was shown in~\cite{Arutyunov:2013ega,Arutyunov:2015qva,Borsato:2016hud} that this matrix coincides with the large string tension limit of the S-matrix invariant under the quantum group of the centrally-extended $\mathfrak{psu}(2|2)^2$ algebra~\cite{Beisert:2008tw,Beisert:2010kk,Hoare:2011wr}. This matrix is thus a $q$-deformation of the light-cone S-matrix of the undeformed superstring on $AdS_5 \times S^5$~\cite{Staudacher:2004tk,Beisert:2005tm,Beisert:2006qh,Janik:2006dc,Arutyunov:2006iu,Beisert:2006ib,Beisert:2006ez,Arutyunov:2006yd} (see also the reviews~\cite{Arutyunov:2009ga,Ahn:2010ka}).

An interesting question about these deformations concerns their geometry, which is in particular important to understand whether or not they define a string theory. As a type IIB string theory, the undeformed model defines a background on $AdS_5\times S^5$ which is a solution of the equations of type IIB supergravity. It is now known~\cite{Borsato:2016ose} that the background of the $\lambda$-deformation is also a solution of the equations of type IIB supergravity (see also~\cite{Borsato:2016zcf,Chervonyi:2016ajp,Chervonyi:2016bfl} and~\cite{Sfetsos:2014cea,Demulder:2015lva}). Thus, the $\lambda$-deformation defines a string theory. However, the Arutyunov-Borsato-Frolov~\cite{Arutyunov:2013ega,Arutyunov:2015qva,Borsato:2016hud} background obtained from the (inhomogeneous) Yang-Baxter deformation is not a solution of type IIB supergravity equations~\cite{Arutyunov:2015qva} but is a solution of generalised type IIB supergravity equations~\cite{Arutyunov:2015mqj,Wulff:2016tju} (see also~\cite{Hoare:2015wia}).

Another aspect of interest of these models concerns the AdS/CFT correspondence. As the undeformed superstring model is dual, through this correspondence, to the $\mathcal{N}=4$ super-Yang-Mills theory in four dimensions, it is natural to wonder if the deformed models are dual to a deformation of this gauge theory. This is still an open question (see however~\cite{Kameyama:2016yuv} and~\cite{vanTongeren:2015uha,vanTongeren:2016eeb,Araujo:2017jap} for homogeneous Yang-Baxter deformations).

\subsection[The algebraic structure behind integrable $\s$-models: the twist function]{The algebraic structure behind integrable $\bm\s$-models: the twist function}
\label{SubSec:IntroStructure}

\paragraph{Integrable $\bm\s$-models as models with twist function.} The $\s$-models presented above and their deformations possess a Lax matrix (see Subsection \ref{SubSec:IntroIntegrability} of this introduction). The Hamiltonian analysis of most of these models has been performed, showing their integrability at the Hamiltonian level. More precisely, as explained in details in Chapter \ref{Chap:Models} of this thesis, it was shown that the Poisson bracket of the Lax matrix of most of these models takes the form of a non-ultralocal Maillet bracket (see Subsection \ref{SubSec:IntroIntegrability}), ensuring the existence of an infinite number of conserved charges in involution for these models.

Moreover, it was observed that the $\Rc$-matrices governing the Maillet brackets of all these models share a common algebraic structure. They are encoded in a model-dependent rational function $\vp$ of the spectral parameter, that we call the twist function. More precisely, they are given by
\vspace{-2pt}\begin{equation*}
\Rc\ti{12}(\lambda,\mu) = \Rc^0\ti{12}(\lambda,\mu) \vp(\mu)^{-1}, \vspace{-2pt}
\end{equation*}
where $\Rc^0$ is a standard $\Rc$-matrix on a loop algebra. This observation thus enables one to interpret integrable $\s$-models as part of a larger class of integrable field theories, the models which possess such a twist function.

\paragraph{Twist function and deformations of $\bm\s$-models.} An interesting aspect of this common algebraic structure concerns the deformations of $\s$-models. Indeed, one sees that the known deformations of $\s$-models do not affect the dependence of the Lax matrix of the model on the spectral parameter. The effect of the deformation is in fact to modify the Poisson bracket of this Lax matrix by modifying the twist function of the model. More precisely, one observes that the deformations of $\s$-models modify the poles of the twist function.

Let us note here that homogeneous Yang-Baxter type deformations exhibit a slightly different behaviour than other deformations in this context. Studying their Hamiltonian structure, one shows that these deformations do not affect the Poisson bracket of the Lax matrix of the model and thus its twist function. In this case, the effect of the deformation is purely to modify the way the Lax matrix depends on the fundamental fields of the model. However, these homogeneous deformations are part of the class of models with twist function and thus, all results which apply to this whole class can be applied to them.

\paragraph{Models with twist function as Affine Gaudin Models.} The observation that $\s$-models are models with twist function led recently to their reinterpretation in terms of so-called classical Affine Gaudin Models (AGM). Gaudin models form a general class of integrable systems which are constructed from certain Lie algebras. In particular, the AGM are the Gaudin models associated with affine Kac-Moody algebras. They were studied recently by Vicedo in~\cite{Vicedo:2017cge}. More precisely, it was shown in~\cite{Vicedo:2017cge} that AGM can be seen as integrable field theories with twist function. Moreover, in the framework of AGM, the twist function appears quite naturally as part of the data defining the model (in particular, the poles of the twist function are the so-called sites of the Gaudin model).

Conversely, Vicedo also proved in~\cite{Vicedo:2017cge} that all known models with twist function (including integrable $\s$-models and their deformations but also affine Toda field theories) are realisations of some particular AGM. This sheds the light on the algebraic origin of integrable $\s$-models and exhibits the common mathematical structure behind their integrability.

\subsection{Goal of this thesis}

One of the main goals of my PhD projects was to develop general methods to study integrable $\s$-models in a model-independent way, using their common algebraic structure as models with twist function. Such methods would show the key role played by the twist function in the study of integrable $\s$-models and the interest of treating these models as a whole class. Below, we list the main questions addressed in this thesis following this direction. They form mainly the first part of this thesis.

\paragraph{Local charges in involution.} It is known~\cite{Evans:1999mj,Evans:2000qx} that some undeformed integrable $\s$-models (the PCM and the symmetric space $\s$-model) possess an infinity of local charges in involution. It is natural to ask whether this is a unique feature of these particular models or if it is shared by other integrable $\s$-models, as for example deformed ones. We address this question in Chapter \ref{Chap:LocalCharges} of this thesis and show that it is in fact a result shared by all models with twist function, under certain assumptions on this twist function. As such, it applies to all integrable $\s$-models and their deformations but also more generally to AGM.

\paragraph{Deformed symmetries of Yang-Baxter type deformations.} As mentioned in Subsection \ref{SubSec:IntroSigma}, the Yang-Baxter type deformations are characterised by the breaking of some global symmetry and the appearance of some $q$-deformed Poisson-Hopf algebra of conserved charges. However, it is not known whether these deformed conserved charges are associated with a symmetry of the model. We address this question in Chapter \ref{Chap:PLie} and exhibit a Poisson-Lie deformed symmetry associated with these charges. The approach used to derive this result is model-independent and uses the common features of Yang-Baxter type deformations. In particular, a key point of this method is the fact that these deformations have for effect to split a double pole of the undeformed twist function into two simple poles, showing the key role played by the twist function in these deformations.

\paragraph{Twist function of the Bi-Yang-Baxter model.} One of my other PhD project presented in this thesis does not concern the development of general results for $\s$-models based on the twist function. It consists rather of proving that a particular deformed $\s$-model, the Bi-Yang-Baxter model, is also described by a twist function, as other deformed $\s$-models. We prove this fact in Chapter \ref{Chap:Models}, Section \ref{Sec:BYB} of this thesis (modulo some technical subtleties), thus showing that the Bi-Yang-Baxter model also belongs to the more general class of models with twist function.

\section{Quantum Gaudin models}

A second part of this thesis is devoted to quantum aspects of integrable models, and more precisely of Gaudin models, either finite or affine.

\subsection{Quantum integrability}
\label{SubSec:IntroQuant}

\paragraph{Quantum integrable mechanical systems and spin chains.} For physical systems with a finite number of degrees of freedom, one can define quantum integrability similarly to classical integrability, replacing conserved charges in involution by conserved commuting operators. Such systems can be quantised mechanical systems, as the Toda chain, or spin chains, as the XXX or XXZ Heisenberg chains.

Given a quantum model, one of the main steps towards its resolution is to find its spectrum, \textit{i.e.} to find the eigenvalues of its Hamiltonian. For an integrable model, this Hamiltonian commutes with a large number of commuting operators. Provided that these operators are diagonalisable, there exists a common basis of eigenvectors for all these operators and one can aim to study the whole spectrum of all these operators at once. For certain quantum integrable models, this can be achieved by various methods. Example of such methods are the coordinate Bethe ansatz or the algebraic Bethe ansatz, which proved quite efficient for the study of quantum integrable spin chains.

\paragraph{Integrable quantum field theory.} Defining the notion of integrability for quantum field theories is a subtle question, which does not admit a universally accepted answer. In this thesis, we will choose to define integrability for a quantum field theory as the existence of an infinity of commuting conserved operators. An important feature of models which admit such a property is that their scattering matrix factorises. More precisely, one shows that there is no possibility of creation and annihilation of particles and that all scattering processes can be reduced to combination of scattering of two particles. Moreover, one can often constrain the form of these two by two scattering matrices using the symmetries of the model. Once these two by two scattering matrices are known, one can extract more results on the theory, such as its spectrum, \textit{via} the Thermodynamic Bethe Ansatz. This is the so-called Factorised Scattering Theory~\cite{Zamolodchikov:1978xm}.

If one is given a classical integrable field theory, a natural goal is to find a quantisation of this theory which leads to an integrable quantum field theory. The usual approach of quantum field theory based on second quantisation of the canonical bracket and perturbative expansions is in general not well adapted for such a goal. Alternative quantisation methods which preserve integrability by construction have been developed. The main idea of these methods is to discretise the space coordinate (while preserving the integrability), to write the Poisson brackets of the evaluations of the Lax matrix on this discretisation and to quantise these Poisson brackets. These methods of quantisation lead to the so-called Quantum Inverse Scattering Method, allowing the use of the algebraic Bethe ansatz on this theory. One then has to take a continuum limit to obtain a quantum field theory.

These methods were mostly developed~\cite{Faddeev:1979gh} (see also the review~\cite{Faddeev:1982} and the book~\cite{Korepin:1997}) for ultralocal theories, satisfying the Sklyanin bracket (see Subsection \ref{SubSec:IntroIntegrability} of this introduction). Developing a similar method for non-ultralocal theories first necessitates to find a discrete integrable Poisson algebra whose continuum limit coincides with the Maillet bracket. Such an algebra was proposed in~\cite{Freidel:1991jx,Freidel:1991jv}. However, the corresponding $\Rc$-matrix arising from its continuum limit satisfies very particular conditions. A similar construction for an arbitrary $\Rc$-matrix is still lacking at the time.

\paragraph{Quantum integrable $\bm\s$-models ?} In the previous section of this introduction, we presented particular examples of classical integrable field theories, the integrable $\s$-models. The methods of quantisation by discretisation mentioned above cannot be applied to these models, as the Poisson bracket of their Lax matrix is non-ultralocal and does not satisfy the condition to apply the results of~\cite{Freidel:1991jx,Freidel:1991jv}. Hence, there is still no proof of the quantum integrability of these models from first principles and one cannot apply the Quantum Inverse Scattering Method to solve them.

However, quantum $\s$-models and the quantum superstring on $AdS_5 \times S^5$ have been extensively studied in the literature. These works rest on several assumptions, in particular the hypothesis that one can apply the Factorised Scattering Theory to these models, allowing the description of their spectrum through the Thermodynamic Bethe Ansatz. A way to prove this hypothesis would be to show the quantum integrability of these models, which is not done yet (note however that the factorisation of the scattering matrix was proven for particular $\s$-models~\cite{Luscher:1977uq}). Such a program turned out to be extremely successful in the context of the AdS/CFT correspondence (see for example the review~\cite{Beisert:2010jr}). It led to the study of the so-called Quantum Spectral Curve~\cite{Gromov:2013pga}.

\paragraph{Quantum affine Gaudin models.} As mentioned in Subsection \ref{SubSec:IntroStructure} of this introduction, classical integrable $\s$-models have been re-interepreted as realisations of affine Gaudin models in~\cite{Vicedo:2017cge}. It was proposed in the same article that this re-interpretation can be used to prove, in the long-term, the integrability of quantum $\s$-models and to determine their spectrum. This proposition is based on the many results known about another type of quantum Gaudin models, the finite ones (associated with finite Kac-Moody algebras instead of affine ones). Let us now discuss these results.

\subsection{Quantum finite Gaudin models}
\label{SubSec:IntroFinite}

\paragraph{The models.} Finite Kac-Moody algebras are the semi-simple Lie algebras of finite dimension. The associated finite Gaudin models possess a finite number of degrees of freedom. More precisely, they describe integrable spin chains. They were introduced by Gaudin, first in~\cite{Gaudin_76a} for the Lie algebra $\sld$, as a limit of the XXX spin chain, and later in~\cite{Gaudin_book83} for an arbitrary semi-simple Lie algebra.

In addition to this underlying Lie algebra $\g$, a Gaudin model is constructed from the data of $N$ representations of $\g$, which describe the sites of the spin chain, attached to $N$ points of the complex plane, which represent the positions of these sites.

Using the invariant non-degenerate bilinear form on the semi-simple Lie algebra, one defines a set of $N$ commuting Hamiltonians of the Gaudin models. These Hamiltonians encode quadratic spin-spin interactions between the sites. A specificity of these Hamiltonians, compared to the ones of most other integrable spin chains, is that these interactions are long-range: every site of the chain interacts with all other sites.

\paragraph{Bethe ansatz.} An important step towards the resolution of Gaudin models is to diagonalise these Hamiltonians. As they commute, they are simultaneously diagonalisable and one can look for a basis of common eigenvectors of all Hamiltonians. When the representations at the sites of the model are highest-weight representations of $\g$, one can look for this basis using the algebraic Bethe ansatz. This is a general method used to diagonalise integrable spin chains which admit a monodromy matrix, satisfying certain commutation relations. The Hamiltonian of such a spin chain is extracted from the monodromy matrix. The Bethe ansatz method then constructs eigenvectors of this Hamiltonian, starting from the vacuum state of the model and acting by suitable lowering operators, extracted from the monodromy matrix.

The Bethe ansatz was applied to the Gaudin model on $\g=\sld$ by Gaudin himself, in its original article~\cite{Gaudin_76a}. For spin chains associated with Lie algebras of higher rank, one usually works recursively on the rank by successive applications of the Bethe ansatz for $\sld$: this is the so-called nested Bethe ansatz. This method has been applied to Gaudin models~\cite{Jurco:1989mg}, resulting in a rather intricate algorithm constructing the eigenvectors. Another approach to the Bethe ansatz, which does not use a recursive algorithm, was developed by Babujian and Flume in~\cite{Babujian:1993ts}, based on the results~\cite{Varchenko:1991} of Schechtman and Varchenko about the Knizhnik-Zamolodchikov equation (see also~\cite{Reshetikhin:1994qw}). 

\paragraph{Higher Gaudin Hamiltonians.} In their pioneering work~\cite{Feigin:1994in} on Gaudin models, Feigin, Frenkel and Reshetikhin have proved that the quadratic Gaudin Hamiltonians are part of a larger commutative algebra. This larger commutative algebra, called the Gaudin algebra or Bethe algebra~\cite{mukhin_2009a}, contains Hamiltonians of degrees greater than two (for $\g$ of rank at least two), which then commute between themselves and with the quadratic ones. In~\cite{Feigin:1994in}, this subalgebra is constructed abstractly, using modules of an affine algebra at the critical level. The higher degree Gaudin Hamiltonians were later constructed explicitly for simple algebras of type A~\cite{Talalaev:2004qi,Chervov:2006xk,Tarasov:2006} and of type B, C and D~\cite{Molev:2013} (see also~\cite{Rybnikov:2008} for the uniqueness of these quantum Hamiltonians, proving that these abstract and explicit constructions coincide).

\paragraph{FFR approach.} As these higher degree Hamiltonians commute with the quadratic ones and between themselves, they can be simultaneously diagonalised. In the article~\cite{Feigin:1994in}, Feigin, Frenkel and Reshetikhin also proposed a method to describe the eigenvalues of these Hamiltonians. This approach, developed further in~\cite{Frenkel:1995zp,Frenkel:2003qx,Mukhin:2005,Frenkel:2004qy,Chervov:2004ty,Chervov:2009ck,Rybnikov:2016} and to which we shall refer as the FFR approach, does not require the use of the Bethe ansatz and in fact recovers its results. It describes the spectrum of the Gaudin Hamiltonians (including higher degree ones) in terms of differential operators, called opers. These opers are constructed from the so-called Langlands dual of the inital Lie algebra $\g$, which is also a semi-simple and finite dimensional Lie algebra.

The FFR approach is related~\cite{Frenkel:2004qy} to a deep mathematical result called the Geometric Langlands Correspondence. It is part of the so-called Langlands program, which aims at finding relations between various domains of mathematics: geometry, group theory, Galois theory, ... This places the FFR approach at the frontier between physics and pure mathematics.

\paragraph{Generalisations.} The results mentioned above concern the ``usual'' finite Gaudin model, as introduced by Gaudin in~\cite{Gaudin_book83}. This model can be generalised in various ways.

For example, the Lax matrix of the usual Gaudin model depends rationally on the spectral parameter and has simple poles at the sites of the model. One can consider a generalisation of this model where the Lax matrix possesses poles of arbitrary multiplicities, as introduced by Feigin, Frenkel and Toledano Laredo in \cite{Feigin:2006xs}. In the same article, they prove that this model with multiplicities admits quadratic and higher degree Hamiltonians and describe their spectrum in terms of opers. As for the case with no multiplicities, these opers are differential operators, associated with the Langlands dual of $\g$, the main difference being that they now possess so-called irregular singularities.

One can also construct generalised Gaudin models on a finite algebra $\g$, associated with arbitrary skew-symmetric solutions of the classical Yang-Baxter equation on the loop algebra of $\g$. The usual Gaudin model then corresponds to the standard rational solution of this equation. The nested Bethe ansatz for these models has been developed for rational and trigonometric solutions of the classical Yang-Baxter equation in~\cite{Jurco:1989mg} (for classical Lie algebras). The Bethe ansatz for the Lie algebra $\sld$ and elliptic solutions was described in~\cite{Sklyanin:1996pr}.

A similar construction for non-skew symmetric solutions of the classical Yang-Baxter equation has been proposed by Skrypnyk in~\cite{Skrypnyk:2006}. In particular, one can construct such solutions from automorphisms of the Lie algebra $\g$ of finite order. The corresponding generalised Gaudin models are called cyclotomic Gaudin models (particular examples of these models also appear in~\cite{Crampe:2007cj}). In~\cite{Vicedo:2014zza}, Vicedo and Young proved the existence for cyclotomic models of higher degree Hamiltonians, using a method similar to the one of~\cite{Feigin:1994in} for the usual Gaudin model. Moreover, they developed the (non-nested) algebraic Bethe ansatz for these models, describing the eigenvectors and eigenvalues of these Hamiltonians (some first results using the nested Bethe ansatz for particular cyclotomic Gaudin models were also found in~\cite{Skrypnyk:2013usa}). The same authors constructed the Hamiltonians of cyclotomic Gaudin models with multiplicities in~\cite{Vicedo_161109059} (including higher degree Hamiltonians) and described their spectrum through the Bethe ansatz (under some restrictions on the multiplicities).

The cyclotomic and non-cyclotomic Gaudin models are complex spin chains. In~\cite{Vicedo:2017cge}, Vicedo considered similar models at the classical level, called dihedral Gaudin models, which admit additional reality conditions, ensuring in particular that the Hamiltonian of the model is real. So far, dihedral Gaudin models were not considered at the quantum level. It is natural to expect that these models can be quantised into quantum integrable spin chains with self-adjoint Hamiltonian.

\subsection{Quantum affine Gaudin models}

As explained above, the article~\cite{Vicedo:2017cge} offers hope that quantum integrable $\s$-models can be studied as quantum affine Gaudin models. At the moment, such a goal is however quite distant, as not much is known about quantum affine Gaudin models. In particular, in~\cite{Vicedo:2017cge}, classical integrable $\s$-models are identified with realisations of dihedral affine Gaudin models, which thus include many generalisations compared to the simplest affine Gaudin models (multiple poles, cyclotomy, reality conditions, ...). These generalisations were never studied at the quantum level for affine Gaudin models (see previous subsection for finite ones). However, there exist promising results about quantum affine Gaudin models with no multiplicities, no cyclotomy and no reality conditions.

\paragraph{Hamiltonians of quantum affine Gaudin models.} The work~\cite{Varchenko:1991} on the Knizhnik-Zamolodchikov equation concerned arbitrary (symmetrisable) Kac-Moody algebras. In particular, it applies also to affine algebras, showing that quantum affine Gaudin models possess quadratic commuting Hamiltonians. Moreover, the Bethe ansatz for these Hamiltonians can be derived from the results of~\cite{Varchenko:1991}, allowing the diagonalisation of these Hamiltonians on tensor product of highest-weight representations of the affine algebra.

These results rely on the algebraic structure shared by all Kac-Moody algebras. In particular, it treats in a very similar way the finite Gaudin models and the affine ones. Considering the extensive number of results one can find on finite Gaudin models, one can hope that some of these results can also be extended to the affine case. For example, it was conjectured by Feigin and Frenkel~\cite{Feigin:2007mr}, in the context of the study of the quantum KdV equation, that quantum affine Gaudin models also possess higher degree commuting Hamiltonians (more precisely an infinity of such Hamiltonians).

\paragraph{Affine FFR approach and ODE/IM correspondence.} In the article~\cite{Feigin:2007mr}, it was also conjectured that the FFR approach describing the spectrum of finite Gaudin models (see previous subsection) generalises to affine Gaudin models. This is supported by the existence of a notion of affine opers, which is an affine equivalent of the finite opers used in the FFR approach for finite Gaudin models. On the long term, such an affine FFR approach could allow the description of the spectrum of integrable field theories.

One of the motivation for such a program is the so-called ODE/IM correspondence. This correspondence establishes a link between the eigenvalues of Integrals of Motion (IM) of some integrable quantum field theories and the solutions of some Ordinary Differential Equations (ODE). The first example of such a correspondence concerns the quantum KdV equation and was developed by Bazhanov, Lukyanov and Zamolodchikov in~\cite{Bazhanov:1998wj,Bazhanov:2003ni}, following the first insight of Dorey and Tateo~\cite{Dorey:1998pt}. It was shown in~\cite{Feigin:2007mr} that the ODE appearing in this example can be encoded in affine opers associated with the affine algebra $\widehat{\mathfrak{sl}}(2,\C)$. Following this observation, Feigin and Frenkel proposed in that article that the ODE/IM finds its origin in the (conjectural) affine FFR approach, as it would also relate the spectrum of integrable models with affine opers.

Other examples of ODE/IM correspondences have been proposed in the last two decades, including for some integrable field theories with twist function. As these models were reinterpreted as affine Gaudin models in~\cite{Vicedo:2017cge}, it was proposed, in the same spirit as the conjectures of Feigin and Frenkel, that the ODE/IM correspondence for these models originates from the affine FFR approach. We will discuss this idea in more details in the conclusion of this thesis, as one of its main perspective. Thus, we also refer to the discussion in Section \ref{Sec:Perspectives} for references about the ODE/IM.

\subsection{Goal of this thesis}

During my PhD, I have worked on both finite and affine quantum Gaudin models and the FFR approach. The results I obtained in these projects will be the main subjects of the second part of this thesis.

\paragraph{Quantum affine Gaudin models and affine opers.} One of my long-term goals is to prove the existence of higher degree Hamiltonians for quantum affine Gaudin models. In this thesis, I will present several results in this direction. The first one, discussed in Chapter \ref{Chap:GaudinClass}, is actually a result at the classical level: applying the results of Chapter \ref{Chap:LocalCharges}, we prove the existence of such Hamiltonians for classical affine Gaudin models (under slight restrictions but including real cyclotomic models with arbitrary multiplicities).

In Chapter \ref{Chap:QuantumAffine}, we develop further the theory of affine opers. Using this result and having in mind the idea of generalising the FFR approach to affine Gaudin models, we give and support conjectures on the form of the higher degree quantum Hamiltonians and their eigenvalues. The main result supporting these conjectures is the explicit construction of the first higher degree Hamiltonian, the cubic one (for untwisted affine algebras of type A).

\paragraph{Towards a cyclotomic FFR approach.} In Chapter \ref{Chap:QuantumFinite}, we discuss a conjectured generalisation of the FFR approach to cyclotomic finite Gaudin models (see Subsection \ref{SubSec:IntroFinite}). For that, we define a notion of cyclotomic finite opers and extend known results about non-cyclotomic opers to this new setting.

\section{Plan of this thesis}

This thesis is separated into two parts.\\

The first one is devoted to the class of integrable classical field theories with twist function. We start by reviewing the Lax formalism of classical integrable field theories in Chapter \ref{Chap:Lax}. In particular, we discuss the Maillet bracket and non-ultralocal field theories. We end this chapter with the definition of the models with twist function, which serves as a general formalism for the first part of the thesis.

In Chapter \ref{Chap:Models}, we present the main examples of integrable field theories with twist function, the integrable $\s$-models and their deformations. Most of this chapter is a review of known results about these models, in particular their Lax matrix and its Hamiltonian structure. The last section of the chapter presents some new results about the Bi-Yang-Baxter model. More precisely, we prove that this model is integrable at the Hamiltonian level and belongs to the class of models with twist function. This section is based on the article~\cite{Delduc:2015xdm}, that I wrote during my PhD with F. Delduc, M. Magro and B. Vicedo, and a few subsequent results. In this section, we will give a less technical presentation of the content of~\cite{Delduc:2015xdm}, which is meant to be readable independently.

In Chapter \ref{Chap:LocalCharges}, we show the existence of an infinity of local charges in involution for all integrable models with twist function, under certain assumptions on the twist function. In particular, we apply this result to all integrable $\s$-models and their deformations. We go on to show that these charges generate compatible integrable equations and thus form an infinite integrable hierarchy. This chapter is mainly similar to the article~\cite{Lacroix:2017isl} of M. Magro, B. Vicedo and myself.

Chapter \ref{Chap:PLie} concerns the deformed symmetries of Yang-Baxter type deformations. Considering the poles of the twist function, we show that all these models admit a $q$-deformed Poisson-Hopf algebra of conserved charges. We then show that these charges are associated with Poisson-Lie symmetries of the model and construct explicitly these symmetries. In particular, we show that these symmetries are non-local. The content of this chapter is almost equivalent to the one of my PhD publication~\cite{Delduc:2016ihq} with F. Delduc, M. Magro and B. Vicedo.\\

The second part of this thesis concerns Gaudin models, either finite or affine, and both at the classical and the quantum levels.

Chapter \ref{Chap:GaudinClass} is a review about classical Gaudin models. We start by discussing these models for an arbitrary Lie algebra (with a non-degenerate invariant bilinear form). We then specialise to classical affine Gaudin models and explain how they can be interpreted as integrable field theories with twist function. Conversely, we discuss on the example of the Yang-Baxter model how integrable $\s$-models can be seen as realisations of affine Gaudin models. This chapter is almost entirely a review of known results (mostly the article~\cite{Vicedo:2017cge} on affine Gaudin models). However, we present a new result in Subsection \ref{SubSec:HierarchyAGM} by applying the results of Chapter \ref{Chap:LocalCharges} to affine Gaudin models, proving that they possess an integrable hierarchy.

Chapter \ref{Chap:QuantumFinite} concerns quantum finite Gaudin models. We first review the construction of these models and their Hamiltonians. We then discuss their spectrum, presenting two approaches for its description: the Bethe ansatz and the FFR approach. Subsection \ref{SubSec:Cyclo} concerns the generalisation of the FFR approach to cyclotomic Gaudin models and is a short summary of the article~\cite{Lacroix:2016mpg}, that I wrote during my PhD with B. Vicedo.

We discuss quantum affine Gaudin models and affine opers in Chapter \ref{Chap:QuantumAffine}, based on the preprints~\cite{Lacroix:2018fhf} and~\cite{Lacroix:2018cag} of B. Vicedo, C.A.S. Young and myself. We first show that functions on affine opers take the form of hypergeometric integrals. Based on this result, we give a series of conjectures about quantum higher order Hamiltonians of affine Gaudin models and their spectrum. We support these conjectures by various observations, in particular the construction of the cubic Hamiltonian. The content of this chapter is mostly a less technical summary of~\cite{Lacroix:2018fhf}, which is meant to be readable independently (although some of the results of~\cite{Lacroix:2018fhf} are not discussed in this thesis). For brevity, the main result of the preprint~\cite{Lacroix:2018cag} (the construction of the cubic Hamiltonian) is simply mentioned in Chapter \ref{Chap:QuantumAffine}, without any details or proofs.\\

We conclude this thesis by a brief overview and some perspectives in Chapter \ref{Chap:Conclusion}. Technical appendices about Lie algebras, Poisson and symplectic geometries and $\Rc$-matrices form the last chapters \ref{App:Lie}, \ref{App:Poisson} and \ref{App:RMat}.

\cleardoublepage
\part{Integrable field theories with twist function}

\cleardoublepage
\chapter{Lax formalism and twist function}
\label{Chap:Lax}
 
This chapter is a self-contained review about classical integrable hamiltonian field theories. The goal is to introduce the class of non-ultralocal integrable field theories with twist function, through the key notion of Lax matrix. We consider a two-dimensional Minkowski space-time, with time coordinate $t$ and space coordinate $x$, which can be taken on the real line $\R$ or on the circle $\mathbb{S}^1$. We are interested in classical field theories on this space-time, given as a set of dynamical fields $\phi_i$, depending on the coordinates $(x,t)$, together with partial differential equations describing the time evolution of these fields. If $x$ is taken in $\R$, we suppose that the fields $\phi_i(x,t)$ decrease sufficiently fast when $x$ goes to $\pm \infty$. If $x$ is taken on the circle $\mathbb{S}^1 \simeq \lsb 0,2\pi\rsb$, we suppose that the fields $\phi_i(x,t)$ satisfy periodic boundary conditions $\phi_i(x=0,t)=\phi_i(x=2\pi,t)$.

Such a theory is said to be integrable if it admits an infinite number of conserved charges, in involution in the hamiltonian sense. We will discuss the hamiltonian framework and the involution properties in section \ref{Sec:Ham}. We first focus on a condition of existence of an infinite number of conserved charges through the Lax formalism.

\section{Lax matrices and monodromies}
\label{Sec:Lax}

Let us consider a pair of matrices $\Lc(\lambda,x,t)$ and $\M(\lambda,x,t)$, depending on the fields $\phi_i(x,t)$ of the model and on a complex auxiliary parameter $\lambda\in\C$ called the spectral parameter. More precisely, we will consider that $\Lc$ and $\M$ are valued in some finite dimensional Lie algebra $\g$ with Lie bracket $\lsb \cdot, \cdot \rsb$ (see the Appendix \ref{App:Lie} for conventions and definitions on Lie algebras). We say that $(\Lc,\M)$ forms a \textbf{Lax pair} of the model if the equations of motion of the fields $\phi_i$ can be rewritten as the \textbf{Lax equation}~\cite{Lax:1968fm}
\begin{equation}\label{Eq:Lax}
\p_t \Lc(\lambda,x,t) - \p_x \M(\lambda,x,t) + \lsb \M(\lambda,x,t), \Lc(\lambda,x,t) \rsb = 0, \;\; \forall \lambda\in\C.
\end{equation}
Note that this equation can be reformulated more geometrically as the zero curvature equation
\begin{equation*}
\lsb \nabla_t, \nabla_x \rsb = 0
\end{equation*}
of the two-dimensional $\g$-connection
\begin{equation*}
\nabla = \lrb \nabla_x, \nabla_t \rrb = \lrb \p_x + \Lc(\lambda,x,t), \p_t + \M(\lambda,x,t) \rrb.
\end{equation*}

Let $G$ be a connected Lie group with Lie algebra $\g$ (for example the adjoint group of $\g$). We define the transfer matrices of the connection $\nabla$ between the points $x$ and $y$ as the path-ordered exponential
\begin{equation*}
T(\lambda \, ; x,y\, ; t) = \Pexp \lrb - \int_y^x \dd z \; \Lc(\lambda,z,t) \rrb,
\end{equation*}
valued in the group $G$. We refer to the Appendix \ref{App:Pexp} for the definition and properties of path-ordered exponentials. In particular, it is proven in this appendix that the Lax equation \eqref{Eq:Lax} implies that
\begin{equation}\label{Eq:DerTempT}
\p_t T(\lambda\, ; x,y;t) = T(\lambda \, ; x,y;t) \M(\lambda,y,t) - \M(\lambda,x,t) T(\lambda \, ; x,y;t),
\end{equation}
for all $\lambda\in\C$. We will now distinguish the cases where $x$ is taken on the real line or on the circle.

\paragraph{Real line coordinate.} Let us suppose that $x$ is taken on the real line $\R$. We define the \textbf{monodromy matrix} $T(\lambda,t)$ as the transfer matrix from $-\infty$ to $+\infty$, \textit{i.e.} as
\begin{equation*}
T(\lambda,t) = T(\lambda\, ; + \infty, -\infty \, ; t).
\end{equation*}
If we suppose that the field $\M(\lambda,x,t)$ decreases sufficiently fast at $x = \pm \infty$, we get from equation \eqref{Eq:DerTempT} that
\begin{equation*}
\p_t T(\lambda,t) = 0, \;\; \forall \lambda\in\C.
\end{equation*}
Thus, a consequence of the Lax equation \eqref{Eq:Lax} is that the whole monodromy matrix is conserved. Note that this equation is true for any value of the spectral parameter $\lambda\in\C$. Thus, varying $\lambda$, we get an infinite number of conserved charges. This was the reason for the introduction of the spectral parameter dependence in the Lax pair $(\Lc,\M)$: as $G$ is a finite dimensional manifold, a monodromy matrix without a dependence on an arbitrary auxiliary parameter $\lambda$ would produce only a finite number of conserved quantities.

\paragraph{Circle coordinate.} Let us now suppose that the coordinate $x$ is on the circle $\mathbb{S}^1 \simeq [0,2\pi]$. We then define the monodromy matrix as
\begin{equation*}
T(\lambda,t) = T(\lambda\, ; 0, 2\pi \, ; t).
\end{equation*}
Assuming periodic boundary conditions $\M(\lambda,0,t) = \M(\lambda,2\pi,t)$, the equation \eqref{Eq:DerTempT} becomes
\begin{equation*}
\p_t T(\lambda,t) = \lsb T(\lambda,t), \M(\lambda,0,t) \rsb, \;\; \forall\lambda\in\C.
\end{equation*}
In contrast with the case of a coordinate on the real line, the monodromy matrix $T(\lambda,t)$ is then not conserved. However, as the time evolution of $T(\lambda,t)$ is given by a commutator, one have
\begin{equation*}
\p_t \Phi \bigl( T(\lambda,t) \bigr) = 0, \;\; \forall \lambda\in\C,
\end{equation*}
for any function $\Phi : G\rightarrow \R $ which is invariant under conjugacy. Thus, the quantity $\Phi \bigl( T(\lambda,t) \bigr)$ is conserved along the time evolution of the model. Here also, this equation of conservation is true for any value of $\lambda\in\C$, hence the existence of an infinite number of conserved quantities.\\

We now come back to the general case, with $x$ either on the real line or on the circle. In both cases, the quantities $\Phi \bigl( T(\lambda,t) \bigr)$, for $\Phi$ a conjugacy invariant function on $G$, are conserved (for $x\in\R$, the whole monodromy $T(\lambda,t)$ is conserved, hence so is any function of $T(\lambda,t)$). A generic way of constructing such invariant functions $\Phi$ on $G$ is to consider traces of powers in a matricial representation $\rho: G \rightarrow GL(d)$ of $G$. Indeed, the function
\begin{equation*}
\begin{array}{rccl}
\Phi^\rho_n : & G & \longrightarrow & \R \\
              & g & \longmapsto     & \Tr \bigl( \rho(g)^n \bigr)
\end{array}
\end{equation*}
for $n\in\N$, is invariant under conjugacy transformation $g \mapsto hgh^{-1}$. Moreover, if $\g$ is semi-simple (which will almost always be the case in the following chapters) and if we fix a representation $\rho$ of $G$, it is known that the functions $\Phi^\rho_n$ generate all invariant functions on $G$. We can then consider the conserved charges
\begin{equation*}
\Q_n(\lambda,t) = \Phi^\rho_n \bigl( T(\lambda,t) \bigr) = \Tr \bigl( \rho(T(\lambda,t))^n \bigr).\vspace{5pt}
\end{equation*}

Let us finish this section by discussing light-cone coordinates. For relativistic field theories, it can be useful to work with the light-cone coordinates $x^\pm = \frac{1}{2} \lrb t \pm x \rrb$ instead of the usual space-time coordinates $(x,t)$. The derivatives with respect to $x^\pm$ are then given by $\p_\pm = \p_t \pm \p_x$. Introducing the light-cone components
\begin{equation}\label{Eq:DefLaxLC}
\Lc_\pm(\lambda,x^\pm) = \M(\lambda,x,t) \pm \Lc(\lambda,x,t)
\end{equation}
of the Lax pair $(\Lc,\M)$, the Lax equation \eqref{Eq:Lax} can be rewritten as
\begin{equation}\label{Eq:LaxLC}
\p_+ \Lc_-(\lambda,x^\pm) - \p_- \Lc_+(\lambda,x^\pm) + \lsb \Lc_+(\lambda,x^\pm), \Lc_-(\lambda,x^\pm) \rsb = 0.
\end{equation}
For relativistic field theories, it will be often simpler to find a Lax pair in light-cone coordinates. However, to construct conserved quantities, one then has to come back to the space-time Lax pair $(\Lc,\M)$ (more precisely, one needs the Lax matrix $\Lc$, \textit{i.e.} the spatial component of the Lax pair).

\section{Hamiltonian integrability and non-ultralocal Poisson brackets}
\label{Sec:Ham}

As explained above, we say that a model is integrable if it possesses an infinite number of conserved quantities in involution (in the hamiltonian sense). In the previous section, we have seen how one can construct an infinite number of conserved charges using Lax pairs and monodromy matrices. It is then natural to ask under which conditions these quantities are in involution (as in this case we have a proof of the integrability of the model). To discuss this, we first need to discuss the hamiltonian formulation of classical field theories.

\paragraph{Hamiltonian field theories.} A hamiltonian field theory is given by a phase space $M$, a Poisson bracket $\lwb \cdot, \cdot \rwb$ and a Hamiltonian $\Hc$. The phase space $M$ describes the possible space configurations of the dynamical fields of the model. As we are in the hamiltonian formalism, we consider the fields as depending on the space coordinate $x$ (the time $t$ is no longer a coordinate of the field but is considered as a hamiltonian flow, as explained below). The phase space $M$ is a Poisson manifold, meaning that the space $\F[M]$ of functionals on $M$ is equipped with a Poisson bracket
\vspace{-5pt}\begin{equation*}
\begin{array}{rccl}
\lwb \cdot, \cdot \rwb : & \F[M] \times \F[M] & \longrightarrow & \F[M] \\
 & (f, g) & \rightarrow & \lwb f, g \rwb
\end{array}.
\end{equation*}
The Poisson bracket is a skew-symmetric bilinear derivation satisfying the Jacobi identity:
\begin{equation*}
\forall f,g,h \in \F[M], \;\;\; \lwb f, \lwb g, h \rwb \rwb + \lwb g, \lwb h, f \rwb \rwb + \lwb h, \lwb f, g \rwb \rwb = 0.
\end{equation*}
The dynamic of the model on $M$ is encoded in a functional $\Hc \in \F[M]$, called the Hamiltonian. More precisely, for any functional $f \in \F[M]$ (we restrict here to functionals which do not depend explicitly on the time $t$), its time evolution is given by the hamiltonian flow of $\Hc$:
\begin{equation}\label{Eq:DynHam}
\p_t f = \lwb \Hc, f \rwb.
\end{equation}
We say that two functionals $f$ and $g$ in $\F[M]$ are in involution if their Poisson bracket vanishes, \textit{i.e.} if
\begin{equation*}
\lwb f, g \rwb = 0.
\end{equation*}
In particular, by equation \eqref{Eq:DynHam}, $f$ is in involution with the Hamiltonian $\Hc$ if and only if $f$ is conserved, \textit{i.e.} if $\p_t f =0$.

Let us also introduce here the momentum $\Pc$ of the theory. It is also a functional in $\F[M]$, whose Hamiltonian flow generates the derivative with respect to the space coordinate $x$, \textit{i.e.} such that
\begin{equation*}
\forall f \in \F[M], \;\;\; \lwb \Pc, f \rwb = \p_x f.
\end{equation*}

\paragraph{Poisson bracket of the Lax matrix.} We will now use the definitions and results of Section \ref{Sec:Lax}. In particular, we suppose that we have a Lax pair $\bigl(\Lc(\lambda,x),\M(\lambda,x)\bigr)$ satisfying the Lax equation \eqref{Eq:Lax}, which can be rewritten in the hamiltonian formalism as
\begin{equation}\label{Eq:ZCEH}
\lwb \Hc, \Lc(\lambda,x) \rwb - \lwb \Pc, \M(\lambda,x) \rwb + \lsb \M(\lambda,x), \Lc(\lambda,x) \rsb =0.
\end{equation}
One can then construct the monodromy matrix $T(\lambda)$ associated with this Lax pair and obtain an infinity of conserved charges $\Phi\bigl(T(\lambda)\bigr)$, with $\Phi$ invariant functions on $G$.

As we are interested in integrable models, it is natural to ask under which condition these conserved charges are in involution. They are defined from the monodromy matrix $T(\lambda)$, which itself is defined from the Lax matrix $\Lc(\lambda,x)$. Thus, the Poisson brackets of these charges depend on the Poisson brackets between the components of the Lax matrix.

To discuss Poisson brackets of Lie algebra-valued functionals in a compact way, we will use tensorial notations. For $X$ a functional valued in $U(\g)$, the universal enveloping algebra of $\g$, we define
\begin{equation*}
X\ti{1} = X \otimes \Id \;\;\;\; \text{ and } \;\;\;\; X\ti{2} = \Id \otimes X,
\end{equation*}
which belong to $U(\g)\otimes U(\g)\otimes\F[M]$, \textit{i.e.} are $U(\g)\otimes U(\g)$-valued functionals.\newpage

Let $\lwb I^a \rwb$ be a basis of the Lie algebra $\g$ and $X,Y\in\g\otimes\F[M]$ be $\g$-valued functionals. We decompose $X$ and $Y$ in this basis as
\begin{equation*}
X = X_a I^a \;\;\;\; \text{ and } \;\;\;\; Y = Y_a I^a,
\end{equation*}
where we used an implicit summation convention on repeated indices. The coefficients $X_a$ and $Y_a$ are then scalar-valued functionals in $\F[M]$. Let us define the Poisson bracket
\begin{equation*}
\lwb X\ti{1}, Y\ti{2} \rwb = \lbrace X_a, Y_b \rbrace \, I^a \otimes I^b,
\end{equation*}
valued in $\g \otimes \g$. One checks that it is independent of the choice of basis $\lwb I^a \rwb$ of $\g$. In particular, one can consider
\begin{equation}\label{Eq:PBLaxT}
\lwb \Lc\ti{1}(\lambda,x), \Lc\ti{2}(\mu,y) \rwb,
\end{equation}
which encodes all Poisson brackets between the components of the Lax matrix.

\paragraph{Ultralocal Sklyanin bracket.} As explained above, we are looking for conditions on the Poisson bracket \eqref{Eq:PBLaxT} for the conserved charges $\Phi \bigl( T(\lambda) \bigr)$ and $\Psi \bigl( T(\mu) \bigr)$ to be in involution for all spectral parameters $\lambda,\mu\in\C$ and all invariant functions $\Phi$ and $\Psi$. A first sufficient condition for this has been found by Sklyanin in~\cite{Sklyanin:1982tf} (see also~\cite{SemenovTianShansky:1983ik} and~\cite{Babebook}). It requires the bracket \eqref{Eq:PBLaxT} to be of the form
\begin{equation}\label{Eq:UL}
\lwb \Lc\ti{1}(\lambda,x), \Lc\ti{2}(\mu,y) \rwb = \lsb \Rc\ti{12}(\lambda,\mu), \Lc\ti{1}(\lambda,x) + \Lc\ti{2}(\mu,x) \rsb \delta_{xy},
\end{equation}
where $\delta_{xy}$ is the Dirac $\delta$-distribution and $\Rc\ti{12}$ is a matrix in $\g \otimes \g$ depending on the spectral parameters $\lambda,\mu\in\C$. The skew-symmetry of the Poisson bracket \eqref{Eq:UL} requires this matrix to be skew-symmetric, in the sense that
\begin{equation}\label{Eq:SkewR}
\Rc\ti{12}(\lambda,\mu) = - \Rc\ti{21}(\mu,\lambda).
\end{equation}
The Poisson bracket \eqref{Eq:UL} is said to be ultralocal, as it contains only a Dirac $\delta$-distribution and not its derivatives. As we shall see in chapter \ref{Chap:Models}, this is not enough to describe the hamiltonian integrability of various models, for which there exists a Lax matrix whose Poisson bracket also contains derivatives of $\delta$-distribution.

\paragraph{Non-ultralocal Maillet bracket.} A generalisation of the ultralocal Sklyanin bracket \eqref{Eq:UL} has been presented by Maillet in~\cite{Maillet:1985fn,Maillet:1985ek}. It is still a sufficient condition on the Poisson bracket of the Lax matrix for the conserved charges extracted from the monodromy to be in involution. However, this condition allows for this Poisson bracket to contain a derivative $\delta'_{xy}$ of the Dirac distribution. It takes the form
\begin{align}\label{Eq:PBR}
\hspace{-60pt}\lwb \Lc\ti{1}(\lambda,x), \Lc\ti{2}(\mu,y) \rwb
&= \lsb \Rc\ti{12}(\lambda,\mu), \Lc\ti{1}(\lambda,x) \rsb \delta_{xy} - \lsb \Rc\ti{21}(\mu,\lambda), \Lc\ti{2}(\mu,x) \rsb \delta_{xy}\\
& \hspace{30pt} - \left( \Rc\ti{12}(\lambda,\mu) + \Rc\ti{21}(\mu,\lambda) \right) \delta'_{xy}, \notag
\end{align}
where here also $\Rc$ is a $\g\otimes\g$-valued function of two spectral parameters $\lambda$ and $\mu$. This bracket is called \textbf{Maillet bracket}, or \textbf{non-ultralocal bracket} (due to the presence of the $\delta'$ term), and will play a major role in this thesis.

Let us note that here, we supposed that $\Rc$ was not dynamical, \textit{i.e.} that it did not depend on $x$ trough a dependence on the fields of the model. The original articles~\cite{Maillet:1985fn,Maillet:1985ek} of Maillet also included the case of dynamical $\Rc$ matrices. However, as we will not need this generalisation in this thesis, we will restrain to the case of non-dynamical $\Rc$ for simplicity.

It is worth noticing that the skew-symmetry of the Poisson bracket \eqref{Eq:PBR} is automatically satisfied, without requiring any further condition on the matrix $\Rc\ti{12}(\lambda,\mu)$. If one considers the case of a skew-symmetric $\Rc$, \textit{i.e.} which satisfies equation \eqref{Eq:SkewR}, the Maillet bracket \eqref{Eq:PBR} reduces to the Sklyanin bracket \eqref{Eq:UL}. In general, the matrix $\Rc$ of a Maillet bracket can possess a skew-symmetric part $r$ and a symmetric part $s$:
\begin{equation*}
r\ti{12}(\lambda,\mu) = \frac{1}{2} \left( \Rc\ti{12}(\lambda,\mu) - \Rc\ti{21}(\mu,\lambda) \right) \;\;\;\; \text{ and } \;\;\;\; s\ti{12}(\lambda,\mu) = \frac{1}{2} \left( \Rc\ti{12}(\lambda,\mu) + \Rc\ti{21}(\mu,\lambda) \right).
\end{equation*}
The bracket \eqref{Eq:PBR} then takes the following form (this is in fact the original form proposed in the article~\cite{Maillet:1985fn}):
\begin{align}\label{Eq:PBrs}
\hspace{-60pt}\lwb \Lc\ti{1}(\lambda,x), \Lc\ti{2}(\mu,y) \rwb
&= \lsb r\ti{12}(\lambda,\mu), \Lc\ti{1}(\lambda,x) + \Lc\ti{2}(\mu,x) \rsb \delta_{xy} \\
& \hspace{30pt} + \lsb s\ti{12}(\lambda,\mu), \Lc\ti{1}(\lambda,x) - \Lc\ti{2}(\mu,x) \rsb \delta_{xy} - 2 s\ti{12}(\lambda,\mu) \, \delta'_{xy}, \notag
\end{align}
somtimes called a $r/s$-system. The non-ultralocality of the bracket, \textit{i.e.} the term proportional to $\delta'_{xy}$, is then controlled by the symmetric part $s$ of $\Rc$.

\section{Yang-Baxter equation, $\Rc$-matrices and twist function}
\label{Sec:CYBE}

\paragraph{Classical Yang-Baxter equation.}
As we have seen in the previous sections, a sufficient condition to have an integrable field theory is the existence of a Lax matrix $\Lc(\lambda,x)$ satisfying a non-ultralocal Poisson bracket \eqref{Eq:PBR}, controlled by a $\g\otimes\g$-valued matrix $\Rc\ti{12}(\lambda,\mu)$. A natural question to ask at this point is whether this matrix can be anything or if there are some constraints on it, coming from the properties that a Poisson bracket must satisfy. As we have seen, the skew-symmetry of the bracket \eqref{Eq:PBR} is automatically verified, without requiring any additional constraint on $\Rc$. However, this bracket must also satisfy the Jacobi identity, \textit{i.e.} we must have
\begin{align}
&\lwb \Lc\ti{1}(\lambda_1,x_1), \lwb \Lc\ti{2}(\lambda_2,x_2), \Lc\ti{3}(\lambda_3,x_3) \rwb \rwb + \lwb \Lc\ti{2}(\lambda_2,x_2), \lwb \Lc\ti{3}(\lambda_3,x_3), \Lc\ti{1}(\lambda_1,x_1) \rwb \rwb \\
& \hspace{200pt} + \lwb \Lc\ti{3}(\lambda_3,x_3), \lwb \Lc\ti{1}(\lambda_1,x_1), \Lc\ti{2}(\lambda_2,x_2) \rwb \rwb = 0, \notag
\end{align}
which is to be understood as an identity in $\g\otimes\g\otimes\g$. Using the expression \eqref{Eq:PBR}, one can check that this is equivalent to
\begin{equation}\label{Eq:CondJac}
\lsb \YB{1}{2}{3}, \Lc\ti{1}(\lambda_1,x) \rsb + \lsb \YB{2}{3}{1}, \Lc\ti{2}(\lambda_2,x) \rsb + \lsb \YB{3}{1}{2}, \Lc\ti{3}(\lambda_3,x) \rsb = 0,
\end{equation}
where
\begin{equation*}
\YB{1}{2}{3} = \lsb \Rc\ti{12}(\lambda_1,\lambda_2), \Rc\ti{13}(\lambda_1,\lambda_3) \rsb + \lsb \Rc\ti{12}(\lambda_1,\lambda_2), \Rc\ti{23}(\lambda_2,\lambda_3) \rsb + \lsb \Rc\ti{32}(\lambda_3,\lambda_2), \Rc\ti{13}(\lambda_1,\lambda_3) \rsb.
\end{equation*}
Thus, a sufficient condition for the Jacobi identity to be verified is the so-called \textbf{Classical Yang-Baxter Equation (CYBE)}:
\begin{equation}\label{Eq:CYBE}
\lsb \Rc\ti{12}(\lambda_1,\lambda_2), \Rc\ti{13}(\lambda_1,\lambda_3) \rsb + \lsb \Rc\ti{12}(\lambda_1,\lambda_2), \Rc\ti{23}(\lambda_2,\lambda_3) \rsb + \lsb \Rc\ti{32}(\lambda_3,\lambda_2), \Rc\ti{13}(\lambda_1,\lambda_3) \rsb = 0.
\end{equation}
The matrices that satisfy this equation are called $\bm{\Rc}$\textbf{-matrices}. Although it is not a necessary condition to have a Jacoby identity, we shall restrict ourselves from now on to matrices $\Rc$ which satisfy this stronger condition. Indeed, the CYBE \eqref{Eq:CYBE} is much more easy to manipulate than the general condition \eqref{Eq:CondJac}, as it depends only on $\Rc$ and not on the Lax matrix $\Lc$ anymore. Moreover, the CYBE is an algebraic, non-dynamical, equation in $\g\otimes\g\otimes\g$, which only depends on the Lie bracket over $\g$. As we shall see in the next paragraph, this allows general schemes of constructions of solutions of the CYBE. For more details on the algebraic interpretations of the CYBE, we refer to the appendix \ref{App:RMat}.

\paragraph{Standard $\bm{\Rc}$-matrices.}
In this paragraph, we present some particular solutions of the CYBE called the standard $\Rc$-matrices. Let us suppose that $\g$ is a semi-simple Lie algebra (see appendix \ref{App:SemiSimple} for definitions and details). In particular, it possesses a non-degenerate invariant bilinear form $\kappa$. Let us consider the so-called split Casimir of $\g$:
\begin{equation*}
C\ti{12} = \kappa_{ab} I^a \otimes I^b \in \g\otimes\g,
\end{equation*}
which is independent of the choice of basis $\lbrace I^a \rbrace$ of $\g$. It is symmetric ($C\ti{12}=C\ti{21}$) and satisfies the following identity (see Appendix \ref{App:Casimir}):
\begin{equation*}
\lsb C\ti{12}, X\ti{1}+X\ti{2} \rsb = 0, \;\; \forall X\in\g.
\end{equation*}
In particular, this implies that
\begin{equation}\label{Eq:QuadCas}
\lsb C\ti{12}, C\ti{23} \rsb = \lsb C\ti{32}, C\ti{13} \rsb = -\lsb C\ti{12}, C\ti{13} \rsb.
\end{equation}
Using this identity in $\g\otimes\g\otimes\g$, together with the circle lemma
\begin{equation}\label{Eq:CircleLemma}
\frac{1}{(\lambda_2-\lambda_1)(\lambda_3-\lambda_1)} - \frac{1}{(\lambda_2-\lambda_1)(\lambda_3-\lambda_2)} - \frac{1}{(\lambda_2-\lambda_3)(\lambda_3-\lambda_1)} = 0,
\end{equation}
we get that
\begin{equation}\label{Eq:R0NonTwisted}
\Rc^0\ti{12} \left( \lambda,\mu \right) = \frac{C\ti{12}}{\mu-\lambda}
\end{equation}
is a solution of the CYBE \eqref{Eq:CYBE}. We refer the reader to the appendix \ref{App:RMat} for a more algebraic interpretation of this solution of the CYBE, related to the non-twisted loop algebra $\Lc(\g)$. We call it \textbf{the standard (non-twisted) $\bm{\Rc}$-matrix on $\bm{\Lg}$}. Let us note that, as the quadratic Casimir is symmetric, the standard non-twisted $\Rc$-matrix $\Rc^0$ is skew-symmetric. \\

Suppose now that we are given an automorphism $\s$ of $\g$, of finite order $T$ (see the appendix \ref{App:Torsion} for more details on these automorphisms). It is a standard result that it preserves the Killing form $\kappa$. As a consequence, the quadratic Casimir satisfies
\begin{equation}\label{Eq:CasInvSigma}
\s\ti{1} \s\ti{2} C\ti{12} = C\ti{12},
\end{equation}
where $\s\ti{1}$ and $\s\ti{2}$ act respectively on the first and second tensor factor of $\g\otimes\g$. Let us chose a primitive $T^{\rm th}$ root of unity $\omega$. We then define the so-called \textbf{standard twisted $\bm{\Rc}$-matrix on $\bm{\Lg}$} as
\begin{equation}\label{Eq:RCyc}
\Rc^0\ti{12}(\lambda,\mu) = \frac{1}{T}\sum_{k=0}^{T-1} \frac{\s\ti{1}^k C\ti{12}}{\mu-\omega^{-k}\lambda}.
\end{equation}
Using the invariance \eqref{Eq:CasInvSigma} of the Casimir under $\s$ and the fact that $\s$ is an automorphism of $\g$, one checks that $\Rc^0$ is a solution of the CYBE \eqref{Eq:CYBE}. We refer to appendix \ref{App:RMat} for a more algebraic interpretation of this solution in terms of the twisted loop algebra $\Lc(\g,\s)$. Let us note that, contrarily to the non-twisted $\Rc$-matrix, the twisted standard $\Rc$-matrix is not skew-symmetric.

Both the twisted and non-twisted standard $\Rc$-matrices are singular at $\lambda=\mu$. Moreover, the asymptotic behaviour of the twisted matrix around this singularity is the one of the non-twisted one (divided by $T$), as we have
\begin{equation*}
\Rc^0\ti{12}(\lambda,\mu) = \frac{1}{T} \frac{C\ti{12}}{\mu-\lambda} + O\bigl( (\lambda-\mu)^0 \bigr).
\end{equation*}

As $\s^T=\Id$, the eigenvalues of $\s$ are of the form $\omega^p$ where, by convention, we take $p\in\lbrace 0, \ldots, T-1 \rbrace$. We denote by $\g^{(p)}$ the corresponding eigenspace and by $\pi^{(p)}$ the projection on $\g^{(p)}$ in the direct sum $\g=\bigoplus_{p=0}^{T-1} \g^{(p)}$ (see the appendix \ref{App:Torsion} for more details about these eigenspaces). Defining $C\ti{12}^{(p)}=\pi^{(p)}\ti{1}C\ti{12}=C\ti{21}^{(-p)}$, we can rewrite $\Rc^0$ as
\begin{equation}\label{Eq:RCas}
\Rc^0\ti{12}(\lambda,\mu) = \sum_{p=0}^{T-1} \frac{\lambda^p\mu^{T-1-p}}{\mu^T-\lambda^T} C\ti{12}^{(p)}.
\end{equation}

\paragraph{Twist function.} We end this section by a quick but crucial remark~\cite{Maillet:1985ec,Sevostyanov:1995hd,Vicedo:2010qd}. Let us fix a solution $\Rc^0$ of the CYBE \eqref{Eq:CYBE}. Typically, in this thesis, we will choose $\Rc^0$ to be a non-twisted or twisted standard $\Rc$-matrix on $\Lg$, as described in the previous paragraph. It is easy to check that for any function $\vp$ from $\C$ to itself, the matrix $\Rc\ti{12}(\lambda,\mu) = \Rc^0\ti{12}(\lambda,\mu) \vp(\mu)^{-1}$ is also a solution of the CYBE \eqref{Eq:CYBE}. Indeed, the left-hand side of the CYBE for $\Rc$ is simply the one for $\Rc^0$ multiplied by $\vp(\lambda_2)^{-1}\vp(\lambda_3)^{-1}$, so the CYBE for $\Rc^0$ implies the one for $\Rc$. This function, that is often supposed to be a rational function of the spectral parameter, is called \textbf{the twist function}~\cite{Vicedo:2010qd}. We refer the reader to the appendix \ref{App:RMat} for the loop algebra interpretation of the corresponding $\Rc$-matrix.

\section{Integrable models with twist function}
\label{Sec:ModelsTwist}

We define an integrable model with twist function to be a Hamiltonian field theory such that:
\begin{itemize}
\item its equations of motion admit a \textbf{Lax pair} $\bigl(\Lc(\lambda,x),\M(\lambda,x)\bigr)$, valued in $\g$ and depending rationally on the spectral parameter ;
\item the Lax matrix $\Lc$ satisfies a \textbf{non-ultralocal Maillet bracket} of the form \eqref{Eq:PBR} ;
\item the $\Rc$-matrix underlying the Maillet bracket is of the form \begin{equation}\label{Eq:DefR}
\Rc\ti{12}(\lambda,\mu) = \Rc^0\ti{12}(\lambda,\mu) \vp(\mu)^{-1}
\end{equation}
with $\Rc^0$ a \textbf{standard $\bm{\Rc}$-matrix} on $\Lg$ and $\vp$ a rational function, the \textbf{twist function} ;
\item if the standard matrix $\Rc^0$ is twisted by an automorphism $\s$, the Lax matrix and the twist function satisfy some \textbf{equivariance properties}, described below ;
\item the Lax matrix and the automorphism $\s$ satisfy some \textbf{reality conditions}, described below.
\end{itemize}

\paragraph{Equivariance properties.} As explained above, if the standard matrix $\Rc^0$ is twisted by an automorphism $\s$ of order $T$, we require some additional equivariance condition on $\Lc$ and $\vp$. In this case, we speak of a \textbf{cyclotomic model}.

As the order of $\s$ is $T$, it defines an action of the cyclic group $\Z_T=\Z/T\Z$ on $\g$. On the other hand, $\Z_T$ can be seen as acting on the complex numbers $\C$ \textit{via} multiplication by $\omega$. We then remark that the first tensor factor of the matrix $\Rc^0$ is equivariant under these two actions, in the sense that
\begin{equation}\label{Eq:EquiR}
\s\ti{1} \Rc^0\ti{12}(\lambda,\mu) = \Rc^0\ti{12}(\omega\lambda,\mu).
\end{equation}
We will suppose that the Lax matrix $\Lc$ possesses a similar equivariance property, namely
\begin{equation}\label{Eq:EquiL}
\s\bigl( \Lc(\lambda,x) \bigr) = \Lc(\omega\lambda,x).
\end{equation}
Moreover, using the invariance \eqref{Eq:CasInvSigma} of the Casimir under $\s$, one gets, on the second tensor factor,
\begin{equation}\label{Eq:EquiR2}
\s\ti{2} \Rc^0\ti{12}(\lambda,\mu) = \omega\Rc^0\ti{12}(\lambda,\omega\mu).
\end{equation}
The compatibility of these properties with the Poisson bracket \eqref{Eq:PBR} imposes that
\begin{equation}\label{Eq:TwistEqui}
\varphi(\omega\lambda)=\omega^{-1}\varphi(\lambda),
\end{equation}
from which we deduce that $\lambda\varphi(\lambda)$ is invariant under the action of $\Z_T$. Thus, there exists a rational function $\zeta$ such that
\begin{equation}\label{Eq:DefZeta}
\lambda \varphi(\lambda) = \zeta(\lambda^T).
\end{equation}
Let us note that the equivariance properties \eqref{Eq:EquiR} and \eqref{Eq:EquiR2} of $\Rc^0$, together with the one of the twist function \eqref{Eq:TwistEqui}, implies
\begin{equation*}
\s\ti{1} \Rc\ti{12}(\lambda,\mu) = \Rc\ti{12}(\omega\lambda,\mu) \;\;\;\; \text{ and } \;\;\;\; \s\ti{2} \Rc\ti{12}(\lambda,\mu) = \Rc\ti{12}(\lambda,\omega\mu).
\end{equation*}
Thus, the matrix $\Rc$ is equivariant both in the first and the second tensor factors.

\paragraph{Reality condition.} In non-ultralocal models with twist function, we suppose that the Lax matrix depends rationally on the spectral parameter $\lambda$, which is a complex number. Thus, the Lie algebra $\g$ in which the Lax matrix is valued is a complex Lie algebra. However, many field theories are described by real valued fields: we keep track of this in the Lax matrix through reality conditions.

Let $\g_0$ be a real form of the complex Lie algebra $\g$ (see the Appendix \ref{App:RealForms} for definitions and conventions on real forms). It can be seen as the algebra of fixed points under an anti-automorphism involution $\tau$ of $\g$. We will suppose that the Lax matrix satisfies the following \textbf{reality condition}:
\begin{equation}\label{Eq:Reality}
\tau \bigl( \Lc(\lambda,x) \bigr) = \Lc\lrb \bar{\lambda}, x \rrb,
\end{equation}
where $\bar{\lambda}$ is the complex conjugate of the spectral parameter $\lambda$.

If the model is cyclotomic, \textit{i.e.} the standard matrix 
$\Rc^0$ is twisted by an automorphism $\s$, we will assume that $\s$ satisfies the following property:
\begin{equation}\label{Eq:Dihedral}
\s \circ \tau = \tau \circ \sigma^{-1},
\end{equation}
that we call the \textbf{dihedrality condition}. The quadratic Casimir satisfies
\begin{equation*}
\tau\ti{1} C\ti{12} = \tau\ti{2} C\ti{12} = C\ti{12}.
\end{equation*}
Combining this equation with the dihedrality condition \eqref{Eq:Dihedral}, we get the following reality conditions on the standard $\Rc$-matrix:
\begin{equation}\label{Eq:RealR0}
\tau\ti{1} \Rc^0\ti{12}(\lambda,\mu) = \tau\ti{2} \Rc^0\ti{12}(\lambda,\mu) = \Rc^0\ti{12} \lrb \bar{\lambda},\bar{\mu} \rrb.
\end{equation}
The compatibility of the Poisson bracket \eqref{Eq:PBR} with the reality conditions \eqref{Eq:Reality} and \eqref{Eq:RealR0} requires that the twist function is a rational function with real coefficients, \textit{i.e.} that
\begin{equation}\label{Eq:TwistReal}
\overline{\varphi(\lambda)}=\varphi(\bar{\lambda}).
\end{equation}
Combining the two conditions \eqref{Eq:RealR0} and \eqref{Eq:TwistReal}, we find
\begin{equation}\label{Eq:RealR}
\tau\ti{1} \Rc\ti{12}(\lambda,\mu) = \tau\ti{2} \Rc\ti{12}(\lambda,\mu) = \Rc\ti{12} \lrb \bar{\lambda},\bar{\mu} \rrb.
\end{equation}

\paragraph{Dihedrality.} Let us end this section by explaining the name of the dihedrality condition \eqref{Eq:Dihedral}. We consider the abstract group $\Gamma_T$ defined by the following presentation:
\begin{equation}\label{Eq:DihedralGroup}
\Gamma_T = \left\langle \mathrm{r},\mathrm{s} \; \bigl| \; \mathrm{r}^T=\Id, \; \mathrm{s}^2=\Id, \; \mathrm{r}\mathrm{s}=\mathrm{s}\mathrm{r}^{-1} \right\rangle,
\end{equation}
\textit{i.e.} the group generated by two abstract generators $\mathrm{r}$ and $\mathrm{s}$ satisfying the relations above. It is a classical result from group theory that $\Gamma_T$ is then of order $2T$ and is isomorphic to the \textbf{dihedral group} $D_T$, defined as the group of symmetries of the regular polygon with $T$ edges ($\mathrm{r}$ then represents the rotation of angle $\frac{2\pi}{T}$ and $\mathrm{s}$ the symmetry with respect to an axis of the polygon).\\

Let $\omega=\exp\left(\frac{2i\pi}{T}\right)$ be a primitive $T^{\rm th}$-root of unity. There is a natural action of the dihedral group $\Gamma_T$ on the complex plan as the multiplication by $\omega$ and the complex conjugation:
\begin{equation}\label{Eq:ActionDihedralPlan}
\forall \,\lambda\in\C, \;\;\;\; \rm r.\lambda = \omega \lambda \;\;\;\; \text{ and } \;\;\;\; \rm s.\lambda = \overline{\lambda}.
\end{equation}
In the same way, there is an action of $\Gamma_T$ on the Lie algebra $\g$, defined from the automorphism $\s$ of order $T$ and the antilinear involutive automorphism $\tau$ introduced above:
\begin{equation*}
\forall\, X\in\g, \;\;\;\; \rm r.X = \s(X) \;\;\;\; \text{ and } \;\;\;\; \rm s.X = \tau(X),
\end{equation*}
as $\s$ and $\tau$ satisfy the dihedrality condition \eqref{Eq:Dihedral}. The cyclotomic property \eqref{Eq:EquiL} and the reality condition \eqref{Eq:Reality} imposed on the Lax matrix $\Lc$ can be understood as the equivariance of $\Lc$ under the action of $\Gamma_T$:
\begin{equation*}
\forall \, {\rm u} \in \Gamma_T, \;\;\;\; {\rm u}. \Lc(\lambda,x) = \Lc(  {\rm u}. \lambda, x).
\end{equation*}

\cleardoublepage
\chapter[Examples of models with twist function: integrable $\s$-models]{Examples of models with twist function: integrable $\bm{\s}$-models}
\label{Chap:Models}

\vspace{-6pt} In the previous chapter, we introduced the class of integrable models with twist function. In the present one, we discuss various examples of such models. All the examples considered here are integrable $\s$-models but there exist other theories with twist function, such as the Korteweg-de Vries (KdV) equation or affine Toda field theories, that we shall not discuss here (see~\cite{Vicedo:2017cge}).

The first three sections of this chapter are reviews about $\s$-models which are known to possess a twist function. The first section discusses the general framework of integrable $\s$-models. The second section concerns the Principal Chiral Model (PCM), which is the simplest example of integrable $\s$-models, and the $\Z_T$-coset models. We review here their Lax formulation and the Poisson bracket of their Lax matrix, exhibiting the structure of a non-ultralocal model with twist function. 

The third section concerns integrable deformations of the PCM and the $\Z_T$-coset models. Indeed, it has been discovered in the past decades that these models admit continuous deformations (controlled by real parameters) which still admit a Lax pair. Moreover, the hamiltonian analysis of these models has been conducted and it has been found that the Poisson bracket of their Lax matrix is also governed by a twist function. More precisely, it has been observed that the effect of the deformation is to deform the poles of the twist function of the model. In the third section, we review different results on this subject that exist in the literature and present an overall view of the panorama of integrable deformed $\s$-models and their twist function.

The fourth and final section is about a particular two-parameter deformation of the PCM, the so-called Bi-Yang-Baxter model. This model was introduced by Klim{\v{c}}ik in~\cite{Klimcik:2008eq}, who also proved in~\cite{Klimcik:2014bta} that its equations of motion can be recast as a Lax equation, thus proving the existence of an infinite number of conserved charges for this model. However, the hamiltonian analysis of the Bi-Yang-Baxter model was never carried out. My first PhD project was the study of the Poisson bracket of the Lax matrix of the Bi-Yang-Baxter model. In Section \ref{Sec:BYB}, after recalling the construction of the model and of its Lax pair, I present the results I found in~\cite{Delduc:2015xdm} concerning the hamiltonian integrability of the Bi-Yang-Baxter model. In particular, we shall see that the Bi-Yang-Baxter model also enters the class of models with twist function.

\section{Generalities about integrable $\s$-models}

\subsection{Lie group valued field and currents}
\label{SubSec:FieldSigma}

Before going into the details of the different models, let us discuss some general aspects regarding integrable $\s$-models. A $\s$-model (not necessarily integrable) is a two-dimensional field theory with dynamical fields $\phi : \Sigma \rightarrow M$, where $\Sigma$ is the two-dimensional Minkowski space-time with coordinates $(x,t)$, called the worldsheet, and $M$ is a Riemannian manifold, which we call the target space. We are interested in some of these models, which have the additional property of being integrable. These share the property that their target space is a real Lie group $G_0$ or one of its cosets, \textit{i.e.} the quotient $G_0/H$ of $G_0$ by a particular subgroup $H$. We shall suppose here that $G_0$ is a connected semi-simple Lie group. We will denote $G=G_0^\C$ the complexification of $G_0$, which is then a complex semi-simple Lie group (see Appendix \ref{App:RealForms}).

As we will see in more details in this chapter, we shall not describe the coset models with fields directly valued in $G_0/H$. Instead, we shall consider an equivalent formulation with a field valued in $G_0$, together with a gauge symmetry in $H$, which eliminates the additional degrees of freedom. As a result, all the models we shall consider are defined by a field $g : \Sigma \rightarrow G_0$, valued in the Lie group $G_0$. We shall use extensively the so-called \textbf{left currents}
\begin{equation*}
j^L_0 = g^{-1} \p_t g = -\bigl( \p_t g^{-1} \bigl) g  \;\;\;\; \text{ and } \;\;\;\; j^L_1 = g^{-1} \p_x g = -\bigl( \p_x g^{-1} \bigl) g,
\end{equation*}
which are $\g_0$-valued fields, with $\g_0$ the Lie algebra of $G_0$. These are called left currents as they are invariant under the left multiplication
\begin{equation*}
L_h : g \longmapsto hg
\end{equation*}
of $g$ by a constant element $h$ of $G_0$. There are also \textbf{right currents}
\begin{equation*}
j^R_0 = g \p_t g^{-1} = - \bigl( \p_t g) g^{-1} \;\;\;\; \text{ and } \;\;\;\; j^R_1 = g \p_x g^{-1} = - \bigl( \p_x g) g^{-1},
\end{equation*}
invariant under the right multiplication $R_h:g \mapsto gh$. They are related to the left currents by a conjugacy transformation:
\begin{equation}\label{Eq:LeftToRight}
j^R_\mu = -g j^L_\mu g^{-1}, \;\;\; \text{for } \mu=0,1.
\end{equation}
The left and right currents are flat, \textit{i.e.} they satisfy the zero curvature equation:
\begin{equation}\label{Eq:MaurerCartan}
\p_t j^{L,R}_1 - \p_x j^{L,R}_0 + \lsb j^{L,R}_0,j^{L,R}_1 \rsb = 0,
\end{equation}
called \textbf{the Maurer-Cartan equation}. Finally let us also introduce the light-cone currents:
\begin{equation}\label{Eq:DefJpm}
j^L_\pm = g^{-1} \p_\pm g = j^L_0 \pm j^L_1 \;\;\;\; \text{ and } \;\;\;\; j^R_\pm = g \p_\pm g^{-1} = j^R_0 \pm j^R_1.
\end{equation}
In light-cone coordinates, the Maurer-Cartan equation reads
\begin{equation}\label{Eq:MaurerCartanLC}
\p_+ j^{L,R}_- - \p_- j^{L,R}_+ + \lsb j^{L,R}_+,j^{L,R}_- \rsb = 0.
\end{equation}

\subsection{Hamiltonian formulation}
\label{SubSec:SigmaModelHam}

\paragraph{Conjugate momenta and phase space.} The $\s$-models are naturally defined as Lagrangian field theories with an action $S[g]$ depending on the field $g(x,t)$. As we are interested in the integrability properties of $\s$-models, we will need to consider them in the Hamiltonian formalism. We thus pass from the field $g$ valued in $G_0$ and depending on $(x,t)$ to fields depending only on the space coordinate $x$ and valued in the cotangent bundle $T^*G_0$. These fields form the phase space of the model and contain the $G_0$-valued coordinate-field $g(x)$, together with some conjugate momentum fields.

Let us be more explicit about that. We fix some local coordinates $\psi^i : G_0 \rightarrow \R$ ($i=1,\cdots,n$) of $G_0$, where $n$ is the dimension of $G_0$ (more precisely, the coordinates $\psi^i$ could be defined only on an open subset of $G_0$). Locally, one can describe the field $g$ as $n$ real-valued fields $\phi^i(x)=\psi^i\bigl(g(x)\bigr)$. The Lagrangian density of the model can then be written as a function $L\bigl(\phi^i,\p_\mu \phi^i\bigr)$ of the fields $\phi^i$'s and their space-time derivatives. In the Hamiltonian formalism, these coordinate fields are paired with conjugate momenta $\pi_i(x)$, obtained from the Lagrangian density $L$ as
\vspace{-3pt}\begin{equation*}
\pi_i = \frac{\p L}{\p (\p_t \phi^i)}.\vspace{-3pt}
\end{equation*}
The fields $\phi^i$'s and $\pi_i$'s then describe the whole phase space of the model and satisfy the canonical Poisson brackets
\begin{subequations}\label{Eq:CanPBCoord}
\begin{align}
\lwb \phi^i(x), \phi^j(y) \rwb & = 0, \\
\lwb \pi_i(x), \pi_j(y) \rwb & = 0, \\
\lwb \pi_i(x), \phi^j(y) \rwb & = \delta^j_{\;i} \delta_{xy},
\end{align}
\end{subequations}
where $\delta^j_{\;i}$ is the Kronecker symbol and $\delta_{xy}$ is the Dirac $\delta$-distribution. \newpage

We now want a coordinate-free description of this phase-space. The coordinate fields $\phi^i$ are equivalent to the $G_0$-valued field $g$. Thus, there is left to find a way to encode the conjugate momenta $\pi_i$ in a coordinate-free way. Let $\p_i$ denote the derivative with respect to the local coordinate $\psi^i$: $g^{-1} \p_i g$ then belongs to the Lie algebra $\g_0$. Let us fix a basis $\lwb I_a \rwb_{a=1,\cdots,n}$ of $\g_0$ and define $L^a_{\;\,i}$ such that
\begin{equation*}
g^{-1} \p_i g = L^a_{\;\,i} I_a.
\end{equation*}
The matrix $\bigl( L^a_{\;\,i} \bigr)_{i,a=1,\cdots,n}$ is then invertible. We shall write $\bigl( L^i_{\;\,a} \bigr)_{i,a=1,\cdots,n}$ its inverse. It verifies
\begin{equation*}
L^a_{\;\,i} L^i_{\;\,b} = \delta^a_{\;b} \;\;\;\; \text{ and } \;\;\;\; L^i_{\;\,a} L^a_{\;\,j} = \delta^i_{\;j},
\end{equation*}
where the summation on repeated indices is implied. We then define the $\g_0$-valued field
\begin{equation}\label{Eq:DefX}
X = L^i_{\;\,a} \pi_i \, \kappa^{ab} I_b,
\end{equation}
where $\kappa^{ab}$ is the Killing form of $\g_0$ in the basis $\lwb I^a \rwb$, dual to $\lwb I_a \rwb$. One can then check that this field is invariant under a change of the coordinates $\lwb\psi^i\rwb$ on $G_0$ or of basis $\lwb I^a\rwb$ of $\g_0$.

Thus, the phase-space of the model is parametrised by the $G_0$-valued field $g(x)$ and the $\g_0$-valued field $X(x)$. Together, they form a field valued in the cotangent bundle $T^*G_0$. One can check that the canonical brackets \eqref{Eq:CanPBCoord} can be rewritten in terms of the fields $g$ and $X$ as
\begin{subequations}\label{Eq:PBTstarG}
\begin{align}
\left\lbrace g\ti{1}(x), g\ti{2}(y) \right\rbrace & = 0, \\
\left\lbrace X\ti{1}(x), g\ti{2}(y) \right\rbrace & = g\ti{2}(x) C\ti{12} \delta_{xy},\label{Eq:PBXg} \\
\left\lbrace X\ti{1}(x), X\ti{2}(y) \right\rbrace & = -\left[ C\ti{12}, X\ti{2}(x) \right] \delta_{xy},\label{Eq:PBXX}
\end{align}
\end{subequations}
where $C\ti{12}$ is the quadratic Casimir of $\g_0$ (see Appendix \ref{App:Lie}).

\paragraph{Left and right multiplication.} Let us investigate what is the Hamiltonian flow generated by the integral of the field $X$. Let us define
\begin{equation}\label{Eq:RightMomentMap}
m^R = \int \dd x \; X(x),
\end{equation}
where the integral is either on the real line $\R$ or on the circle $\mathbb{S}^1$, depending on the space-time of the theory. This is a $\g_0$-valued quantity, which can be seen as $\g^*_0$-valued \textit{via} the duality between $\g_0$ and $\g^*_0$ induced by the Killing form $\kappa$. Moreover, we deduce from the Poisson bracket \eqref{Eq:PBXX} that it satisfies the Kostant-Kirillov bracket (see Appendix \ref{App:KK}).
\begin{equation*}
\lwb m^R\ti{1}, m^R\ti{2} \rwb = -\lsb C\ti{12}, m^R\ti{2} \rsb.
\end{equation*}
Thus, it can be chosen as the moment map of an infinitesimal action of $\g_0$ (see Appendix \ref{App:HamAction}). For $\epsilon\in\g_0$ infinitesimal, the corresponding action on an observable $\mathcal{O}\in\F[M]$ is given by
\begin{equation*}
\delta^R_\epsilon \mathcal{O} = \kappa \bigl( \epsilon, \lwb m^R, \mathcal{O} \rwb \bigr).
\end{equation*}
Using the completeness relation \eqref{Eq:CasComp}, one finds that the action $\delta^R_\epsilon$ on the phase space parametrised by the fields $g$ and $X$ is given by
\begin{equation*}
\delta^R_\epsilon g(x) = g(x) \epsilon \;\;\;\; \text{ and } \;\;\;\; \delta^R_\epsilon X(x) = -\left[ \epsilon, X(x) \right].
\end{equation*}
Thus, the action of $\delta^R_\epsilon$ on $g$ is the infinitesimal transformation induced by the right multiplication $g \mapsto gh$ by constant elements $h$ of $G_0$. One can also describe the left multiplication action by considering $gXg^{-1}$ instead of $X$. Indeed, the canonical bracket can be also written as
\begin{subequations}\label{Eq:PBconjg}
\begin{align}
\left\lbrace g\ti{1}(x), g\ti{2}(y) \right\rbrace & = 0, \\
\left\lbrace \bigl( gXg^{-1} \bigr)\ti{1}(x), g\ti{2}(y) \right\rbrace & = C\ti{12}\, g\ti{2}(x) \delta_{xy}, \\
\left\lbrace \bigl( gXg^{-1} \bigr)\ti{1}(x), \bigl( gXg^{-1} \bigr)\ti{2}(y) \right\rbrace & = \left[ C\ti{12}, \bigl( gXg^{-1} \bigr)\ti{2}(x) \right] \delta_{xy}.
\end{align}
\end{subequations}
The moment map
\begin{equation}\label{Eq:LeftMomentMap}
m^L = \int \dd x \; g(x)X(x)g(x)^{-1}
\end{equation}
generates the infinitesimal left multiplication on $g$:
\begin{equation*}
\delta^L_\epsilon g(x) = \epsilon g(x) \;\;\;\; \text{ and } \;\;\;\; \delta^L_\epsilon \bigl( gXg^{-1} \bigr)(x) = \left[ \epsilon, \bigl( gXg^{-1} \bigr)(x) \right].
\end{equation*}
One can remark that $X$ and $gXg^{-1}$ are respectively invariant under the left and right actions:
\begin{equation*}
\delta^L_\epsilon X(x) = 0 \;\;\;\; \text{ and } \;\;\;\; \delta^R_\epsilon \bigl(gXg^{-1}\bigr)(x) = 0.
\end{equation*}

\paragraph{Momentum and Hamiltonian.} Let us consider the current $j^L=j^L_1=g^{-1} \p_x g$ (when working in Hamiltonian formalism, we shall drop the indices $\mu=1$ as we dropped out the time dependence and kept only the space dependence). We will use the following Poisson brackets, which are a consequence of \eqref{Eq:PBTstarG}:
\begin{subequations}\label{Eq:PBj}
\begin{align}
\lwb g\ti{1}(x), j^L\ti{2}(y) \rwb &= 0,\\
\lwb j^L\ti{1}(x), j^L\ti{2}(y) \rwb &= 0, \\
\lwb X\ti{1}(x), j^L\ti{2}(y) \rwb &= - \bigl[ C\ti{12}, j^L\ti{2}(x) \bigr] \delta_{xy} - C\ti{12} \delta'_{xy}.
\end{align}
\end{subequations}
Using these Poisson brackets, we find that the total momentum of the theory is given by
\begin{equation*}
\Pc = \int \dd x \; \kappa \bigl( X, j^L \bigr).
\end{equation*}
Indeed, one checks that
\begin{equation*}
\lwb \Pc, g(x) \rwb = g(x) j^L(x) = \p_x g(x) \;\;\;\; \text{ and } \;\;\;\; \lwb \Pc, X(x) \rwb = \p_x X(x).
\end{equation*}

To conclude this subsection, let us say a few words about the Hamiltonian of the system. If we choose a set of coordinate fields $\phi^i$, as we have done above, the Hamiltonian $\Hc$ is related to the Lagragian density $L$ by the Legendre transformation
\begin{equation*}
\Hc = \int \dd x \; \bigl( \pi_i \p_t \phi^i - L \bigr),
\end{equation*}
with $\pi_i$ the conjugate momenta. These momenta are encoded in a coordinate-free way in the current $X$ defined as \eqref{Eq:DefX}. One can then rewrite the Hamiltonian also in a coordinate-free way as
\begin{equation}\label{Eq:HamDensity}
\Hc = \int \dd x \; \Bigl( \kappa \bigl(X, g^{-1} \p_t g \bigr) - L \Bigr).
\end{equation}

\section{Undeformed integrable $\s$-models}
\label{Sec:Undef}

\subsection{The Principal Chiral Model}
\label{SubSec:PCM}

\paragraph{Action and equations of motion.} The Principal Chiral Model (PCM) is the simplest example of an integrable $\s$-model. Its target space is a real Lie group $G_0$, which we assume to be connected and semi-simple, equipped with the Killing metric. It is defined by the action
\begin{equation*}
S_{\text{PCM}}[g] = \frac{K}{2} \int_\Sigma \dd x \, \dd t \; \Bigl( \kappa( g^{-1} \p_t g, g^{-1} \p_t g ) - \kappa( g^{-1} \p_x g, g^{-1} \p_x g ) \Bigr),
\end{equation*}
where $K$ is a global constant factor and $\kappa$ is the Killing form on $\g_0$. This can be reexpressed in terms of the left or right currents $j^{L,R}_\mu$ as
\begin{equation}\label{Eq:PCMCurrent}
S_{\text{PCM}}[g] = \frac{K}{2} \int_\Sigma \dd x \, \dd t \; \Bigl( \kappa( j^L_0, j^L_0 ) - \kappa( j^L_1, j^L_1 ) \Bigr) = \frac{K}{2} \int_\Sigma \dd x \, \dd t \; \Bigl( \kappa( j^R_0, j^R_0 ) - \kappa( j^R_1, j^R_1 ) \Bigr),
\end{equation} 
where the expression with the right currents $j^R_\mu$ follows from equation  \eqref{Eq:LeftToRight} and the invariance of $\kappa$ under conjugacy transformation.

It will be useful to express the action of the PCM using light-cone coordinates $x^\pm$ and the light-cone currents $j^{L,R}_\pm$. One then finds
\begin{equation}\label{Eq:PCMLC}
S_{\text{PCM}}[g] = K \int_\Sigma \dd x^+ \, \dd x^- \; \kappa\bigl( j^L_+, j^L_- \bigr) = K \int_\Sigma \dd x^+ \, \dd x^- \; \kappa\bigl( j^R_+, j^R_- \bigr)
\end{equation}
Varying this action with respect to $g$, one finds that the equations of motion of the PCM can be written as the conservation equations
\begin{equation}\label{Eq:EoMPcm}
\p_t j^{L,R}_0 - \p_x j^{L,R}_1 = \frac{1}{2} \left(\p_+ j^{L,R}_- + \p_- j^{L,R}_+\right) = 0.
\end{equation}
The equation of conservation of $j^L_\mu$ is equivalent to the one of $j^R_\mu$, using the equation \eqref{Eq:LeftToRight}.

\paragraph{Global symmetries.} As we have just observed, the equations of motion of the PCM take the form of the conservation equations of the $\g_0$-valued currents $j^{L,R}_\mu$. Thus, the quantities
\begin{equation}\label{Eq:ChargesMult}
K \int \dd x \; j^L_0(x,t) \;\;\;\; \text{ and } \;\;\;\; K \int \dd x \; j^R_0(x,t)
\end{equation}
are conserved (the global factor $K$ is here for future convenience). By the Noether theorem, these conserved charges are associated with global symmetries of the model.

Let us recall that the right and left currents are respectively invariant under the right multiplication $g\mapsto gh$ and the left multiplication $g\mapsto hg$ of $g$ by a constant element $h$ of $G_0$. As the Lagrangian density of the PCM can be entirely written in terms of either one of these currents, it is clear that these transformations are global symmetries of the PCM. Applying the Noether theorem, one finds that the conserved quantities associated with these symmetries are the ones of equation \eqref{Eq:ChargesMult}. It is worth noticing that the charge associated with the right multiplication is the integral of the left current and vice-versa.

\paragraph{Lax equation.} Let us now investigate the integrability properties of the PCM, following the ideas of Chapter \ref{Chap:Lax}. In particular, let us start by exhibiting the Lax representation of the equations of motion of the PCM, using a general scheme developped by Zakharov and Mikhailov~\cite{Zakharov:1973pp} (some first results were also found by Pohlmeyer in~\cite{Pohlmeyer:1975nb}). The latter relies on the existence of a conserved and flat current (\textit{i.e.} satisfying a conservation equation and a zero curvature equation). In the case of the PCM, we have seen that the equations of motion \eqref{Eq:EoMPcm} take the form of the conservation equations of the currents $j^L_\mu$ and $j^R_\mu$. Moreover, according to the Maurer-Cartan equation \eqref{Eq:MaurerCartan}, these currents are flat. Let us then define the following light-cone Lax pair
\begin{equation}\label{Eq:ZakMik}
\Lc^{\text{PCM}}_\pm (\lambda,x,t) = \frac{j^L_\pm}{1 \mp \lambda},
\end{equation}
depending on the spectral parameter $\lambda\in\C$ (one could also have considered a Lax matrix associated with right currents, but we shall focus here on this particular choice). One then has
\begin{equation*}
\p_+ \Lc^{\text{PCM}}_-(\lambda) - \p_- \Lc^{\text{PCM}}_+(\lambda) + \lsb \Lc^{\text{PCM}}_+(\lambda), \Lc^{\text{PCM}}_-(\lambda) \rsb = \frac{1}{1-\lambda^2} \Bigl( \p_+ j^L_- - \p_- j^L_+ + \lsb j^L_+,j^L_- \rsb - \lambda \left( \p_+ j^L_- + \p_- j^L_+ \right) \Bigr).
\end{equation*}
The constant term in the bracket exactly vanishes according to the (light-cone) Maurer-Cartan equation \eqref{Eq:MaurerCartan}. It is worth noticing that this equation is true off-shell, \textit{i.e.} without using the equation of motion of the model. The above equation then reduces to
\begin{equation*}
\p_+ \Lc^{\text{PCM}}_-(\lambda) - \p_- \Lc^{\text{PCM}}_+(\lambda) + \lsb \Lc^{\text{PCM}}_+(\lambda), \Lc^{\text{PCM}}_-(\lambda) \rsb = \frac{\lambda}{\lambda^2-1} \left( \p_+ j^L_- + \p_- j^L_+ \right).
\end{equation*}
Thus, we see that the equations of motion \eqref{Eq:EoMPcm} of the PCM are equivalent to the (light-cone) Lax equation \eqref{Eq:LaxLC}. This proves the existence of a Lax pair representation of the PCM. Moreover, we see that this procedure for constructing a Lax pair generalises to every model which possesses a flat and conserved current.

Let us end this paragraph by exhibiting the Lax pair $(\Lc_{\text{PCM}},\M_{\text{PCM}})$ corresponding to the space-time coordinates $(x,t)$. Inverting the relation \eqref{Eq:DefLaxLC} and using the expression \eqref{Eq:ZakMik} of $\Lc^{\text{PCM}}_\pm$, one finds
\begin{equation}\label{Eq:LaxPcmLag}
\Lc_{\text{PCM}}(\lambda) = \frac{j^L_1 + \lambda j^L_0}{1-\lambda^2} \;\;\;\; \text{ and } \;\;\;\; \M_{\text{PCM}}(\lambda) = \frac{j^L_0 + \lambda j^L_1}{1-\lambda^2}. 
\end{equation}

\paragraph{Hamiltonian analysis.} We aim to show that the PCM belongs to the class of non-ultralocal models with twist function: thus, we want to compute the Poisson bracket of the Lax matrix with itself. For that, we first need to pass from the Lagrangian to the Hamiltonian formalism. The description of the phase space of the PCM, which is common to all integrable $\s$-models, was presented in the subsection \ref{SubSec:SigmaModelHam}. It is encoded in the $G_0$-valued field $g(x)$ and the $\g_0$-valued current $X(x)$, satisfying the canonical brackets \eqref{Eq:PBTstarG}. Choosing a set of coordinates $\phi^i$ and computing the corresponding conjugate momenta $\pi_i$, we compute the expression of the field $X$, as defined in \eqref{Eq:DefX}. For the PCM, one simply finds
\begin{equation}\label{Eq:XPcm}
X = K g^{-1} \p_t g = K j^L_0.
\end{equation}
As expected, one finds that $X$ is independent of the choice of coordinates $\phi^i$. We note that the right moment map \eqref{Eq:RightMomentMap} can then be reexpressed as
\begin{equation*}
m^R = K \int \dd x \; j^L_0(x),
\end{equation*}
which is equal to the Noether charge \eqref{Eq:ChargesMult} associated with the right multiplication symmetry of the PCM. This is natural, as the moment map $m^R$ generates the infinitesimal right multiplication, as seen in the subsection \ref{SubSec:SigmaModelHam}. In the same way, the left multiplication moment map $m^L$ coincides with the second Noether charge in equation \eqref{Eq:ChargesMult}.\\

Using the equation \eqref{Eq:HamDensity} together with the expression \eqref{Eq:XPcm} of $X$, we express the Hamiltonian of the PCM in terms of $X$ and $j^L=j^L_1$:
\begin{equation*}
\Hc_{\text{PCM}} = \frac{1}{2} \int \dd x \; \left( \frac{1}{K} \kappa\bigl(X,X\bigr) + K \kappa\bigl(j^L,j^L\bigr) \right).
\end{equation*}
As a consistency check, one can compute that
\begin{equation*}
\p_t j^L_0 = \frac{1}{K} \lwb \Hc_{\text{PCM}}, X \rwb = \p_x j^L,
\end{equation*}
so that we recover the Lagrangian equation of motion \eqref{Eq:EoMPcm}.

\paragraph{Maillet bracket and twist function.} Finally, let us compute the Poisson bracket of the Lax matrix of the PCM and show that it is governed by a twist function, as done by Maillet in~\cite{Maillet:1985ec}. We start by re-expressing the Lax matrix \eqref{Eq:LaxPcmLag} in terms of the currents $X$ and $j^L$, whose Poisson bracket we know. One simply finds
\begin{equation}\label{Eq:LaxPcm}
\Lc_{\text{PCM}}(\lambda,x) = \frac{j^L(x) + \lambda K^{-1} X(x)}{1-\lambda^2}.
\end{equation}
We now have all the ingredients to compute the Poisson bracket of the Lax matrix with itself, starting from the above expression and the brackets \eqref{Eq:PBTstarG} and \eqref{Eq:PBj}. After a few computations, involving also the circle lemma \eqref{Eq:CircleLemma}, one finds that the Lax matrix satisfies a non-ultralocal Maillet bracket \eqref{Eq:PBR}. The corresponding $\Rc$-matrix is given by
\begin{equation*}
\Rc^{\text{PCM}}\ti{12}(\lambda,\mu) = \Rc^0\ti{12}(\lambda,\mu)\vp_{\text{PCM}}(\mu)^{-1},
\end{equation*}
where $\Rc^0$ is the non-twisted standard $\Rc$-matrix on $\Lg$ and the twist function is given by
\begin{equation}\label{Eq:TwistPCM}
\vp_{\text{PCM}}(\lambda) = K \left( \frac{1}{\lambda^2} - 1 \right) = K\frac{1-\lambda^2}{\lambda^2}.
\end{equation}
As announced, this proves that the PCM is a non-ultralocal model with twist function. According to the definition of such a model given in Section \ref{Sec:ModelsTwist}, we also have to verify the reality condition \eqref{Eq:Reality}: it is obvious from Equation \eqref{Eq:LaxPcm}, as $j^L$ and $X$ belongs to $\g_0$ and thus are invariant under the anti-involution $\tau$ (note that we do not have to verify the equivariance conditions of Section \ref{Sec:ModelsTwist}, as the standard $\Rc$-matrix is non-twisted).\\

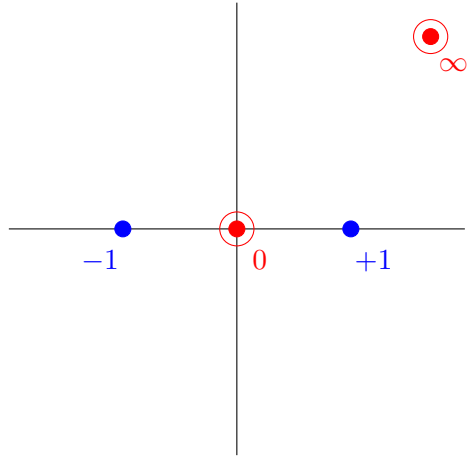
\begin{figure}[H]
\begin{center}
	\begin{tikzpicture}[scale=1.5]
 		\draw (-2,0) to (2,0);
		\draw (0,-2) to (0,2);
		\draw[blue,fill=blue] (1,0) circle (0.07);
		\draw[blue,fill=blue] (-1,0) circle (0.07);	
		\node[blue,below] at (1.2,-0.1) {$+1$}; 
		\node[blue,below] at (-1.2,-0.1) {$-1$}; 
		\draw[red] (1.7,1.7) circle (0.15);
		\draw[red,fill=red] (1.7,1.7) circle (0.07);
		\node[red,below] at (1.9,1.6) {$\infty$};
		\draw[red] (0,0) circle (0.15);
		\draw[red,fill=red] (0,0) circle (0.07);
		\node[red,below] at (0.2,-0.1) {$0$};
 	\end{tikzpicture}
\end{center}\vspace{-10pt}
\caption{{\color{red}Poles} and {\color{blue}{zeros}} of the twist function of the PCM.}
\label{Fig:PolesZerosPcm}
\end{figure}

Although it is not clear yet why we are interested in this, let us study the analytical properties of the twist function $\vp_{\text{PCM}}$. More precisely, let us look at the poles and the zeros of the 1-form $\vp_{\text{PCM}}(\lambda) \dd \lambda$. It is clear that it possesses two simple zeros at $+1$ and $-1$. Moreover, it possesses a double pole at $0$. Under the inversion of parameter $\alpha=\lambda^{-1}$, this 1-form transforms as
\begin{equation*}
\vp_{\text{PCM}}(\lambda)\, \dd \lambda = - \vp_{\text{PCM}}\left(\frac{1}{\alpha}\right) \frac{\dd \alpha}{\alpha^2} = K \left( \frac{1}{\alpha^2} - 1 \right) \dd \alpha.
\end{equation*}
This has a double pole in $\alpha=0$, so the 1-form $\vp_{\text{PCM}}(\lambda) \dd \lambda$ has a double pole at infinity. As a conclusion, the twist function of the PCM has double poles at zero and infinity and simple zeros at $+1$ and $-1$, as represented in Figure \ref{Fig:PolesZerosPcm}. Let us note also that the two zeros $\pm 1$  of the twist function are also poles of the Lax matrix $\Lc(\lambda,x)$ (we shall use this property later).\\

\subsection[$\Z_T$-coset $\s$-models]{$\bm{\Z_T}$-coset $\bm{\s}$-models}
\label{SubSec:ZT}

\paragraph{$\bm{\Z_T}$-grading.} We shall now review the $\Z_T$-coset $\s$-models and their integrable structure. They are $\s$-models with target space $G_0/H$, where $G_0$ is a connected semi-simple Lie group and $H$ is a particular subgroup of $G_0$. More precisely, we suppose that $H$ is such that the corresponding subalgebra $\h$ of $\g_0$ is equal to the $0^{\rm th}$-grade $\g^{(0)}_0$ of a $\Z_T$-grading of $\g_0$
\begin{equation*}
\g_0 = \bigoplus_{p=0}^{T-1} \g_0^{(p)}, \;\;\;\; \text{ with } \;\;\;\; \bigl[ \g^{(p)}_0, \g^{(q)}_0 \bigr] \subset \g^{(p+q \; \text{mod} \; T)}.
\end{equation*}
According to Corollary \ref{Cor:RealGradings}, such gradings are in one-to-one correspondence with automorphisms $\s\in\Aut(\g)$ of order $T$ satisfying the dihedrality condition \eqref{Eq:Dihedral}. We will use the notations introduced in Section \ref{Sec:CYBE} for the construction of the standard $\Rc$-matrix twisted by $\s$ (we refer the reader to the Appendix \ref{App:Torsion} for more details about finite order automorphisms of Lie algebras). In particular, $\omega$ is a primitive $T^{\rm{th}}$-root of unity and $\g^{(p)}$ ($p=0,\cdots,T-1$) is the eigenspace of $\s$ of eigenvalue $\omega^p$. For $X\in\g$, we define $X^{(p)} = \pi^{(p)} X$ the projection of $X$ on $\g^{(p)}$ in the decomposition $\bigoplus_{p=0}^{T-1} \g^{(p)}$.

For simplicity, we shall only detail here the case $T=2$ and just summarise at the end of the subsection how the results on the integrability of the $\Z_2$-coset model generalises to $\Z_T$-cosets for arbitrary $T$. When $T=2$, $\s$ is an involution (\textit{i.e.} $\s^2=\Id$), $\omega=-1$ and the eigenspaces
\begin{equation*}
\g^{(0)} = \lwb X \in\g \; \bigl| \; \s(X)=X \rwb \;\;\;\; \text{ and } \;\;\;\; \g^{(1)} = \lwb X \in\g \; \bigl| \; \s(X)=-X \rwb
\end{equation*}
satisfy
\begin{equation}\label{Eq:GradingZ2}
\bigl[ \g^{(0)}, \g^{(0)} \bigr] \subset \g^{(0)}, \;\;\; \bigl[ \g^{(0)}, \g^{(1)} \bigr] \subset \g^{(1)} \;\;\; \text{and} \;\;\; \bigl[ \g^{(1)}, \g^{(1)} \bigr] \subset \g^{(0)}.
\end{equation}
In this case, the Lie algebra $\h$ of the subgroup $H$ is identified with the subalgebra $\g^{(0)} \cap \g_0$ of elements of the real form $\g_0$ stabilised by $\s$. In this case $T=2$, the coset space $G_0/H$ is called a symmetric-space.

\paragraph{Action and equations of motion.} As explained briefly in Subsection \ref{SubSec:FieldSigma}, we will describe the $\s$-model on $G_0/H$ as a model on a $G_0$-valued field $g$ together with a gauge symmetry in $H$ to eliminate the redundant degrees of freedom. More explicitly, the symmetric-space $\s$-model on $G_0/H$ is given by the action
\begin{equation*}
S_{\Z_2}[g] = \frac{K}{2} \int_\Sigma \dd x \, \dd t \; \left( \kappa \Bigl( \bigl( g^{-1} \p_t g \bigr)^{(1)}, \bigl( g^{-1} \p_t g \bigr)^{(1)} \Bigr) -  \kappa \Bigl( \bigl( g^{-1} \p_x g \bigr)^{(1)}, \bigl( g^{-1} \p_x g \bigr)^{(1)} \Bigr) \right).
\end{equation*}
One can reexpress this action in terms of left currents $j^L_\mu$ as
\begin{equation*}
S_{\Z_2}[g] = \frac{K}{2} \int_\Sigma \dd x \, \dd t \; \left( \kappa \Bigl( j_0^{L\,(1)}, j_0^{L\,(1)} \Bigr) -  \kappa \Bigl( j_1^{L\,(1)}, j_1^{L\,(1)} \Bigr) \right).
\end{equation*}
In the same way, one can also use the light-cone left currents $j^L_\pm$ and the light-cone coordinates $x^\pm$:
\begin{equation}\label{Eq:AcionZTLC}
S_{\Z_2}[g] = K \int_\Sigma \dd x^+ \, \dd x^- \; \kappa \Bigl( j_+^{L\,(1)}, j_-^{L\,(1)} \Bigr).
\end{equation}
Let us consider the local right multiplication
\begin{equation}\label{Eq:GaugeZT}
g(x,t) \longmapsto g(x,t) h(x,t), \;\;\; h(x,t) \in H,
\end{equation}
by a field $h$ valued in $H=G^{(0)}\cap G_0$. Under this transformation, the left-current transforms as
\begin{equation*}
j^L_\pm \longmapsto  h^{-1} j^L_\pm h + h^{-1} \p_\pm h.
\end{equation*} 
By construction, $h$ is valued in $G^{(0)}$ so $h^{-1} \p_\pm h$ belongs to $\g^{(0)}$. Moreover, according to the grading relations \eqref{Eq:GradingZ2}, the conjugacy transformation by $h \in G^{(0)}$ preserves the graded subspaces $\g^{(0)}$ and $\g^{(1)}$. Thus, we deduce that the graded components of $j^L_\pm$ transform simply as
\begin{equation}\label{Eq:CurrentGauge}
j^{L\,(0)}_\pm \longmapsto h^{-1} j^{L\,(0)}_\pm h + h^{-1} \p_\pm h \;\;\;\; \text{ and } \;\;\;\; j^{L\,(1)}_\pm \longmapsto h^{-1} j^{L\,(1)}_\pm h.
\end{equation}
In particular, as the Killing form $\kappa$ is invariant under conjugacy transformations, we see that the action \eqref{Eq:AcionZTLC} is invariant under the local transformation \eqref{Eq:GaugeZT}. Thus, the model possesses a gauge symmetry under the right multiplication by elements of $H$. The physical degrees of freedom of the model are then in the quotient $G_0/H$. One can then recover the usual $\s$-model on the symmetric-space $G_0/H$ by gauge fixing the theory.\\

Let us end this paragraph by expressing the equations of motion of the model, obtained by varying the field $g$ in the action \eqref{Eq:AcionZTLC}. They read
\begin{equation}\label{Eq:EomZ2}
D_+ j^{L\,(1)}_- + D_- j^{L\,(1)}_+ = 0,
\end{equation}
where $D_\pm$ denotes the covariant derivative
\begin{equation*}
D_\pm = \p_\pm + \bigl[ j^{L\,(0)}_\pm, \cdot \bigr].
\end{equation*}
As a consistency check, we can verify that this equation of motion is invariant under the gauge transformation \eqref{Eq:GaugeZT}. According to equation \eqref{Eq:CurrentGauge}, the currents $j^{L\,(1)}_\pm$ transform covariantly under this transformation and the currents $j^{L\,(0)}_\pm$ transform as gauge fields (which justifies the name of covariant derivative for $D_\pm$). Thus, the left-hand side of the equation of motion \eqref{Eq:EomZ2} is covariant under the gauge transformation, hence the gauge symmetry at the level of the equations of motion.

\paragraph{Global left symmetry.} In addition to the right gauge symmetry described above, the $\Z_2$-coset model possesses a global left symmetry. Indeed, the action \eqref{Eq:AcionZTLC} is expressed only in terms of the left-invariant currents $j^L_\pm$. Thus, the action is invariant under the left multiplication $L_h : g \mapsto hg$ by a constant element $h$ of the group $G_0$. By the Noether theorem, this global symmetry is associated with the following equation of conservation:
\begin{equation*}
\p_+ \Bigl( g j^{L\,(1)}_- g^{-1} \Bigr)  + \p_- \Bigl( g j^{L\,(1)}_+ g^{-1} \Bigr) = 0,
\end{equation*}
which is equivalent to the equation of motion \eqref{Eq:EomZ2}. Thus, the charge
\begin{equation*}
\int \dd x \; g(x,t) j^{L\;(1)}_0(x,t) g^{-1}(x,t)
\end{equation*}
is conserved, \textit{i.e.} does not depend on $t$.

\paragraph{Lax equation.} Let us show that the equations of motion of the $\Z_2$-coset $\s$-model can be recast as a Lax equation \eqref{Eq:LaxLC} (in light-cone coordinates), as proved originally by Eichenherr and Forger in~\cite{Eichenherr:1979ci} (first results on particular examples were found by Pohlmeyer in~\cite{Pohlmeyer:1975nb}). As in the case of the PCM (subsection \ref{SubSec:PCM}), we will need the flatness of the left-current $j^L_\pm$, \textit{i.e.} the Maurer-Cartan equation \eqref{Eq:MaurerCartanLC}. As the $\Z_2$-coset model is expressed in terms of the graded components $j^{L\,(0)}_\pm$ and $j^{L\,(1)}_\pm$, let us decompose the Maurer-Cartan equation along this grading. Using the grading relation \eqref{Eq:GradingZ2}, one finds the two following equations:
\begin{subequations}\label{Eq:MCZ2}
\begin{align}
\p_+ j^{L\,(0)}_- - \p_- j^{L\,(0)}_+ + \Bigl[ j^{L\,(0)}_+, j^{L\,(0)}_- \Bigr] + \Bigl[ j^{L\,(1)}_+, j^{L\,(1)}_- \Bigr] &= 0, \\
\p_+ j^{L\,(1)}_- - \p_- j^{L\,(1)}_+ + \Bigl[ j^{L\,(0)}_+, j^{L\,(1)}_- \Bigr] + \Bigl[ j^{L\,(1)}_+, j^{L\,(0)}_- \Bigr] &= 0.
\end{align}
\end{subequations}
We define the light-cone Lax pair
\begin{equation}\label{Eq:LaxZ2LC}
\Lc^{\Z_2}_\pm(\lambda) = j^{L\,(0)}_\pm + \lambda^{\pm 1} j^{L\,(1)}_\pm, 
\end{equation}
depending on the spectral parameter $\lambda\in\C$. One has
\begin{align}\label{Eq:ZCEZ2}
&\hspace{-40pt}\p_+ \Lc^{\Z_2}_-(\lambda) - \p_- \Lc^{\Z_2}_+(\lambda) + \bigl[ \Lc^{\Z_2}_+(\lambda), \Lc^{\Z_2}_-(\lambda) \bigr] \\
& = \;\;\; \p_+ j^{L\,(0)}_- - \p_- j^{L\,(0)}_+ + \Bigl[ j^{L\,(0)}_+, j^{L\,(0)}_- \Bigr] + \Bigl[ j^{L\,(1)}_+, j^{L\,(1)}_- \Bigr] \notag \\
& \hspace{35pt} + \frac{1}{2} \left( \lambda + \frac{1}{\lambda} \right) \Bigl( \p_+ j^{L\,(1)}_- - \p_- j^{L\,(1)}_+ + \Bigl[ j^{L\,(0)}_+, j^{L\,(1)}_- \Bigr] + \Bigl[ j^{L\,(1)}_+, j^{L\,(0)}_- \Bigr] \Bigr) \notag \\
& \hspace{35pt} + \frac{1}{2} \left( \lambda - \frac{1}{\lambda} \right) \Bigl( D_+ j^{L\,(1)}_- + D_- j^{L\,(1)}_+ \Bigr). \notag
\end{align}
The first two lines vanish off-shell (without the equations of motion) due to the Maurer-Cartan equations \eqref{Eq:MCZ2}. Thus, we get
\begin{equation*}
\p_+ \Lc^{\Z_2}_-(\lambda) - \p_- \Lc^{\Z_2}_+(\lambda) + \bigl[ \Lc^{\Z_2}_+(\lambda), \Lc^{\Z_2}_-(\lambda) \bigr] = \frac{1}{2} \left( \lambda - \frac{1}{\lambda} \right) \Bigl( D_+ j^{L\,(1)}_- + D_- j^{L\,(1)}_+ \Bigr).
\end{equation*}
The equation of motion \eqref{Eq:EomZ2} of the $\Z_2$-coset is then equivalent to the Lax equation \eqref{Eq:LaxLC}, as expected. To end this paragraph, let us give the corresponding Lax pair $(\Lc_{\Z_2},\M_{\Z_2})$ in space-time coordinates:
\begin{subequations}
\begin{align}
\Lc_{\Z_2}(\lambda) &= j^{L\,(0)}_1 + \frac{1}{2} \left(\lambda-\frac{1}{\lambda}\right) j^{L\,(1)}_0 + \frac{1}{2} \left(\lambda+\frac{1}{\lambda}\right) j^{L\,(1)}_1 \label{Eq:LaxZ2Lag} \\
\M_{\Z_2}(\lambda) &= j^{L\,(0)}_0 + \frac{1}{2} \left(\lambda+\frac{1}{\lambda}\right) j^{L\,(1)}_0 + \frac{1}{2} \left(\lambda-\frac{1}{\lambda}\right) j^{L\,(1)}_1.
\end{align}
\end{subequations}
Using the fact that $\s\left( j^{L\,(0)}_\mu \right) = j^{L\,(0)}_\mu$ and $\s\left( j^{L\,(1)}_\mu \right) = -j^{L\,(1)}_\mu$, we can note that
\begin{equation*}
\s \bigl( \Lc_{\Z_2}(\lambda) \bigr) = \Lc_{\Z_2}(-\lambda) \;\;\;\; \text{ and } \;\;\;\; \s \bigl( \M_{\Z_2}(\lambda) \bigr) = \M_{\Z_2}(-\lambda).
\end{equation*}
Thus, the Lax matrix $\Lc_{\Z_2}(\lambda)$ satisfies the equivariance condition \eqref{Eq:EquiL} (as $\omega=-1$ for $T=2$).

\paragraph{Hamiltonian analysis.} Let us discuss the Hamiltonian formulation of the $\Z_2$-coset model. The phase space of the model is described in Subsection \ref{SubSec:SigmaModelHam}. It contains the $G_0$-valued field $g(x)$ and the $\g_0$-valued field $X(x)$. Choosing coordinate fields $\phi^i$ on $G_0$, one can compute the corresponding conjugate momenta $\pi_i$ and deduce the Lagrangian expression \eqref{Eq:DefX} of $X$. One finds
\begin{equation}\label{Eq:XZ2}
X \approx K \bigl( g^{-1} \p_t g \bigr)^{(1)} \approx K \, j^{L\,(1)}_0,
\end{equation}
where the notation $\approx$ instead of $=$ will be justified in what follows. A direct consequence of this expression is that
\begin{equation*}
X^{(0)} \approx 0.
\end{equation*}
This can seem surprising at first, as $X^{(0)}=0$ is for example incompatible with the Poisson bracket \eqref{Eq:PBXg}. This is a consequence of the gauge symmetry of the $\Z_2$-coset model. Indeed, according to Dirac theory~\cite{dirac1964lectures}, when passing from a Lagrangian system with gauge symmetry to the Hamiltonian formalism, one encounters constraints on the phase space. These are relations between the coordinates $\phi^i$ and the conjugate momentum $\pi_i$. In the case of the $\Z_2$-coset model, these relations are encoded in the fact that $X^{(0)}$ vanishes.

We shall use here the Dirac terminology and say that an equation is weak when it is true only with the constraint (we then use the notation $\approx$ instead of $=$). At the contrary, we will say that an equation is strong when it is true without imposing the constraint (we shall then keep the usual equality sign $=$ for those). The full treatment of the $\Z_2$-coset model as a constrained Hamiltonian system would require the application of the Dirac procedure. We shall not enter into too much details here and will just refer to the textbooks~\cite{dirac1964lectures,Henneaux:1992ig} when we have to use results on constrained systems.

Let us finish this discussion about the constraint by a quick consistency check. For constraints coming from a gauge symmetry, it is a standard result that the Hamiltonian flow of this constraint generates the corresponding gauge transformation. Let us check this fact here. For $\epsilon(x)$ an infinitesimal field valued in $\h=\g^{(0)}\cap\g_0$, one can check from the Poisson bracket \eqref{Eq:PBXg} that
\begin{equation}\label{Eq:GaugeHam}
\delta_\epsilon g(x) = \lwb \int \dd y \; \kappa\Bigl(X^{(0)}(y),\epsilon(y)\Bigr), g(x) \rwb = g(x) \epsilon(x).
\end{equation}
This is the infinitesimal version of the gauge transformation
\begin{equation*}
g(x) \mapsto g(x)h(x), \;\;\; h(x) \in H.
\end{equation*}
Thus, we recover that the constraint $X^{(0)}$ generates the gauge symmetry of the model, as expected.\\

Let us discuss the Hamiltonian of the model, using the formula \eqref{Eq:HamDensity} obtained in the general discussion. As we are considering a constrained system, we have the freedom of adding to the Hamiltonian of equation \eqref{Eq:HamDensity} any term proportional to the constraint. Doing the explicit computation for the case of the $\Z_2$-model, one finds
\begin{equation}\label{Eq:HamZ2}
\Hc_{\Z_2} = \frac{1}{2} \int \dd x \; \left( \frac{1}{K} \kappa\bigl(X,X\bigr) - K \kappa\bigl(j^{L\,(1)},j^{L\,(1)}\bigr) + \kappa \bigl( X^{(0)}(x), \mu(x) \bigr) \right),
\end{equation}
where $j^L=j^L_1$ and $\mu$ is an arbitrary $\h$-valued field, called the Lagrange multiplier. According to equation \eqref{Eq:GaugeHam}, the Hamiltonian flow generated by the term containing $\mu$ is purely a gauge term and thus does not change the physics of the model.

\paragraph{Maillet bracket and twist function.} Let us now determine the Poisson bracket of the Lax matrix with itself and show that the $\Z_2$-coset model possesses a twist function~\cite{Delduc:2012qb,Sevostyanov:1995hd}. We first need to reexpress the Lax matrix \eqref{Eq:LaxZ2Lag} in terms of the Hamiltonian fields $g$ and $X$. Moreover, as $X^{(0)}$ is a constraint, one has the freedom to add to the Lax matrix a term proportional to $X^{(0)}$. Using the expression \eqref{Eq:XZ2} of $X$, one then gets
\begin{equation}\label{Eq:LaxZ2Ham}
\Lc_{\Z_2}(\lambda) = j^{L\,(0)} + \frac{1}{2} \left(\lambda+\frac{1}{\lambda}\right) j^{L\,(1)} + \frac{1}{2K} \left(\lambda-\frac{1}{\lambda}\right) X^{(1)} + f(\lambda) X^{(0)},
\end{equation}
where $f(\lambda)$ is an arbitrary function of the spectral parameter. Starting from the Poisson brackets \eqref{Eq:PBTstarG} and \eqref{Eq:PBj}, we compute the bracket of the Lax matrix \eqref{Eq:LaxZ2Ham} with itself. We find that it takes the form of a non-ultralocal Maillet bracket, strongly (\textit{i.e.} without using the constraint $X^{(0)}\approx 0$), if we choose the function $f$ to be
\begin{equation}\label{Eq:fCont}
f(\lambda) = \frac{\lambda^2-1}{2K}.
\end{equation}
In this case, the $\Rc$-matrix governing the Maillet bracket takes the form
\begin{equation*}
\Rc^{\Z_2}\ti{12}(\lambda,\mu) = \Rc^0\ti{12}(\lambda,\mu) \vp_{\Z_2}(\mu)^{-1}.
\end{equation*}
The matrix $\Rc^0$ is given by
\begin{equation*}
\Rc^0\ti{12}(\lambda,\mu) = \frac{\mu C\ti{12}^{(0)}}{\mu^2-\lambda^2} + \frac{\lambda C\ti{12}^{(1)}}{\mu^2-\lambda^2},
\end{equation*}
with $C\ti{12}^{(p)}=\pi^{(p)}\ti{1} C\ti{12}$ the partial quadratic Casimirs. We recognize here the twisted standard $\Rc$-matrix \eqref{Eq:RCas}, for $T=2$. The twist function takes the form
\begin{equation}\label{Eq:TwistZ2}
\vp_{\Z_2}(\lambda) = \frac{2K \lambda}{(\lambda^2-1)^2}.
\end{equation}
With the function $f$ chosen as in \eqref{Eq:fCont}, the Lax matrix then becomes
\begin{equation}\label{Eq:LaxZ2}
\Lc_{\Z_2}(\lambda) = j^{L\,(0)} + \frac{1}{2} \left(\lambda+\frac{1}{\lambda}\right) j^{L\,(1)} + \frac{1}{2K} \left(\lambda^2-1\right) X^{(0)} + \frac{1}{2K} \left(\lambda-\frac{1}{\lambda}\right) X^{(1)}.
\end{equation}
This Lax matrix satisfies the equivariance condition \eqref{Eq:EquiL} and the reality condition \eqref{Eq:Reality}. Moreover, the twist function also satisfies the equivariance property \eqref{Eq:TwistEqui} and the reality condition \eqref{Eq:TwistReal}, as expected from the consistency of the Maillet bracket. This proves that the $\Z_2$-coset $\s$-model belongs to the class of non-ultralocal models with twist function. Following the nomenclature of Section \ref{Sec:ModelsTwist}, we say that it is a cyclotomic model of order $T=2$, as it is associated with the standard $\Rc$-matrix on $\Lg$ twisted by the involution $\s$. \\

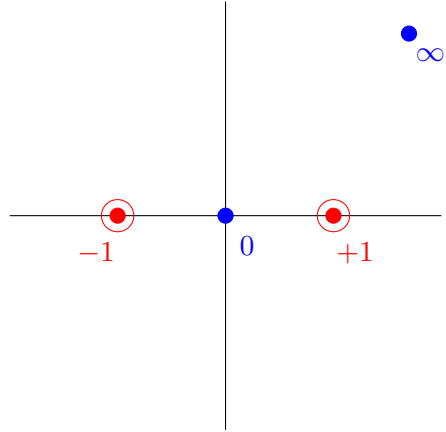
\begin{figure}[H]
\begin{center}
	\begin{tikzpicture}[scale=1.42]
 		\draw (-2,0) to (2,0);
		\draw (0,-2) to (0,2);
		\draw[red] (1,0) circle (0.15);
		\draw[red,fill=red] (1,0) circle (0.07);
		\draw[red] (-1,0) circle (0.15);
		\draw[red,fill=red] (-1,0) circle (0.07);	
		\node[red,below] at (1.2,-0.15) {$+1$}; 
		\node[red,below] at (-1.2,-0.15) {$-1$}; 
		\draw[blue,fill=blue] (1.7,1.7) circle (0.07);
		\node[red,blue] at (1.9,1.5) {$\infty$};
		\draw[blue,fill=blue] (0,0) circle (0.07);
		\node[blue,below] at (0.2,-0.1) {$0$};
 	\end{tikzpicture}
\end{center}\vspace{-10pt}
\caption{{\color{red}Poles} and {\color{blue}{zeros}} of the twist function of the $\Z_2$-coset $\s$-model.}\vspace{-4pt}
\label{Fig:PolesZerosZ2}
\end{figure}

As for the PCM, let us study the analytical properties of the twist function and more precisely, the poles and zeros of the 1-form $\vp_{\Z_2}(\lambda)\dd \lambda$. One sees that it has two double poles at $+1$ and $-1$ and two simple zeros at $0$ and infinity. These poles and zeros are shown in figure \ref{Fig:PolesZerosZ2}. \\

\paragraph{$\bm{\Z_T}$-cosets.} We close this subsection by discussing briefly general $\Z_T$-coset $\s$-models, for arbitrary $T$ (we will use the notations introduced at the beginning of this subsection, with $\s$ an automorphism of order $T$). These models were introduced by Young in~\cite{Young:2005jv}, together with their Lax pair representation (see also~\cite{Beisert:2012ue} and~\cite{Bykov:2017vsm}). The Hamiltonian analysis of these models and the computation of the Poisson bracket of their Lax matrix was done by Ke, Li, Wang and Yue in \cite{Ke:2011zzb}. In particular, they showed that the $\Z_T$-coset models are non-ultralocal models with twist function, cyclotomic of order $T$. The Lax matrix of these models reads
\begin{equation}\label{Eq:LaxZT}
\Lc_{\Z_T}(\lambda,x) = \sum_{k=1}^{T} \frac{(T-k) + k\lambda^{-T}}{T}\lambda^k j^{L\,(k)}(x)  + \sum_{k=1}^{T} \frac{1-\lambda^{-T}}{T} \lambda^k X^{(k)}(x).
\end{equation}
Note that in this equation, and in general, we consider the exponents $(k)$ only modulo $T$, so that $X^{(T)}=X^{(0)}$ for example. All $\Z_T$-coset models possess a gauge symmetry under the action of the subgroup $H=G^{(0)}\cap G_0$ of $G_0$. As in the case of the $\Z_2$-coset presented above, the field $X^{(0)}$ is the constraint associated with this gauge symmetry. It is clear that equation \eqref{Eq:LaxZT} reduces to the Lax matrix \eqref{Eq:LaxZ2} of the $\Z_2$-coset model for $T=2$ (and for the choice of the global factor $K=1$). Note that the Lax matrix \eqref{Eq:LaxZT} satisfies the equivariance condition \eqref{Eq:EquiL} for the automorphism $\s$ of order $T$ introduced above.

The Poisson bracket of the Lax matrix \eqref{Eq:LaxZT} takes the form 
of a Maillet non-ultralocal bracket \eqref{Eq:PBR}. The associated $\Rc$-matrix is given by a twist function, as in \eqref{Eq:DefR}. More precisely, the matrix $\Rc^0$ is the standard $\Rc$-matrix on $\Lc(\g)$ twisted by the automorphism $\s$ of order $T$ and the twist function is given by
\begin{equation}\label{Eq:TwistZT}
\vp_{\Z_T}(\lambda) = \frac{T \lambda^{T-1}}{(1-\lambda^T)^2}.
\end{equation}
As for the Lax matrix, the twist function \eqref{Eq:TwistZT} reduces to the one \eqref{Eq:TwistZ2} of the $\Z_2$-coset model when $T=2$ (and $K=1$). As expected, this twist function satisfies the equivariance condition \eqref{Eq:TwistEqui}. The 1-form $\vp_{\Z_T}(\lambda) \dd \lambda$ has two zeros in $0$ and infinity and $T$ double poles, located at each $T^{\rm{th}}$-root of the unity $\omega^p$. \\

Let us say a few words about supercoset models. Instead of a Lie algebra, one can consider a super-Lie algebra, equipped with a $\Z_{2T}$-grading (for a super-Lie algebra, we only consider even gradings, as the definition of a super-Lie algebra already contains a $\Z_2$-grading). The construction mentioned above for $\Z_T$-coset then generalises and one can write a $\Z_{2T}$-supercoset $\s$-model (this was also considered by Young in~\cite{Young:2005jv}). The Hamiltonian integrable structure of these models (Lax matrix, Maillet bracket, twist function) is the same as the one of the $\Z_{2T}$-coset models described above, when replacing Lie algebras by super-Lie algebras (this was also shown in~\cite{Ke:2011zzb}).

A slightly different construction allows to consider the Green-Schwarz superstring on $AdS_5 \times S^5$: in particular, one has to take into account the worldsheet diffeomorphism gauge symmetry of the model, as well as the $\kappa$-symmetry (the latter comes from the fact that the Green-Schwarz model, contrarily to the supercoset model mentioned above, has no kinetic terms involving fermions). The action of this model was worked out by Metsaev and Tseytlin in~\cite{Metsaev:1998it}. Its interpretation as a $\Z_4$-supercoset model may be seen from~\cite{Berkovits:1999zq} (see also the review~\cite{Arutyunov:2009ga}). This model was extensively studied in the context of the AdS/CFT correspondence, as its holographic dual is the $\mathcal{N}=4$ super-Yang-Mills theory in four dimensions (see for example~\cite{Beisert:2010jr}). It was shown to admit a Lax pair by Bena, Polchinski and Roiban in~\cite{Bena:2003wd}. Its hamiltonian integrability and its Maillet structure were discovered by Magro in~\cite{Magro:2008dv}. The underlying algebraic structure was understood in~\cite{Vicedo:2009sn,Vicedo:2010qd} and the twist function determined in~\cite{Vicedo:2010qd}. This twist function coincides with the one of a $\Z_4$-coset model, given by \eqref{Eq:TwistZT} with $T=4$.

\section[Deformed integrable $\s$-models]{Deformed integrable $\bm{\s}$-models}
\label{Sec:DefModel}

In the past decades, several integrable deformations of the PCM and of the $\Z_T$-coset $\s$-models were discovered. These are models which depend on one or more deformation parameters, such that they reduce to the undeformed model when these parameters are equal to zero and such that they still admit a Lax pair formulation. Moreover, for all models for which the hamiltonian analysis of the Lax matrix has been carried out, it was found that they also belong to the class of models with twist function, as the undeformed one. Interestingly, as we shall see, the effect of the deformations is to deform the poles of the twist function.

We will not detail the construction and the analysis of all these models here. In the first subsection, we will develop the first known example of such a deformed model, the so-called Yang-Baxter model, which is a one-parameter deformation of the PCM that we will often use as an example in this thesis. In the second subsection, we will review briefly the whole panorama of deformations of the PCM, focusing in particular on their twist functions. Finally, in the third section, we shall discuss the deformations of $\Z_2$-coset models.

\subsection{The Yang-Baxter model}
\label{SubSec:YB}

\paragraph{Action.} The Yang-Baxter model is a one-parameter integrable deformation of the PCM, introduced by Klim\v{c}ik in~\cite{Klimcik:2002zj}. We will use the notation of the subsection \ref{SubSec:PCM} on the PCM. We will suppose moreover that we are given a linear map $R:\g\rightarrow\g$, stabilising the real form $\g_0$ and skew-symmetric with respect to the Killing form $\kappa$, which satisfies the so-called modified Classical Yang-Baxter Equation (mCYBE):
\begin{equation}\label{Eq:mCYBE}
\forall \, X,Y \in \g, \; \; [RX,RY]-R\bigl([RX,Y]+[X,RY]\bigr) = -c^2 [X,Y],
\end{equation}
with $c=1$ (split case) or $c=i$ (non-split case). Although this equation is called the modified CYBE, it should not be confused with the CYBE \eqref{Eq:CYBE} (note in particular that there is no dependence on some spectral parameter in the above equation). The link between these equations and their algebraic interpretation are explained in Appendix \ref{App:RMat}. In most examples and applications, we will consider the so-called standard (split or non-split) matrix $R$, as described in Appendix \ref{App:RMat}, although the results of the present subsection hold for any solution of the mCYBE.

The Yang-Baxter model is then defined by the following action, depending on the $G_0$-valued field $g$, a real parameter $\eta\in\R$ and a global factor $K$:
\begin{equation}\label{Eq:ActionYBR}
S_\eta[g] = K \int_\Sigma \dd x^+ \, \dd x^- \; \kappa\lrb j^R_+, \frac{1}{1-\eta R} j^R_- \rrb =  K \int_\Sigma \dd x^+ \, \dd x^- \; \kappa\lrb j^R_-, \frac{1}{1+\eta R} j^R_+ \rrb,
\end{equation}
where the second equality is obtained using the skew-symmetry of $R$ with respect to $\kappa$. Note here that we use the notation $\eta$ for the deformation parameter, following the conventions of~\cite{Delduc:2013fga}, although this parameter is often denoted $\varkappa$ in the literature. Comparing this equation to equation \eqref{Eq:PCMLC}, it is clear that
\begin{equation*}
S_{\eta=0}[g] = S_{\text{PCM}}[g].
\end{equation*}
Thus, this action defines a one-parameter deformation of the PCM.

Let us discuss briefly the invertibility of the operators $1 \pm \eta R$, which is necessary for the good definition of the action \eqref{Eq:ActionYBR}. This invertibility is always ensured for $\eta$ close to $0$. In the case of a standard matrix $R$, the eigenvalue of $R$ are $0$ and $\pm 1$ for the split case and $0$ and $\pm i$ for the non-split one. Thus, $1\pm \eta R$ is invertible for any value of $\eta\in\R$ in the non-split case and invertible for $\eta \neq \pm 1$ for the split case.

The above action is expressed in terms of the right currents $j^R_\pm$. Using the invariance of $\kappa$ under conjugacy transformation and the relation \eqref{Eq:LeftToRight} between the right currents and the left ones, one can re-express this action as
\begin{equation}\label{Eq:ActionYBL}
S_\eta[g] = K \int_\Sigma \dd x^+ \, \dd x^- \; \kappa\lrb j^L_+, \frac{1}{1-\eta R_g} j^L_- \rrb =  K \int_\Sigma \dd x^+ \, \dd x^- \; \kappa\lrb j^L_-, \frac{1}{1+\eta R_g} j^L_+ \rrb,
\end{equation}
where
\begin{equation*}
R_g = \Ad^{-1}_g \circ R \circ \Ad_g.
\end{equation*}
Later in this subsection, when discussing the Hamiltonian formalism of the model, we will need the expression of the action in terms of space-time coordinates $(x,t)$ instead of light-cone ones. This is given by
\begin{align}\label{Eq:ActionYBtx}
S_\eta[g] & = \frac{K}{2} \int_\Sigma \dd x \, \dd t \left[ \kappa \left( g^{-1} \p_t g, \frac{1}{1-\eta^2 R^2_g} g^{-1} \p_t g \right) - \kappa \left( g^{-1} \p_x g, \frac{1}{1-\eta^2 R^2_g} g^{-1} \p_x g \right) \right. \\
 &\hspace{230pt}- \left. \kappa \left( g^{-1} \p_t g, \frac{2 \eta R_g}{1-\eta^2 R^2_g} g^{-1} \p_x g \right) \right]. \notag
\end{align}

\paragraph{(Preserved and broken) Symmetries.} As already observed, the Yang-Baxter action \eqref{Eq:ActionYBR} is expressed only in terms of the right currents $j^R_\pm$. As these currents are invariant under the right multiplication $g \mapsto gh$ by a constant element of $G_0$, it is clear that the Yang-Baxter model is invariant under this transformation, as the PCM is (we will come back on this fact later).

As explained in Subsection \ref{SubSec:PCM}, the PCM is also invariant under the left multiplication $g \mapsto hg$. This left symmetry is broken by the Yang-Baxter deformation. Indeed, under the transformation $g \mapsto hg$, the action \eqref{Eq:ActionYBR} transforms as
\begin{equation*}
S_\eta[hg] = K \int_\Sigma \dd x^+ \, \dd x^- \; \kappa\lrb j^R_+, \Ad_h^{-1} \circ \frac{1}{1-\eta R} \circ \Ad_h \; j^R_- \rrb.
\end{equation*}
As $R$ does not commute with $\Ad_h$, this is different from $S_\eta[g]$ when $\eta \neq 0$. We shall discuss again this breaking of symmetry in Chapter \ref{Chap:PLie}.

Let us now come back to the right symmetry, which is not broken by the Yang-Baxter deformation. By the Noether theorem, it is associated with the conservation equation
\begin{equation}\label{Eq:ConservationYB}
\p_+ K_- + \p_- K_+ = 0
\end{equation}
of a $\g_0$-valued current $K_\pm$. One finds that
\begin{equation}\label{Eq:CurrentYB}
K_\pm = \frac{1-c^2\eta^2}{1 \pm \eta R_g} j^L_\pm = \Oc_\pm j^L_\pm.
\end{equation}
This conservation equation is one way to write the whole set of equations of motion of the Yang-Baxter model. A conserved current is only defined up to a global constant factor: here, we introduced the factor $1-c^2 \eta^2$ for future convenience. We note that in the non-split case $c=i$, this factor is equal to $1+\eta^2$ and thus is non-zero for any value of $\eta$. In the split case $c=1$, we see that this factor is not zero, except for the values $\eta = \pm 1$: as discussed above, these correspond also to the values where the the operator $1 \pm \eta R$ is non-invertible, so we shall exclude them for the present discussion. One easily checks that the current $K_\pm$ reduces to the left-current $j^L_\pm$ in the undeformed case $\eta=0$.\\

As we have just seen, the Yang-Baxter deformation of the PCM preserves the right symmetry of the model but breaks the left symmetry. One could have also introduced a Yang-Baxter deformation which preserves the left symmetry and breaks the right one, by simply replacing the right current $j^R_\pm$ by the left current $j^L_\pm$ in the action \eqref{Eq:ActionYBR}. This model is equivalent to the one studied here and all the computations we present also apply to it by exchanging ``left and right objects''. Thus, we shall focus only on the model defined by \eqref{Eq:ActionYBR}.

\paragraph{Lax equation.} Let us discuss the Lax representation of the equations of motion of the Yang-Baxter model, as first introduced by Klim\v{c}ik in~\cite{Klimcik:2008eq}. Recall that in the undeformed (PCM) case, we constructed a Lax pair following the scheme of Zakharov and Mikhailov, based on the existence of a conserved and flat current. In the case of the Yang-Baxter model, we already found a conserved current $K_\pm$, associated with the global right symmetry. It is thus natural to wonder whether $K_\pm$ is flat, \textit{i.e.} whether we have or not
\begin{equation}\label{Eq:FlatK}
\p_+ K_- - \p_- K_+ + \lsb K_+, K_- \rsb = 0.
\end{equation}
In the undeformed case, this flatness condition reduces to the Maurer-Cartan equation \eqref{Eq:MaurerCartanLC} of $j^L_\pm$. Thus, we should expect to also use the Maurer-Cartan equation in the deformed case. To compute the derivatives $\p_\pm K_\mp$, one needs to take into account the fact that the operators $\Oc_\pm$ in the definition \eqref{Eq:CurrentYB} of $K_\pm$ also depend on the field $g$. In general, for a $g$-dependent field $Y$ and an infinitesimal variation $\delta g$ of $g$, one finds that the induced variation of $\Oc_\pm Y$ is
\begin{equation*}
\delta \bigl( \Oc_\pm Y \bigr) = \Oc_\pm \bigl( \delta Y \bigr) + \Oc_\pm \Bigl( \left[g^{-1}\delta g, Y \right] \Bigr) - \left[ g^{-1}\delta g, \Oc_\pm Y \right].
\end{equation*}
This allows to compute the derivatives $\p_+ K_-$ and $\p_- K_+$. Combining these with the Maurer-Cartan equation of $j^L_\pm$ and the conservation equation \eqref{Eq:ConservationYB} of $K_\pm$, one checks that the flatness equation \eqref{Eq:FlatK} indeed holds. In the derivation of this equation, one has to use extensively the mCYBE equation \eqref{Eq:mCYBE} (or more precisely the fact that $R_g$ also satisfies the mCYBE). Thus, it is the fact that $R$ is a solution of the mCYBE that ensures the flatness of $K_\pm$ and hence the integrability of the model. Note also that the zero curvature equation \eqref{Eq:FlatK} is not invariant under the multiplication of $K_\pm$ by a global constant factor: this is the origin of the factor $1-c^2\eta^2$ that we introduced in the definition \eqref{Eq:CurrentYB} of $K_\pm$ (another factor would spoil the flatness of $K_\pm$).

Following Zakharov and Mikhailov, the Lax pair of the Yang-Baxter model is given by
\begin{equation}\label{Eq:LaxYBLag}
\Lc_\eta(\lambda) = \frac{K_1+\lambda K_0}{1 - \lambda^2} \;\;\;\; \text{ and } \;\;\;\; \M_\eta(\lambda) = \frac{K_0+\lambda K_1}{1 - \lambda^2},
\end{equation}
with $K_0 = \frac{1}{2}(K_++K_-)$ and $K_1 = \frac{1}{2}(K_+-K_-)$. Although it is not equal to the Lax pair found in the original paper~\cite{Klimcik:2008eq} of Klim\v{c}ik, it is related to it by a formal gauge transformation by $g$. It is clear that this Lax pair reduces to the one \eqref{Eq:LaxPcmLag} we considered for the PCM in the undeformed limit:
\begin{equation*}
\Lc_{\eta=0}(\lambda) = \Lc_{\text{PCM}}(\lambda) \;\;\;\; \text{ and } \;\;\;\; \M_{\eta=0}(\lambda) = \M_{\text{PCM}}(\lambda).
\end{equation*}

\paragraph{Hamiltonian analysis.} Let us discuss the Hamitlonian formalism of the Yang-Baxter model. The phase-space of the model is described by the $G_0$-valued field $g$ and the $\g_0$-valued field $X$, as explained in Subsection \ref{SubSec:SigmaModelHam}. Choosing local coordinates $\phi^i$ on $G_0$, one can compute the corresponding conjugate momenta $\pi_i$ from the action \eqref{Eq:ActionYBtx} and deduce the Lagrangian expression of $X$. One then finds
\begin{equation}\label{Eq:XYB}
X = \frac{K}{1-\eta^2 R_g^2} j^L_0 - \frac{K \eta R_g}{1-\eta^2 R_g^2} j^L_1,
\end{equation}
which reduces to $X=Kj^L_0$ in the undeformed limit, as expected from \eqref{Eq:XPcm}. From the action \eqref{Eq:ActionYBtx} and the equation \eqref{Eq:HamDensity}, we compute the Hamiltonian of the Yang-Baxter model and find
\begin{equation}\label{Eq:HamYB}
\Hc_\eta = \int \dd x \left( \frac{1}{2K} \kappa\left( (1-\eta^2 R_g^2)X, X \right) + \frac{K}{2} \kappa\left(j^L,j^L\right) + \kappa\left(j^L,\eta R_g X\right) \right),
\end{equation}
with $j^L=j^L_1$ as usual. Using the relation \eqref{Eq:XYB} between $X$ and $j^L_0$, we can express the current $K_0$ and $K_1$ in terms of $X$ and $j^L$:
\begin{equation}\label{Eq:KHam}
K_0 = \frac{1-c^2\eta^2}{K} X \;\;\;\; \text{ and } \;\;\;\; K_1 = (1-c^2\eta^2) \left( j^L - \frac{\eta}{K} R_g X \right).
\end{equation}
We then note that the Hamiltonian \eqref{Eq:HamYB} can be rewritten as
\begin{equation*}
\Hc_\eta = \frac{K}{2(1-\eta^2 c^2)^2} \int \dd x \, \Bigl( \kappa(K_0,K_0) + \kappa(K_1,K_1) \Bigr).
\end{equation*} 

\paragraph{Maillet non-ultralocal bracket.} Using the Hamiltonian expression \eqref{Eq:KHam} of the current $K_\mu$, we rewrite the Lax matrix \eqref{Eq:LaxYBLag} as
\begin{equation}\label{Eq:LaxYBHam}
\Lc_\eta (\lambda) = \frac{1-c^2\eta^2}{1-\lambda^2} \left( j^L - \frac{\eta}{K} R_g X  + \frac{\lambda}{K} X \right)
\end{equation}
We now want the Poisson bracket of the Lax matrix with itself, as computed by Delduc, Magro and Vicedo in~\cite{Delduc:2013fga}. This necessitates the computation of some intermediate Poisson brackets involving the field $R_g X$. We will not enter into more details here, as some similar computations will be presented in the section \ref{Sec:BYB} on the Bi-Yang-Baxter model. In the end, using the mCYBE \eqref{Eq:mCYBE} (or more precisely the mCYBE for $R_g$), one finds that the Poisson bracket of the Lax matrix takes a Maillet non-ultralocal form \eqref{Eq:PBR}.

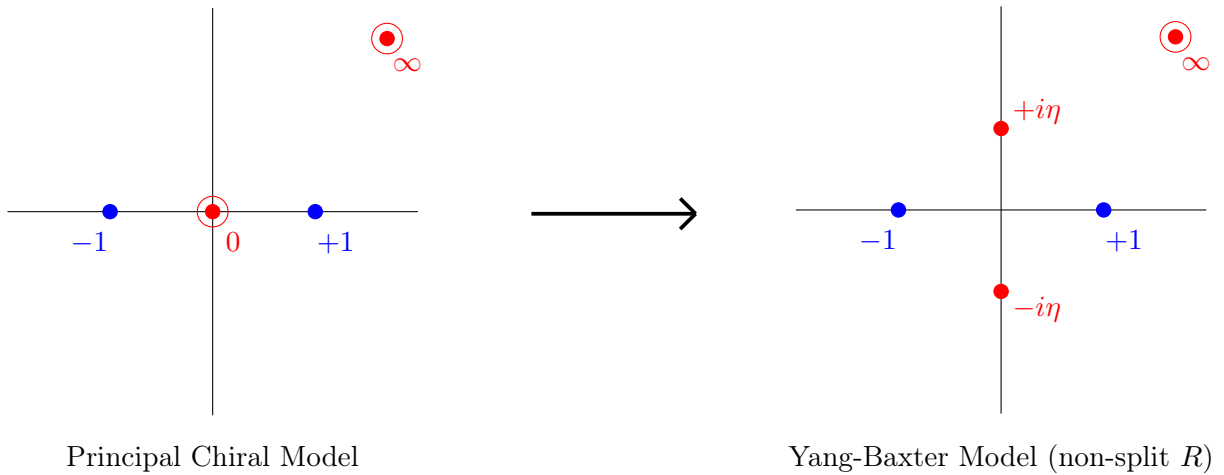
\begin{figure}[H]
\begin{minipage}{0.4\textwidth}
\begin{center}
	\begin{tikzpicture}[scale=1.35]
 		\draw (-2,0) to (2,0);
		\draw (0,-2) to (0,2);
		\draw[blue,fill=blue] (1,0) circle (0.07);
		\draw[blue,fill=blue] (-1,0) circle (0.07);	
		\node[blue,below] at (1.2,-0.1) {$+1$}; 
		\node[blue,below] at (-1.2,-0.1) {$-1$}; 
		\draw[red] (1.7,1.7) circle (0.15);
		\draw[red,fill=red] (1.7,1.7) circle (0.07);
		\node[red,below] at (1.9,1.6) {$\infty$};
		\draw[red] (0,0) circle (0.15);
		\draw[red,fill=red] (0,0) circle (0.07);
		\node[red,below] at (0.2,-0.1) {$0$};
		\node[below] at (0,-2.2) {Principal Chiral Model};
 	\end{tikzpicture}
\end{center}
\end{minipage}
\begin{minipage}{0.2\textwidth}
\begin{center}
	\begin{tikzpicture}[scale=1.35]
		\node[below] at (0,1.8) {};
 		\draw[line width=0.5mm] (-0.8,0) to (0.8,0);
 		\draw[line width=0.5mm] (0.65,0.15) to (0.8,0);
 		\draw[line width=0.5mm] (0.65,-0.15) to (0.8,0); 		
		\node[below] at (0,-2.2) {};
 	\end{tikzpicture}
\end{center}
\end{minipage}
\begin{minipage}{0.4\textwidth}
\begin{center}
	\begin{tikzpicture}[scale=1.35]
 		\draw (-2,0) to (2,0);
		\draw (0,-2) to (0,2);
		\draw[blue,fill=blue] (1,0) circle (0.07);
		\draw[blue,fill=blue] (-1,0) circle (0.07);	
		\node[blue,below] at (1.2,-0.1) {$+1$}; 
		\node[blue,below] at (-1.2,-0.1) {$-1$}; 
		\draw[red,fill=red] (1.7,1.7) circle (0.07);
		\draw[red] (1.7,1.7) circle (0.15);
		\node[red,below] at (1.9,1.6) {$\infty$};
		\draw[red,fill=red] (0,0.8) circle (0.07);
		\draw[red,fill=red] (0,-0.8) circle (0.07);
		\node[red,above] at (0.35,0.75) {$+i\eta$};
		\node[red,below] at (0.35,-0.75) {$-i\eta$};
		\node[below] at (0,-2.2) {Yang-Baxter Model (non-split $R$)};
 	\end{tikzpicture}
\end{center}
\end{minipage}
\caption{Effect of the Yang-Baxter deformation on the {\color{red}poles} and {\color{blue} zeros} of the twist function.}
\label{Fig:PolesZerosYB}
\end{figure}

More precisely, the $\Rc$-matrix underlying this Maillet bracket is given by the standard non-twisted $\Rc$-matrix on $\Lg$ and the twist function:
\begin{equation}\label{Eq:TwistYB}
\vp_\eta(\lambda) = \frac{K}{1-c^2\eta^2} \frac{1-\lambda^2}{\lambda^2-c^2\eta^2}.
\end{equation}
It is clear that this twist function reduces to the one \eqref{Eq:TwistPCM} of the PCM in the undeformed limit $\eta=0$. Let us now study the analytical properties of the 1-form $\vp_\eta(\lambda) \dd \lambda$, summarised on Figure \ref{Fig:PolesZerosYB}. As for the PCM, it possesses two simple zeros at $\lambda=+1$ and $\lambda=-1$. It also possesses a double pole at infinity and two simple poles at $\pm c \eta$ (\textit{i.e.} at $\pm \eta$ for a split matrix $R$ and at $\pm i \eta$ for a non-split one). This is in contrast with the PCM case, for which there was a double pole at infinity and a double pole at $0$. The effect of the Yang-Baxter deformation is thus to split the double pole at $0$ into two simple poles at $\pm c \eta$, without deforming the zeros. We shall see that this is a common feature of all integrable deformations of $\s$-models.

\paragraph{Lax matrix at the pole of the twist function.} As we have seen above, the twist function of the Yang-Baxter model possesses two simple poles at $\lambda_\pm = \pm c \eta$. We shall come back on the interpretation of these poles later, in the second part of this thesis. At this stage, let us just make the following observation, which will be useful for Chapter \ref{Chap:PLie}.

It is a classical result that a Lax pair of a model is not defined uniquely. In particular, one always has a freedom of performing a formal gauge transformation on the Lax pair $(\Lc,\M)$ by a $G$-valued field $u(x,t)$ (this comes from the fact that the Lax equation is a zero curvature equation):
\begin{equation}\label{Eq:FormalGauge}
\Lc^u(\lambda) = u \Lc(\lambda) u^{-1} + u \p_x u^{-1} \;\;\;\; \text{ and } \;\;\;\; \M^u(\lambda) = u \M(\lambda) u^{-1} + u \p_t u^{-1}.
\end{equation}
The Lax equation \eqref{Eq:Lax} for the pair $(\Lc,\M)$ is then equivalent to the one for the Lax pair $(\Lc^u,\M^u)$. For the Yang-Baxter model, we can then consider the Lax pair $\lrb\Lc^g_\eta,\M^g_\eta\rrb$, obtained by a gauge transformation with $u=g$.

We will be interested in the evaluation of this new Lax matrix $\Lc^g_\eta$ at the poles of the twist function. Starting from the expression \eqref{Eq:LaxYBHam}, we get
\begin{equation}\label{Eq:LgYB}
\Lc^g_\eta(\lambda_\pm) = -\gamma R^\mp \bigl( gXg^{-1} \bigr),
\end{equation}
with (see Appendix \ref{App:RMat} for more informations about the operators $R^\pm$)
\begin{equation*}
R^\pm = R \pm c \; \Id \;\;\;\; \text{ and } \;\;\;\; \gamma = \frac{\eta}{K}.
\end{equation*}
The equation \eqref{Eq:LgYB} will be the starting point for the studies of Poisson-Lie symmetries of the Yang-Baxter model in Chapter \ref{Chap:PLie}. To conclude this paragraph, let us also note that the factor $\gamma$ satisfies
\begin{equation}\label{Eq:GammaYB}
\frac{1}{\gamma} = \pm 2c \; \res_{\lambda=\lambda_\pm} \, \vp_\eta(\lambda) \; \dd \lambda.
\end{equation}

\subsection{Deformations of the PCM}
\label{SubSec:dPCM}

In the previous subsection, we presented in details the Yang-Baxter model, which is a particular example of an integrable deformation of the PCM. In this subsection, we will review briefly the whole landscape of these deformations.\\

As we have seen in the previous subsection, the Yang-Baxter model consists in breaking the left symmetry of the PCM, while preserving the right one. In~\cite{Klimcik:2008eq}, Klim\v{c}ik proposed a deformation of the Yang-Baxter model itself, depending on another parameter $\etat$, which also breaks the right symmetry while keeping the existence of a Lax pair. This model is called the Bi-Yang-Baxter model. We shall not discuss it further here: it will be the subject of the following section, as part of the presentation of my PhD works.\\

Another possibility to deform the PCM is to add~\cite{Witten:1983ar} to its action a so-called Wess-Zumino term (multiplied by a constant parameter $k$). We will not explain here how this term is constructed and refer to~\cite{Witten:1983ar,Wess:1971yu,Novikov:1982ei} for more details on this matter. It was shown in~\cite{Abdalla:1982yd} (see also~\cite{deVega:1979zy}) that this model admits a Lax pair for any value of $k$. Its twist function has been studied in~\cite{Delduc:2014uaa} and proved to be
\begin{equation}\label{Eq:TwistWZk}
\vp_k(\lambda) = \frac{1-\lambda^2}{(\lambda-k)^2}.
\end{equation}
In the undeformed limit $k$ goes to $0$, one recovers the twist function \eqref{Eq:TwistPCM} of the PCM (for the global factor $K=1$). An interesting remark is that, as for the Yang-Baxter case, the deformation modifies the poles of the 1-form $\vp_k(\lambda)\dd \lambda$, without modifying its zeros. More precisely, the deformation moved the double pole at $\lambda=0$ of the PCM to a double pole at $\lambda=k$. The double pole at infinity and the two simple zeros at $\pm 1$ of the PCM are not affected by this deformation.\\

Some deformations of the PCM can be combined together to form multi-parameter deformations. In fact, it was shown recently by Delduc, Hoare, Kameyama and Magro~\cite{Delduc:2017fib} that there exists a three-parameter integrable deformation of the PCM combining the left and right deformations $\eta$ and $\etat$ of the Bi-Yang-Baxter model (see above) and the addition of a Wess-Zumino term (with factor $k$). More precisely, they have constructed such a model which admits a Lax pair representation. The Hamiltonian integrability of this model, however, has not been studied yet although it is expected that it enters the class of models with a twist function. Although this might not be a simple thing to prove, this three-parameter deformation is expected to be the maximal way of deforming the twist function of the PCM.

As the twist function of the three-parameter deformation of the PCM has not been computed yet, we will restrict ourselves to its two-parameter limits. One of them is the Bi-Yang-Baxter model mentioned above. The other one is a combination of the Yang-Baxter deformation (with parameter $\eta$) and of the addition of a Wess-Zumino term (with parameter $k$). It was first proposed in~\cite{Balog:1993es} and further studied in~\cite{Kawaguchi:2011mz,Kawaguchi:2013gma} for the group $G_0=SU(2)$. The model for a general group was constructed and shown to be integrable by Delduc, Magro and Vicedo in~\cite{Delduc:2014uaa}. Its twist function takes the form
\begin{equation}\label{Eq:TwistdPCM}
\vp_{\text{dPCM}}(\lambda) = \frac{1-\lambda^2}{(\lambda-k)^2+\Ac^2},
\end{equation}
where the subscript dPCM stands for deformed PCM and
\begin{equation*}
\Ac = \eta \sqrt{1-\frac{k^2}{1+\eta^2}}.
\end{equation*}
Note that this model was considered in the litterature only for a Yang-Baxter deformation associated with a non-split solution $R$ of the mCYBE (\textit{i.e.} $c=i$ in \eqref{Eq:mCYBE}). One easily checks that $\Ac=0$ for $\eta=0$, so that the twist function \eqref{Eq:TwistdPCM} coincides with \eqref{Eq:TwistWZk} in this limit. In the same way, $\Ac=\eta$ for $k=0$, so \eqref{Eq:TwistdPCM} reduces to the twist function \eqref{Eq:TwistYB} of the Yang-Baxter model (for $c=i$ and an appropriate choice of $K$) in this case.\\

The integrable structure of the general deformed PCM shares some similarities with the one of the PCM. In particular, its Lax matrix takes the ``model-independent'' form
\begin{equation}\label{Eq:LaxdPCM}
\Lc_{\text{dPCM}}(\lambda,x) = \frac{A(x) + \lambda \, \Pi(x)}{1-\lambda^2},
\end{equation}
where $A$ and $\Pi$ are $\g_0$-valued fields. The deformation is contained in the (model-dependent) expression of $A$ and $\Pi$ in terms of the fields $g$ and $X$. In particular, according to Subsection \ref{SubSec:PCM}, one has $A=j^L$ and $\Pi=X$ in the undeformed case $k=\eta=0$. In the same way, in the Yang-Baxter case $k=0$ and $\eta\neq 0$, $A$ and $\Pi$ coincide with the currents $K_1$ and $K_0$ of Subsection \ref{SubSec:YB}.

The expression of the currents $A$ and $\Pi$ for the general case $k\neq 0$ is more involved: it is easier to express them in terms of $g$ and another field $Y$ instead of $X$. The relation between $(g,X)$ and $(g,Y)$ is difficult to state explicitly due to some non-locality issues coming with Wess-Zumino terms. We shall not enter into this matter here as we will not need the explicit definition of $A$ and $\Pi$ in this PhD (see~\cite{Delduc:2014uaa} for details).

The form of the Lax matrix \eqref{Eq:LaxdPCM}, specific to a Zakharov-Mikhailov scheme, reflects the existence for all these models of a flat and conserved current. This current is associated by the Noether theorem with the right multiplication symmetry of the model, which is conserved both by the Yang-Baxter deformation and the addition of the Wess-Zumino term.\\

Another common feature shared by the two-parameter dPCM and its (less deformed) limits is the expression of the Hamiltonian and total momentum of the model. They are given in terms of the fields $A$ and $\Pi$ by
\begin{subequations}\label{Eq:HamMomPCM}
\begin{align}
\Hc_{\text{dPCM}} &= \frac{\mathcal{B}}{2} \int \dd x \; \Bigl( (\Ac^2+k^2+1) \left( \kappa(\Pi,\Pi)+\kappa(A,A) \right) + 4k \, \kappa(\Pi,A) \Bigr), \\
\Pc_{\text{dPCM}} &= \mathcal{B} \int \dd x \; \Bigl( k\left( \kappa(\Pi,\Pi)+\kappa(A,A) \right) + (\Ac^2+k^2+1) \kappa(\Pi,A) \Bigr),
\end{align}
\end{subequations}
with $\mathcal{B}$ a global factor depending on $\Ac$ and $k$ \textit{via} the relation $\mathcal{B}=-\dfrac{1}{4}\varphi'_{\text{dPCM}}(1)\varphi'_{\text{dPCM}}(-1)$. Interestingly, one can always rewrite the above expressions as
\begin{equation}\label{Eq:HamMomdPCM}
\Hc_{\text{dPCM}} = \Hc_+ - \Hc_- \;\;\;\ \text{ and } \;\;\;\; \Pc_{\text{dPCM}} = \Hc_+ + \Hc_-,
\end{equation}
with
\begin{equation*}
\Hc_\pm = -\frac{1}{2} \; {\large \res_{\lambda=\pm 1}} \; \vp_{\text{dPCM}}(\lambda) \int \dd x \; \kappa \bigl( \Lc_{\text{dPCM}}(\lambda,x),\Lc_{\text{dPCM}}(\lambda,x) \bigr).
\end{equation*}
We shall use this fact in Chapter \ref{Chap:LocalCharges}.

There exists another type of integrable deformation of the PCM, called the $\lambda$-deformation~\cite{Sfetsos:2013wia}. It consists more of a deformation of the non-abelian T-dual of the PCM, which is a model equivalent to the PCM. We shall not present here the results on this model. The twist function of this model turns out to be of the same form than the one of a split Yang-Baxter deformation~\cite{Hollowood:2014rla,Vicedo:2015pna}. This is due to a deeper relation between the Yang-Baxter model and the $\lambda$-deformation~\cite{Vicedo:2015pna,Hoare:2015gda,Klimcik:2015gba}, involving a Poisson-Lie duality~\cite{Klimcik:1995jn,Klimcik:1995dy,Klimcik:1996br,Klimcik:2002zj}.

\subsection[Deformations of $\Z_2$-coset models]{Deformations of $\bm{\Z_2}$-coset models}
\label{SubSec:dZ2}

We shall now discuss the integrable deformations of the $\Z_
2$-coset $\s$-model. We will use the notations of Subsection \ref{SubSec:ZT}, in which we described the undeformed theory. There are two types of deformations of the $\Z_2$-coset model: the Yang-Baxter deformation and $\lambda$-deformation. As for the PCM, the two give a similar integrable structure and in particular a twist function of the same form. We shall focus on the Yang-Baxter deformation and refer to~\cite{Hollowood:2014rla} for the $\lambda$-deformation.

The deformed model we are going to study here is part of a general scheme of integrable deformations called Yang-Baxter type deformations~\cite{Vicedo:2015pna}, or $\eta$-deformation, which applies to a certain class of models with twist function. In particular, the $\eta$-deformation of the PCM is the Yang-Baxter model, that we studied in Subsection \ref{SubSec:YB}. In this subsection, we shall skip most technical details and insist on the common features shared by this model and the Yang-Baxter one. They actually are common features of general Yang-Baxter type deformations and will be the starting key points for the chapter \ref{Chap:PLie} of this thesis. Moreover, in this subsection, we will present a common formalism describing the integrable structure of both the undeformed $\Z_2$-coset and its $\eta$-deformation, as we did for the PCM and its deformations in the previous subsection.

\paragraph{Action and symmetries.} The $\Z_2$-coset $\eta$-deformation has been constructed by Delduc, Magro and Vicedo in~\cite{Delduc:2013fga}. As for the Yang-Baxter model, it is based on a skew-symmetric solution $R$ of the mCYBE \eqref{Eq:mCYBE}. Its action is given by
\begin{equation}\label{Eq:ActionZ2eta}
S_{\Z_2,\eta}[g] = K \int_\Sigma \dd x^+ \, \dd x^- \; \kappa \left( j_+^{L\,(1)}, \frac{1}{1-\eta R_g \circ \pi^{(1)}} j_-^{L\,(1)} \right),
\end{equation}
where $R_g=\Ad_g^{-1} \circ R \circ \Ad_g$, as in the PCM case, and $\pi^{(1)}$ is the projection on the eigenspace $\g^{(1)}$ of $\s$. Note the similarity between this action and the one of the Yang-Baxter model, written as \eqref{Eq:ActionYBL}.

We expect a deformation of the $\Z_2$-coset model to possess the same degrees of freedom as the undeformed model. Thus, we expect the gauge symmetry \eqref{Eq:GaugeZT} of the undeformed model to also leave invariant the action \eqref{Eq:ActionZ2eta}. This can be easily verified using the expression \eqref{Eq:CurrentGauge} of the transformation of $j^{L\,(1)}_\pm$ under a gauge transformation and the fact that $R_g$ transforms as $\Ad_h^{-1} \circ R \circ \Ad_h$. Thus, the $\eta$-deformation preserves the right gauge symmetry of the $\Z_2$-coset model.

Recall that the undeformed model also possesses a global left symmetry (see Subsection \ref{SubSec:ZT}). Due to the presence of the operator $R_g$ in the action \eqref{Eq:ActionZ2eta}, this symmetry is broken by the $\eta$-deformation. This is similar to what happens for the Yang-Baxter model, \textit{i.e.} the $\eta$-deformation of the PCM, which preserves the right symmetry of the PCM but breaks the left one. This symmetry breaking is one of the common feature shared by the Yang-Baxter type deformations.

\paragraph{Lax matrix.} It was found in~\cite{Delduc:2013fga} that the model \eqref{Eq:ActionZ2eta} admits a Lax pair. In particular, the Lax matrix can be expressed as
\begin{equation}\label{Eq:LaxdZ2}
\Lc_{\dd\Z_2}(\lambda)  = A^{(0)} + \frac{1}{2}\left(\frac{1}{\lambda}+\lambda\right)A^{(1)} + \frac{1}{2}(\lambda^2-1) \Pi^{(0)} + \frac{1}{2}\left(\lambda - \frac{1}{\lambda}\right) \Pi^{(1)},
\end{equation}
where $A$ and $\Pi$ are $\g_0$-valued fields. Similarly to the deformations of the PCM in Subsection \ref{SubSec:dPCM}, this structure is common to the deformed and undeformed $\Z_2$-models. The deformation is purely contained in the dependence of $A$ and $\Pi$ in terms of $g$ and $X$. According to equation \eqref{Eq:LaxZ2}, we have in the undeformed limit $\eta=0$ that $\Pi=K^{-1}X$ and $A=j^L$. We refer to~\cite{Delduc:2013fga} for the expressions of $A$ and $\Pi$ in the deformed case. In both the undeformed and the deformed models, the field $\Pi^{(0)}$ is the constraint associated with the gauge symmetry \eqref{Eq:GaugeZT}. Finally, let us note that the Lax matrix satisfies the equivariance and reality conditions \eqref{Eq:EquiL} and \eqref{Eq:Reality}.

\paragraph{Twist function.} The computation of the Poisson bracket of the Lax matrix is also carried out in~\cite{Delduc:2013fga}. It takes the form of a Maillet non-ultralocal bracket. The $\Rc$-matrix of the latter is given by the standard matrix $\Rc^0$, twisted by the automorphism $\s$, and the twist function:
\begin{equation}\label{Eq:TwistdZ2}
\vp_{\dd\Z_2}(\lambda) = \frac{2 K \lambda}{(\lambda^2-1)^2-c^2\eta^2 (\lambda^2+1)^2}.
\end{equation}
This twist function clearly reduces to the one \eqref{Eq:TwistZ2} of the undeformed $\Z_2$-coset model when $\eta=0$. Note that it also satisfies the equivariance and reality conditions \eqref{Eq:TwistEqui} and \eqref{Eq:TwistReal}.

\begin{figure}[H]
\begin{minipage}{0.4\textwidth}
\begin{center}
	\begin{tikzpicture}[scale=1.5]
 		\draw (-2,0) to (2,0);
		\draw (0,-2) to (0,2);
		\draw[dashed] (0,0) circle (1);
		\draw[red] (1,0) circle (0.15);
		\draw[red,fill=red] (1,0) circle (0.07);
		\draw[red] (-1,0) circle (0.15);
		\draw[red,fill=red] (-1,0) circle (0.07);	
		\node[red,below] at (1.2,-0.15) {$+1$}; 
		\node[red,below] at (-1.2,-0.15) {$-1$}; 
		\draw[blue,fill=blue] (1.7,1.7) circle (0.07);
		\node[red,blue] at (1.9,1.5) {$\infty$};
		\draw[blue,fill=blue] (0,0) circle (0.07);
		\node[blue,below] at (0.2,-0.1) {$0$};
		\node[below] at (0,-2.2) {$\Z_2$-coset $\s$-model};
 	\end{tikzpicture}
\end{center}
\end{minipage}
\begin{minipage}{0.2\textwidth}
\begin{center}
	\begin{tikzpicture}[scale=1.35]
		\node[below] at (0,1.8) {};
 		\draw[line width=0.5mm] (-0.8,0) to (0.8,0);
 		\draw[line width=0.5mm] (0.65,0.15) to (0.8,0);
 		\draw[line width=0.5mm] (0.65,-0.15) to (0.8,0); 		
		\node[below] at (0,-2.2) {};
 	\end{tikzpicture}
\end{center}
\end{minipage}
\begin{minipage}{0.4\textwidth}
\begin{center}
	\begin{tikzpicture}[scale=1.5]
 		\draw (-2,0) to (2,0);
		\draw (0,-2) to (0,2);
		\draw[dashed] (0,0) to (0.9,0.436);
		\draw[dashed] (0,0) to (0.9,-0.436);
		\draw[dashed] (0,0) to (-0.9,0.436);
		\draw[dashed] (0,0) to (-0.9,-0.436);
		\draw[dashed] (0,0) circle (1);
		\draw[red,fill=red] (0.9,0.436) circle (0.07);
		\draw[red,fill=red] (0.9,-0.436) circle (0.07);
		\node[red,right] at (1,0.6) {$e^{i\theta}$};
		\node[red,right] at (1,-0.45) {$e^{-i\theta}$}; 
		\draw[red,fill=red] (-0.9,0.436) circle (0.07);
		\draw[red,fill=red] (-0.9,-0.436) circle (0.07);
		\node[red,left] at (-1,0.52) {$-e^{i\theta}$};
		\node[red,left] at (-1,-0.45) {$-e^{-i\theta}$};
		\draw[blue,fill=blue] (1.7,1.7) circle (0.07);
		\node[blue] at (1.9,1.5) {$\infty$};
		\draw[blue,fill=blue] (0,0) circle (0.07);
		\node[blue,below] at (0.2,-0.1) {$0$};
		\draw[thick,myGreen,->] ([shift=(0:0.7cm)]0,0) arc (0:26:0.7cm);
		\node[myGreen,right] at (0.65,0.18) {$\theta$}; 
		\node[below] at (0,-2.2) {$\eta$-deformation (non-split $R$)};
 	\end{tikzpicture}
\end{center}
\end{minipage}
\caption{Effect of the Yang-Baxter deformation on the {\color{red}poles} and {\color{blue} zeros} of the twist function.}
\label{Fig:PolesZerosZ2eta}
\end{figure}
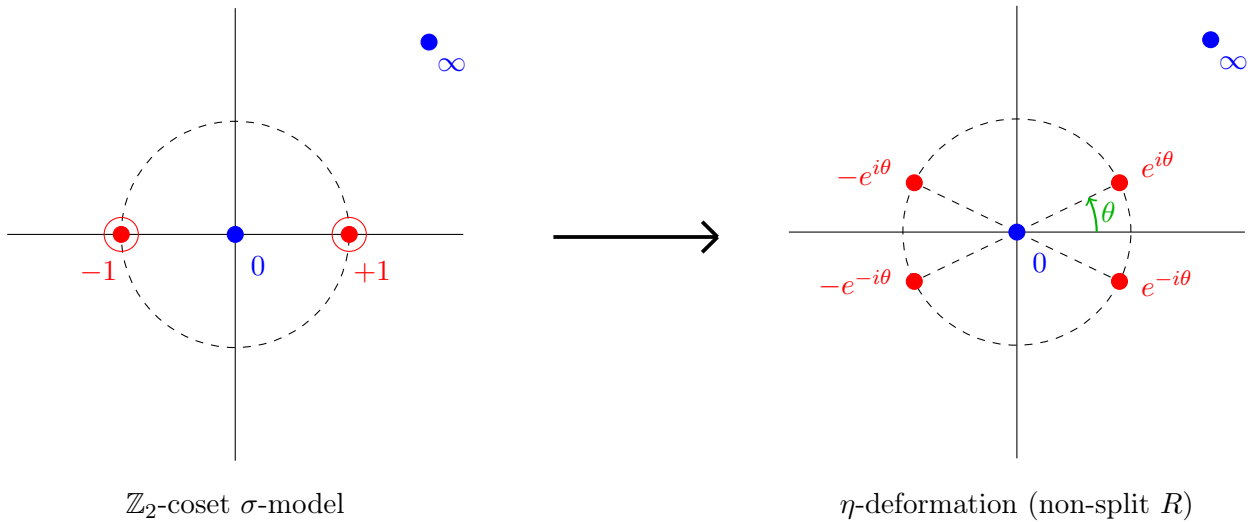

The 1-form $\vp_{\dd\Z_2}(\lambda)\dd\lambda$ possesses two simple zeros at $0$ and infinity, as for the undeformed $\Z_2$-coset model. However, the two double poles $+1$ and $-1$ of the undeformed model have been split into simple poles $\lambda_\pm$ and $-\lambda_\pm$, given by
\begin{equation*}
\lambda_\pm = \sqrt{\frac{1 \pm c\eta}{1 \mp c \eta}} = \frac{1}{\lambda_\mp}.
\end{equation*}
For a split matrix $R$ ($c=1$), these poles are situated on the real axis. For a non-split $R$ ($c=i$), they belong to the unit circle of the complex plane. We can rewrite them as $\lambda_\pm = e^{\pm i\theta}$, with the angle $\theta$ defined by $\eta = \tan\theta$, as represented in Figure \ref{Fig:PolesZerosZ2eta}.

We observe that the two poles $+1$ and $-1$ of the undeformed model split in a symmetric way under the reflection with respect to the origin. This is forced by the equivariance property \eqref{Eq:TwistEqui} of the twist function, which in this case implies that
\begin{equation*}
\vp_{\dd\Z_2}(-\lambda) = -\vp_{\dd\Z_2}(\lambda).
\end{equation*}
Indeed, for every pole $\lambda_\infty$, there must then also be a pole at $-\lambda_\infty$. In some sense (which will be made clearer in the second part of this thesis), one can consider these two poles as a unique set of poles related by the transformation $\lambda \mapsto -\lambda$. Thus, we see that the effect of the $\eta$-deformation is to split a (set of) double pole(s) into two (sets of) simple poles. This is similar to the case of the Yang-Baxter model and is one of the common feature of the Yang-Baxter type deformations.\\

We end this paragraph by a last observation. Performing the gauge transformation of the Lax matrix \eqref{Eq:LaxdPCM} by $g$ and evaluating it at $\lambda_\pm$, one finds
\begin{equation}\label{Eq:LgdZ2}
\Lc^g_{\dd\Z_2}(\lambda_\pm) = -\gamma\, R^\mp \bigl( g X g^{-1} \bigr), \;\;\;\; \text{ with } \;\;\;\; \frac{1}{\gamma} = \pm 4c \; \res_{\lambda=\lambda_\pm} \, \vp_{\dd\Z_2}(\lambda) \; \dd \lambda,
\end{equation}
and $R^\pm = R \pm c\,\Id$. This is to be compared with equation \eqref{Eq:LgYB}, derived for the Yang-Baxter model. This is the main common feature of Yang-Baxter type models, that we shall use extensively in Chapter \ref{Chap:PLie}.

\paragraph{Deformed $\bm{\Z_T}$-coset models.} Let us end this section by discussing the deformations of $\Z_T$-coset $\s$-models. It is expected that the $\Z_T$-coset models admit an integrable deformation of Yang-Baxter type. The effect of this deformation would be to break the left symmetry of the model and to split the double poles of the twist function \eqref{Eq:TwistZT} into simple poles. However, this deformed model has never been constructed explicitly. This construction would be an interesting project, to complete the study of the whole landscape of deformations of $\s$-models. We expect the Lax matrix of this deformed model to be of the form
\begin{equation}\label{Eq:LaxdZT}
\Lc_{\dd\Z_T}(\lambda,x) = \sum_{k=1}^{T} \frac{(T-k) + k\lambda^{-T}}{T}\lambda^k A^{(k)}(x)  + \sum_{k=1}^{T} \frac{1-\lambda^{-T}}{T} \lambda^k \Pi^{(k)}(x),
\end{equation}
with $A$ and $\Pi$ some $\g_0$-valued field depending on $g$, $X$ and $\eta$.

It is worth mentioning that there exists a Yang-Baxter type deformation of the Green-Schwarz superstring on $AdS_5 \times S^5$, which was discovered by Delduc, Magro and Vicedo in~\cite{Delduc:2013qra,Delduc:2014kha}. Its Lax matrix is of the form \eqref{Eq:LaxdZT} for $T=4$. The undeformed model possesses the same twist function as the $\Z_4$-coset $\s$-model, \textit{i.e.} equation \eqref{Eq:TwistZT} with $T=4$. The deformation modifies this twist function as
\begin{equation*}
\vp_{\text{dGS}}(\lambda) = \frac{4 \alpha \lambda^3}{(\lambda^4-1)^2+\beta(\lambda^4+1)^2},
\end{equation*}
where $\alpha$ and $\beta$ are constants depending on the deformation parameter $\eta$, satisfying $\alpha=1$ and $\beta=0$ when $\eta=0$ (thus, for $\eta=0$, we recover the twist function \eqref{Eq:TwistZT} for $T=4$). This twist function has 8 simple poles at $i^k e^{\pm i \theta}$, $k=0,1,2,3$, with $\eta=\tan \theta $. Thus, the effect of the deformation is to split the double poles $\lbrace 1,i,-1,-i \rbrace$ into pairs of simple poles, as expected for a Yang-Baxter type deformation. The deformation of the Green-Schwarz model also exhibits other properties of Yang-Baxter type deformations: the left symmetry of the model is broken by the deformation and the Lax matrix of the model satisfies the relation \eqref{Eq:LgdZ2} for the simple poles $\lambda_\pm = e^{\pm i \theta}$ (see~\cite{Delduc:2014kha}).

\section{Bi-Yang-Baxter model}
\label{Sec:BYB}

In this section, I will present some results I obtained during my PhD about the Bi-Yang-Baxter model. These results are part of the publication~\cite{Delduc:2015xdm}: here, I will summarize the main ideas of~\cite{Delduc:2015xdm} (and a few subsequent results) without entering into too much technical details. We refer to the original article for those details. Note that some conventions and notations (in particular the sign of the Lax matrix $\Lc$) are different in the core text of this thesis and in the article~\cite{Delduc:2015xdm}. In the present section, we will keep the conventions that we used in the previous chapter and sections and will transpose the results of~\cite{Delduc:2015xdm} in those.

The Bi-Yang-Baxter model (BYBM) is a double deformation of the PCM, introduced by Klim\v{c}ik in~\cite{Klimcik:2008eq}, which depends on two deformation parameters $\eta$ and $\etat$. The limits $\eta=0$ and $\etat=0$ respectively correspond to the right and left Yang-Baxter model, \textit{i.e.} the one-parameter deformations of the PCM which break respectively the right and left symmetries of the model. The BYBM is then a combination of these two Yang-Baxter deformations, which breaks both the left and the right symmetries.

The BYBM was shown to possess a Lax pair representation by Klim\v{c}ik in~\cite{Klimcik:2014bta}, hence proving the existence of an infinite number of conserved charges. However, the Hamiltonian integrability of the model, \textit{i.e.} the fact that these conserved charges are in involution one with another, was not proved. This was the subject of my first PhD project, which resulted in the article~\cite{Delduc:2015xdm}. We showed that the BYBM enters the class of non-ultralocal models with twist function, thus proving its Hamiltonian integrability. More precisely, as we shall see, we have shown this for a reformulation of the BYBM as a $\Z_2$-coset model on $\Dz/\Gdz$. We shall start by introducing and studying this formulation and will end the section by discussing the more usual formulation as a deformation of the PCM.

\subsection{The Bi-Yang-Baxter model in Lagrangian formulation}
\label{SubSec:BYBLag}

\paragraph{Action and symmetries.} We shall use the notations of the previous chapter and sections. In particular, we consider a real Lie group $G_0$ with Lie algebra $\g_0$. Let $R$ and $\Rt$ be two skew-symmetric solutions of the mCYBE \eqref{Eq:mCYBE} on $\g_0$ (we shall restrict here to the case of the non-split mCYBE, \textit{i.e.} with $c=i$). We define the Bi-Yang-Baxter model (BYBM) by the following action, for a field $(g,\gt)$ valued in the double group $G_0\times G_0$ 
\cite{Hoare:2014oua}
\begin{equation}\label{Eq:ActionBYB}
S_{\eta,\etat}[g,\gt]= \frac{K}{2} \int \dd x^+ \dd x^- \; \kappa\left(j^L_+-\jt^L_+,\left(1-\frac{\eta}{2} R_g-\frac{\etat}{2}\Rt_{\gt} 
\right)^{-1} (j^L_--\jt^L_-)\right).
\end{equation}
$K$, $\eta$ and $\etat$ are real parameters, the currents $j^L_\pm=g^{-1} \p_\pm g$ and $\jt^L_\pm = \gt^{-1} \p_\pm \gt$ are the left currents of $g$ and $\gt$, as introduced in Subsection \ref{SubSec:FieldSigma} and finally, we have
\begin{equation*}
R_g = \Ad_g^{-1} \circ R \circ \Ad_g \;\;\;\; \text{ and } \;\;\;\;
\Rt_{\gt} = \Ad_{\gt}^{-1} \circ \Rt \circ \Ad_{\gt}.
\end{equation*}
Let us notice here that $R_g$ and $\Rt_{\gt}$ are also skew-symmetric solutions of the mCYBE \eqref{Eq:mCYBE}.\\

Let us consider the subgroup $\Gdz$ of $\Dz$ composed by elements of the form $(h,h)$, $h\in G_0$. It acts on $(g,\gt)\in\Dz$ by the right multiplication
\vspace{-2pt}\begin{equation}\label{Eq:GaugeBYB}
g \longmapsto gh \;\;\;\; \text{ and } \;\;\;\; \gt \longmapsto \gt h.\vspace{-2pt}
\end{equation}
One easily checks that the action \eqref{Eq:ActionBYB} is invariant under the transformation \eqref{Eq:GaugeBYB}, for $h$ an arbitrary field valued in $G_0$. Thus, the action \eqref{Eq:ActionBYB} possesses a gauge symmetry under the right action of $\Gdz$.

Let us also consider the action on $(g,\gt)$ of the right multiplication by constant elements $(h,\hti)$ in $\Dz$:
\begin{equation}\label{Eq:LeftBYBCoset}
g \longmapsto hg \;\;\;\; \text{ and } \;\;\;\; \gt \longmapsto \hti\gt.
\end{equation}
One checks that the action \eqref{Eq:ActionBYB} is invariant under this transformation if and only if $\eta=\etat=0$. We shall come back on this remark later.

\paragraph{Gauge fixing and deformed ``PCM'' formulation.} As we observed above, the action \eqref{Eq:ActionBYB} possesses a gauge symmetry under the right multiplication by $\Gdz$. The real degrees of freedom of this model then belong to the quotient $\Dz/\Gdz$. One can identify this quotient with the group $G_0$, by observing that each orbit contains a unique element of the form $(g',\Id)$. Thus, one should be able to reformulate the model in terms of a $G_0$-valued field. This is done at the level of the action by performing a gauge transformation \eqref{Eq:GaugeBYB} with $h=\gt^{-1}$. Doing so, one finds a new action on the gauge-invariant field $g'=g \gt^{-1}$, which reads
\vspace{-3pt}\begin{equation}\label{Eq:ActionBYB2}
S_{\eta,\etat}[g'] = \frac{K}{2} \int \dd x^+ \dd x^- \; \kappa\left(j'^L_+,\left(1-\frac{\eta}{2} R_{g'}-\frac{\etat}{2}\Rt\right)^{-1} j'^L_-\right).
\end{equation}
This is the original action proposed by Klim\v{c}ik in~\cite{Klimcik:2008eq}. In this formulation, it is clear that the BYBM is a two-parameter deformation of the PCM \eqref{Eq:PCMLC}.

Let us note that the action \eqref{Eq:LeftBYBCoset} of $\Dz$ acts on the field $g'$ by the following combination of the left and right multiplications
\begin{equation*}
g' \mapsto h g \hti^{-1}.
\end{equation*}
In particular, both the left and right multiplication are not symmetries of the action \eqref{Eq:ActionBYB2} if $\eta$ and $\etat$ are non zero. When $\eta=\etat=0$, these symmetries are restored, as the model then reduces to the undeformed PCM model.

It is clear in this gauge-fixed formulation that the BYBM \eqref{Eq:ActionBYB2} reduces to the Yang-Baxter model \eqref{Eq:ActionYBL} when $\etat=0$ (with a change from $\eta$ to $\frac{\eta}{2}$ and from $K$ to $\frac{K}{2}$). In this case, the global right symmetry $g'\mapsto g'h$ is restored and the left symmetry $g' \mapsto hg'$ stays broken. In the same way, the limit $\eta=0$ is also the Yang-Baxter model, but with the left symmetry preserved and the right symmetry broken. The BYBM is thus a combination of both the left and the right Yang-Baxter deformations. According to Subsection \ref{SubSec:YB}, the one-parameter limits $\eta=0$ and $\etat=0$ are thus both integrable models with twist function.

\paragraph{$\bm{\Z_2}$-coset Yang-Baxter limit.}  As we shall see now, there exist two other one-parameter limits of the model which also admit a twist function. They arise naturally from the non-gauge fixed formulation \eqref{Eq:ActionBYB}, once we identify it with a deformation of a $\Z_2$-coset $\s$-model.

Let us consider the double group $DG_0=\Dz$. It is equipped with an involutive automorphism $\delta$ defined as the exchange of the two $G_0$ factors in $DG_0$:
\begin{equation*}
\begin{array}{rccc}
\delta : & DG_0=\Dz & \longrightarrow & DG_0=\Dz \\
         &  (g,\gt) &   \longmapsto   & (\gt,g) 
\end{array}.
\end{equation*}
Let us note here that the subgroup $DG_0^{(0)}$ of $\delta$-fixed-points is equal to $\Gdz$: thus, we can identify the coset $\Dz/\Gdz$ with the $\Z_2$-coset $DG_0/DG_0^{(0)}$. The automorphism $\delta$ induces an involutive automorphism of the Lie algebra $D\g_0=\g_0\times\g_0$, that we shall still denote $\delta$. We will use the notations of Subsection \ref{SubSec:ZT} and Appendix \ref{App:Torsion} for finite order automorphisms. In particular, the eigenspaces of $\delta$ are
\begin{equation*}
D\g_0^{(0)} = \lwb (X,X),\, X \in\g_0 \rwb = \g_0^{\text{diag}} \;\;\;\; \text{ and } \;\;\;\; D\g_0^{(1)} = \lwb (X,-X), \, X\in\g_0 \rwb.
\end{equation*}
The projectors $\pi^{(p)}$ of the decomposition $D\g_0=D\g_0^{(0)}\oplus D\g_0^{(1)}$ are then given by
\begin{equation*}
\begin{array}{rccc}
\pi^{(0)} : & D\g_0 & \longrightarrow & D\g_0^{(0)} \\
            & (X,Y) &   \longmapsto   & \frac{1}{2} (X+Y,X+Y) 
\end{array} \;\;\;\; \text{ and } \;\;\;\; \begin{array}{rccc}
\pi^{(1)} : & D\g_0 & \longrightarrow & D\g_0^{(1)} \\
            & (X,Y) &   \longmapsto   & \frac{1}{2} (X-Y,Y-X) 
\end{array}.
\end{equation*}
We will encode the two fields $g$ and $\gt$ in the $DG_0$-valued field $f=(g,\gt)$. Note that the grade (1) part of the left current $f^{-1}\p_\pm f$ is given by
\begin{equation*}
\bigl( f^{-1}\p_\pm f\bigr)^{(1)} = \frac{1}{2} \bigl( j^L_\pm - \jt^L_\pm, \jt^L_\pm - j^L_\pm \bigr).
\end{equation*}
The Killing form on the double algebra $D\g_0$ is given by $\kappa_D\bigl( (X,Y), (X',Y') \bigr) = \kappa(X,X') + \kappa(Y,Y')$, where $\kappa$ is the Killing form on $\g_0$.
Let us define $\mathfrak{R}=\bigl(R,\Rt\bigr)$. It is a skew-symmetric solution of the mCYBE \eqref{Eq:mCYBE} on $D\g_0$. In the limit $\eta=\etat$, we can rewrite the action \eqref{Eq:ActionBYB} as
\begin{equation*}
S_{\eta=\etat} [f] = K \int \dd x^+ \dd x^- \; \kappa_D\left( \bigl(f^{-1}\p_+ f \bigr)^{(1)},\frac{1}{1-\eta \, \mathfrak{R}_f \circ \pi^{(1)}} \bigl( f^{-1} \p_- f \bigr)^{(1)} \right).
\end{equation*}
This is the Yang-Baxter type deformation of the $\Z_2$-coset $\s$-model on $DG_0/DG_0^{(0)}$, as described in Subsection \ref{SubSec:dZ2}. Thus, the one-parameter limit $\eta=\etat$ is also an integrable model with twist function.

In the same way, we check that the limit $\eta=-\etat$ is also a Yang-Baxter type deformation of the $\Z_2$-coset $\s$-model on $DG_0/DG_0^{(0)}$, but with the solution $\mathfrak{R}=\bigl(R,-\Rt\bigr)$ of the mCYBE on $D\g_0$. Thus, this is also a one-parameter limit which admits a twist function. Moreover, let us note that the twist function of the limit $\eta=\etat$ is the same as the one of the limit $\eta=-\etat$, as the twist function of a Yang-Baxter type deformation does not depend on the choice of the matrix $\mathfrak{R}$.

\subsection{Lax pair of the BYBM}
\label{SubSec:BYBLax}

In this section, we shall show that the BYBM equations of motion can be recast into a Lax equation. This was first done by Klim\v{c}ik in~\cite{Klimcik:2014bta}, using the gauge-fixed action \eqref{Eq:ActionBYB2}. Here, we shall start from the non-gauge-fixed one \eqref{Eq:ActionBYB} and follow a construction close to the one for the Lax pair of the $\Z_2$-coset model (see Subsection \ref{SubSec:ZT}). We will explain the main step of this construction and refer to our article~\cite{Delduc:2015xdm} for more details.

\paragraph{Equations of motion.} Varying the action \eqref{Eq:ActionBYB} with respect to the field $g$, one finds the following equations of motion:
\begin{equation}\label{Eq:BYBeom}
EOM=\p_+J_-+[a_+,J_-] + \p_-J_++[a_-,J_+] = 0,
\end{equation}
where we introduced
\begin{equation}\label{Eq:BYBJ}
J_\pm = \left( 1\pm\dfrac{\eta}{2}R_g\pm\dfrac{\etat}{2}\Rt_{\gt} \right)^{-1} (j^L_\pm-\jt^L_\pm)
\end{equation}
and a ``gauge field''
\begin{equation}\label{Eq:BYBa}
a_\pm = j^L_\pm \mp \frac{\eta}{2} R_gJ_\pm = \left(1\pm \frac{\etat}{2}\Rt_{\gt}\right)J_\pm+\jt^L_\pm .
\end{equation}
We note that the equation of motion is invariant under the redefinition $a_\pm \mapsto a_\pm + \rho J_\pm$ of $a_\pm$, where $\rho$ is a constant. Varying the field $\gt$ in the action \eqref{Eq:ActionBYB}, one finds some other equations of motion. Using the freedom in the definition of the gauge field $a_\pm$ mentioned above, one checks that this equation of motion is equivalent to the one \eqref{Eq:BYBeom} on $g$. This redundancy in the equations of motion is due to the gauge symmetry \eqref{Eq:GaugeBYB} of the model.

\paragraph{Search for a Lax pair.} The equation of motion \eqref{Eq:BYBeom} takes the form of a covariant conservation equation, for the current $J_\pm$, with the covariant derivative $\p_\pm + \left[a_\pm,\cdot\right]$. This was also the form of the equations of motion of the $\Z_2$-coset $\s$-model in Subsection \ref{SubSec:ZT}. We shall use the results about the Lax representation of the $\Z_2$-coset model as a guide for the search of a Lax pair for the BYBM. In particular, following equation \eqref{Eq:LaxZ2LC}, one could try to define the Lax pair $\Lc^{\text{BYB}}_\pm(\lambda) = a_\pm + \lambda^{\pm 1} J_\pm$.

However, let us note that the covariant conservation equation on $J_\pm$ is also verified by the current $\frac{\zeta}{2}J_\pm$, for any constant $\zeta$. Moreover, recall that the ``gauge-field'' $a_\pm$ can be redefined by adding to it a term proportional to $J_\pm$. For now on, we shall then consider the most general gauge-field
\begin{equation}\label{Eq:BYBA}
A_\pm = a_\pm + \rho J_\pm,
\end{equation}
with $\rho$ an arbitrary constant, and define the Lax pair to be
\begin{equation}\label{Eq:LaxPairBYB}
\Lc^{\text{BYB}}_\pm (\lambda) = A_\pm + \frac{\zeta}{2}\lambda^{\pm 1} J_\pm.
\end{equation}
Following equation \eqref{Eq:ZCEZ2}, we then find
\begin{align}\label{Eq:ZCELaxBYB}
&\hspace{-40pt}\p_+ \Lc^{\text{BYB}}_-(\lambda) - \p_- \Lc^{\text{BYB}}_+(\lambda) + \bigl[ \Lc^{\text{BYB}}_+(\lambda), \Lc^{\text{BYB}}_-(\lambda) \bigr] \\
& = \;\;\; \frac{\zeta}{4} \left( \lambda + \frac{1}{\lambda} \right) EOM + \left( \p_+ A_- - \p_- A_+ + \bigl[ A_+, A_- \bigr] + \frac{\zeta^2}{4}\bigl[ J_+, J_- \bigr] \right) \notag \\
& \hspace{35pt} + \frac{\zeta}{4} \left( \lambda + \frac{1}{\lambda} \right) \Bigl( \p_+ J_- - \p_- J_+ + \bigl[ A_+, J_- \bigr] + \bigl[ J_+, A_- \bigr] \Bigr). \notag
\end{align}
The first term on the right-hand side obviously vanishes on-shell, \textit{i.e.} when the equations of motion $EOM=0$ are satisfied. Using the fact that $R_g$ and $\Rt_{\gt}$ are solutions of the mCYBE \eqref{Eq:mCYBE}, combined with the Maurer-Cartan equation \eqref{Eq:MaurerCartanLC} on $j^L_\pm$ (and the one for $\jt^L_\pm$), one finds that
\begin{equation*}
\p_+ J_- - \p_- J_+ + \bigl[ A_+, J_- \bigr] + \bigl[ J_+, A_- \bigr]  = \left( 2\rho + 1 - \frac{\eta^2-\etat^2}{2} \right) \bigl[J_+, J_-\bigr] + \frac{1}{2} \bigl( \eta R_g + \etat \Rt_{\gt} \bigr) (EOM). 
\end{equation*}
Thus, if we choose
\begin{equation}\label{Eq:BYBrho}
\rho = - \frac{1}{2}\left( 1 - \frac{\eta^2-\etat^2}{4} \right),
\end{equation}
we see that the last term in equation \eqref{Eq:ZCELaxBYB} vanishes on shell. In the same way, using the mCYBE and the Maurer-Cartan equations, one gets
\begin{equation*}
\p_+ A_- - \p_- A_+ + \bigl[ A_+, A_- \bigr] + \frac{\zeta^2}{4}\bigl[ J_+, J_- \bigr] = - \frac{1}{4}\left( 1 - \frac{\eta^2-\etat^2}{2} \right) \bigl( \eta R_g + \etat \Rt_{\gt} \bigr) (EOM) + \eta R_g(EOM),
\end{equation*}
for
\begin{equation}\label{Eq:BYBzeta}
\zeta = \sqrt{\left(1+\frac{(\eta+\etat)^2}{4}\right)\left(1+\frac{(\eta-\etat)^2}{4}\right)}.
\end{equation}
Thus, for the choice \eqref{Eq:BYBrho} and \eqref{Eq:BYBzeta} of $\rho$ and $\zeta$, one finds that the vanishing of \eqref{Eq:ZCELaxBYB} is equivalent to the one of $EOM$. In other words, we found a Lax pair formulation of the equations of motion of the BYBM.

\paragraph{Lax pair in the double algebra.} As explained above, the action \eqref{Eq:ActionBYB} is interpreted as a deformation of the $\Z_2$-coset $\s$-model on $DG_0/DG_0^{(0)}$. Thus, the Lax pair of the model should naturally belong to the double algebra $D\g=\g\times\g$ (where $\g$ is the complexification of $\g_0$). However, in the previous paragraph, we constructed a Lax pair $\Lc_\pm^{\text{BYB}}(\lambda)$ in a single copy of $\g$. We shall now see how this Lax pair can be ``extended'' to a Lax pair $\Lc^D_\pm$ valued in the double algebra $\g\times\g$.

It is clear that the Lax equation \eqref{Eq:LaxLC} for $\Lc^D_\pm$ is equivalent to the Lax equations for the two $\g$-valued factors of $\Lc^D_\pm$. Thus, we shall choose the left factor to be $\Lc^{\text{BYB}}_\pm$. Moreover, as we are considering a deformation of the $\Z_2$-coset model with involution $\delta$ (the exchange automorphism), it is natural to ask the Lax pair $\Lc^D_\pm$ to  satisfy the equivariance relation
\begin{equation*}
\delta \left( \Lc^D_\pm(\lambda) \right) = \Lc^D_\pm(-\lambda).
\end{equation*}
This fixes the right factor of $\Lc^D_\pm(\lambda)$ to be $\Lc_\pm^{\text{BYB}}(-\lambda)$. We then define
\begin{equation*}
\Lc^D_\pm(\lambda) = \left( \Lc_\pm^{\text{BYB}}(\lambda), \Lc_\pm^{\text{BYB}}(-\lambda) \right) \in D\g.
\end{equation*}
It is clear that the Lax equation for $\Lc^D_\pm$ is satisfied, as it is satisfied by $\Lc^{\text{BYB}}_\pm$.\\

In the undeformed limit $\eta=\etat=0$, we have $J_\pm=j^L_\pm-\jt^L_\pm$, $A_\pm = \frac{1}{2}(j^L_\pm+\jt^L_\pm)$ and $\zeta=1$. Thus, in this limit, one gets
\begin{equation*}
\Lc^D_\pm(\lambda) = \frac{1}{2} \bigl(j^L_\pm+\jt^L_\pm,j^L_\pm+\jt^L_\pm\bigr) + \frac{\lambda^{\pm 1}}{2} \bigl(j^L_\pm-\jt^L_\pm,j^L_\pm-\jt^L_\pm\bigr) = \bigl(f^{-1}\p_\pm f \bigr)^{(0)} + \lambda^{\pm 1} \bigl(f^{-1}\p_\pm f \bigr)^{(1)},
\end{equation*} 
with the $DG_0$-valued field $f=(g,\gt)$. Thus, for $\eta=\etat=0$, $\Lc^D_\pm$ reduces to the Lax matrix of the $\Z_2$-coset model on $DG_0/DG^{(0)}_0$, as constructed in equation \eqref{Eq:LaxZ2LC}.

As we explained in the last paragraph of Subsection \ref{SubSec:BYBLag}, the one-parameter limits $\eta=\pm\etat$ coincide with the Yang-Baxter type deformation of the $\Z_2$-coset on $DG_0/DG^{(0)}_0$. One can also check that in this limit, the Lax pair $\Lc^D_\pm$ reduces to the Lax pair considered in Subsection \ref{SubSec:dZ2} for the deformed $\Z_2$-coset model.

\paragraph{Lax matrix.} In this subsection, we have constructed light-cone Lax pairs $\Lc^{\text{BYB}}_\pm$ and $\Lc^D_\pm$ for the BYBM, valued respectively in $\g$ and $D\g$. From these pairs, one can construct the corresponding Lax matrices
\begin{equation}\label{Eq:LaxBYBLag}
\Lc_{\text{BYB}}(\lambda)=\frac{1}{2} \left( \Lc^{\text{BYB}}_+(\lambda) - \Lc^{\text{BYB}}_-(\lambda) \right) \;\;\;\; \text{ and } \;\;\;\; \Lc_D(\lambda) = \frac{1}{2} \left( \Lc^D_+(\lambda) - \Lc^D_-(\lambda) \right).
\end{equation}
One then has
\begin{equation*}
\Lc_{\text{BYB}}(\lambda) = \frac{A_+-A_-}{2} + \frac{\zeta\lambda}{4} J_+ - \frac{\zeta}{4\lambda} J_-
\end{equation*}
and
\begin{equation}\label{Eq:BYBLd}
\Lc_D(\lambda) = \left( \Lc_{\text{BYB}}(\lambda), \Lc_{\text{BYB}}(-\lambda) \right).
\end{equation}

\subsection{$\Rc$-matrices on the double algebra}

As explained in the previous subsection, the Lax matrix $\Lc_D$ of the BYBM in its non-gauged-fixed formulation is naturally valued in the double algebra $D\g$. As we will prove later, the Hamiltonian integrability of the BYBM is ensured by the fact that this Lax matrix satisfies a Maillet non-ultralocal bracket \eqref{Eq:PBR}, with a $\Rc$-matrix $\Rc^D\ti{12}$ valued in $D\g \otimes D\g$. As the BYBM is a deformation of the $\Z_2$-coset model on $DG_0/DG_0^{(0)}$, one could expect $\Rc^D\ti{12}$ to be of the form \eqref{Eq:DefR}, with a twist function $\vp$ and $\Rc^0$ the standard $\Rc$-matrix on $\Lc(D\g)$ twisted by the automorphism $\delta$. However, in the case of the BYBM, the matrix $\Rc^0$ takes a slightly more general form, that we shall explain here.

\paragraph{The standard construction.} Let us recall the main steps of the construction of the standard twisted $\Rc$-matrix on $\Lc(D\g)$ (see the Appendix \ref{App:RMat} for the general theory). Let us consider the loop algebra $\Lc(D\g)=D\g (\!(\lambda)\!) =D \g \otimes \C(\!(\lambda)\!)$ of Laurent series in a complex parameter $\lambda$ valued in the double algebra $D\g$ and equipped with the natural Lie bracket. The exchange automorphism $\delta$ on $D\g$ induces an automorphism $\hat{\delta}$ on $D\g(\!(\lambda)\!)$ defined for all $M\in D\g(\!(\lambda)\!)$ by
\vspace{-4pt}\begin{equation*}
\hat{\delta}(M)(\lambda)=\delta\bigl(M(-\lambda)\bigr).\vspace{-5pt}
\end{equation*}
Denote by $D\g(\!(\lambda)\!)^{\hat{\delta}}$ the twisted loop algebra, \textit{i.e.} the subalgebra of $D\g(\!(\lambda)\!)$ formed by the fixed points of $\hat{\delta}$. It admits a natural vector space decomposition
\vspace{-5pt}\begin{equation} \label{Eq:DgDecomp}
D \g(\!(\lambda)\!)^{\hat{\delta}} = D \g[[\lambda]]^{\hat{\delta}} \oplus \left(\lambda^{-1} D \g[\lambda^{-1}]\right)^{\hat{\delta}} \vspace{-3pt}	
\end{equation}
into subalgebras of positive and strictly negative powers of the loop parameter $\lambda$, respectively. Let $\pi_+$ and $\pi_-$ denote the projection operators relative to this decomposition. The operator
\vspace{-4pt}\begin{equation} \label{Eq:ROp}
\Rc^D = \pi_+ - \pi_-\vspace{-4pt}
\end{equation}
defines a solution of the mCYBE on $D \g(\!(\lambda)\!)^{\hat{\delta}}$. Suppose now that we are given an invariant inner product $\langle \cdot, \cdot \rangle$ on the twisted loop algebra $D \g(\!(\lambda)\!)^{\hat{\delta}}$. We define the kernel $\Rc^D\ti{12}(\lambda, \mu)$ of the operator $\Rc^D$ in \eqref{Eq:ROp}, with respect to $\langle \cdot, \cdot \rangle$, as the rational function $\Rc^D\ti{12}(\lambda, \mu)$ of two complex variables and valued in $D\g \otimes D\g$, such that for all $M \in D \g(\!(\lambda)\!)^{\hat{\delta}}$ we have
\begin{equation*}
\langle \Rc^D\ti{12}(\lambda, \mu),M\ti{2}(\mu) \rangle \ti{2} = (\Rc^D M)(\lambda).
\end{equation*}
This matrix is then a solution of the CYBE \eqref{Eq:CYBE} (see Appendix \ref{App:RMat} for more details about this construction). The usual construction for $\Rc$-matrices with twist function is based on the inner product
\vspace{-2pt}\begin{equation}\label{Eq:psOdd}
\langle M, N \rangle_\varphi = 2 \; \res_{\lambda=0} \; \kappa_D \bigl( M(\lambda),N(\lambda) \bigr) \varphi(\lambda) \dd\lambda,\vspace{-3pt}
\end{equation}
for any $M, N \in D\g(\!(\lambda)\!)$. It is easy to check that this inner product is invariant under $\hat{\delta}$ and thus induces an inner product on the twisted loop algebra $D\g(\!( \lambda )\!)^{\hat \delta}$, if and only if $\varphi$ is an odd function. This yields a kernel $\Rc^D\ti{12}(\lambda,\mu)$ of the form \eqref{Eq:DefR}, with twist function $\vp$ and $\Rc^0$ the standard matrix on $\Lc(D\g)$ twisted by $\delta$. The parity condition on $\vp$ is then the equivariance condition \eqref{Eq:TwistEqui}.

\paragraph{Inner product for the BYBM.} We shall now present a more general construction, which allows for a non-odd twist function $\vp$. As we are considering the double Lie algebra $D\g$, one can define an even more general inner product invariant under $\hat{\delta}$, by separating explicitly the left and right part of $D\g$. That is, for any $M=(m,\tilde{m})$ and $N=(n,\tilde{n})$ in $D\g(\!(\lambda)\!)$ we define
\begin{equation}\label{Eq:ps}
\langle M, N \rangle_\varphi = 2\;\res_{\lambda=0} \; \kappa \bigl( m(\lambda),n(\lambda) \bigr) \varphi_{\text{BYB}}(\lambda) \dd\lambda - 2\;\res_{\lambda=0} \; \kappa \bigl( \tilde{m}(\lambda),\tilde{n}(\lambda) \bigr) \varphi_{\text{BYB}}(-\lambda) \dd\lambda,
\end{equation}
where $\kappa$ is the Killing form on $\g$. When $\varphi_{\text{BYB}}$ is odd, we recover the twisted inner product \eqref{Eq:psOdd}. This construction allows to consider twist functions without parity constraint. The kernel of $\Rc^D$ with respect to the inner product \eqref{Eq:ps} is given by
\vspace{-3pt}\begin{equation}\label{Eq:RDouble}
\Rc^D\ti{12}(\lambda, \mu) = \frac{1}{2}\left( \frac{C^{LL}\ti{12}}{\mu - \lambda} + \frac{C^{RL}\ti{12}}{\mu + \lambda} \right) \varphi_{\text{BYB}}(\mu)^{-1} - \frac{1}{2}\left( \frac{C^{RR}\ti{12}}{\mu - \lambda} + \frac{C^{LR}\ti{12}}{\mu + \lambda} \right) \varphi_{\text{BYB}}(- \mu)^{-1},\vspace{-4pt}
\end{equation}
where we defined the partial split Casimirs ($\kappa^{ab}$ is the Killing form of $\g$ in the basis $\lbrace I^a \rbrace$)
\vspace{-3pt}\begin{alignat*}{2}
C^{LL}\ti{12} &= \kappa^{ab}(I_a,0) \otimes (I_b,0), &\qquad
C^{RR}\ti{12} &= \kappa^{ab}(0,I_a) \otimes (0,I_b), \\
C^{LR}\ti{12} &= \kappa^{ab}(I_a,0) \otimes (0,I_b), &\qquad
C^{RL}\ti{12} &= \kappa^{ab}(0,I_a) \otimes (I_b,0).
\end{alignat*}
If $\vp_{\text{BYB}}$ were an odd function, we would get
\vspace{-6pt}\begin{equation*}
\Rc^D\ti{12}(\lambda,\mu) = \Rc^{0,D}\ti{12}(\lambda,\mu)\vp_{\text{BYB}}(\mu)^{-1}, \;\;\;\; \text{ with } \;\;\;\; \Rc^{0,D}\ti{12}(\lambda,\mu) = \frac{1}{2} \left( \frac{C\ti{12}^D}{\mu-\lambda} + \frac{\delta\ti{1}C^D\ti{12}}{\mu+\lambda} \right),
\end{equation*}
where $C^D\ti{12} = C^{LL}\ti{12} + C^{RR}\ti{12}$ is the split Casimir of the double algebra $D\g$. $\Rc^{0,D}$ is the standard $\Rc$-matrix on $\Lc(D\g)$ twisted by $\delta$. We thus recover the usual notion of $\Rc$-matrix with twist function $\vp_{\text{BYB}}$ in this case. The twist function $\vp_{\text{BYB}}$ of the BYBM that we will find is not an odd function in general, so we will need the matrix \eqref{Eq:RDouble}. However, in the limit $\eta=\pm\etat$, the twist function $\vp_{\text{BYB}}$ will be odd and we will recover the usual setting, as expected from the fact that this limit is simply the $\eta$-deformation of the $\Z_2$-coset model on $DG_0/DG^{(0)}_0$.

\paragraph{Projection on $\bm{\g}$.} Let us suppose that the $D\g$-valued Lax matrix $\Lc_D$ \eqref{Eq:BYBLd} satisfies a Maillet bracket \eqref{Eq:PBR} with $\Rc$-matrix \eqref{Eq:RDouble}. Projecting this Poisson bracket on the left factor of $D\g=\g\times\g$, one finds that the Lax matrix $\Lc_{\text{BYB}}$ also satisfies a Maillet bracket with the $\Rc$-matrix:
\begin{equation}\label{Eq:RBYB}
\Rc^{\text{BYB}}\ti{12} = \frac{1}{2}\frac{C\ti{12}}{\mu-\lambda} \vp_{\text{BYB}}(\mu)^{-1},
\end{equation}
with $C\ti{12} = \kappa^{ab} I_a \otimes I_b$ the split Casimir of $\g$. Conversely, if $\Lc_{\text{BYB}}$ satisfies the Maillet bracket with $\Rc$-matrix \eqref{Eq:RBYB}, one checks that the matrix $\Lc^D$ in the double algebra satisfies the bracket with $\Rc$-matrix \eqref{Eq:RDouble}. For the rest of this section, we shall then focus on the Poisson bracket of $\Lc_{\text{BYB}}$ and find the corresponding twist function $\vp_{\text{BYB}}$.

\subsection{Hamiltonian analysis and twist function of the BYBM}

In Subsection \ref{SubSec:BYBLax}, we have presented the Lax pair of the BYBM. We will now compute the Poisson bracket of the Lax matrix with itself. For that, we first need to go from the Lagrangian formulation \eqref{Eq:ActionBYB} of the BYBM to its Hamiltonian formulation.

\paragraph{Phase space, constraint and Hamiltonian.} The BYBM in its non-gauge-fixed formulation \eqref{Eq:ActionBYB} is defined on two $G_0$-valued fields $g$ and $\gt$. Following Subsection \ref{SubSec:SigmaModelHam}, the phase space of the BYBM is thus parametrised in terms of the $G_0$-valued fields $g$ and $\gt$ and the $\g_0$-valued fields $X$ and $\Xt$. The Poisson bracket of $g$ and $X$ is then \eqref{Eq:PBTstarG}. The fields $\gt$ and $\Xt$ satisfy exactly the same brackets as $g$ and $X$. Moreover, the fields $g$ and $X$ have zero Poisson brackets with the fields $\gt$ and $\Xt$.

Starting from the action \eqref{Eq:ActionBYB}, one can compute the conjugate momenta of the BYBM and deduce the Lagrangian expression of $X$ and $\Xt$. One then finds
\begin{equation}\label{Eq:BYBX}
X \approx -\Xt \approx \frac{K}{4} (J_+ + J_-),
\end{equation}
where the current $J_\pm$ was defined in \eqref{Eq:BYBJ} and the use of the symbol $\approx$ instead of $=$ will be explained in what follows. We deduce from equation \eqref{Eq:BYBX} that 
\begin{equation}\label{Eq:BYBConst}
X+\Xt \approx 0.
\end{equation}
This is a constraint on the phase space, which is due to the existence of a gauge symmetry in the model. We refer to the discussion of constraints in coset models in Subsection \ref{SubSec:ZT} and will use the notations introduced there in the rest of the section (in particular, the symbol $\approx$ holds for weak equations, true only on the constrained phase space).\\

In Subsection \ref{SubSec:BYBLag}, we explained that we can see the BYBM as a deformation of a $\Z_2$-coset model, with dynamical field $f=(g,\gt)$ in the double group $DG_0$. Thus, one can also parametrise the phase space of the model with the $DG_0$-valued field $f$ and a $D\g_0$-valued field $Z$ (equivalent of the field $X$ for $f$). One then finds
\begin{equation*}
Z = (X,\Xt) \approx K \bigl( J^D_0 \bigr)^{(1)},
\end{equation*}
with
\begin{equation*}
J^D_0 = \frac{J^D_++J^D_-}{2} \;\;\;\; \text{ and } \;\;\;\; J^D_\pm = \left( \left( 1\pm\dfrac{\eta}{2}R_g\pm\dfrac{\etat}{2}\Rt_{\gt} \right)^{-1} j^L_\pm, \left( 1\pm\dfrac{\eta}{2}R_g\pm\dfrac{\etat}{2}\Rt_{\gt} \right)^{-1} \jt^L_\pm \right).
\end{equation*}
We thus re-express the constraint \eqref{Eq:BYBConst} as $Z^{(0)}\approx 0$ (as we would have in the undeformed limit). Moreover, in the undeformed limit $\eta=\etat=0$, we have $J^D_0 = f^{-1} \p_t f$. Thus, we recover $Z \approx K \bigl( f^{-1} \p_t f \bigr)^{(1)}$, as we expect from equation \eqref{Eq:XZ2}.\\

Let us end this subsection by expressing the Hamiltonian of the BYBM. Starting from equation \eqref{Eq:ActionBYB} and performing the Legendre transformation, we express the Hamiltonian in terms of the current $J_\pm$ as
\begin{equation*}
\Hc_{\text{BYB}} = \frac{K}{8} \int \dd x \; \Bigl( \kappa(J_+,J_+) + \kappa(J_-,J_-) + \kappa(\Lambda,X+\Xt) \Bigr),
\end{equation*}
where $\Lambda$ is a Lagrange multiplier associated with the gauge constraint $X+\Xt$. In terms of the $D\g_0$-valued field currents $Z=K\bigl(J^D_0\bigr)^{(1)}$ and $J^D_1=\frac{1}{2}(J^D_+-J^D_-)$, one finds
\begin{equation*}
\Hc_{\text{BYB}} = \frac{1}{2}\int \dd x \; \left( \frac{1}{K} \kappa_D( Z, Z ) + K \, \kappa_D \Bigl( J^{D\,(1)}_1,J^{D\,(1)}_1 \Bigr) + \kappa_D \Bigl(\mu, Z^{(0)} \Bigr)\right),
\end{equation*}
with $\mu$ a $D\g_0^{(0)}$-valued Lagrange multiplier. In the undeformed limit $\eta=\etat=0$, one has $J^D_1 = (j^L_1,\jt^L_1) = f^{-1} \p_x f$ and we then recover the Hamiltonian \eqref{Eq:HamZ2} for the $\Z_2$-coset $DG_0/DG_0^{(0)}$.

\paragraph{Hamiltonian Lax matrix.} Let us consider the Lax matrix \eqref{Eq:LaxBYBLag}. Using the expression \eqref{Eq:BYBX} of $X$ and $\Xt$, one can re-express the currents $J_\pm$ and $A_\pm$ in terms of $g$, $\gt$, $j=j^L_1$, $\jt=\jt^L_1$, $X$ and $\Xt$. Thus, we can express the Lax matrix $\Lc_{\text{BYB}}(\lambda)$ in terms of these fields.

Moreover, because of the constraint \eqref{Eq:BYBConst}, one can add to the expression obtained this way a term $f(\lambda)(X+\Xt)$, with $f$ an arbitrary function of the spectral parameter. One could potentially add other extra terms, for instance 
proportional to $R_g(X+\Xt)$ and $\Rt_{\gt}(X+\Xt)$. These allow to change terms like $R_g\Xt$ or $\Rt_{\gt}X$ into $R_gX$ and $\Rt_{\gt}\Xt$. For reasons of symmetry and simplicity, we will 
use this freedom to keep only terms proportional to $R_gX$ and $\Rt_{\gt}\Xt$. Similarly, when dealing with terms proportional to $X$ or $\Xt$ (without operators $R$ or $\Rt$), we shall always transform those in terms proportional to $X-\Xt$ (this can always be done by redefining the function $f(\lambda)$ introduced above).

Following this prescription, we obtain an expression for the Lax matrix $\Lc_{\text{BYB}}(\lambda)$ which is linear in the set of fields $\Oc=\lwb j, \jt, X, \Xt, R_gX, \Rt_{\gt}\Xt \rwb$:
\begin{equation}\label{Eq:LaxHamBYB}
\Lc_{\text{BYB}}(\lambda,x) = \sum_{Q\in\Oc} C_Q(\lambda) Q(x),
\end{equation}
where the $C_Q(\lambda)$'s are non-dynamical functions of the spectral parameter $\lambda$. We shall not detail the computation and refer to~\cite{Delduc:2015xdm}, Section 2.2.3 for more details. The expressions we obtained for the coefficients $C_Q$'s are
\begin{subequations}
\begin{align}
C_j(\lambda) &= \frac{1}{2}\left(1+\frac{\eta^2-\etat^2}{4}+\frac{\zeta}{2}\left(\lambda+\frac{1}{\lambda}\right)\right), \\
C_X(\lambda) &= \frac{\zeta}{4K}\left(\lambda-\frac{1}{\lambda}\right) + f(\lambda), \\
C_{R_gX}(\lambda) &= -\frac{\eta}{2K}\left(1+\frac{\eta^2-\etat^2}{4}+\frac{\zeta}{2}\left(\lambda+\frac{1}{\lambda}\right)\right), \\
C_{\jt}(\lambda) &= \frac{1}{2}\left(1+\frac{\etat^2-\eta^2}{4}-\frac{\zeta}{2}\left(\lambda+\frac{1}{\lambda}\right)\right),\\
C_{\Xt}(\lambda) &= -\frac{\zeta}{4K}\left(\lambda-\frac{1}{\lambda}\right) +f(\lambda), \\
C_{\RgXt}(\lambda) &= -\frac{\etat}{2K}\left(1+\frac{\etat^2-\eta^2}{4}-\frac{\zeta}{2}\left(\lambda+\frac{1}{\lambda}\right)\right).
\end{align}
\end{subequations}

\paragraph{Maillet bracket and twist function.} Starting from the expressions \eqref{Eq:LaxHamBYB}, we can compute the Poisson bracket of the Lax matrix $\Lc_{\text{BYB}}$ with itself. For that, we need some intermediary Poisson brackets, between the different fields $Q\in\Oc$. The brackets between $j,X,R_gX$ and $\jt,\Xt,\RgXt$ are all zero. The brackets of $j$ and $X$ are given by \eqref{Eq:PBXX} and \eqref{Eq:PBj} and $\jt$ and $\Xt$ satisfy similar ones.

The brackets involving the fields $R_gX$ and $\RgXt$ can be computed from the brackets \eqref{Eq:PBTstarG} and \eqref{Eq:PBj}, using the fact that the action of any derivation $\delta$ on $R_gX$ is given by
\begin{equation}\label{Eq:DeltaRgX}
\delta \bigl( R_gX \bigr) = R_g \bigl( \delta X \bigr) + R_g \bigl( \bigl[ g^{-1} \delta g, X \bigr] \bigr) - \bigl[ g^{-1}\delta g, R_g X \bigr].
\end{equation}
In particular, let us mention the following bracket
\begin{equation*}
\left\lbrace (R_gX)\ti{1}(x), (R_gX)\ti{2}(y) \right\rbrace = \left[ C\ti{12}, X\ti{2}(x) \right] \delta_{xy},
\end{equation*}
which is derived using equation \eqref{Eq:DeltaRgX} and the mCYBE \eqref{Eq:mCYBE} on $R_g$ (with $c=i$).\\

We will not show the details of the computation of the Poisson bracket of the Lax matrix \eqref{Eq:LaxHamBYB} with itself here. The key steps can be found in the Sections 3.2 and 3.3 of~\cite{Delduc:2015xdm}. One then finds that this Poisson bracket takes the form of a Maillet bracket \eqref{Eq:PBR} if we choose the function $f$ to be
\begin{equation*}
f(\lambda)= \frac{\zeta^2}{4K}(1+\lambda^2) - \frac{1}{2K}\left(1-\frac{(\eta^2-\etat^2)^2}{16}\right) + \frac{\zeta(\eta^2-\etat^2)}{16K}\left(3\lambda+\frac{1}{\lambda}\right).
\end{equation*}
Moreover, the $\Rc$-matrix $\Rc^{\text{BYB}}\ti{12}(\lambda,\mu)$ of this Maillet bracket is of the form \eqref{Eq:RBYB}, with the twist function
\begin{equation}\label{Eq:TwistBYB}
\vp_{\text{BYB}}(\lambda) = \frac{1}{\zeta^2} \frac{2 K\lambda}{\lambda^4+\dfrac{\eta^2-\etat^2}
{\zeta}\lambda^3+\left(2+\dfrac{(\eta^2-\etat^2)^2-16}{4\zeta^2}\right)\lambda^2+\dfrac{\eta^2-\etat^2}{\zeta}\lambda+1}.
\end{equation}

\paragraph{Analysis of the twist function.} Let us first note that we can rewrite \eqref{Eq:TwistBYB} as
\begin{equation}\label{Eq:TwistBYB2}
\vp_{\text{BYB}}(\lambda) = \frac{2 K\lambda}{(\lambda^2-1)^2+(\zeta^2-1)(\lambda^2+1)^2 + (\eta^2-\etat^2) \zeta \lambda \left(\lambda^2 + \dfrac{\eta^2-\etat^2}{4\zeta} \lambda + 1\right)}.
\end{equation}
This expression is appropriate to discuss the one-parameter limits $\eta=\pm\etat$. Indeed, in this case, we have $\zeta=\sqrt{1+\eta^2}$ so it is clear that the twist function \eqref{Eq:TwistBYB2} reduces to the one \eqref{Eq:TwistdZ2} of the Yang-Baxter deformation of a $\Z_2$-coset model (with $c=i$, as here).\\ 

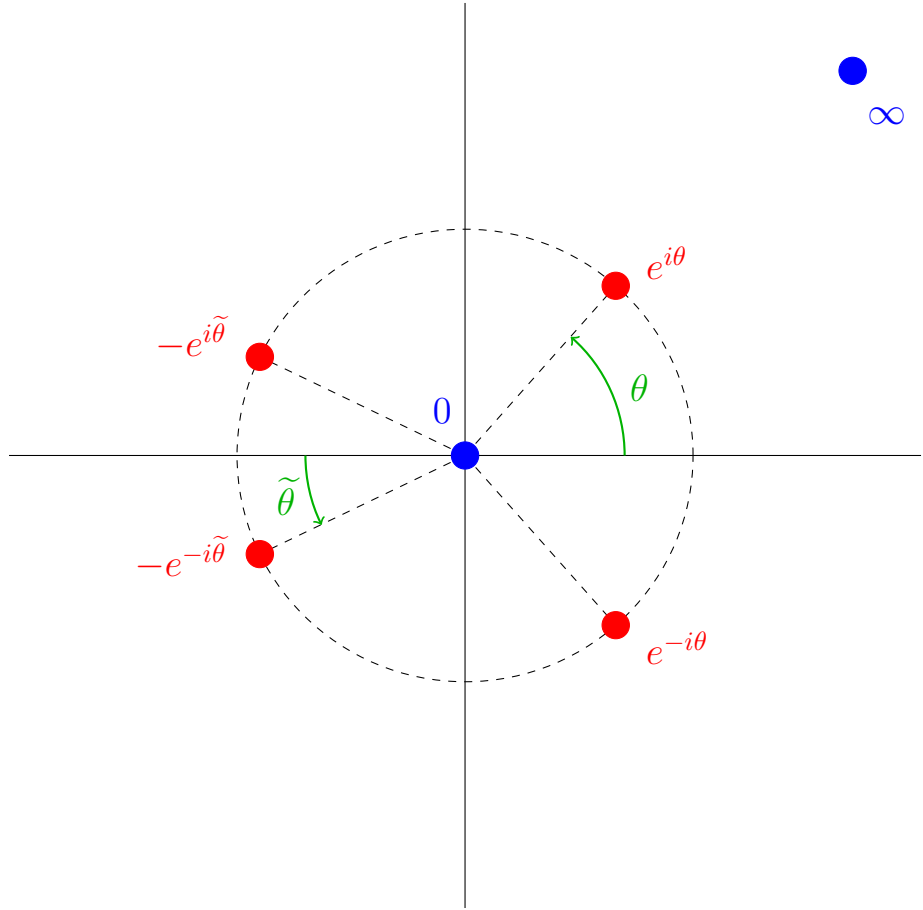
\begin{figure}[H]
\begin{center}
	\begin{tikzpicture}[scale=3]
 		\draw (-2,0) to (2,0);
		\draw (0,-2) to (0,2);
		\draw[dashed] (0,0) to (0.661,0.75);
		\draw[dashed] (0,0) to (0.661,-0.75);
		\draw[dashed] (0,0) to (-0.9,0.436);
		\draw[dashed] (0,0) to (-0.9,-0.436);
		\draw[dashed] (0,0) circle (1);
		\draw[red,fill=red] (0.661,0.75) circle (0.06);
		\draw[red,fill=red] (0.661,-0.75) circle (0.06);
		\node[red,right] at (0.75,0.85) {\Large $e^{i\theta}$};
		\node[red,right] at (0.75,-0.85) {\Large $e^{-i\theta}$}; 
		\draw[red,fill=red] (-0.9,0.436) circle (0.06);
		\draw[red,fill=red] (-0.9,-0.436) circle (0.06);
		\node[red,left] at (-1,0.52) {\Large $-e^{i\widetilde\theta}$};
		\node[red,left] at (-1,-0.45) {\Large $-e^{-i\widetilde\theta}$};
		\draw[blue,fill=blue] (1.7,1.7) circle (0.06);
		\node[blue] at (1.85,1.5) {\Large $\infty$};
		\draw[blue,fill=blue] (0,0) circle (0.06);
		\node[blue,above] at (-0.1,0.1) {\Large $0$};
		\draw[thick,myGreen,->] ([shift=(0:0.7cm)]0,0) arc (0:48.5:0.7cm);
		\node[myGreen,right] at (0.68,0.3) {\Large $\theta$}; 
		\draw[thick,myGreen,->] ([shift=(0:-0.7cm)]0,0) arc (0:26:-0.7cm);
		\node[myGreen,right] at (-0.87,-0.18) {\Large $\widetilde\theta$}; 
 	\end{tikzpicture}
\end{center}
\caption{{\color{red}Poles} and {\color{blue} zeros} of the twist function of the BYBM.}
\label{Fig:PolesZerosBYB}
\end{figure}

Let us now study the analytical properties of the 1-form $\vp_{\text{BYB}}(\lambda) \, \dd\lambda$. It possesses two simple zeros at $0$ and infinity, similarly to the undeformed model. The double poles $+1$ and $-1$ of the undeformed model have been splited into two pairs $(\lambda_+,\lambda_-)$ and $(\lt_+,\lt_-)$ of simple poles by the Bi-Yang-Baxter deformation. The expressions of these poles in terms of the parameters $\eta$ and $\etat$ are given by
\begin{subequations}
\begin{align}
\lambda_\pm & =  \frac{1-\frac{1}{4}(\eta^2-\etat^2) \pm i  \eta}{\zeta} = \lambda_\mp^* \\
\lt_\pm & =-\frac{1+\frac{1}{4}(\eta^2-\etat^2) \pm i  \etat}{\zeta}= \lt_\mp^*.
\end{align}
\end{subequations}
These poles belong to the unit circle. They can be written in a trigonometric form $\lambda_\pm = e^{\pm i \theta}$ and $\lt_\pm=-e^{\pm i \widetilde{\theta}}$, with the angles $\theta$ and $\widetilde{\theta}$ defined by
\begin{equation*}
\tan\theta = \frac{\eta}{1 - \frac{1}{4}(\eta^2-\etat^2)} \;\;\;\; \text{ and } \;\;\;\; \tan\widetilde{\theta} = \frac{\etat}{1 + \frac{1}{4}(\eta^2-\etat^2)}.
\end{equation*}
The zeros and poles of the 1-form $\vp_{\text{BYB}}(\lambda)\,\dd\lambda$ are represented in Figure \ref{Fig:PolesZerosBYB}.

In the one-parameter limit $\eta=\etat$, we recover $\theta=\widetilde\theta=\arctan\eta$. The poles of the twist function are then as in Figure \ref{Fig:PolesZerosZ2eta}. In particular, in this limit we recover the symmetry of the poles with respect to the symmetry $\lambda\mapsto-\lambda$, as the twist function reduces to the odd one in \eqref{Eq:TwistdZ2}. The Bi-Yang-Baxter deformation allows to split the double poles $\pm 1$ with different angles, by breaking the equivariance relation $\vp(-\lambda)=-\vp(\lambda)$. In the gauged-fixed formulation \eqref{Eq:ActionBYB2}, this is related to the fact that $\eta$ and $\etat$ control the breaking of respectively the left and the right symmetry of the PCM. We shall explain this in Chapter \ref{Chap:PLie}.

\paragraph{Yang-Baxter type deformations.} As we just observed, the Bi-Yang-Baxter deformation has for effect to split the double poles $\pm 1$ into pairs of simple poles and to break the left and right symmetries of the undeformed model. In the previous Section \ref{Sec:DefModel}, we have observed that these effects were some of the features shared by the Yang-Baxter type deformations. Let us consider the gauge transformation of the Lax matrix $\Lc_{\text{BYB}}(\lambda)$ by the $G_0$-valued field $g$ and evaluate it at the poles $\lambda=\lambda_\pm$ of the twist function. One then finds
\begin{equation}\label{Eq:LgBYBz2}
\Lc^g_{\text{BYB}}(\lambda_\pm) = -\gamma R^\mp \bigl( gXg^{-1} \bigr), \;\;\;\; \text{ with } \;\;\;\; \frac{1}{\gamma} =  \frac{K}{\eta} = \pm 4i \; \res_{\lambda=\lambda_\pm} \, \vp_{\text{BYB}}(\lambda) \; \dd \lambda,
\end{equation}
with $R^\pm = R \pm i \,\Id$. Similarly, one finds
\begin{equation}\label{Eq:LgtBYBz2}
\Lc^{\gt}_{\text{BYB}}(\lt_\pm) = -\widetilde\gamma \Rt^\mp \bigl( \gt\Xt\gt^{-1} \bigr), \;\;\;\; \text{ with } \;\;\;\; \frac{1}{\widetilde\gamma} = \frac{K}{\etat} = \pm 4i \; \res_{\lambda=\lt_\pm} \, \vp_{\text{BYB}}(\lambda) \; \dd \lambda. 
\end{equation}
This is to be compared with the expressions \eqref{Eq:LgYB} and \eqref{Eq:LgdZ2}. It is also one of the characteristic feature of Yang-Baxter type deformations and will allow us to apply the results of Chapter \ref{Chap:PLie} to the BYBM.

\subsection{Gauge-fixing and Lax matrix}
\label{SubSec:BYBgauge-fixed}

As explained in Subsection \ref{SubSec:BYBLag}, the gauge-fixed BYBM is given by the action \eqref{Eq:ActionBYB2} and is obtained by performing a gauge transformation \eqref{Eq:GaugeBYB} with $h=\gt^{-1}$. We shall use the results for the non-gauge-fixed model to describe the Lax matrix of the gauge-fixed one and the corresponding Maillet bracket.

\paragraph{Lax pair and gauge transformation.} We first need to understand how a gauge transformation \eqref{Eq:GaugeBYB} acts on the Lax pair \eqref{Eq:LaxPairBYB} of the non-gauge-fixed model. From equation \eqref{Eq:BYBJ}, one finds that the currents $J_\pm$ transform covariantly under the gauge transformation \eqref{Eq:GaugeBYB}:
\begin{equation*}
J_\pm \longmapsto h^{-1} J_\pm h.
\end{equation*}
In the same way, using the expression \eqref{Eq:BYBa} and \eqref{Eq:BYBA} of $a_\pm$ and $A_\pm$, one shows that they transform as gauge fields:
\begin{equation*}
a_\pm \longmapsto h^{-1} a_\pm h + h^{-1}\p_\pm h \;\;\;\; \text{ and } \;\;\;\; A_\pm \longmapsto h^{-1} A_\pm h + h^{-1}\p_\pm h.
\end{equation*}
We note in passing that these two properties ensure the invariance of the equation of motion \eqref{Eq:BYBeom} under the gauge transformation, as expected.

A consequence of these transformation properties is that the Lax pair \eqref{Eq:LaxPairBYB} of the non-gauged-fixed BYBM transforms under a gauge transformation as
\begin{equation*}
\Lc^{\text{BYB}}_\pm(\lambda) \mapsto h^{-1} \Lc^{\text{BYB}}_\pm(\lambda) h + h^{-1} \p_\pm h = \bigl(\Lc^{\text{BYB}}_\pm \bigr)^{h^{-1}} (\lambda).
\end{equation*}
Thus, the action of a gauge transformation \eqref{Eq:GaugeBYB} on the Lax pair coincides with the formal gauge transformation by $h^{-1}$, as defined in equation \eqref{Eq:FormalGauge}.\\

As a consequence, one performs the gauge-fixing of the model at the level of the Lax pair by considering the formal gauge transformation by $\gt$. Moreover, we shall consider the following change of spectral parameter:
\begin{equation}\label{Eq:ParamCosetToPCM}
\lambda = \frac{1+\xi}{1-\xi}.
\end{equation}
We then define the gauge-fixed Lax pair of the BYBM as
\begin{equation*}
\Lc^{\text{GF}}_\pm(\xi) = \bigl(\Lc^{\text{BYB}}_\pm \bigr)^{\gt}\left(\lambda=\frac{1+\xi}{1-\xi}\right).
\end{equation*}
This gauge-fixed Lax pair can be expressed in terms of the gauge-invariant field $g'=g \gt^{-1}$ introduced in Subsection \ref{SubSec:BYBLag}. Using
\begin{equation*}
j'^L_\pm = g'^{-1} \p_\pm g' = \gt \left( j^L_\pm - \jt^L_\pm \right) \gt^{-1},
\end{equation*}
one finds
\begin{equation}\label{Eq:LaxGF}
\Lc^{\text{GF}}_\pm(\xi) = \frac{K_\pm}{1 \mp \xi} + B_\pm,
\end{equation}
with
\begin{subequations}
\begin{align}
K_\pm &= \zeta \left(1 \pm \frac{\eta}{2} R_{g'} \pm \frac{\etat}{2} \Rt \right)^{-1} j'^L_\pm \label{Eq:BYBK} \\
B_\pm &= \frac{1}{2\zeta} \left( 1 + \frac{\eta^2-\etat^2}{4} - \zeta \pm \etat \Rt \right) K_\pm. \label{Eq:BYBB}
\end{align}
\end{subequations}
The Lax pair \eqref{Eq:LaxGF} is the one found by Klim\v{c}ik in~\cite{Klimcik:2014bta}, up to a change of spectral parameter. It is to be compared with the one \eqref{Eq:ZakMik} of the PCM. In particular, due to the presence of the field $B_\pm$, this Lax pair is not of Zakharov-Mikhailov type. This is natural as the Bi-Yang-Baxter deformation breaks both left and right symmetries of the PCM: indeed, one then cannot find a flat and conserved current to construct a Zakharov-Mikhailov Lax pair, as we did for the PCM and the Yang-Baxter model. 

Let us finish by discussing the one-parameter limit $\etat=0$. Recall from Subsection \ref{SubSec:BYBLag}, equation \eqref{Eq:ActionBYB2}, that this limit corresponds to the Yang-Baxter model \eqref{Eq:ActionYBL} (where the deformation parameter $\eta$ is replaced by $\frac{\eta}{2}$). In this case, one finds $\zeta=1+\frac{\eta^2}{4}$, so in particular, the field $B_\pm$ in \eqref{Eq:BYBB} vanishes. We are then left with a Lax pair of Zakharov-Mikhailov type. Moreover, the current $K_\pm$ in \eqref{Eq:BYBK} then coincides with the current $K_\pm$ given for the Yang-Baxter model in \eqref{Eq:CurrentYB} (with $c=i$ and $\eta$ replaced by $\frac{\eta}{2}$). Thus, the gauge-fixed Lax pair of the BYBM reduces to the one of the Yang-Baxter model in the limit $\etat=0$.

One can also take the limit $\eta=0$. In this case, one finds the Yang-Baxter model obtained by breaking the right symmetry instead of the left. As explained in Subsection \ref{SubSec:YB}, this model is equivalent to the one deformed on the left by considering $g'^{-1}$ instead of $g'$, so we shall not discuss it further here.

\paragraph{Maillet bracket and gauge transformation.} As we have explained above, one performs the gauge-fixing of the BYBM on the Lax pair by considering a formal gauge transformation by $\gt$. This can also be done on the Hamiltonian Lax matrix \eqref{Eq:LaxHamBYB} and one can then express the result in terms of the gauge-invariant fields
\begin{equation*}
g' = g \gt^{-1}, \;\;\;\; j' = \gt \left( j - \jt \right) \;\;\;\; \text{ and } \;\;\;\; X' = \gt X \gt^{-1},
\end{equation*}
using the gauge constraint $X+\Xt=0$. One then finds
\begin{equation*}
\Lc^{\gt}_{\text{BYB}}(\lambda) = C_j(\lambda) j' + \bigl( C_X(\lambda) - C_{\Xt}(\lambda) \bigr) X' + C_{R_gX}(\lambda) R_{g'} X' - C_{\RgXt}(\lambda) \Rt X'.
\end{equation*}
Note here that the spectral parameter $\lambda$ is the same as in the non-gauge-fixed Lax matrix \eqref{Eq:LaxHamBYB}. One can also perform the change of spectral parameter \eqref{Eq:ParamCosetToPCM} and define
\begin{equation*}
\Lc_{\text{GF}}(\xi) = \Lc^{\gt}_{\text{BYB}}\left(\lambda=\frac{1+\xi}{1-\xi}\right).
\end{equation*}
This coincides with the Lax matrix one would obtain starting from the Lax pair \eqref{Eq:LaxGF}, and expressing the corresponding Lax matrix in terms of the phase space fields ($g'$ and $X'$) of the gauge-fixed model. One could then compute the Poisson bracket of the Lax matrix with itself using the canonical Poisson brackets in terms of $g'$ and $X'$. Here, we shall use a different method to obtain this result.\\

In the Hamiltonian formalism, one performs the gauge fixing at the level of the Poisson bracket using the Dirac bracket~\cite{dirac1964lectures}. The quantities $g'$ and $X'$ being gauge-invariant, their Dirac bracket coincides with their non-gauged-fixed bracket. Thus, one can compute the Poisson bracket of the gauge-fixed Lax matrix $\Lc_{\text{GF}}$ simply by computing the one of the non-gauged-fixed matrix $\Lc^{\gt}_{\text{BYB}}$.

For that, we shall use a general result on the transformation of Maillet brackets under a formal gauge transformation of the Lax matrix. We will not recall this general result here and refer to our article~\cite{Delduc:2015xdm} for details. Let us summarise the outcome of this procedure: starting form the Maillet bracket of $\Lc_{\text{BYB}}$ with the $\Rc$-matrix \eqref{Eq:RBYB}, one finds that the gauge transformed matrix $\Lc^{\gt}_{\text{BYB}}$ also satisfies a Maillet bracket with $\Rc$-matrix
\begin{equation}\label{Eq:BYBRg}
\Rc^{\gt}\ti{12}(\lambda,\mu) = \frac{1}{2} \frac{C\ti{12}}{\mu-\lambda} \vp_{\text{BYB}}(\mu)^{-1} - C_{\Xt}(\mu) C\ti{12} + C_{\RgXt}(\mu) \Rt\ti{12},
\end{equation}
where
\begin{equation*}
\Rt\ti{12} = \Rt\ti{1}C\ti{12} = -\Rt\ti{21}
\end{equation*}
is the kernel of the operator $\Rt$ (the last equality is a consequence of the skew-symmetry of $\Rt$).

As a consequence, the Lax matrix $\Lc_{\text{GF}}(\xi)$ also satisfies a Maillet non-ultralocal bracket. The corresponding $\Rc$-matrix is simply obtained from \eqref{Eq:BYBRg} by the change of spectral parameter \eqref{Eq:ParamCosetToPCM}:
\begin{equation*}
\Rc^{\text{GF}}\ti{12}\left( \xi,\upsilon \right) = \Rc^{\gt}\ti{12} \left( \frac{1+\xi}{1-\xi}, \frac{1+\upsilon}{1-\upsilon} \right).
\end{equation*}
One then finds
\begin{equation*}
\Rc^{\text{GF}}\ti{12}\left( \xi,\upsilon \right) = \Rc^{0,\text{GF}}\ti{12}\left( \xi,\upsilon \right) \vp_{\text{GF}}(\upsilon)^{-1}
\end{equation*}
with
\begin{align}
\Rc^{0,\text{GF}}\ti{12}\left( \xi,\upsilon \right) &= \frac{C\ti{12}}{\upsilon-\xi} - \frac{\alpha_4 \upsilon}{\alpha_3+\alpha_4\upsilon^2}C\ti{12} + \frac{1}{2}\frac{\etat}{\alpha_3+\alpha_4\upsilon^2}\Rt\ti{12}, \label{Eq:R0GF} \\
\vp_{\text{GF}}(\xi) &= \frac{K}{2} \frac{1-\xi^2}{(\alpha_1+\alpha_2\xi^2)(\alpha_3+\alpha_4\xi^2)}, \label{Eq:TwistGF}
\end{align}
and
\begin{subequations}
\begin{align}
\alpha_1 &= \frac{1}{2}\left(\zeta-1+\frac{\eta^2-\etat^2}{4}\right), & \alpha_2 &= \frac{1}{2}\left(\zeta+1-\frac{\eta^2-\etat^2}{4}\right), \\
\alpha_3 &= \frac{1}{2}\left(\zeta+1+\frac{\eta^2-\etat^2}{4}\right), & \alpha_4 &= \frac{1}{2}\left(\zeta-1-\frac{\eta^2-\etat^2}{4}\right).
\end{align}
\end{subequations}
The gauge-fixed BYBM does not exactly enter the class of models with twist function, as the matrix $\Rc^{0,GF}$ is not a standard $\Rc$-matrix on $\g$. However, it shares some similar structures with these models. In particular, the matrix $\Rc^{0,GF}$ satisfies the CYBE \eqref{Eq:CYBE} and the asymptotic condition
\begin{equation}\label{Eq:RBYBAsymp}
\Rc^{0,\text{GF}}\ti{12}(\xi,\upsilon) = \frac{C\ti{12}}{\upsilon-\xi} + O\bigl( (\xi-\upsilon)^0 \bigr),
\end{equation}
as the standard matrix $\Rc^0$. This will allow us to apply the results of Chapter \ref{Chap:LocalCharges} to the gauge-fixed BYBM. With a slight abuse of notation, we will call $\vp_{\text{GF}}$ the twist function of the gauge-fixed BYBM. We note the following relation between $\vp_{\text{BYB}}(\lambda)$ and $\vp_{\text{GF}}(\xi)$:
\begin{equation*}
2\vp_{\text{BYB}}(\lambda)\, \dd\lambda = \vp_{\text{GF}}(\xi) \,\dd\xi.
\end{equation*}

Let us consider the one-parameter limit $\etat=0$, which coincides with the Yang-Baxter model. In this case, we find $\zeta=\alpha_3=1+\frac{\eta^2}{4}$, $\alpha_1=\frac{\eta^2}{4}$, $\alpha_2=1$ and $\alpha_4=0$. Thus, we see that the matrix $\Rc^{0,\text{GF}}$ reduces to the standard matrix on $\g$ and $\vp_{\text{GF}}$ reduces to the twist function \eqref{Eq:TwistYB} of the Yang-Baxter model, with $\eta$ and $K$ replaced by $\frac{\eta}{2}$ and $\frac{K}{2}$.

\paragraph{Poles and zeros of the twist function.} Finally, let us study the analytical properties of the 1-form $\vp_{\text{GF}}(\xi)\,\dd\xi$. It possesses two simple zeros at $+1$ and $-1$, as for the undeformed PCM or the Yang-Baxter model. Recall that the Yang-Baxter deformation had for effect to split the double pole at $0$ of the PCM into two simples poles on the imaginary axis, without modifying the double pole at infinity. In the case of the BYBM, both the double poles at $0$ and infinity are split into pairs of simple poles, given by
\begin{equation*}
\xi_\pm = \pm i \sqrt{\frac{\alpha_1}{\alpha_2}} \;\;\;\; \text{ and } \;\;\;\; \widetilde{\xi}_\pm = \pm i \sqrt{\frac{\alpha_3}{\alpha_4}}.
\end{equation*}
They belong to the imaginary axis as one can show that the $\alpha_i$'s are all positive numbers. Note that in the one-parameter limit $\etat=0$, we get $\alpha_4=0$: we thus see that in this limit, the two poles $\widetilde{\xi}_\pm$ recombine as a double pole at infinity, as expected from the Yang-Baxter model case. Note that these poles are related to the ones of $\vp_{\text{BYB}}$ by the relation
\begin{equation*}
\lambda_\pm = \frac{1+\xi_\mp}{1-\xi_\mp} \;\;\;\; \text{ and } \;\;\;\; \lt_\pm = \frac{1+\widetilde{\xi}_\mp}{1-\widetilde{\xi}_\mp}.
\end{equation*}

As for the Yang-Baxter model, it is useful to evaluate the Lax matrix or its gauge transformation by $g'$ at the poles of the twist function. Indeed, one then finds
\begin{subequations}
\begin{align}
\Lc_{\text{GF}}^{g'}(\xi_\pm) &= -\gamma R^\mp \bigl( g' X' g'^{-1} \bigr) & \text{ with } \;\;\;\; \frac{1}{\gamma} = \frac{K}{\eta} = \pm 2i \; \res_{\xi=\xi_\pm} \, \vp_{\text{GF}}(\xi) \; \dd \xi, \label{Eq:LgBYB} \\
\Lc_{\text{GF}}(\widetilde\xi_\pm) &= -\widetilde\gamma \Rt^\mp X'  & \text{ with } \;\;\;\; \frac{1}{\widetilde\gamma} = \frac{K}{\etat} = \pm 2i \; \res_{\xi=\widetilde\xi_\pm} \, \vp_{\text{GF}}(\xi) \; \dd \xi. \label{Eq:LpmBYB}
\end{align}
\end{subequations}
We shall use these relations in Chapter \ref{Chap:PLie}.\\

One can also extract informations on the model from the zeros $+1$ and $-1$ of the twist function $\vp_{\text{GF}}(\xi)$. Indeed, one can reconstruct the Hamiltonian $\Hc_{\text{GF}}$ and total momentum $\Pc_{\text{GF}}$ of the gauge-fixed model from the Lax matrix:
\begin{equation*}
\Hc_{\text{GF}} = \Hc_+ - \Hc_- \;\;\;\; \text{ and } \;\;\;\; \Pc_{\text{GF}} = \Hc_+ + \Hc_-,
\end{equation*}
with
\begin{equation*}
\Hc_\pm = - \frac{1}{2} \; {\large \res_{\xi=\pm 1}} \; \vp_{\text{GF}}(\xi) \int \dd x \; \kappa \bigl( \Lc_{\text{GF}}(\xi,x),\Lc_{\text{GF}}(\xi,x) \bigr).
\end{equation*}
Note that the same equation was already satisfied by the other deformations of the PCM, as seen in equation \eqref{Eq:HamMomdPCM}. We will use this fact in Chapter \ref{Chap:LocalCharges}.

\cleardoublepage
\chapter{Local charges in involution and integrable hierarchies}
\label{Chap:LocalCharges}

This chapter is based on the article~\cite{Lacroix:2017isl}, that I wrote during my PhD with M. Magro and B. Vicedo. The content of this chapter is the same as the one of~\cite{Lacroix:2017isl} and is made to be read independently.\\

In Chapter \ref{Chap:Lax}, we introduced the general Lax formalism for integrable fields theories. This formalism ensures the existence for these theories of an infinite number of conserved charges in involution, which are constructed from the monodromy matrix of the theory. In particular, these charges are said to be non-local. Before going further, let us briefly recall what we mean by local and non-local quantities.

We consider a Hamiltonian field theory which describes the time evolution of some fundamental fields $\phi_i(x)$'s, depending on the spatial coordinate $x$. A quantity $\K(x)$ is said to be a local field of the theory if it is a function of the evaluation of the $\phi_i$'s and their derivatives at a unique point $x$ of space. For example, $\phi_1(x)^2$ and $\phi_1(x)+2\phi'_2(x)$ are local fields but $\int_0^x \phi_1(y)\,\dd y$ or $\phi_1(x)-\phi_2(0)$ are not. A local charge is the integral
\begin{equation*}
\int \dd x \; \K(x),
\end{equation*}
of a local field $\K(x)$ on the whole space (the real line $\R$ or the circle $\mathbb{S}^1$). The field $\K(x)$ is then called the density of the charge. In particular, we always consider here local field theories, for which the Hamiltonian is a local charge, ensuring that the time evolution of the $\phi_i$'s is encoded in local partial differential equations.

The charges extracted from the monodromy matrix are non-local, in the sense that they cannot be written as integrals of some local fields (more precisely, the monodromy matrix is expressed as a series of nested integrals). This makes the computation and the manipulation of these charges quite involved. Let us illustrate that on an example. Let $\Q$ be a conserved charge extracted from the monodromy and consider its Hamiltonian flow $\p_\Q=\lbrace \Q,\cdot \rbrace$ on the phase space of the model: it is an interesting object for the study of the model as it commutes with the time flow $\p_t$ (because the charge $\Q$ is conserved). However, as $\Q$ is non-local, this flow $\p_\Q$ does not take the form of partial differential equations on the fields $\phi_i$'s and is thus quite difficult to study.\\

It is thus interesting to find conserved charges in involution which are local (another motivation for that will be explained in the second part of this thesis). For integrable $\s$-models (see Chapter \ref{Chap:Models}), an infinity of such charges were found in two particular examples. They were constructed for the Principal Chiral Model by Evans, Hassan, MacKay and Mountain in~\cite{Evans:1999mj} and for the $\Z_2$-coset $\s$-model by Evans and Moutain in~\cite{Evans:2000qx}. For completeness, let us also mention the papers~\cite{Evans:2000hx} by Evans, Hassan, MacKay and Mountain and~\cite{Evans:2005zd} by Evans and Young for super-symmetric generalisations of the above-mentioned publications.

However, the existence of such charges for other integrable $\s$-models, such as deformed models or $\Z_T$-coset models for $T>2$, was not proved. In~\cite{Lacroix:2017isl}, I have shown, together with my collaborators, that all integrable $\s$-models, including integrable deformations, possess an infinite number of local conserved charges in involution. One of the main characteristic of our approach in~\cite{Lacroix:2017isl} is that we do not treat each model individually: at the contrary, we develop a general construction which applies to the whole family of integrable $\s$-models at once. This construction relies on the common mathematical structure shared by these models: the existence of a twist function (see Chapters \ref{Chap:Lax} and \ref{Chap:Models}). As such, it applies to all models with a twist function, provided the latter possesses what we call a regular zero (see the precise definition below).

Given the length of this chapter, and to avoid going directly into technicalities, we start by a summary of the results presented in more details in the subsequent sections.

\section{Summary}
\label{Sec:SummaryLoc}

The purpose of the present chapter is to provide another application of the general formalism of Maillet brackets with twist function (see Chapter \ref{Chap:Lax}). Specifically, we will describe how, in this general framework, infinite towers of local charges can be associated with certain zeros of the twist function, all of which are in pairwise involution. Following the same spirit as recalled above, the starting point of our approach was to reinterpret the construction of local charges in the principal chiral model due to Evans, Hassan, MacKay and Mountain \cite{Evans:1999mj} in the present language of twist functions. In fact, this construction had soon been generalised to include also the (supersymmetric) principal chiral model with a Wess-Zumino term in \cite{Evans:2000hx}, symmetric space $\sigma$-models in \cite{Evans:2000qx} as well as supersymmetric coset $\sigma$-models in \cite{Evans:2005zd}. Each of these generalisations can be regarded as further evidence that such a construction should hold for any integrable field theory with twist function, while at the same time providing indications on how to do so. In this section, we will briefly summarise the main results of the chapter.

Let us first note that in all of the integrable $\sigma$-models with twist function described above, every zero of $\varphi(\lambda)$ is such that $\varphi(\lambda) \mathcal L(\lambda, x)$ is regular there. In a general integrable field theory with twist function $\varphi(\lambda)$ we shall say that any \textbf{zero} of $\varphi(\lambda)$ with this property is \textbf{regular}. We denote by $\mathcal Z$ the set of regular zeros of $\varphi(\lambda)$ in $\C$. We shall further distinguish between two types of zeros: cyclotomic ones and 
non-cyclotomic ones. This notion depends on the order $T$ of the automorphism $\sigma$ appearing in the $\Rc$-matrix of the system (see  Chapter \ref{Chap:Lax}). 
In a model with $T=1$, every point is by definition non-cyclotomic, 
whereas in  a model with $T>1$, every point is non-cyclotomic except for the origin and infinity. As explained in subsection \ref{Sec:Infinity}, throughout our analysis the point at infinity will be treated in much the same way as the origin by using an inversion of the spectral parameter.\\

With every $\lambda_0 \in \mathcal Z$, or every $\lambda_0 \in \mathcal Z \cup \{ \infty \}$ if infinity is also a regular zero, we will associate a subset of integers $\mathcal E_{\lambda_0} \subset \Z_{\geq 2}$ and a corresponding tower of local charges $\mathcal Q^{\lambda_0}_n$ labelled by $n \in \mathcal E_{\lambda_0}$. The first main property of these charges which we will establish is that any two such charges $\mathcal Q^{\lambda_0}_n$ and $\mathcal Q^{\mu_0}_m$ for any $\lambda_0, \mu_0 \in \mathcal Z$ and $n \in \mathcal E_{\lambda_0}$, $m \in \mathcal E_{\mu_0}$ are in involution. Moreover, if infinity is a regular zero and either $\lambda_0$ or $\mu_0$ is taken to be the point at infinity, the corresponding local charges will only Poisson commute up to a certain field $\mathcal C(x)$ which will coincide with the coset constraint in $\Z_T$-coset $\sigma$-models. Following the standard terminology from the theory of constrained Hamiltonian systems, we will refer to equalities as being \emph{weak} when they hold only after setting this particular field to zero, see subsection \ref{Sec:AlgebraLoc}. Furthermore, we show that in every example of integrable $\sigma$-model considered, the Hamiltonian can be expressed as a particular linear combination of the collection of quadratic local charges $\mathcal Q^{\lambda_0}_2$ for $\lambda_0 \in \mathcal Z \cup \{ \infty \}$ and the momentum of the model. It then follows that all of the local charges are conserved.

Let us briefly outline the construction of the local charges by considering first the case when $\lambda_0 \in \mathcal Z$ is non-cyclotomic. If the Lie algebra $\g$ is of type B, C or D then the density of the local charge $\mathcal Q^{\lambda_0}_n$ is obtained simply by evaluating
\begin{equation} \label{tr phi L intro}
\Tr \big( \varphi(\lambda)^n \mathcal L(\lambda, x)^n \big)
\end{equation}
at the regular zero $\lambda_0$. When $\g$ is of type A, on the other hand, the density of the local charge $\mathcal Q^{\lambda_0}_n$ is given instead by a certain polynomial in the above expressions, determined as in \cite{Evans:1999mj} with the help of a generating function. In either case, $\mathcal E_{\lambda_0}$ is given here by the set of \textbf{exponents of the affine Kac-Moody algebra $\bm{\widehat{\g}}$} associated with $\g$, shifted by one (we do not treat the case of the Pfaffian in type D). In the example of the principal chiral model on a real Lie group $G_0$ (see Subsection \ref{Sec:PCM}) treated in \cite{Evans:1999mj}, the twist function has simple zeros at $\pm 1$ and the evaluation of $\varphi(\lambda) \mathcal L(\lambda, x)$ at $\lambda = \pm 1$ produces the light-cone currents $j^L_\pm = g^{-1} \partial_\pm g$ of the theory (see Subsection \ref{SubSec:FieldSigma}). We recover in this way the higher spin local charges in involution of the principal chiral model constructed in \cite{Evans:1999mj}.

When the regular zero $\lambda_0 \in \mathcal Z$ is cyclotomic, \emph{i.e.} $\lambda_0 = 0$, it may 
happen, as a result of the equivariance properties of both the Lax matrix and twist function, that 
the evaluation of \eqref{tr phi L intro} at the point $\lambda_0$ vanishes identically. More precisely, 
the first non-vanishing term in the power series expansion of \eqref{tr phi L intro} around 
$\lambda = 0$ is of order $\lambda^{r_n}$ for some $0 \leq r_n \leq T-1$. If the Lie algebra 
$\g$ is of type B, C or D, or also of type A with an inner automorphism $\sigma$, then we 
define the density of the local charge $\mathcal Q^0_n$ as the coefficient of this leading term. The 
case when $\g$ is of type A and the 
automorphism $\sigma$ is not inner is treated in a similar fashion 
to the case of a non-cyclotomic point in type A, with the densities of the local charges $\mathcal Q^0_n$ being obtained by means of a generating function. In each case it turns out that we need to restrict attention to indices $n$ such that $0 \leq r_n < T-1$. As a result, and in contrast to the case of a non-cyclotomic regular zero, some exponents of the affine Kac-Moody algebra $\widehat{\g}$ are `dropped' in the construction of the subset $\mathcal E_0$, specifically those such that $r_n = T-1$. In the case of a symmetric space $\sigma$-model, for which $T=2$ so that only charges for which $r_n = 0$ are kept, we recover in this way the local charges found in \cite{Evans:2000qx}.\\

The collection of local charges $\Q^{\lambda_0}_n$, $\lambda_0 \in \mathcal Z$, $n \in \mathcal E_{\lambda_0}$ in involution generates an infinite set of Poisson commuting Hamiltonian flows $\left\lbrace \Q^{\lambda_0}_n, \cdot \right\rbrace$ on the phase space of the model. With every such flow we then associate a corresponding $\g$-valued connection
$\nabla^{\lambda_0}_n = \left\lbrace \Q^{\lambda_0}_n, \cdot \right\rbrace + \M^{\lambda_0}_n (\lambda,x)$ for some $\g$-valued matrix $\M^{\lambda_0}_n (\lambda,x)$ depending on the spectral parameter $\lambda$.
The second main property of the local charges $\Q^{\lambda_0}_n$, $\lambda_0 \in \mathcal Z$, $n \in \mathcal E_{\lambda_0}$ which we establish is that the connection $\nabla^{\lambda_0}_n$ for any $\lambda_0 \in \mathcal Z$ and $n \in \mathcal E_{\lambda_0}$ commutes with the connection $\nabla_x = \partial_x + \mathcal L(\lambda, x)$. In this sense, the local charges generate a \textbf{hierarchy of integrable equations}. We use this result to deduce that the local charges $\Q^{\lambda_0}_n$, $\lambda_0 \in \mathcal Z$, $n \in \mathcal E_{\lambda_0}$ are in involution with the non-local charges extracted from the monodromy of $\Lc(\lambda,x)$.
Moreover, we go on to show that when $\g$ is of type B, C or D, any two such connections $\nabla^{\lambda_0}_n$ and $\nabla^{\mu_0}_m$ for $\lambda_0, \mu_0 \in \mathcal Z$ and $n \in \mathcal E_{\lambda_0}$, $m \in \mathcal E_{\mu_0}$ also commute with one another. Finally, we have also checked these results in the case of type A for low values of $n$ and $m$ and on this basis we conjecture it to hold in general. If infinity is a regular zero then the majority of these results still hold in the weak sense when we consider also the local charges associated with infinity.\\

This chapter is organised as follows. The general framework which we employ throughout the chapter is introduced in Section \ref{Sec:FraandGenRes}. In particular, we introduce the notion of a regular zero in the complex plane which plays 
a central role in our analysis. In subsection \ref{Sec:Infinity}, we define the notion of a regular zero at infinity and relate it to that of a regular zero at the origin by inversion of the spectral parameter. Finally, we establish some general results in subsection \ref{subsec-pbtrpow}. Section \ref{Sec:NonCycZero} is devoted to the procedure for extracting local charges in involution in the case of a non-cyclotomic regular zero. In particular, we present in subsection \ref{Sec:GenNonCyc} an explicit construction of the currents $\K_n^{\lambda_0}$ for type A algebras using generating functions in the spirit of \cite{Evans:1999mj}. Section \ref{Sec:CycZero} deals with charges at cyclotomic zeros. We explain how the equivariance properties of the various objects affect the construction of local conserved charges in involution. Here the Lie algebras of type B, C and D can still be treated 
uniformly but in type A we need to consider separately the cases when the 
automorphism $\sigma$ is inner or not. The generating function for Lie algebras of type A 
with non-inner automorphism is presented in subsection \ref{Sec:CycGenerating}. A list of properties of these local charges is collated in section \ref{Sec:PrOfLocCha}, including the fact that the local charges extracted from different regular zeros Poisson commute (weakly when the point at infinity is involved). Moreover, we show that all the local charges commute with the field $\mathcal C(x)$ which will play the role of the constraint in $\Z_T$-coset $\sigma$-models, therefore showing that they are gauge invariant. We also discuss the reality conditions of all the local charges. The Hamiltonian flows of the local charges $\Q^{\lambda_0}_n$ are studied in detail in 
section \ref{Sec:IntHierZeroCurv}. The main result that any two of the $\g$-valued connections $\nabla^{\lambda_0}_n$ and $\nabla^{\mu_0}_m$ satisfy a zero curvature equation is established in subsection \ref{sec: ZC eq}. Finally, in section \ref{Sec:Applications} we apply all these results to the whole family of integrable $\s$-models described in Chapter \ref{Chap:Models}. Some technical appendices, specific to the present chapter, are given here as sections \ref{App:ExtSigma} and \ref{App:Xi}.

\section{Framework and general results} \label{Sec:FraandGenRes}

\subsection{Framework: regular zeros of a model with twist function}
\label{Sec:Model}

In this chapter, we consider an integrable model with twist function, as described in Section \ref{Sec:ModelsTwist}. We will use the notations of this section and more generally of Chapter \ref{Chap:Lax}. In particular, let us consider the twist function $\vp(\lambda)$ and the Lax matrix $\Lc(\lambda,x)$ of the model, which are rational functions of the spectral parameter $\lambda$. Let $\lambda_0\in\C$ be a \textbf{zero} of the twist function, so that $\vp(\lambda_0) = 0$. We will say that this zero is \textbf{regular} if $\vp(\lambda)\Lc(\lambda,x)$ is holomorphic at $\lambda=\lambda_0$.

Let us suppose that the model we are considering is cyclotomic with respect to an automorphism $\s$ of order $T$ (note that a non-cyclotomic model can be considered as a cyclotomic one with $\s=\Id$ and $T=1$). As in the Section \ref{Sec:ModelsTwist}, we consider the action of the cyclic group $\Z_T$ on the complex plane $\C$ by the multiplication by $\omega$, a $T^{\rm th}$-root of unit. The Lax matrix and the twist function then satisfy the equivariance properties \eqref{Eq:EquiL} and \eqref{Eq:TwistEqui}. Due to these properties, if $\lambda_0\in\C$ is a regular zero, all points of the orbit $\Z_T\lambda_0$ are also regular zeros. Let us pick (arbitrarily) one of them. We then form a set $\Zc$ of regular zeros of $\varphi$ such that for every pair of distinct points $\lambda_0$ and $\mu_0$ in $\Zc$, the orbits $\Z_T\lambda_0$ and $\Z_T\mu_0$ are disjoint.

As explained in subsection \ref{Sec:Infinity}, we will also be interested in the case where the differential form $\varphi(\lambda)\dd \lambda$ has a zero at infinity, \textit{i.e.} where
\begin{equation}\label{Eq:Psi}
\psi(\alpha) = - \frac{1}{\alpha^2}\varphi\left(\frac{1}{\alpha}\right)
\end{equation}
has a zero at $\alpha=0$. We will also see that the appropriate notion of a regular zero at infinity corresponds to requiring that $\frac{1}{\alpha}\varphi\left(\frac{1}{\alpha}\right)\Lc\left(\frac{1}{\alpha},x\right)$ be holomorphic at $\alpha=0$.\\

As an example, let us determine the regular zeros of the PCM. Its Lax matrix and twist function are respectively given by \eqref{Eq:LaxPcm} and \eqref{Eq:TwistPCM}. As mentioned in Subsection \ref{SubSec:PCM}, the twist function admits two simple zeros, at $+1$ and $-1$. Moreover, the Lax matrix has simple poles at $\lambda=\pm 1$, \textit{i.e.} at these zeros. Thus, the product $\vp_{\text{PCM}}(\lambda)\Lc_{\text{PCM}}(\lambda,x)$ is regular at $\lambda=\pm 1$, which are then regular zeros. Let us note here that the evaluation of $\vp_{\text{PCM}}(\lambda)\Lc_{\text{PCM}}(\lambda,x)$ at these zeros gives the light-cone currents $j^L_\pm$ (see Subsection \ref{SubSec:FieldSigma}).

\subsection{Infinity and inversion of the spectral parameter}
\label{Sec:Infinity}

In this chapter we will construct a tower of local charges associated with each regular zero of the twist function. As mentioned in the previous subsection, the set of regular zeros can include the point at infinity, although the sense in which infinity can be a regular zero is slightly different from the definition of finite regular zeros. In this subsection, we show how the notion of a regular zero at infinity is related to that of a regular zero at the origin through inversion of the spectral parameter, \textit{i.e.} by the change of parameter $\lambda \mapsto \alpha=\lambda^{-1}$.
Under such a change of spectral parameter we have
\begin{equation*}
\varphi(\lambda) \dd\lambda = \psi(\alpha)\dd\alpha,
\end{equation*}
where $\psi(\alpha)$ is defined in equation \eqref{Eq:Psi}. Suppose that infinity is a zero of the twist function, \textit{i.e.} that $\psi(0)=0$, and define
\begin{equation}\label{Eq:DefP}
P(\alpha,x)=\frac{1}{\alpha}\varphi\left(\frac{1}{\alpha}\right)\Lc\left(\frac{1}{\alpha},x\right).
\end{equation}
We will say that infinity is a \textbf{regular zero} if $P(\alpha,x)$ is regular at $\alpha=0$. In the remainder of this subsection we will assume this to be the case. We then set
\begin{equation}\label{Eq:DefC}
\Cc(x) = P(0,x).
\end{equation}
From the equivariance properties \eqref{Eq:EquiL} and \eqref{Eq:TwistEqui} of $\Lc$ and $\varphi$, we deduce that $\Cc$ is valued in the grading $\g^{(0)}$. Let us note here that in the $\Z_T$-coset models, described in Subsection \ref{SubSec:ZT}, this field $\Cc$ coincides with the gauge constraint $X^{(0)}$.

Starting from the Poisson bracket \eqref{Eq:PBR} and using the form \eqref{Eq:DefR} of the $\Rc$-matrix, we find
\begin{align*}
 \left\lbrace \Lc(\lambda,x)\ti{1}, P(\alpha,y)\ti{2} \right\rbrace & =  \left[\alpha^{-1}\Rc^0\ti{12}\left(\lambda,\alpha^{-1}\right), \Lc(\lambda,x)\ti{1} \right] \delta_{xy} - \left[ \Rc^0\ti{21}\left(\alpha^{-1},\lambda \right)\varphi(\lambda)^{-1}, P(\alpha,x)\ti{2} \right] \delta_{xy} \\
 & \hspace{50pt} - \Bigl( \alpha^{-1}\Rc^0\ti{12}\left(\lambda,\alpha^{-1}\right) - \alpha\psi(\alpha)\Rc^0\ti{21}\left(\alpha^{-1},\lambda \right) \varphi(\lambda)^{-1} \Bigr)  \delta'_{xy}
\end{align*}
Using the expression \eqref{Eq:RCyc} of $\Rc^0$, we have
\begin{equation}\label{Eq:RAsymptoticInfinity}
\alpha^{-1}\Rc^0\ti{12}\left(\lambda,\alpha^{-1}\right) \;\xrightarrow{\alpha\to 0}\; \frac{1}{T} \sum_{k=0}^{T-1} \s^k\ti{1}C\ti{12} = C^{(0)}\ti{12}, \;\;\;\;\; \Rc^0\ti{21}\left(\alpha^{-1},\lambda \right) \;\xrightarrow{\alpha\to 0}\; 0.
\end{equation}
As $P(\alpha,x)$ and $\alpha\psi(\alpha)$ are regular at 0, taking the limit $\alpha\to 0$ in the above Poisson bracket, we then obtain
\begin{equation}\label{Eq:PBLC}
\left\lbrace \Lc(\lambda,x)\ti{1}, \Cc(y)\ti{2} \right\rbrace = \bigl[ C^{(0)}\ti{12}, \Lc(\lambda,x)\ti{1} \bigr] \delta_{xy} - C^{(0)}\ti{12}\delta'_{xy}.
\end{equation}
Applying the same kind of reasoning we also find
\begin{equation}\label{Eq:PBCC}
\left\lbrace \Cc(x)\ti{1}, \Cc(y)\ti{2} \right\rbrace = \bigl[ C^{(0)}\ti{12}, \Cc(x)\ti{1} \bigr] \delta_{xy}.
\end{equation}
Let us define a new Lax matrix
\begin{equation*}
\Lct(\lambda,x) = \Lc(\lambda,x) - \lambda^{-1}\varphi(\lambda)^{-1}\Cc(x).
\end{equation*}
From the fact that $[ C^{(k)}\ti{12}, Z\ti{1} ] = -[ C^{(k)}\ti{12}, Z\ti{2} ]$ for any $Z\in\g^{(0)}$, we find that
\begin{equation*}
\lambda \left[ \Rc^0\ti{21}(\mu,\lambda), Z\ti{2} \right] - \mu \left[ \Rc^0\ti{12}(\lambda,\mu), Z\ti{1} \right]  = \left[ C^{(0)}\ti{12}, Z\ti{2} \right].
\end{equation*}
Using this identity and the Poisson brackets \eqref{Eq:PBR}, \eqref{Eq:PBLC} and \eqref{Eq:PBCC}, we prove that the Poisson bracket of $\Lct$ with itself is also of the Maillet form, namely
\begin{align}\label{Eq:PBRt}
\left\lbrace \Lct(\lambda,x)\ti{1}, \Lct(\mu,y)\ti{2} \right\rbrace &=
\left[ \Rct\ti{12}(\lambda,\mu), \Lct(\lambda,x)\ti{1} \right] \delta_{xy} - \left[ \Rct\ti{21}(\mu,\lambda), \Lct(\mu,y)\ti{2} \right] \delta_{xy} \\
 &\hspace{50pt}  - \; \Bigl( \Rct\ti{12}(\lambda,\mu) + \Rct\ti{21}(\mu,\lambda) \Bigr) \delta'_{xy}, \notag
\end{align}
where $\Rct\ti{12}(\lambda,\mu)=\Rct^0\ti{12}(\lambda,\mu)\varphi(\mu)^{-1}$ and
\begin{equation}\label{Eq:DefRct}
\Rct^0\ti{12}(\lambda,\mu) = \Rc^0\ti{12}(\lambda,\mu) - \mu^{-1} C^{(0)}\ti{12}.
\end{equation}
We now define
\begin{equation*}
\Lc^\infty(\alpha,x) = \Lct\left(\frac{1}{\alpha},x\right).
\end{equation*}
The following theorem is the main result of this subsection.
\begin{theorem}\label{Thm:PBLcI}
The Poisson bracket of $\Lc^\infty$ with itself reads
\begin{align}\label{Eq:PBLcI}
\left\lbrace \Lc^\infty(\alpha,x)\ti{1}, \Lc^\infty(\beta,y)\ti{2} \right\rbrace &=
\left[ \Rc^\infty\ti{12}(\alpha,\beta), \Lc^\infty(\alpha,x)\ti{1} \right] \delta_{xy} - \left[ \Rc^\infty\ti{21}(\beta,\alpha), \Lc^\infty(\beta,y)\ti{2} \right] \delta_{xy} \\
 &\hspace{50pt}  - \; \Bigl( \Rc^\infty\ti{12}(\alpha,\beta) + \Rc^\infty\ti{21}(\beta,\alpha) \Bigr) \delta'_{xy}, \notag
\end{align}
where
\begin{equation*}
\Rc^\infty\ti{12}(\alpha,\beta) = \Rc^0\ti{21}(\alpha,\beta)\psi(\beta)^{-1}
\end{equation*}
satisfies the classical Yang-Baxter equation \eqref{Eq:CYBE}.
\end{theorem}
\begin{proof}
Using equation \eqref{Eq:RCas}, we find that
\begin{equation}\label{Eq:RtCas}
\Rct^0\ti{12}(\lambda,\mu) = \sum_{k=1}^{T} \frac{\lambda^k\mu^{T-1-k}}{\mu^T-\lambda^T} C\ti{12}^{(p)}.\end{equation}
The theorem follows from the Poisson bracket \eqref{Eq:PBRt} and the identity
\begin{equation}\label{Eq:RtInvR}
\Rct^0\ti{12}\left(\frac{1}{\alpha},\frac{1}{\beta}\right)= - \beta^2 \Rc^0\ti{21}(\alpha,\beta),
\end{equation}
which is a consequence of equation \eqref{Eq:RtCas}.
\end{proof}

To interpret Theorem \ref{Thm:PBLcI}, let us note that the matrix $\Rc^0\ti{21}$ is nothing but the matrix $\Rc^0\ti{12}$ for the automorphism $\s^{-1}$. Moreover, from the equivariance properties \eqref{Eq:EquiL} and \eqref{Eq:TwistEqui}, we find that the corresponding properties of $\Lc^\infty$ and $\psi$ are
\begin{equation}\label{Eq:EquiInfinity}
\s^{-1}\bigl(\Lc^\infty(\alpha,x)\bigr) = \Lc^\infty(\omega\alpha,x)
\;\;\;\;\;\;\; \text{ and } \;\;\;\;\;\;\;
\psi(\omega\alpha)=\omega^{-1}\psi(\alpha).
\end{equation}
The Poisson bracket of $\Lc^\infty$ is thus the one of a model with twist function $\psi$, automorphism $\s^{-1}$ and spectral parameter $\alpha=\lambda^{-1}$. Moreover, the point $\alpha=0$ is a regular zero of this model. Indeed, we supposed that $\alpha$ was a zero of $\psi(\alpha)$ and one can check explicitly that $\psi(\alpha)\Lc^\infty(\alpha,x)$ is regular at $\alpha=0$.\\

It is worth noting that the procedure just described is involutive, in the following sense. If $\varphi(\lambda)\Lc(\lambda,x)$ is regular at $\lambda=0$, one can check that $\alpha=\infty$ (which corresponds to $\lambda=0$) is a regular zero of the model with Lax matrix $\Lc^\infty$ and, moreover, that the corresponding field $\Cc^\infty$ obtained by evaluating $\lambda^{-1}\psi(\lambda^{-1})\Lc^\infty(\lambda^{-1},x)$ at $\lambda=0$ is equal to $\Cc$. Re-inverting the spectral parameter $\alpha$ to $\lambda=\alpha^{-1}$, we can thus construct a ``new'' Lax matrix $\Lc^\infty(\lambda^{-1},x)-\lambda \psi(\lambda^{-1})^{-1}\Cc(x)$. According to Theorem \ref{Thm:PBLcI}, this Lax matrix should satisfy a Maillet bracket with twist function $\varphi$ and automorphism $\s$. A direct computation reveals that this Lax matrix is actually equal to the initial Lax matrix $\Lc$.\\

Let us end this subsection by illustrating the inversion of spectral parameter on the example of $\Z_T$-coset models. As noted above, for these models the field $\Cc$ coincides with the constraint $X^{(0)}$. After performing the change of spectral parameter $\lambda \mapsto \alpha=\lambda^{-1}$, we find a twist function
\begin{equation*}
\psi_{\Z_T}(\alpha) = -\frac{T\alpha^{T-1}}{(1-\alpha^T)^2} = -\varphi_{\Z_T}(\alpha).
\end{equation*}
Note that the property $\psi(\alpha)=-\varphi(\alpha)$ is also true for the twist function \eqref{Eq:TwistZ2} of the $\eta$-deformed $\Z_2$-model. The new Lax matrix is
\begin{equation*}
\Lc^\infty_{\Z_T}(\alpha,x) = \sum_{k=1}^{T} \frac{(T-k) + k\alpha^{-T}}{T}\alpha^k j^{L\,(T-k)}(x) - \sum_{k=1}^{T} \frac{1-\alpha^{-T}}{T} \alpha^k X^{(T-k)}(x).
\end{equation*}
Comparing this to the initial Lax matrix \eqref{Eq:LaxZT}, we see that it simply corresponds (up to a minus sign on terms involving $X^{(k)}$) to changing every grading $(k)$ to $(T-k)$, which is equivalent to considering the automorphism $\s^{-1}$ instead of $\s$.

\subsection[Poisson brackets of traces of powers of $\Lc$]{Poisson brackets of traces of powers of $\bm{\Lc}$}
\label{subsec-pbtrpow}

In this chapter, we will focus on the case where the Lie algebra $\g$ is simple. More precisely, we will restrict to the classical types A, B, C and D of the Cartan classification (see Appendix \ref{App:SemiSimple}), seen in their defining representations\footnote{Here $J_n$ is the standard symplectic structure on $\mathbb{C}^{2n}$ given by $J_n=\bigg( \begin{matrix} 0 & \text{Id}\\ - \text{Id} & 0 \end{matrix} \bigg)$ and $\null^t M$ denotes the transpose of $M$.}:
\begin{table}[h]
\begin{center}
\begin{tabular}{lcl}
Type & ~~ & Algebra \\
\hline \hline
A    &    & $\sl(n,\C)=\left\lbrace M \in M_n(\C) \; | \; \Tr(M)=0 \right\rbrace$ \\
B,D  &    & $\so(n,\C)=\left\lbrace M \in M_n(\C) \; | \; \null^t M+M=0 \right\rbrace$ \\
C    &    & $\spc(2n,\C)=\left\lbrace M \in M_{2n}(\C) \; | \; \null^t M J_n + J_n M =0 \right\rbrace$
\end{tabular}
\caption{Defining representations of classical Lie algebras.\label{Tab:Alg}}
\end{center}\vspace{-16pt}
\end{table}

We may therefore take powers of elements of $\g$ and traces of these matrices. In the following sections, we will extract local charges in involution from the traces of powers of the Lax matrix $\Lc$. In this subsection, we will establish general results on the Poisson brackets of powers of $\Lc$ and their traces.\\

For simplicity, we will change the definition of the Casimir $C\ti{12}$ for this chapter, by considering the bilinear form $\Tr(XY)$ in the defining representation $\g$, instead of $\Tr(\ad_X\ad_Y)$ in the adjoint representation (see Appendix \ref{App:Casimir}). This only changes the normalisation of the Casimir, as these two bilinear forms are proportional. Thus it does not affect the results described here but simplifies greatly the presentation, as all computations of this chapter are done in the defining representation. In particular, the completeness relation \eqref{Eq:CasComp} has to be changed into
\begin{equation}\label{Eq:CompRel}
\Tr\ti{2}( C\ti{12} X\ti{2} ) = X, \;\;\;\;\; \forall \; X\in\g.
\end{equation}

\begin{lemma}\label{Lem:PBPow}
Suppose that $X$ and $Y$ are $\g$-valued quantities such that
\begin{equation*}
\left\lbrace X\ti{1}, Y\ti{2} \right\rbrace = \left[a\ti{12},X\ti{1}\right] + \left[b\ti{12},Y\ti{2} \right] + c\ti{12}.
\end{equation*}
Then the Poisson brackets of powers of $X$ and $Y$ are
\begin{equation*}
\left\lbrace X^n\ti{1}, Y^m\ti{2} \right\rbrace = \left[a^{(nm)}\ti{12},X\ti{1}\right] + \left[b^{(nm)}\ti{12},Y\ti{2} \right] + c^{(nm)}\ti{12},
\end{equation*}
where, for $t=a,b,c$, we defined
\begin{equation*}
t^{(nm)}\ti{12} = \sum_{k=0}^{n-1} \sum_{l=0}^{m-1} X\ti{1}^k Y\ti{2}^l \, t\ti{12} \, X\ti{1}^{n-1-k} Y\ti{2}^{m-1-l}.
\end{equation*}
\end{lemma}
\begin{proof}
The Poisson bracket being a derivation, we can use the Leibniz rule yielding
\begin{equation*}
\left\lbrace X^n\ti{1}, Y^m\ti{2} \right\rbrace = \sum_{k=0}^{n-1} \sum_{l=0}^{m-1} X\ti{1}^k Y\ti{2}^l \, \left\lbrace X\ti{1}, Y\ti{2} \right\rbrace \, X\ti{1}^{n-1-k} Y\ti{2}^{m-1-l}.
\end{equation*}
We conclude observing that $ X\ti{1}^k Y\ti{2}^l $ and $X\ti{1}^{n-1-k} Y\ti{2}^{m-1-l}$ commute with $X\ti{1}$ and $Y\ti{2}$ and using the identity
\begin{equation*}
M_1[M_2,N]M_3 = [M_1M_2M_3,N],
\end{equation*}
true for any matrices $M_1$, $M_2$, $M_3$ and $N$ such that $[M_1,N]=[M_3,N]=0$.
\end{proof}

\begin{corollary}\label{Cor:PBTr}
Suppose that $X$ and $Y$ are $\g$-valued quantities such that
\begin{equation*}
\left\lbrace X\ti{1}, Y\ti{2} \right\rbrace = \left[a\ti{12},X\ti{1}\right] + \left[b\ti{12},Y\ti{2} \right] + c\ti{12}.
\end{equation*}
Then we have
\begin{subequations}
\begin{align*}
\bigl\lbrace \emph{\Tr}(X^n), \emph{\Tr}(Y^m) \bigr\rbrace &= nm \, \emph{\Tr}\ti{12} \bigl( c\ti{12}X^{n-1}\ti{1}Y^{m-1}\ti{2} \bigr), \\
\bigl\lbrace X, \emph{\Tr}(Y^m) \bigr\rbrace &= m\left[ \emph{\Tr}\ti{2}\bigl(a\ti{12}Y^{m-1}\ti{2}\bigr), X \right]  + m \, \emph{\Tr}\ti{2}\bigl(c\ti{12}Y^{m-1}\ti{2}\bigr).
\end{align*}
\end{subequations}
\end{corollary}
\begin{proof}
Starting with Lemma \ref{Lem:PBPow}, the corollary follows from the cyclicity of the trace and the vanishing of traces of commutators.
\end{proof}

Let us now apply these results to the Lax matrix $\Lc$. We work in the framework described in Chapter \ref{Chap:Lax}. We define
\begin{equation}\label{Eq:DefS}
S_n(\lambda,x) = \varphi(\lambda)^n \Lc(\lambda,x)^n
\end{equation}
and
\begin{equation}\label{Eq:DefT}
\Tc_n(\lambda,x) = \Tr\bigl( S_n(\lambda,x) \bigr).
\end{equation}
Starting with the Poisson bracket \eqref{Eq:PBR} and the expression \eqref{Eq:DefR} of the $\Rc$-matrix, we apply Corollary \ref{Cor:PBTr}. We find that
\begin{equation}\label{Eq:PBT}
\left\lbrace \Tc_n(\lambda,x), \Tc_m(\mu,y) \right\rbrace = -nm \, \Tr\ti{12} \Bigl( U\ti{12}(\lambda,\mu) S_{n-1}(\lambda,x)\ti{1}S_{m-1}(\mu,y)\ti{2} \Bigr) \delta'_{xy},
\end{equation}
with
\begin{equation}\label{Eq:DefU}
U\ti{12}(\lambda,\mu) = \varphi(\lambda)\Rc^0\ti{12}(\lambda,\mu) + \varphi(\mu)\Rc^0\ti{21}(\mu,\lambda).
\end{equation}

\section{Charges at non-cyclotomic zeros}
\label{Sec:NonCycZero}

The purpose of this section is to describe the procedure for extracting local charges in involution from non-cyclotomic regular zeros of the twist function $\varphi$. Let us first explain what we mean here by a \emph{non-cyclotomic} point. If $T=1$, \textit{i.e.} if $\s=\Id$ and there 
is no cyclotomic invariance, we define any point as being non-cyclotomic. If $T\in\Z_{>1}$, a non-cyclotomic point is a point which is not fixed by the action of the cyclic group $\Z_T$, \textit{i.e.} which is not the origin or infinity.\\

Throughout this section we fix a non-cyclotomic regular zero $\lambda_0$. We will focus here on the case where $\lambda_0$ is different from infinity. The case $\lambda_0=\infty$ is treated by the same method, just replacing $\Lc$ by $\Lc^\infty$ and $\varphi$ by $\psi$ (cf. subsection \ref{Sec:Infinity}). The fact that $\lambda_0$ is a regular zero implies that $S_n(\lambda,x)$ and $\Tc_n(\lambda,x)$, defined in equations \eqref{Eq:DefS} and \eqref{Eq:DefT}, are both holomorphic at $\lambda=\lambda_0$. Thus, we can define the current
\begin{equation}\label{Eq:DefJNonCyc}
\J_n^{\lambda_0}(x) = \Tc_n(\lambda_0,x).
\end{equation}

Let us briefly comment on the explicit expression of these currents in the case of the PCM. As explained in paragraph \ref{Sec:Model}, the PCM has two regular zeros at $+1$ and $-1$. The corresponding currents are
\begin{equation*}
\J_{n,\text{PCM}}^{\pm 1}(x) = \Tr\bigl(j_\pm^{L}(x)^n\bigr).
\end{equation*}
These currents are the one investigated in \cite{Evans:1999mj}, from which local charges in involution for the PCM are constructed. In this section, we will follow the method developed in \cite{Evans:1999mj}, generalising it to any current \eqref{Eq:DefJNonCyc} associated with a non-cyclotomic regular zero $\lambda_0$ of the model.

\subsection{Poisson algebra of the currents}

We begin by computing the Poisson bracket of the currents $\J_n^{\lambda_0}(x)$ and $\J_m^{\lambda_0}(y)$. Specifically, we would like to evaluate equation \eqref{Eq:PBT} at $\lambda=\mu=\lambda_0$. Since $\lambda_0$ is a regular zero, $S_{n-1}(\lambda_0,x)$ and $S_{m-1}(\lambda_0,y)$ are well defined. Thus, it remains to determine $U\ti{12}(\lambda_0,\lambda_0)$. Starting with the definition \eqref{Eq:DefU} of $U$ and using $\varphi(\lambda_0)=0$, one has
\begin{equation*}
U\ti{12}(\lambda,\lambda_0) = \varphi(\lambda)\Rc^0\ti{12}(\lambda,\lambda_0).
\end{equation*}
Recall from equation \eqref{Eq:RCyc} that $\Rc^0\ti{12}(\lambda,\lambda_0)$ is not regular at $\lambda=\lambda_0$, so that we cannot simply evaluate the above equation at $\lambda=\lambda_0$. However, as $\lambda_0$ is a non-cyclotomic point, the matrix $\Rc^0$ has the following local behaviour
\begin{equation}\label{Eq:RAsymptotic}
\Rc^0\ti{12}(\lambda,\lambda_0) = -\frac{1}{T}\frac{C\ti{12}}{\lambda-\lambda_0} + A^{\lambda_0}\ti{12}(\lambda),
\end{equation}
where $A^{\lambda_0}\ti{12}(\lambda)$ is regular at $\lambda=\lambda_0$. Using again $\varphi(\lambda_0)=0$, we then obtain
\begin{equation*}
U\ti{12}(\lambda_0,\lambda_0) = -\frac{\varphi'(\lambda_0)}{T} C\ti{12},
\end{equation*}
where $\varphi'$ denotes the derivative of $\varphi$ with respect to the spectral parameter $\lambda$. Thus, one has
\begin{equation}\label{Eq:PBJ}
\left\lbrace \J^{\lambda_0}_n(x), \J^{\lambda_0}_m(y) \right\rbrace = \frac{nm}{T} \varphi'(\lambda_0) \, \Tr\ti{12} \Bigl( C\ti{12} S_{n-1}(\lambda_0,x)\ti{1}S_{m-1}(\lambda_0,y)\ti{2} \Bigr) \delta'_{xy}.
\end{equation}
Recall the completeness relation \eqref{Eq:CompRel}. We cannot directly apply this identity to equation \eqref{Eq:PBJ} as $S_{m-1}(\lambda_0,y)$ does not belong to $\g$ in general (recall that $S_{m-1}$ is defined as the $(m-1)^{\rm st}$ power of a matrix in $\g$).

Following~\cite{Evans:1999mj}, we will show in the next subsections how to circumvent this difficulty. We will treat separately the case where $\g$ is of type B, C or D and the case where $\g$ is of type A.

\subsection{Type B, C and D algebras}
\label{Sec:NonCycZeroBCD}

Let us first consider the case where $\g$ is of type B, C or D, \textit{i.e.} where $\g$ is an orthogonal or a symplectic algebra (cf. Table \ref{Tab:Alg}). One can check that, for these algebras, if $X$ belongs to $\g$, $X^n$ also belongs to $\g$ if $n$ is odd. Moreover, all matrices in $\g$ are traceless. We then deduce that the currents $\J_n^{\lambda_0}$ are zero for $n$ odd. Thus, we will only extract local charges from the traces of even powers of $\Lc$, \textit{i.e.} from the currents $\J^{\lambda_0}_{2n}$.

The Poisson bracket of such currents is given by equation \eqref{Eq:PBJ}. The right hand side contains $\Tr\ti{2}\bigl(C\ti{12} S_{2m-1}(\lambda_0,y)\ti{2} \bigr)$, and since $2m-1$ is odd we have $S_{2m-1}(\lambda_0,y) \in \g$. Hence, we can apply the completeness relation \eqref{Eq:CompRel}, which yields
\begin{equation*}
\left\lbrace \J^{\lambda_0}_{2n}(x), \J^{\lambda_0}_{2m}(y) \right\rbrace = 4nm\frac{\varphi'(\lambda_0)}{T} \, \Tr \Bigl( S_{2n-1}(\lambda_0,x) S_{2m-1}(\lambda_0,y) \Bigr) \delta'_{xy}.
\end{equation*}
Using the definition \eqref{Eq:DefS} of $S$, one has
\begin{equation}\label{Eq:DerS}
\Tr\bigl(S_p(\lambda,x)\p_xS_q(\lambda,x)\bigr) = \frac{q}{p+q} \p_x \Tc_{p+q}(\lambda,x).
\end{equation}
Using the identities $f(y)\delta'_{xy}=\p_x \bigl(f(x)\bigr)\delta_{xy}+f(x)\delta'_{xy}$ and \eqref{Eq:DerS}, we obtain
\begin{equation}\label{Eq:PBJTypeBCD}
\left\lbrace \J^{\lambda_0}_{2n}(x), \J^{\lambda_0}_{2m}(y) \right\rbrace = 4nm\frac{\varphi'(\lambda_0)}{T} \left( \J^{\lambda_0}_{2n+2m-2}(x) \delta'_{xy} + \frac{2m-1}{2n+2m-2} \p_x \bigl( \J^{\lambda_0}_{2n+2m-2}(x) \bigr) \delta_{xy} \right).
\end{equation}
Define the local charges
\begin{equation}\label{Eq:DefQJ}
\Q^{\lambda_0}_{2n} = \int \dd x \; \J^{\lambda_0}_{2n}(x),
\end{equation}
where the integration is over the whole domain of the spatial coordinate $x$ (\textit{i.e.} the real line $\R$ or the circle $\mathbb{S}^1$).
Once integrated over $y$, the right hand side of \eqref{Eq:PBJTypeBCD} is a total derivative with respect to $x$. Assuming the periodicity of the fields if $x\in \mathbb{S}^1$ or that they decrease at infinity if $x\in\R$, we then conclude that
\begin{equation*}
\left\lbrace \Q^{\lambda_0}_{2n}, \Q^{\lambda_0}_{2m} \right\rbrace = 0.
\end{equation*}
In conclusion, we have constructed a tower of local charges $\Q^{\lambda_0}_{2n}$ in involution, as integrals of the currents $\J^{\lambda_0}_{2n}(x)$. These currents are polynomials in the fields appearing in the Lax matrix $\Lc(\lambda,x)$. More precisely, the current $\J^{\lambda_0}_{2n}$ is a homogeneous polynomial of degree $2n$.\\

Up to a global factor, the Poisson bracket \eqref{Eq:PBJTypeBCD} is the same as the bracket (4.16) of~\cite{Evans:1999mj}. Thus, we can apply the methods developed in~\cite{Evans:1999mj}. In particular, this allows to construct a more general tower of local charges $\Q^{\lambda_0}_{2n}(\xi)$ in involution, depending on a free parameter $\xi\in\R$. These charges are defined as integrals
\begin{equation*}
\Q^{\lambda_0}_{2n}(\xi) = \int \dd x \; \K_{2n}^{\lambda_0}(\xi,x)
\end{equation*}
of some currents $\K_{2n}^{\lambda_0}(\xi)$. These currents are given by homogeneous polynomials in the $\J^{\lambda_0}_{2k}$'s, depending on the free parameter $\xi\in\R$. In particular, the first currents $\K_{2n}^{\lambda_0}(\xi)$ are given by:
\begin{align}\label{Eq:KAlpha}
&\K^{\lambda_0}_2(\xi) = \J^{\lambda_0}_2, \;\;\;\;\;\; \K^{\lambda_0}_4(\xi)=\J^{\lambda_0}_4-\frac{3\xi}{2}(\J^{\lambda_0}_2)^2,  \notag \\
&\K^{\lambda_0}_6(\xi) = \J^{\lambda_0}_6 - \frac{15\xi}{4}\J^{\lambda_0}_2\J^{\lambda_0}_4 + \frac{25\xi^2}{8}(\J^{\lambda_0}_2)^3.
\end{align}
The expression of the current $\K_{2n}^{\lambda_0}(\xi)$ is determined (up to a global factor) recursively from equation \eqref{Eq:PBJTypeBCD} by demanding that the charge $\Q_{2n}^{\lambda_0}(\xi)$ be in involution with all the charges $\Q_{2m}^{\lambda_0}(\xi)$ ($m=2,\ldots,n-1$) constructed thus far. It can also be found without recursion with the help of a generating function, which allows a general proof of the involution of the charges $\Q_{2n}^{\lambda_0}(\xi)$: we refer the reader to the subsection \ref{Sec:GenNonCyc} for more details.

Taking $\xi=0$ in equation \eqref{Eq:KAlpha}, we get $\K_{2n}^{\lambda_0}(\xi=0)=\J_{2n}^{\lambda_0}$. Hence, we recover the local charges $\Q^{\lambda_0}_{2n}$ introduced in equation \eqref{Eq:DefQJ} as a special case of this one-parameter family of local charges. For different parameters $\xi$ and $\xi'$, the towers of charges $\Q_{2n}^{\lambda_0}(\xi)$ and $\Q_{2n}^{\lambda_0}(\xi')$ 
do not commute with one another. We thus have to work with a fixed value of $\xi$: in the rest of this chapter, we will mainly focus on the simplest case $\xi=0$. This choice is justified first by simplicity, but also because the proof of the existence of an integrable hierarchy associated to the charges $\Q_{2n}^{\lambda_0}(\xi)$, presented in section \ref{Sec:IntHierZeroCurv}, works only for the case $\xi=0$.

\subsection{Type A algebras}
\label{Sec:TypeANonCyc}

Let us now consider the case where $\g$ is of type A, \textit{i.e.} where $\g=\sl(d,\C)$ for some $d \in \Z_{\geq 2}$ (see Table \ref{Tab:Alg}). If $X\in\g$, we have $\Tr(X)=0$ by definition, but in general $X^n\notin\g$ and $\Tr(X^n)\neq 0$ for $n \geq 2$. Thus, we consider the currents $\J^{\lambda_0}_n$ for $n\geq 2$. The Poisson bracket between two such currents is given by equation \eqref{Eq:PBJ}. Since in general $S_{m-1}(\lambda_0,y)$ does not belong to $\g$, we cannot use the completeness relation \eqref{Eq:CompRel} to simplify this equation. However, a variant of the identity \eqref{Eq:CompRel} exists for any matrix $Z\in M_d(\C)$. Indeed, using the facts that $Z-\frac{1}{d}\Tr(Z)\Id$ belongs to $\g$ and that $\Tr\ti{2}(C\ti{12})=0$, we find that
\begin{equation}\label{Eq:CompRelSl}
\Tr\ti{2}\bigl(C\ti{12}Z\ti{2}\bigr) = Z - \frac{1}{d}\Tr(Z)\Id.
\end{equation}
Applying this relation to equation \eqref{Eq:PBJ} and using the identities $f(y)\delta'_{xy}=\p_x \bigl(f(x)\bigr)\delta_{xy}+f(x)\delta'_{xy}$ and \eqref{Eq:DerS}, we obtain
\begin{align}\label{Eq:PBJTypeA}
\left\lbrace \J_n^{\lambda_0}(x), \J_m^{\lambda_0}(y) \right\rbrace
&=  nm \frac{\varphi'(\lambda_0)}{T} \left( \J_{n+m-2}^{\lambda_0}(x) \delta'_{xy}  - \frac{1}{d}\J_{n-1}^{\lambda_0}(x)\J_{m-1}^{\lambda_0}(x)\delta'_{xy} \right.  \\
 & \hspace{30pt} \left. + \frac{m-1}{n+m-2} \p_x \left( \J_{n+m-2}^{\lambda_0}(x) \right) \delta_{xy} - \frac{1}{d}\J_{n-1}^{\lambda_0}(x) \p_x \left( \J_{m-1}^{\lambda_0}(x) \right) \delta_{xy} \right). \notag
\end{align}
Integrating both sides over $x$ and $y$, we see that the right hand side does not vanish identically as it did in subsection \ref{Sec:NonCycZeroBCD}. Nevertheless, following the method of~\cite{Evans:1999mj} we will be able to construct new currents $\K_n^{\lambda_0}$ such that the charges
\begin{equation}\label{Eq:DefQK}
\Q^{\lambda_0}_n = \int \dd x \; \K_n^{\lambda_0}(x)
\end{equation}
Poisson commute with one another.\\

The Poisson bracket \eqref{Eq:PBJTypeA} is to be compared to equation (4.5) of~\cite{Evans:1999mj}, from which it differs only by an overall factor. We can therefore directly apply the procedure developed in~\cite{Evans:1999mj} to the present case so as to construct the desired currents $\K^{\lambda_0}_n$'s. The expression for the first $\K^{\lambda_0}_n$'s read
\begin{align}\label{Eq:KJ}
&\K^{\lambda_0}_2 = \J^{\lambda_0}_2, \;\;\; \K^{\lambda_0}_3=\J^{\lambda_0}_3, \;\;\; \K^{\lambda_0}_4=\J^{\lambda_0}_4-\frac{3}{2d}(\J^{\lambda_0}_2)^2,  \;\;\; \K^{\lambda_0}_5 = \J^{\lambda_0}_5 - \frac{10}{3d} \J^{\lambda_0}_2\J^{\lambda_0}_3, \notag \\
&\K^{\lambda_0}_6 = \J^{\lambda_0}_6 - \frac{5}{3d}(\J^{\lambda_0}_3)^2 - \frac{15}{4d}\J^{\lambda_0}_2\J^{\lambda_0}_4 + \frac{25}{8d^2}(\J^{\lambda_0}_2)^3.
\end{align}
These currents are similar to the currents $\K^{\lambda_0}_n(\xi)$ described in \eqref{Eq:KAlpha} for $\g$ of type B, C or D. More precisely, the current \eqref{Eq:KJ} coincide with the currents $\K^{\lambda_0}_n\left(\frac{1}{d}\right)$, recalling that for type B, C and D, the $\J^{\lambda_0}_{2k+1}$'s vanish. As for $\K^{\lambda_0}_n(\xi)$ in type B, C and D, the expression of the current $\K_n^{\lambda_0}$ for type A is determined (up to a global factor) recursively from equation \eqref{Eq:PBJTypeA} by demanding that the charge $\Q_n^{\lambda_0}$ be in involution with all the charges $\Q_m^{\lambda_0}$ ($m=2,\ldots,n-1$) constructed thus far. However, in the present case, one does not have the freedom of a free parameter $\xi$ in the definition of $\K_n^{\lambda_0}$: there is a unique tower of charges in involution $\Q^{\lambda_0}_n$.

As in the case of type B, C and D algebras, the current $\K_n^{\lambda_0}(x)$ is a homogeneous polynomial of degree $n$ in the fields appearing in the Lax matrix $\Lc(\lambda,x)$. And as explained in~\cite{Evans:1999mj}, the degrees $n$ for which the current $\K_n^{\lambda_0}(x)$ is non-zero are the exponents of the untwisted affine Kac-Moody algebra $\widehat{\g}$ plus one.

At this stage, we do not have a proof that the recursive algorithm described above can be applied indefinitely. We shall now recall from~\cite{Evans:1999mj} how to construct explicitly the current $\K_n^{\lambda_0}$ without a recursive algorithm, using generating functions.

\subsection{Generating functions}
\label{Sec:GenNonCyc}

In the previous subsections \ref{Sec:NonCycZeroBCD} and \ref{Sec:TypeANonCyc}, we introduced currents $\K_n^{\lambda_0}(\xi)$ (for types B, C and D) and $\K_n^{\lambda_0}$ (for type A), constructed recursively from the currents $\J^{\lambda_0}_n$ (and which depended on a free parameter $\xi$ for types B, C and D). In this subsection, we will show how to construct these currents using generating functions.

We will mainly focus on the case where $\g$ is of type A and will briefly comment on types B, C and D at the end of the subsection. Let us then suppose that $\g=\sl(d,\C)$, so that we can use the notations and results of subsection \ref{Sec:TypeANonCyc}. We introduce
\begin{equation}\label{Eq:DefF}
F(\lambda,\mu,x)= \Tr \log\bigl( \Id- \mu \varphi(\lambda)\Lc(\lambda,x) \bigr)
\end{equation}
and
\begin{equation}\label{Eq:DefA}
A(\lambda,\mu,x) = \det\bigl( \Id- \mu \varphi(\lambda)\Lc(\lambda,x) \bigr),
\end{equation}
so that $A(\lambda,\mu,x) = \exp\bigl(F(\lambda,\mu,x)\bigr)$. By expanding the matricial logarithm in \eqref{Eq:DefF} as a power series in $\mu$ one finds
\begin{equation} \label{Eq:PowF}
F(\lambda,\mu,x) = - \sum_{k=2}^{\infty} \frac{\mu^k}{k} \Tc_k(\lambda,x),
\end{equation}
with $\Tc_n(\lambda,x)$ defined in equation \eqref{Eq:DefT}. We are interested in the evaluations of $F(\lambda,\mu,x)$ and $A(\lambda,\mu,x)$ at $\lambda=\lambda_0$, which are well defined as $\lambda_0$ is a regular zero. Following~\cite{Evans:1999mj}, we look for $\K_n^{\lambda_0}(x)$ in the form of
\begin{equation}\label{Eq:KGen1}
\K^{\lambda_0}_n(x) = A(\lambda_0,\mu,x)^{p_n} \Bigr|_{\mu^{n}}
\end{equation}
for some rational number $p_n$, where $f(\mu)|_{\mu^n}$ denotes the coefficient of $\mu^n$ in the power series expansion of $f(\mu)$.

The Poisson brackets of the currents $\Tc_n(\lambda_0,x)=\J_n^{\lambda_0}(x)$ are given by equation \eqref{Eq:PBJTypeA}. This allows one to compute $\left\lbrace F(\lambda_0,\mu,x), F(\lambda_0,\nu,y) \right\rbrace$ and $\left\lbrace A(\lambda_0,\mu,x), A(\lambda_0,\nu,y) \right\rbrace$. As equation \eqref{Eq:PBJTypeA} coincides with the equation (4.5) of~\cite{Evans:1999mj} up to a global factor, these Poisson brackets are the same as in~\cite{Evans:1999mj} (equations (4.13) and (4.14)), still up to the global factor. Thus, the procedure of~\cite{Evans:1999mj} applies and we conclude that the Poisson bracket of the local charges \eqref{Eq:DefQK} defined in terms of the currents \eqref{Eq:KGen1} is
\begin{equation*}
\left\lbrace \Q^{\lambda_0}_n, \Q^{\lambda_0}_m \right\rbrace
=  p_n p_m \mu\nu \frac{\varphi'(\lambda_0)}{T} \int \dd x \; A(\lambda_0,\mu,x)^{p_n} \p_x \bigl( A(\lambda_0,\nu,x)^{p_m} \bigr) h_{mn}(\mu,\nu) \Big|_{\mu^n \nu^m},
\end{equation*}
where
\begin{equation} \label{Eq:hnm def}
h_{nm}(\mu,\nu) = \left[ \left( \frac{n-1}{p_n}\nu - \frac{m-1}{p_m}\mu  \right) \frac{1}{\mu-\nu} + \frac{1}{d}\frac{(n-1)(m-1)}{p_n p_m} \right].
\end{equation}
It follows that the charges $\Q_n^{\lambda_0}$ are in involution if we choose, for any $k\in\Z_{\geq 2}$, $p_k=\frac{k-1}{d}$. The corresponding currents are given by
\begin{equation}\label{Eq:KGen2}
\K_n^{\lambda_0}(x) = \left. \exp \left( - \frac{n-1}{d} \sum_{k=2}^{\infty} \frac{\mu^k}{k} \J_k^{\lambda_0}(x) \right) \right|_{\mu^n}.
\end{equation}
One can check that the first currents defined by this generating function are given by equation \eqref{Eq:KJ}, up to overall global factors. The current $\K_n^{\lambda_0}(x)$ is the evaluation at $\lambda=\lambda_0$ of the more general current
\begin{equation}\label{Eq:DefW}
\W_n(\lambda,x) = A(\lambda,\mu,x)^{(n-1)/d} \Bigr|_{\mu^{n}},
\end{equation}
which we will need later. The equation
\begin{equation}\label{Eq:SerieW}
\W_n(\lambda,x) = \left. \exp \left( - \frac{n-1}{d} \sum_{k=2}^{\infty} \frac{\mu^k}{k} \Tc_k(\lambda,x) \right) \right|_{\mu^n}
\end{equation}
allows one to compute $\W_n(\lambda,x)$ as a polynomial in the $\Tc_k(\lambda,x)$. More precisely, $\W_n$ is related to the $\Tc_k$'s in the same way that $\K^{\lambda_0}_n$ is related to the $\J^{\lambda_0}_k$'s.\\

We end this subsection by saying a few words on Lie algebras $\g$ of type B, C or D. In this case, we saw in subsection \ref{Sec:NonCycZeroBCD} that the local charges in involution can be taken as integrals of currents $\K_{2n}^{\lambda_0}(\xi)$, depending on a free parameter $\xi$ (see equation \eqref{Eq:KAlpha}). These currents can be obtained from the $\J^{\lambda_0}_{2k}$'s using a generating function, similar to the one presented above for type A. We will not enter into details here and will just present the final result, based on reference~\cite{Evans:1999mj}. The current $\K^{\lambda_0}_{2n}(\xi)$ can be computed as:
\begin{equation}
\K_{2n}^{\lambda_0}(\xi,x) = \left. \exp \left( - \frac{\xi(2n-1)}{2} \sum_{k=1}^{\infty} \frac{\mu^k}{k} \J_{2k}^{\lambda_0}(x) \right) \right|_{\mu^n}.
\end{equation}
Starting from the Poisson bracket \eqref{Eq:PBJTypeBCD}, one can show that the corresponding charges $\Q^{\lambda_0}_{2n}(\xi)$ are in involution, using similar techniques as above for type A. We refer the interested reader to reference~\cite{Evans:1999mj} for details on the proof. An explicit computation shows that the first currents $\K^{\lambda_0}_{2n}(\xi)$ obtained from the above equation are given by equation \eqref{Eq:KAlpha}, up to overall global factors.

\subsection{Summary}
\label{Sec:SummaryNonCyc}

To conclude this section, let us summarise the results that we obtained. In particular, we will use this as an opportunity to extend the notations $\K_n^{\lambda_0}$ and $\W_n$, defined for a type A algebra in the previous subsections, to other types. This will serve to uniformise the notation in the rest of the chapter.

When $\g$ is of type A, the currents $\K_n^{\lambda_0}(x)$ are given in subsection \ref{Sec:GenNonCyc} through equation \eqref{Eq:KGen2}. We also defined a current $\W_n(\lambda,x)$ depending on the spectral parameter $\lambda$ in equation \eqref{Eq:SerieW}. For a Lie algebra $\g$ of type B, C or D (as treated in subsection \ref{Sec:NonCycZeroBCD}), we introduced currents $\K^{\lambda_0}_n(\xi)$, depending on a free parameter $\xi$. However, as explained at the end of susbection \ref{Sec:NonCycZeroBCD}, we will only use the currents $\J_n^{\lambda_0}(x)=\K_n^{\lambda_0}(\xi=0,x)$ in the rest of this chapter. In order to employ uniform notations throughout the chapter, we shall define in this case $\K_n^{\lambda_0}(x)=\J_n^{\lambda_0}(x)$ and $\W_n(\lambda,x)=\Tc_n(\lambda,x)$.

With these conventions, independently of the type of $\g$, the current $\K^{\lambda_0}_n(x)$ is the evaluation of $\W_n(\lambda,x)$ at $\lambda=\lambda_0$ and the charge $\Q^{\lambda_0}_n$ is given by
\begin{equation}
\Q_n^{\lambda_0} = \int \dd x \; \K^{\lambda_0}_n(x).
\end{equation}

Recall also that we restrict the degrees $n$ of the currents $\K_n^{\lambda_0}$ to some subset $\E_{\lambda_0}$ of $\Z_{\geq 2}$. In fact, independently of the type of $\g$, $\E_{\lambda_0}$ can (almost) be seen as the set of exponents of the affine algebra $\widehat{\g}$ plus one. This was already observed for type A in subsection \ref{Sec:TypeANonCyc}, based on the results of~\cite{Evans:1999mj}. For types B, C and D, we saw in subsection \ref{Sec:NonCycZeroBCD} that $\E_{\lambda_0}$ is the set of all even numbers, which turns out to coincide with the exponents of $\widehat{\g}$ plus one for types B and C~\cite{Evans:1999mj}. For type D, there are some exponents missing in this construction (the rank modulo the Coxeter number), which are related to the Pfaffian (see~\cite{Evans:1999mj}). Although we do not consider the Pfaffian here, we expect that it should be possible to construct a corresponding local charge in the present framework too.

Having introduced these type-independent notations, we can summarise the results of this section by the following theorem.

\begin{theorem} \label{thm: involution of Qs}
Let $\lambda_0$ be a non-cyclotomic regular zero of the model. Then, for any $m$ and $n$ in $\E_{\lambda_0}$, the charges $\Q^{\lambda_0}_n$ and $\Q^{\lambda_0}_m$ are in involution, \textit{i.e.} we have
\begin{equation*}
\lbrace \Q_n^{\lambda_0}, \Q_m^{\lambda_0} \rbrace = 0.
\end{equation*}
\end{theorem}

The notations and results summarised above will be generalised to the case of 
cyclotomic zeros in the following section. Let us note here that there will be some 
subtlety in the definition of the current $\W_n$ for an algebra of type A in the case 
when the automorphism $\s$ is inner, compared to the definition 
given above. We shall discuss this in subsection \ref{Sec:SummaryCyc}.

\section{Charges at cyclotomic zeros}
\label{Sec:CycZero}

In this section, we explain how to construct towers of local charges in involution attached to cyclotomic regular zeros of the twist function $\varphi$. Recall that a cyclotomic point is a point fixed by the action of the cyclic group $\Z_T$, \textit{i.e.} the origin or infinity. Suppose we are considering a model with a regular zero at infinity. As explained in subsection \ref{Sec:Infinity}, working in the new spectral parameter $\alpha=\lambda^{-1}$ and with the new Lax matrix $\Lc^\infty$ amounts to treating, instead, a model with  a regular zero at $\alpha=0$ and automorphism $\s^{-1}$. Hence it is sufficient to describe the extraction of local charges at the origin.

Throughout this section we therefore consider a  model with $T>1$ and 
a regular zero at the origin. We thus have $\varphi(0)=0$ and $\varphi(\lambda)\Lc(\lambda,x)$ regular at 0. Using the equivariance property \eqref{Eq:TwistEqui}, we see that the smallest power of $\lambda$ in $\varphi$ is of the form $\alpha T-1$, for some $\alpha\in\Z_{\geq 1}$. In terms of the function $\zeta$, defined in equation \eqref{Eq:DefZeta}, this implies that $\zeta(\lambda^T)=O(\lambda^{\alpha T})$. We will mostly need the fact that $\zeta(\lambda^T)=O(\lambda^T)$, \textit{i.e.} that $\varphi(\lambda)=O(\lambda^{T-1})$, and more precisely the asymptotic property
\begin{equation}\label{Eq:ZetaAsymptotic}
\zeta(\lambda^T)=\zeta'(0)\lambda^T + O(\lambda^{2T}).
\end{equation}

Recall that in the previous section we extracted local charges by evaluating the traces of powers of $\varphi(\lambda)\Lc(\lambda,x)$ at the regular zeros. In the case of a cyclotomic point, this method is not sufficient to extract all charges, as such traces can vanish. To understand how to construct the whole algebra of local charges, we will first need to establish equivariance properties of $S_n(\lambda,x)$.

\subsection{Equivariance properties}
\label{Sec:EquivT}

Recall the equivariance properties \eqref{Eq:TwistEqui} and \eqref{Eq:EquiL} of $\varphi$ and $\Lc$. In this subsection, we look for a similar relation for $S_n(\lambda,x)$. In general, $S_n(\lambda,x)$ does not belong to the Lie algebra $\g$ since it is defined as the power of an element of $\g$ seen in the fundamental representation. Thus, one cannot consider directly the action of $\s$ on $S_n(\lambda,x)$.

We refer here to the discussion of appendix \ref{App:ExtSigma}. We will restrict to the case where $\s$ is not one of the special automorphisms of $D_4 = \so(8,\C)$. In this case, we can extend $\s$ to a linear endomorphism on the space $F$ of all matrices acting on the fundamental representation, that we shall still denote $\s$ (see details in appendix \ref{App:ExtSigma}). Note that this new endomorphism $\s$ of $F$ is still of order $T$. We will also need the following properties of $\s$. For any $Z\in F$ we have
\begin{subequations}
\begin{align}
\s(Z^n) &= \epsilon^{n-1} \s(Z)^n, \label{Eq:SigmaPow}\\
\Tr\bigl(\s(Z)\bigr) &= \epsilon \Tr(Z), \label{Eq:SigmaTr}
\end{align}
\end{subequations}
for some $\epsilon$ in $\lbrace 1,-1 \rbrace$. Note that $\epsilon$ is always 1 except when $\g=\sl(d,\C)$ and $\s$ has a non-trivial outer part, in which case $T$ is even. We shall write $\epsilon=\omega^{\frac{\eta T}{2}}$, with $\eta$ in $\lbrace 0,1 \rbrace$.

From equations \eqref{Eq:TwistEqui} and \eqref{Eq:EquiL} and the identity \eqref{Eq:SigmaPow}, we deduce that $S_n$ satisfies the equivariance property
\begin{equation}\label{Eq:EquiS}
\s\bigl(S_n(\lambda,x)\bigr) = \omega^{\kappa(n-1)+1} S_n(\omega\lambda,x),
\end{equation}
with $\kappa=1+\frac{\eta T}{2}$. Let us consider the power series expansion
\begin{equation}\label{Eq:PowS}
S_n(\lambda,x) = \sum_{r=0}^{\infty} A_{n,r}(x) \lambda^r.
\end{equation}
We then find
\begin{equation}\label{Eq:EquiA}
\s(A_{n,r})=\omega^{r+\kappa(n-1)+1} A_{n,r}.
\end{equation}
Taking the trace and using equation \eqref{Eq:SigmaTr}, we find
\begin{equation}\label{Eq:EquiTrA}
\Tr(A_{n,r})=\omega^{r+n\kappa} \Tr(A_{n,r}).
\end{equation}
Thus, $\Tr(A_{n,r})$ vanishes except if $r\equiv r_n \, [T]$, where $r_n$ is the remainder of the euclidian division of $-n\kappa$ by $T$. We define
\begin{equation*}
\J^0_n(x) = \Tr \bigl( A_{n,r_n}(x) \bigr).
\end{equation*}
In particular, the first term in the power series expansion of $\Tc_n(\lambda,x)$ is $\lambda^{r_n} \J^0_n(x)$. Note that $\J^0_n(x)$ is the evaluation of $\Tc_n(\lambda,x)$ at $\lambda=0$ if and only if $r_n=0$, \textit{i.e.} if $T$ divides $n\kappa$. Note also, as $-2\kappa \equiv -2 \, [T]$, that $r_2=T-2$. Thus, we find
\begin{equation*}
\J^0_2(x) = \zeta'(0) \res_{\lambda=0} \varphi(\lambda) \Tr \bigl( \Lc(\lambda,x)^2 \bigr),
\end{equation*}
where $\zeta$ was defined in equation \eqref{Eq:DefZeta}. Finally, let us remark that equation \eqref{Eq:EquiTrA} implies that $\Tc_n(\lambda,x)$ has the following equivariance property
\begin{equation}\label{Eq:EquiT}
\Tc_n(\omega\lambda,x) = \omega^{r_n} \Tc_n(\lambda,x).
\end{equation}

\subsection{Poisson algebra of the currents}
\label{Sec:PBCurrentsCyc}

One can extract the Poisson brackets of the currents $\J^0_n(x)$ and $\J^0_m(y)$ as the coefficient of $\lambda^{r_n+r_m}$ in the power series expansion of $\left\lbrace \Tc_n(\lambda,x), \Tc_m(\lambda,y) \right\rbrace$. The latter can be computed from equation \eqref{Eq:PBT}. Specifically, using the identity \eqref{Eq:RCas} we find
\begin{equation*}
\lambda\mu U\ti{12}(\lambda,\mu)
= - \frac{\zeta(\lambda^T)-\zeta(\mu^T)}{\lambda^T-\mu^T} \sum_{k=0}^{T-1} \lambda^k \mu^{T-k} C^{(k)}\ti{12} + \zeta(\mu^T) C^{(0)}\ti{12},
\end{equation*}
with $\zeta$ defined in equation \eqref{Eq:DefZeta}. Taking the limit $\mu \to \lambda$ we obtain
\begin{equation}\label{Eq:UAround0}
U\ti{12} (\lambda,\lambda) = - \lambda^{T-2} \zeta'(\lambda^T) C\ti{12} + \lambda^{-2} \zeta(\lambda^T) C\ti{12}^{(0)},
\end{equation}
so that
\begin{align}\label{Eq:PBTCyc}
\left\lbrace \Tc_n(\lambda,x), \Tc_m(\lambda,y) \right\rbrace
&  = nm \lambda^{T-2} \zeta'(\lambda^T) \Tr\ti{12}\Bigl( C\ti{12} S_{n-1}(\lambda,x)\ti{1} S_{m-1}(\lambda,y)\ti{2} \Bigr) \delta'_{xy} \\
& \hspace{40pt} -nm \lambda^{-2} \zeta(\lambda^T) \Tr\ti{12}\Bigl( C^{(0)}\ti{12} S_{n-1}(\lambda,x)\ti{1} S_{m-1}(\lambda,y)\ti{2} \Bigr) \delta'_{xy}. \notag
\end{align}

The first term of this Poisson bracket has the same structure as the 
Poisson bracket \eqref{Eq:PBJ}. The main difference coming from cyclotomy is thus the second term, which involves the partial Casimir $C\ti{12}^{(0)}$. We recall that we have the partial completeness relation
\begin{equation}\label{Eq:CompRelPart}
\Tr\ti{2}(C^{(0)}\ti{12} Z\ti{2}) = \pi^{(0)} (Z),
\end{equation}
for any $Z\in\g$.\\

The second term in \eqref{Eq:PBTCyc} will therefore involve the projection $S^{(0)}_{n-1}(\lambda,x)$ of $S_{n-1}(\lambda,x)$ onto the grading zero $F^{(0)} = \left\lbrace Z \in F \, | \, \s(Z)=Z \right\rbrace$. To determine these projections, we can make use of the power series expansion \eqref{Eq:PowS} and equation \eqref{Eq:EquiA}. In particular, one finds that $A_{n-1,r+T}$ is in $F^{(0)}$ if and only if $A_{n-1,r}$ also belongs to $F^{(0)}$. Let us then define $q_n$ to be the unique integer between $0$ and $T-1$ such that $A_{n-1,r}$ belongs to $F^{(0)}$ if and only if $r \equiv q_n \, [T]$. Using equation \eqref{Eq:EquiA} we find $q_n \equiv r_n+1 \, [T]$. So $q_n = r_n +1$ if $r_n \leq T-2$ and $q_n=0$ if $r_n=T-1$.\\

\noi To simplify the Poisson bracket \eqref{Eq:PBTCyc}, we will need to distinguish between three cases:
\begin{itemize}
\item $\g$ is of type B, C or D,
\item $\g$ is of type A and $\s$ is inner,
\item $\g$ is of type A and $\s$ is not inner.
\end{itemize}

\subsection{Algebra of type B, C or D}
\label{Sec:CycBCD}

We first consider $\g$ to be of type B, C or D.  Recall that in this case $S_{2n-1}(\lambda,x)$ belongs to the Lie algebra so that $\Tc_{2n-1}(\lambda,x)$ is zero and hence we consider only the currents $\J^0_{2n}(x)$. Moreover, we can use the completeness relations \eqref{Eq:CompRel} and \eqref{Eq:CompRelPart} in \eqref{Eq:PBTCyc}. We then find
\begin{align} \label{Eq:PBTCycBCD}
\left\lbrace \Tc_{2n}(\lambda,x), \Tc_{2m}(\lambda,y) \right\rbrace
& = 4nm \lambda^{T-2} \zeta'(\lambda^T) \Tr\Bigl( S_{2n-1}(\lambda,x) S_{2m-1}(\lambda,y) \Bigr) \delta'_{xy}  \\
& \hspace{40pt} - 4nm \lambda^{-2} \zeta(\lambda^T) \Tr\Bigl( S^{(0)}_{2n-1}(\lambda,x) S^{(0)}_{2m-1}(\lambda,y) \Bigr) \delta'_{xy}. \notag
\end{align}
After integration over $y$, the first term becomes a total derivative with respect to $x$ by virtue of equation \eqref{Eq:DerS} and thus vanishes when integrated over $x$.\\

Recall moreover that the Poisson bracket of $\J^0_{2n}(x)$ with $\J^0_{2m}(y)$ is obtained from \eqref{Eq:PBTCycBCD} by keeping only the term $\lambda^{r_{2n}+r_{2m}}$ in the power series expansion. We note that the smallest power of $\lambda$ in the second term of \eqref{Eq:PBTCycBCD} is $\alpha T-2+q_{2n}+q_{2m}$ (cf. equation \eqref{Eq:ZetaAsymptotic} and above). As we saw in the previous subsection, $q_k=r_k+1$ if $r_k \leq T-2$ and $q_k = 0$ if $r_k=T-1$. In the case where $r_{2n}$ and $r_{2m}$ are different from $T-1$, the smallest power of $\lambda$ is then $r_{2n}+r_{2m}+\alpha T$ so the second term of \eqref{Eq:PBTCycBCD} does not contribute to the Poisson bracket of $\J^0_{2n}(x)$ with $\J^0_{2m}(y)$, as $\alpha \geq 1$.

If $r_{2n}$ or $r_{2m}$ is equal to $T-1$ then there will be a contribution from this term involving other objects than only the $\J
^0_k$'s, preventing us from constructing charges in involution. Thus, we will only consider the currents $\J^0_{2k}(x)$ such that $r_{2k} \neq T-1$. We then have
\begin{equation*}
\left\lbrace \Q^0_{2n}, \Q^0_{2m} \right\rbrace = 0,
\end{equation*}
where $\Q^0_{2k}$ is the integral of the current $\J^0_{2k}(x)$.\\

We have thus extracted a tower of local charges in involution from the Lax matrix around the origin. Just as in the non-cyclotomic case, these charges are integrals of some polynomials of even degrees in the fields appearing in the Lax matrix. The main difference with the non-cyclotomic case is the fact that, in general, we do not have a current of any even degree. More precisely, we `dropped' the currents of degree $2n$, for all $n$ such that $r_{2n}= T-1$. Recall from appendix \ref{App:ExtSigma} that in the case of an algebra of type B, C or D, we have $\epsilon=1$ and $\kappa=1$. Thus $r_{2n}$ is the remainder of the euclidian division of $-2n$ by $T$, which means that $r_{2n}=T-1$ if and only if $2n \equiv 1 \, [T]$. In particular, we see that there is no drop of any degrees if $T$ is even.

\subsection{Algebra of type A and $\s$ inner}
\label{Sec:TypeATrivial}

Let us now suppose that $\g$ is $\sl(d,\C)$ and $\s$ is inner. In this case, we have the generalised completeness relation 
\eqref{Eq:CompRelSl}. Moreover, we also have a similar identity for the partial Casimir $C\ti{12}^{(0)}$, derived as follows. Recall that for any $Z\in F$, $Z-\frac{1}{d}\Tr(Z)\Id$ 
belongs to $\g$. Moreover, we note that the identity $\Id$ is in the 
grading zero $F^{(0)}$ for $\s$ inner (cf. appendix \ref{App:ExtSigma}). Using equation \eqref{Eq:CompRelPart}, we then have
\begin{equation}\label{Eq:CompRelPartSl}
\Tr\ti{2} \left( C^{(0)}\ti{12} Z\ti{2} \right) = \pi^{(0)}(Z) - \frac{1}{d} \Tr(Z) \Id.
\end{equation}
Using equations \eqref{Eq:CompRelSl} and \eqref{Eq:CompRelPartSl} in the Poisson bracket \eqref{Eq:PBTCyc}, we obtain
\begin{align} \label{Eq:PBTCycA1}
& \left\lbrace \Tc_n(\lambda,x), \Tc_m(\lambda,y) \right\rbrace
 = nm \lambda^{T-2} \zeta'(\lambda^T) \Tr\Bigl( S_{n-1}(\lambda,x) S_{m-1}(\lambda,y) \Bigr) \delta'_{xy} \\
& \hspace{6pt} - nm \lambda^{-2} \zeta(\lambda^T) \Tr\Bigl( S^{(0)}_{n-1}(\lambda,x) S^{(0)}_{m-1}(\lambda,y) \Bigr) \delta'_{xy}
 + \frac{nm}{d} \frac{ \zeta(\lambda^T) - \lambda^{T} \zeta'(\lambda^T)}{\lambda^2} \Tc_{n-1}(\lambda,x) \Tc_{m-1}(\lambda,y) \delta'_{xy} \notag
\end{align}

The Poisson bracket of $\J^0_n(x)$ with $\J^0_m(y)$ is obtained by extracting the coefficient of $\lambda^{r_n+r_m}$ in the above equation.
To treat the second term on the right hand side of this equation, we follow the discussion of the previous subsection \ref{Sec:CycBCD}. The smallest power of $\lambda$ appearing in this term is $\alpha T-2+q_n+q_m$ and if we restrict to $n$ and $m$ such that $r_n$ and $r_m$ are different from $T-1$, this power is strictly greater than $r_n+r_m$. The term then does not contribute to the Poisson bracket $\left\lbrace \J^0_n(x), \J^0_m(y) \right\rbrace$. Let us turn to the third term on the right hand side of equation \eqref{Eq:PBTCycA1}. It can be seen from equation \eqref{Eq:ZetaAsymptotic} that $\zeta(\lambda^T)-\lambda^T \zeta'(\lambda^T) = O(\lambda^{2T})$. The smallest power of $\lambda$ that can appear in this term is thus $2T-2+r_{n-1}+r_{m-1}$, which is always greater than $2T-2$ and therefore strictly greater than $r_n+r_m$ if $r_n$ and $r_m$ are different from $T-1$.\\

In conclusion, only the first term of the right hand side of \eqref{Eq:PBTCycA1} contributes to the Poisson bracket $\left\lbrace \J^0_n(x), \J^0_m(y) \right\rbrace$, which then has the same structure as in the previous subsection. Integrating this bracket over $x$ and $y$, we recognise the integral of a total derivative proportional to $\p_x\Tc_{n+m-2}(\lambda,x)$, which then vanishes, assuming appropriate boundary conditions. Thus, for any $n$ and $m$ such that $r_n$ and $r_m$ are different from $T-1$, we have
\begin{equation*}
\left\lbrace \Q^0_n, \Q^0_m \right\rbrace = 0
\end{equation*}
with $\Q^0_k$ the integral of the current $\J^0_k(x)$. As in the subsection 
\ref{Sec:CycBCD}, we have $\epsilon=1$ and $\kappa=1$ for $\s$ inner. It follows that the integers $n$ such that $r_n=T-1$ (for which we do not consider the charge $\Q^0_n$) are the ones equal to 1 modulo $T$.

\subsection{Algebra of type A and $\s$ not inner}
\label{Sec:TypeANonTrivial}

Finally, let us treat the case where $\g=\sl(d,\C)$ and $\s$ not inner. 
In particular, this implies that $T$ is even and we shall write $T=2S$ in this subsection. We still 
have the generalised completeness relation \eqref{Eq:CompRelSl}. As $\s$ is not inner, 
we have $\s(\Id)=-\Id$ and hence $\pi^{(0)}(\Id)=0$. We deduce that in this case, the partial completeness relation \eqref{Eq:CompRelPart} actually holds for any $Z\in F$. Equation \eqref{Eq:PBTCyc} then gives
\begin{align} \label{Eq:PBTCycA2}
& \hspace{-5pt}\left\lbrace \Tc_n(\lambda,x), \Tc_m(\lambda,y) \right\rbrace  = nm \lambda^{T-2} \zeta'(\lambda^T) \Tr\Bigl( S_{n-1}(\lambda,x) S_{m-1}(\lambda,y) \Bigr) \delta'_{xy} \\
& \hspace{28pt} - nm \lambda^{-2} \zeta(\lambda^T) \Tr\Bigl( S^{(0)}_{n-1}(\lambda,x) S^{(0)}_{m-1}(\lambda,y) \Bigr) \delta'_{xy}  - \frac{nm}{d} \lambda^{T-2} \zeta'(\lambda^T) \Tc_{n-1}(\lambda,x) \Tc_{m-1}(\lambda,y) \delta'_{xy}. \notag
\end{align}
We follow the method of the previous subsections and look for the power $r_n+r_m$ of $\lambda$ in the right hand side of this bracket. As explained in subsection \ref{Sec:CycBCD}, the second term does not contribute when we restrict to $r_n$ and $r_m$ different from $T-1$.\\

The first term is treated as in the case of a non-cyclotomic point: using the identity $f(y)\delta'_{xy}=f(x)\delta'_{xy}+\p_x\bigl(f(x)\bigr) \delta_{xy}$ and the equation \eqref{Eq:DerS}, we find
\begin{equation}\label{Eq:SxS}
\Tr\Bigl( S_{n-1}(\lambda,x) S_{m-1}(\lambda,y) \Bigr) \delta'_{xy} = \Tc_{n+m-2}(\lambda,x) \delta'_{xy} + \frac{m-1}{n+m-2} \p_x \bigl( \Tc_{n+m-2}(\lambda,x) \bigr) \delta_{xy}.
\end{equation}
The powers of $\lambda$ appearing in the power series expansion of the first term are then of the form $r_{n+m-2} - 2 + aT$, with $a \in \Z_{\geq 1}$. One has $r_{n+m-2} \equiv r_n +r_m + 2 \, [T]$ and $0 \leq r_{n+m-2} \leq T-1$. Moreover, $r_n+r_m+2$ is always between $0$ and $2T-2$ if we suppose $r_n$ and $r_n$ different from $T-1$. If $ 0 \leq r_n+r_m+2 < T $, we have $r_{n+m-2}=r_n+r_m+2$ and the powers $r_{n+m+2}-2+ aT$ are then all strictly greater than $r_n+r_m$. If $T \leq r_n+r_m+2 \leq 2T-2$ then $r_{n+m-2}=r_n+r_m+2-T$ and the power $r_{n+m-2}-2+aT$ is equal to $r_n+r_m$ if and only if $a=1$.

Finally, let us consider the third term on the right hand side of \eqref{Eq:PBTCycA2}. The powers of $\lambda$ in its power series expansion are of the form $r_{n-1}+r_{m-1}-2+ aT$, with $a \in \Z_{\geq 1}$. Note that $r_{k-1} \equiv r_k + 1 + S \equiv r_k + 1 - S \, [T]$. Thus, $r_{k-1} = r_k + 1 + S$ if $0 \leq r_k < S-1$ and $r_{k-1} = r_k +1 - S$ if $S-1 \leq r_k \leq T -1$. We then conclude that the power $r_{n-1}+r_{m-1}-2+aT$ is equal to $r_n+r_m$ if and only if $r_n + 1 -S \geq 0$, $r_m + 1 - S \geq 0$ and $a=1$.

Combining all the above results, we find a closed expression for the Poisson bracket of the currents $\J^0_n(x)$ and $\J^0_m(y)$ when $r_n$ and $r_m$ are different form $T-1$, specifically
\begin{align}\label{Eq:PBJCycA}
& \left\lbrace \J^0_n(x), \J^0_m(y) \right\rbrace
= \theta_{r_n+r_m+2-T} \, nm\zeta'(0) \left( \J^0_{n+m-2}(x) \delta'_{xy} + \frac{m-1}{n+m-2} \p_x \left( \J_{n+m-2}^{0}(x) \right) \delta_{xy}  \right) \notag \\
 & \hspace{20pt} - \theta_{r_n+1-S}\, \theta_{r_m+1-S} \, \frac{nm}{d}\zeta'(0)  \Bigl(  \J_{n-1}^{0}(x)\J_{m-1}^{0}(x)\delta'_{xy} + \J_{n-1}^{0}(x) \p_x \left( \J_{m-1}^{0}(x) \right) \delta_{xy} \Bigr),
\end{align}
where $\theta_k=1$ if $k\in\Z_{\geq 0}$ and $\theta_k=0$ if $k\in\Z_{<0}$.\\

As in the case of a non-cyclotomic zero, one can construct charges in involution by taking integrals of new currents $\K^0_n(x)$, constructed as polynomials of the currents $\J^0_m(x)$ for $m \leq n$. The method in the present case is similar: one can construct the currents $\K^0_n$ recursively by asking that the corresponding charge $\Q^0_n$ Poisson commutes with $\Q^0_m$ for all $m<n$. One of the main difference with the non-cyclotomic case is the fact that we do not consider a current $\K^0_n$ when $r_n=T-1$ (we say that such a degree $n$ drops from the construction). The second difference is the presence of the terms $\theta_k$ in the Poisson bracket \eqref{Eq:PBJCycA} compared to \eqref{Eq:PBJTypeA}. Since these terms depend on the numbers $r_k$, the construction of the $\K^0_n$'s will depend on $T$. As an illustration, we give here the expression of the first $\K^0_n$'s for $T$ equal to 2 and 4.

In the case $T=2$, we have $\kappa=2$ hence $r_n=0$ for all $n\geq 2$. Thus, there is no drop of any current due to the condition $r_n\neq T-1$ and the current $\J^0_n(x)$ is simply the evaluation of $\Tc_n(\lambda,x)$ at $\lambda=0$. Moreover, since $2-T$ and $1-S$ are both zero for $T=2$, we note that all $\theta_k$ terms in the Poisson bracket \eqref{Eq:PBJCycA} are equal to 1. The construction for the $\K^0_n$'s is therefore the same as in the non-cyclotomic case and their expression is given by \eqref{Eq:KJ}.

Let us now consider the case $T=4$. We find $\kappa=3$ and $r_n \equiv n \, [4]$ and therefore drop the currents $\J^0_{4k+3}(x)$. Constructing the $\K^0_n$'s recursively we find
\begin{align}\label{Eq:KJCyc4}
&\K^0_2 = \J^0_2, \;\;\; \K^0_4=\J^0_4, \;\;\; \K^0_5=\J^0_5, \\
&\K^0_6 = \J^0_6 - \frac{15}{4d}\J^0_2\J^0_4, \;\;\; \K^0_8 = \J^0_8 - \frac{7}{4d}\bigl(\J_4^0\bigr)^2 \notag
\end{align}
where we dropped the currents $\K^0_3$ and $\K^0_7$ and more generally all currents $\K^0_{4k+3}$.\\

Comparing these currents to the ones constructed in the non-cyclotomic case \eqref{Eq:KJ}, one can observe a pattern in the cyclotomic procedure. Here also, the current $\K^0_n$ is constructed by correcting $\J^0_n$ with monomials $\J^0_{m_1} \ldots \J^0_{m_p}$ such that $m_1+\ldots+m_p=n$. We observe that not all the monomials in the non-cyclotomic corrections appear among the cyclotomic ones but the ones that do have the same coefficients (for example $15/4d$ for the $\J_2\J_4$ correction of $\J_6$). Moreover, we note that a monomial $\J^0_{m_1} \ldots \J^0_{m_p}$ appearing in the non-cyclotomic procedure is also present in the cyclotomic expression if and only if $r_{m_1} + \ldots + r_{m_p} = r_n$.

The above observations are still found to hold for larger values of $T$ and $n$ (although we do not include the corresponding expression of $\K^0_n$ for conciseness). This allows one to find the currents $\K^0_n(x)$ without going through the recursive procedure if one already knows the result for the non-cyclotomic case. A more systematic approach to constructing higher conserved charges in involution in the cyclotomic case would be to find a generating function for the $\K^0_n$'s, generalising the results 
of subsection \ref{Sec:GenNonCyc}. This will be the subject of the next subsection.

\subsection{Generating function for type A with $\s$ not inner}
\label{Sec:CycGenerating}

In subsection \ref{Sec:GenNonCyc} we presented, following~\cite{Evans:1999mj}, the generating function for constructing the currents $\K_n(x)$ in the non-cyclotomic setting. In particular, we found that the relation between the $\K_n(x)$'s and the $\J_m(x)$'s is given by equation \eqref{Eq:KGen2}. In the previous subsection we showed how the currents $\K^0_n(x)$ could be constructed in the cyclotomic case from the knowledge of the corresponding result in the non-cyclotomic case. In particular, starting from the expression of $\K_n$ as a polynomial of the $\J_m$'s, we observed that $\K^0_n$ can be constructed in the same way by keeping monomials $\J^0_{m_1} \ldots \J^0_{m_p}$ with the same coefficient if and only if $r_n=r_{m_1}+\ldots+r_{m_p}$. This procedure for going from the non-cyclotomic to the cyclotomic setting has a natural interpretation in terms of equation \eqref{Eq:KGen2}. Indeed, the current $\K^0_n$ constructed above is equal to
\begin{equation}\label{Eq:KGenCyc1}
\hspace{-2pt}\K_n^{0}(x) = \left. \exp \left( - \frac{n-1}{d} \sum_{k=2}^{\infty} \frac{\mu^k}{k} \lambda^{r_k}\J_k^{0}(x) \right) \right|_{\mu^n \, \lambda^{r_n}},
\end{equation}
where the projection onto the term $\lambda^{r_n}$ ensures that we keep only the monomials satisfying the condition $r_n=r_{m_1}+\ldots+r_{m_p}$.\\

Recall that $\lambda^{r_k} \J_k^0(x)$ is the first term in the power series expansion of $\Tc_k(\lambda,x)$. Moreover, the next terms start with a $(r_k+T)^{\rm th}$ power of $\lambda$. Since $r_n < T$, such terms can be added to the exponent in equation \eqref{Eq:KGenCyc1} without changing the left hand side as they cannot contribute to a $\lambda^{r_n}$-term. We may therefore also write
\begin{equation}\label{Eq:KGenCyc2}
\hspace{-2pt}\K_n^{0}(x) = \left. \exp \left( - \frac{n-1}{d} \sum_{k=2}^{\infty} \frac{\mu^k}{k} \Tc_k(\lambda,x) \right) \right|_{\mu^n \, \lambda^{r_n}}.
\end{equation}
In terms of the definitions \eqref{Eq:DefF} and \eqref{Eq:DefA} of $F(\lambda,\mu,x)$ and $A(\lambda,\mu,x)$ and equation \eqref{Eq:PowF}, we can further re-express equation \eqref{Eq:KGenCyc2} as
\begin{equation}\label{Eq:KGenCyc3}
\K_n^{0}(x) = \left. A(\lambda,\mu,x)^{(n-1)/d} \right|_{\mu^n \, \lambda^{r_n}},
\end{equation}
or again
\begin{equation*}
\K_n^{0}(x) = \left. \exp \left( \frac{n-1}{d} F(\lambda,\mu,x) \right) \right|_{\mu^n \, \lambda^{r_n}}.
\end{equation*}
Starting from equation \eqref{Eq:PBTCycA2}, using equation \eqref{Eq:SxS} and the identity $f(y)\delta'_{xy}=f(x)\delta'_{xy}+\p_x\bigl(f(x)\bigr) \delta_{xy}$, we get
\begin{equation}\label{Eq:PBTTypeA}
\hspace{-4pt}\left\lbrace \Tc_n(\lambda,x), \Tc_m(\lambda,y) \right\rbrace = nm \, \Omega_{nm}(\lambda,x,y) + nm \, \Delta_{nm}(\lambda,x,y)
\end{equation}
where
\begin{subequations}
\begin{align}
\Omega_{nm}(\lambda,x,y)
&=  \lambda^{T-2}\zeta'(\lambda) \left( \Tc_{n+m-2}(\lambda,x) \delta'_{xy}  - \frac{1}{d}\Tc_{n-1}(\lambda,x)\Tc_{m-1}(\lambda,x)\delta'_{xy}\right. \label{Eq:DefOmega} \\
 & \hspace{40pt} \left. + \frac{m-1}{n+m-2} \p_x \left( \Tc_{n+m-2}(\lambda,x) \right) \delta_{xy} - \frac{1}{d}\Tc_{n-1}(\lambda,x) \p_x \left( \Tc_{m-1}(\lambda,x) \right) \delta_{xy} \right), \notag \\[2mm]
 \Delta_{nm}(\lambda,x,y) &=\lambda^{-2} \zeta(\lambda^T) \Tr\Bigl( S^{(0)}_{n-1}(\lambda,x) S^{(0)}_{m-1}(\lambda,y) \Bigr) \delta'_{xy}. \label{Eq:DefDelta}
\end{align}
\end{subequations}

We want to compute the Poisson brackets between the charges $\Q^0_n$ defined as integrals of the currents \eqref{Eq:KGenCyc3}. To begin with, note that the Poisson bracket between $F(\lambda,\mu,x)$ and $F(\lambda,\nu,y)$ can be obtained from equations \eqref{Eq:PowF} and \eqref{Eq:PBTTypeA}. We then find that
\begin{equation}\label{Eq:PBApAq}
\left\lbrace A(\lambda,\mu,x)^p, A(\lambda,\nu,y)^q \right\rbrace = pq \, A(\lambda,\mu,x)^p A(\lambda,\nu,y)^q \sum_{k,l=2}^{\infty} \bigl(\Omega_{kl}(\lambda,x,y)+\Delta_{kl}(\lambda,x,y) \bigr) \mu^k \nu^l.
\end{equation}
Up to a global factor and treating the spectral parameter $\lambda$ as an external parameter, $\Omega_{nm}(\lambda,x,y)$ has the same structure as the right hand side of equation \eqref{Eq:PBJTypeA}, which as we saw already had the same structure as equation (4.5) of~\cite{Evans:1999mj}. This equation (4.5) is used in~\cite{Evans:1999mj} to compute the Poisson brackets of the generating functions (equations (4.13) to (4.15)). The method developed in~\cite{Evans:1999mj} for computing these Poisson brackets then applies to the terms $\Omega_{kl}$ in equation \eqref{Eq:PBApAq} and gives a similar result, up to the global factor and the dependence on $\lambda$ that we keep. Specifically, we find
\begin{align} \label{Eq:PBGenCyc}
& \hspace{-30pt}\int \dd x \int \dd y \left\lbrace A(\lambda,\mu,x)^{p_n} \Bigr|_{\mu^n}, A(\lambda,\nu,y)^{p_m} \Bigr|_{\nu^m} \right\rbrace \\
& \hspace{30pt} = \, \left.  p_n p_m \, \lambda^{T-2}\zeta'(\lambda) \, \mu\nu \int \dd x \; A(\lambda,\mu,x)^{p_n} \p_x \bigl( A(\lambda,\nu,x)^{p_m} \bigr) h_{nm}(\mu,\nu) \;\; \right|_{\mu^n\nu^m} \notag  \\
& \hspace{50pt} + \left. p_n p_m \int \dd x \int \dd y  \; \sum_{k=2}^n \sum_{l=2}^m \Delta_{kl}(\lambda,x,y) \, A(\lambda,\mu,x)^{p_n} \Bigl|_{\mu^{n-k}} A(\lambda,\nu,y)^{p_m} \;\; \right|_{\nu^{n-l}}, \notag
\end{align}
where $h_{nm}(\mu,\nu)$ was defined in equation \eqref{Eq:hnm def}.

The first term on the right hand side of \eqref{Eq:PBGenCyc}, proportional to $h_{nm}(\mu,\nu)$, vanishes when $p_k = \frac{k-1}{d}$ for all $k \in \Z_{\geq 2}$, as expected. It therefore remains to show that the second term also does not contribute when we restrict to the $(r_n+r_m)$-th power of $\lambda$. From equation \eqref{Eq:DefDelta}, we see that the powers of $\lambda$ appearing in the power series expansion of $\Delta_{kl}(\lambda,x,y)$ are of the form $q_k+q_l-2+aT$, with $a\,\geq\alpha\geq\, 1$ and $q_n$ defined in subsection \ref{Sec:PBCurrentsCyc}.

The equivariance property \eqref{Eq:EquiT} can be rewritten as $\Tc_n(\omega\lambda,x)=\omega^{-n\kappa} \Tc_n(\lambda,x)$. In terms of the generating function $F(\lambda,\mu,x)$, this can be re-expressed as $F(\omega\lambda,\mu,x)=F(\lambda,\omega^{-\kappa} \mu,x)$. Thus, we also have $A(\omega\lambda,\mu,x)^p=A(\lambda,\omega^{-\kappa} \mu,x)^p$. Finally, we deduce that
\begin{equation}\label{Eq:EquiAGen}
A(\omega\lambda,\mu,x)^p \Bigr|_{\mu^k} = \omega^{r_k} A(\lambda,\mu,x)^p \Bigr|_{\mu^k}
\end{equation}
In particular, this implies that the powers of $\lambda$ appearing in $A(\lambda,\mu,x)^p \bigr|_{\mu^k}$ are of the form $r_k+bT$ with $b\geq 0$. In conclusion, the powers of $\lambda$ in
\begin{equation}\label{Eq:DeltaAA}
\Delta_{kl}(\lambda,x,y) \, A(\lambda,\mu,x)^p \Bigl|_{\mu^{n-k}} A(\lambda,\nu,y)^q \Bigr|_{\nu^{n-l}}
\end{equation}
are of the form $q_k+r_{n-k}+q_l+r_{m-l}-2+cT$, with $c\geq 1$.

Recall from subsection \ref{Sec:PBCurrentsCyc} that $q_k \equiv r_k+1 \, [T]$, and therefore $q_k+r_{n-k} \equiv r_n + 1 \, [T]$. Suppose now that $r_n$ and $r_m$ are different from, and so in particular strictly less than, $T-1$. As $q_k+r_{n-k}$ is always positive and congruent to $r_n+1$ modulo $T$, which is strictly less than $T$, it then follows that $q_k+r_{n-k} \geq r_n +1$. Similarly, we have $q_l+r_{m-l} \geq r_m+1$ and we thus deduce that $q_k+r_{n-k}+q_l+r_{m-l}-2+cT$ is greater than or equal to $r_n + r_m + T$. We deduce that the term \eqref{Eq:DeltaAA} cannot contribute to the $(r_n+r_m)^{\rm th}$ power of $\lambda$, as required.\\

In conclusion, we have found that, for any $n$ and $m$ such that $r_n$ and $r_m$ are different from $T-1$, one has $\left\lbrace \Q^0_n, \Q^0_m \right\rbrace = 0$, where the charge $\Q^0_k$ is defined as the integral of the current $\K_k^0(x)$ given by \eqref{Eq:KGenCyc3}.
Recall that the current $\J^0_n(x)$ is constructed as the coefficient of $\lambda^{r_n}$ in the power series expansion of $\Tc_n(\lambda,x)$. Similarly, one can rewrite equation \eqref{Eq:KGenCyc3} as
\begin{equation}\label{Eq:KW}
\K^0_n(x) = \W_n(\lambda,x)\Bigr|_{\lambda^{r_n}},
\end{equation}
with $\W_n$ defined in equation \eqref{Eq:DefW}.

\subsection{Summary}
\label{Sec:SummaryCyc}

Let us summarise the results of this section, as we did for non-cyclotomic zeros in subsection \ref{Sec:SummaryNonCyc}. In general, we define a charge $\Q_n^0$, associated with a current $\K_n^0(x)$, by
\begin{equation}\label{Eq:SummaryCyc}
\Q_n^0 = \int \dd x \; \K_n^0(x), \hspace{30pt} \text{ with } \hspace{30pt}
\K_n^0(x) = \W_n(\lambda,x) \Bigr|_{\lambda^{r_n}},
\end{equation}
where the definition of $\W_n(\lambda,x)$ and $r_n$ depends on the type of 
$\g$ and whether $\s$ is inner or not.

For $\g$ of type B, C or D (see subsection \ref{Sec:CycBCD}) and for $\g$ of type A and $\s$ 
 inner (see subsection \ref{Sec:TypeATrivial}), we simply choose 
 $\W_n(\lambda,x)=\Tc_n(\lambda,x)$, so that $\K_n^0(x)=\J_n^0(x)$. In this case, 
 we consider $r_n$ as the remainder of the euclidian division of $-n$ by $T$. When 
 $\g$ is of type A and $\s$ is not inner (the case discussed in subsections \ref{Sec:TypeANonTrivial} and \ref{Sec:CycGenerating}), we choose $\W_n(\lambda,x)$ as given by equation \eqref{Eq:SerieW}. In this case, $r_n$ is defined as the remainder of the euclidian division of $-n\big( 1+\frac{T}{2} \big)$ by $T$. \\

The equivariance properties \eqref{Eq:EquiT} and \eqref{Eq:EquiAGen} imply that, 
independently of the type of $\g$ and  of $\s$ being inner or not, we have
\begin{equation}
\W_n(\omega\lambda,x) = \omega^{r_n}\W_n(\lambda,x),
\end{equation}
for $\W_n$ defined as above. It therefore follows that the powers of $\lambda$ appearing in $\W_n(\lambda,x)$ are of the form $r_n+kT$, with $k\in\Z_{\geq 0}$. In particular, the current $\K^0_n(x)$ of equation \eqref{Eq:SummaryCyc} is the coefficient of the smallest power of $\lambda$ in $\W_n(\lambda,x)$.\\

As in the non-cyclotomic case, we restrict the degree $n$ of the currents $\K^0_n(x)$ to some subset $\E_0$ of $\Z_{\geq 2}$. More precisely, $n$ belongs to $\E_0$ if $n-1$ is an exponent of the affine algebra $\widehat{\g}$ and $r_n$ is different from $T-1$ (with the exception of the exponents related to the Pfaffian in type D, as in subsection \ref{Sec:SummaryNonCyc}). The results of this section can be summarised as the following theorem.

\begin{theorem} \label{thm: involution of Q0s}
Let $n,m\in\E_0$. Then the charges $\Q^0_n$ and $\Q^0_m$ are in involution, \textit{i.e.} we have
\begin{equation*}
\left\lbrace \Q^0_n, \Q^0_m \right\rbrace = 0.
\end{equation*}
\end{theorem}

There is a subtlety in the definition of $\W_n(\lambda,x)$ for $\g$ of type A 
and $\s$ inner. Indeed, in this case the current $\K_n^0(x)$ is extracted just from $\Tc_n(\lambda,x)$ as recalled above. Yet in section \ref{Sec:NonCycZero}, the current $\K^{\lambda_0}_n$ at a non-cyclotomic point was extracted instead from $\W_n(\lambda,x)$ which differs from $\Tc_n(\lambda,x)$. Thus, for this case, we  choose the appropriate definition of $\W_n(\lambda,x)$ depending on whether the regular zeros of the considered model are cyclotomic or not.\\

We end this section by an open question. For a non-cyclotomic regular zero $\lambda_0$ and $\g$ of type B, C or D, we considered local charges in involution $\Q_n^{\lambda_0}$ built as the integral of the currents $\J_n^{\lambda_0}(x)$ (see subsection \ref{Sec:NonCycZeroBCD}). However, based on the results of~\cite{Evans:1999mj}, we also exhibited a more general family of local charges in involution $\Q^{\lambda_0}_n(\xi)$, depending on a free parameter $\xi\in\R$ and whose corresponding currents $\K_n^{\lambda_0}(\xi,x)$ are constructed as polynomials in the $\J^{\lambda_0}_k$'s.

In the subsection \ref{Sec:CycBCD} of the present section, where we deal with a cyclotomic regular zero at the origin for $\g$ of type B, C or D, the charges $\Q^0_n$ are also constructed simply as integrals of the currents $\J^0_n(x)$. It is thus natural to ask if there exists in this case a more general family of charges $\Q^0_n(\xi)$, depending on a free parameter $\xi$, as for the non-cyclotomic case. Moreover, for $\g$ of type A and $\s$ inner (as treated in subsection \ref{Sec:TypeATrivial}), the charges $\Q^0_n$ are also integrals of the currents $\J_n^0(x)$ (we do not need to construct more complicated currents to obtain charges in involution). It is thus also natural to look for a more general family of charges $\Q^0_n(\xi)$. This would be an interesting result, as it would exhibit an important qualitative difference between the non-cyclotomic case and the cyclotomic one (with $\s$ inner), for $\g$ of type A.

We expect these one-parameter families (for $\g$ of type B, C and D or for $\g$ of type A with $\s$ inner) to exist. More precisely, we expect them to be given by the first non-zero coefficient in the power series expansions of a suitable generating function, depending on $\xi$, around the cyclotomic regular zero $\lambda=0$. As for the non-cyclotomic case, we will focus in this chapter on the local charges which do not depend on a free parameter $\xi$ (as described at the beginning of this subsection), for the same reasons as the ones discussed at the end of subsection \ref{Sec:NonCycZeroBCD} for a non-cyclotomic zero.

\section{Properties of the local charges} \label{Sec:PrOfLocCha}

\subsection{Algebra of local charges in involution}
\label{Sec:AlgebraLoc}

In the previous sections, we constructed a tower of local charges in involution at every regular zero $\lambda_0 \in \Zc$. More precisely, we constructed currents $\K_n^{\lambda_0}(x)$, with degrees $n$ in some subset $\E_{\lambda_0}$ of $\Z_{\geq 2}$, such that the charges $\Q^{\lambda_0}_n$ defined as the integral of $\K^{\lambda_0}_n(x)$ are in involution with one another. In this subsection, we study the whole algebra of local charges in involution, formed by all the $\Q^{\lambda_0}_n$ for $\lambda_0\in \Zc$ and $n\in\E_{\lambda_0}$. More precisely, we prove that currents $\K_n^{\lambda_0}(x)$ and $\K_m^{\mu_0}(y)$ extracted at different regular zeros are in involution. We establish this result for regular zeros in $\Zc$, excluding the point at infinity. If infinity is a regular zero then one can also extract charges in involution $\Q^\infty_n$, using the results of subsection \ref{Sec:Infinity}. The Poisson brackets of the corresponding currents with the currents at finite regular zeros involve some subtleties and will be treated separately at the end of the subsection.\\

In general, the currents $\K_n^{\lambda_0}(x)$ and $\K_m^{\mu_0}(y)$ are constructed as polynomials of the currents $\J_n^{\lambda_0}(x)$ and $\J_m^{\mu_0}(y)$. It is therefore sufficient to prove that the Poisson bracket of $\J_n^{\lambda_0}(x)$ and $\J_m^{\mu_0}(y)$ is zero. The currents $\J_n^{\lambda_0}(x)$ and $\J_m^{\mu_0}(y)$ are extracted from $\Tc_n(\lambda,x)$ and $\Tc_m(\mu,y)$, whose Poisson bracket is given by equation \eqref{Eq:PBT}. We can suppose that $\mu_0$ is different from $0$ and thus is a non-cyclotomic point, so that $\J^{\mu_0}_m(y) = \Tc_m(\mu_0,y)$. Using the fact that $U\ti{12}(\lambda,\mu_0)=\varphi(\lambda)\Rc^0\ti{12}(\lambda,\mu_0)$ since $\varphi(\mu_0)=0$, we can evaluate equation \eqref{Eq:PBT} at $\mu=\mu_0$ to find
\begin{equation}\label{Eq:PBTJ}
\left\lbrace \Tc_n(\lambda,x), \J^{\mu_0}_m(y) \right\rbrace = -nm \, \varphi(\lambda) \Tr\ti{12} \Bigl( \Rc^0\ti{12}(\lambda,\mu_0) S_{n-1}(\lambda,x)\ti{1}S_{m-1}(\mu_0,y)\ti{2} \Bigr) \delta'_{xy}.
\end{equation}
We will now treat separately the cases $\lambda_0$ cyclotomic or $\lambda_0$ non-cyclotomic.\\

Suppose that $\lambda_0$ is non-cyclotomic so that $\J^{\lambda_0}_n(x)$ is simply the evaluation of $\Tc_n(\lambda,x)$ at $\lambda=\lambda_0$. Recall from paragraph \ref{Sec:Model} that, by construction of the set $\Zc$, as $\lambda_0$ and $\mu_0$ are different elements of $\Zc$, the cyclotomic orbits $\Z_T\lambda_0$ and $\Z_T\mu_0$ are disjoint. In particular, this means that $\Rc^0\ti{12}(\lambda,\mu_0)$ is regular at $\lambda=\lambda_0$. Indeed, by equation \eqref{Eq:RCyc} the poles of $\Rc^0\ti{12}(\lambda,\mu_0)$ are the points of the orbit $\Z_T\mu_0$. Moreover, $S_{n-1}(\lambda,x)$ is regular at $\lambda=\lambda_0$ and we have $\varphi(\lambda_0)=0$. Thus, evaluating equation \eqref{Eq:PBTJ} at $\lambda=\lambda_0$ we find that the currents $\J^{\lambda_0}_n(x)$ and $\J^{\mu_0}_m(y)$ are in involution, as expected.\\

Let us now treat the cyclotomic case where $\lambda_0=0$. In this case, $\J^0_n(x)$ is the coefficient of $\lambda^{r_n}$ in the power series expansion of $\Tc_n(\lambda,x)$ (cf. subsection \ref{Sec:EquivT}). The Poisson bracket of $\J^0_n(x)$ with $\J^{\mu_0}_m(y)$ is then the coefficient of $\lambda^{r_n}$ in equation \eqref{Eq:PBTJ}. Recall from section \ref{Sec:CycZero} that for $n\in\E_0$, we have $r_n < T-1$. Yet, in equation \eqref{Eq:PBTJ}, $\Rc^0\ti{12}(\lambda,\mu_0)$ and $S_{n-1}(\lambda,x)$ are regular at $\lambda=0$ and $\varphi(\lambda)=O(\lambda^{T-1})$, hence the involution of $\J^0_n(x)$ and $\J^{\mu_0}_m(y)$. In conclusion, we have proved

\begin{theorem}\label{Thm:DiffZeros}
Let $\lambda_0,\mu_0 \in \Zc$ and let $n\in\E_{\lambda_0}$ and $m\in\E_{\mu_0}$. Then, if $\lambda_0\neq\mu_0$, we have
\begin{equation*}
\left\lbrace \J^{\lambda_0}_n(x), \J^{\mu_0}_m(y) \right\rbrace = 0.
\end{equation*}
\end{theorem}
Combining this theorem with the results of previous sections, we conclude that the local charges $\Q^{\lambda_0}_n$, for all $\lambda_0\in\Zc$ and $n\in\E_{\lambda_0}$, are in involution with one another.\\

We now turn to the case where one of the regular zeros is the point at infinity. In this case, the current $\J^\infty_m(y)$ is extracted from the Lax matrix $\Lc^\infty(\alpha,y)$. From the Poisson brackets \eqref{Eq:PBR} and \eqref{Eq:PBLC}, we get
\begin{align}\label{Eq:PBLLI}
& \left\lbrace \Lc(\lambda,x)\ti{1}, \Lc^\infty(\alpha,y)\ti{2} \right\rbrace = \left[ \Rct\ti{12}(\lambda,\alpha^{-1}), \Lc(\lambda,x)\ti{1} \right] \delta_{xy} - \left[ \Rc\ti{21}(\alpha^{-1},\lambda), \Lc^\infty(\alpha,y)\ti{2} \right] \delta_{xy} \\
& \hspace{60pt}  - \; \bigl( \Rct\ti{12}(\lambda,\alpha^{-1}) + \Rc\ti{21}(\alpha^{-1},\lambda) \bigr) \delta'_{xy}  + \alpha^{-1}\psi(\alpha)^{-1} \left[ \Rc\ti{21}(\alpha^{-1},\lambda), \Cc(x)\ti{2} \right] \delta_{xy}, \notag
\end{align}
with the matrix $\Rct$ defined in \eqref{Eq:DefRct}.

In the following, we will say that an equation is true \textit{weakly}, and we will then use the symbol $\approx$ instead of $=$, if the equation holds when one puts the field $\Cc(x)$ to zero. This denomination and its interest will be made clear when studying $\Z_T$-coset models, in which case the field $\Cc(x)$ is interpreted as a gauge constraint. Note, in particular, that the last term of equation \eqref{Eq:PBLLI} vanishes weakly. Using Corollary \ref{Cor:PBTr}, we may then compute the Poisson brackets of $\Tc_n(\lambda,x)$ with
\begin{equation*}
\Tc^\infty_m(\alpha,y)=\Tr\bigl(S^\infty_n(\alpha,x)\bigr)=\Tr\bigl( \psi(\alpha)^n \Lc^\infty(\alpha,x)^n \bigr)
\end{equation*}
weakly, to find
\begin{equation}\label{Eq:PBTTI}
\left\lbrace \Tc_n(\lambda,x), \Tc^\infty_m(\alpha,y) \right\rbrace \approx -nm \, \Tr\ti{12} \Bigl( V\ti{12}(\lambda,\alpha) S_{n-1}(\lambda,x)\ti{1}S^\infty_{m-1}(\alpha,y)\ti{2} \Bigr) \delta'_{xy},
\end{equation}
where
\begin{equation*}
V\ti{12}(\lambda,\alpha) = - \alpha^{-2} \varphi(\lambda) \Rct^0\ti{12}(\lambda,\alpha^{-1}) + \psi(\alpha) \Rc^0\ti{21}(\alpha^{-1},\lambda).
\end{equation*}
Suppose first that $\lambda_0\in\Zc$ is non-cyclotomic. We have $\varphi(\lambda_0)=0$, and hence
\begin{equation*}
\left\lbrace \J^{\lambda_0}_n(x), \Tc^\infty_m(\alpha,y) \right\rbrace \approx -nm \, \psi(\alpha)  \Tr\ti{12} \Bigl( \Rc^0\ti{21}(\alpha^{-1},\lambda_0) S_{n-1}(\lambda_0,x)\ti{1}S^\infty_{m-1}(\alpha,y)\ti{2} \Bigr) \delta'_{xy}.
\end{equation*}
The Poisson bracket between $\J^{\lambda_0}_m(x)$ and $\J^\infty_m(y)$ is then obtained, weakly, by extracting the coefficient of $\alpha^{r_m}$ in the equation above. For $m\in\E_\infty$, we have $r_m < T-1$. Yet $\psi(\alpha)=O(\alpha^{T-1})$ and $ \Rc^0\ti{21}(\alpha^{-1},\lambda_0)$ and $S^\infty_{m-1}(\alpha,y)$ are regular at $\alpha=0$. Thus $\J^{\lambda_0}_n(x)$ and $\J^\infty_m(y)$ are weakly in involution.

It remains to consider the case where $\lambda_0=0$. In this case, $\J^0_n(x)$ is the coefficient of $\lambda^{r_n}$ in $\Tc_n(\lambda,x)$ and we restrict to $n$ such that $r_n < T-1$. We have to find the coefficient of $\lambda^{r_n}\alpha^{r_m}$ in equation \eqref{Eq:PBTTI}. Due to the presence of $\varphi(\lambda)$ or $\psi(\alpha)$ in the two terms appearing in $V\ti{12}(\lambda,\alpha)$, we see that either the power of $\lambda$ or the power of $\alpha$ is greater than $T-1$ and thus cannot contribute to the term $\lambda^{r_n}\alpha^{r_m}$. In conclusion, we have the following theorem.

\begin{theorem}\label{Thm:InvolutionInfinity}
Suppose that infinity is a regular zero of the model. Let $\lambda_0\in\Zc$, $n\in\E_{\lambda_0}$ and $m\in\E_{\infty}$. Then we have
\begin{equation*}
\left\lbrace \J^{\lambda_0}_n(x), \J^{\infty}_m(y) \right\rbrace \approx 0.
\end{equation*}
\end{theorem}

Combining this theorem with the results of previous sections, we see that the local charges $\Q^{\lambda_0}_n$, for all $\lambda_0\in\Zc\cup\lbrace\infty\rbrace$ and $n\in\E_{\lambda_0}$, are (at least weakly) in involution with one another.\\

We thus constructed a large algebra of local charges in involution, composed of the charges $\Q^{\lambda_0}_n$, with $\lambda_0$ regular zeros and $n\in\E_{\lambda_0}$. Since these charges are local, they are also in involution with the momentum $\Pc$ of the theory, whose Poisson bracket generates the derivative with respect to the spatial coordinate $x$. We have not yet discussed the conservation properties of these charges. For the models that we will consider as examples in this chapter, we will see in section \ref{Sec:Applications} that the Hamiltonian $\Hc$ of the theory always belongs to the algebra of local charges described above. It therefore follows that all these charges are conserved. More precisely, we will see that $\Hc$ is always a linear combination of the quadratic charges $\Q^{\lambda_0}_2$ and the momentum $\Pc$.

\subsection{Gauge invariance}
\label{Sec:Gauge}

In this subsection, we anticipate the application of the construction developed in the previous sections to integrable $\s$-models on $\Z_T$-coset spaces. In those models, infinity is a regular zero and the corresponding field $\Cc(x)$ (cf. subsection \ref{Sec:Infinity}) is a gauge constraint. We prove here that the currents $\K_n^{\lambda_0}(x)$ constructed at regular zeros $\lambda_0$ in the previous sections are gauge invariant, in the sense that they Poisson commute with the constraint $\Cc(y)$. As the $\K_n^{\lambda_0}$'s are polynomials of the $\J_n^{\lambda_0}$'s, it is enough to prove the following theorem.

\begin{theorem}\label{Thm:Gauge}
Suppose that infinity is a regular zero. Let $\lambda_0 \in \Zc\cup\lbrace \infty \rbrace$ and $n\in\E_{\lambda_0}$. Then, we have
\begin{equation*}
\left\lbrace \J^{\lambda_0}_n(x), \Cc(y) \right\rbrace = 0.
\end{equation*}
\end{theorem}

\begin{proof}
Let us first suppose that $\lambda_0$ is different from infinity. The current $\J^{\lambda_0}_n(x)$ is extracted from $\Tc_n(\lambda,x)=\varphi(\lambda)^n\Lc(\lambda,x)^n$. Recall the Poisson bracket between the Lax matrix $\Lc(\lambda,x)$ and $\Cc(y)$, given by equation \eqref{Eq:PBLC}. By Corollary \ref{Cor:PBTr} we then have
\begin{equation*}
\left\lbrace \Tc_n(\lambda,x), \Cc(y) \right\rbrace = - n\, \varphi(\lambda) \, \Tr\ti{1} \bigl( C^{(0)}\ti{12} S_{n-1}(\lambda,x)\ti{1} \bigr) \delta'_{xy}.
\end{equation*}

If $\lambda_0\neq 0$, $\J_n^{\lambda_0}(x)$ is simply $\Tc_n(\lambda_0,x)$. And since $\lambda_0$ is a regular zero, $S_{n-1}(\lambda,x)$ is regular at $\lambda=\lambda_0$ and $\varphi(\lambda_0)=0$. Evaluating the above Poisson bracket at $\lambda=\lambda_0$, we get the involution of $\J^{\lambda_0}_n(x)$ and $\Cc(y)$. If $\lambda_0=0$, $\J^0_n(x)$ is the coefficient of $\lambda^{r_n}$ in $\Tc_n(\lambda,x)$ and, since $n\in\E_0$, we have $r_n < T-1$. Moreover, since $S_{n-1}(\lambda,x)$ is regular at $\lambda=0$ and $\varphi(\lambda)=O(\lambda^{T-1})$, the $\lambda^{r_n}$-term in the Poisson bracket above is then zero, as required.

Finally, let us treat the case $\lambda_0=\infty$, for which $\J^\infty_n(x)$ is given by the coefficient of $\alpha^{r_n}$ in $\Tc^\infty_n(\alpha,x)= \Tr\big( \psi(\alpha)^n \Lc^\infty(\alpha,x)^n \big)$. Using the definition of $\Lc^\infty$ and the Poisson brackets \eqref{Eq:PBLC} and \eqref{Eq:PBCC}, we find
\begin{equation*}
\left\lbrace \Lc^\infty(\alpha,x)\ti{1}, \Cc(y)\ti{2} \right\rbrace = \bigl[ C^{(0)}\ti{12}, \Lc^\infty(\alpha,x)\ti{1} \bigr] \delta_{xy} - C^{(0)}\ti{12}\delta'_{xy}.
\end{equation*}
This bracket has the same structure as equation \eqref{Eq:PBLC}. Therefore, the case $\lambda_0=\infty$ is treated exactly in the same way than the case $\lambda_0=0$, which ends the proof.
\end{proof}

\subsection{Reality conditions}
\label{Sec:Reality}

To close this section let us discuss the reality conditions on the charges $\Q^{\lambda_0}_n$ extracted at regular zeros in the previous sections. In Chapter \ref{Chap:Models}, we considered integrable $\s$-models with target space $G_0$ or a quotient of $G_0$, where $G_0$ is a real Lie group. If $\g_0$ is the Lie algebra of $G_0$, then the Lax matrix of the model is a $\g$-valued field, where $\g$ is the complexification of $\g_0$. In other words, $\g_0$ is a real form of the complex Lie algebra $\g$: it is thus characterised by a antilinear involutive automorphism $\tau$ (see Appendix \ref{App:RealForms}). The fact that the $\s$-models we consider are on the real form $G_0$ (or one of its quotient) is encoded in the reality conditions \eqref{Eq:Reality} and \eqref{Eq:TwistReal} of the Lax matrix and of the twist function.

In particular, by equation \eqref{Eq:TwistReal}, if $\lambda_0$ is a zero of $\varphi$, its conjugate $\bar{\lambda}_0$ is also a zero of $\varphi$. Combining the reality conditions \eqref{Eq:Reality} and \eqref{Eq:TwistReal}, we also see that if $\lambda_0$ is a regular zero (see paragraph \ref{Sec:Model}), $\bar\lambda_0$ is also a regular zero.
Thus, the regular zeros can be of two types: real ones $\lambda_0\in\R$ and conjugate pairs $\lambda_0$, $\bar\lambda_0$ in $\C\setminus\R$.
\\

We will use the reality condition \eqref{Eq:Reality} in a similar way to the way we used the equivariance property \eqref{Eq:EquiL} in subsection \ref{Sec:EquivT}. In particular, as we consider powers of the Lax matrix, which are not in the Lie algebra $\g$ in general, we will need to extend ``naturally'' the automorphism $\tau$ to the whole algebra $F$ of matrices acting on the defining representation of $\g$. This was done for the automorphism $\s$ in subsection \ref{Sec:EquivT} and appendix \ref{App:ExtSigma}. One can apply similar ideas to $\tau$, using the classification of real forms of the classical Lie algebras A, B, C and D. We do not present the details here and just summarise the results.

There exists an extension of $\tau$ on the whole algebra of matrices $F$, which coincides with $\tau$ when restricted to the Lie algebra $\g$, and that we shall still denote $\tau$. This extension is still an involutive semi-linear map of $F$ to itself. However, it is not in general an algebra homomorphism. The main properties of the extension $\tau$ that we will need are the following. There exists $\gamma\in\lbrace 1, -1 \rbrace$ such that
\begin{subequations}\label{Eq:PropTau}
\begin{align}
\tau(Z^n) &= \gamma^{n-1} \, \tau(Z)^n, \label{Eq:TauPow}\\
\Tr\bigl(\tau(Z)\bigr) &= \gamma\, \overline{\Tr(Z)}, \label{Eq:TauTr}
\end{align}
\end{subequations}
for any $Z\in F$. For every real form $\g_0$ of a classical algebra $\g$ we have $\gamma = 1$, except for the real forms $\mathfrak{su}(p,q,\R)$ of $\sl(d,\C)$ (with $p+q=d$), for which $\gamma=-1$. Using the properties \eqref{Eq:PropTau} with the reality conditions \eqref{Eq:Reality} and \eqref{Eq:TwistReal}, one finds that $\tau\bigl(S_n(\lambda,x)\bigr) = \gamma^{n-1} S_n(\bar{\lambda},x)$ and that
\begin{equation}\label{Eq:TReal}
\overline{\Tc_n(\lambda,x)} = \gamma^n \Tc_n(\bar{\lambda},x).
\end{equation}

Consider a regular zero $\lambda_0$. Suppose first that $\lambda_0$ is complex: its conjugate $\bar{\lambda}_0$ is then also a regular zero. According to the previous sections, we can extract two towers of (possibly complex) currents $\J^{\lambda_0}_n(x)$ and $\J^{\bar\lambda_0}_n(x)$ by evaluating $\Tc_n(\lambda,x)$ at $\lambda=\lambda_0$ or $\lambda=\bar\lambda_0$ (note that $\lambda_0$ cannot be a cyclotomic point as it is complex). However, according to equation \eqref{Eq:TReal}, these currents are not independent. Indeed, they are related by the reality condition
\begin{equation*}
\J^{\bar\lambda_0}_n(x) = \gamma^n \overline{ \J^{\lambda_0}_n(x) }.
\end{equation*}
Thus, considering linear combination of $\Q^{\lambda_0}_n$ and $\Q^{\bar\lambda_0}_n$, we extract from each pair $\lambda_0$, $\bar\lambda_0$ of complex regular zeros two towers of real charges in involution: $\Q^{\lambda_0}_n+\gamma^n\Q^{\bar\lambda_0}_n$ and $i\bigl(\Q^{\lambda_0}_n-\gamma^n\Q^{\bar\lambda_0}_n\bigr)$.\\

Suppose now that $\lambda_0$ is a real and non-cyclotomic regular zero. Equation \eqref{Eq:TReal} then imposes the reality condition
\begin{equation}\label{Eq:RealZero}
\J^{\lambda_0}_n(x) = \gamma^n  \overline{ \J^{\lambda_0}_n(x)}.
\end{equation}
Thus, the current $\J^{\lambda_0}_n$ is either real or pure imaginary. In each case, we can extract only one tower of real local charges.
Consider now the case where $\lambda_0$ is the origin and thus a cyclotomic real point. The current $\J^0_n(x)$ is then the coefficient of $\lambda^{r_n}$ in the power series expansion of $\Tc_n(\lambda,x)$. Yet, this coefficient is also the one of $\bar{\lambda}^{r_n}$ in the power series expansion of $\Tc_n(\bar{\lambda},x)$. The reality condition \eqref{Eq:TReal} then implies that equation \eqref{Eq:RealZero} also holds for $\lambda_0=0$.

Finally, let us discuss the case where $\lambda_0$ is infinity, which we consider as a real point. From the reality conditions \eqref{Eq:Reality} and \eqref{Eq:TwistReal}, we find that the field $\Cc(x)$ defined in subsection \ref{Sec:Infinity} is real, in the sense that $\tau\bigl(\Cc(x)\bigr)=\Cc(x)$. We then obtain reality conditions on the Lax matrix $\Lc^\infty(\alpha,x)$ and the twist function $\psi(\alpha)$ similar to equations \eqref{Eq:Reality} and \eqref{Eq:TwistReal}. As a result we can apply the above discussion, since the point at infinity in the variable $\lambda$ corresponds to the origin in the variable $\alpha$, and conclude that equation \eqref{Eq:RealZero} also holds for $\lambda_0=\infty$.\\

\noi To summarise this subsection, we have shown that one can extract:\vspace{-4pt}
\begin{itemize}\setlength\itemsep{0.1em}
\item one tower of real local charges for each real regular zero $\lambda_0$,
\item two towers of real local charges for each pair $\lambda_0$, $\bar\lambda_0$ of complex regular zeros.\vspace{-4pt}
\end{itemize}
In other words, one can extract as many towers of real charges as there are regular zeros.

\section{Integrable hierarchies and zero curvature equations}
\label{Sec:IntHierZeroCurv}

In the previous sections, we constructed a infinite set of local charges $\Q^{\lambda_0}_n$ in involution, with $\lambda_0$ regular zeros. It induces an infinite set of commuting Hamiltonian flows, defined by $\left\lbrace \Q^{\lambda_0}_n, \cdot \right\rbrace$. In this section, we show that these flows generate a hierarchy of integrable equations. More precisely, we associate with each charge $\Q^{\lambda_0}_n$ a connection
\begin{equation*}
\nabla^{\lambda_0}_n = \left\lbrace \Q^{\lambda_0}_n, \cdot \right\rbrace + \M^{\lambda_0}_n (\lambda,x)
\end{equation*}
which commutes with the connection $\nabla_x = \p_x+\Lc(\lambda,x)$. We show that the connections $\nabla^{\lambda_0}_n$ also commute with one another for finite regular zeros $\lambda_0$. The commutativity of these connections takes the form of zero curvature equations. In particular, we will use the zero curvature equations involving $\Lc(\lambda,x)$ and the $\M^{\lambda_0}_n(\lambda,x)$'s to prove that the local charges $\Q^{\lambda_0}_n$ are in involution with the non-local charges extracted from the monodromy of $\Lc(\lambda,x)$.

\subsection{Zero curvature equations with $\Lc$}
\label{Sec:ZCEL}

The starting point of this chapter is an integrable system with Lax matrix $\Lc(\lambda,x)$ and Hamiltonian $\Hc$. The dynamical equations of this system are generated by the Poisson bracket with $\Hc$. They are encoded in the form of the zero curvature equation \eqref{Eq:ZCEH}. In this subsection, we study the dynamics of the Lax matrix under the Hamiltonian flows generated by the local charges $\Q^{\lambda_0}_n$ constructed in the previous sections. More precisely, we show that these dynamics also take the form of a zero curvature equation on $\Lc(\lambda,x)$:

\begin{theorem}\label{Thm:ZCEL}
Let $\lambda_0\in\Zc$ and $n\in\E_{\lambda_0}$. There exists a matrix $\M^{\lambda_0}_n(\lambda,x)$ such that we have the zero curvature equation
\begin{equation*}
\left\lbrace \Q^{\lambda_0}_n, \Lc(\lambda,x) \right\rbrace - \p_x \M^{\lambda_0}_n(\lambda,x) + \left[ \M^{\lambda_0}_n(\lambda,x), \Lc(\lambda,x) \right] = 0.
\end{equation*}
\end{theorem}

\begin{proof}
Let us apply the second result of Corollary \ref{Cor:PBTr} to the Maillet bracket \eqref{Eq:PBR}. Using the form \eqref{Eq:DefR} of the $\Rc$-matrix, we find
\begin{align}\label{Eq:PBLT}
& \left\lbrace \Lc(\lambda,x), \Tc_n(\mu,y) \right\rbrace = n \left[ \Tr\ti{2}\Bigl( \Rc^0\ti{12}(\lambda,\mu) S_{n-1}(\mu,y)\ti{2} \Bigr), \Lc(\lambda,x) \right] \delta_{xy}  \\
& \hspace{75pt} - n \, \Tr\ti{2} \Bigl( \Rc^0\ti{12}(\lambda,\mu) S_{n-1}(\mu,y)\ti{2} \Bigr) \delta'_{xy}  - n \frac{\varphi(\mu)}{\varphi(\lambda)} \, \Tr\ti{2} \Bigl( \Rc^0\ti{21}(\mu,\lambda)  S_{n-1}(\mu,y)\ti{2} \Bigr) \delta'_{xy}. \notag
\end{align}

Consider first the case where $\lambda_0$ is a non-cylotomic regular zero. Evaluating the equation above at $\mu=\lambda_0$ and using $\varphi(\lambda_0)=0$, we have
\begin{equation}\label{Eq:ZCEJ}
\left\lbrace \Lc(\lambda,x), \J^{\lambda_0}_n(y) \right\rbrace
 = \left[ \mathcal{N}^{\lambda_0}_n(\lambda,y), \Lc(\lambda,x) \right] \delta_{xy}  - \mathcal{N}^{\lambda_0}_n(\lambda,y) \delta'_{xy},
\end{equation}
where
\begin{equation}\label{Eq:DefNNonCyc}
\mathcal{N}^{\lambda_0}_n(\lambda,x) = n \, \Tr\ti{2} \Bigl( \Rc^0\ti{12} (\lambda,\lambda_0) S_{n-1}(\lambda_0,x)\ti{2} \Bigr).
\end{equation}

Suppose now that $\lambda_0$ is the origin, which is a cyclotomic point, in which case $\J^0_n(y)$ is constructed as the coefficient of $\mu^{r_n}$ in the power series expansion of $\Tc_n(\mu,y)$. Moreover, as $n\in\E_0$, we have $r_n<T-1$ (see section \ref{Sec:CycZero}). 
The Poisson bracket $\left\lbrace \Lc(\lambda,x), \J^0_n(y) \right\rbrace$ is thus the $\mu^{r_n}$-term in equation \eqref{Eq:PBLT}. We have $\varphi(\mu)=O(\mu^{T-1})$ and $r_n <T-1$, thus the last term of equation \eqref{Eq:PBLT} cannot contribute to $\mu^{r_n}$. Thus, we also have equation \eqref{Eq:ZCEJ} for $\lambda_0=0$, with
\begin{equation*}
\Nc^0_n(\lambda,x) = n \, \Tr\ti{2} \Bigl( \Rc^0\ti{12} (\lambda,\mu) S_{n-1}(\mu,x)\ti{2} \Bigr) \Bigr|_{\mu^{r_n}}.
\end{equation*}

We will say that $\mathcal{N}^{\lambda_0}_n$ is the Lax matrix associated with the charge defined as the integral of the current $\J_n^{\lambda_0}$. Equation \eqref{Eq:ZCEJ} implies a zero curvature equation for the evolution of $\Lc(\lambda,x)$ under the Hamiltonian flow of this charge. In general, the charge $\Q^{\lambda_0}_n$ is not the integral of $\J_n^{\lambda_0}$ but of $\K_n^{\lambda_0}$ (see previous sections). Recall that $\K_n^{\lambda_0}$ is a polynomial in the $\J_m^{\lambda_0}$'s. We construct the Lax matrix $\M^{\lambda_0}_n(\lambda,x)$ associated with $\Q_n^{\lambda_0}$ by assigning any monomial $\J^{\lambda_0}_{m_1} \ldots \J^{\lambda_0}_{m_p}$ in this polynomial to the matrix
\begin{equation*}
\sum_{k=1}^p \Bigl( \prod_{j\neq k} \J_{m_j}^{\lambda_0}(x) \Bigr) \mathcal{N}^{\lambda_0}_{m_k} (\lambda,x).
\end{equation*}
Using the fact that the Poisson bracket is a derivation, we find from equation \eqref{Eq:ZCEJ} that
\begin{equation}\label{Eq:ZCEK}
\left\lbrace \Lc(\lambda,x), \K^{\lambda_0}_n(y) \right\rbrace
 = \left[ \M^{\lambda_0}_n(\lambda,y), \Lc(\lambda,x) \right] \delta_{xy} - \M^{\lambda_0}_n(\lambda,y) \delta'_{xy}.
\end{equation}
After integration over $y$, we get the required zero curvature equation.
\end{proof}

Thus, the Hamiltonian flows of the charges $\Q^{\lambda_0}_n$ generate dynamical equations that can be recast in the form of zero curvature equations. In conclusion, we have constructed a hierarchy of integrable systems with Lax matrix $\Lc(\lambda,x)$ and Hamiltonians $\Q^{\lambda_0}_n$. The zero curvature equations of Theorem \ref{Thm:ZCEL} can be seen as the commutativity of the connections
\begin{equation}\label{Eq:Nabla}
\nabla^{\lambda_0}_n = \left\lbrace \Q^{\lambda_0}_n, \cdot \right\rbrace + \M^{\lambda_0}_n (\lambda,x)
\end{equation}
with $\nabla_x = \p_x + \Lc(\lambda,x)$. This connection $\nabla_x$ can be thought of as the connection associated with the local momentum $\Pc$ of the theory. As already mentioned, we will see in section \ref{Sec:Applications} that for the models we consider, the Hamiltonian is given by a linear combination $\Hc=\sum_{\lambda_0\in\Zc} a_{\lambda_0} \Q^{\lambda_0}_2 + b \Pc$ of the quadratic charges $\Q^{\lambda_0}_2$ and the momentum $\Pc$. Therefore, the matrix $\M(\lambda,x)$ of equation \eqref{Eq:ZCEH} can be constructed as $\sum_{\lambda_0\in\Zc} a_{\lambda_0} \M^{\lambda_0}_2(\lambda,x) + b \Lc(\lambda,x)$.\\

Theorem \ref{Thm:ZCEL} only treats the case of finite regular zeros $\lambda_0$. Let us also briefly discuss what happens when $\lambda_0=\infty$. In this case, $\J^\infty_n(x)$ is extracted from the Lax matrix $\Lc^\infty(\alpha,x)$. Since this matrix satisfies a Maillet bracket with twist function $\psi(\alpha)$, one can apply the method developed here. Doing so we find that the dynamics of $\Lc^\infty(\alpha,x)$ under the Hamiltonian flow of $\Q^\infty_n$ takes the form of a zero curvature equation. Moreover, starting with the Poisson bracket \eqref{Eq:PBLLI} and working weakly, we also find a weak curvature equation
\begin{equation*}
\left\lbrace \Q^{\infty}_n, \Lc(\lambda,x) \right\rbrace - \p_x \M^{\infty}_n(\lambda,x) + \left[ \M^{\infty}_n(\lambda,x), \Lc(\lambda,x) \right] \approx 0,
\end{equation*}
where the matrix $\M^{\infty}_n(\lambda,x)$ is constructed from
\begin{equation*}
\Nc^\infty_n(\lambda,x) = - n \, \alpha^{-2} \, \Tr\ti{2} \Bigl( \Rct^0\ti{12} (\lambda,\alpha^{-1}) S^\infty_{n-1}(\alpha,x)\ti{2} \Bigr) \Bigr|_{\alpha^{r_n}}
\end{equation*}
in the same way as $\M^{\lambda_0}_n(\lambda,x)$ was built from $\Nc^{\lambda_0}_n(\lambda,x)$ for a finite regular zero $\lambda_0$. In other words, Theorem \ref{Thm:ZCEL} also applies for $\lambda_0=\infty$ when Poisson brackets are considered weakly.\\

Let us end this subsection by stating a few properties of the Lax matrix $\M^{\lambda_0}_n(\lambda,x)$. Using the equivariance property \eqref{Eq:EquiR}, we find that
\begin{equation*}
\s \left( \M^{\lambda_0}_n(\lambda,x) \right) = \M^{\lambda_0}_n(\omega\lambda,x).
\end{equation*}
The Lax matrix $\M^{\lambda_0}_n$ thus satisfies the same equivariance property \eqref{Eq:EquiL} as $\Lc$. Recall that the Lax matrix $\Nc^0_n(\lambda,x)$ is extracted as the $\mu^{r_n}$-term in
\begin{equation}\label{Eq:DefN}
\Nc_n(\mu \,;\lambda,x) =  n \, \Tr\ti{2} \Bigl( \Rc^0\ti{12} (\lambda,\mu) S_{n-1}(\mu,x)\ti{2} \Bigr).
\end{equation}
Consider the equivariance properties \eqref{Eq:EquiS} and
\begin{equation*}
\s\ti{2} \Rc^0\ti{12}(\lambda,\mu) = \omega\Rc^0\ti{12}(\lambda,\omega\mu).
\end{equation*}
Combining it with the fact that $\Tr\bigl(\s(Y)\s(Z)\bigr)=\Tr(YZ)$ for any matrices $Y,Z\in F$ (see appendix \ref{App:ExtSigma}), we find that
\begin{equation}\label{Eq:EquiN}
\Nc_n(\omega\mu\,;\lambda,x) = \omega^{r_n} \Nc_n(\mu\,;\lambda,x).
\end{equation}
Therefore, the power series expansion of $\Nc_n(\mu\,; \lambda,x)$ in $\mu$ contains powers of the form $r_n+kT$, with $k\in\Z_{\geq 0}$. In particular, $\Nc^0_n(\lambda,x)$ is the coefficient of the smallest power in this expansion, in the same way as $\J^0_n(x)$ is in the expansion of $\Tc_n(\mu,x)$.

Let us define $\M_n(\mu\,;\lambda,x)$ from $\Nc_n(\mu\,;\lambda,x)$ and $\Tc_n(\mu,x)$ in the same way we constructed $\M_n^{\lambda_0}(\lambda,x)$ from $\Nc^{\lambda_0}_n(\lambda,x)$ and $\J^{\lambda_0}_n(x)$. In particular, $\M^{\lambda_0}_n(\lambda,x)$ is the evaluation of $\M_n(\mu\,;\lambda,x)$ at $\mu=\lambda_0$. From equations \eqref{Eq:EquiTrA} and \eqref{Eq:EquiN}, we find the following equivariance property
\begin{equation}\label{Eq:EquiM}
\M_n(\omega\mu\,;\lambda,x) = \omega^{r_n} \M_n(\mu\,;\lambda,x).
\end{equation}
So $\M^0_n(\lambda,x)$ is the coefficient of the first term $\mu^{r_n}$ in the power series expansion of $\M_n(\mu\,;\lambda,x)$.

\subsection{Involution with non-local charges}
\label{Sec:NonLocal}

In this subsection, we use the result of the previous one to prove that the local charges $\Q^{\lambda_0}_n$ are in involution with the non-local charges extracted from the monodromy of the Lax matrix $\Lc(\lambda,x)$. This monodromy is defined as the path-ordered exponential (see Appendix \ref{App:Pexp})
\begin{equation*}
T(\lambda) = \Pexp \left( -\int \dd z \, \Lc(\lambda,z) \right),
\end{equation*}
where the integral is taken on the real line $\R$ or the circle $S^1$, depending on the coordinate space of the model. Consider also the partial transfer matrices
\begin{equation*}
T(\lambda\;;x,y) = \Pexp \left( -\int_y^x \dd z \, \Lc(\lambda,z) \right).
\end{equation*}
The properties of these matrices are described in Appendix \ref{App:Pexp}. In particular, their variation under an infinitesimal variation of $\Lc$ is given by equation \eqref{dPExp}.

This formula allows one to compute derivatives of $T$ and in particular its Poisson bracket with the local charge $\Q^{\lambda_0}_n$, for $\lambda_0\in\Zc$ and $n\in\E_{\lambda_0}$. Specifically, we have
\begin{equation}\label{Eq:PBPexp}
\left\lbrace \Q^{\lambda_0}_n, T(\lambda\;;x,y) \right\rbrace
 = -\int_y^x \dd z \, T(\lambda\;;x,z) \left\lbrace \Q^{\lambda_0}_n, \Lc(\lambda,z) \right\rbrace T(\lambda\;;z,y).
\end{equation}
The Poisson bracket of $\Q^{\lambda_0}_n$ and $\Lc(\lambda,z)$ is given by Theorem \ref{Thm:ZCEL}. Using this together with the equations \eqref{Eq:DerPexp} and \eqref{Eq:PBPexp}, we find
\begin{equation*}
\left\lbrace \Q^{\lambda_0}_n, T(\lambda\;;x,y) \right\rbrace = T(\lambda\;;x,y) \M^{\lambda_0}_n(\lambda,y) - \M^{\lambda_0}_n(\lambda,x) T(\lambda\;;x,y).
\end{equation*}
If the spatial coordinate is taken on the real line (from $-\infty$ to $\infty$) and the fields are assumed to be decreasing at infinity fast enough, we get
\begin{equation*}
\left\lbrace \Q^{\lambda_0}_n, T(\lambda) \right\rbrace = 0,
\end{equation*}
\textit{i.e.} the whole monodromy $T(\lambda)$ is in involution with $\Q^{\lambda_0}_n$. If the spatial coordinate is taken on the circle (from $0$ to $2\pi$) and the fields are assumed to be periodic, we get
\begin{equation*}
\left\lbrace \Q^{\lambda_0}_n, T(\lambda) \right\rbrace = \left[ T(\lambda), \M^{\lambda_0}_n(\lambda,0) \right].
\end{equation*}
In this case, $\Q^{\lambda_0}_n$ Poisson commutes with any central function of $T(\lambda)$, \textit{e.g.} the traces $\Tr\bigl(T(\lambda)^k\bigr)$ and the determinant $\det\bigl(T(\lambda)\bigr)$. Thus, we have

\begin{theorem}\label{Thm:NonLocal}
The monodromy $T(\lambda)$ (resp. the central functions of $T(\lambda)$) is in involution with the local charges $\Q^{\lambda_0}_n$ for $\lambda_0\in\Zc$ and $n\in\E_{\lambda_0}$, if the spatial coordinate is taken on the real line (resp. the circle). In particular, it is conserved.
\end{theorem}
\begin{proof}
It just remains to prove the conservation of the non-local charges. This follows from the fact that the Hamiltonian $\Hc$ can be expressed as a linear combination of the quadratic charges $\Q^{\lambda_0}_2$ and the momentum $\Pc$.
\end{proof}

Once again, this theorem applies only for finite regular zeros $\lambda_0$. Following a similar argument to the one given in the previous subsection, it also holds for the charges $\Q^\infty_n$ if we consider Poisson brackets only weakly.

\subsection{Zero curvature equations between the $\M^{\lambda_0}_n$'s} \label{sec: ZC eq}

In subsection \ref{Sec:ZCEL}, we showed that the dynamics of the Lax matrix $\Lc(\lambda,x)$ under the Hamiltonian flow of the local charge $\Q^{\lambda_0}_n$ takes the form of a zero curvature equation with a matrix $\M^{\lambda_0}_n(\lambda,x)$. We thus exhibited a hierarchy of integrable equations, corresponding to the commutativity of the connections $\nabla^{\lambda_0}_n$ with $\nabla_x$. This can be seen as the compatibility condition of the two auxiliary linear problems $\nabla_x\Psi=0$ and $\nabla^{\lambda_0}_n \Psi=0$, with $\Psi$ a function on the phase space, valued in the connected and simply connected Lie group with Lie algebra $\g$. In this subsection, we prove that the connections $\nabla^{\lambda_0}_n$ and $\nabla^{\mu_0}_m$ also commute with one another (except when $\lambda_0$ is finite and $\mu_0=\infty$). This can be seen as the simultaneous compatibility of all auxiliary linear problems $\nabla^{\lambda_0}_n \Psi=0$ and it takes the form of zero curvature equations:

\begin{theorem}\label{Thm:ZCEM}
Let $\lambda_0,\mu_0\in\Zc$, $n\in\E_{\lambda_0}$ and $m\in\E_{\mu_0}$. We have the zero curvature equation
\begin{equation*}
\left\lbrace \Q^{\lambda_0}_n, \M^{\mu_0}_m(\lambda,x) \right\rbrace - \left\lbrace \Q^{\mu_0}_m , \M^{\lambda_0}_n(\lambda,x) \right\rbrace  + \left[ \M^{\lambda_0}_n(\lambda,x), \M^{\mu_0}_m(\lambda,x) \right] = 0.
\end{equation*}
\end{theorem}

This subsection is entirely devoted to the proof of Theorem \ref{Thm:ZCEM}. After stating some general results, we will treat separately the cases $\lambda_0\neq\mu_0$ and $\lambda_0=\mu_0$. Note that for the latter, we only have a complete proof for an algebra $\g$ of type B, C and D. For $\g$ of type A, we verified Theorem \ref{Thm:ZCEM} for the first degrees $n$ and $m$ and conjecture that it holds more generally for any $n$ and $m$. To improve the clarity of the subsection, some technical details of the proof are presented in appendix \ref{App:Xi}.

Here also the theorem concerns the finite regular zeros $\lambda_0$ and $\mu_0$. The method presented in this subsection also applies for $\lambda_0=\mu_0=\infty$ as $\Lc^\infty(\lambda,x)$ also satisfies an $r/s$-system with twist function (Theorem \ref{Thm:PBLcI}). However, the theorem does not hold when $\lambda_0$ is finite and $\mu_0=\infty$, even if Poisson brackets are considered only weakly.

\subsubsection{Some general results}
\label{Sec:ZCEMGeneral}

Let us consider the Poisson bracket \eqref{Eq:PBLT}. It can be rewritten
\begin{align*}
& \hspace{-3pt}\left\lbrace S(\lambda,x), \Tc_m(\mu,y) \right\rbrace = - m \, \Tr\ti{2} \Bigl( U\ti{12}(\lambda,\mu) S_{m-1}(\mu,y)\ti{2} \Bigr) \delta'_{xy} \notag \\
& \hspace{140pt} + m \left[ \Tr\ti{2}\Bigl( \Rc^0\ti{12}(\lambda,\mu) S_{m-1}(\mu,x)\ti{2} \Bigr), S(\lambda,x) \right] \delta_{xy},
\end{align*}
with $S(\lambda,x)=S_1(\lambda,x)=\varphi(\lambda)\Lc(\lambda,x)$. Starting from this Poisson bracket, we elevate $S(\lambda,x)$ to the power $n-1$ and find, using the fact that the Poisson bracket and the commutator are derivations, that
\begin{align}\label{Eq:PBSnT}
&\left\lbrace S_{n-1}(\lambda,x), \Tc_m(\mu,y) \right\rbrace = m \left[ \Tr\ti{2}\Bigl( \Rc^0\ti{12}(\lambda,\mu) S_{m-1}(\mu,x)\ti{2} \Bigr), S_{n-1}(\lambda,x) \right] \delta_{xy} \\
&\hspace{150pt} - m \sum_{k=0}^{n-2} S_k(\lambda,x) \Tr\ti{2} \Bigl( U\ti{12}(\lambda,\mu) S_{m-1}(\mu,y)\ti{2} \Bigr)  S_{n-2-k}(\lambda,x) \delta'_{xy}. \notag
\end{align}
Recall the definition \eqref{Eq:DefN} of $\Nc_n(\lambda\, ; \rho,x)$. From the Poisson bracket \eqref{Eq:PBSnT}, using the cyclicity of the trace, we find
\begin{equation*}
\left\lbrace \Nc_n(\lambda\, ; \rho,x), \Tc_m(\mu,y) \right\rbrace = \Gamma^{\lambda\mu}_{nm}(\rho,x) \delta_{xy} + \Xi^{\lambda\mu}_{nm}(\rho,x,y),
\end{equation*}
where
\begin{align}
&\Gamma^{\lambda\mu}_{nm}(\rho,x) = nm \Tr\ti{23} \Bigl( \bigl[ \Rc^0\ti{12}(\rho,\lambda), \Rc^0\ti{23}(\lambda,\mu) \bigr]   S_{n-1}(\lambda,x) \ti{2} S_{m-1}(\mu,y)\ti{3} \Bigr), \label{Eq:DefGamma} \\
&\Xi^{\lambda\mu}_{nm}(\rho,x,y) = nm \Tr\ti{23} \Bigl( \Rc^0\ti{12}(\rho,\lambda) S_{m-1}(\mu,y)\ti{3} \sum_{k=0}^{n-2} S_k(\lambda,x)\ti{2} U\ti{23}(\lambda,\mu) S_{n-2-k}(\lambda,x)\ti{2} \Bigr) \delta'_{xy}, \label{Eq:DefXi}
\end{align}
with $U$ defined in equation \eqref{Eq:DefU}. Let us introduce
\begin{align*}
&\Y^{\lambda\mu}_{nm}(\rho,x,y) = \left[ \Nc_n(\lambda\, ; \rho,x), \Nc_m(\mu\,;\rho,x) \right]\delta_{xy} \\
& \hspace{90pt} + \left\lbrace \Tc_n(\lambda,y), \Nc_m(\mu\,;\rho,x) \right\rbrace  - \left\lbrace \Tc_m(\mu,y), \Nc_n(\lambda\, ; \rho,x) \right\rbrace . \notag
\end{align*}
It contains a term equal to $\delta_{xy}$ times
\begin{equation*}
\left[ \Nc_n(\lambda\, ; \rho,x), \Nc_m(\mu\,;\rho,x) \right] + \Gamma^{\lambda\mu}_{nm}(\rho,x) - \Gamma^{\mu\lambda}_{mn}(\rho,x).
\end{equation*}
One can show from equations \eqref{Eq:DefN} and \eqref{Eq:DefGamma} that this is equal to
\begin{equation*}
\Tr\ti{23}\Bigl( \Upsilon\ti{123}(\rho,\lambda,\mu) S_{n-1}(\lambda,x)\ti{2} S_{m-1}(\mu,x)\ti{3} \Bigr),
\end{equation*}
with
\begin{align*}
&\hspace{-3pt}\Upsilon\ti{123}(\rho,\lambda,\mu) = \left[ \Rc^0\ti{12}(\rho,\lambda), \Rc^0\ti{13}(\rho,\mu) \right]   + \left[ \Rc^0\ti{12}(\rho,\lambda), \Rc^0\ti{23}(\lambda,\mu) \right] + \left[ \Rc^0\ti{32}(\mu,\lambda), \Rc^0\ti{13}(\rho,\mu) \right].
\end{align*}
This terms vanishes as $\Rc^0$ is a solution of the classical Yang-Baxter equation \eqref{Eq:CYBE}. We are therefore simply left with
\begin{equation}\label{Eq:Y}
\Y^{\lambda\mu}_{nm}(\rho,x,y) = \Xi^{\lambda\mu}_{nm}(\rho,x,y) - \Xi^{\mu\lambda}_{mn}(\rho,x,y).
\end{equation}

The currents $\J_k$ are extracted from $\Tc_k$. But in general, the charges are constructed from currents $\K_k$ which are extracted from $\W_k$, where the definition of $\W_k$ depends on $\g$ and $\s$ (see  subsections \ref{Sec:SummaryNonCyc} and \ref{Sec:SummaryCyc}). In particular, we have $\K^{\lambda_0}_k(x)=\W_k(\lambda_0,x)$ for a non-cyclotomic regular zero $\lambda_0$. For the origin, which is cyclotomic, $\K^0_k(x)$ is the coefficient of $\lambda^{r_k}$ in $\W_k(\lambda,x)$. Let us define
\begin{align}
&\Zc^{\lambda\mu}_{nm}(\rho,x,y) = \left[ \M_n(\lambda\, ; \rho,x), \M_m(\mu\,;\rho,x) \right]\delta_{xy} \\
& \hspace{90pt} + \left\lbrace \W_n(\lambda,y), \M_m(\mu\,;\rho,x) \right\rbrace  - \left\lbrace \W_m(\mu,y), \M_n(\lambda\, ; \rho,x) \right\rbrace . \notag
\end{align}
Using the expression of $\W_k$ and $\M_k$ in terms of $\Tc_k$ and $\Nc_k$, we see that $\Zc^{\lambda\mu}_{nm}(\rho,x,y)$ contains several types of terms:
\begin{enumerate}
\item commutators $\left[\Nc_k(\lambda\,;\rho,x),\Nc_l(\mu\,;\rho,x)\right]$, multiplied by polynomials in the $\Tc_j$'s,
\item $\Gamma^{\lambda\mu}_{kl}(\rho,x)$ and $\Gamma^{\mu\lambda}_{lk}(\rho,x)$, multiplied by polynomials in the $\Tc_j$'s,
\item $\Xi^{\lambda\mu}_{kl}(\rho,x,y)$ and $\Xi^{\mu\lambda}_{lk}(\rho,x,y)$, multiplied by polynomials in the $\Tc_j$'s,
\item $\left\lbrace \Tc_k(\lambda,x), \Tc_l(\mu,y) \right\rbrace$ and $\left\lbrace \Tc_k(\mu,x), \Tc_l(\lambda,y) \right\rbrace$, multiplied by polynomials in $\Tc_j$'s and $\Nc_j$'s.
\end{enumerate}
Moreover, the terms of type 1 and 2 are always ultralocal, \textit{i.e.} proportional to $\delta_{xy}$. It can be seen that these terms always combine into polynomials of $\Tc_j$ multiplied by
\begin{equation*}
\left[ \Nc_k(\lambda\, ; \rho,x), \Nc_l(\mu\,;\rho,x) \right] + \Gamma^{\lambda\mu}_{kl}(\rho,x) - \Gamma^{\mu\lambda}_{lk}(\rho,x).
\end{equation*}
As explained above, this vanishes by virtue of the classical Yang-Baxter equation. Therefore, $\Zc^{\lambda\mu}_{nm}(\rho,x,y)$ is composed only of terms of type 3 and 4.

\subsubsection{Zero curvature equation at different regular zeros}

Let us now prove Theorem \ref{Thm:ZCEM} when $\lambda_0$ and $\mu_0$ are different regular zeros. Since we are not considering here the point at infinity (see discussion after Theorem \ref{Thm:ZCEM}), at least one of them is non-cyclotomic, say $\mu_0$. Recall that $U\ti{23}(\lambda,\mu_0)=\varphi(\lambda)\Rc^0(\lambda,\mu_0)$, as $\varphi(\mu_0)=0$.\\

Consider first the case where $\lambda_0$ is also non-cyclotomic. We will prove the that zero curvature equation of Theorem \ref{Thm:ZCEM} holds by showing that $\Zc^{\lambda_0\mu_0}_{nm}(\rho,x,y)$ vanishes. As explained above, it contains two types of terms. The ones of types 4 contain Poisson brackets between currents $\J^{\lambda_0}_k$ and $\J^{\mu_0}_l$. According to Theorem \ref{Thm:DiffZeros}, these brackets are all zeros. As $\lambda_0$ and $\mu_0$ are two distinct elements of $\Zc$, the cyclotomic orbits $\Z_T\lambda_0$ and $\Z_T\mu_0$ are disjoint and thus $\Rc^0(\lambda,\mu_0)$ is regular at $\lambda=\lambda_0$. We then have $U\ti{23}(\lambda_0,\mu_0)=0$, as $\varphi(\lambda_0)=0$. We deduce from this that $\Xi^{\lambda_0\mu_0}_{kl}(\rho,x,y) = 0$ and similarly $\Xi^{\mu_0\lambda_0}_{lk}(\rho,x,y) = 0$, \textit{i.e.} the terms of type 3 also vanish. Thus $\Zc^{\lambda_0\mu_0}_{nm}(\rho,x,y)=0$, as required. \\

Suppose now that $\lambda_0$ is the origin and hence a cyclotomic point. Recall that $\K^0_n(x)$ and $\M^0_n(\rho,x)$ are the coefficients of $\lambda^{r_n}$ in respectively $\W_n(\lambda,x)$ and $\M_n(\lambda\,;\rho,x)$. Thus, it is enough to show that there is no term $\lambda^{r_n}$ in $\Zc^{\lambda\mu_0}_{nm}(\rho,x,y)$ to prove Theorem \ref{Thm:ZCEM} in this case. Recall that $\Tc_k(\lambda,x)$ and $\Nc_k(\lambda\,;\rho,x)$ contain powers of $\lambda$ of the form $r_k + aT$ with $a\in\Z_{\geq 0}$. As $r_n \leq T-2 <T$ for $n\in\E_{\lambda_0}$, the powers with $a\geq 1$ cannot contribute to the $\lambda^{r_n}$-term. Following the discussion at the end of paragraph \ref{Sec:ZCEMGeneral}, the term $\lambda^{r_n}$ of $\Zc^{\lambda\mu_0}_{nm}(\rho,x,y)$ is thus composed of polynomials in the $\J^0_k$'s and $\Nc^0_k$'s times Poisson brackets of $\J^0_k$ with $\J^{\mu_0}_l$ or terms of the form
\begin{equation*}
\Xi^{\lambda\mu_0}_{kl}(\rho,x,y)\Bigr|_{\lambda^{r_k}} \;\;\; \text{ or } \;\;\; \Xi^{\mu_0\lambda}_{lk}(\rho,x,y)\Bigr|_{\lambda^{r_k}},
\end{equation*}
for $k$ such that $r_k < T-1$. According to Theorem \ref{Thm:DiffZeros}, the Poisson brackets of such $\J^0_k$ with $\J^{\mu_0}_l$ vanish. Moreover, $\Xi^{\lambda\mu_0}_{kl}(\rho,x,y)$ is proportional to $\varphi(\lambda)\Rc^0(\lambda,\mu_0)$. Yet, $\varphi(\lambda)=O(\lambda^{T-1})$ and $r_k<T-1$, hence $\Xi^{\lambda\mu_0}_{kl}(\rho,x,y)\Bigr|_{\lambda^{r_k}}=0$. Similarly $\Xi^{\mu_0\lambda}_{lk}(\rho,x,y)\Bigr|_{\lambda^{r_k}}=0$. Thus, the coefficient of $\lambda^{r_n}$ in $\Zc^{\lambda\mu_0}_{nm}(\rho,x,y)$ vanishes, as required. This ends the proof of Theorem \ref{Thm:ZCEM} for different regular zeros $\lambda_0$ and $\mu_0$.

\subsubsection{Zero curvature equations at a non-cyclotomic regular zero}

Let us now prove Theorem \ref{Thm:ZCEM} for $\lambda_0=\mu_0$. We start with the case where $\lambda_0$ is a non-cyclotomic point. We then want to show that $\displaystyle\int \dd y \,\Zc^{\lambda_0\lambda_0}_{nm}(\rho,x,y)=0$.\\

As in section \ref{Sec:NonCycZero}, we treat separately the Lie algebras of type B, C and D and the Lie algebras of type A. Suppose first that $\g$ is of type B, C or D. In this case, the currents $\K^{\lambda_0}_{2n}$ are equal to the currents $\J^{\lambda_0}_{2n}$ (see subsections \ref{Sec:NonCycZeroBCD} and \ref{Sec:SummaryNonCyc}) and the corresponding Lax matrices $\M^{\lambda_0}_{2n}$ are equal to the matrices $\Nc^{\lambda_0}_{2n}$. Thus, $\Zc^{\lambda_0\lambda_0}_{2n\,2m}(\rho,x,y)$ is simply equal to $\Y^{\lambda_0\lambda_0}_{2n\,2m}(\rho,x,y)$ (see paragraph \ref{Sec:ZCEMGeneral}). According to equation \eqref{Eq:Y}, we have
\begin{equation*}
\Y^{\lambda_0\lambda_0}_{2n\,2m}(\rho,x,y) = \Xi^{\lambda_0\lambda_0}_{2n\,2m}(\rho,x,y) - \Xi^{\lambda_0\lambda_0}_{2m\,2n}(\rho,x,y),
\end{equation*}
where $\Xi$ was defined in equation \eqref{Eq:DefXi}. To avoid cluttering the argument in the present paragraph with too many technicalities, we postpone the details of the computation of $\Xi^{\lambda_0\lambda_0}_{2n\,2m}(\rho,x,y)$ in appendix \ref{App:XiNonCyc}. We find
\begin{equation*}
\Xi^{\lambda_0\lambda_0}_{2n\,2m}(\rho,x,y) = \frac{\varphi'(\lambda_0)}{T} \frac{4nm(1-2n)}{2n+2m-2}\Nc^{\lambda_0}_{2n+2m-2}(\rho,x) \delta'_{xy} + f^{\lambda_0}_{2n\,2m}(\rho,x) \delta_{xy},
\end{equation*}
where the function $f^{\lambda_0}_{2n\,2m}$ satisfies $f^{\lambda_0}_{2n\,2m}=f^{\lambda_0}_{2m\,2n}$ (cf. appendix \ref{App:XiNonCyc}). It then follows that
\begin{equation*}
\Y^{\lambda_0\lambda_0}_{2n\,2m}(\rho,x,y) = \frac{\varphi'(\lambda_0)}{T} \frac{8nm(m-n)}{2n+2m-2}\Nc^{\lambda_0}_{2n+2m-2}(\rho,x) \delta'_{xy},
\end{equation*}
from which we deduce that $\displaystyle \int \dd y \; \Y^{\lambda_0\lambda_0}_{2n\,2m}(\rho,x,y) = 0$, as required.\\

Suppose now that $\g$ is of type A. In this case, the currents $\K^{\lambda_0}_n$ are different from the currents $\J^{\lambda_0}_n$ and we therefore have to consider $\Zc^{\lambda_0\lambda_0}_{nm}$ rather than simply $\Y^{\lambda_0\lambda_0}_{nm}$. According to the discussion at the end of paragraph \ref{Sec:ZCEMGeneral}, it contains polynomials in the $\J^{\lambda_0}_p$'s and $\Nc^{\lambda_0}_p$'s, multiplied by either $\Xi^{\lambda_0\lambda_0}_{kl}(\rho,x,y)$ or $\bigl\lbrace \J^{\lambda_0}_k(x), \J^{\lambda_0}_l(y) \bigr\rbrace$. This last Poisson bracket is given by equation \eqref{Eq:PBJTypeA} and is expressed in terms of the $\J^{\lambda_0}_p$'s. As for type B, C and D, we compute the expression of $\Xi^{\lambda_0\lambda_0}_{kl}$ in appendix \ref{App:XiNonCyc}. We find
\begin{align*}
&\Xi^{\lambda_0\lambda_0}_{k\,l}(\rho,x,y) = - \frac{\varphi'(\lambda_0)}{T} \frac{kl(k-1)}{k+l-2} \Nc^{\lambda_0}_{k+l-2}(\rho,x) \delta'_{xy} \\
& \hspace{145pt}+ \frac{\varphi'(\lambda_0)}{d T} kl \J^{\lambda_0}_{l-1}(y) \Nc^{\lambda_0}_{k-1}(\rho,x) \delta'_{xy} + f^{\lambda_0}_{kl}(\rho,x) \delta_{xy},
\end{align*}
for some function $f^{\lambda_0}_{kl}$ such that $f^{\lambda_0}_{kl}=f^{\lambda_0}_{lk}$.

Hence $\Zc^{\lambda_0\lambda_0}_{nm}$ can be expressed in terms of the $\J^{\lambda_0}_p$'s and $\Nc^{\lambda_0}_p$'s, up to terms involving $f^{\lambda_0}_{kl}$. The latter are always of the form
\begin{equation*}
\alpha \J^{\lambda_0}_{p_1}(x) \ldots \J^{\lambda_0}_{p_q}(x) f^{\lambda_0}_{kl}(\rho,x) \delta_{xy},
\end{equation*}
with $\alpha$ a constant. Moreover, one can check that for any such term, there is also a similar one but with an opposite sign and $k$ and $l$ interchanged. Using the symmetry property $f^{\lambda_0}_{kl}=f^{\lambda_0}_{lk}$, one can then conclude that these terms always vanish.

Therefore $\Zc^{\lambda_0\lambda_0}_{nm}$ can be expressed in terms of the $\J^{\lambda_0}_p$'s and $\Nc^{\lambda_0}_p$'s. Using the first explicit expressions \eqref{Eq:KJ} for the current $\K^{\lambda_0}_n$ and the corresponding expressions for the matrices $\M^{\lambda_0}_n$, it can be check directly that $\displaystyle \int \dd y \; \Zc^{\lambda_0\lambda_0}_{nm}(\rho,x,y)=0$ for the first few degrees $n, m$. Specifically, we have checked this for degrees $n$ and $m$ up to 7. In particular, we observed that we could not have chosen different coefficients in equation \eqref{Eq:KJ} for these zero curvature equations to hold (in the same way that these coefficients were uniquely fixed by requiring the involution of $\Q^{\lambda_0}_n$ and $\Q^{\lambda_0}_m$). Based on these strong observations, we conjecture that it holds for any $n,m\in\E_{\lambda_0}$.

\subsubsection{Zero curvature equations at a cyclotomic regular zero}

Finally, let us prove Theorem \ref{Thm:ZCEM} for $\lambda_0=\mu_0=0$, which is a cyclotomic point. Remember that $\K^0_n(x)$ and $\M^0_n(\rho,x)$ are extracted as the coefficient of $\lambda^{r_n}$ in the power series expansion of $\W_n(\lambda,x)$ and $\M_n(\lambda\,;\rho,x)$ where $r_n$ is the smallest power appearing in these expansions. That is, Theorem \ref{Thm:ZCEM} for $\lambda_0=\mu_0=0$ is equivalent to the statement that
\begin{equation*}
\int \dd y \; \Zc^{\lambda\lambda}_{nm} (\rho,x,y) \Bigr|_{\lambda^{r_n+r_m}} = 0.
\end{equation*}

Let us start with the case of a Lie algebra $\g$ of type B, C or D, for which $\Zc^{\lambda\lambda}_{2n\,2m}=\Y^{\lambda\lambda}_{2n\,2m}$. According to equation \eqref{Eq:Y}, we have
\begin{equation*}
\Y^{\lambda\lambda}_{2n\,2m}(\rho,x,y) = \Xi^{\lambda\lambda}_{2n\,2m}(\rho,x,y) - \Xi^{\lambda\lambda}_{2m\,2n}(\rho,x,y).
\end{equation*}
The computation of $\Xi^{\lambda\lambda}_{2n\,2m}\bigr|_{\lambda^{r_n+r_m}}$ is performed in appendix \ref{App:XiCyc}. The final result is
\begin{align*}
&\Xi^{\lambda\lambda}_{2n\,2m}(\rho,x,y)\Bigr|_{\lambda^{r_{2n}+r_{2m}}} = f^{(0)}_{2n\,2m}(\rho,x) \delta_{xy} - \theta_{r_{2n}+r_{2m}+2-T}\, \zeta'(0) \frac{4nm(2n-1)}{2n+2m-2}\Nc^0_{2n+2m-2}(\rho,x) \delta'_{xy}, \notag
\end{align*}
with $f^{(0)}_{2n\,2m}$ a function symmetric under the exchange of $n$ and $m$. By virtue of this symmetry we find that the terms involving $f$ disappear in $\Y^{\lambda\lambda}_{2n\,2m}(\rho,x,y)\bigr|_{\lambda^{r_n+r_m}}$, while the other terms vanish when integrated over $y$, as required.\\

Consider now $\g=\sl(d,\C)$ of type A. The construction of the currents 
$\K^0_k$ depends on $\s$ being inner or not 
(see subsections \ref{Sec:TypeATrivial}, \ref{Sec:TypeANonTrivial} and \ref{Sec:SummaryCyc}). If 
 $\s$ is inner, then the currents $\K^0_k$ and $\W_k$ are equal to the currents $\J^0_k$ and $\Tc_k$. In this case, we have
\begin{equation*}
\Zc^{\lambda\lambda}_{nm}(\rho,x,y)=\Y^{\lambda\lambda}_{nm}(\rho,x,y) = \Xi^{\lambda\lambda}_{nm}(\rho,x,y) - \Xi^{\lambda\lambda}_{mn}(\rho,x,y).
\end{equation*}
The expression for $\Xi^{\lambda\lambda}_{nm}(\rho,x,y)\bigr|_{\lambda^{r_n+r_m}}$ is given by equation \eqref{Eq:XiATrivial} of appendix \ref{App:XiCyc}. It has the same structure as in the case of types B, C and D: the same arguments then apply and we conclude that the integration of $\Y^{\lambda\lambda}_{nm}(\rho,x,y)\bigr|_{\lambda^{r_n+r_m}}$ over $y$ vanishes.\\

Finally, consider $\g=\sl(d,\C)$ of type A with $\s$ not inner. In this case, the currents $\K^0_k(x)$ and $\W_k(\lambda,x)$ are constructed as polynomials of respectively $\J^0_k(x)$ and $\Tc_k(\lambda,x)$. The corresponding structure of $\Zc^{\lambda\mu}_{nm}$ is discussed at the end of paragraph \ref{Sec:ZCEMGeneral}. In particular, $\Zc^{\lambda\lambda}_{nm}(\rho,x,y)$ is composed of two types of terms:
\begin{itemize}
\item $\Xi^{\lambda\lambda}_{kl}(\rho,x,y)$, multiplied by polynomials in the $\Tc_j(\lambda,\bm{\cdot})$'s,
\item $\bigl\lbrace \Tc_k(\lambda,x), \Tc_l(\lambda,y) \bigr\rbrace$, multiplied by polynomials in the $\Tc_j(\lambda,\bm{\cdot})$'s and $\Nc_j(\lambda \, ; \rho,\bm{\cdot})$'s.
\end{itemize}

We want to extract the coefficient of $\lambda^{r_n+r_m}$ in $\Zc^{\lambda\lambda}_{nm}(\rho,x,y)$. Recall that the powers of $\lambda$ appearing in $\Tc_j(\lambda,\bm{\cdot})$ and $\Nc_j(\lambda \, ; \rho,\bm{\cdot})$ are of the form $r_j+aT$, with $a\in\Z_{\geq 0}$, and that the coefficients corresponding to $a=0$ are $\J^0_j(\bm{\cdot})$ and $\Nc^0_j(\rho,\bm{\cdot})$. In the two types of terms mentioned above, one can check that the terms with $a>0$ will not contribute to the coefficient of $\lambda^{r_n+r_m}$. More precisely, $\Zc^{\lambda\lambda}_{nm}(\rho,x,y)\bigr|_{\lambda^{r_n+r_m}}$ is composed of two types of terms:
\begin{itemize}
\item $\Xi^{\lambda\lambda}_{kl}(\rho,x,y)\bigr|_{\lambda^{r_k+r_l}}$, multiplied by polynomials in the $\J^0_j(\bm{\cdot})$'s,
\item $\bigl\lbrace \J^0_k(x), \J^0_l(y) \bigr\rbrace$, multiplied by polynomials in the $\J^._j(\bm{\cdot})$'s and $\Nc^0_j(\rho,\bm{\cdot})$'s.
\end{itemize}
The Poisson brackets $\bigl\lbrace \J^0_k(x), \J^0_l(y) \bigr\rbrace$ are given by equation \eqref{Eq:PBJCycA}. The expression for $\Xi^{\lambda\lambda}_{kl}(\rho,x,y)\bigr|_{\lambda^{r_k+r_l}}$ is worked out in appendix \ref{App:XiCyc} and reads
\begin{align*}
\Xi^{\lambda\lambda}_{kl}(\rho,x,y)\Bigr|_{\lambda^{r_{k}+r_{l}}} & = f^{(0)}_{kl}(\rho,x) \delta_{xy} - \theta_{r_{k}+r_{l}+2-T}\, \zeta'(0) \frac{kl(k-1)}{k+l-2}\Nc^0_{k+l-2}(\rho,x) \delta'_{xy} \\
& \hspace{65pt} - \theta_{r_{k}+1-S}\theta_{r_{l}+1-S}\, \frac{\zeta'(0)}{d}\, kl \J^0_{l-1}(y) \Nc^0_{k-1}(\rho,x) \delta'_{xy}, \notag
\end{align*}
where $f^{(0)}_{kl}$ is a function invariant under the interchange of $k$ and $l$.

The rest of the argument follows closely that given in the non-cyclotomic case. Specifically, the terms containing $f^{(0)}_{kl}$ are seen to vanish by virtue of this symmetry property. We can thus express $\Zc^{\lambda\lambda}_{nm}(\rho,x,y)\bigr|_{\lambda^{r_n+r_m}}$ in terms of the $\J^0_k$'s and $\Nc^0_k$'s only. One then can check explicitly that this expression vanishes when integrated over $y$, as required. We verified this for the first few degrees $n$ and $m$ (up to 8) and different values of $T$ (from 2 to 6). We therefore conjecture that this is also true for any $n,m \in \E_0$ and any $T$.

\section{Applications}
\label{Sec:Applications}

In Chapter \ref{Chap:Models}, we gave a list of integrable $\s$-models which fit in the framework of models with twist function. In this section, we apply the methods developed in the previous sections to these particular examples, analyse the results and compare them to some existing work in the literature. These models were recently re-interpreted as particular examples of so-called dihedral affine Gaudin models~\cite{Vicedo:2017cge}. We will explain in second part of this thesis how the framework of dihedral affine Gaudin models is particularly suited to apply the methods of the present chapter.

\subsection{Principal chiral model and its deformations}
\label{Sec:PCM}

Let us start with the simplest integrable $\s$-model, the Principal Chiral Model (PCM). The study of local charges of the PCM is already well known and was treated in the reference~\cite{Evans:1999mj}: these results were the principal motivation and guideline for the present chapter. In particular, one of the aims was to generalise the construction of~\cite{Evans:1999mj} to a wider class of models, among which are the integrable two-parameters deformations of the PCM (dPCM). We shall discuss the latter in this subsection.\\

The integrable structure of the dPCM was discussed in subsection \ref{SubSec:dPCM}. 
In this case, $\s=\Id$ so that  $T=1$. Their Lax matrix and twist function are given by equations \eqref{Eq:LaxdPCM} and \eqref{Eq:TwistdPCM} respectively. In the language of this chapter, the regular zeros of these deformed models are $+1$ and $-1$. The evaluation of $\varphi(\lambda)\Lc(\lambda,x)$ at these zeros gives the fields
\begin{equation}\label{Eq:ChiralFieldsdPCM}
J_\pm(x) = \frac{A(x) \pm \Pi(x)}{(k\pm 1)^2+\mathcal{A}^2},
\end{equation}
which reduce to the light-cone currents $j^L_\pm$ in the undeformed case.

The local charges $\Q^{\pm 1}_n$ constructed in the present chapter are related to the traces of powers of these fields. In the undeformed case, \textit{i.e.} for the PCM, these fields coincide with the fields $j^L_\pm$ used in reference~\cite{Evans:1999mj} to construct the local charges. Thus, the method presented in this chapter gives back the results of~\cite{Evans:1999mj} for the PCM, as expected. In the deformed case, it generalises these results, while keeping a similar structure in the construction: in particular, we obtain two towers of local charges in involution, corresponding to the two chiralities of the model, and the spin of these charges is still related to the exponents of the affine Kac-Moody algebra $\widehat{\g}$, as it was in the PCM case~\cite{Evans:1999mj}.\\

The Hamiltonian and momentum of the deformed PCM are given by equations \eqref{Eq:HamMomPCM}. One can check that these are related to the quadratic charges $\Q^{\pm 1}_2$ as follows
\begin{subequations}
\begin{align*}
\Hc_{\text{dPCM}} & = - \frac{\Q^{+1}_2}{2\varphi'_{\text{dPCM}}(+1)} + \frac{\Q^{-1}_2}{2\varphi'_{\text{dPCM}}(-1)} , \\
\Pc_{\text{dPCM}} & = - \frac{\Q^{+1}_2}{2\varphi'_{\text{dPCM}}(+1)} -\frac{\Q^{-1}_2}{2\varphi'_{\text{dPCM}}(-1)} .
\end{align*}
\end{subequations}
In particular, the Hamiltonian belongs to the algebra of local charges in involution so that these charges are conserved (see also the discussion at the end of subsection \ref{Sec:AlgebraLoc}).

This observation also allows one to recover the temporal component $\M(\lambda,x)$ of the Lax pair of the model (see the paragraph below equation \eqref{Eq:Nabla}). More precisely, the equation of motion of the dPCM can be recast as the Lax equation \eqref{Eq:ZCEH}, where
\begin{equation*}
\M_{\text{dPCM}}(\lambda_,x)= - \frac{\M^{+1}_2(\lambda,x)}{2\varphi'_{\text{dPCM}}(+1)} + \frac{\M^{-1}_2(\lambda,x)}{2\varphi'_{\text{dPCM}}(-1)} = \frac{\Pi(x) + \lambda A(x)}{1-\lambda^2}.
\end{equation*}
This zero curvature equation \eqref{Eq:ZCEH} is the first among a whole hierarchy of integrable equations generated by the local charges $\Q^{\pm 1}_n$ (cf. subsection \ref{Sec:ZCEL}).\\

In particular, this result was used in subsection \ref{Sec:NonLocal} to show that the local charges $\Q^{\pm 1}_n$ are in involution with the non-local charges extracted from the monodromy of the Lax matrix $\Lc(\lambda,x)$ (see Theorem \ref{Thm:NonLocal}). In~\cite{Evans:1999mj}, it was shown that the local charges of the undeformed PCM Poisson commute with the non-local charges generating the classical Yangian symmetry of the model. In the framework of this chapter, if we consider the model on the real line $\R$, we expect these non-local charges to be extracted from the expansion of (a gauge transformation of) the monodromy around the pole $\lambda=0$ of the twist function of the PCM.

For the Yang-Baxter model ($k=0$ and $\eta\neq 0$, see paragraph \ref{SubSec:YB}), this Yangian symmetry gets deformed to a quantum affine symmetry~\cite{Delduc:2017brb} (see also Chapter \ref{Chap:PLie} of this thesis for the finite part of the deformed symmetry). In particular, studying the monodromy around the poles $\pm i \eta$ of the twist function of the Yang-Baxter model, one can extract a $q$-deformed affine Poisson-Hopf algebra $\mathscr U_q(\widehat{\g})$. We have therefore proved that this algebra of non-local charges is in involution with the algebra of local charges consisting of the $\Q^{\pm 1}_n$'s.\\

As explained in the chapter \ref{Chap:Models}, the PCM and its deformation are defined on a real Lie group $G_0$, whose Lie algebra $\g_0$ is a real form of $\g$. This real form is characterised by a semi-linear involutive automorphism $\tau$. The Lax matrix \eqref{Eq:LaxdPCM} of these models satisfies the reality condition \eqref{Eq:Reality}. Moreover, the twist function \eqref{Eq:TwistdPCM} verifies the reality condition \eqref{Eq:TwistReal} and the regular zeros of the model ($+1$ and $-1$) are real. Thus, the discussion of the subsection \ref{Sec:Reality} applies to these models and the charges $\Q^{\pm 1}_n$ are real (possibly up to a redefinition of some $\Q^{\pm 1}_n$ by a factor of $i$, depending on $\tau$).

\subsection[Bi-Yang-Baxter model]{Bi-Yang-Baxter model}

There exists another two-parameter deformation of the PCM, the Bi-Yang-Baxter (BYB) $\s$-model \cite{Klimcik:2008eq,Klimcik:2014bta}, which is the subject of the Section \ref{Sec:BYB} of this thesis. Here, we choose to treat the BYB model in its gauge fixed formulation, as described in Subsection \ref{SubSec:BYBgauge-fixed}. As we saw, this formulation does not fit exactly within the framework of models with twist function, but we will explain how one can overcome this difficulty by further relaxing the general assumptions we made. Even though this generalisation could have been done throughout the entire chapter, we chose here to work in a more restricting but more common framework for clarity and simplicity. In this regard, the present subsection is also used to illustrate, \textit{a posteriori}, how the methods and results we found apply under the generalised conditions.\\

The Lax pair of the gauge-fixed BYB model takes the form \eqref{Eq:LaxGF}. The corresponding Lax matrix then reads
\begin{equation}\label{Eq:LaxBYB}
\Lc_{\text{GF}}(\xi,x) = \frac{K_1(x) + \lambda K_0(x)}{1-\xi^2} + K_\infty(x),
\end{equation}
for some $\g$-valued fields $K_0$, $K_1$ and $K_\infty$. The Hamiltonian analysis of the BYB model was done in Section \ref{Sec:BYB}. In particular, the Poisson bracket of the Lax matrix \eqref{Eq:LaxBYB} was computed in Subsection \ref{SubSec:BYBgauge-fixed}. It takes the form of a Maillet bracket with twist function $\vp_{\text{GF}}$ defined in \eqref{Eq:TwistGF} but with the standard $\Rc$-matrix $\Rc^0$ replaced by the matrix $\Rc^{0,\text{GF}}$ defined in \eqref{Eq:R0GF}. \\

As $\Rc^{0,\text{GF}}$ is different from $\Rc^0$, we cannot directly apply the results of this chapter. However, going through the details of the proofs of these results for a non-constrained model with $T=1$, we see that the only properties of the matrix $\Rc^0$ that we used are the CYBE (for the zero curvature equations), the fact that $\Rc^0\ti{12}(\lambda,\mu)$ is holomorphic at pairs $(\lambda_0,\mu_0)$ of distinct regular zeros and the asymptotic property \eqref{Eq:RAsymptotic} near a regular zero $\mu = \lambda_0$. The matrix $\Rc^{0,\text{GF}}$ also satisfies the CYBE, as explained in Subsection \ref{SubSec:BYBgauge-fixed}. Moreover, one easily checks that it also verifies the holomorphy condition and the asymptotic property mentioned above. Thus, the results we found in this chapter also apply to the BYB model.

This is a general observation: we can also treat the models where the matrix $\Rc^0$ is replaced by a matrix $\Rc'$ satisfying some similar properties. More precisely, we require that $\Rc'$\vspace{-4pt}
\begin{itemize}\setlength\itemsep{0.1em}
\item obeys the CYBE \eqref{Eq:CYBE},
\item is holomorphic at $(\lambda_0,\mu_0)$ with $\lambda_0$ and $\mu_0$ different regular zeros in $\Zc$,
\item verifies the asymptotic property \eqref{Eq:RAsymptotic} around non-cyclotomic regular zeros,
\item satisfies the equation \eqref{Eq:UAround0} for $U\ti{12}(\lambda,\lambda)$ around a cyclotomic regular zero, up to a term $O(\lambda^{2T-3})$ (which would not contribute to some $(r_n+r_m)^{\rm th}$ power of $\lambda$ in \eqref{Eq:PBTCyc}).\vspace{-4pt}
\end{itemize}
In particular, let us consider a matrix $\Rc'$ of the form
\begin{equation}\label{Eq:R'}
\Rc'\ti{12}(\lambda,\mu) = \Rc^0\ti{12}(\lambda,\mu) + \mathcal{D}\ti{12}(\mu),
\end{equation}
like the matrix $\Rc^{0,\text{GF}}$. Then $\Rc'\ti{12}(\lambda,\mu)$ is holomorphic for $\lambda$ and $\mu$ going to different regular zeros if $\mathcal{D}$ is holomorphic at any regular zero (this is for example the case for the BYB model). This condition also ensures that the asymptotic property \eqref{Eq:RAsymptotic} is satisfied by $\Rc'$. In the same way the condition on $U\ti{12}(\lambda,\lambda)$ is satisfied by $\Rc'$ if $\mathcal{D}\ti{12}(\lambda)+\mathcal{D}\ti{21}(\lambda)=O(\lambda^{T-2})$.

Let us note, however, that these conditions do not allow to treat the case where infinity is a regular zero in the same way that we did in this chapter (subsections \ref{Sec:Infinity} and \ref{Sec:CycZero}). This would require, among other conditions, that the asymptotic properties \eqref{Eq:RAsymptoticInfinity} at infinity are also satisfied by the matrix $\Rc'$. One can check that a matrix $\Rc'$ of the form \eqref{Eq:R'} can never satisfy the second property of equation \eqref{Eq:RAsymptoticInfinity}.\\

As explained above, we can apply the construction of local charges in involution to the BYB model. These local charges will be very similar to the ones of the PCM and its deformations, described in the previous subsection, so we shall not enter into much details here. Let us note that these charges are related to traces of powers of $K_0(x)\pm K_1(x)$, where $K_0$ and $K_1$ are the fields appearing in the Lax matrix \eqref{Eq:LaxBYB}. As in the case of the PCM (see previous subsection), the Hamiltonian and the momentum of the BYB model are related to the quadratic charges $\Q^{\pm 1}_2$ by the relation
\begin{subequations}
\vspace{-6pt}\begin{align*}
\Hc_{\text{GF}} & = - \frac{\Q^{+1}_2}{2\varphi'_{\text{GF}}(+1)} + \frac{\Q^{-1}_2}{2\varphi'_{\text{GF}}(-1)} , \\
\Pc_{\text{GF}} & = - \frac{\Q^{+1}_2}{2\varphi'_{\text{GF}}(+1)} -\frac{\Q^{-1}_2}{2\varphi'_{\text{GF}}(-1)} .
\end{align*}
\end{subequations}
In particular, the local charges constructed above are all conserved.

\subsection[$\Z_T$-coset models and their deformations]{$\bm{\Z_T}$-coset models and their deformations}

In this subsection, we discuss the construction of local charges in involution 
for $\Z_T$-coset models (and the deformations of $\Z_2$-coset models). These models were described in Subsections \ref{SubSec:ZT} and \ref{SubSec:dZ2}. Their twist function and Lax matrix are given by equations \eqref{Eq:TwistZT} and \eqref{Eq:LaxZT} for undeformed $\Z_T$-coset and by \eqref{Eq:TwistdZ2} and \eqref{Eq:LaxdZ2} for the deformed $\Z_2$-coset. Local charges in involution were constructed for symmetric spaces, \textit{i.e.} $\Z_2$-cosets, in references~\cite{Evans:2000qx} and~\cite{Evans:2005zd}: we shall compare these results with the ones of this chapter.

The regular zeros of these models are the origin and infinity, which are both cyclotomic points. We shall therefore apply here the construction of section \ref{Sec:CycZero}. Moreover, all these models possess a gauge constraint $X^{(0)}$ (see Chapter \ref{Chap:Models}), which is identified with the field at infinity $\Cc(x)$ described in subsection \ref{Sec:Infinity}. The results of subsection \ref{Sec:Gauge} ensure that the densities of the local charges that we construct here are gauge invariant. Indeed, by Theorem \ref{Thm:Gauge}, these densities Poisson commute with the constraint $\Cc$.\\

As in the case of the PCM (see subsection \ref{Sec:PCM}), the degrees of the local charges are related to the exponents of the affine Kac-Moody algebra $\widehat{\g}$ plus one (here also, we do not consider the exponents corresponding to the Pfaffian for type D). However, as explained in section \ref{Sec:CycZero}, the fact that the regular zeros of the model are cyclotomic makes some of the exponents `drop out', in the sense that we cannot construct a charge of the corresponding degree. Recall that a degree $n$ (corresponding to an exponent $n-1$) drops out if $r_n$ is equal to $T-1$ (where $r_n$ was defined in subsection \ref{Sec:EquivT}).

Let us study this in more detail for the case of $\Z_2$-cosets. In particular, we shall compare this phenomenon of exponents dropping out with some results of reference~\cite{Evans:2000qx}. Indeed, in this reference, some local charges in involution were constructed for symmetric spaces (\textit{i.e.} $\Z_2$-cosets). These symmetric spaces correspond to quotients $G_0/G^\s_0$ of the real Lie group $G_0$ by the subgroup of fixed points under the involutive automorphism $\s$. Such spaces were classified, up to isomorphism, for classical compact groups $G_0$.

In particular, the possible exponents (\textit{i.e.} the degrees minus one) of the local charges for each symmetric space of this classification were listed in Table 1 of~\cite{Evans:2000qx}: they form a (potentially proper) subset of the exponents of $\widehat{\g}$. A simple case by case computation of the integers $r_n$ for these symmetric spaces, and thus these automorphisms $\s$, shows that the exponents of $\widehat{\g}$ which do not appear in this list are exactly the exponents that drop out in the formalism of the present chapter. We therefore recover the structure of the degrees of local charges found in~\cite{Evans:2000qx} (except for the integer $h$ of~\cite{Evans:2000qx}, which we could not interpret in the present formalism).\\

An explicit computation of the traces of powers of $\varphi_{\Z_2}(\lambda)\Lc_{\Z_2}(\lambda,x)$ around the origin $\lambda=0$ shows that the charges constructed in this chapter coincide, up to some factors, with the ones constructed in reference~\cite{Evans:2000qx}. The two regular zeros $0$ and $\infty$ correspond to the two chiralities of the model. The article~\cite{Evans:2000qx} focused on one particular chirality. Here, we also have the Poisson brackets between the two towers of local charges constructed in this way. Indeed, according to Theorem \ref{Thm:InvolutionInfinity}, we show that these two towers of charges Poisson commute weakly.

This chapter also generalises the results of~\cite{Evans:2000qx} in different directions. First of all, the present formalism also allows to treat the integrable deformations of the $\Z_2$-coset model. Indeed, as explained in Subsection \ref{SubSec:dZ2}, the regular zeros of these models are still $0$ and $\infty$ and so the methods developed here still apply. The main generalisation is that this chapter does not restrict to (compact) symmetric spaces and also generalises the construction to any $\Z_T$-coset model. Finally, in this chapter we have also studied the hierarchy of equations induced by the flow of these local charges.

Recall the supercoset models discussed at the end of Subsection \ref{SubSec:ZT}, which have a similar structure than the $\Z_T$-coset models but with $\g$ a super-Lie algebra. Although we did not consider such models in this chapter, we expect the construction to extend to these theories, by working with the Grassmann envelope of $\g$ and replacing all traces by supertraces. One should however be careful about how the automorphism $\s$ is extented to the whole matrix algebra (see appendix \ref{App:ExtSigma}) in these supercoset cases. Such considerations could allow the construction of local charges in involution for supercoset $\s$-models whose target space includes AdS manifolds, with possible applications to the Green-Schwarz formulation of string theory.\\

We end this subsection by observing that the Hamiltonian of the $\Z_T$-coset model is related to the quadratic charge $\Q^0_2$ at the origin and the momentum $\Pc_{\Z_T}$ of the theory by
\begin{equation*}
\Hc_{\Z_T} = \frac{\Q^0_2}{\zeta'(0)} + \Pc_{\Z_T},
\end{equation*}
where $\zeta$ was defined in equation \eqref{Eq:DefZeta} (note that this expression also holds for the deformed $\Z_2$ model). Thus, we conclude that the local charges constructed above are conserved, as they commute (at least weakly) with the Hamiltonian.

As the $\Z_T$-coset models are constrained models, their Hamiltonian is defined up to a term $\Tr\bigl( \mu(x) \Cc(x) \bigr)$, where $\mu$ is a $\g$-valued Lagrange multiplier. In this sense, $\Hc_{\Z_T}$ defined above is a particular choice of such a Hamiltonian, which generates a strong zero curvature equation \eqref{Eq:ZCEH} on the Lax matrix $\Lc$. Another choice of Hamiltonian involves the quadratic charge $\Q^\infty_2$ extracted at infinity, namely
\begin{equation*}
\widetilde{\Hc}_{\Z_T} = -\frac{\Q^\infty_2}{\zeta_\infty'(0)} - \Pc_{\Z_T},
\end{equation*}
where $\zeta_\infty$ is defined in the same way than $\zeta$ by $\zeta_\infty(\alpha^T)=\alpha \psi(\alpha)$. This Hamiltonian is weakly equal to $\Hc_{\Z_T}$ and generates a strong curvature equation on the Lax matrix $\Lc^\infty$.

\begin{subappendices}

\section{Extension of the automorphism to the whole space of matrices}
\label{App:ExtSigma}

We consider a Lie algebra $\g$ of classical type A, B, C or D, in its defining matricial representation. We therefore regard elements of $\g$ as acting linearly on a vector space $V$, \textit{i.e.} as element of the space $F$ of endomorphisms of $V$. Note that we can consider some connected matrix group $G \subset F$ whose Lie algebra is $\g$. Let $\s$ be an automorphism of $\g$, of finite order $T$. In this chapter, we are considering powers of elements of $\g$, which do not belong to $\g$ in general but are elements of $F$. Thus, we want to extend the automorphism $\s$ to the whole space of matrices $F$, in a ``natural way''.

\subsection{The conjugacy case}

Let us begin with the case where $\s$ is inner, \textit{i.e.} when $\s : X \in \g \mapsto Q X Q^{-1}$ for some $Q \in G$. Then the extension of $\s$ to $F$, which by a slight abuse of notation we still denote as $\s$, can be naturally defined as
\begin{equation}\label{Eq:ExtInner}
\begin{array}{rccc}
\s : & F & \longmapsto & F \\
                 & X & \longrightarrow & Q X Q^{-1}
\end{array}
\end{equation}
This covers the case of types B and C, as they do not have any non trivial diagram automorphism.\\

Let us now consider the algebra D$_n$, \textit{i.e.} $\g=\so(2n,\C)$, for $n\geq 5$. In this case, there 
always exists one non trivial diagram automorphism. However, this automorphism can be realised on the defining representation as an external conjugation: $\s: X \in \g \mapsto Q X Q^{-1}$ where $Q$ is not in the group $SO(2n,\C)$ but belongs to $O(2n,\C)$. In this case, the endomorphism $\s$ as defined in equation \eqref{Eq:ExtInner} still naturally extends $\s$ on $F$.

Let us say a few words on the algebra D$_4=\so(8,\C)$. It is known to have 6 diagram automorphisms, forming the \textit{triality}, isomorphic to the symmetric group $S_3$. One of them, of order 2, can also be realised as conjugation by a matrix $Q\in O(8,\C)$ and so extends to $F=M_8(\C)$ by equation \eqref{Eq:ExtInner}. The other non-trivial ones cannot be realised in any ``natural way'' on the defining representation and thus cannot easily be extended to the whole space $M_8(\C)$. We shall not consider them in this chapter. \\

We now describe the properties of the extension $\s$ defined in \eqref{Eq:ExtInner}. The fact that the automorphism $\s$ of $\g$ is of order $T$ is equivalent to the fact that $Q^T$ belongs to the centraliser of $\g$ in $F$,
\begin{equation*}
Z_F(\g) = \left\lbrace X \in F \; \text{s.t.} \; [X,Y]=0, \; \forall \, Y\in\g \right\rbrace.
\end{equation*}
By Schur's lemma, this implies that $Q^T = \lambda\,\Id$ for some $\lambda\in\C$. Therefore $\s$ on $F$ defined by \eqref{Eq:ExtInner} is also of order $T$. We shall make extensive use of the following five obvious properties of $\s$ as defined in \eqref{Eq:ExtInner}:
\begin{align*}
\s(XY) = \s(X)\s(Y), \;\;\;\;\;
\s(X^n) = \s(X)^n, \;\;\;\;\;
\s(\Id) = \Id, \\
\Tr\bigl(\s(X)\bigr) = \Tr(X), \;\;\;\;\;
\Tr\bigl(\s(X)\s(Y)\bigr) = \Tr(XY), \;\;\;
\end{align*}
for any $X,Y$ in $F$.

\subsection{The transpose case}

The last case that we have to treat is the one of a type A algebra, with 
$\s$ being not inner. We thus consider the defining representation $\g=\sl(d,\C)$. The action of $\s$ on $\g$ can then always be expressed as $\s : X\in\g \mapsto - Q X\Tp Q^{-1}$, where $X\Tp$ is the transpose of $X$ and $Q$ is a matrix in $SL(d,\C)$. Here also we can naturally extend $\s$ to an endomorphism of $F=M_d(\C)$, which we still denote $\s$, by letting
\begin{equation} \label{sigma F type A}
\begin{array}{rccc}
\s : & F & \longmapsto & F \\
                 & X & \longrightarrow & - Q \, X\Tp Q^{-1}
\end{array}.
\end{equation}

Once again, let us investigate the properties of $\s$. As the automorphism $\s$ of $\g$ is 
not inner, its order $T$ must be even, and we shall write $T=2S$. We note that $\s^2$ acts as conjugation by $R=Q (Q\Tp)^{-1}$. The fact that $\s^T=(\s^2)^S=\left. \Id \right|_{\g}$ is thus equivalent to the fact that $R^S$ belongs to the centraliser $Z_F(\g)$. Thus $R^S = \lambda \, \Id$ for some $\lambda\in\C$ and so $\s$ defined in \eqref{sigma F type A} is also of order $T$. We end the subsection by noting the following five properties of $\s$:
\begin{align*}
\s(XY) = - \s(Y)\s(X), \;\;\;\;\;
\s(X^n) = (-1)^{n-1}\s(X)^n, \;\;\;\;\;
\s(\Id) = - \Id, \\
\Tr\bigl(\s(X)\bigr) = -\Tr(X), \;\;\;\;\;
\Tr\bigl(\s(X)\s(Y)\bigr) = \Tr(XY), \hspace{40pt}
\end{align*}
for any $X,Y$ in $F$.

\section[Computation of $\Xi$]{Computation of $\bm{\Xi}$}
\label{App:Xi}

In this appendix, we give the details of the computation in some particular cases of the term $\Xi^{\lambda\mu}_{nm}(\rho,x,y)$, defined by \eqref{Eq:DefXi}.

\subsection{At a non-cyclotomic regular zero}
\label{App:XiNonCyc}

We first suppose that $\lambda_0$ is a non-cyclotomic regular zero and that $\g$ is of type B, C or D. Recall that in this case, we constructed currents $\K^{\lambda_0}_{2n}=\J^{\lambda_0}_{2n}$ and the associated Lax matrices $\M^{\lambda_0}_{2n}=\Nc^{\lambda_0}_{2n}$. We want to compute $\Xi^{\lambda_0\lambda_0}_{2n\,2m}(\rho,x,y)$, starting from equation \eqref{Eq:DefXi}. Recall from section \ref{Sec:NonCycZero} that for a non-cyclotomic zero $\lambda_0$, one has $U\ti{23}(\lambda_0,\lambda_0)=-\frac{1}{T}\varphi'(\lambda_0)C\ti{12}$. Recall also from subsection \ref{Sec:NonCycZeroBCD} that $S_{2m-1}(\lambda_0,x)$ belongs to the Lie algebra $\g$, as it is an odd power of a matrix in $\g$. Using the completeness relation \eqref{Eq:CompRel}, we find
\begin{align}\label{Eq:XiNonCyc1}
\Xi^{\lambda_0\lambda_0}_{2n\,2m}(\rho,x,y) = - \frac{4nm\varphi'(\lambda_0)}{T} \Tr\ti{2} \left( \Rc^0\ti{12}(\rho,\lambda_0)  \sum_{k=0}^{2n-2} S_k(\lambda_0,x)\ti{2} S_{2m-1}(\lambda_0,y)\ti{2} S_{2n-2-k}(\lambda_0,x)\ti{2} \right) \delta'_{xy},
\end{align}
Using the identity $f(y)\delta'_{xy}=f(x)\delta'_{xy}+\bigl(\p_x f(x) \bigr) \delta_{xy}$ and the fact that $S_p(\lambda,x)S_q(\lambda,x)=S_{p+q}(\lambda,x)$, we get
\begin{equation*}
\Xi^{\lambda_0\lambda_0}_{2n\,2m}(\rho,x,y) = f^{\lambda_0}_{2n\,2m}(\rho,x) \delta_{xy} - \frac{4nm\varphi'(\lambda_0)(2n-1)}{T} \Tr\ti{2} \Bigl( \Rc^0\ti{12}(\rho,\lambda_0) S_{2n+2m-3}(\lambda_0,x)\ti{2} \Bigr) \delta'_{xy},
\end{equation*}
where
\begin{equation*}
f^{\lambda_0}_{2n\,2m}(\rho,x) = - \frac{4nm\varphi'(\lambda_0)}{T} \Tr\ti{2} \Bigl( \Rc^0\ti{12}(\rho,\lambda_0)  \sum_{k=0}^{2n-2} S_k(\lambda_0,x)\ti{2} \p_x\bigl(S_{2m-1}(\lambda_0,x)\ti{2}\bigr) S_{2n-2-k}(\lambda_0,x)\ti{2} \Bigr).
\end{equation*}
Recalling the definition \eqref{Eq:DefNNonCyc} of $\Nc^{\lambda_0}_p$, we obtain
\begin{equation}\label{Eq:XiNonCyc2}
\Xi^{\lambda_0\lambda_0}_{2n\,2m}(\rho,x,y) = \frac{\varphi'(\lambda_0)}{T} \frac{4nm(1-2n)}{2n+2m-2}\Nc^{\lambda_0}_{2n+2m-2}(\rho,x) \delta'_{xy} + f^{\lambda_0}_{2n\,2m}(\rho,x) \delta_{xy}.
\end{equation}
As $\p_x S_p(\lambda,x)=\sum_{l=0}^{p-1} S_{l}(\lambda,x)\p_x \bigl( S(\lambda,x) \bigr) S_{p-1-l}(\lambda,x)$, one can rewrite the function $f^{\lambda_0}_{2n\,2m}$ as
\begin{align*}
& f^{\lambda_0}_{2n\,2m}(\rho,x) = - \frac{4nm\varphi'(\lambda_0)}{T} \sum_{k=0}^{2n-2}\sum_{l=0}^{2m-2} \Tr\ti{2} \Bigl( \Rc^0\ti{12}(\rho,\lambda_0)  \\
& \hspace{150pt} S_{k+l}(\lambda_0,x)\ti{2} \p_x\bigl(S(\lambda_0,x)\ti{2}\bigr) S_{2n+2m-4-k-l}(\lambda_0,x)\ti{2} \Bigr).
\end{align*}
In particular, note that $f^{\lambda_0}_{2n\,2m}=f^{\lambda_0}_{2m\,2n}$.\\

Let us now compute $\Xi^{\lambda_0\lambda_0}_{nm}$ for a non-cyclotomic regular zero $\lambda_0$ and an algebra $\g$ of type A. As in the case of type B, C or D, we have $U\ti{23}(\lambda_0,\lambda_0)=-\frac{1}{T}\varphi'(\lambda_0)C\ti{23}$. Using the generalised completeness relation \eqref{Eq:CompRelSl} and the fact that $\J^{\lambda_0}_p(x)=\Tr\bigl( S_p(\lambda_0,x)\bigr)$, we find from equation \eqref{Eq:DefXi} that
\begin{align*}
\Xi^{\lambda_0\lambda_0}_{n\,m}(\rho,x,y) & = - \frac{nm\varphi'(\lambda_0)}{T} \Tr\ti{2} \Bigl( \Rc^0\ti{12}(\rho,\lambda_0)  \sum_{k=0}^{n-2} S_k(\lambda_0,x)\ti{2} S_{m-1}(\lambda_0,y)\ti{2} S_{n-2-k}(\lambda_0,x)\ti{2} \Bigr) \delta'_{xy} \\
& \hspace{40pt} + \frac{nm(n-1)\varphi'(\lambda_0)}{dT} \J^{\lambda_0}_{m-1}(y) \Tr\ti{2} \Bigl( \Rc^0\ti{12}(\rho,\lambda_0) S_{n-2}(\lambda_0,x) \Bigr) \delta'_{xy}.
\end{align*}
From the identity $f(y)\delta'_{xy}=f(x)\delta'_{xy}+\bigl(\p_x f(x) \bigr) \delta_{xy}$ and equation \eqref{Eq:DefNNonCyc}, we find
\begin{align*}
\Xi^{\lambda_0\lambda_0}_{n\,m}(\rho,x,y) & = - \frac{\varphi'(\lambda_0)}{T} \frac{nm(n-1)}{n+m-2} \Nc^{\lambda_0}_{n+m-2}(\rho,x) \delta'_{xy} \\
& \hspace{70pt}+ \frac{\varphi'(\lambda_0)}{dT} nm \J^{\lambda_0}_{m-1}(y) \Nc^{\lambda_0}_{n-1}(\rho,x) \delta'_{xy} + f^{\lambda_0}_{nm}(\rho,x) \delta_{xy},
\end{align*}
with
\begin{equation*}
f^{\lambda_0}_{nm}(\rho,x) = - \frac{nm\varphi'(\lambda_0)}{T} \Tr\ti{2} \Bigl( \Rc^0\ti{12}(\rho,\lambda_0)  \sum_{k=0}^{n-2} S_k(\lambda_0,x)\ti{2} \p_x\bigl(S_{m-1}(\lambda_0,x)\ti{2}\bigr) S_{n-2-k}(\lambda_0,x)\ti{2} \Bigr).
\end{equation*}
As in the case of type B, C or D, we can re-express $f^{\lambda_0}_{nm}$ as
\begin{equation*}
f^{\lambda_0}_{nm}(\rho,x) = - \frac{nm\varphi'(\lambda_0)}{T} \sum_{k=0}^{n-2}\sum_{l=0}^{m-2} \Tr\ti{2} \Bigl( \Rc^0\ti{12}(\rho,\lambda_0) S_{k+l}(\lambda_0,x)\ti{2} \p_x\bigl(S(\lambda_0,x)\ti{2}\bigr) S_{n+m-4-k-l}(\lambda_0,x)\ti{2} \Bigr).
\end{equation*}
In particular, note that $f^{\lambda_0}_{nm}=f^{\lambda_0}_{mn}$.

\subsection{Around a cyclotomic regular zero}
\label{App:XiCyc}

This subsection is devoted to the computation of $\Xi^{\lambda\lambda}_{nm}(\rho,x,y)$ around the origin $\lambda=0$ and more precisely to the computation of the coefficient of $\lambda^{r_n+r_m}$ in its series expansion. Our starting point is the definition \eqref{Eq:DefXi} of $\Xi^{\lambda\mu}_{nm}$. To evaluate this equation at $\mu=\lambda$,	we will need the expression of $U\ti{12}(\lambda,\lambda)$. We saw in section \ref{Sec:CycZero} that this is given by equation \eqref{Eq:UAround0}.\\

The presence of the partial Casimir $C^{(0)}\ti{12}$ in this equation will gives rise to projections of $S_{m-1}(\lambda,y)$ onto the grading $F^{(0)}=\lbrace Z\in F \, | \, \s(Z)=Z \rbrace$ of the matrix algebra. More precisely, the calculations will involve
\begin{equation}\label{Eq:Defg}
g^\lambda_{mn}(\rho,x,y) = nm\lambda^{-2} \zeta(\lambda^T) \Tr\ti{2} \Bigl( \Rc^0\ti{12}(\rho,\lambda) \sum_{k=0}^{n-2} S_k(\lambda,x)\ti{2} S^{(0)}_{m-1}(\lambda,y)\ti{2} S_{n-2-k}(\lambda,x)\ti{2} \Bigr) \delta'_{xy},
\end{equation}
where $S^{(0)}_{p}$ denotes the projection of $S_{p}$ on $F^{(0)}$. As we are computing the coefficient of $\lambda^{r_n+r_m}$ in $\Xi^{\lambda\lambda}_{nm}$, we will consider the $\lambda^{r_n+r_m}$-term of $g^\lambda_{nm}$. Let us show that this term is actually always zero. Using the conventions and results of the subsections \ref{Sec:EquivT} and \ref{Sec:PBCurrentsCyc}, in particular the integers $\alpha$ and $q_m$, we see that the smallest power of $\lambda$ appearing in $g^\lambda_{nm}$ is $a=\alpha T-2+q_m$, as the $S_p(\lambda,x)$'s are regular at $\lambda=0$. Recall that $r_n$ and $r_m$ are both strictly less than $T-1$ when $n,m\in\E_{0}$ and that in this case, we have $q_m=r_m+1$ (see subsection \ref{Sec:PBCurrentsCyc}). Thus $a = \alpha T-1+r_m$ and hence $a > r_n+r_m$ since $\alpha\geq 1$ and $T-1>r_n$. We can then conclude that the coefficient of $\lambda^{r_n+r_m}$ in $g^\lambda_{nm}$ vanishes, as announced.\\

We will also need the function
\begin{equation*}
f^{\lambda}_{nm}(\rho,x) = - nm\lambda^{T-2}\zeta'(\lambda) \sum_{k=0}^{n-2}\sum_{l=0}^{m-2} \Tr\ti{2} \Bigl( \Rc^0\ti{12}(\rho,\lambda)   S_{k+l}(\lambda,x)\ti{2} \p_x\bigl(S(\lambda,x)\ti{2}\bigr) S_{n+m-4-k-l}(\lambda,x)\ti{2} \Bigr),
\end{equation*}
similar to the function $f^{\lambda_0}_{nm}$ defined in the non-cyclotomic case (see previous subsection) and which possesses the same symmetry property $f^\lambda_{nm}=f^{\lambda}_{mn}$. As for $g^\lambda_{nm}$, we will use more precisely the function $f^{(0)}_{nm}=f^\lambda_{nm}\bigr|_{\lambda^{r_n+r_m}}$, which is also symmetric under the exchange of $n$ and $m$.\\

To go further in the computation, we will need to distinguish between the algebras of type B, C and D and the ones of type A. Let us start with types B, C and D. In this case, we restrict to degrees $2n$ and $2m$ (see subsections \ref{Sec:NonCycZeroBCD} and \ref{Sec:CycBCD}) and thus compute $\Xi^{\lambda\lambda}_{2n\,2m}$. Recall that $S_{2m-1}(\lambda,y)$ belongs to the Lie algebra $\g$, so that we can apply the completeness relations \eqref{Eq:CompRel} and \eqref{Eq:CompRelPart} to it. One then gets
\begin{equation*}
\Xi^{\lambda\lambda}_{2n\,2m}(\rho,x,y) = \widetilde{\Xi}^{\lambda\lambda}_{2n\,2m}(\rho,x,y) + g^\lambda_{2n\,2m}(\rho,x,y),
\end{equation*}
with $g^\lambda_{2n\,2m}$ defined in equation \eqref{Eq:Defg} and
\begin{equation*}
\widetilde{\Xi}^{\lambda\lambda}_{2n\,2m}(\rho,x,y) = - 4nm\lambda^{T-2} \zeta'(\lambda^T) \Tr\ti{2} \Bigl( \Rc^0\ti{12}(\rho,\lambda)  \sum_{k=0}^{2n-2} S_k(\lambda,x)\ti{2} S_{2m-1}(\lambda,y)\ti{2} S_{2n-2-k}(\lambda,x)\ti{2} \Bigr) \delta'_{xy}.\notag
\end{equation*}
The first term $\widetilde{\Xi}^{\lambda\lambda}_{2n\,2m}$ has the same structure as $\Xi^{\lambda_0\lambda_0}_{2n\,2m}$ studied in the previous subsection (see equation \eqref{Eq:XiNonCyc1}). Thus, the calculations of that subsection apply here and we get to an equation similar to \eqref{Eq:XiNonCyc2} for $\widetilde{\Xi}^{\lambda\lambda}_{2n\,2m}$. Namely, we have
\begin{align}\label{Eq:XiLambdaBCD}
&\Xi^{\lambda\lambda}_{2n\,2m}(\rho,x,y) = f^{\lambda}_{2n\,2m}(\rho,x) \delta_{xy} + g^{\lambda}_{2n\,2m}(\rho,x,y)  - \lambda^{T-2}\zeta'(\lambda^T) \frac{4nm(2n-1)}{2n+2m-2}\Nc_{2n+2m-2}(\lambda\,;\rho,x) \delta'_{xy},
\end{align}
where $f^\lambda_{2n\,2m}$ is defined above.

We now compute the coefficient of $\lambda^{r_{2n}+r_{2m}}$ in this expression. We showed above that $g^\lambda_{2n\,2m}$ does not contribute to this term and we have defined $f^{(0)}_{2n\,2m}$ as its contribution from $f^\lambda_{2n\,2m}$. Recall also that $\Nc_k(\lambda\,;\rho,x)$ has the same equivariance property as $\Tc_k(\lambda,x)$ (equation \eqref{Eq:EquiN}) so that its power series expansion starts with $\lambda^{r_k}$. Thus, the smallest power of $\lambda$ in the second line of equation \eqref{Eq:XiLambdaBCD} is greater than or equal to $T-2+r_{2n+2m-2}$. We have shown in subsection \ref{Sec:TypeANonTrivial} that this is equal to $r_{2n}+r_{2m}$ or $r_{2n}+r_{2m}+T$, depending on whether $r_{2n}+r_{2m}$ is greater than or strictly less than $T-2$. We find
\begin{equation}\label{Eq:XiBCD}
\Xi^{\lambda\lambda}_{2n\,2m}(\rho,x,y)\Bigr|_{\lambda^{r_{2n}+r_{2m}}} = f^{(0)}_{2n\,2m}(\rho,x) \delta_{xy} - \theta_{r_{2n}+r_{2m}+2-T}\, \zeta'(0) \frac{4nm(2n-1)}{2n+2m-2}\Nc^0_{2n+2m-2}(\rho,x) \delta'_{xy}.
\end{equation}~

Finally, let us study the case of a  Lie algebra of type A, \textit{i.e.} of $\g=\sl(d,\C)$. The term in $\Xi^{\lambda\lambda}_{nm}$ involving the Casimir $C\ti{12}$ is treated with the generalised completeness relation \eqref{Eq:CompRelSl}. In the same way, one has a generalised completeness relation for the partial Casimir $C^{(0)}\ti{12}$. This relation depends on whether the extension of $\s$ to the whole algebra of matrices fixes 
the identity $\Id$ or not, and thus whether  $\s$ is inner or not (see appendix \ref{App:ExtSigma} and subsections \ref{Sec:TypeATrivial} and \ref{Sec:TypeANonTrivial}). In general, one can write
\begin{equation*}
\Tr\ti{12}\left(C\ti{12}^{(0)}Z\ti{2}\right) = \pi^{(0)}(Z) - \frac{a}{d} \Tr(Z),
\end{equation*}
for any $Z\in M_d(\C)$, with $a=1$ if  $\s$ is inner and $a=0$ if not. Using this relation and the relation \eqref{Eq:CompRelSl}, one finds
\begin{equation*}
\Xi^{\lambda\lambda}_{nm}(\rho,x,y) = \widetilde{\Xi}^{\lambda\lambda}_{nm}(\rho,x,y) + g^\lambda_{nm}(\rho,x,y),
\end{equation*}
with $g^\lambda_{nm}$ defined in equation \eqref{Eq:Defg} and
\begin{align*}
\widetilde{\Xi}^{\lambda\lambda}_{nm}(\rho,x,y) & = - nm\lambda^{T-2} \zeta'(\lambda^T) \Tr\ti{2} \Bigl( \Rc^0\ti{12}(\rho,\lambda)  \sum_{k=0}^{n-2} S_k(\lambda,x)\ti{2} S_{m-1}(\lambda,y)\ti{2} S_{n-2-k}(\lambda,x)\ti{2} \Bigr) \delta'_{xy} \notag \\
& \hspace{20pt} + \frac{nm(n-1)}{d} \frac{\lambda^{T} \zeta'(\lambda^T)-a\zeta(\lambda^T)}{\lambda^2} \Tc_{m-1}(\lambda,y) \Tr\ti{2} \Bigl( \Rc^0\ti{12}(\rho,\lambda) S_{n-2}(\lambda,x)\ti{2} \Bigr)\delta'_{xy}.
\end{align*}
The first term in this expression is treated in the same way as in the case of types B, C and D. Moreover, we recognise in the second term the definition of $\Nc_{n-1}(\lambda\,;\rho,x)$. Finally, we obtain
\begin{align}\label{Eq:XiLambdaA}
&\hspace{-20pt}\Xi^{\lambda\lambda}_{nm}(\rho,x,y) = f^{\lambda}_{nm}(\rho,x) \delta_{xy} + g^{\lambda}_{nm}(\rho,x,y) - \lambda^{T-2}\zeta'(\lambda^T) \frac{nm(n-1)}{n+m-2}\Nc_{n+m-2}(\lambda\,;\rho,x) \delta'_{xy},\\
& \hspace{130pt} + \frac{nm}{d} \frac{\lambda^{T} \zeta'(\lambda^T)-a\zeta(\lambda^T)}{\lambda^2} \Tc_{m-1}(\lambda,y) \Nc_{n-1}(\lambda\,;\rho,x)\delta'_{xy}. \notag
\end{align}

We now compute the coefficient of $\lambda^{r_n+r_m}$ in $\Xi^{\lambda\lambda}_{nm}$. As explained at the beginning of this subsection, $g^\lambda_{nm}$ does not contribute to this term and the contribution of $f^\lambda_{nm}$ is defined as $f^{(0)}_{nm}$. The contribution from the third term is calculated as in the case of types B, C and D. In particular, it vanishes when $r_n+r_m$ is strictly less than $T-2$.

Finally, let us discuss the contribution of the last term. First of all, we note that if $a=1$, $\lambda^T\zeta'(\lambda^T)-a\zeta(\lambda^T)=O(\lambda^{2T})$. Thus the powers of $\lambda$ in this term are greater than $2T-2$. Yet, we have $r_n+r_m<2T-2$ for $n,m\in\E_0$, so this term does not contribute to the $\lambda^{r_n+r_m}$-term in this case. Hence 
for $\s$ inner, we have
\begin{equation}\label{Eq:XiATrivial}
\Xi^{\lambda\lambda}_{nm}(\rho,x,y)\Bigr|_{\lambda^{r_{n}+r_{m}}} = f^{(0)}_{nm}(\rho,x) \delta_{xy}  - \theta_{r_{n}+r_{m}+2-T}\, \zeta'(0) \frac{nm(n-1)}{n+m-2}\Nc^0_{n+m-2}(\rho,x) \delta'_{xy},
\end{equation}
as in the case of types B, C and D.

Suppose now that $\s$ is not inner, so that $a=0$. Then the smallest power of $\lambda$ in the last term of equation \eqref{Eq:XiLambdaA} is greater than or equal to $T-2+r_{n-1}+r_{m-1}$. In subsection \ref{Sec:TypeANonTrivial}, we have shown that this is equal to $r_n+r_m$ if both $r_n$ and $r_m$ are greater than $S-1$ and that it is strictly greater than $r_n+r_m$ otherwise. In conclusion, we find
\begin{align}\label{Eq:XiANonTrivial}
\Xi^{\lambda\lambda}_{nm}(\rho,x,y)\Bigr|_{\lambda^{r_{n}+r_{m}}} & = f^{(0)}_{nm}(\rho,x) \delta_{xy} - \theta_{r_{n}+r_{m}+2-T}\, \zeta'(0) \frac{nm(n-1)}{n+m-2}\Nc^0_{n+m-2}(\rho,x) \delta'_{xy} \\
& \hspace{45pt} + \theta_{r_{n}+1-S}\theta_{r_{m}+1-S}\, \zeta'(0)\, \frac{nm}{d} \J^0_{m-1}(y) \Nc^0_{n-1}(\rho,x) \delta'_{xy}. \notag
\end{align}

\end{subappendices}

\cleardoublepage
\chapter{Deformed symmetries of Yang-Baxter deformations as Poisson-Lie symmetries}
\label{Chap:PLie}

This chapter is based on the article~\cite{Delduc:2016ihq}, that I wrote during my PhD with F. Delduc, M. Magro and B. Vicedo. The content of this chapter is the same as the one of~\cite{Delduc:2016ihq} and is made to be read independently.\\

In the chapter \ref{Chap:Models}, we introduced the Yang-Baxter type deformations of $\s$-models, which include the (one-parameter) $\eta$-deformations of the PCM (Subsection \ref{SubSec:YB}) and of the $\Z_2$-coset $\s$-model (Subsection \ref{SubSec:dZ2}), as well as the BYB model (Section \ref{Sec:BYB}), which is a combination of two Yang-Baxter deformations. One of the characteristics of these deformations is that it breaks a global symmetry of the model: the global left multiplication symmetry for the one-parameter deformations of the PCM and $\Z_2$-coset model and both the left and right multiplication symmetries for the BYB model.

This global symmetry of the undeformed model is associated with a set of conserved charges (see Section \ref{Sec:Undef}), which satisfies the Kirillov-Kostant bracket of the underlying Lie algebra $\g_0$ (see appendices \ref{App:KK} and \ref{App:HamAction}). It was already observed in previous articles about Yang-Baxter type deformations \cite{Delduc:2013fga,Delduc:2014kha,Delduc:2015xdm,Kawaguchi:2011pf,Kawaguchi:2012gp,Hoare:2014oua} that the deformed models admit a set of deformed conserved charges, which satisfy a $q$-deformed algebra (technically the semi-classical limit $\U_q(\g_0)$ of a quantum group). In the article~\cite{Delduc:2016ihq}, I have shown, together with my collaborators, that this is a general feature of all Yang-Baxter type deformations. The originality of the approach we developed in~\cite{Delduc:2016ihq} is that it is model independent: it relies mostly on the effect of the deformation on the twist function of the model.

A natural question emerging from the existence of these deformed conserved charges is whether they are associated with a symmetry of the model. This question is made difficult by the fact that the charges do not satisfy a Kirillov-Kostant algebra but rather a $q$-deformed algebra $\U_q(\g_0)$. Indeed, charges satisfying the Kirillov-Kostant bracket of the Lie algebra $\g_0$ form the moment map of an infinitesimal Hamiltonian action of $\g_0$ (see Appendix \ref{App:HamAction}). However, this is not the case for charges satisfying a $q$-deformed algebra $\U_q(\g_0)$. Instead, these charges form the so-called non-abelian moment map of a Poisson-Lie action of $\g_0$, which is a generalisation of a Hamiltonian action.

Using the theory of Poisson-Lie actions, we constructed in~\cite{Delduc:2016ihq} the symmetry associated with the deformed conserved charges of Yang-Baxter type deformations, in a model-independent way. It is a transformation of the fields of the theory, depending on the deformation parameter, which is a symmetry of the deformed model and which reduces to the global symmetry of the undeformed one when the deformation parameter goes to zero. This transformation is the main subject of this chapter.\\

The chapter is constructed as follows. Sections \ref{sec: PL Dr double} to \ref{Sec:Cobound} are a general review of the theory of Poisson-Lie groups and their actions. Section \ref{Sec:qAlgebra} establishes the link between Poisson-Lie actions and the $q$-deformed algebra $\U_q(\g_0)$ (which is a generalisation of the link between Hamiltonian actions and the Kirillov-Kostant bracket). Finally, Section \ref{sec: YB models} is the application of the Poisson-Lie formalism to the study of deformed symmetries in Yang-Baxter type deformations. Some technical results, specific to this chapter, are gathered in Appendix \ref{App:ThmPB}. 

\section{Poisson-Lie groups and Drinfel'd doubles} \label{sec: PL Dr double}

In this section, we recall the main points of the general theory of Poisson-Lie groups\footnote{There are many references on Poisson-Lie groups. 
For the aspects reviewed in the present article, we mainly refer to the articles 
\cite{Drinfeld:1986in,Drinfeld:1983ky,SemenovTianShansky:1985my,Semenov-nonv,luweinstein1990_a,lu_1990_phd,Babelon:1991ah,
Falceto:1992bf,Feher:2002fx} and to the books 
\cite{Babebook,Chari_Pressley_1994}. Further 
references may be found in \cite{Kosmann-Schwarzbach2004}.} and their link to Lie bialgebras, including the formulation in terms of Drinfel'd doubles.

\subsection{Poisson-Lie groups and Lie bialgebras}

A Poisson-Lie group is a real Lie group $G_0$ equipped with a Poisson bracket $\lbrace \cdot, \cdot \rbrace_{G_0}$ which is compatible with the multiplication $G_0\times G_0 \to G_0$ in the sense that the latter is a Poisson map.

Consider the dual space $\g_0^*$ of the Lie algebra $\g_0$. As $\g_0 \simeq T_eG_0$, any element in $\g_0^*$ can be realised as the differential $\dd_e f:T_eG_0 \rightarrow \mathbb{R}$ of a smooth function $f : G_0 \to \mathbb{R}$, taken at the identity $e$. Using this, we define a skew-symmetric product on $\g_0^*$ by
\begin{equation}\label{LieDual}
\left[ d_ef, d_eg \right]_* = d_e \left\lbrace f,g \right\rbrace_{G_0}.
\end{equation}
One can show that this product is well defined, \emph{i.e.} that the results only depend on $d_ef$ and $d_eg$ and not on the choice of $f$ and $g$. Using the Jacobi identity of the Poisson bracket, one finds that $\LB_*$ also satisfies the Jacobi identity, so that
\begin{equation*}
(\g_0^*,[\cdot,\cdot]_*)
\end{equation*}
is a Lie algebra. The Lie bracket $[\cdot,\cdot]_*$ can be seen as a skew-symmetric map $\delta^*:\g_0^* \otimes \g_0^* \rightarrow \g_0^*$. Using the compatibility of the Poisson bracket $\PB_{G_0}$ with the multiplication on $G_0$, one can show that the dual map 
$\delta:\g_0 \rightarrow \g_0 \otimes \g_0$ is a 1-cocycle, \textit{i.e.} that it verifies
\begin{equation}
\delta([X,Y]) = \left( \ad_X \otimes \Id + \Id \otimes \ad_X \right) \delta(Y) - \left( \ad_Y \otimes \Id + \Id \otimes \ad_Y \right) \delta(X),
\end{equation}
with the adjoint actions $\ad_X$ and $\ad_Y$ as defined in Appendix \ref{App:LieGen}. This proves that $(\g_0,\g_0^*)$ is a Lie bialgebra (see for instance  \cite{Chari_Pressley_1994}). Conversely, from any Lie bialgebra, one can define a unique connected and simply connected Poisson-Lie group.

\subsection{Drinfel'd doubles}
\label{SubSec:Drinfeld}

Let $G_0$ be a Poisson-Lie group, with Lie bialgebra $(\g_0,\g_0^*)$. We define the Drinfel'd double of $\g_0$ as the vector space direct sum
\begin{equation*}
D\g_0 = \g_0 \oplus \g_0^*.
\end{equation*}
We will write $\iota$ and $\iota^*$ for the natural embeddings of $\g_0$ and $\g_0^*$ into $D\g_0$ and we will denote elements of $D\g_0$ as $(X,\lambda)$, where $X$ is in $\g_0$ and $\lambda$ is a linear form in $\g_0^*$. One can define a non-degenerate bilinear form on the double $D\g_0$ by
\begin{equation}\label{PairingDg}
\big\langle (X,\lambda) | (Y,\mu) \big\rangle = \langle X, \mu \rangle + \langle Y, \lambda \rangle = \mu(X) + \lambda(Y)
\end{equation}
for any $X, Y \in \g_0$ and $\lambda, \mu \in \g_0^*$, where $\langle \cdot, \cdot \rangle$ denotes the canonical pairing between $\g_0$ and $\g_0^*$. One then has the following result~\cite{Chari_Pressley_1994}:

\begin{theorem}
There exists a unique Lie bracket $[\cdot,\cdot]_D$ on $D\g_0$ such that $\iota$ and $\iota^*$ are Lie homomorphisms from $\g_0$ and $\g_0^*$ to $D\g_0$, and such that the bilinear form $\langle \cdot | \cdot \rangle$ is $\ad$-invariant.
\end{theorem}

The decomposition $D\g_0=\g_0 \oplus \g_0^*$ satisfies the conditions for the application of the Adler-Kostant-Symes (AKS) scheme, described in Appendix \ref{App:AKS}. Thus, if $\pi_{\g_0}$ and $\pi_{\g_0^*}$ denote the projections along this decomposition, the operator $\Rc^D=\pi_{\g_0}-\pi_{\g_0^*}$ of $D\g_0$ is a solution of the split mCYBE on $D\g_0$ (Theorem \ref{Thm:AKS}). As $\langle\cdot|\cdot\rangle$ defines a non-degenerate bilinear form on $D\g_0$, one can consider the kernel $\Rc^D\ti{12}$ of $\Rc^D$ with respect to this form (see Appendix \ref{App:AKS}). It is clear from the definition \eqref{PairingDg} that the subspaces $\iota(\g_0)$ and $\iota^\ast(\g_0^*)$ of $D\g_0$ are both isotropic with respect to $\langle\cdot|\cdot\rangle$: $\Rc^D\ti{12}$ is thus skew-symmetric (see Proposition \ref{Prop:SkewAKS}). By Proposition \ref{Prop:Rop2Rmat}, $\Rc\ti{12}^D$ is a solution of the matricial split mCYBE on $D\g_0$, which reads
\begin{equation*}
\left[ \Rc^D\ti{12},\Rc^D\ti{13} \right]_D + \left[ \Rc^D\ti{12},\Rc^D\ti{23} \right]_D + \left[ \Rc^D\ti{13},\Rc^D\ti{23} \right]_D = \left[ C^D\ti{12},C^D\ti{13} \right]_D,
\end{equation*}
with $C^D\ti{12}$ the quadratic Casimir in $D\g_0 \otimes D\g_0$. Let us consider a basis $\lbrace I^a \rbrace$ of $\g_0$ and the dual basis $\lbrace I_a \rbrace$ of $\g_0^*$. Then by Proposition \ref{Prop:SkewAKS}, $\Rc\ti{12}^D$ reads
\begin{equation}\label{RD}
\Rc^D\ti{12} = \iota(I^a) \otimes \iota^*(I_a) - \iota^*(I_a) \otimes \iota(I^a).
\end{equation}
In the same way, the quadratic Casimir $C^D\ti{12}$ is given by
\begin{equation*}
C^D\ti{12} = \iota(I^a) \otimes \iota^*(I_a) + \iota^*(I_a) \otimes \iota(I^a).
\end{equation*}

\section{Poisson-Lie actions}

In this section we study actions of Poisson-Lie groups on Poisson manifolds \textit{via} the non-abelian moment map formulation.

\subsection{Non-abelian moment map} \label{subsec: namm}

Let $G_0$ be a Poisson-Lie group and $M$ be a Poisson manifold, with Poisson bracket $\lbrace \cdot, \cdot \rbrace$. Let
\begin{equation*}
\rho : G_0 \times M \longrightarrow M
\end{equation*}
be a smooth group action of $G_0$ on $M$. We say that $\rho$ is a Poisson-Lie action if it is a Poisson map, from $G_0 \times M$, with the direct product Poisson structure, to $M$. The map $\rho$ can alternatively be seen as a group homomorphism from $G_0$ to $\Diff{M}$, the group of diffeomorphisms of $M$. Its differential at the identity induces a Lie algebra action
\begin{equation*}
\delta : \g_0 \longrightarrow T\bigl(\Diff{M}\bigr) = \X[M],
\end{equation*}
on the space $\X[M]$ of vector fields on $M$ (see Appendix \ref{App:HamAction}). For $\epsilon\in\g_0$, the vector field $\delta_\epsilon$ acts naturally on any smooth function $f:M\rightarrow\mathbb{R}$. We consider the case where there exists a map
\begin{equation*}
\Gamma : M \longrightarrow G_0^*,
\end{equation*}
where the dual group $G_0^*$ is the connected and simply connected Lie group with Lie algebra $\g_0^*$, such that
\begin{equation}\label{MomentMap}
\delta_\epsilon f = -\left\langle \epsilon, \Gamma^{-1}\left\lbrace \Gamma, f \right\rbrace \right\rangle.
\end{equation}
The map $\Gamma$ is called the non-abelian moment map of the action of $G$ on $M$. If $M$ is symplectic and simply connected, then such a map always exists. We can note here that $\Gamma$ is defined up to a left multiplication by a constant element in $G_0^*$. Conversely, every transformation of the form \eqref{MomentMap} preserves the Poisson bracket if the parameter $\epsilon$ has a non-zero bracket with itself, coming from the Poisson-Lie structure on $G_0$.\\

To illustrate this concept, let us investigate here the case of a usual Hamiltonian action of $G_0$ on $M$. For any fixed $g$, the action $\rho(g,\cdot)$ is then a canonical transformation on $M$. In other words, $\rho$ is a Poisson map for the trivial Poisson structure on $G_0$. The induced Lie bracket on $\g_0^*$ defined by \eqref{LieDual} is then trivial, so that the dual group $G_0^*$ is abelian. We write $\Gamma = \exp(-Q)$, with $Q: M \rightarrow \g_0^*$. As $G_0^*$ is abelian, the transformation \eqref{MomentMap} simply becomes
\begin{equation}\label{CanonicalCase}
\delta_\epsilon f = \left\langle \epsilon, \left\lbrace Q, f \right\rbrace \right\rangle.\vspace{-2pt}
\end{equation}
We recognize here the usual expression \eqref{Eq:DeltaMomentMap} for a Hamiltonian action of $G_0$ on $M$, with $Q$ the moment map (see Appendix \ref{App:HamAction}). When this action is a symmetry of a Hamiltonian system, decomposing $Q$ with respect to the dual basis of $\g_0^*$ as $Q=Q^a I_a$, we obtain $\text{dim}\,G$ conserved charges $Q^a$.

\subsection{Poisson brackets of the non-abelian moment map}
\label{SubSec:PBGamma}

Let us recall that since $\rho$ is a Lie group action, $\delta$ is a Lie algebra action. In other words, $\delta$ is a Lie homomorphism which is to say that
\begin{equation}\label{ActionLie}
\left[ \delta_\epsilon,\delta_{\epsilon'} \right] = \delta_{[\epsilon,\epsilon']},
\end{equation}
for any $\epsilon, \epsilon' \in \g_0$.
In the case of a usual Hamiltonian action of $G_0$ on $M$, where $\delta_\epsilon$ is given by equation \eqref{CanonicalCase}, it is a well-known fact that the homomorphism condition \eqref{ActionLie} implies that the Poisson algebra of the charges $Q^a$ takes the form
\begin{equation*}
\{ Q^a, Q^b \} = \fs{ab}{c} Q^c + N^{ab}
\end{equation*}
of the Lie algebra relations in $\g_0$ up to central charges $N^{ab}$, where $\ft{ab}{c}$ are structure constants of $\g_0$ with respect to the basis $\{ I^a \}$, \emph{i.e.}
\begin{equation*}
[ I^a, I^b ] = \fs{ab}{c} I^c.
\end{equation*}
It is therefore natural to ask whether we can extract from equation \eqref{ActionLie} some informations on the Poisson bracket of $\Gamma$ with itself. One important step is to note that, from equation \eqref{MomentMap}, using the Jacobi and Leibniz identities on $\lbrace\cdot,\cdot\rbrace$, the action of $[\delta_\epsilon,\delta_{\epsilon'}]$ on any function $f$ takes the following rather simple form
\begin{equation}\label{ComDelta}
\left[ \delta_\epsilon,\delta_{\epsilon'} \right]f = \left\langle \epsilon\ti{1} \epsilon'\ti{2}, \Gamma\ti{1}^{-1} \Gamma\ti{2}^{-1} \left\lbrace \lbrace \Gamma\ti{1}, \Gamma\ti{2} \rbrace \Gamma\ti{1}^{-1} \Gamma\ti{2}^{-1}, f \right\rbrace \Gamma\ti{1} \Gamma\ti{2} \right\rangle\ti{12}.
\end{equation}
In order to treat the right hand side of equation \eqref{ActionLie}, we pass to the Drinfel'd double formulation recalled in section \ref{SubSec:Drinfeld}. Indeed, we can write
\begin{equation*}
\delta_{[\epsilon,\epsilon']}f = -\left\langle  [\iota(\epsilon),\iota(\epsilon')]_D \; \bigr| \, \iota^*(\Gamma)^{-1} \left\lbrace \iota^*(\Gamma), f \right\rbrace \right\rangle,
\end{equation*}
where, by abuse of notation, we still denote by $\iota^*$ the lift of $\iota^*:\g_0^* \hookrightarrow D\g_0$ to the group embedding $G_0^* \hookrightarrow DG_0$. Using the expression \eqref{RD} of $\Rc^D\ti{12}$ we note that
\begin{equation*}
\iota(\epsilon')\ti{1} = \langle \iota(\epsilon')\ti{2} | \Rc^D\ti{12} \rangle\ti{2} \quad \text{and} \quad \iota(\epsilon)\ti{2} = -\langle \iota(\epsilon)\ti{1} | \Rc^D\ti{12} \rangle\ti{1}.
\end{equation*}
From these two equations and the fact that the pairing $\langle\cdot|\cdot\rangle$ is invariant with respect to the $[\cdot,\cdot]_D$ bracket, we obtain
\begin{equation*}
\delta_{[\epsilon,\epsilon']}f = -\frac{1}{2}\left\langle \iota(\epsilon)\ti{1} \iota(\epsilon')\ti{2} \, \Bigr|  \left[ \Rc^D\ti{12}, \iota^*(\Gamma)^{-1}\ti{1} \left\lbrace \iota^*(\Gamma)\ti{1}, f \right\rbrace + \iota^*(\Gamma)^{-1}\ti{2} \left\lbrace \iota^*(\Gamma)\ti{2}, f \right\rbrace \right]_D \right\rangle\ti{12}.
\end{equation*}
This expression can be rewritten as
\begin{equation}\label{DeltaCom}
\delta_{[\epsilon,\epsilon']}f = \frac{1}{2} \left\langle \iota(\epsilon)\ti{1} \iota(\epsilon')\ti{2} \, \Bigr| \, \iota^*(\Gamma)^{-1}\ti{1} \iota^*(\Gamma)^{-1}\ti{2} \left\lbrace \iota^*(\Gamma)\ti{1}\iota^*(\Gamma)\ti{2} \Rc^D\ti{12} \iota^*(\Gamma)^{-1}\ti{1} \iota^*(\Gamma)^{-1}\ti{2}, f \right\rbrace \iota^*(\Gamma)\ti{1}\iota^*(\Gamma)\ti{2} \right\rangle\ti{12}.
\end{equation}
Using the fact that the pairing $\langle\cdot|\cdot\rangle$ is non-degenerate between $\g_0$ and $\g_0^*$, by equating \eqref{ComDelta} and \eqref{DeltaCom} we arrive at
\begin{equation}\label{PBGammaP}
\left\lbrace \iota^*(\Gamma)\ti{1}, \iota^*(\Gamma)\ti{2} \right\rbrace \iota^*(\Gamma)^{-1}\ti{1} \iota^*(\Gamma)^{-1}\ti{2} = \frac{1}{2}
\iota^*(\Gamma)\ti{1}\iota^*(\Gamma)\ti{2} \Rc^D\ti{12} \iota^*(\Gamma)^{-1}\ti{1} \iota^*(\Gamma)^{-1}\ti{2} + P\ti{12}
\end{equation}
where the element $P\ti{12} \in D\g_0 \otimes D\g_0$ is a central charge of the Poisson bracket $\lbrace\cdot,\cdot\rbrace$. Let us study the properties that must fulfil $P\ti{12}$. First of all, it must be skew-symmetric. Moreover, it should 
be such that the right hand side of \eqref{PBGammaP} lives in $\iota^*(\g_0^*)\otimes\iota^*(\g_0^*)$. It is a well-known 
consequence of the Adler-Kostant-Symes construction that, for any $y \in \iota^*(G_0^*)$, we have 
$y\ti{1}y\ti{2}\Rc^D\ti{12}y\ti{1}^{-1}y\ti{2}^{-1}-\Rc^D\ti{12} \in \iota^*(\g_0^*)\otimes\iota^*(\g_0^*)$. Thus, 
defining $N\ti{12}=P\ti{12}+\frac{1}{2}\Rc^D\ti{12}$, we can write
\begin{equation}\label{PBGamma}
\left\lbrace \iota^*(\Gamma)\ti{1}, \iota^*(\Gamma)\ti{2} \right\rbrace  =
-\frac{1}{2} \left[ \Rc^D\ti{12}, \iota^*(\Gamma)\ti{1}\iota^*(\Gamma)\ti{2} \right] + N\ti{12} 
\iota^*(\Gamma)\ti{1}\iota^*(\Gamma)\ti{2},
\end{equation}
with $N\ti{12} \in \iota^*(\g_0^*) \otimes \iota^*(\g_0^*)$ skew-symmetric. The last requirement on 
$N\ti{12}$ is that the Poisson bracket \eqref{PBGamma} must satisfy the Jacobi identity. Let us first remark that this is the case when $N\ti{12}=0$, as $\Rc^D\ti{12}$ verifies the mCYBE. We will see in the next sections why this case is of particular interest.

More generally, we recognise in \eqref{PBGamma} a quadratic algebra of $ad$-type, in the 
nomenclature of \cite{Freidel:1991jx,Freidel:1991jv}. In this case, a necessary and sufficient condition for the Jacobi identity to hold is that $N\ti{12}-\frac{1}{2}\Rc^D\ti{12}$ satisfies the mCYBE. In particular, this is the case if
\begin{equation*}
N\ti{12}=\frac{1}{2}\Rc^D\ti{12}-\frac{1}{2}\iota^*(C)\ti{1}^{-1}\iota^*(C)\ti{2}^{-1}\Rc^D\ti{12}\iota^*(C)\ti{1}\iota^*(C)\ti{2},
\end{equation*}
for some constant $C \in G_0^*$. For this $N\ti{12}$ we define $\tilde{\Gamma}=C\Gamma$. As noted in subsection \ref{subsec: namm}, $\tilde{\Gamma}$ is still a good non-abelian moment map. Moreover, the Poisson bracket of $\tilde{\Gamma}$ becomes
\begin{equation*}
\big\{ \iota^*(\tilde{\Gamma})\ti{1}, \iota^*(\tilde{\Gamma})\ti{2} \big\}  =
-\frac{1}{2} \big[ \Rc^D\ti{12}, \iota^*(\tilde{\Gamma})\ti{1}\iota^*(\tilde{\Gamma})\ti{2}\big].
\end{equation*}
Conversely, if the Poisson brackets of $\Gamma$ are of the form \eqref{PBGamma}, then the transformation \eqref{MomentMap} is a Lie algebra action of $\g_0$, \textit{i.e.} it satisfies \eqref{ActionLie}.

\section{Coboundary Poisson-Lie groups and $R$-matrices}
\label{Sec:Cobound}

One important class of Lie bialgebras are the so-called coboundary ones, which are given by $R$-matrices, solutions of the mCYBE (see Appendix \ref{App:CYBE}). In this section, we recall their properties and apply the abstract result \eqref{PBGamma} of the previous section to this particular case.

\subsection[$R$-matrices, Sklyanin bracket and $\g_R$ dual algebra]{$\bm{R}$-matrices, Sklyanin bracket and $\bm{\g_R}$ dual algebra}
\label{SubSec:GR}

Let $\g_0$ be a Lie algebra and $R:\g_0\rightarrow\g_0$ a linear map solution of the mCYBE on $\g_0$, namely
\begin{equation}\label{mCYBE}
[RX,RY]-R\bigl([RX,Y]+[X,RY]\bigr) = -c^2 [X,Y],
\end{equation}
for all $X,Y \in \g_0$ with $c=1$ (split case) or $c=i$ (non-split case). We define $R^\pm = R \pm c \, \Id$, and introduce the $R$-bracket
\begin{equation*}
[X,Y]_R = \gamma \big( [RX,Y]+[X,RY] \big) = \gamma \big( [R^\pm X, Y] + [X, R^\mp Y] \big),
\end{equation*}
with $\gamma$ a real constant. An important consequence of the mCYBE is that the vector space $\g_0$ equipped with the $R$-bracket is also a Lie algebra. We therefore have two Lie algebra structures on the vector space $\g_0$: the usual one $(\g_0,[\cdot,\cdot])$, that we shall still note $\g_0$ and
\begin{equation*}
\g_R = (\g_0,[\cdot,\cdot]_R).
\end{equation*}

This construction is related to Poisson-Lie groups. Suppose now that $\g_0$ is semi-simple and let $\kappa$ denote its Killing form. Let $R$ be a skew-symmetric solution of the mCYBE on $\g_0$. We denote by $R\ti{12}\in\g_0\otimes\g_0$ its kernel with respect to $\kappa$. One can then define a Poisson-Lie structure on $G_0$ with the Sklyanin bracket
\begin{equation*}
\left\lbrace x\ti{1}, x\ti{2} \right\rbrace_{G_0} = \gamma \left[R\ti{12}, x\ti{1}x\ti{2} \right].
\end{equation*}
Since $\g_0$ is semi-simple, its Killing form $\kappa$ is non-degenerate. This allows us to define a natural pairing $\pi$ between $\g_0$ and its dual $\g_0^*$ by considering, for any $X\in\g_0$, the linear form
\begin{equation*}
\begin{array}{rccc}
\pi(X) : &\g_0 & \longrightarrow & \mathbb{R} \\
         & Y & \longmapsto    & \kappa(X,Y)
\end{array}
\end{equation*}
As a vector space, $\g_R$ is equal to $\g_0$, so $\pi$ can be seen as a linear isomorphism from $\g_R$ to $\g_0^*$. The following lemma then gives a concrete realisation of the dual Lie algebra $\g_0^*$.

\begin{lemma}
Let $G_0$ be a Poisson-Lie group, with the Sklyanin bracket associated with a solution $R$ of the mCYBE on $\g_0$. Equip $\g_0^*$ with the Lie bracket \eqref{LieDual}. Then the map
\begin{equation*}
\pi : \g_R \longrightarrow \g_0^*
\end{equation*}
is a Lie algebra isomorphism.
\end{lemma}

\subsection[Real and complex doubles and the dual Lie algebras $\g_{DR}$ and $\g_\pm$]{Real and complex doubles and the dual Lie algebras $\bm{\g_{DR}}$ and $\bm{\g_\pm}$}
\label{SubSec:Doubles}

In this section, we study separately the split and non-split cases.

\paragraph{Split case.}
Define the real double of $\g_0$ as the Lie algebra direct sum
$\g_0 \oplus \g_0$. We introduce the subspaces
\begin{equation*}
\gd = \bigl\lbrace (X,X), \, X\in\g_0 \bigr\rbrace, \qquad
\g_{DR} = \bigl\lbrace (R^+X,R^-X), \, X\in\g_0 \bigr\rbrace.
\end{equation*}
It is clear that, for any endomorphism $R$, $\gd$ and $\g_{DR}$ form a direct sum decomposition of $\g_0 \oplus \g_0$. Moreover, $\gd$ is a Lie subalgebra of $\g_0 \oplus \g_0$. One shows that, when $R$ is a split solution of the mCYBE, $\g_{DR}$ is also a Lie subalgebra of $\g_0 \oplus \g_0$ isomorphic to $\g_R$. More precisely,

\begin{lemma}
If $R$ is a solution of the split mCYBE on $\g_0$, then
\begin{equation*}
\begin{array}{rccc}
\Delta : &\g_R & \longrightarrow & \g_{DR} \\
          & X  & \longmapsto     & \gamma(R^+X,R^-X)
\end{array}
\end{equation*}
is a Lie algebra isomorphism, whose inverse is given for all $(X,Y) \in \g_{DR} \subset \g_0 \oplus \g_0$ by
\begin{equation}\label{DeltaInv}
\Delta^{-1}(X,Y) = \frac{1}{2\gamma}(X-Y).
\end{equation}
\end{lemma}

We have obtained yet another realisation of $\g_0^*$, this time in the real double. Moreover the subalgebra $\gd$ is isomorphic to $\g_0$. Hence we have realised both $\g_0$ and $\g_0^*$ as subalgebras of the real double $\g_0 \oplus \g_0$. In fact, by the following lemma the real double $\g_0 \oplus \g_0$ itself is a realisation of the abstract Drinfel'd double $D\g_0=\g_0\oplus\g_0^*$ (cf. section \ref{SubSec:Drinfeld}).

\begin{lemma}\label{Lemma:PhiSplit}
If $R$ is a skew-symmetric solution of the split mCYBE on $\g_0$, then
\begin{equation*}
\begin{array}{rccc}
\Phi :& D\g_0 & \longrightarrow &\g_0 \oplus \g_0 \\
       &    (X,\lambda)  & \longmapsto   &  (X,X) + \Delta\circ\pi^{-1}(\lambda)
\end{array}
\end{equation*}
is a Lie algebra isomorphism, such that $\Phi\bigl(\iota(\g_0)\bigr)=\gd$ and $\Phi\bigl(\iota^*(\g_0^*)\bigr)=\g_{DR}$. Its inverse is given for every $(X,Y) \in \g_0 \oplus \g_0$ by
\begin{equation*}
\Phi^{-1}(X,Y) = \frac{1}{2} \left( R^+Y-R^-X, \frac{1}{\gamma} \pi(X-Y) \right).
\end{equation*}
Moreover, $\Phi$ sends the pairing \eqref{PairingDg} on $D\g_0$ to the non-degenerate $\ad$-invariant bilinear form on $\g_0 \oplus \g_0$ defined, for all $X_1, X_2, Y_1, Y_2 \in \g_0$, by
\begin{equation*}
\big\langle (X_1,Y_1)|(X_2,Y_2) \big\rangle = \frac{1}{2\gamma} \big( \kappa(X_1,X_2) - \kappa(Y_1,Y_2) \big).
\end{equation*}
\end{lemma}

\paragraph{Non-split case.}
 One can perform a similar analysis in the case of a non-split solution of the mCYBE ($c=i$). Here we introduce the complex double $\g_0^\C$ as the complexification of $\g_0$ (see Appendix \ref{App:RealForms}), namely
\begin{equation*}
\g_0^\C = \lbrace X+iY, \, X,Y \in \g_0 \rbrace
\end{equation*}
We define the complex conjugation relative to the real form $\g_0$ as
\begin{equation*}
\begin{array}{rccc}
\tau : &\g_0^\C & \longrightarrow & \g_0^\C \\
       &  X+iY  & \longmapsto     & X-iY
\end{array}
\end{equation*}
This is a semi-linear involutive automorphism of $\g_0^\C$ and $\g_0$ itself can be seen as a Lie subalgebra of $\g_0^\C$, viewed as a real Lie algebra. More precisely, $\g_0$ is the subalgebra of $\g_0^\C$ fixed by $\tau$ (see appendix \ref{App:RealForms}).

We introduce the subspaces
\begin{equation*}
\g_\pm = \lbrace R^\pm X, \, X\in\g_0 \rbrace
\end{equation*}
of $\g_0^\C$. Note that $\g_\pm=\tau(\g_\mp)$. We have the vector space decompositions $\g_0^\C=\g_0\oplus\g_+ = \g_0\oplus\g_-$. Moreover, as a consequence of the mCYBE, $\g_\pm$ are Lie subalgebras of $\g_0^\C$ isomorphic to $\g_R$.

\begin{lemma}
If $R$ is a solution of the non-split mCYBE on $\g_0$, then
\begin{equation*}
\Delta_\pm = \gamma R^\pm : \g_R \longrightarrow \g_\pm
\end{equation*}
is a Lie algebra isomorphism, whose inverse is given for each $X \in \g_{\pm} \subset \g_0^\C$ by
\begin{equation}\label{DeltaInvNS}
\Delta_\pm^{-1}(X) = \pm\frac{X-\tau(X)}{2i\gamma}.
\end{equation}
\end{lemma}

 As in the split case, we realised $\g_0$ and $\g_0^*$ as subalgebras of $\g_0^\C$. Moreover, the complex double $\g_0^\C$ provides another realisation of the abstract Drinfel'd double $D\g_0$ by the following result.

\begin{lemma}\label{Lemma:PhiNonSplit}
If $R$ is a skew-symmetric solution of the non-split mCYBE on $\g_0$, then
\begin{equation*}
\begin{array}{rccc}
\Phi_\pm : &D\g_0 & \longrightarrow &\g_0^\C \\
           &(X,\lambda)  & \longmapsto  &   X + \gamma R^\pm \circ\pi^{-1}(\lambda)
\end{array}
\end{equation*}
is a Lie algebra isomorphism, such that $\Phi_\pm\bigl(\iota(\g_0)\bigr)=\g_0$ and $\Phi_\pm\bigl(\iota^*(\g_0^*)\bigr)=\g_\pm$. Its inverse is given for any $X \in \g_0^\C$ by
\begin{equation*}
\Phi_\pm^{-1}(X) = \frac{1}{2i} \left( R^+\big( \tau(X) \big) - R^-(X), \pm \frac{1}{\gamma} \pi \big( X - \tau(X) \big) \right).
\end{equation*}

Moreover, $\Phi_\pm$ sends the pairing \eqref{PairingDg} on $D\g_0$ to the non-degenerate $\ad$-invariant bilinear form on $\g_0^\C$ defined, for all $X, Y \in \g_0^\C$, by
\begin{equation*}
\left\langle X|Y \right\rangle = \pm\frac{1}{\gamma} \,\textup{Im} \big( \kappa(X,Y)\big).
\end{equation*}
\end{lemma}

\subsection[Poisson-Lie action of $G_0$: Semenov-Tian-Shansky brackets]{Poisson-Lie action of $\bm{G_0}$: Semenov-Tian-Shansky brackets}
\label{SubSec:PL2STS}

In the previous subsections, we provided concrete realisations of both the dual Lie algebra $\g_0^*$ and the Drinfel'd double $D\g_0$ for (split and non-split) coboundary Poisson-Lie groups. By abuse of notation, we will denote by the same symbols the lift of these realisations to the dual group $G_0^*$ and the Drinfel'd double group $DG_0$. In section \ref{SubSec:PBGamma}, we found an abstract expression \eqref{PBGamma} for the Poisson bracket of the non-abelian moment map viewed in the Drinfel'd double. We will now investigate what this Poisson bracket becomes in the concrete realisations of $DG_0$.

\paragraph{Split case.}
The non-abelian moment map $\Gamma$ can be seen as a map to the group $G_R$ \textit{via} the Killing pairing $\pi$, namely
\begin{equation*}
\Gamma_R = \pi^{-1}(\Gamma) \in G_R,
\end{equation*}
and in turn as an element of the group $G_{DR}$ \textit{via} the morphism $\Delta$,
\begin{equation*}
(\Gamma^+,\Gamma^-) = \Delta(\Gamma_R) = \Delta\circ\pi^{-1}(\Gamma) \in G_{DR} \subset G\times G.
\end{equation*}
The real double $G_0 \times G_0$ is related to the Drinfel'd double $DG_0$ by the morphism $\Phi$ (cf lemma \ref{Lemma:PhiSplit}). Let us remark here that:
\begin{equation*}
\Phi\bigl(\iota^*(\Gamma)\bigr)=(\Gamma^+,\Gamma^-).
\end{equation*}
The Poisson bracket \eqref{PBGamma} then becomes, under the action of $\Phi\ti{1}\Phi\ti{2}$:
\begin{equation}\label{PBRealDouble}
\left\lbrace (\Gamma^+,\Gamma^-)\ti{1}, (\Gamma^+,\Gamma^-)\ti{2} \right\rbrace = -\frac{1}{2} \left[ \Phi\ti{1} \Phi\ti{2} \Rc^D\ti{12}, (\Gamma^+,\Gamma^-)\ti{1} (\Gamma^+,\Gamma^-)\ti{2} \right] + \tilde{N}\ti{12} (\Gamma^+,\Gamma^-)\ti{1} (\Gamma^+,\Gamma^-)\ti{2}
\end{equation}
with $\widetilde{N}\ti{12} = \Phi\ti{1}\Phi\ti{2} N\ti{12}$ the central charge.

The objects in the above formula belongs to $(\g_0 \oplus \g_0) \otimes (\g_0 \oplus \g_0)$. Such objects can be written as vectors with four $\g_0\otimes\g_0$-valued components as
\begin{equation*}
(X,Y) \otimes (X',Y') =
\begin{pmatrix}
X \otimes X' \\ X \otimes Y' \\ Y \otimes X' \\ Y \otimes Y'
\end{pmatrix}.
\end{equation*}
Let us now compute $\Phi\ti{1} \Phi\ti{2} \Rc^D\ti{12}$. We have
\begin{equation*}
\Phi\circ\iota(I^a) = (I^a,I^a)
\end{equation*}
and
\begin{equation*}
\Phi\circ\iota^*(I_a) = \Delta\circ\pi^{-1}(I_a) =  \kappa_{ab} \Delta(I^b) = \gamma \kappa_{ab} (R^+I^b,R^-I^b)
\end{equation*}
where $\kappa_{ab}$ is the inverse of the Killing form written in the basis $\lbrace I^a \rbrace$. We therefore find that
\begin{equation}\label{PhiRD}
\Phi\ti{1} \Phi\ti{2} \Rc^D\ti{12}
= -2\gamma \begin{pmatrix}
 R\ti{12} \\ R^+\ti{12} \\ R^-\ti{12} \\ R\ti{12}
 \end{pmatrix}.
\end{equation}
Let us write
\begin{equation*}
\widetilde{N}\ti{12} = \begin{pmatrix}
N^{++}\ti{12} \\ N^{+-}\ti{12} \\ N^{-+}\ti{12} \\ N^{--}\ti{12}
\end{pmatrix}.
\end{equation*}
The four components of the Poisson bracket \eqref{PBRealDouble} then read
\begin{subequations}
\begin{align}
\left\lbrace \Gamma^+\ti{1}, \Gamma^+\ti{2} \right\rbrace &= \gamma \left[ R\ti{12}, \Gamma^+\ti{1} \Gamma^+\ti{2} \right] + N^{++}\ti{12} \Gamma^+\ti{1} \Gamma^+\ti{2}, \\[5pt]
\left\lbrace \Gamma^+\ti{1}, \Gamma^-\ti{2} \right\rbrace &= \gamma \left[ R^+\ti{12}, \Gamma^+\ti{1} \Gamma^-\ti{2} \right] + N^{+-}\ti{12} \Gamma^+\ti{1} \Gamma^-\ti{2}, \\[5pt]
\left\lbrace \Gamma^-\ti{1}, \Gamma^+\ti{2} \right\rbrace &= \gamma \left[ R^-\ti{12}, \Gamma^-\ti{1} \Gamma^+\ti{2} \right] + N^{-+}\ti{12} \Gamma^-\ti{1} \Gamma^+\ti{2}, \\[5pt]
\left\lbrace \Gamma^-\ti{1}, \Gamma^-\ti{2} \right\rbrace &= \gamma \left[ R\ti{12}, \Gamma^-\ti{1} \Gamma^-\ti{2} \right] + N^{--}\ti{12} \Gamma^-\ti{1} \Gamma^-\ti{2}.
\end{align}
\end{subequations}
When the central charges $N^{\pm\pm}\ti{12}$ vanish, these are the Semenov-Tian-Shansky brackets
\begin{subequations}\label{STSpb}
\begin{align}
\left\lbrace \Gamma^\pm\ti{1}, \Gamma^\pm\ti{2} \right\rbrace &= \gamma \left[ R\ti{12}, \Gamma^\pm\ti{1} \Gamma^\pm\ti{2} \right], \\
\left\lbrace \Gamma^\pm\ti{1}, \Gamma^\mp\ti{2} \right\rbrace &= \gamma \left[ R^\pm\ti{12}, \Gamma^\pm\ti{1} \Gamma^\mp\ti{2} \right].
\end{align}
\end{subequations}

Finally, let us emphasise that the transformation law \eqref{MomentMap} can be re-expressed in terms of the non-abelian moment map $(\Gamma^+,\Gamma^-)$ \textit{via} the morphism $\Delta\circ\pi^{-1}$, giving explicitly
\begin{equation} \label{ActionSplit}
\delta_\epsilon f = -\frac{1}{2\gamma} \kappa \Bigl( \epsilon, (\Gamma^+)^{-1}\left\lbrace \Gamma^+, f \right\rbrace - (\Gamma^-)^{-1}\left\lbrace \Gamma^-, f \right\rbrace \Bigr).
\end{equation}

\paragraph{Non-split case.}
From the non-abelian moment map $\Gamma_R=\pi^{-1}(\Gamma)$ seen in the group $G_R$, we can construct two different realisations of $\Gamma$ in the complex double $G_0^\C$:
\begin{equation*}
\Gamma^+ = \Delta_+ (\Gamma_R) \in G_+ \quad \text{and} \quad \Gamma^- = \Delta_- (\Gamma_R) \in G_-.
\end{equation*}
These are not independent. They are related by the semi-linear automorphism $\tau$ as $\Gamma^\pm = \tau(\Gamma^\mp)$. Note that
\begin{equation*}
\Phi_\pm\circ\iota^*(\Gamma) = \Gamma^\pm.
\end{equation*}
For any $\eta,\varepsilon \in \lbrace +,- \rbrace$, applying $\Phi_\eta\null\ti{1}\Phi_\varepsilon\null\ti{2}$ to the bracket \eqref{PBGamma}, we obtain
\begin{equation*}
\left\lbrace \Gamma^\eta\ti{1}, \Gamma^\varepsilon\ti{2} \right\rbrace = -\frac{1}{2} \left[ \Phi_\eta\null\ti{1}\Phi_\varepsilon\null\ti{2}\Rc^D\ti{12}, \Gamma^\eta\ti{1} \Gamma^\varepsilon\ti{2} \right] + N^{\eta\varepsilon}\ti{12} \Gamma^\eta\ti{1} \Gamma^\varepsilon\ti{2},
\end{equation*}
with the central charges $N^{\eta\varepsilon}\ti{12} = \Phi_\eta\null\ti{1}\Phi_\varepsilon\null\ti{2} \tilde{N}\ti{12}$. We have
\begin{equation*}
\Phi_\pm\circ\iota(I^a)=I^a \; \; \; \; \text{and} \; \; \; \; \Phi_\pm\circ\iota^*(I_a)=\gamma R^\pm \bigl( \pi^{-1}(I_a) \bigr) =\gamma\sum_b \kappa_{ab} R^\pm I^b,
\end{equation*}
so that
\begin{equation*}
\Phi_\eta\null\ti{1}\Phi_\varepsilon\null\ti{2} \Rc^D\ti{12} = \gamma \bigl(R^\varepsilon\ti{21}-R^\eta\ti{12}\bigr)= -\gamma \bigl(R^{\eta}\ti{12}+R^{-\varepsilon}\ti{12} \bigr).
\end{equation*}
Thus, in the non-split case, the non-abelian moment maps $\Gamma^+$ and $\Gamma^-$ also satisfy, up to central charges, the Semenov-Tian-Shansky Poisson brackets \eqref{STSpb}.\\

Applying $\Phi_\pm$ to equation \eqref{MomentMap} yields the transformation law in terms of the non-abelian moment map $\Gamma^\pm$ which reads
\begin{equation}\label{ActionNonSplit}
\delta_\epsilon f = \mp \frac{1}{\gamma} \, \textup{Im} \, \Bigl( \kappa \bigl( \epsilon, (\Gamma^\pm)^{-1}\left\lbrace \Gamma^\pm, f \right\rbrace  \bigr) \Bigr) =  -\frac{1}{2i\gamma} \kappa \Bigl( \epsilon, (\Gamma^+)^{-1}\left\lbrace \Gamma^+, f \right\rbrace - (\Gamma^-)^{-1}\left\lbrace \Gamma^-, f \right\rbrace \Bigr).
\end{equation}

\subsection[Poisson-Lie action of $G_0^*$: Sklyanin bracket on $U$]{Poisson-Lie action of $\bm{G_0^*}$: Sklyanin bracket} \label{subsec: Sklyanin bracket}

In this section we consider the case of a coboundary Lie bialgebra specified by a split $R$-matrix. We have canonical isomorphisms $\g_{DR}^*\simeq\g_0^{**}\simeq\g_0$ of vector spaces. Moreover, the dual Lie group $G_0^* \simeq G_{DR}$ is a Poisson-Lie group when equipped with the Semenov-Tian-Shansky bracket
\begin{subequations}
\begin{align}
\left\lbrace x^\pm\ti{1}, x^\pm\ti{2} \right\rbrace &= \gamma \left[ R\ti{12}, x^\pm\ti{1} x^\pm\ti{2} \right], \\
\left\lbrace x^\pm\ti{1}, x^\mp\ti{2} \right\rbrace &= \gamma \left[ R^\pm\ti{12}, x^\pm\ti{1} x^\mp\ti{2} \right].
\end{align}
\end{subequations}
The induced Lie structure on $\g_{DR}^*$ is isomorphic to $\g_0$, so that the isomorphism $\g_0^{**}\simeq\g_{DR}^*\simeq\g_0$ also holds at the level of Lie algebras. Thus the Drinfel'd double $D\g_0^*$ of the dual Lie algebra $\g_0^*$ is isomorphic to the Drinfel'd double $D\g_0$ of the original Lie algebra $\g_0$.

As a consequence, the formalism developed in the previous sections can also be used to treat a Poisson-Lie action of the dual group $G_0^* \simeq G_{DR}$ on a Poisson manifold $M$. In this case, the non-abelian moment map is an application
\begin{equation*}
U : M \rightarrow G_0.
\end{equation*}
We can see it as an element $\iota(U)$ of the Drinfel'd double $DG_0$. The results presented in section \ref{SubSec:PBGamma} still apply in this case and the Poisson bracket of $U$ is then given by
\begin{equation*}
\left\lbrace \iota(U)\ti{1}, \iota(U)\ti{2} \right\rbrace  =
-\frac{1}{2} \left[ \Rc^D\ti{12}, \iota(U)\ti{1}\iota(U)\ti{2} \right] + M\ti{12} \iota(U)\ti{1}\iota(U)\ti{2},
\end{equation*}
where $M\ti{12}$ is a central charge valued in $\iota(\g_0) \otimes \iota(\g_0)$. Applying the morphism $\Phi$ to this equation, noting that $\Phi\bigl(\iota(U)\bigr)=(U,U)$ and recalling equation \eqref{PhiRD}, we obtain
\begin{equation}\label{PBU}
\left\lbrace U\ti{1}, U\ti{2} \right\rbrace = \gamma\left[ R\ti{12}, U\ti{1} U\ti{2} \right] + \widetilde{M}\ti{12} U\ti{1}U\ti{2},
\end{equation}
with $\widetilde{M}\ti{12}=\Phi\ti{1} \Phi\ti{2} \bigl( M\ti{12} \bigr)$. This is, up to central charges, the Sklyanin bracket.

\section[Link with $q$-deformed algebras]{Link with $q$-deformed algebras}
\label{Sec:qAlgebra}

In this section we suppose that $\g_0$ is either the split real form or a non-split real form of the semi-simple complexification $\g=\g_0^\C$. The definitions and basic properties of semi-simple complex Lie algebras are recalled in appendix \ref{App:SemiSimple} and those of (non-)split real forms in appendix \ref{App:RealSS}.

\subsection{Real forms and standard $R$-matrices}
\label{SubSec:Stand}

Recall the construction of the standard $R$-matrix on $\g$, explained in Appendix \ref{App:StandardFinite}. If $\pi_\pm$ and $\pi_\h$ denotes the projections on $\n_\pm$ and $\h$ in the Cartan-Weyl decomposition $\g=\n_+\oplus\n_-\oplus\n_+$, the standard $R$-matrix on $\g$ reads
\begin{equation}
R=c(\pi_+-\pi_-),
\end{equation}
with $c=1$ (split case) or $c=i$ (non-split case). It is a solution of the operator mCYBE \eqref{mCYBE} on $\g$. Moreover, as explained in Appendix \ref{App:StandardFinite}, it defines a $R$-matrix on the real form $\g_0$ (for $c=1$ for the split real form and for $c=i$ for the non-split real form).

As shown in Appendix \ref{App:StandardFinite}, the kernel of $R$ with respect to the Killing form is
\begin{equation*}
R\ti{12} = c\sum_{\alpha > 0} \left(E_\alpha \otimes F_{\alpha} - F_{\alpha} \otimes E_\alpha \right)
\end{equation*}
Likewise, the kernel of $R^\pm$ is $R^\pm\ti{12}=R\ti{12}\pm c  \,C\ti{12}$, where the quadratic Casimir is given
\begin{equation*}
C\ti{12} = H\ti{12} + \sum_{\alpha > 0} \left(E_\alpha \otimes F_{\alpha} + F_{\alpha} \otimes E_\alpha \right),
\end{equation*}
with $H\ti{12} \in \g \otimes \g$ as introduced in Appendix \ref{App:Casimir}.

\subsection{Extraction of charges}

We saw in section \ref{SubSec:PL2STS} that, in the split case, the non-abelian moment map can be regarded as an element $(\Gamma^+,\Gamma^-)$ of $G_{DR} \subset G_0\times G_0$. In the non-split case, it can be represented as either $\Gamma^+ \in G_+ \subset G$ or $\Gamma^- \in G_- \subset G$ (with $\Gamma^+$ and $\Gamma^-$ related by $\Gamma^\pm=\tau(\Gamma^\mp)$, where $\tau$ is the semi-linear involutive automorphism of $G$ defining the real subgroup $G_0$). Here we will treat the two cases together.

In the split (resp. non-split) case, the Lie algebras $R^\pm(\g_0)$ are the positive and negative Borel subalgebras of $\g_0$ (resp. $\g$), with opposite Cartan parts. Therefore $\Gamma^+$ and $\Gamma^-$ are elements of the positive and negative Borel subgroups of $G_0$ (resp. $G$), with Cartan parts inverses of one another. We choose to parametrise them as follows
\begin{equation}\label{DefDM}
\Gamma^+ = M^+D, \; \; \; \Gamma^-=D^{-1}M^-, \; \; \; M^\pm \in N_\pm, \; \; \; D \in H.
\end{equation}

We now extract scalar charges from $D$ and $M^\pm$. Starting with the Cartan part $D$, we choose a decomposition with respect to the basis of fundamental weight, recalled in appendix \ref{App:BasesCartan},
\begin{equation}\label{DecompoD}
D = \exp\left(ic \gamma \sum_{i=1}^\ell Q^H_i P_i \right) = \prod_{i=1}^\ell \exp\left(i c\gamma Q^H_i P_i \right).
\end{equation}
The order of the product has no importance since the Cartan subgroup is abelian. We define
\begin{equation}\label{DefZ}
Z = i c \gamma\sum_{i=1}^\ell Q^H_i P_i \in \h,
\end{equation}
so that $D=\exp(Z)$.\\

The extraction of suitable charges from $M^+$ and $M^-$ is more involved. Let us fix a labelling $\beta_1,\ldots,\beta_n$ of the positive roots, where $n$ is the number of positive roots. We can parametrise $M^\pm$ by scalar charges $Q^E_\beta$ as
\begin{equation}\label{DecompoM}
M^\pm = \prod_{i=1}^n \exp\left( \pm ic\gamma A_{\pm\beta_i} Q^E_{\pm\beta_i} E_{\pm\beta_i} \right),
\end{equation}
where the $A_\beta$'s are normalisation constants to be fixed later. Define
\begin{equation} \label{DefUV}
u_{(i)} = \exp\left( ic\gamma A_{\beta_i} Q^E_{\beta_i} E_{\beta_i} \right) \;\;\;\; \text{ and } \;\;\;\; v_{(i)} = \exp\left( -ic\gamma A_{-\beta_i} Q^E_{-\beta_i} E_{-\beta_i} \right)
\end{equation}
so that we can write
\begin{equation}\label{DecompoUV}
M^+ = u_{(1)} \ldots u_{(n)}  \;\;\;\; \text{ and } \;\;\;\; M^- = v_{(1)} \ldots v_{(n)}.\vspace{8pt}
\end{equation}
Since the Lie groups $N_\pm$ are not abelian, these products depend on the choice of the ordering $\beta_1,\ldots,\beta_n$ of the positive roots. We choose an ordering such that
\begin{equation}\label{ConditionOrder}
\text{if } i < j \text{ and } \beta_i + \beta_j \text{ is a root, then } \beta_i + \beta_j = \beta_k \text{ with } i < k < j.
\end{equation}
Such an ordering can be constructed from the (partial) normal order described in~\cite{Delduc:2013fga}. We label the simple roots $\alpha_1, \ldots, \alpha_\ell$ in a way which is compatible with the ordering $\beta_1,\ldots,\beta_n$, \textit{i.e.} such that $\alpha_i=\beta_{k_i}$ with $1=k_1 \leq \ldots \leq k_\ell=n$.

\subsection[Semenov-Tian-Shansky brackets and $\U_q(\g_0)$ algebra]{Semenov-Tian-Shansky brackets and $\bm{\U_q(\g_0)}$ algebra}
\label{SubSec:Uqg}

We will now start from the Semenov-Tian-Shansky brackets \eqref{STSpb} for the non-abelian moment map $\Gamma^\pm$ and extract from it the corresponding Poisson brackets between the charges $Q^H_i$ and $Q^E_\beta$, as defined above. We will make extensive use of two theorems for the extraction of Poisson brackets that we present in appendix \ref{App:ThmPB}. We shall denote by $\pi_k$ and $\pi_{-k}$ the projections onto $\C E_{\beta_k}$ and $\C E_{-\beta_k}$ with respect to the Cartan-Weyl decomposition
\begin{equation*}
\g = \bigoplus_{k=1}^n \left( \C E_{\beta_k} \oplus \C E_{-\beta_k} \right) \oplus \h.
\end{equation*}

\subsubsection[Poisson brackets of $D$ and $M^\pm$]{Poisson brackets of $\bm{D}$ and $\bm{M^\pm}$}

Consider the decomposition \eqref{DefDM} of $\Gamma^+$ and $\Gamma^-$. Using Theorem \ref{TheoremPB1}, we obtain
\begin{subequations}\label{PBDMM}
\begin{align}
\left\lbrace D\ti{1}, D\ti{2} \right\rbrace &= 0, \label{PBDD}\\
\left\lbrace D\ti{1}, M^\pm\ti{2} \right\rbrace &= c\gamma D\ti{1} \left[H\ti{12}, M^\pm\ti{2} \right], \label{PBDM}\\
\left\lbrace M^+\ti{1}, M^-\ti{2} \right\rbrace &= \gamma \Bigl( D\ti{2} R^{++}\ti{12} D^{-1}\ti{2} M^+\ti{1}M^-\ti{2} - M^+\ti{1}M^-\ti{2} D^{-1}\ti{2}R^{++}\ti{12}D\ti{2} \Bigr), \label{PBMpMm}\\
\left\lbrace M^\pm\ti{1}, M^\pm\ti{2} \right\rbrace &= \gamma \Bigl( \left[ R\ti{12}, M^\pm\ti{1}M^\pm\ti{2} \right] \mp c \,M^\pm\ti{1} H\ti{12} M^\pm\ti{2} \pm c\,M^\pm\ti{2} H\ti{12} M^\pm\ti{1} \Bigr). \label{PBMM} 
\end{align}
\end{subequations}
where we have introduced (see Appendix \ref{App:StandardFinite})
\begin{subequations}
\begin{align}
R^{++}\ti{12} &= R^+\ti{12}-c\,H\ti{12} = 2c\sum_{\alpha > 0} E_\alpha \otimes F_{\alpha}, \\
R^{--}\ti{12} & = R^-\ti{12}+c\,H\ti{12} = - 2c\sum_{\alpha > 0} F_{\alpha} \otimes E_\alpha.
\end{align}
\end{subequations}
We also made use of the following identity, valid for any $h\in H$ and $\epsilon\in\lbrace \emptyset, +, ++, -, -- \rbrace$,
\begin{equation*}
h\ti{1}h\ti{2}R^\epsilon\ti{12}h\ti{1}^{-1}h\ti{2}^{-1}=R^\epsilon\ti{12}.
\end{equation*}
From the Poisson bracket \eqref{PBDD}, one simply finds
\begin{equation}\label{PBQHQH}
\left\lbrace Q^H_i, Q^H_j \right\rbrace = 0.
\end{equation}

\subsubsection[Poisson bracket between $Q^H_i$ and $Q^E_\beta$]{Poisson bracket between $\bm{Q^H_i}$ and $\bm{Q^E_\beta}$}
\label{SubSubSec:QHQE}

The partial Casimir tensor $H\ti{12}$ on the Cartan subalgebra can be expressed in terms of the dual bases of weights $P_i$ and co-roots $\ach_i$ (cf. Appendices \ref{App:BasesCartan} and \ref{App:Casimir}) as
\begin{equation*}
H\ti{12} = \sum_{i=1}^\ell P_i \otimes \ach_i.
\end{equation*}
This allows us to extract the Poisson bracket between $Q^H_i$ and $M^\pm$ by projecting equation \eqref{PBDM} onto $P_i$ in the first tensor factor, namely
\begin{equation*}
i \lbrace Q^H_i, M^\pm \rbrace = \left[ \ach_i, M^\pm \right].
\end{equation*}

 We will now treat the bracket with $M^+$, the case of $M^-$ being similar. We introduce
\begin{equation*}
w_{(k)} = u_{(k)} \ldots u_{(n)},
\end{equation*}
such that $M^+=w_{(1)}$ and $w_{(k)}=u_{(k)}w_{(k+1)}$. Using this decomposition and Theorem \ref{TheoremPB1}, one shows by induction on $k$ that, for every $k\in\lbrace1,\ldots,n\rbrace$, we have
\begin{subequations}
\begin{align}
i u_{(k)}^{-1} \lbrace Q^H_i, u_{(k)} \rbrace &= u_{(k)}^{-1} \ach_i u_{(k)} - \ach_i, \label{PBQHu}\\
i \lbrace Q^H_i, w_{(k)} \rbrace w_{(k)}^{-1} &= \ach_i - w_{(k)} \ach_i w_{(k)}^{-1}. 
\end{align}
\end{subequations}
This induction relies on the fact that for any $k$, the adjoint action of $w_{(k+1)}$ on $\ach_i$ only creates nilpotent generators $E_\gamma$ coresponding to roots of the form $\gamma=a_{k+1}\beta_{k+1} + \ldots + a_n\beta_n$, with $a_{k+1},\ldots,a_n \in \N$. One can show from the ordering condition \eqref{ConditionOrder} that these roots are always strictly superior to the root $\beta_k$, which allows to perform the projection needed in Theorem \ref{TheoremPB1}. Using the definition \eqref{DefUV} of $u_{(k)}$, equation \eqref{PBQHu} becomes
\begin{equation*}
i \lbrace Q^H_i, Q^E_{\beta_k} \rbrace = \beta_k(\ach_i) Q^E_{\beta_k}.
\end{equation*}
Applying the same method to the Poisson bracket with $M^-$, we find that this equation holds for any root $\beta$, positive or negative. In the case of a simple root (or its opposite) $\beta=\pm\alpha_j$, we have $\beta(\ach_i)=\pm\alpha_j(\ach_i)=\pm a_{ij}$ (cf. appendix \ref{App:BasesCartan}). We therefore obtain
\begin{equation}\label{PBQEQH}
i \lbrace Q^H_i, Q^E_{\pm\alpha_j} \rbrace = \pm a_{ij} Q^E_{\pm\alpha_j}.
\end{equation}

\subsubsection[Poisson bracket between $Q^E_{\alpha_i}$ and $Q^E_{-\alpha_j}$]{Poisson bracket between $\bm{Q^E_{\alpha_i}}$ and $\bm{Q^E_{-\alpha_j}}$}

Fixing two simple roots $\alpha_i$ and $\alpha_j$, we want to compute the Poisson bracket between $Q^E_{\alpha_i}$ and $Q^E_{-\alpha_j}$. Recall that $\alpha_i=\beta_{k_i}$ and $\alpha_j=\beta_{k_j}$. Considering the decomposition \eqref{DecompoUV} of $M^\pm$, we need to extract the Poisson bracket of $u_{(k_i)}$ with $v_{(k_j)}$. Define
\begin{align*}
x &=u_{(1)} \ldots u_{(k_i)}, & \tilde{x} &= v_{(1)} \ldots v_{(k_j)},\\
y &=u_{(k_i+1)} \ldots u_{(n)}, & \tilde{y} &= v_{(k_j+1)} \ldots v_{(n)}.
\end{align*}
By Theorem \ref{TheoremPB2}, applied on both tensor factors, we may write
\begin{equation*}
u_{(k_i)\,}^{-1}\null\ti{1} v_{(k_j)\,}^{-1}\null\ti{2} \left\lbrace u_{(k_i)\,}\null\ti{1}, v_{(k_j)\,}\null\ti{2} \right\rbrace = \pi_{k_i} \otimes \pi_{-k_j} \bigl( \Pc\ti{12} \bigr),
\end{equation*}
where
$\Pc\ti{12} = x^{-1}\ti{1}\tilde{x}^{-1}\ti{2} \left\lbrace M^+\ti{1}, M^-\ti{2} \right\rbrace y^{-1}\ti{1} \tilde{y}^{-1}\ti{2}$.
On the other hand, from equation \eqref{PBMpMm} we find
\begin{equation*}
\Pc\ti{12} = \gamma \Bigl( x^{-1}\ti{1}\tilde{x}^{-1}\ti{2} D\ti{2} R^{++}\ti{12} D^{-1}\ti{2} x\ti{1}\tilde{x}\ti{2} - y\ti{1} \tilde{y}\ti{2} D^{-1}\ti{2}R^{++}\ti{12}D\ti{2} y^{-1}\ti{1} \tilde{y}^{-1}\ti{2} \Bigr).
\end{equation*}
Recalling from \eqref{DefZ} that $D=\exp(Z)$ with $Z\in\h$, we have
\begin{equation*}
D^{\pm 1}\ti{2}R^{++}\ti{12}D^{\mp 1}\ti{2} = 2c \sum_{\alpha > 0} \exp\bigl(\mp \alpha(Z)\bigr) E_\alpha \otimes E_{-\alpha},
\end{equation*}
so that
\begin{equation*}
\Pc\ti{12} = 2c \gamma \sum_{\alpha > 0}  \Bigl( \exp\bigl(- \alpha(Z)\bigr) \left( x^{-1} E_\alpha x \right) \otimes \left( \tilde{x}^{-1} E_{-\alpha} \tilde{x} \right) - \exp\bigl(\alpha(Z)\bigr) \left( y E_\alpha y^{-1} \right) \otimes \left( \tilde{y} E_{-\alpha} \tilde{y}^{-1} \right) \Bigr).
\end{equation*}
The adjoint action of any $E_\beta$ (appearing in $x$ or $y$) on $E_\alpha$ cannot create the simple root generator $E_{\alpha_i}$ and similarly for $E_{-\alpha_j}$ on the second space. It follows that
\begin{equation*}
\pi_{k_i} \otimes \pi_{-k_j} \bigl( \Pc\ti{12} \bigr) = 2c \gamma \delta_{ij} \Bigl( \exp\bigl(- \alpha_i(Z)\bigr) - \exp\bigl(\alpha_i(Z)\bigr) \Bigr)  E_{\alpha_i} \otimes E_{-\alpha_j}.
\end{equation*}
Yet, by definition \eqref{DefUV} of the $u_{(k)}$'s and $v_{(k)}$'s, we find
\begin{equation*}
\pi_{k_i} \otimes \pi_{-k_j} \bigl( \Pc\ti{12} \bigr) = u_{(k_i)\,}^{-1}\null\ti{1} v_{(k_j)\,}^{-1}\null\ti{2} \left\lbrace u_{(k_i)\,}\null\ti{1}, v_{(k_j)\,}\null\ti{2} \right\rbrace = c^2 \gamma^2 A_{\alpha_i} A_{-\alpha_j} \left\lbrace Q^E_{\alpha_i}, Q^E_{-\alpha_j} \right\rbrace E_{\alpha_i} \otimes E_{-\alpha_j},
\end{equation*}
so that
\begin{equation*}
i \big\{ Q^E_{\alpha_i}, Q^E_{-\alpha_j} \big\} = \frac{2i}{c\gamma A_{\alpha_i} A_{-\alpha_j}} \delta_{ij} \Bigl( \exp\bigl(-\alpha_i(Z)\bigr) - \exp\bigl(\alpha_i(Z)\bigr) \Bigr).
\end{equation*}
From equation \eqref{DefZ}, one has (cf. appendix \ref{App:BasesCartan})
\begin{equation*}
\alpha_i(Z) = ic\gamma \sum_{k=1}^\ell Q_k^H \alpha_i(P_k) = i c\gamma \sum_{k=1}^\ell Q_k^H d_i \delta_{ik} = i c\gamma d_i Q^H_i.
\end{equation*}
Introducing the deformation parameter
\begin{equation}\label{Defq}
q=e^{-ic\gamma},
\end{equation}
we therefore have
\begin{equation*}
i \big\{ Q^E_{\alpha_i}, Q^E_{-\alpha_j} \big\} = \frac{2i}{\gamma c A_{\alpha_i} A_{-\alpha_j}} \delta_{ij} \Bigl( q^{d_iQ^H_i}-q^{-d_iQ^H_i} \Bigr).
\end{equation*}
Finally, if we fix the normalisation $A_{\pm \alpha}$ for simple roots as
\begin{equation}\label{Aalphai}
A_{\pm\alpha_i} = \left( \frac{4 \sinh(i c \gamma d_i)}{ic \gamma} \right)^{\frac{1}{2}},
\end{equation}
then we may rewrite the above Poisson brackets as
\begin{equation}\label{PBQEpQEm}
i \big\{ Q^E_{\alpha_i}, Q^E_{-\alpha_j} \big\} = \delta_{ij} \frac{q^{d_iQ^H_i}-q^{-d_iQ^H_i}}{q^{d_i}-q^{-d_i}}.
\end{equation}

\subsubsection[$q$-Poisson-Serre relations]{$\bm{q}$-Poisson-Serre relations}
\label{SubSubSec:qSerre}

The Poisson brackets \eqref{PBQHQH}, \eqref{PBQEQH} and \eqref{PBQEpQEm} obtained so far are part of the defining relations of the semiclassical limit $\U_q(\g_0)$ of the quantum group $U_{\widehat q}(\g_0)$ with $\widehat q = q^\hbar$, as given in \cite{Delduc:2013fga} (see also~\cite{Ballesteros_2009}). The complete set of relations characterising $\U_q(\g_0)$ also includes the so-called $q$-Poisson-Serre relations. The purpose of the present subsection is to derive these from the Poisson bracket \eqref{PBMM}. We will only treat the case of positive roots, the negative one being handled similarly.

\paragraph{Poisson brackets of $\bm{Q^E_{\alpha_i}}$ with $\bm{M^+}$.}
Let us fix a simple root $\alpha_i$. We recall that $\alpha_i=\beta_{k_i}$, so that $Q^E_{\alpha_i}$ is to be extracted from $u_{(k_i)}$. Introduce
\begin{align*}
x &= u_{(1)} \ldots u_{(k_i)}, \\
y &= u_{(k_i+1)} \ldots u_{(n)},
\end{align*}
so that $M^+=xy$. By Theorem \ref{TheoremPB2} we have
\begin{equation*}
\left(u_{(k_i)}\right)\ti{1}^{-1} \left\lbrace u_{(k_i)}\null\ti{1}, M^+\ti{2} \right\rbrace = \left(\pi_{k_i}\right)\ti{1} \bigl( \Pc\ti{12} \bigr),
\end{equation*}
where $\Pc\ti{12} = x^{-1}\ti{1} \left\lbrace M^+\ti{1}, M^+\ti{2} \right\rbrace y^{-1}\ti{1}$.
On the other hand, from \eqref{PBMM} we have
\begin{equation*}
\Pc\ti{12} = \gamma \Pc^R\ti{12} + \gamma\Pc^H\ti{12} 
\end{equation*}
with
\begin{align*}
\Pc^R\ti{12} &= x^{-1}\ti{1} R^{++}\ti{12} x\ti{1} M^+\ti{2} - M^+\ti{2} y\ti{1} R^{++}\ti{12} y^{-1}\ti{1}, \\
\Pc^H\ti{12} &= c\left( x^{-1}\ti{1} H\ti{12} x\ti{1} - y\ti{1}H\ti{12}y\ti{1}^{-1} \right) M^+\ti{2} + c\,M^+\ti{2} \left( x^{-1}\ti{1} H\ti{12} x\ti{1} - y\ti{1}H\ti{12}y\ti{1}^{-1} \right).
\end{align*}
By writing  $H\ti{12}=\displaystyle \sum_{j=1}^\ell \och_j\otimes H_j$ (cf. appendices \ref{App:BasesCartan} and \ref{App:Casimir}), these can be rewritten
\begin{align*}
\Pc^R\ti{12} &= 2c\sum_{\alpha > 0} \bigl( x^{-1} E_\alpha x \bigr) \otimes \bigl( E_{-\alpha} M^+ \bigr) - 2c\sum_{\alpha > 0} \bigl( yE_\alpha y^{-1} \bigr) \otimes \bigl( M^+ E_{-\alpha} \bigr), \\
\Pc^H\ti{12} &= c\sum_{j=1}^\ell \bigl(x^{-1} \och_j x - y \och_j y^{-1} \bigr) \otimes \bigl( H_j M^+ + M^+ H_j \bigr).
\end{align*}
The adjoint action of any $E_\beta$ (appearing in $x$ or $y$) on $E_\alpha$ cannot create the simple root generator $E_{\alpha_i}$. Thus, we have
\begin{equation*}
\left(\pi_{k_i}\right)\ti{1} \bigl( \Pc^R\ti{12} \bigr) = 2c \, E_{\alpha_i} \otimes \bigl( E_{-\alpha_i} M^+ -  M^+ E_{-\alpha_i} \bigr).
\end{equation*}
In the same way, in the adjoint actions of $E_\beta$'s from $x$ or $y$ on $\och_j$, only a unique adjoint action of $E_{\alpha_i}$, coming from $u_{(k_i)}$ in $x$, can create the simple root generator $E_{\alpha_i}$. Therefore
\begin{equation*}
\pi_{k_i} \bigl(x^{-1} \och_j x - y \och_j y^{-1} \bigr) = -ic\gamma A_{\alpha_i} Q^E_{\alpha_i} \ad_{E_{\alpha_i}} \bigl( \och_j\bigr) = ic\gamma A_{\alpha_i} Q^E_{\alpha_i} \delta_{ij} E_{\alpha_i},
\end{equation*}
and hence
\begin{equation*}
\left(\pi_{k_i}\right)\ti{1} \bigl( \Pc^H\ti{12} \bigr) = ic^2 \gamma A_{\alpha_i} Q^E_{\alpha_i} E_{\alpha_i} \otimes \bigl( H_i M^+ + M^+ H_i \bigr).
\end{equation*}
Putting together all the above we arrive at
\begin{equation*}
\left(u_{(k_i)}\right)\ti{1}^{-1} \left\lbrace u_{(k_i)}\null\ti{1}, M^+\ti{2} \right\rbrace = c\gamma E_{\alpha_i} \otimes \Bigl( 2\bigl[ E_{-\alpha_i}, M^+ \bigr] + ic\gamma A_{\alpha_i} Q^E_{\alpha_i} \bigl( H_i M^+ + M^+ H_i \bigr) \Bigr).
\end{equation*}
Yet, by definition of $u_{(k_i)}$ in \eqref{DefUV} we have
\begin{equation*}
\left(u_{(k_i)}\right)\ti{1}^{-1} \left\lbrace u_{(k_i)}\null\ti{1}, M^+\ti{2} \right\rbrace = ic\gamma A_{\alpha_i} E_{\alpha_i} \otimes \left\lbrace Q^E_{\alpha_i}, M^+ \right\rbrace,
\end{equation*}
and hence
\begin{equation}\label{PBQMp}
i A_{\alpha_i} \left\lbrace Q^E_{\alpha_i}, M^+ \right\rbrace = 2\bigl[ E_{-\alpha_i}, M^+ \bigr] + ic\gamma A_{\alpha_i} Q^E_{\alpha_i} \bigl( H_i M^+ + M^+ H_i \bigr)
\end{equation}

\paragraph{$\bm{\alpha_i}$-string through $\bm{\alpha_j}$.}
Let us now consider another simple root $\alpha_j$. We suppose here that $\alpha_i > \alpha_j$. The $\alpha_i$-string through $\alpha_j$ is then contained between $\alpha_j$ and $\alpha_i$. Specifically, we have
\begin{equation*}
\alpha_j < \alpha_j + \alpha_i < \ldots < \alpha_j - a_{ij} \alpha_i <  \alpha_i,
\end{equation*}
with $a$ the Cartan matrix of $\g$ (cf. appendix \ref{App:BasesCartan}). Let $r \in \lbrace 0, \ldots, -a_{ij} \rbrace$ and $p \in \lbrace 1,\ldots,n \rbrace$ be such that
\begin{equation*}
\beta_p = \alpha_j + r \alpha_i,
\end{equation*}
We define
\begin{align*}
x &= u_{(1)} \ldots u_{(p)}, \\
y &= u_{(p+1)} \ldots u_{(n)},
\end{align*}
and
$\Q = x^{-1} \left\lbrace Q_{\alpha_i}^E, M^+ \right\rbrace y^{-1}$.
By Theorem \ref{TheoremPB2}, we have
\begin{equation} \label{pip up}
u_{(p)}^{-1} \left\lbrace Q^E_{\alpha_i}, u_{(p)} \right\rbrace = \pi_p (\Q).
\end{equation}
On the other hand, from the Poisson bracket \eqref{PBQMp} we get
\begin{equation}\label{Qc}
i A_{\alpha_i} \Q = 2 \bigl( x^{-1} E_{-\alpha_i} x - y E_{-\alpha_i} y^{-1} \bigr) + i c\gamma A_{\alpha_i} Q^E_{\alpha_i} \bigl( x^{-1} H_i x + y H_i y^{-1} \bigr).
\end{equation}

The projection onto $E_{\beta_p}$ of the terms involving $H_i$ on the right hand side of \eqref{Qc} can be computed as follows. We note that $y H_i y^{-1}$ is composed of nilpotent generators $E_\beta$ with $\beta$ a sum of roots superior to $\beta_p$, which therefore cannot be $\beta_p$. In the same way, the adjoint action of $x$ on $H_i$ creates nilpotent generators $E_\beta$ with $\beta$ a sum of roots inferior or equal to $\beta_p$. Such $\beta$ can be either strictly inferior to $\beta_p$ or $\beta_p$ itself. Therefore the only way to have $E_{\beta_p}$ in $x^{-1}H_i x$ is by the simple adjoint action on $H_i$ of the generator $E_{\beta_p}$ (appearing in $u_{(p)}$). Thus, we have
\begin{equation}\label{ProjH}
\pi_p \bigl( x^{-1} H_i x + y H_i y^{-1} \bigr) = -ic\gamma A_{\beta_p} Q^E_{\beta_p} \ad_{E_{\beta_p}}\bigl(H_i\bigr) = ic\gamma A_{\beta_p} Q^E_{\beta_p} (\alpha_i,\beta_p) E_{\beta_p}.
\end{equation}

Next, consider the term $x^{-1} E_{-\alpha_i} x - y E_{-\alpha_i} y^{-1}$ on the right hand side of \eqref{Qc}. It is composed of generators $E_\beta$, with $\beta=\gamma-\alpha_i$ and $\gamma$ a sum of roots either all inferior or equal to $\beta_p$ (for $x$) or all superior (for $y$). We want to project this on $E_{\beta_p}$. Yet, having $\beta=\beta_p$ requires $\gamma=\beta_p+\alpha_i$. As $\beta_p < \alpha_i$, this means that $\beta_p < \gamma$, hence $\gamma$ comes from the adjoint action of $y$. To be more precise, $yE_{-\alpha_i}y^{-1}$ is composed of elements of the form (up to prefactors)
\begin{equation*}
\ad^{a_{p+1}}_{E_{\beta_{p+1}}} \ldots \ad^{a_n}_{E_{\beta_n}} \bigl( E_{-\alpha_i} \bigr), \; \; \; \text{ with } a_{p+1},\ldots,a_n \in \N.
\end{equation*}
Such a term is proportional to $E_{\gamma-\alpha_i}$, with $\gamma=a_{p+1}\beta_{p+1}+\ldots+a_n\beta_n$. In order to get $E_{\beta_p}$, one must have $\gamma=\alpha_i+\beta_p=\alpha_j+(r+1)\alpha_i$. Therefore, we want to solve
\begin{equation*}
a_{p+1}\beta_{p+1}+\ldots+a_n\beta_n = \alpha_j+(r+1)\alpha_i,
\end{equation*}
with $a_{p+1},\ldots,a_n$ non-negative integers. If a root $\beta_q$, for $q>p$, contains a simple root $\alpha_k$ different from $\alpha_i$ and $\alpha_j$, it is clear from the equation above that $a_q$ must be zero, as $\alpha_k$ does not appear in the right hand side of the equation.\\
\indent Moreover, the only roots superior to $\beta_p$ and containing only $\alpha_i$ and $\alpha_j$ as simple roots are $\alpha_j+(r+1)\alpha_i, \alpha_j+(r+2)\alpha_i, \ldots, \alpha_j-A_{ij}\alpha_i$ and $\alpha_i$. The only way that a non-negative integer linear combination of these roots can give $\alpha_j+(r+1)\alpha_i$ is if all the coefficients are zero except for that of the root $\alpha_j+(r+1)\alpha_i$ itself. Thus, the projection of $x^{-1} E_{-\alpha_i} x - y E_{-\alpha_i} y^{-1}$ onto $E_{\beta_p}$ comes from the simple adjoint action of $E_{\alpha_j+(r+1)\alpha_i}$ on $E_{-\alpha_i}$ (if $\alpha_j+(r+1)\alpha_i$ is a root). Hence
\begin{equation}\label{ProjF}
\begin{split}
\pi_p \bigl( x^{-1} E_{-\alpha_i} x - y E_{-\alpha_i} y^{-1} \bigr) &= - ic\gamma A_{\alpha_j+(r+1)\alpha_i} Q^E_{\alpha_j+(r+1)\alpha_i} \bigl[ E_{\alpha_j+(r+1)\alpha_i}, E_{-\alpha_i} \bigr] \\
 &= - ic\gamma A_{\alpha_j+(r+1)\alpha_i} Q^E_{\alpha_j+(r+1)\alpha_i} N_{\beta_p,\alpha_i} E_{\beta_p},
\end{split}
\end{equation}
if $\alpha_j+(r+1)\alpha_i$ is a root, and is zero otherwise.

Applying $\pi_p$ to \eqref{Qc} and using the results \eqref{ProjH} and \eqref{ProjF} gives
\begin{equation*}
iA_{\alpha_i} \pi_p(\Q) = -2ic\gamma A_{\alpha_j+(r+1)\alpha_i} N_{\beta_p,\alpha_i} Q^E_{\alpha_j+(r+1)\alpha_i} E_{\beta_p} + (ic\gamma)^2 A_{\alpha_i} A_{\beta_p} (\alpha_i,\beta_p) Q^E_{\alpha_i} Q^E_{\beta_p} E_{\beta_p}.
\end{equation*}
Yet from \eqref{pip up} together with the definition of $u_{(p)}$ in \eqref{DefUV} we have
\begin{equation*}
\pi_p (\Q) = u_{(p)}^{-1} \big\{ Q^E_{\alpha_i}, u_{(p)} \big\} = ic\gamma A_{\beta_p} \big\{ Q^E_{\alpha_i}, Q^E_{\beta_p} \big\} E_{\beta_p},
\end{equation*}
and hence
\begin{equation*}
\big\{ Q^E_{\alpha_i}, Q^E_{\beta_p} \big\} = \frac{ A_{\alpha_j+(r+1)\alpha_i}}{A_{\alpha_i} A_{\beta_p}} 2 i N_{\beta_p,\alpha_i} Q^E_{\alpha_j+(r+1)\alpha_i} + c \gamma (\alpha_i,\beta_p) Q^E_{\alpha_i} Q^E_{\beta_p}.
\end{equation*}
We define the $q$-bracket of two charges associated with the positive roots $\alpha$ and $\beta$ as
\begin{equation*}
\big\{ Q^E_\alpha, Q^E_\beta \big\}_q = \left\lbrace Q^E_\alpha, Q^E_\beta \right\rbrace + c\gamma (\alpha,\beta) Q^E_\alpha Q^E_\beta.
\end{equation*}
Moreover, if we fix the normalisation constant $A_\alpha$ for $\alpha$ in the $\alpha_i$-string through $\alpha_j$ as
\begin{equation*}
A_{\alpha_j+r\alpha_i} = A_{\alpha_j}A_{\alpha_i}^r,
\end{equation*}
then we deduce that
\begin{equation*}
\big\{ Q^E_{\alpha_j+r\alpha_i}, Q^E_{\alpha_i} \big\}_q =   2i N_{\alpha_i,\alpha_j+r\alpha_i} Q^E_{\alpha_j+(r+1)\alpha_i},
\end{equation*}
if $\alpha_j+(r+1)\alpha_i$ is a root and is zero otherwise.

By induction, we get the $q$-Poisson-Serre relation
\begin{equation}
\bigl\lbrace \bigl\lbrace \ldots \bigl\lbrace Q_{\alpha_j}^E, \underbrace{Q_{\alpha	_i}^E \bigr\rbrace_q, \ldots Q_{\alpha_i}^E \bigr\rbrace_q, Q_{\alpha_i}^E}_{1-a_{ij} \text{ times}} \bigr\rbrace_q = 0.
\end{equation}
One can treat the case $\alpha_i < \alpha_j$ in a similar way. For that, one needs to use a slightly different version of Theorem \ref{TheoremPB2}, involving the quantity
$\left( u_{(1)} \ldots u_{(p-1)} \right)^{-1} \left\lbrace Q^E_{\alpha_i}, M^+ \right\rbrace \left( u_{(p)} \ldots u_{(n)} \right)^{-1}$ instead of
$\left( u_{(1)} \ldots u_{(p)} \right)^{-1} \left\lbrace Q^E_{\alpha_i}, M^+ \right\rbrace \left( u_{(p+1)} \ldots u_{(n)} \right)^{-1}$. This yields the $q$-Poisson-Serre relation
\begin{equation}
\bigl\lbrace \underbrace{Q_{\alpha_i}^E, \bigl\lbrace Q_{\alpha_i}^E, \ldots, \bigl\lbrace Q^E_{\alpha_i}}_{1-a_{ij} \text{ times}}, Q^E_{\alpha_j} \bigr\rbrace_q \ldots \bigr\rbrace_q \bigr\rbrace_q = 0.
\end{equation}

Applying the same method as above to the Poisson bracket in \eqref{PBMM} involving $M^-$, one finds that the charges $Q^E_{-\alpha_i}$ also verifiy $q$-Poisson-Serre relations, but with respect to the deformed bracket $\lbrace\cdot,\cdot\rbrace_{q^{-1}}$, defined for two negative roots $\alpha$ and $\beta$ as
\begin{equation*}
\big\{ Q_\alpha^E, Q_\beta^E \big\}_{q^{-1}} = \big\{ Q^E_\alpha, Q^E_\beta \big\} - c\gamma (\alpha,\beta) Q^E_\alpha Q^E_\beta.
\end{equation*}

\subsubsection{Reality conditions}
\label{SubSubSec:RealCond}

The Poisson brackets \eqref{PBQHQH}, \eqref{PBQEQH} and \eqref{PBQEpQEm}, together with the $q$-Poisson-Serre relations stated above are the defining Poisson bracket relations of the semiclassical limit $\U_q(\g_0)$ of the quantum group $U_{\widehat q}(\g_0)$ with $\widehat q = q^\hbar$. It only remains to check that the required reality conditions are verified by the charges $Q^H_i$ and $Q^E_\alpha$. We shall address this question in the present subsection, first in the split case and then in the non-split one.

\paragraph{Split case.}
When $c=1$, the deformation parameter \eqref{Defq} becomes
\begin{equation*}
q=e^{-i\gamma},
\end{equation*}
so that $|q|=1$, \textit{i.e.} $q$ is a phase. Now the moment map $(\Gamma^+,\Gamma^-)$ takes values in the real double $G_0 \times G_0$, therefore the reality condition is simply
\begin{equation*}
\tau(\Gamma^\pm)=\Gamma^\pm, 
\end{equation*}
with $\tau$ the split semi-linear automorphism described in appendix \ref{App:RealSS} and lifted to the complexified group $G$. As $\tau$ stabilises the Cartan and the nilpotent subgroups and since the decomposition \eqref{DefDM} is unique, one has
\begin{equation*}
\tau(D)=D \; \; \; \; \text{ and } \; \; \; \; \tau(M^\pm)=M^\pm.
\end{equation*}
We recall (cf. appendix \ref{App:RealSS}) that in the split case $\tau(P_i)=P_i$ for $i\in\lbrace 1,\ldots,\ell \rbrace$ (since $P_i$ is a real linear combination of the $H_j$) and $\tau(E_\alpha)=E_\alpha$ for any root $\alpha$. Considering the extraction of charges \eqref{DecompoD} and \eqref{DecompoM} with $c=1$, the above reality condition gives
\begin{align*}
\overline{Q^H_i} &= -Q^H_i, \\
\overline{A_{\pm\alpha_i} Q^E_{\pm\alpha_i}} &= -A_{\pm\alpha_i}Q^E_{\pm\alpha_i}.
\end{align*}
The normalisation constants $A_{\pm\alpha_i}$ are given by equation \eqref{Aalphai}, which in the split case reads
\begin{equation*}
A_{\pm\alpha_i} = \left( \frac{4 \sin(\gamma d_i)}{\gamma} \right)^{\frac{1}{2}}.
\end{equation*}
We will restrict attention to the case where
\begin{equation*}
-\pi \leq \gamma d_i \leq \pi,
\end{equation*}
for any $i$, so that the $A_{\pm\alpha_i}$ are real numbers. As a result, the reality conditions are simply
\begin{equation}
|q|=1, \; \; \; \; \overline{Q^H_i} = -Q^H_i \; \; \; \; \text{and} \; \; \; \; \overline{Q^E_{\pm\alpha_i}} = -Q^E_{\pm\alpha_i}.
\end{equation}
These are the reality conditions for the split real form $\U_q(\g)$, which correspond precisely to the semiclassical counterpart of the reality conditions on $U_{\widehat q}(\g_0)$ as given in \cite{Ruegg:1993eq}.

\paragraph{Non-split case.}
For $c=i$, the deformation parameter \eqref{Defq} now reads
\begin{equation*}
q=e^\gamma,
\end{equation*}
which is a real number. As explained in subsection \ref{SubSec:PL2STS}, the two moment maps $\Gamma^+$ and $\Gamma^-$ are not independent. They are related by the reality condition
\begin{equation*}
\tau(\Gamma^\pm)=\Gamma^\mp,
\end{equation*}
where $\tau$ is the non-split semi-linear automorphism described in appendix \ref{App:RealSS}, lifted to the complexified group $G$. Since $\tau$ stabilises the Cartan subgroup $H$ but exchanges the unipotent subgroups $N_\pm$, applying $\tau$ to the decomposition \eqref{DefDM} we get
\begin{align*}
\tau(\Gamma^+) &= \tau(M^+)\tau(D)=\underbrace{\tau(D)}_{\in H} \underbrace{\tau(D)^{-1}\tau(M^+)\tau(D)}_{\in N_-},\\
\tau(\Gamma^-) &= \tau(D)^{-1}\tau(M^-) = \underbrace{\tau(D)^{-1}\tau(M^-)\tau(D)}_{\in N_+} \underbrace{\tau(D)^{-1}}_{\in H},
\end{align*}
where we used the fact that an adjoint action of a Cartan element on a element of $N_\pm$ is still in $N_\pm$. Equating $\tau(\Gamma^+)$ with $\Gamma^-=D^{-1}M^-$ and $\tau(\Gamma^-)$ with $\Gamma^+=M^+D$ we obtain
\begin{equation} \label{theta D M}
\tau(D) = D^{-1} \; \; \; \; \text{ and } \; \; \; \; \tau(M^\pm) = D^{-1} M^\mp D.
\end{equation}
Recall (cf. appendix \ref{App:RealSS}) that $\tau(P_i)=-P_i$. Using the extraction of Cartan charges \eqref{DecompoD}, with $c=i$, we find
\begin{equation*}
\overline{Q^H_i}=Q^H_i.
\end{equation*}
From the decomposition \eqref{DecompoM}, we have
\begin{equation*}
D^{-1}M^\pm D = \prod_{k=1}^n \exp\left( \mp \gamma A_{\pm\beta_k} Q^E_{\pm\beta_k} D^{-1} E_{\pm\beta_k} D \right).
\end{equation*}
Moreover, since $D=\exp(Z)$ since $Z$ defined in \eqref{DefZ},
\begin{equation*}
D^{-1} E_{\pm\beta_k} D = \exp\bigl(\mp \beta_k(Z)\bigr) E_{\pm\beta_k} = \exp\bigl(\pm \gamma \sum_{j=1}^\ell Q_j^H \beta_k(P_j)\bigr)E_{\pm\beta_k}.
\end{equation*}
In particular, for $k=k_i$, \textit{i.e.} for $\beta_k$ the simple root $\alpha_i$, using \eqref{AlphaP} we get
\begin{equation*}
D^{-1} E_{\pm\alpha_i} D = q^{\pm d_i Q_i^H}E_{\pm\alpha_i}.
\end{equation*}
The term corresponding to the simple root $\alpha_i$ in $D^{-1}M^\mp D$ therefore reads
\begin{equation*}
\exp \Bigl( \pm \gamma A_{\mp\alpha_i} q^{\mp d_i Q^H_i} Q^E_{\mp\alpha_i} E_{\mp\alpha_i} \Bigr).
\end{equation*}
Yet we have $\tau(E_{\pm\alpha_i})=-\lambda_i E_{\mp\alpha_i}$, so that the corresponding term in $\tau(M^\pm)$ is
\begin{equation*}
\exp \Bigl( \pm \gamma \overline{A_{\pm\alpha_i} Q^E_{\pm\alpha_i}} \lambda_i E_{\mp\alpha_i} \Bigr).
\end{equation*}
It now follows from the second equality in \eqref{theta D M} that
\begin{equation*}
\overline{A_{\pm\alpha_i} Q^E_{\pm\alpha_i}} = \lambda_i q^{\mp d_i Q^H_i} A_{\mp\alpha_i} Q^E_{\mp\alpha_i}.
\end{equation*}
The normalisation constants $A_{\pm\alpha_i}$ are given by \eqref{Aalphai}, which for $c=i$ take the form
\begin{equation*}
A_{\pm\alpha_i} = \left( \frac{4 \sinh(\gamma d_i)}{\gamma} \right)^{\frac{1}{2}}
\end{equation*}
and are therefore real numbers. Hence, the reality conditions are
\begin{equation}
q\in\mathbb{R}, \; \; \; \; \overline{Q^H_i}=Q^H_i \; \; \; \; \text{and} \; \; \; \; \overline{ Q^E_{\pm\alpha_i}}=\lambda_i q^{\mp d_i Q^H_i} Q^E_{\mp\alpha_i}.
\end{equation}
According to~\cite{Ruegg:1993eq}, these are the reality conditions of the non-split real form $\U_q(\g_0)$.

\subsection[Sklyanin bracket and $\U_q(\g^*)$ algebra]{Sklyanin bracket and $\bm{\U_q(\g_0^*)}$ algebra}

As in subsection \ref{subsec: Sklyanin bracket}, in what follows we consider only the split case. We start from the Poisson bracket \eqref{PBU} with the central quantity $\widetilde{M}\ti{12}$ set to zero. In other words, $U$ satisfies the Sklyanin Poisson bracket
\begin{equation}\label{PBU2}
\left\lbrace U\ti{1}, U\ti{2} \right\rbrace = \gamma\left[ R\ti{12}, U\ti{1} U\ti{2} \right].
\end{equation}
Let us decompose $U$ as
\begin{equation*}
U=M^-DM^+,
\end{equation*}
with $D \in H$ and $M^\pm \in N_\pm$. Using Theorem \ref{TheoremPB1}, we can extract the Poisson brackets between $D$, $M^+$ and $M^-$ from equation \eqref{PBU2} to find
\begin{subequations}
\begin{align}
\left\lbrace D\ti{1}, D\ti{2} \right\rbrace &= 0,\\
\left\lbrace D\ti{1}, M^\pm\ti{2} \right\rbrace &= \pm c\gamma D\ti{1} \left[H\ti{12}, M^\pm\ti{2} \right], \label{PBDM2}\\
\left\lbrace M^+\ti{1}, M^-\ti{2} \right\rbrace &= 0, \label{PBMpMm2}\\\left\lbrace M^\pm\ti{1}, M^\pm\ti{2} \right\rbrace &= \gamma \Bigl( \left[ R\ti{12}, M^\pm\ti{1}M^\pm\ti{2} \right] \mp c\,M^\pm\ti{1} H\ti{12} M^\pm\ti{2} \pm c\,M^\pm\ti{2} H\ti{12} M^\pm\ti{1} \Bigr). \label{PBMM2}
\end{align}
\end{subequations}
We notice that these Poisson brackets are very similar to \eqref{PBDMM}, the main difference being that $M^+$ and $M^-$ Poisson commute in the present case.

The methods of subsection \ref{SubSec:Uqg} can be applied to this case. For each positive root $\alpha$, we extract the positive nilpotent charge $Q^E_\alpha$ from $M^+$ and negative nilpotent charge $Q^E_{-\alpha}$ from $M^-$ as we did before using the decomposition \eqref{DecompoM}. Likewise, we extract Cartan charges $Q^H_i$ from $D$ as we did in equation \eqref{DecompoD}.

From equation \eqref{PBMM2}, following same the procedure outline in subsection \ref{SubSubSec:qSerre}, we find that the nilpotent charges $Q^E_\alpha$ and $Q^E_{-\alpha}$ satisfy the $q$-Poisson-Serre relations. In other words, these charges span nilpotent $q$-deformed Poisson algebras $\U_q(\n_+)$ and $\U_q(\n_-)$. Moreover, it is clear from \eqref{PBMpMm2} that elements from these two algebras Poisson commute.

Similarly, we can apply the methods of subsection \ref{SubSubSec:QHQE} to the Poisson bracket \eqref{PBDM2}. Doing so, we find that, for any positive root $\alpha$,
\begin{equation*}
i \lbrace Q^H_i, Q^E_\alpha \rbrace = \alpha(\ach_i) Q^E_\alpha \; \; \; \; \text{ and } \; \; \; \; i \lbrace Q^H_i, Q^E_{-\alpha} \rbrace = \alpha(\ach_i) Q^E_{-\alpha}.
\end{equation*}
This implies that the Cartan charges $Q^H_i$ for $i=1,\ldots,n$ together with the charges $Q^E_\alpha$ for $\alpha > 0$ span a $q$-deformed positive Borel algebra $\U_q(\bo_+)$. In the same way, the charges $-Q^H_i$ for $i=1,\ldots,n$ together with the charges $Q^E_{-\alpha}$ for $\alpha > 0$ span a $q$-deformed negative Borel algebra $\U_q(\bo_-)$. The combination of all the charges $Q^H_i$ for $i=1,\ldots,n$ and $Q^E_\alpha$ for all roots $\alpha$ therefore span a $q$-deformed Poisson algebra which we could call $\U_q(\g_{DR})$.

Since $U$ takes value in the split real form $G_0$ we have $\tau(U)=U$, with $\tau$ the split semi-linear automorphism of appendix \ref{App:RealSS}. Moreover, since $\tau$ stabilises the subgroups $H$ and $N_\pm$, we deduce that $\tau(D)=D$ and $\tau(M^\pm)=M^\pm$. The reality conditions are then identical to those of the split case in the subsection \ref{SubSubSec:RealCond}, so that
\begin{equation}
|q|=1, \; \; \; \; \; \overline{Q^H_i}=-Q^H_i \; \; \; \; \text{and} \; \; \; \; \; \overline{Q^E_{\pm\alpha}}=-Q^E_{\pm\alpha}.
\end{equation}

\section{Application to Yang-Baxter type models} \label{sec: YB models}

\subsection{Yang-Baxter type models}

In this section we will apply the formalism of Poisson-Lie groups and non-abelian moment maps to discuss the symmetries of Yang-Baxter type models. The latter can be defined as the result of applying a general procedure for constructing integrable deformations of a broad family of integrable models. We refer to \cite{Vicedo:2015pna} for details on the general construction. To start this section, we will recall the main features of the construction, mostly based on the examples of Yang-Baxter type deformations presented in Chapter \ref{Chap:Models}: the Yang-Baxter model (deformation of the PCM, Subsection \ref{SubSec:YB}), the $\eta$-deformation of the $\Z_2$-coset model (Subsection \ref{SubSec:dZ2}) and the Bi-Yang-Baxter model (Section \ref{Sec:BYB}).

Yang-Baxter type deformations apply to models which possesses a twist function (see Chapter \ref{Chap:Lax}) with a double pole $\lambda_0\in\R$. To construct a Yang-Baxter type model we begin by modifying the twist function $\varphi(\lambda)$ of the undeformed model, by deforming its double pole at $\lambda_0$ to a pair of simple poles which we denote $\lambda_\pm$, while keeping all other poles and zeroes fixed. In order to preserve the reality condition \eqref{Eq:TwistReal} of the twist function, we should take either $\lambda_\pm$ both real or $\lambda_+$ and $\lambda_-$ complex conjugate of one another. We refer to these two cases as the real and complex branches respectively.
Finally, we require also that the deformed twist function $\varphi(\lambda)$ has opposite residues at the simple poles $\lambda_+$ and $\lambda_-$, which allows us to define a single real deformation parameter $\gamma \in \mathbb{R}$ by
\begin{equation*}
\frac{1}{\gamma} = 2cT \res_{\lambda=\lambda_+} \varphi(\lambda) \dd\lambda = - 2cT \res_{\lambda=\lambda_-} \varphi(\lambda) \dd\lambda,
\end{equation*}
with $c = 1$ in the real branch and $c = i$ in the complex branch, and with $T$ the order of cyclotomy of the model (see Section \ref{Sec:ModelsTwist}). This definition of $\gamma$ agrees with the ones given for examples of Yang-Baxter type models, namely in equations \eqref{Eq:GammaYB}, \eqref{Eq:LgdZ2} and \eqref{Eq:LgBYB}.

The phase space of Yang-Baxter type models contains a $G_0$-valued field $g$ and a $\g_0$-valued field $X$ satisfying the Poisson brackets \eqref{Eq:PBTstarG}. In particular, the evaluation of the (gauge transformed) Lax matrix of the model at the poles $\lambda_\pm$ is given by
\begin{equation}\label{Eq:Lgpm2}
\Lc^g(\lambda_\pm) = -\gamma R^\mp \bigl( gXg^{-1} \bigr),
\end{equation}
where $R^\pm=R \pm c\,\Id$, for a certain solution $R$ of the mCYBE on $\g$ (which is part of the definition of the Yang-Baxter deformation). This general statement was verified for the examples mentioned above in  equations \eqref{Eq:LgYB}, \eqref{Eq:LgdZ2}, \eqref{Eq:LgBYB}.

In this section, we will restrict to models whose space coordinate $x$ belongs to the real line $\R$ (and not the circle $\mathbb{S}^1$). Let us then introduce the transfer matrices of the $\g_0$-valued fields $\Lc^g(\lambda_\pm)$, given as the following path-order exponential (see Appendix \ref{App:Pexp}):
\begin{equation*}
T^g(\lambda_\pm\,;\,x,y) = \Pexp \left( - \int_{x}^{y} \Lc^g(\lambda_\pm,z)\,\dd z \right).
\end{equation*}
In particular, we will be interested in the monodromy matrices
\begin{equation*}
T^g(\lambda_\pm) = T^g(\lambda_\pm\,;\,-\infty,+\infty).
\end{equation*}
As $\Lc^g(\lambda_\pm)$ satisfies a zero curvature equation (because $\Lc$ satisfies the Lax equation \eqref{Eq:Lax}) and as the space coordinate is taken on the real line, the matrices $T^g(\lambda_\pm)$ are conserved quantities of the model (see Appendix \ref{App:Pexp}).

\subsection[Poisson brackets of $T^g(\lambda_\pm)$]{Poisson brackets of $\bm{T^g(\lambda_\pm)}$}

In this subsection, we will compute the Poisson bracket of $T^g(\lambda_\pm)$ with itself and with $T^g(\lambda_\mp)$. In general, when a Lax matrix obeys Poisson brackets of Maillet type, the Poisson bracket of its path-ordered exponential is ill-defined, due to the presence of non-ultralocal terms (the calculation then requires a regularisation procedure~\cite{Maillet:1985ek}). As we will see, this issue will not appear for the particular case of $T^g(\lambda_\pm)$.

Recall that $g$ and $X$ satisfy the Poisson brackets \eqref{Eq:PBTstarG}. For this section, we will use another parametrisation of the phase space of the model, by defining the fields
\begin{equation*}
h=g^{-1} \;\;\;\; \text{ and } \;\;\;\; Y = -gXg^{-1}
\end{equation*}
Let us motivate this choice. Considering $Y$ instead of $X$ is natural given the expression \eqref{Eq:Lgpm2} of $\Lc^g(\lambda_\pm)$. Indeed, one then have
\begin{equation}\label{Eq:Lgpm}
\Lc^g(\lambda_\pm) = \gamma R^\mp Y.
\end{equation}
The Poisson brackets of $g$ and $Y$ were considered in equation \eqref{Eq:PBconjg}. Our motivation to consider the field $h$ instead of $g$ is a matter of convention (more precisely it simplifies some changes of conventions between our article~\cite{Delduc:2016ihq} and this thesis): indeed, the Poisson brackets of $h$ and $Y$ are exactly the same as the Poisson brackets of $g$ and $X$:
\begin{subequations}
\begin{align}
\left\lbrace h\ti{1}(x), h\ti{2}(y) \right\rbrace & = 0, \\
\left\lbrace Y\ti{1}(x), h\ti{2}(y) \right\rbrace & = h\ti{2}(x)  C\ti{12} \, \delta_{xy}, \label{Eq:PBhY}\\
\left\lbrace Y\ti{1}(x), Y\ti{2}(y) \right\rbrace & = -\left[ C\ti{12}, Y\ti{2}(x) \right] \delta_{xy}. \label{Eq:PBYY}
\end{align}
\end{subequations}

The variation of the path-ordered exponential of some field under an infinitesimal variation of this field is given by equation \eqref{dPExp}. In particular, for $\varepsilon,\varkappa\in\lbrace +,- \rbrace$, one has
\begin{align}\label{Eq:PBTgTg1}
\left\lbrace T^g(\lambda_\varepsilon)\ti{1}, T^g(\lambda_\varkappa)\ti{2} \right\rbrace & = \displaystyle \int_{-\infty}^{+\infty} \dd x \int_{-\infty}^{+\infty} \dd y \; \,  T^g (\lambda_\varepsilon \, ; \, x, +\infty)\ti{1} T^g (\lambda_\varkappa \, ; \, y, +\infty)\ti{2}\\
 & \hspace{45pt} \left\lbrace \Lc^g(\lambda_\varepsilon,x)\ti{1}, \Lc^g(\lambda_\varkappa,y)\ti{2} \right\rbrace T^g (\lambda_\varepsilon \, ;\, -\infty, x)\ti{1} T^g (\lambda_\varkappa \, ; \, -\infty,y)\ti{2}. \notag
\end{align}
One then needs the bracket $\left\lbrace \Lc^g(\lambda_\varepsilon,x)\ti{1}, \Lc^g(\lambda_\varkappa,y)\ti{2} \right\rbrace$. Starting from equation \eqref{Eq:Lgpm} and using the bracket \eqref{Eq:PBYY}, we get
\begin{equation}\label{Eq:PBLgLg}
\left\lbrace \Lc^g(\lambda_\varepsilon,x)\ti{1}, \Lc^g(\lambda_\varkappa,y)\ti{2} \right\rbrace = \gamma\left[ \Lc^g(\lambda_\varepsilon,x)\ti{1} + \Lc^g(\lambda_\varkappa,x)\ti{2}, R^{\varkappa}\ti{12} \right] \delta_{xy}.
\end{equation}
To obtain this Poisson bracket, we used the mCYBE \eqref{Eq:mCYBE} on $R$ and the skew-symmetry of $R$, in the form
\begin{equation*}
R^{\varepsilon}\ti{12} = R\ti{1}^\varepsilon C\ti{12} = -R\ti{2}^{-\varepsilon} C\ti{12} = - R^{-\varepsilon}\ti{21}.
\end{equation*}
In particular, the Poisson bracket \eqref{Eq:PBLgLg} is ultralocal. Thus, as mentioned above, the computation of the Poisson brackets of the monodromies $T^g(\lambda_\pm)$ does not suffer of non-ultralocality issues. Combining equations \eqref{Eq:PBTgTg1} and \eqref{Eq:PBLgLg} with the property \eqref{Eq:DerPexp} of path-order exponentials, one gets
\begin{eqnarray}
\left\lbrace T^g(\lambda_\pm)\ti{1}, T^g(\lambda_\pm)\ti{2} \right\rbrace &=& \gamma \left[ R\ti{12}, T^g(\lambda_\pm)\ti{1} T^g(\lambda_\pm)\ti{2} \right]. \\
\left\lbrace T^g(\lambda_\mp)\ti{1}, T^g(\lambda_\pm)\ti{2} \right\rbrace &=& \gamma \left[ R^{\pm}\ti{12}, T^g(\lambda_\mp)\ti{1} T^g(\lambda_\pm)\ti{2} \right].
\end{eqnarray}
In conclusion, $T^g(\lambda_-)$ and $T^g(\lambda_+)$ satisfy the Semenov-Tian-Shansky Poisson brackets \eqref{STSpb}.\\

In particular, let us suppose that $R$ is a standard $R$-matrix on $\g$ as described in Subsection \ref{SubSec:Stand} and Appendix \ref{App:StandardFinite}. Then, using the methods of Subsection \ref{SubSec:Uqg}, one can extract from $T^g(\lambda_\pm)$ conserved charges satisfying the $q$-deformed Poisson-Hopf algebra $\U_q(\g_0)$ (note that the reality conditions on $T^g(\lambda_\pm)$ necessary to apply the results of Subsection \ref{SubSec:Uqg} can be proven from the reality condition \eqref{Eq:Reality} of the Lax matrix). These deformed conserved charges were already found in~\cite{Delduc:2013fga} for the $\eta$-deformations of the PCM and the $\Z_2$-coset $\s$-model. The result presented above then generalises this to all Yang-Baxter type models, in a model independent way. The rest of this section is devoted to the study of the (Poisson-Lie) symmetry associated with these deformed charges (note however that this study does not require $R$ to be standard).

Before pursuing, let us make a brief parenthesis. For the undeformed PCM, the conserved charges mentioned above are part of a bigger (actually infinite) algebra, the classical analogue of the Yangian $Y(\g_0)$ of $\g_0$~\cite{MacKay:1992he,Bernard:1992ya}. It was proved recently~\cite{Delduc:2017brb} (after the publication of the results presented here) that for the Yang-Baxter model, this Yangian is deformed in the algebra $\U_q(\,\widehat{\g}_0)$, where $\widehat{\g}_0$ is the affine algebra associated with $\g_0$.

\subsection[Poisson-Lie $G$-symmetry]{Poisson-Lie $\bm{G_0}$-symmetry}

For the remainder of this section we shall restrict attention to the non-split case. The treatment of the split case is completely analogous.

\subsubsection{The non-abelian moment map}

According to \eqref{Eq:Lgpm}, $\Lc^g(\lambda_\mp, \sigma)$ take values in the subalgebras $\g_{\pm}$ of the complex double $\g$ (cf. subsection \ref{SubSec:Doubles}). Hence the path-ordered exponentials $T^g(\lambda_\mp \, ; x,y)$ belong to the subgroups $G_\pm$, which are realisations of the dual group $G_0^*$.
Moreover, we proved in the previous subsection that $T^g(\lambda_\mp)$ satisfies the Semenov-Tian-Shansky bracket. It follows from subsection \ref{SubSec:PBGamma} that $T^g(\lambda_\mp)$ has the right Poisson brackets for being the non-abelian moment map of a Poisson-Lie action of $G_0$. In the notations of the previous sections, we therefore consider
\begin{equation}\label{GammapmYB}
\Gamma^\pm = T^g(\lambda_\mp).
\end{equation}
as the two realisations of a non-abelian moment map in $G_\pm$, embedded in the complex double $G$. It is natural to also look for the expression of this non-abelian moment map in the other realisation of the dual group $G_0^*$, namely the group $G_R$ described in subsection \ref{SubSec:GR}. This is given by $\Gamma_R=\Delta_\pm^{-1} \bigl( T^g(\lambda_\mp) \bigr)$, with $\Delta_\pm : G_R \rightarrow G_\pm$ the automorphisms described in subsection \ref{SubSec:Doubles}. In order to evaluate this explicitly we note that \eqref{Eq:Lgpm} can be written as $\Lc^g(\lambda_\mp, x) = \Delta_\pm Y(x)$. Therefore, according to equation \eqref{AutoPExp}, the non-abelian moment map seen in $G_R$ simply reads
\begin{equation}\label{GammaRYB}
\Gamma_R = \Pexp_{G_R} \left( - \int_{-\infty}^{+\infty} \dd x \; Y(x) \right).
\end{equation}
In this expression, $Y(x)$ is seen as an element of $\g_R$ and the path-ordered exponential is taken in the group $G_R$.

\subsubsection[Transformation law of $g$, $h$ and $Y$]{Transformation law of $\bm{g}$, $\bm{h}$ and $\bm{Y}$}

As motivated in the previous subsection, we consider the Poisson-Lie action of $G_0$ generated by the non-abelian moment map $\Gamma^\pm = T^g(\lambda_\mp) \in G_\pm$. According to equation \eqref{ActionNonSplit}, the infinitesimal form of this action is given by
\begin{equation}\label{TransfoTg}
\delta_\epsilon f = \frac{1}{2i\gamma} \kappa \Bigl( \epsilon, T^g(\lambda_+)^{-1}\left\lbrace T^g(\lambda_+), f \right\rbrace - T^g(\lambda_-)^{-1}\left\lbrace T^g(\lambda_-), f \right\rbrace \Bigr) = -\kappa \bigl( \epsilon, \Gamma_R^{-1} \left\lbrace\Gamma_R,f\right\rbrace \bigr).
\end{equation}

In the undeformed case $\gamma=0$, the group $G_R$ is abelian and, from the expression \eqref{GammaRYB} of the non-abelian moment map $\Gamma_R$, the transformation \eqref{TransfoTg} becomes the usual Hamiltonian action with moment map $\int_{-\infty}^{+\infty} \dd x \; Y(x)$. From the brackets \eqref{Eq:PBhY} and \eqref{Eq:PBYY}, one checks that this is the action of $\g_0$ by right multiplication on $h$:
\begin{equation}\label{TransfogXNonDef}
\delta_\epsilon h(x) = h(x) \epsilon \; \; \; \; \text{ and } \; \; \; \; \delta_\epsilon Y(x) = [Y(x),\epsilon].
\end{equation}
We will see in the rest of this subsection that, for $\gamma \neq 0$, the Poisson-Lie action generated by $T^g(\lambda_\mp)$ is still a right multiplication of $h$, but with a more complicated parameter. Note that in terms of the initial field $g=h^{-1}$, the transformation \eqref{TransfogXNonDef} becomes the left multiplication on $g$.

Since the Poisson bracket of $\Lc^g(\lambda_\mp, x)$ with the fields $h$ and $Y$ is ultralocal, we can compute the Poisson brackets of $T^g(\lambda_\pm)$ with $h$ and $Y$ using equation \eqref{dPExp}, without the need for any regularisation. We find
\begin{align*}
\left\lbrace T^g(\lambda_\mp)\ti{1}, h(x) \ti{2} \right\rbrace &= -\gamma \, T^g(\lambda_\mp \, ; \, x, +\infty)\ti{1}\, h(x)\ti{2}\, R^\pm\ti{12}\, T^g(\lambda_\mp \, ; \, -\infty,x)\ti{1}, \\
\left\lbrace T^g(\lambda_\mp)\ti{1}, Y(x) \ti{2} \right\rbrace &= - \gamma \, T^g(\lambda_\mp \, ; \, x, +\infty)\ti{1}\, \left[Y\ti{2}(x), R^\pm\ti{12} \right] T^g(\lambda_\mp \, ; \, -\infty,x)\ti{1}.
\end{align*}
Inserting these expressions into \eqref{TransfoTg}, we obtain the transformation law of $h$ and $Y$,
\begin{equation}\label{TransfogX}
\delta_\epsilon h(x) = h(x)K(x) \; \; \; \; \text{ and } \; \; \; \; \delta_\epsilon Y(x) = [Y(x), K(x)],
\end{equation}
where we have defined
\begin{align}\label{DefK}
K(x) &= \frac{1}{2i} R^+\Bigl( T^g(\lambda_+ \, ; \, -\infty, x) \, \epsilon \, T^g(\lambda_+ \, ; \, -\infty, x)^{-1} \Bigr) \notag\\
&\qquad\qquad - \frac{1}{2i} R^-\Bigl( T^g(\lambda_- \, ; \, -\infty, x) \, \epsilon \, T^g(\lambda_- \, ; \, -\infty, x)^{-1} \Bigr).
\end{align}
We note that this transformation law has the same structure as the undeformed one \eqref{TransfogXNonDef} but with $\epsilon$ replaced by a more complicated (and non-constant) expression $K(x)$. In particular, this field is non-local, as it contains $T^g(\lambda_\pm \, ; \, -\infty, x)$. Since $T^g(\lambda_\pm \, ; \, -\infty, x)$ becomes equal to the identity when $\gamma=0$, we see that $K$ turns back into $\epsilon$ in the undeformed case. As for the undeformed transformation,  if one goes back to the initial field $g=h^{-1}$, one finds that the transformation \eqref{TransfogX} acts as a left multiplication
\begin{equation}\label{Eq:transfog}
\delta_\epsilon g(x) = -K(x)g(x).
\end{equation}

According to the paragraph following equation \eqref{MomentMap}, the transformation \eqref{TransfogX} must preserve the Poisson brackets on $(h,Y)$ if $\epsilon$ possesses a Poisson bracket with itself, coming from the linearisation of the Sklyanin Poisson bracket on $G_0$:
\begin{equation}\label{SklyaninAlg}
\bigl\lbrace \epsilon\ti{1}, \epsilon\ti{2} \bigr\rbrace = \gamma \bigl[ R\ti{12}, \epsilon\ti{1} + \epsilon\ti{2} \bigr].
\end{equation}
For coherence, one can check this directly from the expression \eqref{DefK} of $K(x)$. This (slightly long) computation involves some algebraic manipulations to simplify the expressions, in particular the identity
\begin{equation*}
\Ad_{T^g(\lambda_\mp)} \circ R^\pm = R^\pm \circ \Ad^{G_R}_{\Gamma_R},
\end{equation*}
which is a consequence of equation \eqref{AutoPExp}, applied to the automorphism $\Delta_\pm$.\\

The transformation law \eqref{TransfogX} may seem complicated because of the non-local expression \eqref{DefK} for $K(x)$. However, it can be re-interpreted in a simpler way by introducing the more adapted variables \cite{Klimcik:2002zj,Delduc:2013fga,Vicedo:2015pna}
\begin{equation*}
\Psi_\pm(x) = h(x) \xi_\pm(x), \; \; \; \; \text{ with } \; \; \; \; \xi_\pm(x) = T^g(\lambda_\mp \, ; \, -\infty, x).
\end{equation*}
In terms of these, the quantity \eqref{DefK} may be then rewritten as
\begin{equation}\label{DefK2}
K(x) = \frac{1}{2i} R^+\Bigl( \xi_-(x) \,\epsilon\,  \xi_-(x)^{-1} \Bigr) - \frac{1}{2i} R^-\Bigl( \xi_+(x) \,\epsilon\,  \xi_+(x)^{-1} \Bigr).
\end{equation}
If we also introduce
\begin{equation*}
Z(x) = \frac{1}{2i} \Bigl( \xi_+(x) \,\epsilon\, \xi_+(x)^{-1} -  \xi_-(x) \,\epsilon\, \xi_-(x)^{-1} \Bigr),
\end{equation*}
then one checks that
\begin{equation*}
\delta_\epsilon Y(x) = \bigl[ Y(x), K(x) \bigr] = -\frac{1}{\gamma} \Bigl( \p_x Z(x) + \bigl[ Y(x),Z(x) \bigr]_R \Bigr).
\end{equation*}
Using this identity and equation \eqref{dPExp}, we find that the transformation law of $\xi_\pm$ reads
\begin{equation*}
\delta_\epsilon \xi_\pm(x) = R^\pm Z(x) \, \xi_\pm(x).
\end{equation*}
Finally, it follows that the pair of fields $\Psi_\pm$ simply transform as
\begin{equation*}
\delta_\epsilon \Psi_\pm(x) = \Psi_\pm(x)\epsilon .
\end{equation*}

It was observed in \cite{Delduc:2013fga} that for Yang-Baxter type deformations with standard $R$-matrices, the Cartan part of the $G_0$-symmetry is preserved. This can be checked here explicitly: indeed, for $\epsilon \in \h$ we find that the definition \eqref{DefK} of $K$ reduces to $\epsilon$, so that the infinitesimal transformation in the Cartan direction remains undeformed, as in \eqref{TransfogXNonDef}.
This fact can also be seen in terms of Poisson-Lie actions. For standard $R$-matrices, the Sklyanin bracket \eqref{SklyaninAlg} vanishes when restricted to the Cartan subalgebra $\h$. The corresponding action is then a usual Hamiltonian symmetry.\\

Let us end this paragraph by saying a few words about the BYB model. As explained in Section \ref{Sec:BYB}, it is the combination of two Yang-Baxter type deformations, breaking both the left and the right multiplication symmetries. The discussion above applies to the BYB model, as it satisfies equation \eqref{Eq:Lgpm}, by equation \eqref{Eq:LgBYB}. The BYB model then admits a deformed left multiplication symmetry \eqref{Eq:transfog} (on $g'$, or equivalently the right symmetry \eqref{TransfogX} on $h'=g'^{\,{-1}}$). This deformed symmetry is associated with conserved charges forming a deformed algebra $\U_q(\g_0)$.

Recall however that the BYB model aslo satisfies equation \eqref{Eq:LpmBYB}, which is similar to equation \eqref{Eq:LgBYB} but with $Y$ replaced by $X'$. This can be understood as the result of the Yang-Baxter type deformation on the right (recall from Subsection \ref{SubSec:SigmaModelHam} that the integral of $X'$ generates the right multiplication on $g'$). All the computation done above can be done with $Y$ replaced by $X'$ and $h$ by $g'$ and we would then find that the BYB model also admits a deformed right multiplication symmetry on $g'$, associated with a deformed algebra $\U_{\tilde{q}}(\g_0)$, where $\tilde{q}=e^{-ic\widetilde{\gamma}}$.

Recall also that the BYB model can be seen as a deformation of a $\Z_2$-coset $\s$-model (this was the approach followed in Section \ref{Sec:BYB}). In particular, in this formulation, we obtained two equations of the form \eqref{Eq:Lgpm}, respectively in \eqref{Eq:LgBYBz2} and \eqref{Eq:LgtBYBz2}. Thus, one can apply the method presented above and find two deformed left multiplication symmetries of the BYB model, acting respectively on $g$ and $\gt$. After gauge fixing, these transformations will reduce to the deformed left and right multiplication on the gauge-invariant field $g'=g\gt^{-1}$.

\subsubsection{Poisson-Lie symmetry: variation of the Hamiltonian and the action}

In this section, we consider the case of the Yang-Baxter $\sigma$-model. The conservation of $T^g(\lambda_\pm)$ can be seen as the fact that it has a vanishing Poisson bracket with the Hamiltonian $\Hc$ of the model. This implies that the Hamiltonian is invariant under the Poisson-Lie action generated by $T^g(\lambda_\pm)$, namely
\begin{equation*}
\delta_\epsilon \Hc = 0.
\end{equation*}
Thus, the transformation \eqref{TransfogX} is a symmetry of the Hamiltonian.

Let us now compute the variation of the action under the transformation. In the case of a Hamiltonian action ($G_0^*$ abelian), the transformation is canonical and the invariance of the Hamiltonian is then equivalent to the invariance of the action. The situation is slightly more involved in the case of a Poisson-Lie action. The first order action is given
\begin{equation*}
S = \int \dd t \,\dd x \, \kappa\bigl( h^{-1} \p_t h, Y \bigr) - \int \dd t \, \Hc.
\end{equation*}
Consider the transformation \eqref{TransfogX} of $h$ and $Y$ and, at first, let us allow the parameter $\epsilon$ to be a function of the time parameter $t$. Since $\Hc$ is invariant under this transformation, the variation of the action becomes
\begin{equation*}
\delta_\epsilon S = \int\dd t \,\dd x \, \delta_\epsilon \Bigl( \kappa\bigl( h^{-1} \p_t h , Y \bigr) \Bigr).
\end{equation*}
We have
$\delta_\epsilon \bigl( h^{-1}\p_t h \bigr) = \p_t K + \bigl[ h^{-1}\p_t h, K \bigr]$,
so that
\begin{equation*}
\delta_\epsilon \Bigl( \kappa\bigl( h^{-1} (\p_t h), Y \bigr) \Bigr) = \kappa \bigl( \p_t K, Y \bigr) = \p_t \bigl( \kappa(K,Y) \bigr) - \kappa\bigl( K, \p_t Y \bigr).
\end{equation*}
Using expression \eqref{DefK2} for $K$, the skew-symmetry of $R$ and the invariance of $\kappa$ under adjoint action, one finds
\begin{equation*}
\kappa\bigl( K, \p_t Y \bigr) = \frac{1}{2i} \kappa \bigl( \epsilon, \xi_+^{-1} \p_t(R^+Y) \xi_+ - \xi_-^{-1} \p_t(R^-Y) \xi_+ \bigr).
\end{equation*}
Discarding the boundary terms at initial and final times, we get
\begin{align*}
\delta_\epsilon S &= \int\dd t \, \frac{1}{2i\gamma} \kappa \left( \epsilon, T^g(\lambda_+)^{-1} \int_{-\infty}^\infty \dd x \, T^g(\lambda_+ \, ; \, x,+\infty) \p_t \Lc^g(\lambda_+,x) T^g(\lambda_+ \, ; \, -\infty,x) \right) \\
&\qquad - \int\dd t \, \frac{1}{2i\gamma} \kappa \left( \epsilon, T^g(\lambda_-)^{-1} \int_{-\infty}^\infty \dd x \, T^g(\lambda_- \, ; \, x,+\infty) \p_t \Lc^g(\lambda_-,x) T^g(\lambda_- \, ; \, -\infty,x) \right).
\end{align*}
Using equation \eqref{dPExp}, this may be rewritten as
\begin{equation*}
\delta_\epsilon S = \frac{1}{2i\gamma} \int \dd\tau \, \kappa \bigl( \epsilon, T^g(\lambda_+)^{-1}\p_t T^g(\lambda_+)-T^g(\lambda_-)^{-1}\p_t T^g(\lambda_-) \bigr).
\end{equation*}
In terms of the ``abstract'' non-abelian moment map $\Gamma$, seen as a $G_0^*$-valued map, this is simply
\begin{equation*}
\delta_\epsilon S = -\int \dd t \,  \bigl\langle \epsilon(t), \Gamma^{-1}\p_t \Gamma \bigr\rangle.
\end{equation*}
By the principle of least action, $\delta_\epsilon S$ must be zero for any function $\epsilon(t)$, as long as the fields are on-shell. Thus, we recover the fact that $\Gamma$ is conserved.\\

It is worth noticing that, if $G_0^*$ is non-abelian, $\Gamma^{-1} \p_t \Gamma$ is not a total time derivative. Thus, when we choose a constant parameter $\epsilon$, we cannot conclude that $\delta_\epsilon S=0$. That is to say, the action is not invariant under the Poisson-Lie symmetry. However, the latter is still a symmetry of the model since the Hamiltonian is invariant.

\begin{subappendices}
\section{Poisson brackets extraction theorems}
\label{App:ThmPB}

In this appendix, we state and prove two theorems allowing to extract the Poisson brackets of the factors of a Lie-group-valued quantity.

\begin{theorem}\label{TheoremPB1}
Let $F_1$ and $F_2$ be two Lie groups, that are decomposable into two subgroups: $F_i=G_iH_i$. This group factorisation corresponds to a direct sum of Lie algebras $\f_i = \g_i \oplus \h_i$. We denote by $\pi_{\g_i}$ and $\pi_{\h_i}$ the associated projections. We consider $A \in F_1$ and $B \in F_2$ that we factorise as
\begin{equation*}
A=uv, \; \; \; \text{and} \: \: \: B=xy,
\end{equation*}
for $(u,v) \in G_1 \times H_1$ and $(x,y) \in G_2 \times H_2$. We define
\begin{equation*}
\Pc\ti{12} = u^{-1}\ti{1}x^{-1}\ti{2} \lbrace A\ti{1}, B\ti{2} \rbrace v^{-1}\ti{1}y^{-1}\ti{2} \; \in \f_1 \otimes \f_2.
\end{equation*}
Then, we have
\begin{align*}
\lbrace u\ti{1}, x\ti{2} \rbrace &= u\ti{1}x\ti{2} \; \pi_{\g_1} \otimes \pi_{\g_2} \bigl( \Pc\ti{12} \bigr), \\
\lbrace u\ti{1}, y\ti{2} \rbrace &= u\ti{1} \; \pi_{\g_1} \otimes \pi_{\h_2} \bigl( \Pc\ti{12} \bigr) \; y\ti{2}, \\
\lbrace v\ti{1}, x\ti{2} \rbrace &= x\ti{2} \; \pi_{\h_1} \otimes \pi_{\g_2} \bigl( \Pc\ti{12} \bigr) \; v\ti{1}, \\\lbrace v\ti{1}, y\ti{2} \rbrace &= \pi_{\h_1} \otimes \pi_{\h_2} \bigl( \Pc\ti{12} \bigr) \; v\ti{1}y\ti{2},
\end{align*}
\begin{proof}
Using the Leibniz rule, we have
\begin{equation*}
\lbrace A\ti{1}, B\ti{2} \rbrace = \lbrace u\ti{1}, x\ti{2} \rbrace v\ti{1}y\ti{2} + x\ti{2}\lbrace u\ti{1}, y\ti{2} \rbrace v\ti{1} + u\ti{1}\lbrace v\ti{1}, x\ti{2} \rbrace y\ti{2} + u\ti{1}x\ti{2}\lbrace v\ti{1}, y\ti{2} \rbrace,
\end{equation*}
so that
\begin{equation*}
\Pc\ti{12} = 
\underbrace{u^{-1}\ti{1}x^{-1}\ti{2} \lbrace u\ti{1}, x\ti{2} \rbrace}_{\displaystyle \in \g_1 \otimes \g_2} + 
\underbrace{u^{-1}\ti{1} \lbrace u\ti{1}, y\ti{2} \rbrace y^{-1}\ti{2}}_{\displaystyle \in \g_1 \otimes \h_2} +
\underbrace{x^{-1}\ti{2} \lbrace v\ti{1}, x\ti{2} \rbrace v^{-1}\ti{1}}_{\displaystyle \in \h_1 \otimes \g_2} +
\underbrace{\lbrace v\ti{1}, y\ti{2} \rbrace v^{-1}\ti{1} y^{-1}\ti{2}}_{\displaystyle \in \h_1 \otimes \h_2},
\end{equation*}
and hence the theorem.
\end{proof}
\end{theorem}

\begin{theorem}\label{TheoremPB2}
Let $G$ be a Lie group that factorises into subgroups as $G=G_1 \ldots G_p$. This group factorisation corresponds to a direct sum of Lie algebras $\g = \g_1 \oplus \ldots \oplus \g_p$, with associated projections $\pi_i$. Suppose this decomposition is such that, for all $i\in\lbrace 1,\ldots,p \rbrace$,
\begin{equation*}
\g_{<i}=\bigoplus_{k=1}^{i-1} \g_k \; \; \; \; \text{ and } \; \; \; \; \g_{>i}=\bigoplus_{k=i+1}^p \g_k
\end{equation*}
are subalgebras of $\g$ and $[\g_i,\g_{<i}] \subseteq \g_{<i}$.

Let $f$ be a $\mathbb{R}$-valued function and $A$ a $G$-valued function, that we factorise as
\begin{equation*}
A=A^{(1)} \ldots A^{(p)}, \; \; \; \text{with } \left(A^{(1)},\ldots,A^{(p)}\right) \in G_1 \times \ldots \times G_p.
\end{equation*}
If we define
\begin{equation*}
\Pc^{(i)} = \bigl( A^{(1)} \ldots A^{(i)} \bigr)^{-1} \left\lbrace A, f \right\rbrace \bigl( A^{(i+1)} \ldots A^{(p)} \bigr)^{-1} \in \g,
\end{equation*}
then we have
\begin{equation*}
\left(A^{(i)}\right)^{-1}\left\lbrace A^{(i)}, f \right\rbrace = \pi_i  \bigl( \Pc^{(i)} \bigr).
\end{equation*}
\begin{proof}
Let
$B = A^{(1)} \ldots A^{(i-1)}$ and
$C = A^{(i+1)} \ldots A^{(p)}$.
Using the Leibniz rule, we have
\begin{equation*}
\Pc^{(i)} = \underbrace{\left(A^{(i)}\right)^{-1} \left\lbrace A^{(i)}, f \right\rbrace}_{\in \g_i} + \left(A^{(i)}\right)^{-1} \underbrace{\bigl( B^{-1} \left\lbrace B, f \right\rbrace \bigr) }_{\in \g_{<i}}A^{(i)} + \underbrace{\lbrace C, f \rbrace C^{-1}}_{\in\g_{>i}}.
\end{equation*}
Since $B^{-1} \left\lbrace B, f \right\rbrace$ belongs to $\g_{<i}$, the assumption on the Lie subalgebras $\g_k$ tells us that the adjoint action
\begin{equation*}
\left(A^{(i)}\right)^{-1}\bigl( B^{-1} \left\lbrace B, f \right\rbrace \bigr)A^{(i)}
\end{equation*}
still belongs to $\g_{<i}$, hence the theorem.
\end{proof}
\end{theorem}

\end{subappendices}

\part{Gaudin models}

\cleardoublepage
\chapter{Classical Gaudin models}
\label{Chap:GaudinClass}

The first part of this thesis concerned integrable models with twist function and in particular integrable $\s$-models. In the recent publication~\cite{Vicedo:2017cge}, Vicedo reinterpreted these field theories as so-called classical Dihedral Affine Gaudin Models (DAGM). Gaudin models, which are the subject of the second part of this thesis, are a particular class of integrable systems, associated with quadratic Lie algebras (Lie algebras with a non-degenerate invariant bilinear form). 

In particular, one can construct Gaudin models associated with Kac-Moody algebras. These are a particular class of quadratic Lie algebras (see for example~\cite{Kac:1990gs}). They are classified in three types: finite, affine and indefinite. The simplest Kac-Moody algebras are the ones of finite type: they exactly coincide with finite-dimensional semi-simple Lie algebras, as described in Appendix \ref{App:SemiSimple}. The associated Gaudin models, that we shall call finite Gaudin models, are thus integrable systems with a finite number of degrees of freedom. They have been extensively studied in the literature and are used to describe mechanical systems and spin chains.

The Kac-Moody algebras of non-finite type are much more complicated: in particular, they are all infinite dimensional. Among them, the affine Kac-Moody algebras are the simplest, as they are related to loop algebras of finite ones. Using this relation, Vicedo has shown in~\cite{Vicedo:2017cge} that classical affine Gaudin models (associated with affine Kac-Moody algebras) can be viewed as two dimensional integrable field theories. More precisely, they are models with twist function, as described in the first part of this thesis.

In this chapter, we will focus on classical Gaudin models. In Section \ref{Sec:ClassicalGaudin}, we will develop the general theory of classical Gaudin models for an arbitrary quadratic Lie algebra. We will also describe briefly finite Gaudin models (illustrated with an example, the unreduced Neumann model). In Section \ref{Sec:AGM}, we will discuss classical affine Gaudin models and their relations with integrable fields theories with twist function. Most of this chapter is a review of the literature on classical Gaudin models, in particular based on the article~\cite{Vicedo:2017cge}. However, we will end Section \ref{Sec:AGM} by a new result about integrable hierarchies of affine Gaudin models, which is part of my article~\cite{Lacroix:2017isl} with M. Magro and B. Vicedo.

\section{Classical Gaudin models}
\label{Sec:ClassicalGaudin}

Let $\g$ be a quadratic Lie algebra, \textit{i.e.} a Lie algebra possessing a non-degenerate invariant bilinear form $\kappa$. For now, we will consider only complex Lie algebras. Real Gaudin models will be discussed later in Subsection \ref{SubSec:RealGaudin}.

Let us consider a basis $\lbrace I^a \rbrace$ of $\g$, the bilinear form $\kappa^{ab}=\kappa(I^a,I^b)$ written in this basis, its inverse $\kappa_{ab}$ and the dual basis $\lbrace I_a = \kappa_{ab} I^b \rbrace$. One can then consider the quadratic Casimir
\begin{equation}\label{Eq:Cas}
C\ti{12} = I^a \otimes I_a = \kappa_{ab} \, I^a \otimes I^b \in \g \otimes \g.
\end{equation}
It is defined in Appendix \ref{App:Casimir} in the particular case of a semi-simple finite dimensional algebra equipped with the Killing form. However, its main properties \eqref{Eq:CasIdentity} and \eqref{Eq:CasComp} generalise easily to any quadratic Lie algebra.

Let us mention here a technical subtlety. If the Lie algebra $\g$ is infinite dimensional, the sum defining $C\ti{12}$ in equation \eqref{Eq:Cas} is infinite. A rigorous definition of $C\ti{12}$ then requires to give a sense to this infinite sum, by considering an appropriate topology on $\g\otimes\g$. The quadratic Casimir then belongs to the completion $\g \hat{\otimes} \g$ with respect to this topology. We shall not enter into these technicalities here: when manipulating infinite sums, we will suppose that these can be well defined in an appropriate completion. These questions of convergence will appear in Section \ref{Sec:AGM}, as we will consider affine Kac-Moody algebras, which are infinite dimensional. We will then give a brief summary of how to treat these infinite sums in this case.\\

In addition to the quadratic Lie algebra $\g$, a Gaudin model is defined from the data of $N$ points $\lambda_1,\cdots,\lambda_N$ of the complex plane $\C$, given with multiplicities $m_1,\cdots,m_N \in \Z_{\geq 1}$. These points are called the sites of the model. For simplicity, we will restrict in this section to the case where all sites have multiplicity one. In the same way, for this first presentation of Gaudin models, we will not consider the so-called cyclotomic and dihedral Gaudin models. For completeness, and as it is important for the link between affine Gaudin models and integrable field theories, we will say a few words about the general formalism with arbitrary multiplicities and dihedrality at the end of this section. We refer to~\cite{Vicedo:2017cge} for an exhaustive and rigorous presentation.

\subsection{Phase space and Lax matrix}
\label{SubSec:GaudinPhaseSpace}

\paragraph{Phase space.} Consider the dual $\g^*$ of the Lie algebra $\g$. Recall that it can be made a Poisson manifold by the so-called Kirillov-Kostant bracket, as described in Appendix \ref{App:KK} (technically the Appendix \ref{App:KK} treats the case of a real Lie algebra but the discussion generalises easily to complex ones, the Poisson bracket then being a $\C$-linear derivation of the complex algebra of functions on $\g^*$). The phase space $M$ of the Gaudin model with sites $\lambda_1,\cdots,\lambda_N$ is the Cartesian product of $N$ copies of the Poisson manifold $\g^*$. Concretely, this means that the algebra $\F[M]$ of functions on $M$ is generated by elements $X^a_{(r)}$ ($r=1,\cdots,N$), satisfying the Poisson bracket
\begin{equation*}
\lwb X^a_{(r)}, X^b_{(s)} \rwb = \delta_{rs} \, \fs{ab}{c} X^c_{(r)},
\end{equation*}
where the $\ft{ab}{c}$'s are the structure constants of $\g$.

Using the non-degenerate form $\kappa$, one can encode all the fundamental functions $X^a_{(r)}$, for a fixed $r\in\lbrace 1,\cdots,N \rbrace$, in a unique object
\begin{equation}\label{Eq:XGaudin}
X^{(r)} = \kappa_{ab}\, I^a \otimes X^b_{(r)} \in \g \otimes \F[M].
\end{equation}
Here also, one would have to consider a completion of $\g \otimes \F[M]$ to consider $X^{(r)}$ if the algebra $\g$ is infinite. As an element of $\g \otimes \F[M]$, $X^{(r)}$ is a $\g$-valued function on $M$. As explained in Appendix \ref{App:KK}, the Poisson bracket of the $X^{(r)}$'s can be written in tensorial notations as
\begin{equation*}
\bigl\lbrace X^{(r)}\ti{1}, X^{(s)}\ti{2} \bigr\rbrace = \delta_{rs} \bigl[ C\ti{12}, X^{(r)}\ti{1} \bigr] = -\delta_{rs} \bigl[ C\ti{12}, X^{(r)}\ti{2} \bigr].
\end{equation*}

\paragraph{Lax matrix.} Let us now define the Lax matrix of the Gaudin model. It is a rational function of a complex parameter $\lambda$ (the spectral parameter) which depends on the positions $\lambda_r$'s of the sites and contains all the $\g$-valued functions $X^{(r)}$:
\begin{equation}\label{Eq:LaxGaudin}
\Ls(\lambda) = \sum_{r=1}^N \frac{X^{(r)}}{\lambda-\lambda_r} + \Omega,
\end{equation}
where $\Omega$ is a constant element in $\g$, in the sense that it has a vanishing Poisson bracket with all functions in $\F[M]$ (this constant element can be seen as a site with multiplicity two at infinity, but we shall not develop further this interpretation here, cf.~\cite{Vicedo:2017cge}). Using the circle lemma \eqref{Eq:CircleLemma}, one checks that the Poisson bracket of the Lax matrix reads
\begin{equation}\label{Eq:PBGaudin}
\lwb \Ls\ti{1}(\lambda), \Ls\ti{2}(\mu) \rwb = \lsb \rc\ti{12}(\lambda,\mu), \Ls\ti{1}(\lambda) + \Ls\ti{2}(\mu) \rsb,
\end{equation}
with
\begin{equation*}
\rc\ti{12}(\lambda,\mu) = \frac{C\ti{12}}{\mu-\lambda}.
\end{equation*}
Note that this Poisson bracket is independent of the constant element $\Omega\in\g$ (to get this result, one needs to use the identity \eqref{Eq:CasIdentity}).
 
In particular, if $\g$ is taken to be a finite dimensional semi-simple algebra, the matrix $\rc$ coincides with $\Rc^0$, the standard non-twisted $\Rc$-matrix on $\Lc(\g)$, as introduced in Section \ref{Sec:CYBE}. As explained in this section, this matrix satisfies the CYBE \eqref{Eq:CYBE}. Although we considered the case of a semi-simple finite algebra $\g$, the proof generalises to any quadratic Lie algebra $\g$ and in general the matrix $\rc\ti{12}(\lambda,\mu)$ satisfies the CYBE.

Note that $\rc$ is skew-symmetric, in the sense that $\rc\ti{12}(\lambda,\mu) = -\rc\ti{21}(\mu,\lambda)$. This property ensures the skew-symmetry of the bracket \eqref{Eq:PBGaudin}. Note that one can rewrite the Poisson bracket \eqref{Eq:PBGaudin} in a form where the skew-symmetry is obvious:
\begin{equation*}
\lwb \Ls\ti{1}(\lambda), \Ls\ti{2}(\mu) \rwb = \lsb \rc\ti{12}(\lambda,\mu), \Ls\ti{1}(\lambda) \rsb - \lsb \rc\ti{21}(\mu,\lambda), \Ls\ti{2}(\mu) \rsb.
\end{equation*}

\subsection{Hamiltonians and Lax equation}
\label{SubSec:GaudinHam}

\paragraph{Quadratic Hamiltonians.} So far, we defined the phase space $M$ of the Gaudin model, whose fundamental coordinates are encoded in the Lax matrix \eqref{Eq:LaxGaudin}. To define entirely the model, one has to specify a time evolution on this phase space, in the form of a Hamiltonian. Recall the bilinear form $\kappa$ on $\g$. We define the so-called quadratic Hamiltonian of the model as
\begin{equation}\label{Eq:GaudinHamLambda}
\Hs(\lambda) = \frac{1}{2}\kappa\bigl( \Ls(\lambda), \Ls(\lambda) \bigr).
\end{equation}
As $\Ls(\lambda)$ is a $\g$-valued function on $M$, $\Hs(\lambda)$ is simply a function on $M$. Using the expression \eqref{Eq:LaxGaudin}, one finds the following partial fraction decomposition of $\Hs(\lambda)$:
\begin{equation}\label{Eq:HamDecomp}
\Hs(\lambda) =  \Delta_\infty + \sum_{r=1}^N \left( \frac{\Delta_r}{(\lambda-\lambda_r)^2} + \frac{\Hs_r}{\lambda-\lambda_r} \right),
\end{equation}
with
\begin{equation}\label{Eq:CasimirsGaudin}
\Delta_r =  \frac{1}{2}\kappa\bigl( X^{(r)}, X^{(r)} \bigr), \;\;\;\;  \Delta_\infty =  \frac{1}{2}\kappa(\Omega,\Omega)
\end{equation}
and
\begin{equation}\label{Eq:HrGaudin}
\Hs_r = \sum_{s\neq r} \frac{\kappa\bigl(X^{(r)},X^{(s)}\bigr)}{\lambda_r-\lambda_s} + \kappa\bigl(X^{(r)},\Omega\bigr).
\end{equation}
Using the Poisson bracket \eqref{Eq:PBGaudin} and the fact that $\kappa$ is invariant (\textit{i.e.} that $\kappa([X,Y],Z)=\kappa(X,[Y,Z])$ for all $X,Y,Z\in\g$), one finds that
\begin{equation}\label{Eq:HlInvo}
\lwb \Hs(\lambda), \Hs(\mu) \rwb = 0, \;\;\;\;\ \forall\, \lambda,\mu\in\C.
\end{equation}
Thus the quadratic Hamiltonian is in involution with itself for any values of the spectral parameter. In particular, one finds that the $\Hs_r$'s, the $\Delta_r$'s and $\Delta_\infty$ are all in involution with one another.

As it turns out, one can show that the $\Delta_r$'s and $\Delta_\infty$ are Poisson Casimirs of the model, \textit{i.e.} that they have a vanishing Poisson bracket with any function in $\F[M]$. However, the $\Hs_r$'s are not Poisson Casimirs in general and their involution is thus a non trivial result. We will define the Hamiltonian of the model as a linear combination
\begin{equation}\label{Eq:HamComb}
\Hs = \sum_{r=1}^N c_r \Hs_r
\end{equation}
of the $\Hs_r$. In particular, the $\Hs_r$'s are conserved charges in involution of the model.

\paragraph{Lax equation.} It is natural to study the time flow of the Gaudin model, defined as the Hamiltonian flow of $\Hs$ on the phase space: $\p_t=\lbrace \Hs, \cdot \rbrace$. For that, let us compute the Hamiltonian flow of the quadratic Hamiltonian $\Hs(\mu)$ on the Lax matrix $\Ls(\lambda)$. Starting from the bracket \eqref{Eq:PBGaudin}, using the invariance of $\kappa$ and the completeness relation \eqref{Eq:CasComp}, one finds that
\begin{equation}\label{Eq:LaxEqGaudinSpectral}
\bigl\lbrace \Hs(\mu), \Ls(\lambda) \bigr\rbrace = \bigl[ \Ms(\mu,\lambda), \Ls(\lambda) \bigr], \;\;\;\; \text{ with } \;\;\;\; \Ms(\mu,\lambda) = \frac{\Ls(\mu)}{\lambda-\mu}.
\end{equation}
In particular, we find
\begin{equation*}
\lbrace \Hs_r, \Ls(\lambda) \rbrace = \left[ \frac{X^{(r)}}{\lambda-\lambda_r}, \Ls(\lambda) \right].
\end{equation*}
Thus, the time evolution of $\Ls(\lambda)$ takes the form of the Lax equation
\begin{equation}\label{Eq:LaxEqGaudin}
\p_t \Ls(\lambda) = \bigl\lbrace \Hs, \Ls(\lambda) \bigr\rbrace = \bigl[ \Ms(\lambda), \Ls(\lambda) \bigr],
\end{equation}
with
\begin{equation*}
\Ms(\lambda) = \sum_{r=1}^N c_r \frac{X^{(r)}}{\lambda-\lambda_r}.
\end{equation*}
We will explain in Section \ref{Sec:AGM} the relation between this Lax equation when $\g$ is an affine Kac-Moody algebra and the Lax equation \eqref{Eq:Lax} for a field theory.

\paragraph{Higher degree Hamiltonians.} In this subsection, we defined the quadratic Hamiltonians of the Gaudin model, using the bilinear form $\kappa$. These quadratic Hamiltonians were then in involution, as a consequence of the invariance of $\kappa$. This construction generalises for any invariant polynomial on $\g$, as we shall explain now.

Let $\Phi$ be a polynomial of degree $d$ on $\g$. One can see $\Phi$ as a totally symmetric $d$-linear form on $\g$, \textit{i.e.} such that, for all permutation $\s$ of $\lbrace 1,\cdots,d \rbrace$, we have
\begin{equation*}
\Phi\bigl(Y_1,\cdots,Y_d\bigr) = \Phi\bigl(Y_{\s(1)},\cdots,Y_{\s(d)}\bigr), \;\;\;\; \forall \; Y_1,\cdots,Y_d \in \g.
\end{equation*}
We say that $\Phi$ is invariant if for any $Y_1,\cdots,Y_d,Z\in\g$ we have
\begin{equation*}
\Phi\bigl([Z,Y_1],Y_2,\cdots,Y_d\bigr) + \cdots + \Phi\bigl(Y_1,Y_2,\cdots,[Z,Y_d]\bigr) = 0.
\end{equation*}
In particular, the bilinear form defines an invariant polynomial of degree two on $\g$.

Let $\Phi$ be an invariant polynomial of degree $d$ on $\g$. Evaluating this polynomial on the Lax matrix of the Gaudin model, we define the charge
\begin{equation}\label{Eq:ChargeInvPol}
\Qs_\Phi(\lambda) = \frac{1}{d} \Phi\bigl( \Ls(\lambda), \cdots, \Ls(\lambda) \bigr).
\end{equation}
In particular, the quadratic Hamiltonian is then $\Hs(\lambda)=\Qs_\kappa(\lambda)$.

Let $\Phi$ and $\Psi$ be two invariant polynomials on $\g$. Starting from the Poisson bracket \eqref{Eq:PBGaudin} of the Lax matrix, one can compute the Poisson bracket of the charges $\Qs_\Phi(\lambda)$ and $\Qs_\Psi(\mu)$. Using the invariance of $\Phi$ and $\Psi$, one finds that
\begin{equation*}
\bigl\lbrace \Qs_\Phi(\lambda), \Qs_\Psi(\mu) \bigr\rbrace = 0.
\end{equation*}
Thus, charges constructed from invariant polynomials are in involution one with another. In particular, they are in involution with $\Hs(\lambda)$ and thus with the Hamiltonian $\Hs$, hence they are conserved. This is a general method for constructing a large number of commuting conserved charges in Gaudin models.

\subsection{Real classical Gaudin models}
\label{SubSec:RealGaudin}

So far, we discussed complex Gaudin models, defined on complex Lie algebras. Let us now consider real Gaudin models. Let $\g_0$ be a real quadratic Lie algebra. If we choose points $\lambda_1,\cdots,\lambda_N$ on the real line $\R$, the construction presented above for a complex Lie algebra easily applies to the real algebra (by also considering the spectral parameter $\lambda$ to be real). We then get a real Gaudin model.

However, one can consider a larger class of real Gaudin models, by allowing the sites $\lambda_k$'s to be complex. In this case, one has to change the construction of the Gaudin model. Let $\g$ be the complexification of $\g_0$: $\g_0$ can then be seen as the real form $\g_0=\g^\tau$ for some antilinear involutive automorphism $\tau$ of $\g$ (see Appendix \ref{App:RealForms}). Note that the quadratic form $\kappa$ on $\g_0$ extends naturally in a non-degenerate form on $\g$ invariant under $\tau$.

As complex ones, a real Gaudin model is defined by the data of sites $\lambda_1,\cdots,\lambda_N$ in $\C$. However, for the real model we shall suppose that if $\lambda_k$ is a site, $\overline{\lambda}_k$ is not. The phase space of the real model is then defined as follows. If $\lambda_k$ is a site of the model in $\R$, then we consider the real dual $\g_0^*$ of $\g_0$, equipped with the Kirillov-Kostant bracket. This space is described by a $\g_0$-valued observable $X^{(k)}$ as defined in equation \eqref{Eq:XGaudin} (bur for $\g_0$ instead of $\g$). If $\lambda_k$ is a site in $\C\setminus\R$, then we consider the complex dual $\g^*$ of the complexification $\g$, but as a real vector space (which is then of dimension $2\dim(\g_0)$). This space is described by an observable $X^{(k)}$ which is $\g$-valued and not $\g_0$-valued. The phase space $M$ of the real Gaudin model is then the Cartesian product of these vectors spaces associated with each site.

We define the Lax matrix of the model as the following $\g$-valued function of the spectral parameter $\lambda\in\C$:
\begin{equation}\label{Eq:LaxRealGaudin}
\Ls(\lambda) = \sum_{ \substack{k=1 \\ \lambda_k\in\R}}^N \frac{X^{(k)}}{\lambda-\lambda_k} + \sum_{ \substack{k=1 \\ \lambda_k\in\C\setminus\R}}^N \left( \frac{X^{(k)}}{\lambda-\lambda_k} + \frac{\tau\bigl(X^{(k)}\bigr)}{\lambda-\overline{\lambda}_k} \right) + \Omega,
\end{equation}
where $\Omega$ is a constant element of $\g_0$. Using the definitions of $X^{(k)}$ as being either $\g_0$-valued or $\g$-valued, depending on whether $\lambda_k$ is real or not, one checks that this Lax matrix satisfies the reality condition
\begin{equation}\label{Eq:RealGaudin}
\tau\bigl( \Ls(\lambda) \bigr) = \Ls(\overline{\lambda}).
\end{equation}
The rest of the construction of the real Gaudin model is then similar to the complex case. The Poisson bracket of the Lax matrix \eqref{Eq:LaxRealGaudin} of the real model is exactly the same as the complex one, \textit{i.e.} the bracket \eqref{Eq:PBGaudin}.

One defines the quadratic Hamiltonian $\Hs(\lambda)$ as in \eqref{Eq:GaudinHamLambda}, which also satisfies the involution equation \eqref{Eq:HlInvo}. Using the fact that the extended form $\kappa$ on $\g$ is $\tau$-invariant, one finds that this Hamiltonian satisfies the reality condition
\begin{equation*}
\overline{\Hs(\lambda)} = \Hs(\overline{\lambda}).
\end{equation*}
One can then write a partial fraction decomposition of $\Hs(\lambda)$ as in \eqref{Eq:HamDecomp} but with also double and simple poles at $\overline{\lambda}_k$ if $\lambda_k\notin\R$, with coefficients $\overline{\Delta_k}$ and $\overline{\Hs_k}$. One can then define a real Hamiltonian $\Hs$ as in \eqref{Eq:HamComb} by a linear combination of the $\Hs_k$'s and $\overline{\Hs_k}$'s with appropriate reality conditions on the coefficients.

If one considers only real sites $\lambda_k$'s, then this definition of the real Gaudin model agrees with the simplest one presented at the beginning of this subsection, as one then considers only $\g_0$-valued $X^{(k)}$'s.

\subsection{Finite Gaudin models and an example}
\label{SubSec:FiniteClassGaudin}

\paragraph{Classical finite Gaudin models.} In this subsection, we will suppose that $\g$ is a Kac-Moody algebra of finite type, \textit{i.e.} a finite dimensional semi-simple Lie algebra (see Appendix \ref{App:SemiSimple}). The associated Gaudin model then possesses a finite number of degrees of freedom and is thus a system of classical mechanics. In this case, the equation \eqref{Eq:LaxEqGaudin} is what is generally called a Lax equation for a mechanical system (see for example~\cite{Babebook}).

Finite algebras admit many invariant polynomials. For example, let us consider a representation of $\g$ (we can then consider the elements of $\g$ as matrices): the polynomial $X\in\g \mapsto \Tr(X^d)$ is then an invariant polynomial of degree $d$ on $\g$. This way, one can construct a large number of commuting conserved charges in involution for finite Gaudin models.

Following the construction explained above, it appears that there are an infinite number of such charges (for example considering all powers $d\in\N$). However, these charges are not all independent: for example, for matrices of size $n$, the traces $\Tr(X^d)$'s for $d > n$ can be expressed in terms of the ones for $d \leq n$. The structure of the algebra of invariant polynomials on a finite Lie algebra $\g$ has been completely described in the literature~\cite{Dixmier:1977}. In particular, it is generated by $\ell$ independent polynomials $\Phi_i$'s, where $\ell$ is the rank of $\g$. These polynomials have degrees $d_i+1$, where the $d_i$'s are the so-called exponents of the Lie algebra $\g$ (we shall come back on that fact in Chapter \ref{Chap:QuantumFinite}). One then finds all independent charges in involution of the Gaudin model by considering the charges \eqref{Eq:ChargeInvPol} associated with the fundamental invariant polynomials $\Phi_i$.

\paragraph{An example: the unreduced Neumann model.} Let us consider the so-called Neumann model~\cite{Neumann1859}. This model describes the movement of a particle in an (anisotropic) harmonic potential and constrained on a $(N-1)$-dimensional sphere $\mathbb{S}^{N-1}$. We describe the particle by its position $(x_1,\cdots,x_N)$ by embedding the sphere $\mathbb{S}^{N-1}$ in $\R^N$. We then have the constraint $\sum_{i=1}^N x_i^2 = 1$. This model was shown to be integrable by Uhlenbeck~\cite{Uhlenbeck:1982}.

We will consider here a slightly different model, that we shall call the unreduced Neumann model. It is a model on the whole space $\R^N$, which once reduced to the sphere $\mathbb{S}^{N-1}$ coincides with the Neumann model. The phase space of the model is given by the positions $x_i$'s and the corresponding momenta $p_i$'s ($i=1,\cdots,N$). The Hamiltonian of the model is
\begin{equation*}
\Hc_N = \frac{1}{2} \sum_{i=1}^N \omega_i^2 x_i^2 + \frac{1}{4} \sum_{i\neq j} J_{ij}^2,
\end{equation*}
with $\omega_i$ the pulsation of the harmonic oscillator in the axis $i$ and
\begin{equation*}
J_{ij} = x_ip_j - x_jp_i.
\end{equation*}
Let us define the so-called Ulhenbeck quantities
\begin{equation}\label{Eq:Ulh}
\mathcal{F}_i = x_i^2 + \sum_{\substack{ j = 1 \\ j \neq i }}^N \dfrac{J_{ij}^2}{\omega_i^2 - \omega_j^2},
\end{equation}
for $i\in\lbrace 1,\cdots,N\rbrace$. One shows that these are independent conserved charges in involution of the unreduced Neumann model (hence proving its Liouville integrability~\cite{Babebook}). They are related to the Hamiltonian by
\begin{equation}\label{Eq:HamUlh}
\Hc_N = \sum_{i=1}^N \omega_i^2 \mathcal{F}_i.
\end{equation}
Let us consider the two by two matrices~\cite{Avan:1991ib} (see also~\cite{Ragnisco1997})
\begin{equation}\label{Eq:LaxNeumann}
\Lc_N(\lambda) = \sum_{k=1}^N \frac{1}{\lambda-\omega_k^2}\left(\begin{array}{cc} x_k p_k &  x_k^2\\ -p_k^2 & -x_k p_k \end{array}\right) +\Omega, \;\;\;\; \M_N(\lambda) = \sum_{k=1}^N \left(\begin{array}{cc} x_k p_k &  x_k^2\\ -p_k^2 & -x_k p_k \end{array}\right) + \lambda\Omega,
\end{equation}
depending on a spectral parameter $\lambda$, with $\Omega=\left(\begin{array}{cc} 0 &  0\\ -1 & 0 \end{array}\right)$. The equations of motion of the unreduced Neumann model can be recast in the form of the Lax equation
\begin{equation*}
\frac{\dd\;}{\dd t} \Lc_N(\lambda) = \left\lbrace \Hc_N, \Lc_N(\lambda) \right\rbrace = \left[ \M_N(\lambda), \Lc_N(\lambda) \right].
\end{equation*}~

During my PhD, I have co-supervised a master thesis about the Neumann model. The main subject of this project was the search of a two by two Lax pair for an integrable deformation of the Neumann model (for $N=3$), introduced recently by Arutyunov and Medina-Rincon in~\cite{Arutyunov:2014cda}. Another part of the project was to understand the reinterpretation of the unreduced Neumann model as a Gaudin model on the Lie algebra $\sl(2,\R)$, as stated in~\cite{Kuznetsov:1992}. We will present this reinterpretation here, as an example of what we mean exactly when saying that a model is realised as a Gaudin model.

We consider the simple real Lie algebra $\g_0=\sl(2,\R)$, equipped with the invariant bilinear form $\kappa(X,Y)=\frac{1}{2}\Tr(XY)$, making it a quadratic Lie algebra. Let us fix a basis
\begin{equation*}
H=\begin{pmatrix}
1 & 0 \\
0 & -1
\end{pmatrix}, \;\;\;\;
E=\begin{pmatrix}
0 & 1 \\
0 & 0
\end{pmatrix} \;\;\;\; \text{and} \;\;\;\;
F=\begin{pmatrix}
0 & 0 \\
1 & 0
\end{pmatrix}
\end{equation*}
of $\g_0$. We consider a classical real Gaudin model associated with $\g_0=\sl(2,\R)$ and with $N$ sites $\lambda_r=\omega_r^2$ ($r=1,\cdots,N$). As these are real sites, the construction of this real Gaudin model is similar to the complex one (see Subsection \ref{SubSec:RealGaudin}). The phase space $M$ of this model (see subsections \ref{SubSec:GaudinPhaseSpace} and \ref{SubSec:RealGaudin}) is described by $3N$ fundamental functions $X^H_{(r)}$, $X^E_{(r)}$ and $X^F_{(r)}$. The non-vanishing Poisson brackets of these functions are
\begin{equation*}
\bigl\lbrace X^H_{(r)}, X^E_{(s)} \bigr\rbrace = 2\delta_{rs} X^E_{(r)}, \;\;\;\; \bigl\lbrace X^H_{(r)}, X^F_{(s)} \bigr\rbrace = -2\delta_{rs} X^F_{(r)} \;\;\;\; \text{and} \;\;\;\; \bigl\lbrace X^E_{(r)}, X^F_{(s)} \bigr\rbrace = \delta_{rs} X^H_{(r)}.
\end{equation*}
The Lax matrix of the Gaudin model is then
\begin{equation*}
\Ls(\lambda) = \sum_{k=1}^N \frac{1}{\lambda-\omega_k^2} \begin{pmatrix}
X^H_{(k)} & 2X^F_{(k)} \\
2X^E_{(k)} & -X^H_{(k)}
\end{pmatrix} + \Omega,
\end{equation*}
where by anticipation we already choose the constant matrix $\Omega$ to be the one appearing in \eqref{Eq:LaxNeumann}. The presence of factors 2 in the above equation is due to the expression of the form $\kappa$ in the basis $\lbrace H,E,F \rbrace$: $\kappa(H,H)=1$, $\kappa(E,F)=\kappa(F,E)=\frac{1}{2}$.\\

The phase space $M$ of the Gaudin model can be seen as the vector space $\R^{3N}$ with coordinates $\bigl(X^H_{(k)}, X^E_{(k)}, X^F_{(k)}\bigr)_{k=1,\cdots,N}$. In the same way, the phase space of the unreduced Neumann model is simply the vector space $\R^{2N}$ with canonical coordinates $(x_k,p_k)_{k=1,\cdots,N}$. We define the following map from $\R^{2N}$ to $M$:
\begin{equation*}
\begin{array}{rcll}
\pi : &    \R^{2N}     & \longrightarrow & M \simeq \R^{3N} \\
      & (x_k,p_k)_{k=1,\cdots,N}  &   \longmapsto  & \bigl(x_kp_k, -\frac{1}{2}p_k^2, \frac{1}{2}x_k^2 \bigr)_{k=1,\cdots,N}
\end{array}.
\end{equation*}
One checks that this is a Poisson map, \textit{i.e.} that $x_kp_k$, $-\frac{1}{2}p_k^2$ and $\frac{1}{2}x_k^2$ satisfy the same Poisson bracket as $X^H_{(k)}$, $X^E_{(k)}$ and $X^F_{(k)}$. The map $\pi$ induces a morphism of algebras $\pi^*$ from $\F[M]$ to $\F[\R^{2N}]$ by pullback:
\begin{equation*}
\begin{array}{rccc}
\pi^* : & \F[M] & \longrightarrow & \F[\R^{2N}] \\
        &   f   & \longmapsto     & f \circ \pi
\end{array}.
\end{equation*}
As $\pi$ is a Poisson map, $\pi^*$ sends the Poisson bracket on $\F[M]$ to the one on $\F[\R^{2N}]$. We will say that $\pi$ is a realisation of the Gaudin phase space $M$.

As we will see, $\pi$ maps the unreduced Neumann model on the $\sl(2,\R)$-Gaudin model. Indeed, it is clear that
\begin{equation*}
\pi^*\bigl( \Ls (\lambda) \bigr) = \Lc_N(\lambda),
\end{equation*}
hence the Lax matrix of the unreduced Neumann model coincides with the one of the Gaudin model through the map $\pi^*$. In the same way, one checks that the Ulhenbeck quantities \eqref{Eq:Ulh} coincide with the quadratic Hamiltonians \eqref{Eq:HrGaudin} of the Gaudin model:
\begin{equation*}
\pi^*\bigl( \Hs_k ) = \mathcal{F}_k = \frac{1}{2} \res_{\lambda=\omega_k^2} \kappa \bigl( \Lc_N(\lambda),\Lc_N(\lambda) \bigr).
\end{equation*}
By equation \eqref{Eq:HamUlh}, we then have
\vspace{-5pt}\begin{equation*}
\pi^*(\Hs) = \Hc_N, \;\;\;\; \text { with } \;\;\;\; \Hs = \sum_{k=1}^N \omega_k^2 \Hs_k \vspace{-5pt}
\end{equation*}
of the form \eqref{Eq:HamComb}. The quadratic Hamiltonians $\Hs_k$ are part of the quadratic charge $\Hs(\lambda)$. Recall from \eqref{Eq:HamDecomp} that $\Hs(\lambda)$ also contains the Casimirs $\Delta_k$, defined in \eqref{Eq:CasimirsGaudin}. It is easy to check that these Casimirs vanishes under the realisation $\pi$:
\begin{equation*}
\pi^* (\Delta_k) = 0.
\end{equation*}
This is to be partially expected: indeed, the canonical phase space $\R^{2N}$ of the model is symplectic, \textit{i.e.} its Poisson bracket does not possess non constant Casimirs. The image of $\Delta_k$ under $\pi^*$ should then be a constant (and in this case is actually equal to zero).

This illustrates more generally what we mean when saying that a model is identified as a Gaudin model: there exists a Poisson map from the phase space of the model to the phase space of the Gaudin model, which maps the Hamiltonian of the model to the Gaudin Hamiltonian. Note however that this map is in general not an isomorphism, so that the phase space of the initial model is not isomorphic to the phase space of the Gaudin model. This is the case for the map $\pi$ in the case of the unreduced Neumann model: it clearly cannot be surjective on dimensional grounds (and the phase spaces $\R^{2N}$ and $M$ cannot be isomorphic as $\R^{2N}$ is symplectic and $M$ is not).

\subsection{Cyclotomic and dihedral Gaudin models with arbitrary multiplicities}
\label{SubSec:GaudinGen}

In this subsection, we say a few words about various generalisations of the classical Gaudin model described previously in this section. The most general formalism of a dihedral Gaudin model with arbitrary singularities is described in the article~\cite{Vicedo:2017cge} (although this article treats the case of an affine Lie algebra, the transcription of the formalism to an arbitrary quadratic Lie algebra is straightforward). These generalisations are necessary to understand entirely the reinterpretation of integrable $\s$-models as (dihedral) affine Gaudin models.

\paragraph{Gaudin models with multiplicities.} For this section, we restricted to Gaudin models whose sites $\lambda_1,\cdots,\lambda_N$ have multiplicity one. Let us say a few words about Gaudin models with arbitrary multiplicities $m_1,\cdots,m_N \in \Z_{\geq 1}$~\cite{Feigin:2006xs}. We will focus on complex Gaudin models, the generalisation to real ones being quite similar to the case with multiplicity one (as discussed in Subsection \ref{SubSec:RealGaudin}).

The phase space of the model is the Cartesian product of $N$ Poisson vector spaces associated with the $N$ sites. The space associated with $\lambda_r$ is the Kirillov-Kostant space $\Ta^{m_r}\g^*$ associated with the so-called Takiff algebra $\Ta^{m_r}\g$ of multiplicity $m_r$~\cite{Takiff:1971}. The Takiff algebra $\Ta^m\g$ is constructed from the Lie algebra $\g$ as follows. We denote by $\C[\varepsilon]$ the algebra of complex polynomials of a formal variable $\varepsilon$. The subspace $\varepsilon^{m}\C[\varepsilon]$ is then an ideal of $\C[\varepsilon]$. The Takiff algebra associated with $\g$ and of multiplicity $m$ is then defined as
\begin{equation*}
\Ta^m \g = \g \otimes \C[\epsilon]/\epsilon^{m}\C[\epsilon].
\end{equation*}
If $m=1$, one has $\C[\epsilon]/\epsilon\C[\epsilon] \simeq \C$, so that the Takiff algebra $\Ta^1\g$ coincide with $\g$. Let us describe more concretely the Takiff algebra $\Ta^m\g$. We consider a basis $\lbrace I^a \rbrace$ of $\g$, with associated structure constants $\ft{ab}{c}$. Then the Takiff algebra has a basis $I^a_{[p]}$, with $p=0,\cdots,m-1$, with bracket
\begin{equation*}
\Bigl[ I^a_{[p]}, I^b_{[q]} \Bigr] = \left\lbrace \begin{array}{ll}
\fs{ab}{c} I^c_{[p+q]} & \text{ if } p+q\leq m-1,\\
0           & \text{ if } p+q > m-1.
\end{array}  \right.
\end{equation*}

The phase space of the Gaudin model is defined as
\begin{equation*}
M = \Ta^{m_1}\g^* \times \cdots \times \Ta^{m_N}\g^*,
\end{equation*}
where each dual $\Ta^m\g^*$ is equipped with the Kirillov-Kostant bracket. This phase space is then parametrized by $\g$-valued observables $X^{(r)}_{[p]}$, with $r=1,\cdots,N$ and $p=0,\cdots,m_r-1$, satisfying the brackets
\begin{equation}\label{Eq:KKTakiff}
\Bigl\lbrace X^{(r)}_{[p]}\null\ti{1}, X^{(s)}_{[q]}\null\ti{2} \Bigr\rbrace = \left\lbrace \begin{array}{ll}
\delta_{rs}\left[ C\ti{12}, X^{(r)}_{[p+q]}\null\ti{1} \right] & \text{ if } p+q\leq m_r-1,\\
0           & \text{ if } p+q > m_r-1.
\end{array}  \right.
\end{equation}
The Lax matrix of the model is defined as~\cite{Feigin:2006xs}
\begin{equation}\label{Eq:LaxMultiple}
\Ls(\lambda) = \sum_{r=1}^N \sum_{p=0}^{m_r-1} \frac{X^{(r)}_{[p]}}{(\lambda-\lambda_r)^{p+1}} + \Omega,
\end{equation}
with $\Omega$ a constant element of $\g$. As the multiplicity $m_r$ appears as the order of the pole at $\lambda_r$ in the Lax matrix, we will sometimes speak of a Gaudin model with simple, double, triple ... poles instead of a model with sites and multiplicities.

The important result is that this Lax matrix satisfies the same Poisson bracket \eqref{Eq:PBGaudin} as the Lax matrix for simple poles, with the same matrix $\rc\ti{12}(\lambda,\mu)$. One can then construct the Gaudin model in a similar fashion as in the multiplicity one case. In particular, we can consider the quadratic Hamiltonian \eqref{Eq:GaudinHamLambda}, which satisfies the involution equation \eqref{Eq:HlInvo}. This Hamiltonian can then be written as a partial fraction decomposition similar to \eqref{Eq:HamDecomp} but with higher order poles (which gives more conserved charges in involution). The Lax equation \eqref{Eq:LaxEqGaudinSpectral} still holds without corrections.

The most general Gaudin model can also have a site at infinity, with an arbitrary multiplicity. As this requires a somehow special treatment, we will not consider this case here and refer to~\cite{Vicedo:2017cge} (note however that the constant element $\Omega$ in \eqref{Eq:LaxMultiple} corresponds to a double pole at infinity). 

\paragraph{Cyclotomic Gaudin models.} Let us now say a few words about cyclotomic Gaudin models. These are a generalisation of the complex Gaudin model presented in this section (which appeared first in~\cite{Skrypnyk:2005} for the case with no multiplicities and in~\cite{Vicedo_161109059} for the case with arbitrary multiplicities), for Lie algebras $\g$ which possess an automorphism $\s$ of finite order $T$ (see appendix \ref{App:Torsion}). We will use here a formalism close to the one we used in subsection \ref{SubSec:ZT} to described $\Z_T$-coset $\s$-models. The action of $\s$ on $\g$ defines an action of the cyclic group $\Z_T$. One can also define an action of $\Z_T$ on the complex plane \textit{via} the multiplication by $\omega$ a primitive $T^{\rm th}$-root of unity. Cyclotomic Gaudin models are models with a Lax matrix of a form similar to \eqref{Eq:LaxMultiple} but which satisfies the equivariance condition
\begin{equation}\label{Eq:EquivGaudin}
\s \bigl( \Ls(\lambda) \bigr) = \omega\Ls(\omega\lambda).
\end{equation}
Before explaining the construction \textit{ex nihilo} of cyclotomic Gaudin models, let us first gain some intuitions about what these models should satisfy. Indeed, the equivariance condition \eqref{Eq:EquivGaudin} imposes several restrictions on the theory.

Let $\lambda_r$ be a pole of $\Ls(\lambda)$ different from the origin (which is then not fixed under the multiplication by $\omega$), then all the points of the orbit $\Z_T.\lambda_r = \lbrace \omega^k \lambda_r, k=0,\cdots,T-1 \rbrace$ must be poles of $\Ls(\lambda)$, with same multiplicity $m_r$. Moreover, the coefficients over the poles at these points are related to the $X^{(r)}_{[p]}$'s by action of powers of $\s$.

In the same way, let us consider the point $0$, which is fixed under the multiplication by $\omega$. If it is a pole of $\Ls(\lambda)$, then one has
\begin{equation*}
\s\left( X^{(0)}_{[p]} \right) = \omega^{-p} X^{(0)}_{[p]},
\end{equation*}
\textit{i.e.} $X^{(0)}_{[p]}$ is in the grading $\g^{(-p)}$ of $\g$ (see $\Z_T$-gradings in Appendix \ref{App:Torsion}).\\

To construct a cyclotomic Gaudin model, one then considers sites $\lambda_0=0,\lambda_1,\cdots,\lambda_N$, with multiplicities $m_0,m_1,\cdots,m_N$, and associate with them some observables $X^{(r)}_{[p]}$ ($p=0,\cdots,m_r-1$), which are $\g$-valued for $r=1,\cdots,N$ and which are in gradings $\g^{(-p)}$ for $r=0$. We suppose that the $\lambda_r$ are not in the same $\Z_T$-orbits. We then define the Lax matrix to be (with $\Omega\in\g^{(1)}$ constant)~\cite{Vicedo_161109059,Vicedo:2017cge}
\begin{equation}\label{Eq:LaxGaudinCyc}
\Ls(\lambda) = \frac{1}{T}\sum_{k=0}^{T-1}\sum_{r=1}^N \sum_{p=0}^{m_r-1} \frac{\omega^{kp} \s^k X^{(r)}_{[p]}}{(\lambda-\omega^k \lambda_r)^{p+1}} + \sum_{p=0}^{m_0-1} \frac{X^{(0)}_{[p]}}{\lambda^{p+1}} + \Omega.
\end{equation}
One then checks that this Lax matrix satisfies the equivariance condition \eqref{Eq:EquivGaudin}. Note that for all sites $\lambda_r$ ($r=1,\cdots,N$), this Lax matrix possesses poles at all $\omega^k\lambda_r$ in the orbit $\Z_T.\lambda_r$, as expected above. However, the coefficients of these poles are not independent of the ones at $\lambda_r$. Thus, in the cyclotomic Gaudin model, all these poles have to be regarded as attached to one unique site.

So far, we specified a Lax matrix but not the phase space of the model. This phase space is parametrised by the observables $X^{(r)}_{[p]}$. For $r=1,\cdots,N$, we will suppose that $X^{(r)}_{[p]}$ satisfies the Kirillov-Kostant bracket \eqref{Eq:KKTakiff} of the Takiff algebra $\Ta^{m_r}\g$. The observables $X^{(0)}_{[p]}$ satisfy more complicated brackets, which take into account the fact that these are restricted to particular gradings. We will not enter into details here and refer to~\cite{Vicedo:2017cge}. These brackets on $X^{(r)}_{[p]}$ are made for the Lax matrix $\Ls(\lambda)$ to satisfy the bracket
\begin{equation}\label{Eq:PBGaudin2}
\lwb \Ls\ti{1}(\lambda), \Ls\ti{2}(\mu) \rwb = \lsb \rc\ti{12}(\lambda,\mu), \Ls\ti{1}(\lambda) \rsb - \lsb \rc\ti{21}(\mu,\lambda), \Ls\ti{2}(\mu) \rsb,
\end{equation}
with
\begin{equation}\label{Eq:rGaudinCyc}
\rc\ti{12}(\lambda,\mu) = \frac{1}{T}\sum_{k=0}^{T-1} \frac{\s^k\ti{1}C\ti{12}}{\mu-\omega^{-k}\lambda}.
\end{equation}
In particular, for a finite algebra $\g$ with an automorphism $\s$, $\rc$ coincides with the standard $\Rc$-matrix on $\Lc(\g)$ twisted by $\s$, as introduced in \eqref{Eq:RCyc}. The construction of quadratic Hamiltonians in involution and the associated Lax equations developed in the non-cyclotomic case then generalise to the cyclotomic case.

\paragraph{Dihedral Gaudin models.} Let us finally introduce dihedral Gaudin models~\cite{Vicedo:2017cge}. These are the real equivalents of the cyclotomic Gaudin models introduced above. Let $\g_0$ be a real Lie algebra with a $\Z_T$-grading $\g_0 = \bigoplus_{p=0}^{T-1} \g_0^{(p)}$. According to Corollary \ref{Cor:RealGradings}, if $\g_0$ is the real form $\g^\tau$ of $\g$, these gradings are in one-to-one correspondence with automorphisms $\s$ of $\g$ of order $T$, which satisfy the \textbf{dihedrality condition}
\begin{equation*}
\s \circ \tau  = \tau\circ\s^{-1}.
\end{equation*}
A dihedral Gaudin model is then a model whose Lax matrix $\Ls(\lambda)$ satisfies both the reality condition \eqref{Eq:RealGaudin} and the equivariance condition \eqref{Eq:EquivGaudin}:
\begin{equation}\label{Eq:DihedGaudin}
\tau\bigl( \Ls(\lambda) \bigr) = \Ls(\overline{\lambda}) \;\;\;\;  \text{ and } \;\;\;\; \s \bigl( \Ls(\lambda) \bigr) = \omega\Ls(\omega\lambda).
\end{equation}
Such a model is constructed by combining the constructions of real Gaudin models (in Subsection \ref{SubSec:RealGaudin}) and the one of cyclotomic Gaudin models above. Let $\Gamma_T$ be the dihedral group of order $2T$, as defined in \eqref{Eq:DihedralGroup}, and consider its action \eqref{Eq:ActionDihedralPlan} on the complex plane \textit{via} multiplication by $\omega$ and complex conjugation. We fix a certain number of sites $\lambda_0=0,\lambda_1,\cdots,\lambda_N$ in $\C$, whose orbits under $\Gamma_T$ are disjoint, with multiplicity $m_0,m_1,\cdots,m_N$. We then attach to these sites certain objects $X^{(r)}_{[p]}$ ($p=0,\cdots,m_r-1$), whose properties depend on the orbit $\Gamma_T.\lambda_r$:
\begin{enumerate}[(i)]\setlength\itemsep{0.1em}
\item if $\Gamma_T.\lambda_r$ is of size one (fixed point), \textit{i.e.} $r=0$, then $X^{(0)}_{[p]} \in \g_0^{(-p)}$,
\item if $\Gamma_T.\lambda_r$ is of size $T$, \textit{i.e.} (at least) one element of $\Gamma_T.\lambda_r$ is real and non-zero, then $X^{(r)}_{[p]} \in \g_0$,
\item if $\Gamma_T.\lambda_r$ is of maximal size $2T$, \textit{i.e.} all elements $\Gamma_T.\lambda_r$ are in $\C\setminus\R$, then $X^{(r)}_{[p]} \in \g$.
\end{enumerate}
The Lax matrix is then constructed from these objects to satisfy the conditions \eqref{Eq:DihedGaudin}. In particular, it has poles at all points of the orbits $\Gamma_T.\lambda_r$, but the coefficients at these poles are related to the ones at $\lambda_r$ by actions of $\tau$ and powers of $\s$. Hence, we see all these poles as a unique site of the model.

The Poisson brackets satisfied by the $X^{(r)}_{[p]}$ are made so that the Lax matrix satisfies the bracket \eqref{Eq:PBGaudin2} with $\rc$ as in \eqref{Eq:rGaudinCyc}. In the cases (ii) and (iii) above, the $X^{(r)}_{[p]}$'s obey the Takiff Kirillov-Kostant brackets \eqref{Eq:KKTakiff} of respectively $\g_0$ and $\g$. The Poisson brackets of the $X^{(0)}_{[p]}$'s are more complicated and we simply refer to~\cite{Vicedo:2017cge} for their expression.

One could also consider a site at infinity. In this case, the associated $X^{(\infty)}_{[p]}$ would have properties similar to the ones attached to the origin, as $\infty$ is a fixed point under the action of $\Gamma_T$ (we refer to~\cite{Vicedo:2017cge} for the full construction).

Let us end this paragraph by an example. The unreduced Neumann model presented in Subsection \ref{SubSec:FiniteClassGaudin} also possesses another Lax matrix, which is of size $N \times N$. This alternative formulation is a realisation of a $\Gamma_2$-dihedral Gaudin model on $\sl(N,\C)$, with automorphism $\s : X \mapsto -\null^t X$. 

\section{Classical AGM as field theories with twist function}
\label{Sec:AGM}

In this section, we will focus on classical Affine Gaudin Models (AGM). In particular, using the description of affine Kac-Moody algebras in terms of loop algebras, we will explain how AGM can be seen as integrable field theories with twist function and with space coordinate $x$ on the circle $\Sc$. This section is mostly a review of the article~\cite{Vicedo:2017cge} of Vicedo.

\subsection[Models with twist function and the Lie algebra of $\g$-connections on the circle]{Models with twist function and the Lie algebra of $\bm{\g}$-connections on the circle}
\label{SubSec:Connections}

Before entering into the formalism of affine algebras, let us try to get intuition on how a Gaudin model can be understood as a model with twist function. Recall that the Lax matrix of a Gaudin model satisfies a Poisson bracket of the form \eqref{Eq:PBGaudin2}. The Lax matrix of a model with twist function, on the other hand, satisfies a Poisson bracket of the form \eqref{Eq:PBR}. At first sight, these two brackets seem quite dissimilar. In this section, we explain how the bracket \eqref{Eq:PBR} can be put in a form similar to the one of Gaudin models, using the Lie algebra of $\g$-connections on the circle.\\

The fundamental dynamical fields of a field theory on the circle are observables-valued distributions on $\Sc$. They can be seen as infinite Fourier series
\begin{equation}\label{Eq:FieldDist}
\phi(x) = \sum_{n\in\Z} c_n e^{inx}, \;\;\;\; x \in \Sc \simeq [0,2\pi],
\end{equation}
where the coefficients $c_n$ belong to $\F[M]$, \textit{i.e.} are functions on the phase space $M$ of the model (here, we consider complex fields: real ones are treated in a similar fashion requiring $\overline{c}_n=c_{-n}$). The $k^{\rm th}$-derivative of the field $\phi$ with respect to $x$ is then
\begin{equation*}
\p_x^k \phi(x) = \sum_{n\in\Z} (in)^k c_n e^{inx}.
\end{equation*}
We will denote by $\Tt(\Sc)$ the space of trigonometric polynomials on $\Sc$, \textit{i.e.} the vector space with basis
\vspace{-2pt}\begin{equation*}
\begin{array}{rccc}
e_n : & \Sc & \longrightarrow & \C \\
      &  x  & \longmapsto     & e^{inx}
\end{array}.\vspace{-4pt}
\end{equation*}
A field $\phi(x)$ of the form \eqref{Eq:FieldDist} is then in the tensor product $\F[M] \otimes \Tt(\Sc)$ (more precisely, in a completion of this tensor product, as the sum in \eqref{Eq:FieldDist} is infinite).

Let $\g$ be the finite Lie algebra underlying the considered model with twist function. We define the space of $\g$-connections on $\Sc$ as
\begin{equation*}
\Co = \C \p \oplus \g \otimes \Tt(\Sc).
\end{equation*}
Elements of $\Co$ are represented by their evaluation at $x\in\Sc$, which sends $\p$ to $\p_x$ and $e_n$ to $e_n(x)=e^{inx}$. An element $\nabla=\kay \p + \Jc$ in $\Co$, with $\kay\in\C$ and $\Jc\in \g \otimes \Tt(\Sc)$ a $\g$-valued trigonometric polynomial, can then be seen as the $\g$-connection
\begin{equation*}
\nabla(x) = \kay \p_x + \Jc(x).
\end{equation*}
One defines a bracket $\LB$ of two $\g$-valued trigonometric polynomials by considering the point-wise bracket on $\g$. We then extend it to $\Co$ as
\begin{equation*}
\bigl[ \kay_1 \p + \Jc_1, \kay_2 \p + \Jc_2 \bigr](x) = [\Jc_1(x), \Jc_2(x)] + \kay_1 \p_x \Jc_2(x) - \kay_2(x) \p_x \Jc_1(x).
\end{equation*}
One checks that $\LB$ is a Lie bracket on $\Co$.\\

Let us come back to the considered field theory with twist function. From its twist function $\vp(\lambda)$ and its Lax matrix $\Lc(\lambda,x)$, we define the following $\g$-connection on $\Sc$, depending on the spectral parameter $\lambda$:
\begin{equation}\label{Eq:LaxConnection}
\nabla(\lambda,x) = \vp(\lambda) \bigl( \p_x + \Lc(\lambda,x) \bigr) \in \Co.
\end{equation}
As $\Lc(\lambda,\cdot)$ is a $\g$-valued field, its coefficients in a basis of $\g$ are thus elements of (the completion of) the tensor factor $\F[M] \otimes \Tt(\Sc)$: thus, $\nabla(\lambda,\cdot)$ is an element of (the completion of) the tensor product $\F[M] \otimes \Co$. The main result of this subsection is the fact that the Poisson bracket \eqref{Eq:PBR} of $\Lc$, with the $\Rc$-matrix \eqref{Eq:DefR}, can be rewritten as
\begin{equation}\label{Eq:PBNabla}
\lwb \nabla\ti{1}(\lambda,x), \nabla\ti{2}(\mu,y) \rwb = \left[ \Rc^0\ti{12}(\lambda,\mu)\delta_{xy}, \nabla\ti{1}(\lambda,x) \right] - \left[ \Rc^0\ti{21}(\mu,\lambda)\delta_{yx}, \nabla\ti{2}(\mu,y) \right].
\end{equation}
This bracket is valued in $\F[M] \otimes \Co \otimes \Co$ (or more technically in its completion with respect to an appropriate topology). An element of $\Co$ is a $\g$-connection on $\Sc$ and can thus be evaluated at any point of $\Sc$. In this bracket, the points $x$ and $y$ are thus to be understood as attached to the first and second tensor factors of $\Co \otimes \Co$. Note that the dependence of the bracket in the twist function is now only contained in the connection $\nabla$.

Before interpreting this bracket, let us briefly prove the equation \eqref{Eq:PBNabla}. From the Poisson bracket \eqref{Eq:PBR} of $\Lc(\lambda,x)$ and the expression \eqref{Eq:DefR} of $\Rc$, one gets
\begin{eqnarray*}
\lwb \nabla\ti{1}(\lambda,x), \nabla\ti{2}(\mu,y) \rwb
&=& \vp(\lambda)\vp(\mu) \lwb \Lc\ti{1}(\lambda,x), \Lc\ti{2}(\mu,y) \rwb \\
&=&  \lsb \Rc^0\ti{12}(\lambda,\mu), \vp(\lambda)\Lc\ti{1}(\lambda,x) \rsb \delta_{xy} - \lsb \Rc^0\ti{21}(\mu,\lambda), \vp(\mu)\Lc\ti{2}(\mu,x) \rsb \delta_{xy}\\
& &\hspace{30pt} - \left( \vp(\lambda)\Rc^0\ti{12}(\lambda,\mu) + \vp(\mu)\Rc^0\ti{21}(\mu,\lambda) \right) \delta'_{xy} \\
&=& \lsb \Rc^0\ti{12}(\lambda,\mu)\delta_{xy}, \vp(\lambda)\Lc\ti{1}(\lambda,x) \rsb - \lsb \Rc^0\ti{21}(\mu,\lambda)\delta_{yx}, \vp(\mu)\Lc\ti{2}(\mu,x) \rsb\\
& & \hspace{30pt}- \left[ \vp(\lambda)\p_x, \Rc^0\ti{12}(\lambda,\mu)\delta_{xy} \right] + \left[ \vp(\mu)\p_y, \Rc^0\ti{21}(\mu,\lambda)\delta_{yx} \right] \\
&=& \left[ \Rc^0\ti{12}(\lambda,\mu)\delta_{xy}, \nabla\ti{1}(\lambda,x) \right] - \left[ \Rc^0\ti{21}(\mu,\lambda)\delta_{yx}, \nabla\ti{2}(\mu,y) \right].
\end{eqnarray*}

The bracket \eqref{Eq:PBNabla} now has a form similar to the bracket \eqref{Eq:PBGaudin2} of a Gaudin model. One could then hope that models with twist function are related to Gaudin models on the Lie algebra $\Co$. However, this turns out to be impossible. Indeed, there is no invariant non-degenerate bilinear form on the Lie algebra $\Co$. As we will see in the next subsection, we will overcome this difficulty by considering affine Kac-Moody algebras.

\subsection{Affine Kac-Moody algebras from loop algebras}
\label{SubSec:Affine}

\paragraph{Loop algebra.}
In this section we present the construction of affine Kac-Moody algebras from loop algebras. We will not use here the general abstract theory of Kac-Moody algebras, defined from generalised Cartan matrices. We refer the reader to~\cite{Kac:1990gs} for a general treatment of Kac-Moody algebras and the link between the two descriptions of affine ones.

Let us consider a finite dimensional complex semi-simple Lie algebra $\g$, with Killing form $\kappa$. The loop algebra of $\g$ is the algebra
\begin{equation*}
\Loop = \g \otimes \C[t,t^{-1}]
\end{equation*}
of $\g$-valued Laurent polynomials in a formal variable $t$ (note that this definition of the loop algebra is slightly different from the one used in Appendix \ref{App:StandardLoop}, where we considered Laurent series and not Laurent polynomials). The space $\Loop$ is a Lie algebra when equipped with the point-wise bracket
\begin{equation*}
[X \otimes t^p, Y \otimes t^q] = [X,Y] \otimes t^{p+q}, \;\;\;\; \forall \, X,Y\in\g, \; \forall \, p,q \in \Z.
\end{equation*}
We define the (untwisted) affine Kac-Moody algebra $\gft$ as a central and cocentral extension of $\Loop$:
\begin{equation*}
\gft = \Loop \oplus \C \Dd \oplus \C \Kd ,
\end{equation*}
with the Lie bracket
\begin{equation}\label{Eq:ComAffine}
\begin{array}{l}
\hspace{-35pt}\bigl[ X\otimes t^p + a \Dd + b \Kd, Y\otimes t^q + c \Dd + d \Kd \bigr] = \\ \hspace{55pt} [X,Y] \otimes t^{p+q} + a q Y \otimes t^q - c p X \otimes t^p + p \, \kappa(X,Y) \delta_{p+q,0} \Kd.
\end{array}
\end{equation}
We define a bilinear form $\fd$ on $\gft$ by letting
\begin{equation}\label{Eq:FormAffine}
\bigl( X\otimes t^p + a \Dd + b \Kd, Y\otimes t^q + c \Dd + d \Kd \bigr) = \delta_{p+q,0} \kappa(X,Y) + a d + b c.
\end{equation}
One checks that $\fd$ defines an invariant and non-degenerate bilinear form on $\gft$.\\

As the affine Kac-Moody algebra $\gft$ is quadratic, one can consider the split quadratic Casimir $\Ct\ti{12}$ of $\gft$, as defined in equation \eqref{Eq:Cas}. If $\lbrace I^a \rbrace$ is a basis of $\g$ and $\lbrace I_a \rbrace$ is the associated dual basis with respect to $\kappa$, then $\lbrace I^a\otimes t^n, \Kd, \Dd \rbrace$ is a basis of $\gft$, whose dual basis with respect to $\fd$ is $\lbrace I_a \otimes t^{-n}, \Dd, \Kd \rbrace$. The split quadratric Casimir of $\gft$ is then
\begin{equation}\label{Eq:CasAffine}
\Ct\ti{12} = \Kd \otimes \Dd + \Dd \otimes \Kd + \sum_{n\in\Z} \bigl( I^a \otimes t^n \bigr) \otimes \bigl( I_a \otimes t^{-n} \bigr).
\end{equation}
One issue regarding this construction is the fact that $\gft$ is infinite dimensional, which makes the sum in equation \eqref{Eq:CasAffine} infinite and thus ill-defined in $\gft \otimes \gft$. To make sense of this sum, one needs to consider a completion $\gft \hat\otimes \gft$ of this tensor product. For completeness, we will say a few words about completions in the following paragraph (note that this paragraph is not necessary for the comprehension of the main ideas introduced in the rest of this section).

\paragraph{Infinite sums and completion.}\label{Para:Completion} Issues of convergence and completion appear frequently when dealing with AGM. In this thesis, we will not treat these in details, in order to focus on the general ideas behind AGM. We refer to~\cite{Vicedo:2017cge} for a thorough rigorous treatment. However, we will illustrate the kind of completions used when manipulating affine algebras and related vector spaces by a simple example, a completion of $\gft$ itself. Consider the subspaces
\begin{equation*}
\mathsf{F}_n\gft = \g \otimes t^n \C[t], \;\;\; n \in \Z_{\geq 0},
\end{equation*}
of $\gft$. They form a descending $\Z_{\geq 0}$-filtration of $\gft$:
\begin{equation*}
\gft \supset \mathsf{F}_0\gft \supset \mathsf{F}_1\gft \supset \cdots \supset \mathsf{F}_n\gft \supset \cdots.
\end{equation*}
Let us consider a sequence $U=(u_k)_{k\in\Z_{\geq 0}}$ in $\gft$. We say that $U$ is a Cauchy sequence with respect to this filtration if
\begin{equation*}
\forall \, n \in \Z_{\geq 0}, \;\;\;\; \exists \, k \in \Z_{\geq 0} \;\; \text{ such that } \;\; u_p-u_q \in \mathsf{F}_n\gft \;\;\; \text{ if } \; p,q\geq k.
\end{equation*}
There exists a general construction of completions of vector spaces with descending $\Z_{\geq 0}$-filtration such that all Cauchy sequences with respect to this filtration converge in the completion (see more details in~\cite{Vicedo:2017cge}). In particular, this construction allows to make sense of infinite series
\begin{equation*}
\sum_{n\in\Z_{\geq 0}} v_n, \;\;\;\; \text{ with } \;\; v_n \in \mathsf{F}_n\gft.
\end{equation*}
Note that the $\Z_{\geq 0}$-filtration is a Lie algebra filtration, \textit{i.e.} satisfies
\begin{equation*}
[\mathsf{F}_n\gft,\mathsf{F}_m\gft] \subset \mathsf{F}_{n+m}\gft.
\end{equation*}
This ensures that the Lie bracket $\gft$ extends to the associated completion, making it a Lie algebra. In the present case, the completion is simply
\begin{equation*}
\g (\!( t )\!) \oplus \C\Dd \oplus \C\Kd, 
\end{equation*}
constructed as $\gft$ but where we consider $\g$-valued Laurent series in $t$ instead of Laurent polynomials.

The rigorous construction of AGM requires to consider similar completions of various vector spaces, with respect to some filtrations constructed from the one above and the ``conjugate'' one (with powers of $t^{-1}$ instead of $t$). In particular, this allows to make sense of the split quadratic Casimir \eqref{Eq:CasAffine}.
 
\paragraph{Affine algebras and connections on the circle.}
Recall the trigonometric functions on the circle, $e_n: x\in\Sc \mapsto e^{inx}$ ($n\in\Z$), introduced in the previous subsection. We define a map
\begin{equation*}
\rho : \gft = \Loop \oplus \C\Dd \oplus \C\Kd \longrightarrow \Co = \g \otimes \Tt(\Sc) \oplus \C\p
\end{equation*}
by
\begin{equation*}
\rho ( X \otimes t^n ) = X \otimes e_n, \;\;\;\; \rho(\Dd) = -i\p \;\;\;\;\; \rho(\Kd)=0,
\end{equation*}
for $X\in\g$ and $n\in\Z$. One checks that $\rho$ is a Lie algebra morphism from the affine algebra $\gft$ to the Lie algebra $\Co$ of $\g$-connections on the circle.\\

Let us consider the quadratic split Casimir \eqref{Eq:CasAffine} of $\gft$. It is clear that its image under the Lie algebra morphism $\rho$ is given by
\begin{equation*}
(\rho \otimes \rho) \Ct\ti{12} = \sum_{n\in\Z} \bigl( I^a \otimes e_n \bigr) \otimes \bigl( I_a \otimes e_{-n} \bigr)
\end{equation*}
and belongs to (a completion of) $\bigl( \g \otimes \Tt(\Sc)\bigr) \otimes \bigl( \g\otimes\Tt(\Sc) \bigr) \subset \Co \otimes \Co$. Recall that in Subsection \ref{SubSec:Connections}, we considered elements of $\Co$ by evaluating them at points of $\Sc$. Let us evaluate the left factor of the tensor product $\Co \otimes \Co$ at $x\in\Sc$ and the second at $y\in\Sc$ (as we did to consider the Poisson bracket \eqref{Eq:PBNabla} in Subsection \ref{SubSec:Connections}). We then get
\begin{equation*}
(\rho \otimes \rho) \Ct\ti{12}(x,y) = C\ti{12} \D(x,y),
\end{equation*}
where $C\ti{12}=I^a \otimes I_a \in \g\otimes\g$ is the split Casimir of the finite algebra $\g$ and
\begin{equation*}
\D(x,y) = \sum_{n\in\Z} e_n(x) e_{-n}(y) = \sum_{n\in\Z} e^{in(x-y)}.
\end{equation*}
The latter cannot be understood as a function of $(x,y) \in \Sc\times\Sc$ because of the infinite sum on $n$. In fact, such infinite Fourier series where we do not require additional convergence conditions can be seen as distributions on $\Tt(\Sc)$. Let
\begin{equation*}
f = \sum_{c_n \in \Z} c_n e_n,
\end{equation*}
be an element of $\Tt(\Sc)$ (we then have a finite number of non-vanishing $c_n$'s). The action of the distribution $\D(x,\cdot)$ on the trigonometric polynomial $f$ is given by
\begin{equation*}
\frac{1}{2\pi}\int_0^{2\pi} \dd y \; \D(x,y) f(y) = \sum_{n,m\in\Z} c_m e^{inx} \underbrace{\frac{1}{2\pi} \int_0^{2\pi} \dd y \; e^{i(m-n)y} }_{\delta_{nm}} = \sum_{n\in\Z} c_n e^{inx} = f(x).
\end{equation*}
Thus the distribution $\D(x,y)$ is in fact the Dirac $\delta$-distribution $\delta_{xy}$ (note that, as we are on the circle, we consider integrals with a measure normalised by $2\pi$). We then have
\begin{equation}\label{Eq:CasRho}
(\rho \otimes \rho) \Ct\ti{12}(x,y) = C\ti{12}\delta_{xy}.\vspace{5pt}
\end{equation}

Note that the morphism $\rho$ is surjective but not injective, as it sends $\Kd$ to zero. Thus, $\Co$ is not isomorphic to $\gft$. In fact, it is isomorphic to its quotient $\gft/\C\Kd$ by the central ideal $\C\Kd$. This prevents us to translate all properties of $\gft$ on $\Co$. In particular, one cannot ``push'' the invariant non-degenerate bilinear form $\fd$ to an invariant non-degenerate form on $\Co$. We already stated the non-existence of such a form on $\Co$ at the end of subsection \ref{SubSec:Connections}. In particular, this prevented us to define a Gaudin model on $\Co$. However, one can define a Gaudin model on $\gft$, whose Lax matrix would then be $\gft$-valued, and apply the morphism $\rho$ to it to get a Lax matrix valued in $\Co$. We will apply this idea in the next subsection. First, let us remark that although one cannot ``push'' the bilinear form $\fd$ from $\gft$ to $\Co$, one can do it when restricting to $\Loop\subset\gft$:
\begin{equation}\label{Eq:FormRho}
\forall \, J_1,J_2 \in\Loop \subset \gft, \;\;\;\; (J_1,J_2) = \frac{1}{2\pi}\int_0^{2\pi} \dd x \; \kappa\bigl(\Jc_1(x),\Jc_2(x)\bigr), \;\;\; \text{ with } \;\;\; \Jc_i=\rho(J_i).
\end{equation}

\subsection{Classical Affine Gaudin models without dihedrality and multiple poles}
\label{SubSec:SimplestAGM}

\paragraph{Lax matrix and change of auxiliary space.} 
For simplicity, we will start with the simplest AGM on $\gft$: complex, without cyclotomy and with sites of multiplicity one. Its Lax matrix $\Ls(\lambda)$ then satisfies the Poisson bracket
\begin{equation}\label{Eq:PBAGM1}
\lwb \Ls\ti{1}(\lambda), \Ls\ti{2}(\mu) \rwb = \lsb \frac{\Ct\ti{12}}{\mu-\lambda}, \Ls\ti{1}(\lambda) + \Ls\ti{2}(\mu) \rsb,
\end{equation}
according to equation \eqref{Eq:PBGaudin}.

The Lax matrix $\Ls(\lambda)$ is valued in (a completion of) $\gft \otimes \F[M]$, where $M$ is the phase space of the Gaudin model. We call the space $\F[M]$ the algebra of observables of the model and $\gft$ the auxiliary space. Let us modify this auxiliary space by applying the morphism $\rho: \gft \rightarrow \Co$:
\begin{equation*}
\Nt(\lambda)=(\rho \otimes \Id) \Ls(\lambda) \in \Co \otimes \F[M],
\end{equation*}
where $\rho$ and $\Id$ acts respectively on the left and right factor of the tensor product $\gft \otimes \F[M]$. The bracket \eqref{Eq:PBAGM1} is valued in $\gft\otimes\gft\otimes\F[M]$. We get the Poisson bracket of $\Nt\ti{1}(\lambda)$ with $\Nt\ti{2}(\mu)$ by applying $\rho\otimes\rho\otimes\Id$ on equation \eqref{Eq:PBAGM1}. As $\rho$ is a Lie algebra morphism, we get
\begin{equation*}
\lwb \Nt\ti{1}(\lambda), \Nt\ti{2}(\mu) \rwb = \lsb \frac{(\rho\otimes\rho)\Ct\ti{12}}{\mu-\lambda}, \Nt\ti{1}(\lambda) + \Nt\ti{2}(\mu) \rsb.
\end{equation*}
Recall that to write the Poisson bracket \eqref{Eq:PBNabla} of $\Co$-valued objects we evaluated the connections at points $x$ and $y$ of $\Sc$. Using equation \eqref{Eq:CasRho}, one then gets
\begin{equation}\label{Eq:PBNablat}
\lwb \Nt\ti{1}(\lambda,x), \Nt\ti{2}(\mu,y) \rwb = \lsb \frac{C\ti{12}}{\mu-\lambda}\delta_{xy}, \Nt\ti{1}(\lambda,x) + \Nt\ti{2}(\mu,x) \rsb.
\end{equation}
This has exactly the form of the Lax bracket \eqref{Eq:PBNabla} (for $\Rc^0$ the non-twisted standard $\Rc$-matrix $C\ti{12}/(\mu-\lambda)$). The Lax matrix of the AGM, once the auxiliary space changed, then has the same bracket as the Lax connection \eqref{Eq:LaxConnection} of a model with twist function. However, to complete the interpretation of the AGM as a field theory, one also has to understand its phase space and in particular the components of $\Nt(\lambda)$.

\paragraph{Phase space and dynamical fields.} As we consider a non-cyclotomic complex AGM with only simple poles, its phase space $M$ is the product of $N$ copies of the dual space $\gft^*$, equipped with the Kirillov-Kostant bracket. This phase space is then encoded in $N$ quantities $X^{(r)}$ in (a completion of) $\gft \otimes \F[M]$ satisfying the bracket
\begin{equation}\label{Eq:PBKacAbs}
\lwb X^{(r)}\ti{1}, X^{(s)}\ti{2} \rwb = \delta_{rs} \lsb \Ct\ti{12}, X^{(r)}\ti{1} \rsb.
\end{equation}
We will write $X^{(r)}$ as
\begin{equation}\label{Eq:AgmXr}
X^{(r)} = D^{(r)} \Kd + i K^{(r)} \Dd + J^{(r)},
\end{equation}
with $D^{(r)},K^{(r)} \in \F[M]$ and $J^{(r)} \in \Loop \otimes \F[M]$. We will focus first on $K^{(r)}$ and $J^{(r)}$ by considering the image of $X^{(r)}$ through the morphism $\rho$:
\begin{equation}\label{Eq:DefJK}
\bigl( \rho \otimes \Id \bigr) X^{(r)}(x) = K^{(r)} \p_x + \Jc^{(r)}(x),
\end{equation}
where $\Jc^{(r)}=\rho\bigl(J^{(r)}\bigr)$ belongs to (a completion of) $\g \otimes \Tt(\Sc) \otimes \F[M]$. In particular, the components of $\Jc^{(r)}$ in a basis $\lbrace I^a \rbrace$ of $\g$ belong to the completion of $\Tt(\Sc) \otimes \F[M]$:
\begin{equation*}
\Jc_a^{(r)}(x) = \sum_{n\in\Z} c_{a,n}^{(r)} e^{inx},
\end{equation*}
with $c_{a,n}^{(r)} \in \F[M]$. Thus, $\Jc_a^{(r)}$ is an observable-valued distribution on $\Sc$. As explained at the beginning of subsection \ref{SubSec:Connections}, such an object is the dynamical field of an Hamiltonian theory on $\Sc$.

\paragraph{Levels and Kac-Moody currents.} Now that we interpreted the observable $\Jc^{(r)}=\rho\bigl(J^{(r)}\bigr)$ as dynamical fields, let us consider the other observables. In particular, the observable $K^{(r)}$ appears as the coefficient of $i\Dd$ in $X^{(r)}$. However, one sees in equation \eqref{Eq:ComAffine} that the element $\Dd$ can never be created by a commutator in $\gft$. Thus, it cannot appear on the left factor of the bracket \eqref{Eq:PBKacAbs}. Therefore, one gets
\begin{equation*}
\lwb K^{(r)}, X^{(s)} \rwb = 0,
\end{equation*}
for all $r,s=1,\cdots,N$. This means that the observable $K^{(r)}$ is a Poisson Casimir of $\F[M]$ (it Poisson commutes with all functions in $\F[M]$). The elements $K^{(r)}$ are the classical equivalent for the Kostant-Kirillov bracket of the central element $\Kd$ of $\gft$.

In a Hamiltonian field theory, there should not be Poisson Casimirs which are not constant functions. In order to interpret the AGM as a field theory, we thus fix the quantities $K^{(r)}$ to constants. For that, we will consider a Poisson map $\pk$ from $\F[M]$ to a new algebra of observables $\Obs$, such that
\begin{equation}\label{Eq:PiK}
\pk \left(K^{(r)} \right) = k_r, \;\;\;\; \text{ with } \;\;\;\;\; \bm k = (k_1,\cdots,k_N) \in \C^N.
\end{equation}
We will explain more precisely how we construct this map later. The numbers in $\bm k$ are called the \textbf{levels} of the AGM. As the quantities $\Jc^{(r)}$ have already been interpreted as dynamical fields on $\Sc$, we shall consider that $\pk$ does not affect these quantities (again, see the precise statement after), so we still write $\Jc^{(r)}$ their image under $\pk$. We then have
\begin{equation*}
(\rho \otimes \pk) X^{(r)}(x) = k_r \p_x + \Jc^{(r)}(x).
\end{equation*}

The bracket \eqref{Eq:PBKacAbs} is valued in $\gft \otimes \gft \otimes \F[M]$. As $\pk$ is a Poisson map, we get the Poisson brackets of $(\rho \otimes \pk) X^{(r)}\ti{1}(x)$ and $(\rho \otimes \pk) X^{(s)}\ti{2}(y)$ by applying the map $\rho \otimes \rho \otimes \pk$ to \eqref{Eq:PBKacAbs} (and evaluate at $x$ and $y$ in $\Sc$). Doing this, and using the expression \eqref{Eq:CasRho} of $(\rho\otimes\rho)\Ct\ti{12}(x,y)$, we obtain
\begin{equation}\label{Eq:KMC}
\lwb \Jc^{(r)}\ti{1}(x), \Jc^{(r)}\ti{2}(y) \rwb = \left[ C\ti{12}, \Jc^{(r)}\ti{1}(x) \right] \delta_{xy} - k_r C\ti{12} \delta'_{xy}
\end{equation} 
and
\begin{equation*}
\lwb \Jc^{(r)}\ti{1}(x), \Jc^{(s)}\ti{2}(y) \rwb = 0 \;\;\;\; \text{ if } \; k\neq s.
\end{equation*}
The bracket \eqref{Eq:KMC} is then the one of a so-called \textbf{Kac-Moody current} of level $k_r$. The algebra of observables $\Obs$ of the model then contains $N$ commuting Kac-Moody currents.

\paragraph{Momentum of the theory.}So far, we understood the interpretation of the components $\Jc^{(r)}$ and $K^{(r)}$ of $X^{(r)}$. However, there is still a set of observables that we did not study. Indeed, recall that the morphism $\rho$ is not injective, as it sends $\Kd$ to zero. Thus there is still to interpret the coefficient $D^{(r)}$ of $\Kd$ in \eqref{Eq:AgmXr}.

Let us compute its Poisson bracket with the Kac-Moody current $\Jc^{(r)}$. For that, we project the bracket \eqref{Eq:PBKacAbs} on $\Kd$ on the second tensor factor. Given the expression \eqref{Eq:CasAffine} of $\Ct\ti{12}$, one gets
\begin{equation*}
\lwb X^{(r)}, D^{(r)} \rwb = \left[ \Dd, X^{(r)} \right].
\end{equation*}
Applying the morphism $\rho$ to this bracket, we get
\begin{equation*}
\lwb \Jc^{(r)}(x), D^{(r)} \rwb = \left[ -i\p_x, K^{(r)}\p_x + \Jc^{(r)}(x) \right],
\end{equation*}
hence
\begin{equation*}
-i \bigl\lbrace D^{(r)}, \Jc^{(r)}(x) \bigr\rbrace = \p_x \Jc^{(r)}(x).
\end{equation*}
Thus, $-i D^{(r)}$ generates the space derivative on the current $\Jc^{(r)}$. Its Hamiltonian flow then coincides with the one of the momentum of the theory. More precisely, it is clear from \eqref{Eq:PBKacAbs} that
\begin{equation*}
\bigl\lbrace D^{(r)}, \Jc^{(s)}(x) \bigr\rbrace = 0 \;\;\;\; \text{ if } \; r\neq s,
\end{equation*}
hence the Hamiltonian flow of $-iD^{(r)}$ coincides with the momentum of the Kac-Moody current $\rp \Jc r p$ associated with the site $\lambda_r$.

Here, $D^{(r)}$ is an abstract generator in $\F[M]$. Yet, in general in a Hamiltonian field theory, the momentum of the model is not an independent quantity: it is expressed in terms of the dynamical fields so that it generates the space derivative. For the algebra $\Obs$ to describe the observables of a field theory, one should then realise the abstract generator $-iD^{(r)}$ as the momentum $\Pc^{(r)}$ of the Kac-Moody current $\Jc^{(r)}$. We thus choose the map $\pk$ to satisfy
\begin{equation}\label{Eq:PiP}
\pk \left( D^{(r)} \right) = i \Pc^{(r)}.
\end{equation}
As for the treatment of the observables $K^{(r)}$'s, the construction of the map $\pk$ satisfying \eqref{Eq:PiP} will be made more precise in the next paragraph.

Above, we introduced the observable $\Pc^{(r)}$ to be the momentum generating the space derivative on the Kac-Moody current $\Jc^{(r)}$. Let us be more concrete. Indeed, one can find an explicit expression for $\Pc^{(r)}$. From the Poisson bracket \eqref{Eq:KMC}, one finds that
\begin{equation}\label{Eq:Pr}
\Pc^{(r)} = \frac{1}{4\pi k_r} \int_0^{2\pi} \dd x \; \kappa \left( \Jc^{(r)}(x), \Jc^{(r)}(x) \right)
\end{equation}
generates the space derivative on $\Jc^{(r)}$. Note that in order to define $\Pc^{(r)}$, we supposed that the level $k_r$ is non-zero. This expression for $\Pc^{(r)}$ is the so-called classical Segal-Sugawara construction (see for instance~\cite{Etingof:1998}). Its algebraic origin will be explained in the next paragraph, together with the precise construction of $\pk$.

The total momentum $\Pc \in \Obs$ of the resulting field theory is then simply given by
\begin{equation}\label{Eq:PGaudin}
\Pc = \sum_{r=1}^N \Pc^{(r)}.
\end{equation}

\paragraph{Local AGM and the Segal-Sugawara map $\bm{\pk}$.} In this section, we will explain how to construct rigorously the map $\pk$ such that it satisfies \eqref{Eq:PiK} and \eqref{Eq:PiP}. In particular, this map sends the element $K^{(r)}-k_r$ of $\F[M]$ (where we see $k_r$ as a constant function on $M$) to zero. Such a map is easily constructed by taking the quotient by the ideal $(K^{(r)}-k_r)\F[M]$ generated by $K^{(r)}-k_r$. By construction, the canonical map from $\F[M]$ to $\F[M]/(K^{(r)}-k_r)\F[M]$ is thus a morphism of algebra. However, we also want this map to induce a Poisson bracket on the quotient. For that, we need the algebra ideal $(K^{(r)}-k_r)\F[M]$ to also be a Poisson ideal, \textit{i.e.} to be stabilised under the Poisson bracket with any element of $\F[M]$. This is simply ensured by the fact that $K^{(r)}-k_r$ is a Poisson Casimir and thus has vanishing Poisson bracket with all elements of $\F[M]$.

Recall from Subsection \ref{SubSec:GaudinHam} that $\F[M]$ contains other Poisson Casimirs than the $K^{(r)}$'s, namely the quantities $\Delta_r$'s defined as
\begin{equation*}
\Delta_r = \frac{1}{2} \left( X^{(r)}, X^{(r)} \right).
\end{equation*}
Given the expression \eqref{Eq:AgmXr} of $X^{(r)}$, the definition \eqref{Eq:FormAffine} of $\fd$ and equation \eqref{Eq:FormRho}, we get
\begin{equation*}
\Delta_r = i K^{(r)} D^{(r)} + \frac{1}{2} \left(J^{(r)},J^{(r)}\right) = i K^{(r)} D^{(r)} + \frac{1}{4\pi} \int_0^{2\pi} \kappa \left( \Jc^{(r)}(x), \Jc^{(r)}(x) \right) = i K^{(r)} D^{(r)} + k_r \Pc^{(r)}.
\end{equation*}
One then sees that a map $\pk$ which would verify \eqref{Eq:PiK} and \eqref{Eq:PiP} should send the Casimir $\Delta_r$ to zero. As $\Delta_r$ is a Poisson Casimir, the algebra ideal $\Delta_r \F[M]$ is also a Poisson ideal of $\F[M]$. Thus the canonical surjection from $\F[M]$ to $\F[M]/\Delta_r\F[M]$ is an algebra morphism preserving the Poisson bracket.

The map $\pk$ is obtained by considering the quotient of $\F[M]$ by all Poisson ideals $(K^{(r)}-k_r)\F[M]$'s and $\Delta_r\F[M]$'s. More precisely, let us define
\begin{equation*}
\mathcal{I}_{\bm k} =  \bigoplus_{r=1}^N  \Bigl( (K^{(r)}-k_r)\F[M] \oplus \Delta_r \F[M] \Bigr).
\end{equation*}
It is a Poisson ideal of $\F[M]$ as all $(K^{(r)}-k_r)$'s and $\Delta_r$'s are Poisson Casimirs. We then define the algebra of observables of the field theory as the quotient
\begin{equation*}
\Obs = \F[M] / \mathcal{I}_{\bm k}
\end{equation*}
and the map $\pk$ as the canonical surjection
\begin{equation*}
\pk : \F[M] \longrightarrow \Obs.
\end{equation*}
This map is called the (classical) Segal-Sugawara map.\\

We shall call the resulting Hamiltonian theory with observables $\Obs$ the local AGM. We will often speak of the formal AGM for the model before quotienting. Recall that the $X^{(r)}$'s generate the algebra $\F[M]$. Thus, as $\pk$ is surjective, $\Obs$ is generated by the images $\pk(X^{(r)})$'s. Recall the expression \eqref{Eq:AgmXr} of $X^{(r)}$. As $\pk$ sends $K^{(r)}$ to a constant $k_r$ and expresses $D^{(r)}$ in terms of $\Jc^{(r)}=(\rho \otimes \pi_k)J^{(r)}$, we find that $\Obs$ is generated by the $\Jc^{(r)}$. Moreover, as $\pk$ does not impose any additional relations, these generators $\Jc^{(r)}$ are independent in $\Obs$ (this is what we meant after equation \eqref{Eq:PiK} when saying that the map $\pk$ does not affect these quantities).

As explained in this subsection, the quantities $\Jc^{(r)}$ are elements of the completion of $\g\otimes\Tt(\Sc) \otimes \F[M]$, \textit{i.e.} observables-valued distributions on $\Sc$, living in $\g$. Thus, they are $\g$-valued dynamical fields of an Hamiltonian field theory. As they are independent generators of $\Obs$, they form the fundamental fields of the theory. This shows that the local AGM is an Hamiltonian field theory. Note that to define this field theory, we required that the levels $k_r$'s are different from $0$.

\paragraph{Lax matrix and twist function.} Let us now turn our attention to the Lax matrix of the model. According to equation \eqref{Eq:LaxGaudin}, the Lax matrix of the formal AGM is given by
\vspace{-5pt}\begin{equation*}
\Ls(\lambda) = \sum_{i=1}^N \frac{X^{(r)}}{\lambda-\lambda_r} + \Omega,\vspace{-5pt}
\end{equation*}
where $\Omega$ is a constant in $\gft$. This Lax matrix is valued in (a completion of) $\gft\otimes\F[M]$. We will write the constant $\Omega$ as
\begin{equation*}
\Omega = i p_\infty \Kd + i k_\infty \Dd + B,
\end{equation*}
with $k_\infty$ and $p_\infty$ complex numbers and $B\in\Loop$.

As explained in the first paragraph of this subsection, to interpret this Lax matrix as the Lax connection of a model with twist function, we made a change of auxiliary space and considered the image $\Nt(\lambda)$ of $\Ls(\lambda)$ under the morphism $(\rho\otimes\Id)$ (where $\Id$ acts on $\F[M]$). Let us now consider the equivalent object for the local AGM and define
\begin{equation}\label{Eq:NablaAGM}
\nabla (\lambda) = \left( \rho \otimes \pk \right) \Ls(\lambda).
\end{equation}
Evaluating at $x\in\Sc$, we then find
\begin{equation}\label{Eq:NablaGaudin}
\nabla (\lambda,x) = \vp(\lambda) \p_x + \Scc(\lambda,x)
\end{equation}
with
\begin{equation}\label{Eq:TwistGaudin}
\vp(\lambda) = \sum_{r=1}^N \frac{k_r}{\lambda-\lambda_r} + k_\infty
\end{equation}
and
\begin{equation}\label{Eq:SGaudin}
\Scc(\lambda,x) = \sum_{r=1}^N \frac{\Jc^{(r)}(x)}{\lambda-\lambda_r} + \mathcal{B}(x), \;\;\;\; \text{ where }  \;\;\;\; \mathcal{B}=(\rho\otimes\pk)B.
\end{equation}

As $\pk$ is a Poisson map, the $\g$-connection $\nabla$ satisfies the same Poisson bracket \eqref{Eq:PBNablat} than $\Nt$. According to equation \eqref{Eq:LaxConnection}, the function $\vp(\lambda)$ is thus the \textbf{twist function} of the model. Moreover, one recovers the usual \textbf{Lax matrix} of the model as
\begin{equation*}
\Lc(\lambda,x) = \vp(\lambda)^{-1} \Scc(\lambda,x).
\end{equation*}
Considering the inverse reasoning of Subsection \ref{SubSec:Connections}, this Lax matrix then satisfies the Maillet bracket \eqref{Eq:PBR} with $\Rc$-matrix
\begin{equation*}
\frac{C\ti{12}}{\mu-\lambda} \vp(\mu)^{-1}.
\end{equation*}
Note that the twist function appears first in the formal AGM as the coefficient of $i\Dd$ in the formal Lax matrix $\Ls(\lambda)$: it is then a fundamental observable of the formal theory and is later realised as a rational function through the Segal-Sugawara map $\pk$. Note also that as $\vp(\lambda)$ and $\Scc(\lambda,x)$ are extracted from the same object $\Ls(\lambda)$, they possess similar partial fraction decompositions.

\paragraph{Hamiltonian.} So far, we did not discuss the dynamic of the AGM. At the level of the formal AGM, we define the quadratic Hamiltonian (still depending on the spectral parameter) as in \eqref{Eq:GaudinHamLambda}:
\begin{equation*}
\Hs(\lambda) = \frac{1}{2} \bigl( \Ls(\lambda), \Ls(\lambda) \bigr) =   \sum_{r=1}^N \left( \frac{\Delta_r}{(\lambda-\lambda_r)^2} + \frac{\Hs_r}{\lambda-\lambda_r} \right) + \frac{1}{2} (\Omega,\Omega).
\end{equation*}
This Hamiltonian is in involution with itself for any values of the spectral parameter, according to equation \eqref{Eq:HlInvo}.

We then define the Hamiltonian of the local AGM from the formal one \textit{via} the Segal-Sugawara map $\pk$:
\begin{equation*}
\Hc(\lambda) = \pk \bigl( \Hs(\lambda) \bigr) \in \Obs.
\end{equation*}
As $\pk$ is a Poisson map, we have
\begin{equation*}
\lwb \Hc(\lambda), \Hc(\mu) \rwb = 0, \;\;\;\; \forall \, \lambda,\mu\in\C.
\end{equation*}
As explained above, $\pk$ sends the Poisson Casimirs $\Delta_r$'s to zero. Thus, the local Gaudin Hamiltonian has only simple poles:
\begin{equation}\label{Eq:HamPoleSimple}
\Hc(\lambda) = \sum_{r=1}^N \frac{\Hc_r}{\lambda-\lambda_r} + \frac{1}{2} (\Omega,\Omega).
\end{equation}
Note that equation \eqref{Eq:AgmXr} implies
\begin{equation*}
(\Id \otimes \pk) \bigl( \Ls(\lambda) \bigr) = i \Pc(\lambda) \Kd + i \vp(\lambda) \Dd + S(\lambda),
\end{equation*}
where $S(\lambda)\in\Loop$ is such that $\rho\bigl( S(\lambda) \bigr) = \Scc(\lambda)$ and
\begin{equation*}
\Pc(\lambda) = \sum_{r=1}^N \frac{\Pc^{(r)}}{\lambda-\lambda_r} + p_\infty.
\end{equation*}
Note that the total momentum \eqref{Eq:PGaudin} of the local AGM can be seen as
\begin{equation*}
\Pc = -\res_{\lambda=\infty} \; \Pc(\lambda) \, \dd\lambda.
\end{equation*}
We then express the local Hamiltonian, using the identity \eqref{Eq:FormRho}, as
\begin{equation}\label{Eq:HamAgmLoc}
\Hc(\lambda) = \frac{1}{2} \bigl( S(\lambda), S(\lambda) \bigr) - \vp(\lambda)\Pc(\lambda) = \frac{1}{4\pi} \int_0^{2\pi} \dd x \; \kappa \bigl( \Scc(\lambda,x), \Scc(\lambda,x) \bigr) - \vp(\lambda)\Pc(\lambda).
\end{equation}
One easily checks that the double pole of $\Hc(\lambda)$ at $\lambda=\lambda_r$ is equal to
\begin{equation*}
\frac{1}{4\pi} \int_0^{2\pi} \dd x \; \kappa \bigl( \Jc^{(r)}(x), \Jc^{(r)}(x) \bigr) - k_r \Pc^{(r)} = 0,
\end{equation*}
by equation \eqref{Eq:Pr}, as expected from equation \eqref{Eq:HamPoleSimple}. In general, we define the Hamiltonian of the model as a linear combination of the $\Hc_r$'s.

\paragraph{Lax equation.} Let us end this subsection by discussing the dynamic of the Lax matrix. Recall the formal Lax equation \eqref{Eq:LaxGaudin} describing the evolution of the formal Lax matrix $\Ls(\lambda)$ under the formal Hamiltonian $\Hs(\mu)$. As usual, we will apply the map $\rho\otimes\pk$ to this equation to get informations on the local AGM. We define $\Nc$ and $\chi$ by
\begin{equation*}
(\rho \otimes \pk) \Ms(\mu,\lambda,x) = \chi(\mu,\lambda) \p_x + \Nc(\mu,\lambda,x).
\end{equation*}
Starting from the formal Lax equation \eqref{Eq:LaxGaudin} and using the fact that $\rho$ is a Lie morphism on $\gft$, we get
\begin{eqnarray*}
(\rho\otimes\pk) \lwb \Hs(\mu), \Ls(\lambda) \rwb (x)
 &=& (\rho\otimes\pk) \lsb \Ms(\mu,\lambda), \Ls(\lambda) \rsb (x) \\
 &=& \lsb (\rho \otimes \pk) \Ms(\mu,\lambda,x), (\rho \otimes \pk) \Ls(\lambda,x)  \rsb \\
 &=& \vp(\lambda)\lsb \chi(\mu,\lambda) \p_x + \Nc(\mu,\lambda,x), \p_x + \Lc(\lambda,x) \bigr] \rsb \\
 &=& \vp(\lambda) \Bigl( \p_x \bigl( \chi(\mu,\lambda)\Lc(\lambda,x) - \Nc(\mu,\lambda,x) \bigr) + \bigl[ \Nc(\mu,\lambda,x), \Lc(\lambda,x) \bigr] \Bigr) \\
 &=& \vp(\lambda) \Bigl( \p_x \M(\mu,\lambda,x) - \bigl[ \M(\mu,\lambda,x), \Lc(\lambda,x) \bigr] \Bigr),
\end{eqnarray*}
with
\begin{equation*}
\M(\mu,\lambda,x) = \chi(\mu,\lambda)\Lc(\lambda,x) - \Nc(\mu,\lambda,x).
\end{equation*}
On the other hand, using the fact that $\pk$ is a Poisson map on $\F[M]$, we have
\begin{eqnarray*}
(\rho\otimes\pk) \lwb \Hs(\mu), \Ls(\lambda) \rwb (x)
&=& \lwb \pk\bigl(\Hs(\mu)\bigr),  (\rho\otimes\pk)\Ls(\lambda)(x) \rwb \\
&=& \vp(\lambda) \lwb \Hc(\mu), \p_x + \Lc(\lambda,x) \rwb \\
&=& \vp(\lambda) \lwb \Hc(\mu), \Lc(\lambda,x) \rwb.
\end{eqnarray*}
Thus, one gets
\begin{equation*}
\lwb \Hc(\mu), \Lc(\lambda,x) \rwb - \p_x \M(\mu,\lambda,x) + \bigl[ \M(\mu,\lambda,x), \Lc(\lambda,x) \bigr] = 0.
\end{equation*}
This is a zero curvature equation, hence proving that the dynamic of the Lax matrix $\Lc(\lambda)$ under the local quadratic Gaudin Hamiltonian $\Hc(\mu)$ takes the form of a Lax equation \eqref{Eq:ZCEH}.

\paragraph{Summary.} Let us summarise what we have done in this subsection by recalling the main steps of the construction of the local AGM.
\begin{enumerate}[1.]
\setlength\itemsep{0.1em}
\item We start with the formal AGM on the affine algebra $\gft$.
\item Using the definition of $\gft$ in terms of loop algebra, we change the auxiliary space $\gft$ in the space of $\g$-connections on $\Sc$, \textit{via} the morphism $\rho$.
\item Using the Segal-Sugawara map $\pk$, we change the algebra of observables of the model so that the new algebra $\Obs$ describes the phase space of a Hamiltonian field theory .
\item We extract from the formal Gaudin Lax matrix $\Ls(\lambda)$ the twist function $\vp(\lambda)$ and the usual Lax matrix $\Lc(\lambda)$ of the local AGM.
\item The Poisson bracket of the formal Gaudin Lax matrix translates to the Maillet bracket with twist function $\vp$ of the matrix $\Lc$.
\item We define the Hamiltonian of the local AGM as the image of the formal one under $\pk$.
\item The Lax equation on $\Ls$ gives the zero curvature equation on $\Lc$ (\textit{i.e.} a Lax equation for a field theory).
\end{enumerate}
As a conclusion, the local AGM is an integrable field theory with twist function.

\subsection{Dihedral Affine Gaudin models with arbitrary multiplicities}
\label{SubSec:DAGM}

In this subsection, we will discuss generalisations of the AGM presented above by adding successively multiplicities of the sites, cyclotomy and reality conditions, hence ending with the most general DAGM. We will follow the general discussion of these generalisations for an arbitrary Gaudin model in Subsection \ref{SubSec:GaudinGen}, but adapting to the particular case of an affine algebra. In particular, we shall explain how these formal AGM can be mapped to local AGM, as in the previous subsection, and how to interpret local AGM as integrable field theories with twist function.

\paragraph{AGM with multiplicities.} Let us start by considering sites $\lambda_1,\cdots,\lambda_N$ with corresponding multiplicities $m_1, \cdots, m_N \in\Z_{\geq 1}$. In this case, recall from Subsection \ref{SubSec:GaudinGen} that the formal Lax matrix $\Ls(\lambda)$ satisfies the same Poisson bracket \eqref{Eq:PBAGM1} than in the case with multiplicities equal to one. Therefore, applying the Lie morphism $\rho : \gft \rightarrow \Co$ to $\Ls(\lambda)$ to change the auxiliary space of the model, one gets a connection satisfying the bracket \eqref{Eq:PBNablat}, characteristic of a model with twist function.

The main change in the study of AGM with multiplicities is then the interpretation of the observables of the model as the observables of a Hamiltonian field theory. Recall from Subsection \ref{SubSec:GaudinGen} that the phase space of the formal AGM with multiplicities is the Cartesian product
\begin{equation*}
M = \Ta^{m_1}\gft^* \otimes \cdots \otimes \Ta^{m_N}\gft^*
\end{equation*}
of Kirillov-Kostant spaces dual to Takiff algebras $\Ta^m\gft$. This phase space is encoded by $\gft$-valued observables $X^{(r)}_{[p]}$ ($p=0,\cdots,m-1$) satisfying the bracket
\begin{equation}\label{Eq:AgmTakiff}
\Bigl\lbrace X^{(r)}_{[p]}\null\ti{1}, X^{(s)}_{[q]}\null\ti{2} \Bigr\rbrace = \left\lbrace \begin{array}{ll}
\delta_{rs}\left[ \Ct\ti{12}, X^{(r)}_{[p+q]}\null\ti{1} \right] & \text{ if } p+q\leq m_r-1,\\
0           & \text{ if } p+q > m_r-1.
\end{array}  \right.
\end{equation}
We will follow the ideas developed in Subsection \ref{SubSec:SimplestAGM} for the AGM with multiplicities one. In particular, we want to find a Poisson map
\begin{equation*}
\pk : \F[M] \longrightarrow \Obs,
\end{equation*}
from the algebra of observables $\F[M]$ of the formal AGM to the one $\Obs$ of a field theory, which we will call the local AGM. We define observables $K^{(r)}_{[p]}$, $D^{(r)}_{[p]}$ (scalars) and $J^{(r)}_{[p]}$ (in $\Loop$) by
\begin{equation*}
X^{(r)}_{[p]} = D^{(r)}_{[p]} \Kd + i K^{(r)}_{[p]} \Dd + J^{(r)}_{[p]},
\end{equation*}
as in \eqref{Eq:AgmXr}. As in the case with no multiplicities, we find that the $\rp{K}{r}{p}$ are Poisson Casimirs of the algebra $\F[M]$ and choose $\pk$ to send them to some constants:
\begin{equation*}
\pk\left( \rp K r p \right) = k_{r,p}, \;\;\;\; \text{ with } \;\;\;\; \bm k = (k_{1,0},\cdots,k_{r,p},\cdots,k_{N,m_N-1}) \in \C^{m_1+\cdots+m_N},
\end{equation*}
that we also call the levels of the local AGM. We also define
\begin{equation*}
\rp\Jc r p = (\rho\otimes\pk) \rp J r p,
\end{equation*}
which belong to (a completion of) $\g \otimes \Obs \otimes \Tt(\Sc)$ and are thus $\g$-valued dynamical fields. Applying $(\rho\otimes\pk)$ to the Poisson bracket \eqref{Eq:AgmTakiff} and using identity \eqref{Eq:CasRho}, we get
\begin{equation*}
\Bigl\lbrace \Jc^{(r)}_{[p]}\null\ti{1}(x), \Jc^{(s)}_{[q]}\null\ti{2}(y) \Bigr\rbrace = \delta_{rs} \left\lbrace \begin{array}{ll}
\left[ C\ti{12}, \Jc^{(r)}_{[p+q]}\null\ti{1}(x) \right] \delta_{xy} - k_{r,p+q} C\ti{12} \delta'_{xy} & \text{ if } p+q\leq m_r-1,\\
0           & \text{ if } p+q > m_r-1.
\end{array}  \right.
\end{equation*}
The treatment of the $\rp D r p$'s is similar to the one of the $D^{(r)}$'s in the case of multiplicities one. We find that $\rp D r p$ acts on the fields $\rp \Jc r p$ (here also, we denote by the same notations the fields before and after taking the realisation $\pk$) as
\begin{equation*}
-i \Bigl\lbrace D^{(r)}_{[p]}, \Jc^{(s)}_{[q]}(x) \Bigr\rbrace = \delta_{rs} \left\lbrace \begin{array}{ll} \p_x \Jc^{(r)}_{[p+q]}(x)  & \text{ if } p+q\leq m_r-1,\\
0           & \text{ if } p+q > m_r-1.
\end{array}  \right.
\end{equation*}
Moreover, for all $r$, $s$, $p$ and $q$, we have
\begin{equation*}
\Bigl\lbrace D^{(r)}_{[p]}, D^{(s)}_{[q]} \Bigr\rbrace = 0.
\end{equation*}
It is possible to construct quantities $\rp \Pc r p$ from the $\rp \Jc r q$'s that satisfy the same Poisson brackets as the $\rp D r p$'s. We then consider a generalised Segal-Sugawara map
\begin{equation*}
\pk \left( \rp D r p \right) = \rp \Pc r p.
\end{equation*}
For a Takiff algebra of multuplicity two, this generalised Segal-Sugawara construction is the classical analogue of a quantum one, found in~\cite{Babichenko:2012uq}. For the most general (classical) construction with arbitrary multiplicity, we refer to the article~\cite{Vicedo:2017cge}. Note that this construction requires that the ``highest'' levels $k_{r,m_r-1}$'s (in the sense of the levels associated with the highest Takiff modes $[p]=[m_r-1]$) are non-zero.

As the map $\pk$ realises the observables $\rp K r p$'s and $\rp D r p$'s in terms of the fields $\rp \Jc r p$'s, the resulting Poisson algebra $\Obs$ describes the observables of a field theory, that we shall call the local AGM. Once this algebra constructed, the rest of the construction of the local AGM presented in Subsection \ref{SubSec:SimplestAGM} for multiplicities one generalises easily to the case of arbitrary multiplicities. One extracts the twist function $\vp$ and the Lax matrix $\Lc$ of the local AGM from the formal Gaudin Lax matrix:
\begin{equation}\label{Eq:LaxTwistMult}
\vp(\lambda) = \sum_{r=1}^N \sum_{p=0}^{m_r-1} \frac{k_{r,p}}{(\lambda-\lambda_r)^{p+1}} + k_\infty \;\;\;\; \text{ and } \;\;\;\; \vp(\lambda)\Lc(\lambda,x) = \sum_{r=1}^N \sum_{p=0}^{m_r-1} \frac{\rp \Jc r p(x)}{(\lambda-\lambda_r)^{p+1}} + \mathcal{B}(x).
\end{equation}
This Lax matrix $\Lc$ satisfies a Maillet bracket \eqref{Eq:PBR} with twist function (and non twisted standard $\Rc$-matrix) and a zero curvature equation, hence proving that the local AGM with arbitrary multiplicities is an integrable field theory with twist function.

As explained already for the Gaudin model on an arbitrary quadratic Lie algebra, one can also consider a site at infinity with arbitrary multiplicity (in fact, the level $k_\infty$ and the non-dynamical field $\mathcal{B}(x)$ in \eqref{Eq:LaxTwistMult} already correspond to a double pole at infinity). The construction of the local AGM as a field theory with twist function also generalises to this case (see~\cite{Vicedo:2017cge}): the effect on the twist function and Lax matrix \eqref{Eq:LaxTwistMult} would then be to add polynomials in the spectral parameter $\lambda$.

\paragraph{Cyclotomic AGM.} Let us now consider a cyclotomic affine Gaudin model, associated with an automorphism $\s$ of $\gft$ of order $T$. We shall restrict to a particular type of automorphism, which is the lift on $\gft$ of an automorphism of $\g$. Let $\s$ be an automorphism of $\g$ of order $T$. We define an automorphism of $\gft = \C[t,t^{-1}] \oplus \C\Kd \oplus \C\Dd$, that we still denote $\s$, by
\begin{equation*}
\s ( X\otimes t^n ) = \s(X) \otimes t^n, \;\;\;\; \s(\Kd) = \Kd, \;\;\;\;\; \text{and} \;\;\;\; \s(\Dd) = \Dd.
\end{equation*}
The formal Gaudin Lax matrix $\Ls(\lambda)$ then satisfies the equivariance condition \eqref{Eq:EquivGaudin}.

The phase space $M$ of the formal cyclotomic AGM is similar to the one of a non-cyclotomic model, taking into account that the observables $\rp X 0 p$ attached to the site $\lambda_0=0$ at the origin belong to appropriate gradings $\gft^{(-p)}$. One can also construct a Segal-Sugawara map $\pk$ from $\F[M]$ to the algebra $\Obs$ of observables of a field theory. In particular, it sends the observables $\rp K r p$ (the coefficient of $i\Dd$ in $\rp X r p$) to complex numbers $k_{r,p}$, the levels of the theory.  We will not enter into the details of this construction here, as it is quite similar to the one presented above, with a particular treatment of the observables $\rp X 0 p$, and refer to~\cite{Vicedo:2017cge} for details.\\

To generalise the construction of the local AGM to the cyclotomic case, we will focus more on the auxiliary space $\gft$ of the model. One easily checks that the Lie morphism $\rho : \gft \rightarrow \Co$ satisfies:
\begin{equation}\label{Eq:SigmaRho}
\rho \circ \s = \s_c \circ \rho,
\end{equation}
where $\s_c$ is an automorphism of $\Co$ defined by
\begin{equation}\label{Eq:SigmaConn}
\s_c(X \otimes e_n) = \s(X) \otimes e_n \;\;\;\ \text{ and } \;\;\;\; \s_c(\p) = \p.
\end{equation}
Recall that the formal Gaudin Lax matrix $\Ls(\lambda)$ of the AGM satisfies the Poisson bracket \eqref{Eq:PBGaudin2}, with the matrix $\rc$ given as in \eqref{Eq:rGaudinCyc} by
\begin{equation*}
\rc\ti{12}(\lambda,\mu) = \frac{1}{T}\sum_{k=0}^{T-1} \frac{\s^k\ti{1}\Ct\ti{12}}{\mu-\omega^{-k}\lambda}.
\end{equation*}
Applying $\rho\otimes\rho$ to this matrix and using the equivariance property \eqref{Eq:SigmaRho}, one finds
\begin{equation*}
(\rho \otimes \rho) \rc\ti{12}(\lambda,\mu) = \frac{1}{T}\sum_{k=0}^{T-1} \frac{\s_c^k\null\ti{1}}{\mu-\omega^{-k}\lambda}(\rho\otimes\rho)\Ct\ti{12}.
\end{equation*}
Evaluating this $\Co\otimes\Co$-valued equation at $x$ and $y$ in $\Sc$ (on the left and right factors) and using the definition \eqref{Eq:SigmaConn} of $\s_c$ and the identity \eqref{Eq:CasRho}, one finds
\begin{equation*}
(\rho \otimes \rho) \rc\ti{12}(\lambda,\mu) (x,y) = \frac{1}{T}\sum_{k=0}^{T-1} \frac{\s^k\ti{1}C\ti{12}}{\mu-\omega^{-k}\lambda} \delta_{xy} = \Rc^0\ti{12}(\lambda,\mu) \delta_{xy},
\end{equation*}
where $\Rc^0$ is the standard $\Rc$-matrix \eqref{Eq:RCyc} on $\Lc(\g)$ twisted by $\s$. Defining the Lax connection $\nabla$ of the local AGM as in \eqref{Eq:NablaAGM}, one then finds that it satisfies the Poisson bracket \eqref{Eq:PBNabla}, characteristic of a model with twist function. We then extract the twist function $\vp(\lambda)$ and the Lax matrix $\Lc(\lambda,x)$ of the local AGM as in equation \eqref{Eq:LaxConnection}. By equation \eqref{Eq:SigmaConn}, the equivariance condition \eqref{Eq:EquivGaudin} translates to
\begin{equation*}
\s_c \bigl( \nabla(\lambda) \bigr) = \omega \nabla(\omega\lambda).
\end{equation*}
We thus get that the Lax matrix and the twist function of the local AGM satisfy the equivariance conditions \eqref{Eq:TwistEqui} and \eqref{Eq:EquiL}.

The construction of the local Hamiltonian and of the zero curvature equation under its flow generalises easily from the non-cyclotomic case to the cyclotomic one. The local cyclotomic AGM is thus a cyclotomic integrable field theory with twist function (as defined in Section \ref{Sec:ModelsTwist}, explaining the denomination cyclotomic used at the time).

\paragraph{Dihedral AGM.}\label{Para:LocalDAGM} Let us end this subsection by discussing briefly the Dihedral AGM (DAGM), \textit{i.e.} cyclotomic AGM with an additional reality condition. As for the automorphism $\s$ in the cyclotomic AGM, we start with an antilinear involutive automorphism $\tau$ of the finite algebra $\g$. We extend it to $\gft = \Loop \oplus \C\Kd \oplus \C\Dd$ as
\begin{equation}\label{Eq:PiReal}
\tau (X \otimes t^n) = \tau(X) \otimes t^{-n}, \;\;\;\; \tau(\Kd)=-\Kd \text{ and } \;\;\;\; \tau(\Dd) = -\Dd.
\end{equation}
One checks that this defines an antilinear involutive automorphism of $\gft$.

We then define a formal DAGM as explained generally in Subsection \ref{SubSec:GaudinGen}. We shall not discuss in detail the algebra of observables $\F[M]$ of the formal model as the discussion is quite similar to what we did previously. One constructs a generalised Segal-Sugawara map $\pk$ from $\F[M]$ to $\Obs$, which is the algebra of observables of a field theory, the local DAGM. The important property of this map is that it is compatible with the conjugacy on the complex algebra $\F[M]$ and $\Obs$, \textit{i.e.}
\begin{equation*}
\forall \, f\in\F[M], \;\;\;\; \pk (\overline{f}) = \overline{\pk(f)}.
\end{equation*}
The treatment of the auxiliary space of the DAGM is also similar to what was done above. We find that
\begin{equation*}
\rho \circ \tau = \tau_c \circ \rho,
\end{equation*}
where $\tau_c$ is the antilinear involutive automorphism of $\Co$ defined by
\begin{equation*}
\tau_c(X\otimes e_n) = X \otimes e_{-n} \;\;\;\; \text{ and } \;\;\;\; \tau_c(\p) = \p.
\end{equation*}
Note that this definition is compatible with the conjugacy \textit{via} the evaluation at $x\in\Sc$, as
\begin{equation*}
\overline{e_n(x)} = e^{-inx} = e_{-n}(x) \;\;\;\;\; \text{ and } \;\;\;\;\; \overline{\p_x f(x)} = \p_x \overline{f(x)}.
\end{equation*}
Using these properties and equation \eqref{Eq:PiReal} on $\pk$, one finds that the Lax matrix $\Lc(\lambda,x)$ and twist function $\vp(\lambda)$ extracted from the formal Gaudin Lax matrix $\Ls(\lambda)$ satisfy the reality conditions \eqref{Eq:Reality} and \eqref{Eq:TwistReal} (using the reality condition \eqref{Eq:RealGaudin} on $\Ls$).

One constructs a local quadratic Hamiltonian of the theory as in the non-dihedral case and finds that its flow on the Lax matrix takes the form of a zero curvature equation. As in equation \eqref{Eq:HamAgmLoc} for the non-dihedral case, this Hamiltonian reads
\begin{equation}\label{Eq:HamDagmLoc}
\Hc(\lambda) = \frac{1}{4\pi} \int_0^{2\pi} \dd x \; \kappa \bigl( \Scc(\lambda,x), \Scc(\lambda,x) \bigr) - \vp(\lambda)\Pc(\lambda),
\end{equation}
where
\begin{equation*}
\Scc(\lambda,x)=\vp(\lambda)\Lc(\lambda,x)
\end{equation*}
and $\Pc$ is the image under $\pk$ of the coefficient of $i\Kd$ in $\Ls(\lambda)$. This Hamiltonian satisfies the reality condition:
\begin{equation*}
\overline{\Hc(\lambda)}=\Hc( \overline{\lambda} ).
\end{equation*}
In general, we can choose the Hamiltonian $\Hc$ of the theory as any quantity extracted linearly from $\Hc(\lambda)$ (evaluation at particular points, residues, coefficients of higher order poles, integral over $\lambda$, any linear combination of these ...). Such an Hamiltonian is then in involution with $\Hc(\lambda)$ for any value of $\lambda$ and generates a zero curvature equation on $\Lc(\lambda,x)$.

\subsection{Integrable field theories with twist function as DAGM}

In the previous subsections, we proved that a formal DAGM can be realised as an integrable field theory with twist function, the local DAGM. Given the generality of this construction, it is natural to ask whether the known models with twist function can be interpreted as DAGM. The answer is yes for all integrable $\s$-models and their deformations, as shown by B. Vicedo in~\cite{Vicedo:2017cge}.

Let us be more precise about what we mean by that. We consider a local DAGM with algebra of observables $\Obs$, as constructed above, and an integrable field theory $\Tt$ with twist function, whose algebra of observables we denote $\Obs_{\Tt}$. The local DAGM possesses a twist function $\vp(\lambda)$, a Lax matrix $\Lc(\lambda,x)$ and a Hamiltonian $\Hc$, as constructed in the previous subsections. In the same way, by definition, the model with twist function is described by a twist function $\vp_{\Tt}(\lambda)$, a Lax matrix $\Lc_{\Tt}(\lambda,x)$ and a Hamitlonian $\Hc_{\Tt}$. We will say that $\Tt$ is a realisation of the local DAGM if:
\begin{itemize}
\setlength\itemsep{0.1em}
\item there exists a Poisson map $\pi : \Obs \mapsto \Obs_{\Tt}$, which realises the Gaudin observables in terms of the observables of the model $\Tt$ ;
\item the twist functions $\vp(\lambda)$ and $\vp_{\Tt}(\lambda)$ are equal ;
\item the Lax matrices and Hamiltonians of the two models are such that:
\begin{equation*}
\pi \bigl( \Lc(\lambda,x) \bigr) = \Lc_{\Tt}(\lambda,x) \;\;\;\; \text{ and } \;\;\;\; \pi \big( \Hc \bigr) = \Hc_{\Tt}.
\end{equation*}
\end{itemize}

\paragraph{Yang-Baxter model as a DAGM.} Let us illustrate this on a simple example, the Yang-Baxter model. It could seem surprising to take the Yang-Baxter model as a first example, as the undeformed PCM seems simpler. As we will see, in terms of their Gaudin representations, the Yang-Baxter model is actually simpler than the PCM, as it is a DAGM with simple poles, whereas the PCM is not.

Recall then the Yang-Baxter model on a real Lie group $G_0$, described in Subsection \ref{SubSec:YB} of this thesis. For simplicity, we will restrict to the split case $c=1$. We consider the complexification $\g$ of $\g_0$ and the antilinear involutive automorphism $\tau$ such that $\g_0=\g^\tau$. Let us check what are the properties of the Yang-Baxter model as a theory with twist function and deduce what would be the characteristic of an affine Gaudin model that would be realised by the Yang-Baxter model.

The Yang-Baxter model possesses a Lax matrix $\Lc(\lambda,x)$ valued in the Lie algebra $\g$ and satisfying the reality condition \eqref{Eq:Reality}. Thus, the corresponding AGM should be constructed from the Lie algebra $\g$ and be real with respect to the antilinear automorpism $\tau$. The Lax matrix does not satisfy an equivariance condition of the form \eqref{Eq:EquiL} and its Maillet bracket is described by a non-twisted standard $\Rc$-matrix \eqref{Eq:R0NonTwisted}. Thus, the AGM should not be cyclotomic, \textit{i.e.} $\s=\Id$. 

One still has to find what would be the sites and the levels of the AGM. For that, let us consider the twist function $\vp_\eta(\lambda)$ of the Yang-Baxter model, given by \eqref{Eq:TwistYB}. We find that its partial fraction decomposition is
\begin{equation*}
\vp_\eta(\lambda) = \frac{1}{\lambda-\eta}\frac{K}{2\eta} - \frac{1}{\lambda+\eta}\frac{K}{2\eta} - \frac{K}{1-\eta^2}. 
\end{equation*}
Comparing to equation \eqref{Eq:TwistGaudin}, one then consider an AGM with sites $\lambda_1=\eta$ and $\lambda_2=-\eta$ and fix the levels to be
\begin{equation*}
k_1 = \frac{K}{2\eta}, \;\;\;\; k_2 = -\frac{K}{2\eta}, \;\;\;\; k_\infty = - \frac{K}{1-\eta^2}.
\end{equation*}
This way, the Gaudin twist function \eqref{Eq:TwistGaudin} agrees with the twist function $\vp_\eta(\lambda)$.\\

Let us now turn to the phase space of the model, as we would like to construct a map $\pi$ from the observables $\Obs$ of the local real AGM (with sites and levels described above) to the observables $\Obs_{\text{YB}}$ of the Yang-Baxter model. For that, we consider the Lax matrix $\Lc_\eta(\lambda,x)$ of the Yang-Baxter model, given by \eqref{Eq:LaxYBHam}, and construct the matrix
\begin{equation*}
\Scc_\eta(\lambda,x) = \vp_\eta(\lambda) \Lc_\eta(\lambda,x).
\end{equation*}
We find that the partial fraction decomposition of this matrix is
\begin{equation*}
\Scc_\eta(\lambda) = \frac{\mathcal{K}_+}{\lambda-\eta} + \frac{\mathcal{K}_-}{\lambda+\eta},
\end{equation*}
where
\begin{equation*}
\mathcal{K}_+ = \frac{1}{2} \left( X - R_gX + \frac{K}{\eta} j^L \right) \;\;\;\; \text{ and } \;\;\;\; \mathcal{K}_- = \frac{1}{2} \left( X + R_gX - \frac{K}{\eta} j^L \right).
\end{equation*}
The fundamental observables of the local AGM (which generate $\Obs$) are the $\g_0$-valued Kac-Moody current $\Jc^{(1)}$ and $\Jc^{(2)}$ in equation \eqref{Eq:SGaudin} (for two sites). Comparing this equation to $\Scc_\eta$ above, one then defines the map
\begin{equation*}
\pi : \Obs \longrightarrow \Obs_{\text{YB}}
\end{equation*}
by
\begin{equation*}
\pi \bigl( \Jc^{(1)} \bigr) = \mathcal{K}_+ \;\;\;\; \text{ and } \;\;\;\; \pi \bigl( \Jc^{(2)} \bigr) = \mathcal{K}_-. 
\end{equation*}
By construction, this map satisfies $\pi\bigl( \Scc(\lambda,x) \bigr) = \Scc_\eta(\lambda,x)$, hence also $\pi\bigl( \Lc(\lambda,x) \bigr) = \Lc_\eta(\lambda,x)$ as the twist functions of the models are equal (note that we choose the free non-dynamical field $\mathcal{B}$ in \eqref{Eq:SGaudin} to be zero). Moreover, one checks explicitly that $\pi$ is a Poisson map, \textit{i.e.} that the currents $\mathcal{K}_\pm$ are commuting Kac-Moody currents with levels $k_1$ and $k_2$ defined above. The fact that Yang-Baxter deformations possess such Kac-Moody currents was already known before their reinterpretation in terms of Gaudin models (see~\cite{Rajeev:1988hq,Hollowood:2014rla,Vicedo:2015pna}).\\

To prove that the Yang-Baxter model is a realisation of the local DAGM constructed here, one still has to find that the Hamiltonian $\Hc_\eta$ of the Yang-Baxter model is realised as the image under $\pi$ of the Hamiltonian $\Hc$ of the local DAGM. The Hamiltonian $\Hc$ can be defined as any linear combination
\begin{equation*}
\Hc = c_1 \Hc_1 + c_2 \Hc_2, \;\;\;\; \text{ with } \;\;\;\; \Hc_r = \res_{\lambda=\lambda_r} \Hc(\lambda) \dd \lambda,
\end{equation*}
with the local quadratic Hamiltonian $\Hc(\lambda)$ defined in \eqref{Eq:HamAgmLoc}. Starting from the expression \eqref{Eq:HamYB} of the Yang-Baxter Hamiltonian $\Hc_\eta$, one finds that there is a choice of $c_1$ and $c_2$ such that $\pi(\Hc)=\Hc_\eta$. Thus, we successfully interpreted the Yang-Baxter model as the realisation of a local DAGM.

\paragraph{Integrable $\bm\s$-models as DAGM.} More generally, it was proved in~\cite{Vicedo:2017cge} that all integrable $\s$-models described in Chapter \ref{Chap:Models} are realisations of local DAGM, following a method close to the one above.

The PCM and its deformations are realisations of DAGM without cyclotomy (\textit{i.e.} $\s=\Id$). For example, the PCM corresponds to a real AGM with a site at 0 of multiplicity two. For the Yang-Baxter model in the non-split case ($c=i$), there is only one site at $i\eta$: the associated Kac-Moody current is then complex (\textit{i.e.} $\g$-valued and not $\g_0$-valued) and the matrix $\Scc$ then contains this current at the pole $i\eta$ and its image under $\tau$ at the pole $-i\eta$.

The $\Z_T$-cosets and their deformations are realisations of diheral AGM, with the automorphism $\s$ of order $T$ appearing in the definition of the considered coset space. For these models, it is also explained in~\cite{Vicedo:2017cge} how to treat the gauge constraint at the level of the local DAGM.

For completeness, let us also mention the fact that affine Toda field theories have been realised as DAGM in~\cite{Vicedo:2017cge}. However, for these models, it is a realisation of a formal DAGM and not of a local one. This is due to the fact that in this case, one of the highest level $k_{r,m_r-1}$ is zero, which prevents the construction of a Segal-Sugawara map and thus of a local DAGM (see Subsections \ref{SubSec:SimplestAGM} and \ref{SubSec:DAGM}).

\subsection{Integrable hierarchies in AGM}
\label{SubSec:HierarchyAGM}

In this subsection, we discuss integrable hierarchies of (local) AGM. Recall indeed from Subsections \ref{SubSec:SimplestAGM} and \ref{SubSec:DAGM} that formal AGM (and DAGM) can be mapped to integrable field theories with twist function (the local models). Thus, the construction of integrable hierarchies for models with twist function, as presented in Chapter \ref{Chap:LocalCharges} of this thesis, applies to local AGM.

\paragraph{Regular zeros.} Recall from Chapter \ref{Chap:LocalCharges} that this construction necessitates the existence of a regular zero of the model, \textit{i.e.} a zero $\lambda_0$ of $\vp(\lambda)$ such that $\Scc(\lambda,x)=\vp(\lambda)\Lc(\lambda,x)$ is regular at $\lambda=\lambda_0$. In our construction of local AGM, the matrix $\Scc(\lambda,x)$ appears quite naturally as a part of the Lax connection \eqref{Eq:NablaGaudin}. Moreover, comparing the expressions \eqref{Eq:LaxTwistMult} of the twist function and the matrix $\Scc(\lambda,x)$, one sees that the poles of $\Scc(\lambda,x)$ are exactly the poles of the twist function. Thus, if $\lambda_0$ is a zero of the twist function (and so is different from a pole), the matrix $\Scc(\lambda,x)$ is regular at $\lambda_0$.

The expressions used here for $\vp$ and $\Scc$ were constructed for the case of a non dihedral AGM. However, the observation that the poles of $\vp$ and $\Scc$ exactly coincide stays true for the most general local DAGM. Therefore, for a local DAGM, \textbf{every zero of $\bm{\vp(\lambda)}$ is regular}. Thus, the construction of local charges in involution presented in Chapter \ref{Chap:LocalCharges} applies naturally to local DAGM.

\paragraph{Hamiltonian and conservation.} So far, we constructed an infinite number of local charges in involution. However, we do not know yet if these charges are conserved. The Hamiltonian of the local AGM is extracted from the charge $\Hc(\lambda)$, given by \eqref{Eq:HamDagmLoc}. Note that the density
\begin{equation*}
\frac{1}{4\pi} \kappa \bigl( \Scc(\lambda,x), \Scc(\lambda,x) \bigr) 
\end{equation*}
appearing in $\Hc(\lambda)$ coincides, up to a global factor $a$, with the density
\begin{equation*}
\W_2(\lambda,x) = \vp(\lambda)^2 \Tr \bigl( \Lc(\lambda,x)^2 \bigr)
\end{equation*}
introduced in Chapter \ref{Chap:LocalCharges}. Recall that the quadratic charge $\Q^{\lambda_0}_2$ at a regular zero is defined as
\begin{equation*}
\Q^{\lambda_0}_2 = \int_0^{2\pi} \W_2(\lambda_0,x) \dd x.
\end{equation*}
As $\lambda_0$ is a zero of the twist function $\vp(\lambda)$, we then find that
\begin{equation*}
\Q^{\lambda_0}_2 = a \Hc(\lambda_0).
\end{equation*}
Thus, the quadratic charge at $\lambda_0$ is extracted from the local quadratic Hamiltonian $\Hc(\lambda)$. At the end of paragraph \ref{Para:LocalDAGM}, we said that the Hamiltonian $\Hc$ of the local AGM can be chosen as any quantity linearly extracted from $\Hc(\lambda)$. In particular, motivated by the discussion above, one can define $\Hc$ as a linear combination
\begin{equation}\label{Eq:HamLinZeros}
\Hc = \sum_{\lambda_0 \text{ zero of } \vp(\lambda)} b_{\lambda_0} \Hc(\lambda_0).
\end{equation}
By construction, such an Hamiltonian is thus part of the infinite algebra of local charges in involution, which are then all conserved.\\

The question whether any choice of local Hamiltonian extracted from $\Hc(\lambda)$ can be written in the form \eqref{Eq:HamLinZeros} is not easy to answer in general. However, let us explain why this is the case for the simplest AGM: complex, non-cyclotomic and with multiplicities one. In Subsection \ref{SubSec:SimplestAGM}, we observed that in this simple case, the Hamiltonian is expressed in \eqref{Eq:HamPoleSimple} as a sum of simple poles (and a constant non-dynamical term). The most general Hamiltonian linearly extracted from $\Hc(\lambda)$ is thus a linear combination
\begin{equation}\label{Eq:HamLinPoles}
\Hc = \sum_{r=1}^N c_r \Hc_r,
\end{equation}
of the residues $\Hc_r$'s of $\Hc(\lambda)$ at the sites $\lambda=\lambda_r$ (recall that the symbol $\lambda_0$ above did not stand for a site of the model but for a zero of the twist function). It is clear that any Hamiltonian of the form \eqref{Eq:HamLinZeros} is of the form \eqref{Eq:HamLinPoles} (up to the non-dynamical term $(\Omega,\Omega)$ in \eqref{Eq:HamPoleSimple}).

On the other hand, the twist function \eqref{Eq:TwistGaudin} of the local DAGM can be written as
\begin{equation*}
\vp(\lambda)=\frac{P(\lambda)}{(\lambda-\lambda_1)\cdots(\lambda-\lambda_N)},
\end{equation*}
where $P(\lambda)$ is a polynomial of order $N$. Thus, the twist function possesses $N$ zeros $\mu_1,\cdots,\mu_N$. Except maybe for degenerate cases (multiples zeros, ...), we then expect the $N$ residues $\Hc_r$'s of $\Hc(\lambda)$ to be linear combinations of the $N$ evaluations $\Hc(\mu_r)$'s. Therefore, an Hamiltonian of the form \eqref{Eq:HamLinPoles} is also of the form \eqref{Eq:HamLinZeros}, ensuring that the local charges in involution constructed above are conserved.

\cleardoublepage
\chapter{Quantum finite Gaudin models}
\label{Chap:QuantumFinite}

In this chapter, we discuss quantum finite Gaudin models. These were introduced by Gaudin, before the classical Gaudin models, in~\cite{Gaudin_76a} for the Lie algebra $\g=\sl(2,\C)$ and in~\cite{Gaudin_book83} for a general semi-simple Lie algebra $\g$. As physical models, they are interesting as they describe quantum spin chains with long range interaction, while being integrable. For simplicity, we will restrict here to complex quantum Gaudin models with sites of multiplicities one. Moreover, for most of this chapter, we will consider non-cyclotomic Gaudin models.

As explained in Chapter \ref{Chap:GaudinClass}, Subsection \ref{SubSec:FiniteClassGaudin}, classical finite Gaudin models possess a large number of conserved charges in involution, given by the evaluation of invariant polynomials of $\g$ on the Lax matrix on the model. There exist quantum analogues of these conserved charges~\cite{Feigin:1994in,Talalaev:2004qi,Chervov:2006xk,Molev:2013,Rybnikov:2008}: they form a large commutative subalgebra of the algebra of operators of the quantum Gaudin model, called the Gaudin~\cite{Feigin:1994in} (or Bethe~\cite{mukhin_2009a}) subalgebra. In particular, this subalgebra contains the quantum analogue of the quadratic Hamiltonians $\Hs_r$ defined for classical Gaudin models in Chapter \ref{Chap:GaudinClass}.

A natural question coming with the existence of conserved commuting charges is their diagonalisation. Indeed, as these operators commute, they can be simultaneously diagonalised. A first step in the resolution of quantum finite Gaudin models is thus to find the common eigenvectors and the spectrum (eigenvalues) of these commuting conserved charges.

This goal is partially achieved by the so-called Bethe ansatz for Gaudin models~\cite{Babujian:1993ts}. The Bethe ansatz is a general method to solve quantum integrable systems by finding common eigenvectors of their commuting charges. The Bethe ansatz for Gaudin models has proven to be quite successful and has led to very rich developments in the study of these models. However, it has some limits, principally in two directions. The first is the fact that it can be applied only for models where the Hilbert space is a tensor product of highest weight representations of the underlying Lie algebra $\g$. In particular, this excludes models which do not possess a vacuum. The second limitation of the Bethe ansatz is its non-completeness: indeed, in some (rather degenerate cases), the family of common eigenvectors obtained by the Bethe ansatz does not form a basis of the Hilbert space of the model~\cite{Mukhin:2007}.

An alternative but more abstract approach for the description of the spectrum of Gaudin models is the so-called Feigin-Frenkel-Reshetikhin (FFR) approach~\cite{Feigin:1994in}. It describes the eigenvalues of the conserved commuting charges of the Gaudin model in terms of some differential operators, called opers, which are associated with the underlying Lie algebra. One of the great assets of this approach is that it gives a theoretical description of all eigenvalues of the Gaudin charges, for any eigenvector, in any Hilbert space. Moreover, it is related~\cite{Frenkel:2004qy} to a very deep mathematical result, the Geometric Langlands Correspondence, and is thus of interest in pure mathematics.

This chapter is mostly an introductory review of known results on quantum finite Gaudin models. Its plan is the following. In Section \ref{Sec:QuantGaud}, we will explain the construction of quantum Gaudin models: the algebra of operators, the Hilbert space, the quadratic Hamiltonians and the Gaudin subalgebra. In Section \ref{Sec:Bethe}, we will recall the basics of the Bethe ansatz for Gaudin models. Finally, in Section \ref{Sec:FFR}, we give an introduction to the FFR approach of Gaudin models. We will end this section with some new results concerning the generalisation of the FFR approach to quantum cyclotomic Gaudin models, based on my PhD work~\cite{Lacroix:2016mpg} with B. Vicedo.

\section{Constructing quantum finite Gaudin models}
\label{Sec:QuantGaud}

In this section, we explain how to quantise classical Gaudin models associated with a finite Lie algebra $\g$~\cite{Gaudin_book83}. As this chapter concerns only finite Gaudin models, we will simply refer to them as Gaudin models for simplicity: one has to bear in mind that the results presented here do not hold in general for more general Gaudin models (see Chapter \ref{Chap:QuantumAffine} for first results on quantum affine Gaudin models). As explained in the introduction, we will also restrict here to the simplest Gaudin model: complex, without cyclotomy and with simple poles. In this section, we will align our notations to the ones more currently used in the literature about Gaudin models and will thus denote by $z$ the spectral parameter instead of $\lambda$. In particular, we will write $\zb=(z_1,\cdots,z_N) \in \C^N$ the sites of the considered Gaudin model.

\subsection{Algebra of quantum operators and Hilbert spaces}
\label{SubSec:OpHilbert}

\paragraph{Quantizing the Kirillov-Kostant bracket.} Recall that the phase space of the classical Gaudin model is a product of $N$ copies of the dual space $\g^*$, equipped with the Kirillov-Kostant bracket. We fix a basis $\lbrace I^a \rbrace$ of $\g$, with structure constants $\ft{ab}{c}$. The algebra of observables on the space $\g^*$ is then generated by $ \dim\g$ observables $X^a$'s, satisfying the bracket
\begin{equation*}
\lbrace X^a, X^b \rbrace = \fs{ab}{c} X^c.
\end{equation*}
A quantisation of this bracket then consists of a (non-commutative) algebra, generated by operators $\widehat{X}^a$, satisfying the commutation relations
\begin{equation}\label{Eq:QuantKK}
\left[ \widehat{X}^a, \widehat{X}^b \right] = \hbar \fs{ab}{c} \widehat{X}^c.
\end{equation}
These commutation relations are thus the same as the Lie relations of the Lie algebra $\g$ itself (up to the factor $\hbar$). In particular, if $V$ is a vector space and $\rho : \g \rightarrow \End(V)$ is a representation of $\g$, the operators
\begin{equation*}
\widehat{X}^a = \hbar \, \rho( I^a )
\end{equation*}
satisfy the commutation relations \eqref{Eq:QuantKK}.\\

The most general (in a sense made precise below) algebra with generators satisfying the commutation relations \eqref{Eq:QuantKK} is the universal enveloping algebra $U_{\hbar}(\g)$ of $\g$~\cite{Dixmier:1977}. This algebra is defined in the following way. Consider the tensor algebra
\begin{equation*}
T(\g) = \bigoplus_{n=0}^\infty \g^{\otimes n} = \C \oplus \g \oplus (\g\otimes\g) \oplus \cdots \oplus \g^{\otimes n} \oplus \cdots,
\end{equation*}
equipped with the tensor product $\otimes$. It is an associative and unital algebra (the identity element being the element $1$ of $\C=\g^{\otimes 0} \subset T(\g)$). We denote by $I_\hbar(\g)$ the ideal of $T(\g)$ generated by all
\begin{equation*}
X \otimes Y - Y \otimes X - \hbar [X,Y],
\end{equation*}
for $X,Y \in \g$. The universal enveloping algebra is then defined as the quotient
\begin{equation*}
U_\hbar(\g) = T(\g)/I_\hbar(\g).
\end{equation*}
This is also an associative and unital algebra. Moreover, it contains a copy of $\g$ (the image of $\g^{\otimes 1}$ under the quotient by $I_\hbar(\g)$) such that for any $X,Y$ in $\g \subset U_\hbar(\g)$, we have
\begin{equation*}
XY-YX = \hbar [X,Y],
\end{equation*}
where $XY$ denotes the product of $X$ and $Y$ in $U_\hbar(\g)$. The elements of the basis $\lbrace I^a \rbrace$, when seen as elements of $U_\hbar(\g)$, satisfy the commutation relation \eqref{Eq:QuantKK}.

Moreover, one can show~\cite{Dixmier:1977} that if $\Ac$ is any associative algebra generated by elements $\widehat{X}_a$ which satisfy these relations, there exists a surjective morphism of algebra from $U_\hbar(\g)$ to $\Ac$ which sends $I^a$ to $\widehat{X}^a$. In particular, there is a one-to-one correspondence between representations of the Lie algebra $\g$ and modules over the algebra $U_\hbar(\g)$. In this sense, $U_\hbar(\g)$ is the most general quantisation of the Kirillov-Kostant bracket.

A simple rescaling of the generators by $\hbar$ shows that $U_\hbar(\g)$ and $U(\g)=U_1(\g)$ are isomorphic. For simplicity, we will then consider the universal enveloping algebra $U(\g)$ and forget about the Planck constant $\hbar$. When we will talk about classical limits, we will then mean re-introducing the constant $\hbar$ and take the limit $\hbar$ goes to zero.

\paragraph{The algebra of operators of the Gaudin model.} Let us come back to the Gaudin model with sites $\zb=(z_1,\cdots,z_N) \in \C^N$. The phase space of the model is the Cartesian products of $N$ copies of the dual space $\g^*$. Thus, the algebra of observables $\Ac_{\zb}(\g)$ of the quantum model is the $N^{\rm th}$-tensor product of the universal enveloping algebra:
\begin{equation}\label{Eq:OpGaudin}
\Ac_{\zb}(\g) = U(\g)^{\otimes N}.
\end{equation}
For any $X\in\g$, one can consider an operator $X_{(r)} \in \Ac_{\zb}(\g)$, which belongs to the copy of $\g$ in the $r^{\rm th}$-factor of the tensor product \eqref{Eq:OpGaudin}. The algebra $\Ac_{\zb}(\g)$ is then generated by the $N \dim\g$ operators $I_{(k)}^a$, satisfying the commutation relations
\begin{equation*}
\left[ I_{(i)}^a, I_{(j)}^b \right] = \delta_{ij}\,\fs{ab}{c} I_{(i)}^c.
\end{equation*}
We will encode these generators in operators depending on the spectral parameter $z$, by defining, for any $X\in\g$:
\begin{equation*}
X(z) = \sum_{k=1}^N \frac{X_{(k)}}{z-z_k}.
\end{equation*}
From the circle lemma \eqref{Eq:CircleLemma}, one finds the following commutation relation:
\begin{equation}\label{Eq:ComGaudin}
\bigl[ X(z), Y(z') \bigr] = - \frac{[X,Y](z)-[X,Y](z')}{z-z'}, \;\;\;\; \forall \, X,Y \in \g, \;\; \forall \, z,z' \in\C.
\end{equation}
In particular, we have
\begin{equation*}
\bigl[ X(z), Y(z) \bigr] = - [X,Y]'(z),
\end{equation*}
where on the right-hand side, the prime denotes the derivative with respect to $z$.

In the classical limit, $I^a_{(i)}$ becomes the observable $X^a_{(i)}$ of the classical Gaudin model. Thus, the classical limit of the operators $I^a(z)$ are
\begin{equation*}
\sum_{i=1}^N \frac{X^a_{(i)}}{z-z_i}.
\end{equation*}
These are the coefficients of the Lax matrix \eqref{Eq:LaxGaudin} in the dual basis $I_a$ (with the constant element $\Omega$ chosen to be zero).

\paragraph{Hilbert space.}\label{Para:Hilbert} So far, we found an algebra $\Ac_{\zb}(\g)$ which is a quantisation of the algebra of observables of the Gaudin model. Let us define the Hilbert space $H$ of the theory, on which the algebra $\Ac_{\zb}(\g)$ acts linearly. We choose $N$ representations $R_1,\cdots,R_N$ of the Lie algebra $\g$. As explained in the first paragraph of this subsection, these representations are also modules of the universal enveloping algebra $U(\g)$. We define the Hilbert space to be
\begin{equation}\label{Eq:HilbertGaudin}
H = R_1 \otimes \cdots \otimes R_N.
\end{equation}
The algebra $\Ac_{\zb}(\g)$ acts on $H$ by letting the $k^{\rm th}$-factor $U(\g)$ of the tensor product \eqref{Eq:OpGaudin} act on the $k^{\rm th}$-factor $R_k$ in $H$.\\

For most of our purposes, we will consider the $R_k$'s to be highest weight representations of $\g$~\cite{Humphreys:1980dw}. For $\lambda \in \h^*$ a weight of $\g$ (where $\h$ is the Cartan subalgebra of $\g$ and $\h^*$ its dual), we will denote by $V_\lambda$ the Verma module of $\g$. If $\lambda$ is an integral and dominant weight, we can also consider $F_\lambda$, the finite dimensional irreducible representation of $\g$, obtained as a quotient of $V_\lambda$. Let us fix a collection $\lb=(\lambda_1,\cdots,\lambda_N) \in \h^{*\,N}$ of weights. We will often consider the Hilbert space
\begin{equation}\label{Eq:HilbertLambda}
H_{\lb} = V_{\lambda_1} \otimes \cdots \otimes V_{\lambda_N}.
\end{equation}

The particularity of the Verma module $V_\lambda$ is that it possesses a highest-weight vector $v_\lambda$. Recall from Appendix \ref{App:CartanWeyl} the definition of the nilpotent elements $E_\alpha$ of $\g$, associated with roots $\alpha\in\Delta$. In particular, we distinguish the positive nilpotent elements $E_\alpha$, for $\alpha\in\Delta_+$ a positive root, and the negative elements $F_\alpha=E_{-\alpha}$. The highest-weight vector satisfies
\begin{equation}\label{Eq:PosHighest}
E_\alpha.v_\lambda = 0, \;\;\;\; \forall \, \alpha\in\Delta_+.
\end{equation}
Moreover, the action of the Cartan subalgebra $\h \subset \g$ on $v_\lambda$ is given by
\begin{equation}\label{Eq:CartanHighest}
X.v_\lambda = \lambda(X) v_\lambda = \langle X, \lambda \rangle v_\lambda, \;\;\;\; \forall \, X\in\h,
\end{equation}
where $\langle \cdot,\cdot\rangle$ represents the canonical pairing on $\h\times\h^*$.

We define the vacuum state of the Hilbert space $H_{\lb}$ as the tensor product of the highest-weight vectors:
\begin{equation*}
\vl = v_{\lambda_1} \otimes \cdots \otimes v_{\lambda_N}.
\end{equation*}
This vacuum state satisfies
\begin{equation*}
E_\alpha(z).\vl = 0, \;\;\;\; \forall \, \alpha\in\Delta_+
\end{equation*}
and
\begin{equation*}
X(z).\vl = \bigl\langle X, \lambda(z) \bigr\rangle \vl,
\end{equation*}
with
\begin{equation}\label{Eq:Lz}
\lambda(z) = \sum_{i=1}^N \frac{\lambda_i}{z-z_i}\, \in \h^*.
\end{equation}

\subsection{Hamiltonians and Gaudin subalgebra}
\label{SubSec:HamQuant}

\paragraph{Quadratic Casimir of $\bm{U(\g)}$.}\label{Para:Cas} Recall the Killing form $\kappa$ (see Appendix \ref{App:SemiSimple}), which is a non-degenerate invariant bilinear form on $\g$. We denote by $\kappa^{ab}=\kappa(I^a,I^b)$ its evaluation in the basis $\lbrace I^a \rbrace$. As $\kappa$ is non-degenerate, we can define the inverse $\kappa_{ab}$ of $\kappa^{ab}$ and the dual basis $I_a = \kappa_{ab} I^b$. We define the following element of $U(\g)$:
\begin{equation*}
\Delta = \kappa_{ab} I^a I^b = I^a I_a.
\end{equation*}
It is independent of the choice of basis $\lbrace I^a \rbrace$. Moreover, by the invariance of $\kappa$, one finds that $\Delta$ is a Casimir of $U(\g)$, \textit{i.e.} that
\begin{equation*}
\left[ \Delta, X \right] = 0, \;\;\;\; \forall \, X\in U(\g).
\end{equation*}

This Casimir $\Delta$ is the quantum equivalent of the quadratic Poisson Casimir $\kappa_{ab} X^a X^b$ of the Kirillov-Kostant bracket, as described in Subsection \ref{SubSec:GaudinHam} on classical Gaudin Hamiltonians.\\

\paragraph{Quadratic Hamiltonians.} We define the quadratic quantum Gaudin hamiltonian as~\cite{Gaudin_book83}
\begin{equation}\label{Eq:QuantHam}
\Hs(z) = \frac{1}{2}\kappa_{ab} I^a(z) I^b(z) \in \Ac_{\zb}(\g).
\end{equation}
As explained in subsection \ref{SubSec:OpHilbert}, in the classical limit, $I^a(z)$ becomes the coefficient of $I_a$ in the classical Lax matrix $\Ls(z)$ (without the constant element $\Omega$). Thus, in this limit, the quantum Hamiltonian $\Hs(z)$ becomes the classical quadratic Hamiltonian \eqref{Eq:GaudinHamLambda}.

We will denote by $\Delta_{(k)}=\kappa_{ab} I^a_{(k)} I^b_{(k)}$ the Casimir of the algebra $\Ac_{\zb}(\g)=U(\g)^{\otimes N}$, constructed as $\Delta$ in the $k^{\rm th}$-tensor factor of $\Ac_{\zb}(\g)$. The partial fraction decomposition of $\Hs(z)$ is then
\begin{equation}\label{Eq:DecompoH}
\Hs(z) = \sum_{i=1}^N \left( \frac{1}{2}  \frac{\Delta_{(i)}}{(z-z_i)^2} + \frac{\Hs_i}{z-z_i} \right),
\end{equation}
with
\begin{equation}\label{Eq:Hi}
\Hs_i = \sum_{j\neq i} \kappa_{ab} \frac{I^a_{(i)} I^b_{(j)}}{z_i-z_j}.
\end{equation}

Recall the involution property \eqref{Eq:HlInvo} of the classical Gaudin Hamiltonian. Using the commutation relation \eqref{Eq:ComGaudin} and the invariance of $\kappa$, one checks that the Hamiltonian \eqref{Eq:QuantHam} satisfies a similar property at the quantum level:
\begin{equation*}
\left[ \Hs(z), \Hs(z') \right] = 0, \;\;\;\; \forall \, z,z' \in \C.
\end{equation*}
As the $\Delta_{(k)}$'s are Casimirs of the algebra $\Ac_{\zb}(\g)$, this is equivalent to
\begin{equation*}
[ \Hs_i, \Hs_j ] = 0, \;\;\;\; \forall \, i,j\in\lbrace 1,\cdots,N \rbrace.
\end{equation*}
In general, we will define the Hamiltonian $\Hs$ of the Gaudin model to be a linear combination
\begin{equation*}
\Hs=\sum_{k=1}^N c_k \Hs_k.
\end{equation*}

\paragraph{Global $\bm{\g_{(\infty)}}$-symmetry.}\label{Para:DiagAction} For $X\in\g$, we define the operator
\begin{equation*}
\Di{X} = \sum_{i=1}^N X_{(i)} \in \Ac_{\zb}(\g).
\end{equation*}
One checks that the map $X \mapsto \Di{X}$ is a Lie algebra homomorphism, \textit{i.e.} that
\begin{equation*}
\bigl[ \Di{X}, \Di{Y} \bigr] = \Di{[X,Y]}, \;\;\;\; \forall \, X,Y\in\g.
\end{equation*}
Thus, the $\Di{X} $'s form a Lie subalgebra $\Di{\g}$ of $\Ac_{\zb}(\g)$, isomorphic to $\g$. On the Hilbert space \eqref{Eq:HilbertGaudin}, the operator $\Di{X}$ acts by $X$ on each representation $R_i$:
\begin{equation}\label{Eq:DiagAct}
\forall \, w = w_1 \otimes \cdots \otimes w_N \in H, \;\;\;\;  \Di{X}.w = \Bigl( (X.w_1) \otimes \cdots \otimes w_N \Bigr) + \cdots  + \Bigl(  w_1 \otimes \cdots \otimes (X.w_N) \Bigr).
\end{equation}
This defines a representation of $\Di{\g}\simeq \g$ on $H$ that we call the diagonal action.\\

As we will see, this diagonal action is an infinitesimal symmetry of the Gaudin model. For that, we will need the following commutation relation, which is straightforward to prove:
\begin{equation}\label{Eq:ComDiag}
\bigl[ \Di{X}, Y(z) \bigr] = \bigl[ X(z), \Di{Y} \bigr] = [X,Y](z), \;\;\;\; \forall \, X,Y \in \g, \;\; \forall \, z \in \C.
\end{equation}
The commutator of the Gaudin hamiltonian $\Hs(z)$ with the diagonal operator $\Di{I^a}$ is then given by
\begin{eqnarray*}
\bigl[ \Di{I^a}, \Hs(z) \bigr]
&=& \frac{1}{2} \kappa_{bc} \Bigl( \bigl[ \Di{I^a}, I^b(z) \bigr] I^c(z) + I^b(z) \bigl[ \Di{I^a}, I^c(z) \bigr] \Big) \\
&=& \frac{1}{2} \kappa_{bc} \Bigl( \ft{ab}{d} I^d(z) I^c(z) + I^b(z) \ft{ac}{e} I^e(z) \Bigr) \\
&=& \frac{1}{2} \bigl( \fs{ab}{d}\kappa_{be} + \fs{ac}{e}\kappa_{dc} \bigr) I^d(z) I^e(z),
\end{eqnarray*}
where in the last equality, we relabelled some of the indices. The parenthesis in the last line vanishes due to the invariance of $\kappa$. We thus find that
\begin{equation*}
\bigl[ \Di{X}, \Hs(z) \bigr] = 0, \;\;\;\; \forall \, X\in\g, \;\; \forall \, z\in\C.
\end{equation*}
The diagonal action of $\g$ then commutes with the Gaudin Hamiltonian $\Hs(z)$, so it is a symmetry of the Gaudin model.

\paragraph{Higher-degrees Hamiltonians and Gaudin subalgebra.} In this subsection, we already found commuting operators in $\Ac_{\zb}(\g)$: the Casimirs $\Delta_{(k)}$ and the Hamiltonians $\Hs_k$, which were parts of the quadratic Hamiltonian $\Hs(z)$.

We discussed the classical limit of the quantum (finite) Gaudin model in Subsection \ref{SubSec:FiniteClassGaudin}. In particular, we said that these classical models admit a large number of conserved charges in involution. These charges were constructed from invariant polynomials on $\g$. These polynomials are generated by $\ell$ elementary polynomials $\Phi_1,\cdots,\Phi_\ell$, of degrees $d_i+1$, where the $d_i$'s are the so-called exponents of $\g$. We will denote by $E=\lbrace d_1,\cdots,d_\ell\rbrace$ the set\footnote{Technically, $E$ is a multiset, as some of the exponents can come with multiplicities, \textit{i.e.} some $d_i$'s can be identical for different $i$'s. In this case, one has to consider the corresponding $d_i$'s as different elements of $E$.} of exponents of $\g$. There are then $\ell$ corresponding independent conserved charges in involution $\Hs^{d}(z)$, $d\in E$, depending on the spectral parameter $z$ and constructed as polynomials of the Lax matrix of degrees $d+1$. The first exponent of a semi-simple Lie algebra is always 1, corresponding to the existence of a quadratic invariant polynomial on $\g$, namely the Killing form. The corresponding Hamiltonian is simply the quadratic Hamiltonian $\Hs^1(z)=\Hs(z)$ constructed above. The other Hamiltonians $\Hs^d(z)$ for $d>1$ are then called the higher-degree Hamiltonians of the classical Gaudin model.\\

The existence of a large number of invariant polynomials on $\g$ can be seen as a classical statement on Poisson Casimirs of the Kirillov-Kostant bracket. The quantum analogue of this statement is the existence of a large number of Casimirs of the universal enveloping algebra $U(\g)$. Together, these Casimirs form the center $\Zc(g)$ of the algebra $U(\g)$. This center is generated by $\ell$ elements $\Delta^d$, labelled by exponents $d\in E$, corresponding to the $\ell$ fundamental invariant polynomials. In particular, the Casimir associated with the first exponent $d=1$ is the quadratic Casimir $\Delta^1=\Delta$ constructed above.

As one can construct a quantum quadratic Hamiltonian $\Hs^1(z)=\Hs(z)$, one can also construct \textbf{higher-degree quantum Hamiltonians} $\Hs^d(z)$ for all exponents $d\in E$ strictly superior to one, whose classical limits are the higher-degree Hamiltonians mentioned above~\cite{Feigin:1994in} (see also~\cite{Talalaev:2004qi,Chervov:2006xk,Molev:2013} for explicit constructions). The quantum analogue of the involution of these classical Hamiltonians is
\begin{equation*}
\bigl[ \Hs^p(z), \Hs^q(z') \bigr] = 0, \;\;\;\; \forall \, p,q \in E, \;\; \forall \, z,z' \in \C.
\end{equation*}
Moreover, one finds that all Hamiltonians $\Hs^d(z)$ are invariant under the global diagonal action of $\Di{\g}$, as described in the previous paragraph for the quadratic one. Concretely, this means that
\begin{equation*}
\bigl[ \Di{X}, \Hs^d(z) \bigr] = 0, \;\;\;\; \forall \, X\in\g, \;\; \forall \, d \in E, \;\; \forall \, z \in \C.
\end{equation*}
The Hamiltonian $\Hs^d(z)$ is a rational function of the spectral parameter $z$, which possesses poles at the $N$ sites $z_i$'s of the Gaudin model. These poles are of order inferior or equal to $d+1$ and the coefficient of the higher order ($d+1$) pole at $z=z_k$ is (proportional to) the Casimir $\Delta^d_{(k)}$ in the $k^{\rm th}$-site of the algebra $\Ac_{\zb}(\g)=U(\g)^{\otimes N}$.

Let us consider the coefficients of all poles of all rational functions $\Hs^d(z)$ ($d\in E$). These are commuting operators in $\Ac_{\zb}(\g)$. We will denote by $\Zc_{\zb}(\g)$ the algebra generated by all these operators and will call it the \textbf{Gaudin subalgebra} of $\Ac_{\zb}(\g)$~\cite{Feigin:1994in} (it is also sometimes called the \textbf{Bethe subalgebra}~\cite{mukhin_2009a}). By construction, this is an abelian subalgebra of $\Ac_{\zb}(\g)$, which contains the Gaudin Hamitlonian $\Hs$ and the center $\Zc(\g)^{\otimes N}$ of $\Ac_{\zb}(\g)$.

\subsection{Realisations of quantum Gaudin models}

Let us consider a quantum system $\Scc$, with algebra of operators $\Ac_{\Scc}$ acting on a Hilbert space $H_{\Scc}$ and with an Hamiltonian $\Hc_{\Scc}$. We say that $\Scc$ is a realisation of the quantum Gaudin model if there exists an algebra morphism
\begin{equation*}
\pi : \Ac_{\zb}(\g) \longrightarrow \Ac_{\Scc},
\end{equation*}
such that $\pi(\Hs)=\Hc_{\Scc}$. The image of the Gaudin subalgebra $\Zc_{\zb}(\g)$ of $\Ac_{\zb}(\g)$ then contains a large number of commuting conserved charges of the system $\Scc$.

As the algebra $\Ac_{\Scc}$ acts on the Hilbert space $H_{\Scc}$, the map $\pi$ makes $H_{\Scc}$ a representation of the algebra $\Ac_{\zb}(\g)$ of Gaudin operators. In particular, if one is able to solve the Gaudin model on any representation of $\Ac_{\zb}(\g)$ (in particular, find the spectrum of the Hamiltonian $\Hs$ on this representation), one can theoretically solve all realisations of the Gaudin model.

\section{The Bethe ansatz for Gaudin models}
\label{Sec:Bethe}

In Section \ref{Sec:QuantGaud}, we explained how to construct a quantum Gaudin model and its algebra of operators $\Ac_{\zb}(\g)$. In particular, we found a large commutative subalgebra $\Zc_{\zb}(\g)$ of $\Ac_{\zb}(\g)$, containing the Hamiltonians $\Hs^d(z)$. The next step in the study of the Gaudin model is to diagonalise these Hamiltonians and find their spectrum (eigenvalues). In particular, as all operators in $\Zc_{\zb}(\g)$ commute, they can be diagonalised simultaneously and thus possess a common basis of eigenvectors.

The Bethe ansatz is a general method to construct these common eigenvectors when the Hilbert space is a product of highest-weight representations of $\g$~\cite{Babujian:1993ts} (see also~\cite{Varchenko:1991,Reshetikhin:1994qw}). We will use the notations of paragraph \ref{Para:Hilbert}, where we already discussed such Hilbert spaces. In particular, we fix a collection $\lb=(\lambda_1,\cdots,\lambda_N)\in\h^{*\,N}$ of weights and consider the Hilbert space $H_{\lb}$ as defined in equation \eqref{Eq:HilbertLambda}.

Although the Bethe ansatz provides common eigenvectors to all Hamiltonians $\Hs^d(z)$, the description of the eigenvalues of the higher-degree Hamiltonians is more difficult than the ones of the quadratic Hamiltonian. In particular there exists a common formalism for the description of the quadratic Hamiltonian, independent of the Lie algebra $\g$, but such a formalism for higher-degree Hamiltonians is difficult to construct due to the fact that the exponents depend on the Lie algebra $\g$. In this section, we will therefore mainly focus on the diagonalisation of the quadratic Hamiltonian $\Hs(z)=\Hs^1(z)$. However, note that the section \ref{Sec:FFR}, which concerns the more abstract Feigin-Frenkel-Reshetikhin approach, will use a formalism independent of $\g$.

\subsection{Preliminary: eigenvalues of the quadratic Casimirs}
\label{SubSec:EigenCas}

Recall from Paragraph \ref{Para:Cas} the quadratic Casimir $\Delta$ of $U(\g)$. Let us consider the Verma module $V_\lambda$ of weight $\lambda\in\h^*$ (or any representation with highest-weight $\lambda$). As we will see, the highest-weight vector $v_\lambda$ of $V_\lambda$ is an eigenvector under the action of $\Delta$. A generating family of $V_\lambda$ is given by successive applications of the simple negative generators $F_i=F_{\alpha_i}$'s (associated with simple roots $\alpha_i$'s). As $\Delta$ commutes with all the $F_i$'s, all vectors of $V_\lambda$ are eigenvectors of $\Delta$, with the same eigenvalue as $v_\lambda$.

Let us compute this eigenvalue. Recall that a basis of $\g$ is given by the $E_\alpha$'s and $F_\alpha$'s, $\alpha\in\Delta_+$, together with a basis $K_i$ ($i=1,\cdots,\ell$) of the Cartan subalgebra $\h$ (where $\ell=\dim\h$ is the rank of $\g$). We will suppose that $K_i$ is orthonormal with respect to the Killing form $\kappa$ (one can always find such a basis as $\kappa$ restricts non-degenerately to $\h$). The dual basis of $\lbrace E_{\alpha}, F_{\alpha}, K_i \rbrace_{\alpha\in\Delta_+, \, i=1,\cdots,\ell}$ with respect to $\kappa$ is then $\lbrace F_{\alpha}, E_{\alpha}, K_i \rbrace_{\alpha\in\Delta_+, \, i=1,\cdots,\ell}$. We thus find that
\begin{equation*}
\Delta = \sum_{i=1}^\ell K_i K_i + \sum_{\alpha\in\Delta_+} \bigl( E_\alpha F_\alpha + F_\alpha E_\alpha \bigr).
\end{equation*}
Let $\zeta : \h^* \rightarrow \h$ be the canonical isomorphism induced by the Killing form (see Appendix \ref{App:BasesCartan}). It is known that the commutator of $E_\alpha$ with $F_\alpha$ is given by
\begin{equation*}
[E_\alpha,F_\alpha] = \zeta(\alpha), \;\;\;\; \forall \, \alpha\in\Delta_+.
\end{equation*}
We then find
\begin{equation}\label{Eq:CasPBW}
\Delta = \sum_{i=1}^\ell K_i K_i + 2\zeta(\rho) + 2\sum_{\alpha\in\Delta_+} F_\alpha E_\alpha,
\end{equation}
where $\rho$ is the so-called Weyl weight of $\g$:
\begin{equation}\label{Eq:WeylW}
\rho = \frac{1}{2} \sum_{\alpha\in\Delta_+} \alpha.
\end{equation}
By equations \eqref{Eq:PosHighest} and \eqref{Eq:CartanHighest}, one then finds
\begin{equation*}
\Delta.v_\lambda = \left( \sum_{i=1}^\ell \langle K_i, \lambda \rangle^2 + 2\langle \zeta(\rho), \lambda \rangle \right)v_\lambda.
\end{equation*}
The restriction of the Killing form on $\h$ induces a bilinear form $\fd$ on $\h^*$ by (see also Appendix \ref{App:BasesCartan})
\begin{equation*}
(\alpha,\beta) = \kappa\bigl( \zeta(\alpha),\zeta(\beta) \bigr) = \bigl\langle \zeta(\alpha), \beta \bigr\rangle.
\end{equation*}
Using the fact that $\lbrace K_i \rbrace_{i=1,\cdots,\ell}$ is an orthonormal basis of $\h$, one finds
\begin{equation*}
\Delta.v_\lambda = ( \lambda, \lambda+2\rho)v_\lambda.
\end{equation*}
Thus, one has
\begin{equation*}
\Delta.w =  ( \lambda, \lambda+2\rho) w, \;\;\;\; \forall \, w\in V_\lambda.\vspace{12pt}
\end{equation*}

Let us consider the quadratic Hamiltonian $\Hs(z)$, acting on the Hilbert space $\Hl$. Recall the partial fraction decomposition \eqref{Eq:DecompoH} of $\Hs(z)$. The Casimir $\Delta_{(k)}$, which is the coefficient of the double pole of $\Hs(z)$ at $z=z_k$, acts as the Casimir $\Delta$ on the $k^{\rm th}$-tensor fact of $\Hl=V_{\lambda_1}\otimes\cdots\otimes V_{\lambda_N}$. According to the discussion above, we then have
\begin{equation}\label{Eq:EigenCas}
\Delta_{(k)}.w = ( \lambda_k, \lambda_k+2\rho)w, \;\;\;\; \forall\, w\in \Hl.
\end{equation}
The diagonalisation of the quadratic Hamiltonian $\Hs(z)$ thus reduces to the diagonalisation of the residues $\Hs_k$ in \eqref{Eq:DecompoH}.

\subsection{Vacuum eigenvalues}
\label{SubSec:VacuumEigen}

Recall from the paragraph \ref{Para:Hilbert} the vacuum state $\vl = v_{\lambda_1}\otimes\cdots\otimes v_{\lambda_N}$ of $\Hl$. Let us study the action of $\Hs_k$ on $\vl$. We will use the notations of the subsection \ref{SubSec:EigenCas}. From equation \eqref{Eq:Hi}, We get
\begin{equation*}
\Hs_i = \sum_{ \substack{j=1 \\ j\neq i}}^N \frac{1}{z_i-z_j} \left( \sum_{k=1}^\ell K_{k\,(i)} K_{k\,(j)} + \sum_{\alpha\in\Delta_+} \bigl( F_{\alpha\,(i)} E_{\alpha\,(j)} + F_{\alpha\,(j)} E_{\alpha\,(i)} \bigr) \right)
\end{equation*}
Using equations \eqref{Eq:PosHighest} and \eqref{Eq:CartanHighest}, we get that the vacuum state $\vl$ is an eigenvector of the $\Hs_i$'s:
\begin{equation*}
\Hs_i.\vl = \left( \sum_{ \substack{j=1 \\ j\neq i}}^N  \sum_{k=1}^\ell \frac{\langle K_k, \lambda_i\rangle \langle K_k, \lambda_j \rangle}{z_i-z_j}  \right)\vl.
\end{equation*}
As in Subsection \ref{SubSec:EigenCas}, by the orthonormality of the basis $\lbrace K_k \rbrace_{k=1,\cdots,\ell}$ of $\h$, one gets
\begin{equation*}
\Hs_i.\vl = \left( \sum_{ \substack{j=1 \\ j\neq i}}^N  \frac{(\lambda_i,\lambda_j)}{z_i-z_j}  \right)\vl.
\end{equation*}
Combining this with equation \eqref{Eq:EigenCas}, we get
\begin{equation*}
\Hs(z).\vl = \left( \frac{1}{2} \sum_{i=1}^N \frac{(\lambda_i,\lambda_i+2\rho)}{(z-z_i)^2} + \sum_{i=1}^N \frac{1}{z-z_i} \sum_{ \substack{j=1 \\ j\neq i}}^N  \frac{(\lambda_i,\lambda_j)}{z_i-z_j}  \right)\vl.
\end{equation*}
This can be rewritten in terms of $\lambda(z)$, defined in \eqref{Eq:Lz}, as
\begin{equation*}
\Hs(z).\vl = \left( \frac{1}{2}\bigl( \lambda(z), \lambda(z) \bigr) - \bigl( \lambda'(z), \rho \bigr) \right)  \vl = \Ev(z) \vl,
\end{equation*}
where $\lambda'(z)$ denotes the derivative of $\lambda(z)$ with respect to $z$. We call $\Ev(z)$ the vacuum eigenvalue of $\Hs(z)$.

\subsection[Diagonal decomposition of $H_\lambda$]{Diagonal decomposition of $\bm{\Hl}$}
\label{SubSec:DiagBethe}

\paragraph{$\bm{\Di{\h}}$-weight decomposition.} Recall from Paragraph \ref{Para:DiagAction} the Lie algebra $\Di{\g} \subset \Ac_{\zb}(\g)$, which acts on $\Hl$ by the diagonal elements $\Di{X}$, $X\in\g$, as in \eqref{Eq:DiagAct}. Consider the Cartan subalgebra $\Di{\h}$ of $\Di{\g}$. One can decompose the Hilbert space $\Hl$ in $\Di{\h}$-weight spaces:
\begin{equation*}
\Hl = \bigoplus_{\mu \in \Ll} W_{\mu},
\end{equation*}
where $\Ll \subset \h^*$ is composed of weights $\mu\in\h^*$ such that the weight space
\begin{equation*}
W_\mu = \left\lbrace w\in\Hl \; | \; \Di{X}.w = \langle X, \mu \rangle w, \; \forall \, X \in \h \right\rbrace
\end{equation*}
is non trivial.

If the $w_i$'s are elements of $V_{\lambda_i}$ of weights $\mu_i\in\h^*$ under the $\h$-weight decomposition of $V_{\lambda_i}$, the vector $w=w_1 \otimes \cdots \otimes w_N$ of $\Hl$ belongs to $W_{\mu_1+\cdots+\mu_N}$. The highest weight in $\Ll$ is the sum $\lambda_\infty=\lambda_1+\cdots+\lambda_N$ and the corresponding weight space is simply $W_{\lambda_\infty} = \C\vl$. More generally, if 
\begin{equation*}
\mu = \lambda_\infty - \sum_{i=1}^\ell p_i \alpha_i,
\end{equation*}
a generating family of the weight space $W_\mu$ is given by vectors of the form $\mathcal{F}.\vl$, where $\mathcal{F}$ is a product of simple negative generators $F_i=F_{\alpha_i}$, appearing $p_i$ times, and acting on any tensor factor of $\Hl=V_{\lambda_1}\otimes\cdots\otimes V_{\lambda_N}$. For example, $W_{\lambda_\infty-\alpha_i}$ is of dimension $N$ and admits the following basis:
\begin{equation*}
\lbrace F_{i\,(k)}.\vl \, \rbrace_{k=1,\cdots,N}.
\end{equation*}
The description of the weight space $W_\mu$ above implies in particular that all the $W_\mu$'s are finite dimensional.\\

Recall that the diagonal action $\Di{\g}$ is a symmetry of the Gaudin model, in the sense that all $\Di{X}$ ($X\in\g$) commute with the Gaudin Hamiltonians $\Hs^d(z)$. In particular, these Hamiltonians commute with the Cartan subalgebra $\Di{\h}$ of $\Di{\g}$. This implies that they stabilise the weight spaces $W_\mu$'s and that one can find a basis of common eigenvectors of the $\Hs^d(z)$'s and $\Di{\h}$.

\paragraph{Decomposition in highest-weight representations of $\bm{\Di{\g}}$.} Let us consider the Hilbert space $\Hl$ as a representation of $\Di{\g}$. In particular, as it is a tensor product of Verma modules, one can decompose $\Hl$ as a direct sum of Verma modules of $\Di{\g}$~\cite{Humphreys:1972}. In general, the weights and the multiplicities of the Verma modules appearing in this decomposition are difficult to describe. We shall then label these modules by abstract indices $s$ in a set $\Il$ and write
\begin{equation}\label{Eq:DecompoVerma}
\Hl = \bigoplus_{s \in \Il} M_s,
\end{equation}
where $M_s$ is a $\Di{\g}$-submodule of $\Hl$, isomorphic to the Verma module $V_{\mu_s}$, for some weight $\mu_s \in \h^*$. In particular, there is a unique (up to scalar multiplication) highest-weight vector $B_s$ in $M_s$, which is then in $W_{\mu_s}$. Note that one of these modules $M_{s_\infty}$ is a Verma module of weight $\mu_{s_\infty}=\lambda_\infty$, whose highest-weight vector is the vacuum state $B_{s_\infty}=\vl$. It is the unique module of highest weight $\lambda_\infty$ in the decomposition \eqref{Eq:DecompoVerma}.\\

Recall that the Gaudin subalgebra $\Zc_{\zb}(\g)$ (\textit{i.e.} the Gaudin Hamiltonians $\Hs^d(z)$) commute with the diagonal action $\Di{\g}$. One can then choose the labelling $\Il$ of the decomposition \eqref{Eq:DecompoVerma} such that the Gaudin subalgebra $\Zc_{\zb}(\g)$ stabilises all $M_s$'s ($s\in\Il$).

Let $\Q$ be an operator in $\Zc_{\zb}(\g)$. As $\Q$ stabilises $M_s$ and commutes with the action of $\Di{\h}$, the state $\Q.B_s$ is a vector of $M_s$ of $\Di{\h}$-weight $\mu_s$. Thus, it is proportional to $B_s$ itself as $\mu_s$ is the highest-weight of $M_s$. This means that there exists $\chi_s(\Q) \in \C$ such that
\begin{equation*}
\Q.B_s = \chi_s(\Q) B_s.
\end{equation*}

As $B_s$ is the highest-weight vector of $M_s$, all other vectors are obtained from $B_s$ by successive actions of the simple lowering operators $F_{i\,(\infty)}$ ($i=1,\cdots,\ell$). As $\Q$ commutes with the diagonal operators $F_{i\,(\infty)}$, all vectors of $M_s$ are also eigenvectors of $\Q$ of eigenvalue $\chi_s(\Q)$. Thus, we have
\begin{equation*}
\Q.v = \chi_s(\Q) v, \;\;\;\; \forall \, v \in M_s. 
\end{equation*}
To diagonalise the operators of the Gaudin subalgebra $\Zc_{\zb}(\g)$ on $\Hl$, it is then enough to find all highest-weight vectors $B_s$, $s\in\Il$. This is the goal of the Bethe ansatz.

As highest-weight vectors under the action of $\Di{\g}$, the states $B_s$'s are $\Di{\g}$-singular, in the sense that
\begin{equation*}
E_{\alpha\,(\infty)}.B_s = 0, \;\;\;\; \forall \, \alpha\in\Delta_+, \;\; \forall \, s\in\Il.
\end{equation*}
There exist other $\Di{\g}$-singular vectors in $\Hl$, which are not highest-weight vectors of one of the $M_s$'s. In fact, the Bethe ansatz also generates these other singular vectors. We shall come back on this fact in Section \ref{Sec:FFR}.

\subsection{Bethe ansatz at one excitation}

\paragraph{Off-shell Bethe vectors.} Let us illustrate the Bethe ansatz method by finding the simplest eigenvectors of the Hamiltonian $\Hs(z)$'s (after the vacuum state), the so-called Bethe vectors with one excitation. Let $w$ be a complex number and $i\in\lbrace 1,\cdots,\ell \rbrace$. We define the so-called off-shell Bethe vector
\begin{equation*}
\Psi_i(w) = F_i(w).\vl = \sum_{k=1}^N \frac{F_{i\,(k)}}{w-z_k}.\vl.
\end{equation*}
It is clear that $\Psi_i(w)$ belongs to $W_{\lambda_\infty-\alpha_i}$, \textit{i.e.} is of weight $\lambda_\infty-\alpha_i$ under the $\Di{\h}$-weight decomposition of $\Hl$. We say that $\Psi_i(w)$ is a Bethe vector with one excitation in the direction $\alpha_i$. 

\paragraph{Action of $\bm{\Hs_k}$ on the off-shell Bethe vector.} Recall from subsection \ref{SubSec:VacuumEigen} that the vacuum state $\vl$ is an eigenvector of $\Hs(z)$ with eigenvalue $\Ev(z)$. We then have
\begin{equation*}
\Hs(z).\Psi_i(w) = F_i(w) \Hs(z).\vl + \bigl[\Hs(z),F_i(w)\bigr].\vl = \Ev(z) \Psi_i(w) + \bigl[\Hs(z),F_i(w)\bigr].\vl.
\end{equation*}
In particular, taking the residue at $z=z_k$, one has
\begin{equation}\label{Eq:HonBethe}
\Hs_k.\Psi_i(w) =\Evk{k} \Psi_i(w) + \bigl[\Hs_k,F_i(w)\bigr].\vl,
\end{equation}
where
\begin{equation*}
\Evk{k} = \res_{z=z_k} \Ev(z) = \sum_{j\neq k} \frac{(\lambda_k,\lambda_j)}{z_k-z_j}.
\end{equation*}
Using the expression \eqref{Eq:Hi} of $\Hs_k$, one gets
\begin{equation*}
\bigl[\Hs_k,F_i(w)\bigr] = \kappa_{ab} \sum_{j\neq k} \frac{\bigl[I^a_{(k)},F_i(w)\bigr]I^b_{(j)} + I^a_{(k)} \bigl[I^b_{(j)},F_i(w)\bigr]}{z_k-z_j}.
\end{equation*}
Note that
\begin{equation*}
\bigl[X_{(k)},Y(w)\bigr] = \frac{[X,Y]_{(k)}}{w-z_k}, \;\;\;\; \forall \, X,Y \in \g.
\end{equation*}
Thus, one gets
\begin{equation}\label{Eq:ComHamF}
\bigl[\Hs_k,F_i(w)\bigr] = \sum_{j\neq k} \left( \frac{[I_a,F_i]_{(k)} I^a_{(j)}}{(w-z_k)(z_k-z_j)} + \frac{[I_a,F_i]_{(j)} I^a_{(k)} }{(w-z_j)(z_k-z_j)} \right).
\end{equation}
Let us now compute the action of this operator on the vacuum state $\vl$. We let $I^a$ run through the basis $\lbrace E_\alpha \rbrace_{\alpha\in\Delta} \cup \lbrace K_m \rbrace_{m=1,\cdots,\ell}$. If $I^a = E_\alpha$ with $\alpha\in\Delta_+$, then $I^a_{(j)}.\vl=I^a_{(k)}.\vl=0$ so this choice does not contribute to the action of \eqref{Eq:ComHamF} on $\vl$. Let now $I^a=F_\alpha$ with $\alpha\in\Delta_+$ of height $h$: we then have $I_a=E_\alpha$ so $[I_a,F_i]$ is of height $h-1$. Thus, if $h>1$, we get $[I_a,F_i]_{(j)}.\vl=[I_a,F_i]_{(k)}.\vl=0$. Therefore, the only contributions to the action of \eqref{Eq:ComHamF} on $\vl$ are $I^a=F_m$ and $I^a=K_m$ ($m\in\lbrace 1,\cdots,\ell\rbrace$). We then get
\begin{eqnarray*}
\bigl[\Hs_k,F_i(w)\bigr].\vl
&=& \sum_{j\neq k} \sum_{m=1}^\ell \left( \frac{[E_m,F_i]_{(k)} F_{m\,(j)}}{(w-z_k)(z_k-z_j)} + \frac{[E_m,F_i]_{(j)} F_{m\,(k)} }{(w-z_j)(z_k-z_j)} \right).\vl \\
& & \hspace{40pt} + \sum_{j\neq k} \sum_{m=1}^\ell \left( \frac{[K_m,F_i]_{(k)} K_{m\,(j)}}{(w-z_k)(z_k-z_j)} + \frac{[K_m,F_i]_{(j)} K_{m\,(k)} }{(w-z_j)(z_k-z_j)} \right).\vl.
\end{eqnarray*}
Recall the canonical isomorphism $\zeta:\h^*\rightarrow\h$ induced by the Killing form (see Appendix \ref{App:BasesCartan} and Subsection \ref{SubSec:EigenCas}). We then have $[E_m, F_i] = \delta_{mi} \, \zeta(\alpha_i)$. Thus, we get
\begin{eqnarray*}
\bigl[\Hs_k,F_i(w)\bigr].\vl
&=& \sum_{j\neq k} \left( \frac{\zeta(\alpha_i)_{(k)} F_{i\,(j)}}{(w-z_k)(z_k-z_j)} + \frac{\zeta(\alpha_i)_{(j)} F_{i\,(k)} }{(w-z_j)(z_k-z_j)} \right).\vl \\
& & \hspace{40pt} - \sum_{j\neq k} \sum_{m=1}^\ell \langle K_m, \alpha_i \rangle  \left(\frac{F_{i\,(k)} K_{m\,(j)}}{(w-z_k)(z_k-z_j)} + \frac{F_{i\,(j)} K_{m\,(k)} }{(w-z_j)(z_k-z_j)} \right).\vl \\
&=& \frac{\langle \zeta(\alpha_i), \lambda_k \rangle }{(w-z_k)} \sum_{j\neq k} \frac{F_{i\,(j)}}{z_k-z_j}.\vl  + \left( \sum_{j \neq k} \frac{\langle \zeta(\alpha_i),\lambda_j \rangle }{(w-z_j)(z_k-z_j)}\right)  F_{i\,(k)}.\vl \\
& & \hspace{40pt} - \sum_{m=1}^\ell \langle K_m, \alpha_i \rangle  \left( \sum_{j\neq k} \frac{\langle K_m, \lambda_j \rangle}{(z_k-z_j)} \right) \frac{F_{i\,(k)}}{(w-z_k)}.\vl \\
& & \hspace{40pt} - \left( \sum_{m=1}^\ell \langle K_m, \alpha_i \rangle \langle K_m, \lambda_k \rangle \right) \left( \sum_{j\neq k} \frac{F_{i\,(j)}  }{(w-z_j)(z_k-z_j)} \right).\vl
\end{eqnarray*}
Recall the bilinear form $\fd$ on $\h^*$ induced by $\kappa$ (see Appendix \ref{App:BasesCartan} and Subsection \ref{SubSec:EigenCas}). As $K_m$ is an orthonormal basis of $\h$, we get
\begin{eqnarray*}
\bigl[\Hs_k,F_i(w)\bigr].\vl
&=& \frac{(\alpha_i, \lambda_k ) }{(w-z_k)} \sum_{j\neq k} \frac{F_{i\,(j)}}{z_k-z_j}.\vl  + \left( \sum_{j \neq k} \frac{(\alpha_i,\lambda_j ) }{(w-z_j)(z_k-z_j)}\right)  F_{i\,(k)}.\vl \\
& & \hspace{40pt} - \left( \sum_{j\neq k} \frac{ (\alpha_i, \lambda_j )}{(z_k-z_j)} \right) \frac{F_{i\,(k)}}{(w-z_k)}.\vl  - (\alpha_i,\lambda_k) \left( \sum_{j\neq k} \frac{F_{i\,(j)}  }{(w-z_j)(z_k-z_j)} \right).\vl \\
&=& (\alpha_i,\lambda_k) \sum_{j\neq k} \left( \frac{1}{(w-z_k)(z_k-z_j)} - \frac{1}{(w-z_j)(z_k-z_j)} \right) F_{i\,(j)}.\vl \\
& & \hspace{40pt} + \left( \sum_{j \neq k} (\alpha_i,\lambda_j)  \left( \frac{1}{(w-z_j)(z_k-z_j)} - \frac{1}{(w-z_k)(z_k-z_j)}\right) \right) F_{i\,(k)}.\vl.
\end{eqnarray*}
Using the circle lemma \eqref{Eq:CircleLemma}, one gets
\begin{eqnarray*}
\bigl[\Hs_k,F_i(w)\bigr].\vl
&=& \frac{(\alpha_i,\lambda_k)}{w-z_k} \left( \sum_{j\neq k}  \frac{F_{i\,(j)}}{w-z_j} \right) .\vl - \left( \sum_{j \neq k} \frac{(\alpha_i,\lambda_j)}{w-z_j} \right) \frac{F_{i\,(k)}}{w-z_k}.\vl \\
&=& \frac{(\alpha_i,\lambda_k)}{w-z_k} \left( \sum_{j=1}^N  \frac{F_{i\,(j)}}{w-z_j} \right) .\vl - \left( \sum_{j=1}^N \frac{(\alpha_i,\lambda_j)}{w-z_j} \right) \frac{F_{i\,(k)}}{w-z_k}.\vl \\
&=& \frac{(\alpha_i,\lambda_k)}{w-z_k} \Psi_i(w) - \bigl(\alpha_i,\lambda(w)\bigr) \frac{F_{i\,(k)}}{w-z_k}.\vl.
\end{eqnarray*}
Combining this equation with \eqref{Eq:HonBethe}, we find
\begin{equation}\label{Eq:HkPsi}
\Hs_k.\Psi_i(w) = \Ewk{k}(w) \Psi_i(w) - \bigl(\alpha_i,\lambda(w)\bigr) \frac{F_{i\,(k)}}{w-z_k}.\vl,
\end{equation}
with
\begin{equation*}
\Ewk{k}(w) = \Evk{k} + \frac{(\alpha_i,\lambda_k)}{w-z_k}.
\end{equation*}

\paragraph{Bethe equation and eigenvalue with excitation.} It is clear from equation \eqref{Eq:HkPsi} that $\Psi_i(w)$ is an eigenvector of all $\Hs_k$'s if $w$ satisfies the \textbf{Bethe equation}:
\begin{equation}\label{Eq:Bethe1}
\bigl( \alpha_i, \lambda(w) \bigr) = \sum_{k=1}^N \frac{(\alpha_i,\lambda_k)}{w-z_k} = 0.
\end{equation}
In this case, we say that the Bethe vector $\Psi_i(w)$ is on-shell and write it as $\Po_i(w)$. The eigenvalue of $\Hs_k$ on $\Po_i(w)$ is then $\Ewk{k}^{\,\text{on}}(w)$, defined as the value of $\Ewk{k}(w)$ for $w$ solution of the Bethe equation. Such a solution is called a \textbf{Bethe root}.

Recall the partial fraction decomposition \eqref{Eq:DecompoH} of $\Hs(z)$. Recall also that the Casimirs $\Delta_{(i)}$ at the double poles of $\Hs(z)$ act as multiples of the identity on the whole Hilbert space $\Hl$. We then get that the action of $\Hs(z)$ on $\Po_i(w)$ is
\begin{equation*}
\Hs(z).\Po_i(w) = \Exo(z,w) \Po_i(w),
\end{equation*}
with
\begin{equation*}
\Exo(z,w) = \Ev(z) + \sum_{k=1}^N \frac{\Ewk{k}^{\,\text{on}}(w)}{z-z_k}.
\end{equation*}
Let us define
\begin{equation*}
\lambda^i(z,w) = \sum_{i=1}^N \frac{\lambda_k}{z-z_k} - \frac{\alpha_i}{z-w} = \lambda(z) - \frac{\alpha_i}{z-w}.
\end{equation*}
One finds
\begin{equation*}\Ev(z) + \sum_{k=1}^N \frac{\Ewk{k}(w)}{z-z_k} = \frac{1}{2} \bigl( \lambda^i(z,w), \lambda^i(z,w) \bigr) - \frac{\p\;}{\p z} \bigl( \lambda^i(z,w), \rho \bigr) + \frac{\bigl(\alpha_i,\lambda(w)\bigr)}{z-w} - \frac{(\alpha_i,\alpha_i-2\rho)}{2(z-w)^2}.
\end{equation*}
It is a standard result that for a simple root $\alpha_i$, the quantity $(\alpha_i,\alpha_i-2\rho)$ vanishes, so that the last term of the equation above disappears. Moreover, the third term cancels on-shell. Thus, we find
\begin{equation*}
\Exo(z,w) = \frac{1}{2} \bigl( \lambda^i(z,w), \lambda^i(z,w) \bigr) - \frac{\p\;}{\p z} \bigl( \lambda^i(z,w), \rho \bigr).
\end{equation*}

\paragraph{$\bm{\Di{\g}}$-singularity of the on-shell Bethe vector.} Recall the discussion of Subsection \ref{SubSec:DiagBethe} about the diagonal action of $\Di{\g}$. For any $j\in\lbrace 1,\cdots,\ell \rbrace$, using the commutation relation \eqref{Eq:ComDiag}, we get
\begin{equation*}
E_{j\,(\infty)}.\Psi_i(w)
= \bigl[ E_{j\,(\infty)}, F_i(w) \bigr].\vl
= [E_j,F_i](w).\vl 
= \delta_{ij} \, \zeta\bigl(\alpha_i(w)\bigr).\vl
= \delta_{ij} \bigl(\alpha_i,\lambda(w)\bigr) \vl.
\end{equation*}
Given the form \eqref{Eq:Bethe1} of the Bethe equation, one sees that the on-shell Bethe vector is $\Di{\g}$-singular:
\begin{equation*}
E_{j\,(\infty)}.\Po_i(w) = 0, \;\;\;\; \forall \, j \in \lbrace 1,\cdots,\ell \rbrace.
\end{equation*}
The on-shell Bethe vector $\Po_i(w)$ is the highest-weight vector $B_s$ of one of the Verma module $M_s$ in the decomposition \eqref{Eq:DecompoVerma}. All vectors of $M_s$ are then eigenvectors of $\Hs(z)$ with eigenvalue $\Exo(z,w)$. The highest weight of $M_s$ is
\begin{equation*}
\mu_s = \lambda_\infty - \alpha_i = \sum_{k=1}^N \lambda_k - \alpha_i = - \res_{z=\infty} \lambda^i(z,w) \dd z.
\end{equation*}

\paragraph{Eigenvectors at one excitation.} Let us describe all eigenvectors of $\Hs(z)$ with one excitation in the direction $\alpha_i$. For that we need to find a basis of eigenvectors of the weight space $W_{\lambda_\infty-\alpha_i}$. As explained in Subsection \ref{SubSec:DiagBethe}, this weight space is of dimension $N$.

After a multiplication by $\prod_{k=1}^N (w-z_k)$, the Bethe equation \eqref{Eq:Bethe1} becomes a polynomial equation of degree $N-1$ on the Bethe root $w$. Thus, for generic values of the $z_i$'s, there exist $N-1$ distinct solutions to this equation. One then finds $N-1$ linearly independent eigenvectors of $\Hs(z)$ as the corresponding on-shell Bethe vectors.

The last eigenvector in $W_{\lambda_\infty-\alpha_i}$ is not a Bethe vector, as it is not a singular vector under the diagonal action. It belongs to the $\Di{\g}$-module $M_{s_\infty}$ and is obtained from the vacuum state as
\begin{equation*}
F_{i\,(\infty)} . \vl.
\end{equation*}
As a vector in $M_{s_\infty}$, it is an eigenvector of $\Hs(z)$ of eigenvalue $\Ev(z)$.

\subsection{Higher excited Bethe vectors}

Let us end this section by saying a few words about the Bethe ansatz with $M$ excitations, for $M>1$. Let us consider $M$ complex numbers $\bm w = (w_1,\cdots,w_M)$, which will be the Bethe roots of the Bethe vector. We associate with each of these Bethe roots $w_j$ ($j=1,\cdots,M$) a ``color'' $c(j) \in \lbrace 1,\cdots,\ell\rbrace$. The off-shell Bethe vector
\begin{equation*}
\Psi_{\bm c}(\bm w) = \Psi_{c(1),\cdots,c(M)} (w_1,\cdots,w_M)
\end{equation*}
is then a function of the numbers $w_j$ ($j=1,\cdots,M$). More precisely, it is obtained from the vacuum state by actions of products of the lowering operators $F_{c(1)}, \cdots, F_{c(m)}$, distributed on the $N$ tensor factors of $\Hl=V_{\lambda_1}\otimes\cdots\otimes V_{\lambda_N}$, with weightings depending rationally on the Bethe roots $w_j$'s ($j=1,\cdots,M$) and the sites $z_k$'s ($k=1,\cdots,N$). In particular, we have
\begin{equation*}
\Psi_{\bm c}(\bm w) \in W_\mu, \;\;\;\; \text{ with } \;\;\;\; \mu=\sum_{i=1}^N \lambda_i - \sum_{j=1}^M \alpha_{c(j)}.
\end{equation*}
We refer to~\cite{Varchenko:1991,Babujian:1993ts} for the precise construction of $\Psi_{\bm c}(\bm w)$.

One then finds that $\Psi_{\bm c}(\bm w)$ is an eigenvector of the quadratic Hamiltonian $\Hs(z)$ if the Bethe roots satisfy the $M$ Bethe equations:
\begin{equation}\label{Eq:Bethe2}
\sum_{i=1}^N \frac{(\alpha_{c(k)},\lambda_i)}{w_k-z_i} - \sum_{\substack{j=1 \\ j\neq k}}^M \frac{(\alpha_{c(k)},\alpha_{c(j)})}{w_k-w_j} = 0, 
\end{equation}
for $k\in\lbrace 1,\cdots,M \rbrace$. In this case, we get the on-shell Bethe vectors $\Psi^{\text{on}}_{\bm c}(\bm w)$, which then satisfy
\begin{equation*}
\Hs(z).\Psi^{\text{on}}_{\bm c}(\bm w) = \mathcal{E}^{\,\text{on}}_{\bm c}(z,\bm{w}) \Psi^{\text{on}}_{\bm c}(\bm w).
\end{equation*}
To express the eigenvalue $\mathcal{E}^{\,\text{on}}_{\bm c}(z,\bm{w})$, let us introduce the $z$-dependent weight
\begin{equation}\label{Eq:LambdaExc}
\lambda^{\bm c}(z,\bm{w}) = \sum_{i=1}^N \frac{\lambda_i}{z-z_i} - \sum_{j=1}^M \frac{\alpha_{c(j)}}{z-w_j}.
\end{equation}
We then have
\begin{equation}\label{Eq:EigenBethe}
\mathcal{E}^{\,\text{on}}_{\bm c}(z,\bm{w}) = \frac{1}{2} \bigl( \lambda^{\bm c}(z,\bm{w}), \lambda^{\bm c}(z,\bm{w}) \bigr) - \frac{\p\;}{\p z} \bigl( \lambda^{\bm c}(z,\bm{w}), \rho \bigr).
\end{equation}
Note that the weight $\mu\in\h^*$ such that $\Psi_{\bm c}(\bm w)$ belongs to $W_\mu$ can be expressed as
\begin{equation*}
\mu = - \res_{\lambda=\infty} \lambda^{\bm c}(z,\bm{w}) \dd z.
\end{equation*}
Note also that if we write
\begin{equation*}
\lambda^{\bm c}(z,\bm{w}) = \overline{\lambda}_k^{\,\bm c}(z,\bm{w}) - \frac{\alpha_{c(k)}}{z-w_k},
\end{equation*}
then $\overline{\lambda}_k^{\,\bm c}(z,\bm{w})$ is regular at $z=z_k$ and the Bethe equation \eqref{Eq:Bethe2} reads
\begin{equation*}
\left( \alpha_{c(k)}, \overline{\lambda}_k^{\,\bm c}(z_k,\bm{w}) \right) = 0
\end{equation*}
Finally, let us mention that because of the Bethe equations, the on-shell Bethe vector $\Psi^{\text{on}}_{\bm c}(\bm w)$ is $\Di{\g}$-singular.

\section{The Feigin-Frenkel-Reshetikhin approach}
\label{Sec:FFR}

This section is devoted to the Feigin-Frenkel-Reshetikhin (FFR) approach of Gaudin models. It consists of a description of the spectrum of Gaudin models in terms of differential operators called opers. We will give an introductory review about this approach and refer to~\cite{Feigin:1994in,Frenkel:1995zp,Frenkel:2003qx,Mukhin:2005,Frenkel:2004qy,Chervov:2004ty,Chervov:2009ck,Rybnikov:2016} for more details. We will end the section with some new results concerning the generalisation of this approach for cyclotomic Gaudin models. Before explaining the general theory, let us gain some intuition about it by considering the simplest Gaudin model, on the Lie algebra $\g=\sld$.

\subsection[The $\sld$ case]{The $\sldb$ case}
\label{SubSec:FFRsl2}

\paragraph{Bethe ansatz for $\sldb$.} In this subsection, we consider a Gaudin model on the Lie algebra $\g=\sld$, with sites $\zb=(z_1,\cdots,z_N) \in \C^N$. The Cartan subalgebra $\h$ of $\sld$ is one dimensional, hence $\sld$ is of rank $\ell=1$. In particular, $\sld$ admits only one exponent, equal to 1. Thus, the Gaudin subalgebra $\Zc_{\zb}(\g)$ is generated by the coefficients of the quadratic Hamiltonian $\Hs(z)$.

As $\h^*$ is one dimensional, any weight $\lambda\in\h^*$ can be written as $\lambda= S \alpha$, with $S\in\C$ and $\alpha\in\h^*$ the unique positive root of $\sld$. In particular, the Weyl vector is equal to $\rho=\frac{1}{2}\alpha$. The bilinear form $\fd$ induced on $\h^*$ by the Killing form is given by $(\alpha,\alpha)=4$.

Let us consider $N$ weights $\lambda_i= S_i \alpha$, with $S_i\in\C$ and $i\in\lbrace 1,\cdots,N \rbrace$, and the associated Hilbert space $\Hl=V_{\lambda_1}\otimes\cdots\otimes V_{\lambda_N}$, as in the previous sections. As $\Hl$ is a tensor product of highest-weight vectors, we can study the spectrum of $\Hs(z)$ on $\Hl$ by the Bethe ansatz, as described in Section \ref{Sec:Bethe}. Let us then consider a Bethe vector with $M$ excitations and Bethe roots $\bm w = (w_1,\cdots,w_M)$. As $\sld$ is of rank 1, there is only one possible color, \textit{i.e.} each Bethe root is associated with an excitation in the direction $\alpha$. The Bethe equations are given by equation \eqref{Eq:Bethe2} and read in this particualar case
\begin{equation}\label{Eq:BetheSl2}
B_k = \sum_{i=1}^N \frac{S_i}{w_k-z_i} - \sum_{\substack{j=1\\j\neq k}}^M \frac{1}{w_k-w_j} = 0, \;\;\;\; \forall \, k\in\lbrace 1,\cdots,M \rbrace.
\end{equation}
The weight \eqref{Eq:LambdaExc} associated with the Bethe vector is then
\begin{equation}\label{Eq:LzSl2}
\lambda^{\bm c}(z,\bm w) = S(z,\bm w) \alpha \;\;\;\ \text{ with } \;\;\;\; S(z,\bm w) = \sum_{i=1}^N \frac{S_i}{z-z_i} - \sum_{j=1}^M \frac{1}{z-w_j}.
\end{equation} 
The eigenvalue of the on-shell Bethe vector with Bethe roots $\bm w$ is then given by \eqref{Eq:EigenBethe}
\begin{equation}\label{Eq:EigenSl2}
\mathcal{\E}^{\text{on}}_{\bm c} (z,\bm w) = 2 \bigl( S(z,\bm w)^2 - S'(z,\bm w) \bigr),
\end{equation}
where $S'(z,\bm w) = \dfrac{\p S(z,\bm w)}{\p z}$.

\paragraph{$\sldb$-opers.} The goal of this subsection is to show a reformulation of the Bethe ansatz described above in terms of objects called $\sld$-opers. Let us start by defining these objects. We will be interested in the space $\text{Conn}_{\sld}\bigl(\mathbb{P}^1\bigr)$ of meromorphic $\sld$-connections on the Riemann sphere $\CP=\C\cup\lbrace\infty\rbrace$. These are differential operators of the form
\begin{equation*}
a\p_z + A(z),
\end{equation*}
where $a$ is a complex number and $A$ is a $\sld$-valued meromorphic function of $z$. We will use the explicit representation of elements of $\sld$ as two by two traceless matrices. If $g(z)$ is a meromorphic function valued in the associated Lie group $SL(2,\C)$ (composed of two by two matrices of determinant 1), we can consider the formal gauge transformation of the connection $a\p_z + A(z)$ by $g$:
\begin{equation*}
g(z)\bigl( a\p_z + A(z)\bigr)g(z)^{-1} = a\p_z + g(z)A(z)g(z)^{-1} - a\bigl( \p_z g(z) \bigr) g(z)^{-1}.
\end{equation*}

Let us now consider a particular type of $\sld$-connections:
\begin{equation}\label{Eq:opSl2}
\op{\sld} = \left\lbrace \nabla = \p_z + \begin{pmatrix}
a(z) & b(z) \\
 1   & -a(z)
\end{pmatrix} \in \text{Conn}_{\sld}\bigl(\mathbb{P}^1\bigr) \right\rbrace.
\end{equation}
We also define a group
\begin{equation}\label{Eq:NSl2}
N = \left\lbrace
g(z) = \begin{pmatrix}
1 & f(z) \\
0 & 1
\end{pmatrix},
\; f \; \text{meromorphic}
\right\rbrace
\end{equation}
of meromorphic functions valued in the group $SL(2,\C)$. It is a straightforward computation that for $\nabla$ and $g$ of the form above, the gauge transformation of $\nabla$ by $g$ is given by
\begin{equation}\label{Eq:GaugeOper}
g \nabla g^{-1} = \p_z + \begin{pmatrix}
a(z) + f(z) & b(z)-f(z)^2-2f(z)a(z)-f'(z) \\
1 & - a(z) - f(z),
\end{pmatrix}.
\end{equation}
In particular, the gauge transformation by elements of $N$ stabilises the space $\op{\sld}$. This defines a group action of $N$ on $\op{\sld}$. We can thus define the quotient
\begin{equation}\label{Eq:OpSl2}
\Op{\sld} = \op{\sld} / N.
\end{equation}
An element of this quotient is called a \textbf{$\sldb$-oper}. If $\nabla$ is an element of $\op{\sld}$, we denote by $[\nabla]\in\Op{\sld}$ the corresponding oper, \textit{i.e.} the equivalence class of $\nabla$ under the gauge transformations in $N$. The connection $\nabla$ is then a representative of the oper $[\nabla]$.

It is clear from equation \eqref{Eq:GaugeOper} that there is a unique element in the equivalence class $[\nabla]$ with no diagonal coefficients: it is obtained by taking $f=-a$. Such an element is called a \textbf{canonical representative} of the oper $[\nabla]$. Through this representative, the space $\Op{\sld}$ is thus parametrised by a unique meromorphic function $c$ of $z$:
\begin{equation}\label{Eq:CanSl2}
\Op{\sld} \simeq \left\lbrace \p_z + \begin{pmatrix}
0 & c(z) \\
 1   & 0
\end{pmatrix} \in \text{Conn}_{\sld}\bigl(\mathbb{P}^1\bigr) \right\rbrace.
\end{equation}

\paragraph{Reformulation of the Bethe ansatz in terms of $\sldb$-opers.} Let us come back to the $\sld$-Gaudin model and the Bethe ansatz described above. We introduce the following $\sld$-connection:
\begin{equation}\label{Eq:MiuraSl2}
\nabla_{\bm w} = \p_z + \begin{pmatrix}
-S(z,\bm w) & 0 \\
1 & S(z,\bm w)
\end{pmatrix},
\end{equation}
with $S(z,\bm w)$ as in \eqref{Eq:LzSl2}. We will call a connection of this form (with any function of $z$ instead of $S$) a \textbf{Miura $\sldb$-oper}. In particular, it belongs to the space $\op{\sld}$ and we can thus consider the associated oper $[\nabla_{\bm w}]$ in $\Op{\sld}$. The canonical representative of $[\nabla_{\bm w}]$ is given explicitly by
\begin{equation*}
[\nabla_{\bm w}]_{\text{can}} = \p_z + \begin{pmatrix}
0 & S(z,\bm w)^2-S'(z, \bm w) \\
1 & 0
\end{pmatrix}.
\end{equation*}
We recognize in the coefficient of this canonical representative the eigenvalue \eqref{Eq:EigenSl2} of $\Hs(z)$ on the Bethe vector with Bethe roots $\bm w$ (up to a global factor).

\paragraph{Bethe equations.} Let us consider the coefficient
\begin{equation*}
\mathcal{C}(z,\bm w) = S(z,\bm w)^2-S'(z, \bm w),
\end{equation*}
as a function of $z$ and $\bm w$, without requiring that the Bethe roots $\bm w$ are on-shell. In particular, as $S(z,\bm w)$ has simple poles at $z=w_j$ ($j\in\lbrace 1,\cdots,N \rbrace$), $\mathcal{C}(z,\bm w)$ might also have poles at $z=w_j$. One finds that due to a cancellation between the term $S(z,\bm w)^2$ and the term $S'(z,\bm w)$, the double pole of $\mathcal{C}(z,\bm w)$ at $z=w_j$ vanishes. Moreover, an explicit computation shows that
\begin{equation*}
\res_{z=w_j} \mathcal{C}(z,\bm w) \dd z = -2B_j,
\end{equation*}
with $B_j$ defined in equation \eqref{Eq:BetheSl2}. Thus, if the Bethe root $w_j$ is on-shell, this residue vanishes and the coefficient $\mathcal{C}(z,\bm w)$ is then regular at $z=w_j$.

\paragraph{Summary.} As a conclusion of this subsection, let us summarise what we observed. We considered a Bethe vector $\Psi(\bm w)$ for the $\sld$-Gaudin model and observed the following.
\begin{enumerate}[1.]\setlength\itemsep{0.1em}
\item From the weight \eqref{Eq:LambdaExc} associated with $\Psi(\bm w)$, we constructed a Miura $\sld$-oper $\nabla_{\bm w}$.
\item We constructed the canonical representative $[\nabla_{\bm w}]_{\text{can}}$ of the corresponding oper $[\nabla_{\bm w}]$.
\item Although the Miura oper $\nabla_{\bm w}$ has simple poles at $z$ equal to a Bethe root, the canonical representative $[\nabla_{\bm w}]_{\text{can}}$ is regular at this Bethe root if and only if this root satisfies the Bethe equation.
\item When all Bethe equations are satisfied, the coefficient of the canonical representative $[\nabla_{\bm w}]_{\text{can}}$ coincides (up to a factor) with the eigenvalue of $\Hs(z)$ on the Bethe vector.
\end{enumerate}
These results are in fact part of the FFR approach for the $\sld$-Gaudin model, as we shall see in the rest of this section.

\subsection{The FFR approach and the spectrum of the Gaudin model}
\label{SubSec:FFR}

Let us now state and motivate the main ideas of the FFR approach for a general $\g$-Gaudin model, before going into more details.

\paragraph{The main theorem.} The main result of the FFR approach can be stated as follows~\cite{Frenkel:2004qy}:

\begin{theorem}\label{Thm:FFR}
There exists an algebraic variety \emph{$\Opz$} of so-called \emph{$\null^L \g$}-opers such that the Gaudin subalgebra \emph{$\Zg$} is isomorphic to the algebra of polynomial functions on \emph{$\Opz$}. We will write
\emph{\begin{equation*}
\Phi_{\zb,\g} : \Zg \longrightarrow \Fun{\Opz}
\end{equation*}}~\vspace{-12pt}\\
this isomorphism.
\end{theorem}

The definition of the space of opers $\Opz$ and of the map $\Phi_{\zb,\g}$ will be explained later in this section, as it requires the introduction of a new mathematical formalism. For now, let us try to understand the consequence of this theorem for the study of quantum Gaudin models.

If $\Zc$ is a commutative algebra over $\C$, we define the space $\Xi(\Zc)$ of algebra homomorphisms from $\Zc$ to $\C$ and we call the elements of $\Xi(\Zc)$ the \textbf{characters} of $\Zc$. Let $V$ be an algebraic variety and $\Fun{V}$ the algebra of polynomial functions on $V$. For $x$ a point of $V$, we define the evaluation at $x$ as
\begin{equation*}
\begin{array}{rccc}
\ev_x : & \Fun{V} & \longrightarrow & \C \\
        &    f    & \longmapsto     & f(x)
\end{array}.
\end{equation*}
It is clear that $\ev_x$ is an algebra homomorphism from $\Fun{V}$ to $\C$ and thus is a character of $\Fun{V}$. It is a fundamental result of algebraic geometry~\cite{Hartshorne:1977} that
\begin{equation*}
\begin{array}{rccc}
\ev : & V & \longrightarrow & \Xi\bigl( \Fun{V} \bigr) \\
      & x & \longmapsto     & \ev_x
\end{array}
\end{equation*}
is a bijection. We then get the following consequence of Theorem \ref{Thm:FFR}:

\begin{corollary}\label{Cor:Charac}
The map
\emph{\begin{equation*}
\begin{array}{rccc}
\eta_{\zb,\g} : & \Opz   & \longrightarrow & \Xi\bigl( \Zg \bigr) \\
                & [\nabla] &   \longmapsto   & \ev_{[\nabla]} \circ \Phi_{\zb,\g}
\end{array}
\end{equation*}}~\vspace{-12pt}\\
is a bijection.
\end{corollary}
\begin{proof}
Theorem \ref{Thm:FFR} states that the commutative $\C$-algebras $\Zg$ and $\Fun{\Opz}$ are isomorphic (through $\Phi_{\zb,\g}$). Thus the spaces of characters of these algebras should be isomorphic. Concretely, this isomorphism is given by the precomposition by $\Phi_{\zb,\g}$:
\begin{equation*}
\begin{array}{rccc}
\etat_{\zb,\g}: & \Xi \left( \Fun{\Opz} \right) & \longrightarrow & \Xi\bigl( \Zg \bigr) \\
               &     \chi             & \longmapsto     & \chi \circ \Phi_{\zb,\g}
\end{array}.
\end{equation*}
We then get the bijection $\eta_{\zb,\g}$ from $\Opz$ to $\Xi \bigl( \Zg \bigr)$ by composing $\etat_{\zb,\g}$ on the right with the bijection $\ev : \Opz \rightarrow \Xi \left( \Fun{\Opz} \right)$.
\end{proof}

The main Theorem \ref{Thm:FFR} of the FFR approach then allows to realise all characters of the Gaudin subalgebra $\Zg$ as an oper in $\Opz$. Let us now explain why such characters are useful for the study of quantum integrable models.

\paragraph{Characters and quantum integrable models.} Let us consider a quantum model, with an algebra of operators $\Ac$ and Hamiltonian $\Hc$. We suppose that this model is integrable, in the sense that it possesses a large number of conserved commuting charges. Let us then consider the subalgebra $\Zc$ of $\Ac$ generated by these conserved charges (in particular, it contains the Hamiltonian $\Hc$ of the system). As the charges are commuting, the algebra $\Zc$ is a commutative subalgebra of $\Ac$ (in the case of the Gaudin model, this is the Gaudin subalgebra $\Zg$).

Let us now consider the Hilbert space $H$ of the model, on which $\Ac$ acts linearly. A first step towards the resolution of the model is to describe the spectrum of the Hamiltonian $\Hc$ on $H$, or even better, the joint spectrum of the commuting charges in $\Zc$. Indeed, as these charges commute, they can be simultaneously diagonalised. Let then $v\in H$ be a common eigenvector of the operators in $\Zc$. For $\Q$ in $\Zc$, we denote by $\chi_v(\Q)\in\C$ the eigenvalue of $\Q$ on $v$, so that we have
\begin{equation*}
\Q.v = \chi_v(\Q) v.
\end{equation*}
If $\Q$ and $\Q'$ are two operators in $\Zc$, we have
\begin{equation*}
\chi_v(\Q\Q') v = (\Q\Q').v = \Q. \bigl( \Q'.v \bigr) = \chi_v(\Q') \, \Q.v = \chi_v(\Q') \chi_v(\Q) v.
\end{equation*}
Thus, we have
\begin{equation*}
\chi_v(\Q\Q') = \chi_v(\Q) \chi_v(\Q'), \;\;\;\; \forall \, \Q,\Q' \in \Zc,
\end{equation*}
\textit{i.e.} \textbf{$\bm{\chi_v}$ is a character of $\bm{\Zc}$}.

Therefore, if one knows all characters of $\Zc$, one knows all possible eigenvalues of the charges $\Q$, \textit{i.e.} the spectrum of the considered integrable model. The description of $\Xi(\Zc)$ is thus a first step towards the resolution of the model.

\paragraph{Spectrum of the Gaudin model.} Let us come back to the particular case of the Gaudin model. The characters $\Xi\bigl( \Zg \bigr)$ of the Gaudin subalgebra are realised in terms of opers in $\Opz$. Thus, one can describe the spectrum of the Gaudin model (\textit{i.e.} the eigenvalues of the operators in $\Zg$) in terms of opers. More precisely, we get

\begin{theorem}\label{Thm:FFRSpectrum}
Let $H$ be a Hilbert space of the Gaudin model, \textit{i.e.} a linear representation of $\Ac$. Let $v\in H$ be a common eigenvector of the Gaudin subalgebra $\Zg$. Then, there exists an oper \emph{$[\nabla_v] \in \Opz$} such that for any $\Q \in \Zg$, the eigenvalue of $\Q$ on $v$ is given by
\begin{equation}\label{Eq:SpectrumFFR}
\bigl( \Phi_{\zb,\g} (\Q) \bigr) \bigl([\nabla_v]\bigr).
\end{equation}
\end{theorem}

\noi \textbf{Remark:} Let us explain exactly what we mean by equation \eqref{Eq:SpectrumFFR}. Recall from Theorem \ref{Thm:FFR} that $\Phi_{\zb,\g}$ is a bijection from $\Zg$ to the algebra of functions on $\Opz$. Thus, $\Phi_{\zb,\g} (\Q)$ is such a function and we get a complex number $\bigl( \Phi_{\zb,\g} (\Q) \bigr) \bigl([\nabla_v]\bigr)$ when evaluating it on the oper $[\nabla_v] \in \Opz$.

\begin{proof}
We use the notations of the previous paragraph and denote by $\chi_v(\Q)$ the eigenvalue of $\Q$ on $v$: $\chi_v$ is then a character in $\Xi \bigl( \Zg \bigr)$. Let us define
\begin{equation}\label{Eq:NablaV}
[\nabla_v] = \eta^{-1}_{\zb,\g} (\chi_v) \in \Opz,
\end{equation}
where $\eta_{\zb,\g} : \Opz \rightarrow \Xi \bigl( \Zg \bigr)$ is the bijection of Corollary \ref{Cor:Charac}. Then, by definition of this bijection, we have
\begin{equation*}
\chi_v(\Q) = \bigl( \eta_{\zb,\g} \bigl([\nabla_v]\bigr) \bigr) (\Q)
           = \ev_{[\nabla_v]} \bigl( \Phi_{\zb,\g} (\Q) \bigr)
           = \bigl( \Phi_{\zb,\g} (\Q) \bigr) \bigl([\nabla_v]\bigr).
           \qedhere
\end{equation*}
\end{proof}

Theorem \ref{Thm:FFRSpectrum} gives a theoretical description of the spectrum of the Gaudin model. In particular, note that in this theorem, we did not put any restriction on the Hilbert space $H$ of the model. Thus, the FFR approach describes the spectrum of the Gaudin model even on Hilbert spaces which are not a tensor product of highest-weight representations, \textit{i.e.} on Hilbert spaces where the Bethe ansatz does not apply. Moreover, as we mentioned in the introduction of this chapter, it can happen in some degenerate cases that the Bethe ansatz is not complete~\cite{Mukhin:2007}, \textit{i.e.} that it does not result in a basis of eigenvectors. In this case, there is at least one eigenvector $v\in H$ whose eigenvalues are not described by the Bethe ansatz. However, by Theorem \ref{Thm:FFRSpectrum}, one knows that these eigenvalues are described by some oper $[\nabla_v]$. These are some of the advantages of the FFR approach.

However, this approach also has downsides. The main one is the fact that it is mainly an existence result, and not a constructive one. Indeed, the Theorem \ref{Thm:FFRSpectrum} only proves the existence of the oper $[\nabla_v]$ associated with the eigenvector $v$, it does not explain how to construct it. In the proof of the theorem, we defined $[\nabla_v]$ through equation \eqref{Eq:NablaV}. However, this is not a constructive result: indeed, this equation supposed that we already knew the eigenvector $v$ and the corresponding eigenvalues $\chi_v$. Similarly, given an oper $[\nabla]$, one can construct an associated character $\chi$ but the theorem \ref{Thm:FFRSpectrum} does not explain how to find an eigenvector $v$ whose eigenvalues would be described by $\chi$.

These downsides can be overcome when one considers a Hilbert space which is a tensor product of highest-weight representations of $\g$, \textit{i.e.} when the Bethe ansatz applies. In this case, one can re-interpret the Bethe ansatz in terms of opers and find the oper associated with each Bethe vector. The FFR approach then contains the Bethe ansatz.\\

The rest of this Section is mainly devoted to the description of the space of opers $\Opz$ and of the isomorphism $\Phi_{\zb,\g}$ of Theorem \ref{Thm:FFR}, together with the reinterpretation of the Bethe ansatz in terms of these opers. Before that, let us come back to the example of the $\sld$-Gaudin model studied in Subsection \ref{SubSec:FFRsl2} and illustrate the notions introduced above in this example.

\paragraph{Back to $\sldb$.}\label{Para:BackSl2} We will use the notations of Subsection \ref{SubSec:FFRsl2}. In particular, we introduced $\sld$-opers $\Op{\sld}$ as gauge equivalence class of connections of the form
\begin{equation*}
\nabla = \p_z + \begin{pmatrix}
a(z) & b(z) \\
 1   & -a(z)
\end{pmatrix}.
\end{equation*}
We will say that such a connection has regular singularities at $x\in\C$ if $a$ possesses at most simple poles at $z=x$ and $b$ at most double poles. We will also say that it is regular at $x$ if $a$ and $b$ are regular at $z=x$.\\

For a $\sld$-Gaudin model with sites $\zb=(z_1,\cdots,z_n)$, we define the space of opers $\Opsld$ appearing in the FFR approach as opers in $\Op{\sld}$ which possess a representative that has regular singularities at the $z_i$'s, is regular elsewhere and has no constant term. An oper $[\nabla]\in\Op{\sld}$ is in $\Opsld$ if and only if its canonical representative is of the form
\begin{equation}\label{Eq:CanRSSL2}
[\nabla]_\text{can} = \p_z + \begin{pmatrix}
0 & c(z) \\
 1   & 0
\end{pmatrix} \;\;\;\; \text{ with } \;\;\;\; c(z) = \sum_{i=1}^N \left( \frac{c_{i,0}}{2(z-z_i)} + \frac{c_{i,1}}{4(z-z_i)^2} \right).
\end{equation}
We define $2N$ functions $\Gamma_{i,p}$ ($i=1,\cdots,N$ and $p=0,1$) on $\Opsld$ by
\begin{equation*}
\Gamma_{i,p} : [\nabla] \in \Opsld \rightarrow c_{i,p}.
\end{equation*}
As any oper in $\Opsld$ has a canonical representative of the form \eqref{Eq:CanRSSL2}, it is clear that these functions generate all functions on $\Opsld$.

For a Gaudin model on $\sld$, there is only the quadratic Hamiltonians $\Hs(z)$ and no higher-order ones (see Subsection \ref{SubSec:FFRsl2}). The Gaudin algebra $\Zc_{\zb}\bigl( \sld \bigr)$ of the model is then generated by the Casimirs $\Delta_{(i)}$'s and the Hamiltonians $\Hs_i$'s appearing in the partial fraction decomposition \eqref{Eq:DecompoH} of $\Hs(z)$. In this case, the algebra isomorphism $\Phi_{\zb,\sld}$ of Theorem \ref{Thm:FFR} is such that
\begin{equation}\label{Eq:PhiSl2}
\Phi_{\zb,\sld} \bigl( \Hs_i \bigr) = \Gamma_{i,0} \;\;\;\; \text{ and } \;\;\;\; \Phi_{\zb,\sld} \bigl( \Delta_{(i)} \bigr) = \Gamma_{i,1}
\end{equation}
and extends to the whole algebra $\Zc_{\zb}\bigl( \sld \bigr)$ by linearity and multiplication.\\

Let us now come back to the Bethe ansatz for this model. We consider $\Psi^{\text{on}}(\bm w)$ an on-shell Bethe vector, which is then an eigenvector of the Gaudin algebra $\Zc_{\zb}(\sld)$. Theorem \ref{Thm:FFRSpectrum} implies the theoretical existence of an oper $[\nabla_{\bm w}]$ in $\Opsld$ encoding the eigenvalues of $\Zc_{\zb}(\sld)$ on $\Psi^{\text{on}}(\bm w)$. Let us show that we can construct this oper explicitly.

We will use the notations of Subsection \ref{SubSec:FFRsl2}. Let us first consider the off-shell Bethe vector $\Psi(\bm w)$. We define an associated Miura oper $\nabla_{\bm w}$ as in \eqref{Eq:MiuraSl2}. As explained in Subsection \ref{SubSec:FFRsl2}, the canonical representative of the oper $[\nabla_{\bm w}]$ is
\begin{equation*}
[\nabla_{\bm w}]_{\text{can}} = \p_z + \begin{pmatrix}
0 & \Cc(z,\bm w) \\
1 & 0
\end{pmatrix}, \;\;\;\; \text{ with } \;\;\;\; \Cc(z,\bm w) = S(z,\bm w)^2-S'(z, \bm w).
\end{equation*}
In particular, $\Cc(z,\bm w)$ has simple and double poles at all $z=z_i$'s. Moreover, as remarked in Subsection \ref{SubSec:FFRsl2}, it is regular at $z=w_j$ if and only if the Bethe root $w_j$ satisfies the Bethe equation \eqref{Eq:BetheSl2}. Thus, we have
\begin{equation*}
[\nabla_{\bm w}] \in \Opsld \Longleftrightarrow \bm w \text{ satisfy the Bethe equations} \Longleftrightarrow \Psi(\bm w) \text{ is on-shell}.
\end{equation*}
If this is the case, as explained in Subsection \ref{SubSec:FFRsl2}, the on-shell Bethe vector $\Psi^{\text{on}}(w)$ is an eigenvector of $\Hs(z)$ with eigenvalue $2\Cc(z,\bm w)$. In particular, given the expression \eqref{Eq:DecompoH} of $\Hs(z)$, the eigenvalues of $\Hs_i$ and $\Delta_{(i)}$ are
\begin{equation*}
\chi_{\bm w}(\Hs_i) = 2\res_{z=z_i} \Cc(z,\bm w) \;\;\;\; \text{ and } \;\;\;\; \chi_{\bm w}(\Delta_{(i)}) = 4\res_{z=z_i} (z-z_i)\Cc(z,\bm w).
\end{equation*}
Recall the functions $\Gamma_{i,p}$ on $\Opsld$ defined above. As $\Cc(z,\bm w)$ is the coefficient of the canonical representative $[\nabla_{\bm w}]_{\text{can}}$, the equation above can be rewritten
\begin{equation*}
\chi_{\bm w}(\Hs_i) = \Gamma_{i,0} \bigl( [\nabla_{\bm w}] \bigr) \;\;\;\; \text{ and } \;\;\;\; \chi_{\bm w}(\Delta_{(i)}) = \Gamma_{i,1} \bigl( [\nabla_{\bm w}] \bigr).
\end{equation*}
Given the definition \eqref{Eq:PhiSl2} of the FFR isomorphism $\Phi_{\zb,\sld}$, we find that for all $\Q\in\Zc_{\zb}\bigl( \sld \bigr)$, the eigenvalue of $\Q$ on $\Psi^{\text{on}}(\bm w)$ is
\begin{equation*}
\chi_{\bm w} (\Q) = \bigl( \Phi_{\zb,\sld}(\Q) \bigr) \bigl( [\nabla_{\bm w}] \bigr).
\end{equation*}
This illustrates Theorem \ref{Thm:FFRSpectrum} about the spectrum of the $\sld$-Gaudin model. In particular, it shows that for the eigenvector $\Psi^{\text{on}}(\bm w)$ constructed from the Bethe ansatz, one can give an explicit construction of the associated oper $[\nabla_{\bm w}]$ (through a Miura oper).

\subsection[Principal $\sld$ subalgebra and exponents]{Principal $\sldb$ subalgebra and exponents}
\label{SubSec:Princ}

In this subsection, we introduce the notions of principal gradation, of principal $\sld$ subalgebra and of exponents of a semi-simple Lie algebra $\g$, which are necessary for the construction of opers. Most of these notions were introduced by Kostant in~\cite{Kostant:1959,Kostant:1963,Kostant:1978}. We refer to these articles for details and proofs and mention just the results we will need in this section.

\paragraph{Principal gradation.} Let us consider the root system $\Delta$ of $\g$. A root $\alpha$ can be written in the basis of simple roots $\lbrace \alpha_i \rbrace_{i=1,\cdots,\ell} \subset \h^*$:
\begin{equation*}
\alpha = \sum_{i=1}^\ell m_i \alpha_i.
\end{equation*}
The numbers $m_i$'s are either all non-negative integers or all non-positive integers (we then speak of a positive or negative root). We define the height of $\alpha$ as the integer
\begin{equation*}
\htg{\alpha} = \sum_{i=1}^\ell m_i.
\end{equation*}
Recall the Cartan-Weyl decomposition of $\g$:
\begin{equation*}
\g = \h \oplus \left( \bigoplus_{\alpha\in\Delta} \C E_\alpha \right).
\end{equation*}
We define the \textbf{principal gradation} of $\g$ as
\begin{equation*}
\g_0 = \h \;\;\;\;\;\;\; \text{ and } \;\;\;\;\;\;\; \g_d = \bigoplus_{ \substack{ \alpha\in\Delta \\ \htg{\alpha} = d } } \C E_\alpha, \;\; \text{for } d\in\Z \setminus \lbrace 0 \rbrace.
\end{equation*}
If $X\in\g_d$, we say that $X$ is of degree $d$. This defines a $\Z$-gradation of $\g$, \textit{i.e.}
\begin{equation}\label{Eq:PrincGrad}
\g = \bigoplus_{d \in \Z} \g_d, \;\;\;\; \text{ and } \;\;\;\; \bigl[ \g_p, \g_q \bigr] \subset \g_{p+q}.
\end{equation}
Note that the positive nilpotent subalgebra $\n_+$ and Borel subalgebra $\bo_+$ of $\g$ are given by
\begin{equation}\label{Eq:PrincNB}
\n_+ = \bigoplus_{\alpha\in\Delta_+} \C E_\alpha = \bigoplus_{d\in\Z_{\geq 1}} \g_d \;\;\;\; \text{ and } \;\;\;\; \bo_+ = \h \oplus \n_+ = \bigoplus_{d\in\Z_{\geq 0}} \g_d.
\end{equation}
The roots of $\g$ have a maximal (resp. minimal) height $h-1$ (resp. $-h+1$), where $h$ is the so-called Coxeter number of $\g$. Thus, we have
\begin{equation*}
\g_{\pm d} = 0 \;\;\;\; \text{ for } \;\;\;\; d \geq h.
\end{equation*}

\paragraph{Principal $\sldb$ subalgebra.} Let us consider the Cartan subalgebra $\h$ of $\g$ and its basis of coweights $\lbrace \och_i \rbrace_{i=1,\cdots,\ell}$ (see appendix \ref{App:BasesCartan}). This is the dual basis of the simple roots $\lbrace \alpha_i \rbrace_{i=1,\cdots,\ell} \subset \h^*$:
\begin{equation*}
\langle \och_i, \alpha_j \rangle = \delta_{ij}.
\end{equation*}
Let us then introduce the Weyl co-vector
\begin{equation*}
\rch = \sum_{i=1}^\ell \och_i.
\end{equation*}
By definition, it satisfies
\begin{equation*}
\langle \rch, \alpha \rangle = \htg{\alpha}, \;\;\;\; \forall \, \alpha\in\Delta.
\end{equation*}
One can interpret the principal gradation \eqref{Eq:PrincGrad} as a decomposition of $\g$ in eigenspaces of $\ad_{\rch}$:
\begin{equation*}
\g_d = \bigl\lbrace X\in\g \; \bigl| \; [\rch,X] = d \, X \bigr\rbrace, \;\;\;\;\; \forall \, d\in\Z.
\end{equation*}
Recall the isomorphism $\zeta:\h^*\rightarrow\h$ induced by the Killing form. We define numbers $b_i$'s ($i=1,\cdots,\ell$) by the decomposition of the Weyl co-weight in the basis $\lbrace \zeta(\alpha_i) \rbrace_{i=1,\cdots,\ell}$ of $\h$:
\begin{equation*}
\rch = \sum_{i=1}^\ell b_i \zeta(\alpha_i).
\end{equation*}
Let $E_i=E_{\alpha_i}$'s and $F_i=E_{-\alpha_i}$'s ($i\in\lbrace 1,\cdots,\ell\rbrace$) be the Chevalley generators of $\g$ (see Appendix \ref{App:SemiSimple}). We introduce the so-called negative and positive principal nilpotent elements as:
\begin{equation*}
\pn = \sum_{i=1}^\ell F_i \in \g_{-1} \;\;\;\;\;\;\; \text{ and } \;\;\;\;\;\;\; p_1 = \sum_{i=1}^\ell b_iE_i \in \g_1.
\end{equation*}
The three elements $\lbrace \rch,p_1,p_{-1} \rbrace$ of $\g$ satisfy the commutation relations:
\begin{equation*}
\bigl[ \rch, p_{\pm 1} \bigr] = \pm \, p_{\pm 1} \;\;\;\;\;\;\; \text{ and } \;\;\;\;\;\;\; \bigl[ p_1, \pn \bigr] = \rch.
\end{equation*}
The subalgebra $\Span(\rch,p_1,\pn)$ of $\g$ is thus isomorphic to $\sld$. We call it the \textbf{principal $\sldb$ subalgebra} of $\g$.

\paragraph{Exponents of $\bm{\g}$.} Let us consider the centraliser of $p_1$:
\begin{equation*}
\a = \Ker(\ad_{p_1}) = \bigl\lbrace X \in \g \; \bigl| \; [p_1,X]=0 \bigr\rbrace.
\end{equation*}
As shown in~\cite{Kostant:1959}, one has $\dim\,\a = \ell$, where $\ell$ is the rank of $\g$. Moreover, for any $X\in\a$, using the Jacoby identity, we have
\begin{equation*}
\bigl[ p_1, [\rch,X] \bigr] = \bigl[ [p_1,\rch], X \bigr] + \bigl[ \rch, \Ccancel[red]{[p_1,X]} \bigr] = - [p_1,X] = 0.
\end{equation*}
Thus, $\ad_{\rch}$ stabilises the centraliser $\a$. In particular, this implies that one can find a basis $\lbrace q_1, \cdots, q_\ell \rbrace$ of $\a$ which is composed of eigenvectors of $\ad_{\rch}$. Note that we can choose the first element of this basis to be $q_1=p_1$ itself, as it belongs to $\a$. Thus, there exist numbers $d_i \in \Z$ such that
\begin{equation*}
q_i \in \a \cap \g_{d_i}.
\end{equation*} 
The first $d_1$ is the degree of $p_1$, hence $d_1=1$. One shows that the $d_i$'s are positive integers and that the largest $d_i$ is equal to the highest degree $h-1$. These integers $d_i$'s are the \textbf{exponents} $\bm E$ of $\g$. We can order the basis such that
\begin{equation*}
1 = d_1 \leq d_2 \leq \cdots \leq d_\ell = h-1.
\end{equation*}
We will now label the elements $q_i$'s ($i=1,\cdots,\ell$) by their corresponding exponent $d_i\in E$ and hence obtain a basis\footnote{As in Subsection \ref{SubSec:HamQuant}, we define $E$ as the multiset of exponents. Therefore, if two exponents $d_i$'s coincide, they appear both in $E$ and one then has two corresponding elements $p_{d_i}$'s.} $\lbrace p_d \rbrace_{d\in E}$ of $\a$. Note that as $d_1=1$, the element $p_1=q_1$ indeed coincides with $p_1$, justifying the notation. One shows~\cite{Kostant:1959} that for every $d \in \lbrace 1, \cdots, h-1 \rbrace$, we have
\begin{equation} \label{Eq:DecoExpo}
\g_d = \left\{ \begin{array}{ll} 
[p_{-1}, \g_{d+1}] \qquad & \text{if}\quad d \not\in E,\\
{[p_{-1}, \g_{d+1}]} \oplus \C p_d \qquad & \text{if}\quad d \in E.
\end{array} \right.
\end{equation}

\subsection[Opers and Miura Opers associated with $\g$]{Opers and Miura Opers associated with $\bm{\g}$}
\label{SubSec:Opers}

In this subsection, we define the opers and Miura opers associated with an arbitrary semi-simple Lie algebra $\g$. These are generalisations of the opers and Miura opers defined in Subsection \ref{SubSec:FFRsl2} for $\g=\sld$. They were first introduced by Drinfeld and Sokolov in~\cite{Drinfeld:1984qv}, for the study of the KdV equation and its generalisations. For a more complete introduction to opers, see for instance~\cite{Frenkel:book,Beilinson:1995}.

\paragraph{Meromorphic $\bm{\g}$-connections.} We will denote by $\M$ the algebra of meromorphic functions on the Riemann sphere $\CP$. We define the space of $\g$-connections on $\CP$ as
\begin{equation*}
\Cog = \left\lbrace a\p_z + A(z), \; a\in\C, \; A\in \g \otimes \M \right\rbrace.
\end{equation*}
In this definition, the element $A\in\g \otimes \M$ is a $\g$-valued meromorphic function of $z$.

Let $G$ be a connected Lie group with Lie algebra $\g$\footnote{The rest of this section will not depend on the choice of the group $G$. One can for example choose the adjoint group $\Ad(\g)$ or the unique connected simply-connected group with Lie algebra $\g$.}. Let us consider the group $G(\M)$ of $G$-valued meromorphic functions. A rigorous definition of this group can become quite involved (we refer for example to the article~\cite{Lacroix:2016mpg}, in which a construction of $G(\M)$ in terms of affine group schemes is presented). For the purpose of this Subsection, one can consider a matricial representation of $G$ and think of $G(\M)$ as matrices in this representation with meromorphic entries. For $g\in G(\M)$, we define the gauge transformation of $\nabla = a\p_z + A(z)$ in $\Cog$ as
\begin{equation*}
g\nabla g^{-1} = a\p_z + g(z)A(z)g(z)^{-1} - a\bigl( \p_z g(z) \bigr) g(z)^{-1}.
\end{equation*}
This defines an action of $G(\M)$ on $\Cog$.

\paragraph{Opers and Miura opers.} Recall the positive nilpotent subalgebra $\n_+$ and Borel subalgebra $\bo_+$ of $\g$ (see Appendix \ref{App:CartanWeyl} and Subsection \ref{SubSec:Princ}). We define the following subspace of $\Cog$:
\begin{equation}\label{Eq:op}
\op{\g} = \bigl\lbrace \p_z + p_{-1} + A(z), \; A \in \bo_+ \otimes \M \bigr\rbrace.
\end{equation}
There is a unique connected subgroup $N_+$ of $G$ with Lie algebra $\n_+$. This subgroup can be seen as the exponentials of elements of $\n_+$ in $G$:
\begin{equation*}
N_+ = \lbrace \exp(X), \; X\in\n_+ \rbrace \subset G.
\end{equation*}
Let us consider the group $N_+(\M)$ of $N_+$-valued meromorphic functions:
\begin{equation}\label{Eq:NMero}
N_+(\M) = \bigl\lbrace \exp\bigl(m(z)\bigr), \; m\in\n_+\otimes\M \bigr\rbrace,
\end{equation}
which is then a subgroup of $G(\M)$.

Let $g=\exp(m)$ be an element of $N_+(\M)$ and $\nabla=\p_z + p_{-1} + A(z)$ be an element of $\op{\g}$. Then the gauge transformation of $\nabla$ by $g$ reads
\begin{equation*}
g\nabla g^{-1} = \p_z + p_{-1} + \bigl( g(z) p_{-1} g(z)^{-1} - p_{-1} \bigr) + g(z)A(z)g(z)^{-1} - \bigl( \p_z g(z) \bigr) g(z)^{-1}.
\end{equation*}
As $g$ is $N_+$-valued, $\bigl( \p_z g(z) \bigr) g(z)^{-1}$ belongs to $\n_+ \subset \bo_+$. In the same way, $A(z)$ belongs to $\bo_+$ and $g(z)$ to $N_+ \subset B_+$ (the subgroup of $G$ with Lie algebra $\bo_+$), thus the conjugacy $g(z)A(z)g(z)^{-1}$ belongs to $\bo_+$. Finally, using the fact that $\Ad_{g(z)} = \exp\bigl(\ad_{m(z)}\bigr)$ (see Appendix \ref{App:Lie}), we get
\begin{equation*}
g(z) p_{-1} g(z)^{-1} - p_{-1} = \sum_{n=1}^{+\infty} \frac{1}{n!}\ad_{m(z)}^n \bigl( \pn \bigr).
\end{equation*}
Note that this sum is finite as $m(z)$ is a nilpotent element of $\g$. The element $\pn$ is of degree -1 in the principal gradation and $m$ is composed of elements of degrees at least 1, thus $\ad_{m(z)}\bigl(\pn\bigr)$ is composed of elements of degrees at least 0, $\ad_{m(z)}^2\bigl(\pn\bigr)$ of elements of degrees at least 1, and so on. Thus, $g(z) p_{-1} g(z)^{-1} - p_{-1}$ belongs to $\bo_+$.

As a conclusion, we see that $g\nabla g^{-1}$ belongs to $\op{\g}$. Thus, the group $N_+(\M)$ acts by gauge transformations on the space $\op\g$. Let us consider the quotient of this action
\begin{equation}\label{Eq:Op}
\Op\g = \op\g / N_+(\M).
\end{equation}
We call en element of $\Op\g$ a $\bm\g$\textbf{-oper}. If $\nabla$ belongs to $\op\g$, we denote by $[\nabla]$ the equivalence class of $\nabla$ in the quotient $\Op\g$: $\nabla$ is then a representative of $[\nabla]$.\\

We define the space of $\bm\g$\textbf{-Miura opers} as
\begin{equation}\label{Eq:mOp}
\mop{\g} = \bigl\lbrace \p_z + p_{-1} + A(z), \; A \in \h \otimes \M \bigr\rbrace,
\end{equation}
where $\h$ is the Cartan subalgebra of $\g$. As $\h$ is included in the Borel algebra $\bo_+$, $\mop{\g}$ is a subspace of $\op\g$. Thus, for any Miura oper $\nabla\in\mop\g$, one can define the associated oper $[\nabla]\in\Op\g$ (note however that a Miura oper is not an oper, as it is not an equivalence class). 

\paragraph{Back to $\sldb$.} Let us come back to our first example $\g=\sld$, for which we defined opers and Miura opers in Subsection \ref{SubSec:FFRsl2}. The negative principal element $\pn$ and the Borel subalgebra of $\sld$ are respectively
\begin{equation*}
\begin{pmatrix}
0 & 0 \\ 1 & 0
\end{pmatrix} \;\;\;\;\;\;\; \text{ and } \;\;\;\;\;\;\; \left\lbrace \begin{pmatrix}
a & b \\ 0 & -a
\end{pmatrix}, \; a,b\in\C \right\rbrace.
\end{equation*}
Thus, the definition \eqref{Eq:opSl2} of the space $\op\sld$ agrees with the general definition \eqref{Eq:op} for an arbitrary Lie algebra $\g$.

In subsection \ref{SubSec:FFRsl2}, we consider the group $SL(2,\C)$, which is a connected (and simply-connected) Lie group with Lie algebra $\sld$. The connected subgroup of $SL(2,\C)$ with Lie algebra the positive nilpotent subalgebra of $\sld$ reads matricially
\begin{equation*}
\left\lbrace \begin{pmatrix}
1 & a \\ 0 & 1
\end{pmatrix}, \; a\in\C \right\rbrace.
\end{equation*}
Thus, the group $N_+(\M)$ defined in general by \eqref{Eq:NMero} coincides with the group $N$ defined in \eqref{Eq:NSl2} for $G=SL(2,\C)$. As a consequence, the definition \eqref{Eq:OpSl2} of $\sld$-opers agrees with the general definition \eqref{Eq:Op}.

Finally, the Cartan subalgebra of $\sld$ is composed by traceless diagonal matrices. Thus the general definition \eqref{Eq:mOp} of Miura oper reduces to the definition \eqref{Eq:MiuraSl2} for $\g=\sld$.

\paragraph{Canonical representatives of $\bm\g$-opers.} Recall that, for $\sld$-opers, we defined a notion of unique canonical representative. Such a notion exists also for opers associated with any semi-simple Lie algebra $\g$. It involves the centraliser $\a$ of $p_1$ introduced in Subsection \ref{SubSec:Princ}.

\begin{theorem}\label{Thm:Can}
Let \emph{$[\nabla]\in\Op\g$} be a $\g$-oper. There exists a unique representative of $[\nabla]$ of the form
\emph{\begin{equation}\label{Eq:Can}
\p_z  + \pn + A(z), \;\;\;\;\; \text{ with } \;\;\;\; A\in\a\otimes\M.
\end{equation}~\vspace{-12pt} \\}
It is called the \textbf{canonical representative} of $[\nabla]$ and is written $[\nabla]_{\text{can}}$.
\end{theorem}
\begin{proof}
We will not detail the construction of this canonical representative here. As for the construction of the group $G(\M)$ above, we refer to the article~\cite{Lacroix:2016mpg} for a detailed proof. Let us first note that the centraliser $\a$ of $p_1$ is included in $\bo_+$: thus, a connection of the form \eqref{Eq:Can} indeed belongs to the space $\op\g$.

The proof rests on a constructive algorithm, which, from any representative $\nabla\in\op\g$ of the oper, constructs $g\in G(\M)$ such that the gauge transformation $g\nabla g^{-1}$ is of the form \eqref{Eq:Can}. The proof of this algorithm works recursively on the degrees (in the principal gradation) of the elements of $\nabla$. It is mostly based on the property \eqref{Eq:DecoExpo} of the basis $\lbrace p_d \rbrace_{d\in E}$ of $\a$. One sees in the proof that the constructed element $g$ is uniquely determined by the initial representative $\nabla$, hence proving the uniqueness of the canonical representative.

An important point in the proof is the following. Although the gauge transformation $g\nabla g^{-1}$ involves derivatives of $g$, fixing the components of $g$ such that $g\nabla g^{-1}$ is of the form \eqref{Eq:Can} does not require any integration with respect to $z$. The construction is purely algebraic, which ensures that the element $A(z) \in \a$ obtained in the end is a meromorphic function. In the same way, it ensures that the algorithm does not create new singularities: if the initial representative is regular at a point $x\in\C$, so is the canonical representative (the algorithm can however increase the order of existing poles of the initial representative).
\end{proof}

Using the basis $\lbrace p_d \rbrace_{d \in E}$ of $\a$, we can express the canonical representative \eqref{Eq:Can} as
\begin{equation*}
[\nabla]_{\text{can}} = \p_z + p_{-1} + \sum_{d\in E} c^d(z) p_d, \;\;\;\;\; \text{ with } \;\;\;\; c^d(z) \in \M.
\end{equation*}
The canonical element of a $\g$-oper is thus uniquely described by $\ell$ meromorphic functions $c^d(z)$'s, labelled by exponents $d\in E$.\\

Let us end this paragraph by comparing this general notion of canonical element with the one for $\g=\sld$, introduced in Subsection \ref{SubSec:FFRsl2}. The rank of $\sld$ is one, so there is only one element $p_d$, which is $p_1$. An explicit computation of $p_1$ shows that it is matricially given by
\begin{equation*}
p_1 = \frac{1}{2} \begin{pmatrix}
0 & 1 \\
0 & 0
\end{pmatrix}.
\end{equation*}
Thus, the notion of canonical representative of Theorem \ref{Thm:Can} coincides with the one \eqref{Eq:CanSl2} for $\sld$.

\paragraph{Regularities and singularities of opers.} Let $x$ be a point of $\C$. We say that a $\g$-oper $[\nabla]$ is regular at $x$ if it possesses a representative which is regular at $x$. 

\begin{lemma}
Let $[\nabla]$ be a $\g$-oper. Then, $[\nabla]$ is regular at $x\in\C$ if and only if its canonical representative $[\nabla]_{\text{can}}$ is regular at $x$.
\end{lemma}
\begin{proof}
If $[\nabla]_{\text{can}}$ is regular at $x$, the oper $[\nabla]$ admits a representative which is regular at $x$ and is thus by definition regular at $x$.

Conversely, let us suppose that $[\nabla]$ possesses a representative $\nabla\in\op\g$ regular at $x$. Apply then the algorithm of Theorem \ref{Thm:Can} to find the canonical representative $[\nabla]_{\text{can}}$. As explained in the proof of the theorem, this algorithm does not create singularities, hence $[\nabla]_{\text{can}}$ is also regular at $x$. 
\end{proof}

Let $\nabla$ be a connection in $\op\g$. Using the decomposition \eqref{Eq:PrincNB} of $\bo_+$ in terms of the principal gradation, one can write $\nabla$ as
\begin{equation*}
\nabla = \p_z + p_{-1} + \sum_{d=0}^{h-1} A_d(z), \;\;\;\;\; \text{ with } \;\;\;\; A_d \in \g_d \otimes \M.
\end{equation*}
We say that $\nabla$ has (at most) a regular singularity at $x\in\C$ if the functions $A_d(z)$ have poles of order (at most) $d+1$. An oper in $\Op\g$ is said to have a regular singularity at $x$ if it possesses a representative with a regular singularity at $x$. Considering the algorithm constructing the canonical representative of an oper one proves the following lemma~\cite{Frenkel:2003qx}, similar to the one above for regularity.

\begin{lemma}
Let $[\nabla]$ be a $\g$-oper. Then, $[\nabla]$ has a regular singularity at $x\in\C$ if and only if its canonical representative $[\nabla]_{\text{can}}$ has a regular singularity at $x$.
\end{lemma}

One easily checks that these notions of regularity and regular singularity coincide with the ones defined in Paragraph \ref{Para:BackSl2} for $\g=\sld$.

\subsection{The FFR isomorphism and the case of Bethe vectors}

We now have almost all tools to express the FFR isomorphism $\Phi_{\zb,\g}$ of Theorem \ref{Thm:FFR}. We will then study how one can reformulate the Bethe ansatz in terms of the FFR approach and of opers.

\paragraph{The Langlands dual $\bm{\Lang}$.} Let us consider the Cartan matrix $A$ of the Lie algebra $\g$. Its transpose $\null^t A$ is also a Cartan matrix. The simple roots $\alpha_i \in \h^*$ and coroots $\ach_i \in \h$ associated with $A$ are respectively the simple coroots and roots associated with $\null^t A$. We denote by $\Lang$ the semi-simple Lie algebra with Cartan matrix $\null^t A$. The algebra $\Lang$ is called the \textbf{Langlands dual} of $\g$. For a simple Lie algebra $\g$ of type A, D, E, F or G in Cartan's classification, the Langlands dual is isomorphic to $\g$ itself (as $A$ is symmetric or a permutation of its transpose). The Langlands duality exchanges Lie algebras of types B and C.

As the simple coroots of $\Lang$ are the simple roots of $\g$, the Cartan subalgebra of $\Lang$ is naturally identified with the dual $\h^*$ of the Cartan subalgebra of $\g$. The algebras $\g$ and $\Lang$ then have the same rank $\ell$. One can define the Cartan-Weyl decomposition and Chevalley generators of $\Lang$ as for $\g$. One then also constructs the principal $\sld$ subalgebra of $\Lang$, generated by $\rho$, $\pch_1$ and $\pch_{-1}$. In particular, the Weyl coweight $\rho$ of $\Lang$ is defined as for $\g$ and is thus an element of $\null^L\h=\h^*$. In fact, it coincides with the Weyl weight of $\g$~\cite{Humphreys:1980dw}, as defined in equation \eqref{Eq:WeylW}. One then defines the centraliser $\null^L\a$ of $\pch_1$ and its basis $\lbrace\pch_d\rbrace$, with $d$ running through the exponents of $\Lang$. One can show~\cite{Kac:1990gs} that the latter are equal to the exponents of $\g$, so we can label the $\pch_d$'s by $d\in E$. This allows to define the space of $\Lang$-opers $\Op{\Lang}$ and their canonical representative, as for $\g$ in the previous Subsection.

\paragraph{The FFR isomorphism.} Let us consider the Gaudin model on $\g$ with sites $\bm z=(z_1,\cdots,z_N)$. Recall that the Gaudin Hamiltonian $\Hs^d(z)$ associated with the exponent $d\in E$ is a rational function of $z$, with poles of order $d+1$ at all sites $z_i$'s. We will then write $\Hs^d(z)$ as
\begin{equation}\label{Eq:DecompoHd}
\Hs^d(z) = \sum_{i=1}^N \sum_{p=0}^{d} \frac{1}{(p+1)!}\frac{\Q^d_{i,p}}{(z-z_i)^ {p+1}}.
\end{equation}
In particular, we have $\Q^1_{i,0}=\Hs_i$ (residue at $z=z_i$ of the quadratic Hamiltonian $\Hs(z)$) and $\Q^1_{i,1}=\Delta_{(i)}$ (the quadratic Casimir). The Gaudin algebra $\Zc_{\zb}(\g)$ is then generated by the operators $\Q^d_{i,p}$, for $i\in\lbrace 1,\cdots,N \rbrace$, $d\in E$ and $p\in\lbrace 0,\cdots,d\rbrace$.\\

We define the space $\Opz$ as the space of $\Lang$-opers with regular singularities at the $z_i$'s and regular elsewhere. Following the results of Subsection \ref{SubSec:Opers}, if $[\nabla]$ is such an oper, its canonical representative can be written as
\begin{equation}\label{Eq:RSCan}
\nc = \p_z + \pch_{-1} + \sum_{d\in E} A_d \, c^d(z)\, \pch_d,
\end{equation}
where the $A_d$'s are some normalisation constants, independent of $[\nabla]$, and
\begin{equation*}
c^d(z) = \sum_{i=1}^N \sum_{p=0}^{d} \frac{1}{(p+1)!}\frac{c^d_{i,p}}{(z-z_i)^ {p+1}},
\end{equation*}
with $c^d_{i,p}$ complex numbers. We then define the following functions $\Gamma^d_{i,p}$ on $\Opz$, for $i\in\lbrace 1,\cdots,N \rbrace$, $d\in E$ and $p\in\lbrace 0,\cdots,d\rbrace$:
\begin{equation}\label{Eq:Gammaip}
\begin{array}{rccc}
\Gamma^d_{i,p} : & \Opz     & \longrightarrow &   \C \\
                 & [\nabla] &   \longmapsto   & c^d_{i,p}
\end{array}.
\end{equation}
These functions generate the algebra of functions on $\Opz$.\\

We now finally have all the conventions necessary to express the FFR isomorphism $\Phi_{\zb,\g}$ of Theorem \ref{Thm:FFR}. We characterize it by specifying its images on the generators $\Q^d_{i,p}$ of $\Zc_{\zb}(\g)$ and extending it by linearity and multiplication:
\begin{equation}\label{Eq:PhiFFR}
\begin{array}{rccc}
\Phi_{\zb,\g} : & \Zc_{\zb}(\g) & \longrightarrow & \Fun{\Opz} \\
                &   \Q^d_{i,p}  &   \longmapsto   & \Gamma^d_{i,p}  
\end{array}.
\end{equation}
This expression coincides with the one \eqref{Eq:PhiSl2} of $\Phi_{\zb,\sld}$ for the $\sld$-Gaudin model, recalling that the Langlands dual of $\sld$ is $\sld$ itself.

\paragraph{The opers associated with Bethe vectors.} Let us now consider the Gaudin model on the Hilbert space $\Hl$, tensor product of highest-weight representations. One can then apply the Bethe ansatz. If $\Pcwo$ is an on-shell Bethe vector and thus an eigenvector of $\Zg$, Theorem \ref{Thm:FFRSpectrum} ensures the existence of an oper $[\nabla^{\bm c}_{\bm w}]$ in $\Opz$ which encodes the eigenvalues of $\Zg$ on $\Pcwo$. In this paragraph, we will show how one constructs this oper explicitly.

Let us start with an off-shell Bethe vector $\Psi_{\bm c}(\bm w)$, with colors $\bm c =(c(1),\cdots,c(M))$ and Bethe roots $\bm w = (w_1,\cdots,w_M)$. Recall the weight $\lambda^{\bm c}(z,\bm w)$ associated with the Bethe vector $\Pcw$ by equation \eqref{Eq:LambdaExc}. It belongs to the dual $\h^*$ of the Cartan subalgebra of $\g$. Recall from the first paragraph of this subsection that $\h^*$ is naturally identified with the Cartan subalgebra of $\Lang$. We then define the following $\Lang$-Miura oper
\begin{equation*}
\nabla^{\bm c}_{\bm w} = \p_z + \pch_{-1} - \lambda^{\bm c}(z,\bm w) \; \in \; \mop{\Lang}
\end{equation*}
and the associated oper $[\nabla^{\bm c}_{\bm w}]$. The FFR reformulation of the Bethe ansatz can be summarised as follows.

\begin{theorem}\label{Thm:FFRBethe}
The following points are equivalent.
\begin{enumerate}[(i)]
\setlength\itemsep{0.1em}
\item The oper \emph{$[\nabla^{\bm c}_{\bm w}]$} belongs to \emph{$\Opz$} (\textit{i.e.} has regular singularities at the $z_i$'s and is regular elsewhere).
\item The Bethe roots $\bm w$ satisfy the Bethe equations.
\item The Bethe vector \emph{$\Pcw$} is on-shell and thus an eigenvector of $\Zc_{\zb}(\g)$.
\end{enumerate}
In this case, the canonical representative of \emph{$[\nabla^{\bm c}_{\bm w}]$} can be written as
\emph{\begin{equation*}
[\nabla^{\bm c}_{\bm w}]_{\text{can}} = \p_z + \pch_{-1} + \sum_{d\in E} A_d \, \Cc^d(z,\bm w)\, \pch_d,
\end{equation*}\vspace{-12pt}\\}
with $A_d$'s the normalisation constant introduced in equation \eqref{Eq:RSCan} and $\Cc^d(z,\bm w)$'s some meromorphic functions. Then, the eigenvalue of $\Hs^d(z)$ on the on-shell Bethe vector \emph{$\Pcwo$} is equal to $\Cc^d(z,\bm w)$.
\end{theorem}

We will not prove this theorem here and refer to~\cite{Frenkel:2003qx} for a demonstration. Let us analyse the different aspects  and consequences of this theorem. The Miura oper $\nabla^{\bm c}_{\bm w}$ is a representative of the oper $[\nabla^{\bm c}_{\bm w}]$. Its component $\lambda^{\bm c}(z,\bm w)$ is valued in the Cartan subalgebra $\h^*$ of $\Lang$. In terms of the principal gradation $\Lang=\bigoplus_{d\in\Z} \Lang_d$, this Cartan subalgebra corresponds to the grade zero, \textit{i.e.} $\h^*=\Lang_0$ (see Subsection \ref{SubSec:Princ}). Yet, the component $\lambda^{\bm c}(z,\bm w) \in \Lang_0$ has simple poles at the $z_i$'s. Thus, the representative $\nabla^{\bm c}_{\bm w}$ has regular singularities at the $z_i$'s and so has the corresponding oper $[\nabla^{\bm c}_{\bm w}]$.

The oper $[\nabla^{\bm c}_{\bm w}]$ then belongs to $\Opz$ if and only if it is regular at points different from the $z_i$'s. Given the expression \eqref{Eq:LambdaExc} of $\lambda^{\bm c}(z,\bm w)$, the oper can have only singularities (other than at the $z_i$'s) at the Bethe roots $w_j$'s. The first main point of Theorem \ref{Thm:FFRBethe} is that the oper is regular at $w_j$ if and only the Bethe root $w_j$ satisfies the Bethe equation. We already observed this phenomenon in Subsections \ref{SubSec:FFRsl2} and \ref{SubSec:FFR} for the case where $\g=\sld$. In particular, the oper is regular at all $w_j$'s if and only if all Bethe roots satisfy the Bethe equations, hence if and only if the Bethe vector $\Pcw$ is on-shell.

Let us now consider the second part of Theorem \ref{Thm:FFRBethe}, stating that the eigenvalue of $\Hs^d(z)$ on the $\Pcwo$ is $\Cc^d(z,w)$. Here also we observed it explicitly for the case $\g=\sld$ in Subsections \ref{SubSec:FFRsl2} and \ref{SubSec:FFR}. Given the decomposition \eqref{Eq:DecompoHd} of $\Hs^d(z)$ and the definition \eqref{Eq:Gammaip} of the functions $\Gamma_{i,p}^d$ on $\Opz$, this can be reinterpreted as the fact that
\begin{equation*}
\chi_{\bm c, \bm w} \bigl( \Q^d_{i,p} \bigr) = \Gamma^d_{i,p} \bigl( [\nabla^{\bm c}_{\bm w}] \bigr),
\end{equation*}
where $\chi_{\bm c, \bm w}(\Q)$ denotes the eigenvalue of $\Q \in \Zg$ on $\Pcwo$. Thus, according to the definition \eqref{Eq:PhiFFR} of the FFR isomorphism $\Phi_{\zb,\g}$, we have
\begin{equation*}
\chi_{\bm c,\bm w}(\Q) = \bigl( \Phi_{\zb,\g} (\Q) \bigr) \bigl( [\nabla^{\bm c}_{\bm w}] \bigr).
\end{equation*}
This result illustrates the Theorem \ref{Thm:FFRSpectrum} for the eigenvector $\Pcwo$. In particular, it shows that the oper associated with this eigenvector (whose existence is predicted by Theorem \ref{Thm:FFRSpectrum}) can be constructed explicitly \textit{via} the Miura oper $\nabla^{\bm c}_{\bm w}$.

As a conclusion, we see that the FFR approach then contains the Bethe ansatz. As explained in Subsection \ref{SubSec:FFR}, one of the assets of this approach is that it also works when the Bethe ansatz fails (for example on other types of Hilbert spaces). Another asset of the FFR approach that one can observe in the present paragraph is that it gives a general formalism to describe the eigenvalues of both the quadratic Hamiltonian and the higher-degree ones.

\paragraph{Reproduction procedure.} Let us end this subsection by saying a few words about the so-called reproduction procedure. This procedure was first introduced and studied in~\cite{Mukhin_2002a} outside of the FFR formalism. Here, we present the reinterpretation of this procedure in the FFR approach, as understood in~\cite{Frenkel:2003qx,Mukhin:2005}. More details about this can be found in the article~\cite{Lacroix:2016mpg} (indeed, part of the subject of this publication is the generalisation of this procedure to cyclotomic Gaudin models, as explained in the following subsection).

Let us consider an on-shell Bethe vector $\Pcwo$ and the associated $\Lang$-Miura oper $\nabla^{\bm c}_{\bm w}$. The corresponding oper $[\nabla^{\bm c}_{\bm w}]$ is thus in $\Opz$, as the Bethe roots $\bm w$ satisfy the Bethe equations. Let $g$ be an element of the group $N_+(\M)$ and let us consider the gauge transformation $\widetilde{\nabla}^{\bm c}_{\bm w}=g \nabla^{\bm c}_{\bm w} g^{-1}$. This is an other representative of the oper $[\nabla^{\bm c}_{\bm w}]$. Suppose that one can choose $g$ such that $\widetilde{\nabla}^{\bm c}_{\bm w}$ is also a $\Lang$-Miura oper, \textit{i.e.} that
\begin{equation*}
\widetilde{\nabla}^{\bm c}_{\bm w} = \p_z + \pch_{-1} - \widetilde{\mu}(z),
\end{equation*}
with $\widetilde{\mu}$ a $\h^*$-valued meromorphic function. This condition takes the form of a differential equation on $g(z)$. It can be made into the form of a Ricatti equation and can then be solved formally (under some assumptions on the weights $\lambda_i$'s defining the Hilbert space $\Hl$). The Miura oper $\widetilde{\nabla}^{\bm c}_{\bm w}$ is then called a reproduction of the initial one $\nabla^{\bm c}_{\bm w}$.

For generic solutions of this reproduction procedure, we obtain a weight $\widetilde{\mu}(z)$ of the same form as the weight $\lambda^{\bm{\widetilde{c}}}(z,\bm{\widetilde{w}})$ associated with a Bethe vector $\Pcwt$, for other sets of Bethe roots $\bm{\widetilde w}$ and colors $\bm{\widetilde{c}}$. By construction, the oper $[\widetilde{\nabla}^{\bm c}_{\bm w}] = [{\nabla}^{\bm c}_{\bm w}]$ belongs to $\Opz$. Thus, by Theorem \ref{Thm:FFRBethe}, the Bethe vector $\Pcwt$ is on-shell and is therefore an eigenvector of $\Zg$. As the two Miura opers $\nabla^{\bm c}_{\bm w}$ and $\widetilde{\nabla}^{\bm c}_{\bm w}$ correspond to the same oper, they have the same canonical representative. Theorem \ref{Thm:FFRBethe} then implies that the eigenvalues of $\Zg$ on $\Pcwo$ and $\Pcwto$ are identical.

The generic solutions of the reproduction procedure then start from an on-shell Bethe vector $\Pcwo$ and generate other on-shell Bethe vectors $\Pcwto$ with the same eigenvalues. These different Bethe vectors have an interpretation in terms of the diagonal action of $\Di\g$. We discussed this action in Paragraph \ref{Para:DiagAction}: we shall use the notation of this paragraph here. Recall in particular that we decomposed $\Hl$ as a sum of Verma modules $M_s$ of the diagonal action, on which elements of $\Zg$ have a unique eigenvalue. These Verma modules contain several $\Di\g$-singular vectors (including the highest-weight ones $B_s$). The different Bethe vectors obtained above correspond in fact to different singular vectors with the same eigenvalue, including non highest-weight ones.

Outside of these generic solutions, the obtained Miura opers describe eigenvectors which are not Bethe vectors. We shall not enter into details about this here. An interesting result is that the set of all possibles solutions of the reproduction procedure possesses a nice geometric structure. Indeed, it is isomorphic to the flag manifold $G/B_-$, obtained as the quotient of the Lie group $G$ by its Borel subgroup $B_-$.

\subsection{Cyclotomic opers and a conjectural FFR approach for cyclotomic Gaudin models}
\label{SubSec:Cyclo}

This subsection is based on the article~\cite{Lacroix:2016mpg}, that I wrote during my PhD with B. Vicedo. Here, we will simply give a brief summary of its content.

In Chapter \ref{Chap:GaudinClass}, Subsection \ref{SubSec:GaudinGen}, we introduced a generalisation of classical Gaudin models that are called cyclotomic models and which possess additional equivariance properties under an automorphism $\s$ of the underlying Lie algebra $\g$. Quantum cyclotomic Gaudin models for finite algebras can be seen as a particular example of generalised quantum Gaudin model introduced by Skrypnyk in~\cite{Skrypnyk:2006}. Vicedo and Young proved~\cite{Vicedo:2014zza} that cyclotomic Gaudin models possess, similarly to non-cyclotomic ones, a large family of conserved commuting Hamiltonians, whose degrees are labelled by the exponents of the Lie algebra. These Hamiltonians generate a commutative algebra $\Zgc$, which is the cyclotomic equivalent of the Gaudin subalgebra $\Zg$. The Bethe ansatz for cyclotomic Gaudin models has also been developed in~\cite{Vicedo:2014zza} (see also~\cite{Skrypnyk:2013usa} for particular cases) and allows to diagonalise $\Zgc$ on a tensor product of highest-weight representations.

The goal of one of my PhD project with B. Vicedo, which resulted in the article~\cite{Lacroix:2016mpg}, was to lay down the foundations for a generalisation of the FFR approach to cyclotomic models. We summarise briefly the main results of the article here and refer to the whole publication for details.

\paragraph{Cyclotomic opers.} We introduce a notion of cyclotomic $\g$-opers as gauge equivalence classes of particular $\g$-connections. The main difference with non-cyclotomic opers is that these connections must satisfy some equivariance condition with respect to a certain automorphism $\tau$ of $\g$. We generalise most definitions and results about non-cyclotomic opers to the cyclotomic setting (Miura opers, canonical representatives, singularities, ...).

\paragraph{Cyclotomic reproduction procedure.} We study cyclotomic reproduction procedure (see previous Subsection for the non-cyclotomic case). More precisely we characterise the space of cyclotomic Miura opers which correspond to the same cyclotomic oper. The main difference with the non-cylcotomic case is that the possible reproduction transformations must satisfy some invariance condition under an automorphism $\theta$ of $\g$ (which is related to the automorphism $\tau$ above but is not identical). In particular, we show that this space is isomorphic to a subspace of the flag variety $G/B_-$, invariant under the automorphism $\theta$.

\paragraph{Conjectural cyclotomic FFR approach.} We conjecture that the cyclotomic Gaudin algebra $\Zgc$ is isomorphic to the algebra of functions on a certain type of cyclotomic opers $\Opzc$ on the Langlands dual $\Lang$ (the automorphism $\tau$ of $\Lang$ which appears in the definition of these cyclotomic $\Lang$-opers is constructed from the automorphism $\s$ characterising the cyclotomic Gaudin model).

We motivate this conjecture by several observations coming from the cyclotomic Bethe ansatz. As in the non-cyclotomic case, we associate cyclotomic Bethe vectors with particular cyclotomic Miura opers. We then prove that the cyclotomic Bethe equations can be seen as some regularity conditions on the associated oper, as in the non-cyclotomic setting. We also prove that our conjecture gives the correct eigenvalue of the quadratic cyclotomic Gaudin Hamiltonian.

Assuming that the conjecture is true, we use the results on cyclotomic reproduction mentionned above to get some informations about reproduction of cyclotomic Bethe vectors. \textit{Via} the constraints on these reproduction imposed by the automorphism $\theta$, we recover (and to some extent generalise) some results already obtained in~\cite{Varchenko:2015} using other techniques.

\cleardoublepage
\chapter{Quantum affine Gaudin models: towards a quantum hierarchy}
\label{Chap:QuantumAffine}

This chapter is based on the preprints~\cite{Lacroix:2018fhf} and~\cite{Lacroix:2018cag}, that I wrote during my PhD with B. Vicedo and C.A.S. Young. Here, we will summarise the main ideas developed in them.\\

This chapter is devoted to quantum Affine Gaudin models. In general, for any Kac-Moody algebra $\g$, one can define a quantum Gaudin model (complex, non cyclotomic and with simple poles) on $\g$. The algebra of observables of this model is the tensor product $U(\g)^{\otimes N}$, where $N$ is the number of sites of the model and $U(\g)$ is the universal enveloping algebra of $\g$ (more precisely the algebra of operators is a completion of this tensor product). We already introduced the universal enveloping algebra $U(\g)$ as a natural quantisation of the Kirillov-Kostant phase space $\g^*$ for a finite Lie algebra in Subsection \ref{SubSec:OpHilbert}: this discussion generalises straightforwardly to Kac-Moody algebras (modulo the treatment of infinite sums by appropriate completions), justifying this choice of algebra of operators.

One then defines~\cite{Varchenko:1991} a quadratic Hamiltonian $\Hs(z)$ in $U(\g)^{\otimes N}$ using the non-degenerate invariant bilinear form on the Kac-Moody algebra $\g$ (recall that in this thesis, we consider Kac-Moody algebras to be associated with symmetrisable Cartan matrices, ensuring that these algebras possess such a form). When the Gaudin model is considered on a tensor product of highest-weight representations of $\g$, this quadratic Hamiltonian has been diagonalised by Schechtman and Varchenko in~\cite{Varchenko:1991}, using the Bethe ansatz.\\

In particular, this construction applies to affine Kac-Moody algebras and one can thus consider quantum AGM. Classical AGM were discussed in Section \ref{Sec:AGM} of this thesis. In particular, we explained how an AGM can be realised as an integrable field theory with twist function, that we called the local AGM. Using this fact, we constructed in Subsection \ref{SubSec:HierarchyAGM} an infinite hierarchy of local conserved charges in involution for these models, which contains the quadratic Hamiltonian mentioned above. A natural question is then whether this hierarchy can be quantised, \textit{i.e.} whether one can find commuting conserved operators in the quantum AGM which reduce to these local charges in the classical limit. This is the main question addressed in this chapter.

An important property of the classical hierarchy is that the charges are labelled by the positive exponents of the affine Kac-Moody algebra. This bears a striking resemblance with the finite case. Indeed, for finite Gaudin models, one constructs a hierarchy of conserved charges of degrees labelled by the exponents of the underlying finite algebra (see Subsection \ref{SubSec:FiniteClassGaudin}). Moreover, as explained in Chapter \ref{Chap:QuantumFinite}, this hierarchy can be quantised and its spectrum can be described by the Bethe ansatz and the FFR approach. The latter relates the spectrum of the conserved charges with opers associated with the finite algebra, whose description also involves the exponents of the finite algebra.

One can also define a notion of oper associated with an affine algebra. Moreover, these opers are described in terms of the positive exponents of the affine algebra. It is then natural to conjecture that there exists a quantum hierarchy of AGM and that its spectrum is described in terms of affine opers. This was first proposed by Feigen and Frenkel in~\cite{Feigin:2007mr} for the study of the quantum KdV hierarchy.\\

As explained above, the first charge of the hierarchy of AGM, the quadratic Hamiltonian, has already been constructed at the quantum level by Schechtman and Varchenko in~\cite{Varchenko:1991}. This construction was made possible by the existence of a quadratic Casimir of the affine algebra. In the same way, the existence of higher-degree quantum hamiltonians for finite Gaudin models was ensured by the existence of Casimirs of the corresponding degrees for the finite algebra. However, it is known~\cite{Chari:1984} that affine algebras do not possess Casimirs of order greater than two. The construction of higher-order quantum Hamiltonians for AGM thus cannot be done in a similar way than the quadratic one.

This issue can already be seen at the classical level. As explained in Subsection \ref{SubSec:GaudinHam}, Hamiltonians in involution for a classical Gaudin model on a Lie algebra $\g$ can be constructed by evaluating invariant polynomials of $\g$ on the Lax matrix. One then obtains an Hamiltonian depending freely on the spectral parameter of the theory. For instance, the quadratic Hamiltonian of the model is constructed from an invariant polynomial of degree two, the invariant bilinear form. The classical analogue of the non-existence of higher-order Casimirs of the affine algebra is the non-existence of higher order invariant polynomials on this algebra.

In the article~\cite{Lacroix:2017isl} (see also Chapter \ref{Chap:LocalCharges} and Subsection \ref{SubSec:HierarchyAGM} of this thesis), we overcame this difficulty and constructed higher-degree classical Hamiltonians by considering some observables depending on the Lax matrix and evaluating the spectral parameter at particular points, the zeros of the twist function. To find a quantisation of these charges, one then has to find an analogue of this procedure at the quantum level.\\

In the preprint~\cite{Lacroix:2018fhf}, B. Vicedo, C.A.S. Young and myself propose a conjecture for such a construction, guided by the idea that the spectrum of these quantum charges should be described by affine opers. We develop further the study of affine opers and construct functions on these opers which take the form of hypergeometric integrals in the spectral parameter. Based on the idea that such functions should describe the eigenvalues of quantum Hamiltonians of AGM, we conjecture that these Hamiltonians also take the form of such integrals. We support this conjecture by reasoning on the classical limit of the AGM: indeed in this limit, these hypergometric integrals in the spectral parameter become, by a saddle point approximation, evaluations of the spectral parameter at the zeros of the twist function. Considering these integrals is thus a quantum analogue of the procedure used to construct the classical hierarchy.

The plan of the chapter is the following. In Section \ref{Sec:QuantumAGM}, we recall some generalities about affine Lie algebras and quantum affine Gaudin models. In the rest of this chapter, we will summarise the main ideas developed in the preprint~\cite{Lacroix:2018fhf} and state some of the important results without proving them. We refer to the complete preprint for more details and proofs. In Section \ref{Sec:AffineOper}, we discuss the theory of affine opers and more precisely how to define functions on affine opers by hypergeometric integrals. In Section \ref{Sec:AffineHam}, we conjecture the existence of a quantum hierarchy for Gaudin models and the description of its spectrum by affine opers. The classical limit of this hierarchy will also be discussed in this section. Some first results on the cubic quantum Hamiltonian supporting these conjectures are the subject of our second preprint~\cite{Lacroix:2018cag} and are also mentioned in Section \ref{Sec:AffineHam}.

\section{Affine algebras and quantum affine Gaudin models}
\label{Sec:QuantumAGM}

\subsection{Affine Kac-Moody algebras}
\label{SubSec:AffineKac}

Let us consider an affine algebra $\g$. In Chapter \ref{Chap:GaudinClass}, Subsection \ref{SubSec:Affine}, we used a description of this algebra in terms of loop algebras, as it was useful for the reinterpretation of classical AGM as field theories. In this chapter, we will need the description of $\g$ as a Kac-Moody algebra. This can be found for example in~\cite{Kac:1990gs}. The structure theory of the affine algebra $\g$ is similar to the one of a semi-simple Lie algebra: $\g$ possesses a Cartan subalgebra $\h$, a root system $\Delta\subset \h^*$ and some Chevalley generators $E_i$'s and $F_i$'s associated with simple roots $\alpha_i\in\Delta$ and coroots $\ach_i\in\h$ ($i=0,\cdots,\ell$). A basis of $\h$ is given by $\lbrace \ach_0,\cdots,\ach_\ell,\Dd \rbrace$, where $\Dd$ is a so-called derivation element of $\g$. The algebra $\g$ is equipped with a non-degenerate invariant bilinear form $\fd$. In particular, this form restricts to a non-degenerate form on $\h$, which induces a non-degenerate form on $\h^*$, that we shall still write $\fd$.

Let $a_i$ and $a_i^\vee$ be the Kac labels and dual labels of $\g$ (these are coprime positive integers defined from the Cartan matrix of $\g$, see~\cite{Kac:1990gs}). The Coxeter and dual Coxeter numbers of $\g$ are defined as
\begin{equation}\label{Eq:DefCoxeter}
h = \sum_{i=0}^\ell a_i \;\;\;\;\; \text{ and } \;\;\;\;\; h^\vee = \sum_{i=0}^\ell a^\vee_i.
\end{equation}
We define the elements $\delta$ of $\h^*$ and $\Kd$ of $\h$ as
\begin{equation}\label{Eq:DefKd}
\delta = \sum_{i=0}^\ell a_i \alpha_i \;\;\;\;\; \text{ and } \;\;\;\;\; \Kd = \sum_{i=0}^\ell a^\vee_i \ach_i.
\end{equation}
They satisfy
\begin{equation*}
\langle \ach_i, \delta \rangle = \langle \Kd, \alpha_i \rangle = 0, \;\;\;\;\;\; \forall \, i\in\lbrace 0,\cdots,\ell \rbrace.
\end{equation*}
The element $\Kd$ is the central element of $\g$ (which commutes with every $X\in\g$). Let $\rho$ be the unique element of $\h^*$ such that $(\rho,\rho)=0$ and
\begin{equation}\label{Eq:DefRho}
\langle \ach_i, \rho \rangle = 1, \;\;\;\;\; \forall \, i\in\lbrace 0,\cdots,\ell \rbrace.
\end{equation}
Then a basis of $\h^*$ is given by $\lbrace \alpha_0,\cdots,\alpha_\ell,\rho \rbrace$.

One constructs a quadratic Casimir $\Delta$ of $\g$ using the invariant bilinear form $\fd$~\cite{Kac:1990gs}. This Casimir is an element of the center $\Zc(\g)$ of the (completed) universal enveloping algebra $U(\g)$. Let $\lambda\in\h^*$ be a weight of $\g$ and $V_\lambda$ be the Verma module with highest-weight $\lambda$. Then all elements of $V_\lambda$ are eigenvectors of the Casimir $\Delta$, with a unique eigenvalue:
\begin{equation*}
\Delta.v = \bigl( \lambda, \lambda+2\rho \bigr)v, \;\;\;\;\;  \forall \, v\in V_\lambda.
\end{equation*}

\subsection{Quantum affine Gaudin model}
\label{SubSec:QuantAGM}

\paragraph{Generalities.} Let us briefly introduce the quantum affine Gaudin model on $\g$ with sites $\zb=(z_1,\cdots,z_N)$. Its algebra of operators $\Ac_{\zb}$ is given by a completion of the tensor product $U(\g)^{\otimes N}$ (this completion is constructed from the homogeneous gradation of $\g$: cf. Paragraph \ref{Para:Completion} in the classical case). As for the finite case, if $X$ is an element of $\g$, we will write $X_{(k)}$ the copy of $X$ in the $k^{\rm th}$-tensor factor of $\Ac_{\zb}$.

Let $I^a$ be a basis of $\g$, with dual basis $I_a$ with respect to the bilinear form $\fd$. We define the Gaudin Hamiltonian at site $k$ to be
\begin{equation*}
\Hs_k = \sum_{ \substack{j=1 \\ j\neq k}}^N \frac{I^a_{(k)}I_{a\,(j)}}{z_k - z_j} \; \in \Ac_{\zb}. 
\end{equation*}
We define the spectral parameter dependent quadratic Hamiltonian to be
\begin{equation*}
\Hs(z) = \sum_{k=1}^N \left( \frac{1}{2} \frac{\Delta_{(k)}}{(z-z_k)^2} + \frac{\Hs_k}{z-z_k} \right).
\end{equation*}
One then has the commutation relation
\begin{equation*}
\bigl[ \Hs(z), \Hs(z') \bigr] = 0, \;\;\;\;\; \forall \, z,z' \in \C.
\end{equation*}

\paragraph{Hilbert space and levels.} As in the finite case, the tensor product $U(\g)^{\otimes N}$ acts naturally on any tensor product of $N$ representations of $\g$. However, as $\Ac_{\zb}$ is a completion of $U(\g)^{\otimes N}$, the extension of this action to $\Ac_{\zb}$ requires additional constraints on the considered representations. We will not enter into more details about which representations can be considered or not. We shall just need the fact that one can choose a tensor product of highest-weight representations of $\g$. Let then $\lb=(\lambda_1,\cdots,\lambda_N)$ be a collection of $N$ weights of $\g$. As in the finite case, we define the Hilbert space $\Hl$ as the tensor product of the Verma modules $V_{\lambda_i}$, which is then a representation of the algebra of operators $\Ac_{\zb}$.

Recall from Subsection \ref{SubSec:AffineKac} that a basis of $\h^*$ is given by $\lbrace \alpha_0,\cdots,\alpha_\ell,\rho \rbrace$. Let us then decompose the weights $\lambda_i$'s as
\begin{equation}\label{Eq:WeightsLevel}
\lambda_i = \widetilde{\lambda}_i + \frac{k_i}{h^\vee} \rho, \;\;\;\; \text{ with } \;\;\;\; \widetilde{\lambda}_i \in \text{Span}(\alpha_0,\cdots,\alpha_\ell) \;\; \text{ and } \;\; k_i\in\C,
\end{equation}
where the dual Coxeter number $h^\vee$ has been introduced for later convenience.

The central element $\Kd_{(i)}$ acts on the Hilbert space $\Hl$ by the multiplication by $\langle \Kd, \lambda_i \rangle$. Recall that $\langle \Kd,\alpha_j \rangle = 0$, for all $j\in\lbrace 0,\cdots,\ell\rbrace$. Thus, we have $\langle \Kd, \widetilde{\lambda}_i \rangle = 0$. From equations \eqref{Eq:DefCoxeter}, \eqref{Eq:DefKd} and \eqref{Eq:DefRho}, we get
\begin{equation*}
\langle \Kd, \rho \rangle = \sum_{i=0}^\ell a_i^\vee = h^\vee.
\end{equation*}
Thus, we get that $\langle \Kd, \lambda_i \rangle = k_i$. Therefore, the operator $\Kd_{(i)}$ acts on the Hilbert space $\Hl$ as a complex number $k_i$. This is the quantum equivalent of the fact that for classical AGM, we realised the abstract observables $K_{(i)}$'s as complex numbers (see Section \ref{Sec:AGM}). The numbers $k_i$'s are thus the \textbf{levels} of the quantum AGM.

\paragraph{Bethe ansatz.} The Bethe ansatz allows the diagonalisation of the quadratic Hamiltonian $\Hs(z)$ on $\Hl$. It can be derived from the results~\cite{Varchenko:1991} of Schechtman and Varchenko, for any Kac-Moody algebra. In particular, it is quite similar to the Bethe ansatz for the finite Gaudin model presented in Section \ref{Sec:Bethe}, as it is based on the common structure shared by all Kac-Moody algebras. Hence, we will not enter into more details about the Bethe ansatz here and just recall the result.

The off-shell Bethe vector $\Pcw$ is defined from Bethe roots $\bm w = (w_1,\cdots,w_M)\in\C^M$ and associated colors $\bm c =\bigl( c(1),\cdots,c(M)\bigr)\in\lbrace 0,\cdots,\ell \rbrace^M$. It is on-shell (and thus an eigenvector of $\Hs(z)$) if the Bethe roots satisfy the Bethe equations
\begin{equation*}
\sum_{i=1}^N \frac{(\alpha_{c(k)},\lambda_i)}{w_k-z_i} - \sum_{\substack{j=1 \\ j\neq k}}^M \frac{(\alpha_{c(k)},\alpha_{c(j)})}{w_k-w_j} = 0, 
\end{equation*}
for $k\in\lbrace 1,\cdots,M \rbrace$. The eigenvalue of $\Hs(z)$ on $\Pcwo$ is given by
\begin{equation}\label{Eq:EigenAffine}
\mathcal{E}^{\,\text{on}}_{\bm c}(z,\bm{w}) = \frac{1}{2} \bigl( \lambda^{\bm c}(z,\bm{w}), \lambda^{\bm c}(z,\bm{w}) \bigr) - \frac{\p\;}{\p z} \bigl( \lambda^{\bm c}(z,\bm{w}), \rho \bigr),
\end{equation}
where we define $\lambda^{\bm c}(z,\bm{w})$ in $\h^*$ by
\begin{equation}\label{Eq:LambdaAffine}
\lambda^{\bm c}(z,\bm{w}) = \sum_{i=1}^N \frac{\lambda_i}{z-z_i} - \sum_{j=1}^M \frac{\alpha_{c(j)}}{z-w_j}.
\end{equation}

An important point for the rest of this section is the following. Recall the decomposition \eqref{Eq:WeightsLevel} of the weights $\lambda_i$'s. The weight \eqref{Eq:LambdaAffine} can then be written as
\begin{equation}\label{Eq:LambdaTwist}
\lambda^{\bm c}(z,\bm{w}) = \widetilde{\lambda}^{\bm c}(z,\bm{w}) + \frac{\vp(z)}{h^\vee} \rho,
\end{equation}
where $\widetilde{\lambda}^{\bm c}(z,\bm{w})$ belongs to $\text{Span}(\alpha_0,\cdots,\alpha_\ell)$ and
\begin{equation}\label{Eq:TwistAGM}
\vp(z) = \sum_{i=1}^N \frac{k_i}{z-z_i}.
\end{equation}
As the $k_i$'s are the levels of the AGM, the function $\vp(z)$ is the \textbf{twist function} of the model (see equation \eqref{Eq:TwistGaudin} in the classical case). The twist function then appears as the coefficient of $\rho$ in the weight $\lambda^{\bm c}(z,\bm w)$ (note in particular that, although this weight depends on the considered Bethe vector $\Pcw$, the coefficient of $\rho$ does not and is always given by the twist function).

\subsection{Langlands dual and exponents}
\label{SubSec:AffineLangExpo}

\paragraph{Langlands dual.} Our goal in this chapter is the generalisation of the FFR approach to affine Gaudin models. Recall from Section \ref{Sec:FFR} that this approach for finite models involves opers on the Langlands dual of the underlying finite algebra. Thus, we shall need the Langlands dual $\Lang$ of the affine algebra $\g$. As in the finite case, it is the Kac-Moody algebra with Cartan matrix the transpose of the one of $\g$. Thus, it is also an affine algebra. Its Cartan subalgebra $\null^L\h$ is naturally identified with the dual $\h^*$ of the one of $\g$. In particular, the coroots of $\Lang$ are the roots $\alpha_i$'s and its central element is the root $\delta$ defined in equation \eqref{Eq:DefKd}. As a Kac-Moody algebra, $\Lang$ possesses a Cartan-Weyl decomposition
\begin{equation*}
\Lang = \null^L\h \oplus \null^L\n_+ \oplus \null^L\n_- = \h^* \oplus \null^L\n_+ \oplus \null^L\n_-. 
\end{equation*}
We define the positive Borel subalgebra of $\Lang$ as $\bl = \h^* \oplus \null^L\n_+$.

\paragraph{Heisenberg principal subalgebra and exponents.} Let $\check{E}_i$'s and $\check{F}_i$'s ($i=0,\cdots,\ell$) be the positive and negative Chevalley generators of $\Lang$. We define the positive and negative principal elements of $\Lang$ as
\begin{equation*}
\pch_{1} = \sum_{i=0}^\ell a_i \check{E}_i \;\;\;\;\; \text{ and } \;\;\;\;\; \pch_{-1} = \sum_{i=0}^\ell \check{F}_i.
\end{equation*}
We will need the following theorem, which can be found for example in~\cite{Kac:1990gs}.

\begin{theorem}\label{Thm:Heisenberg}
There exist elements $\pch_d$'s in $\Lang$, labelled by the exponents $d\in E$ of $\g$, such that $\pch_1$ and $\pch_{-1}$ are given as above and
\begin{eqnarray}
\bigl[ \pch_d, \pch_e \bigr] &=& d\, \delta_{d+e,0}\, \delta \\
\bigl[ \rho, \pch_d \bigr]   &=& d \, \pch_d.
\end{eqnarray}
\end{theorem}
\noi These elements are the affine equivalent of the $\pch_d$'s introduced in Subsection \ref{SubSec:Princ} for finite algebras. Together with the central element $\delta$, they form the so-called principal Heisenberg subalgebra of $\Lang$. In the rest of this chapter, we will denote by $E_+$ the set of positive exponents.

\section{Affine opers and hypergeometric integrals}
\label{Sec:AffineOper}

In this section, we develop the theory of affine opers. As we have in mind the application to Gaudin models, we will consider here $\Lang$-opers instead of $\g$-opers. Affine opers were already studied by Drinfeld and Sokolov in~\cite{Drinfeld:1984qv} (see also~\cite{Feigin:2007mr,Bazhanov:2013cua}). In the preprint~\cite{Lacroix:2018fhf}, we develop further this theory and in particular define hypergeometric functions on affine opers.

\subsection[Opers and Miura opers associated with $\Lang$]{Opers and Miura opers associated with $\bm{\Lang}$}
\label{SubSec:AffineOpers}

\paragraph{$\bm{\Lang}$-opers.} We will introduce $\Lang$-opers in a way similar to the finite case, discussed in Subsection \ref{SubSec:Opers}. We consider the space $\text{Conn}_{\Lang}\bigl(\CP\bigr)$ of $\Lang$-connections on the Riemann sphere. As for a finite algebra, it is composed of differential operators of the form
\begin{equation*}
a \p_z + A(z), \;\;\;\; \text{ with } \; a\in\C \;\; \text{ and } \;\; A\in \Lang \otimes \M,
\end{equation*}
where $\M$ denotes the algebra of meromorphic functions on $\CP$ ($A(z)$ is thus a $\Lang$-valued meromorphic function of $z$).

Recall the positive Borel subalgebra $\bl$ of $\Lang$ introduced in Subsection \ref{SubSec:AffineLangExpo}. We shall be interested in the following particular subspace of $\text{Conn}_{\Lang}\bigl(\CP\bigr)$:
\begin{equation*}
\op{\Lang} = \bigl\lbrace \p_z + \pch_{-1} + A(z), \, A\in\bl \otimes \M \bigr\rbrace.
\end{equation*}
This is the equivalent of the space \eqref{Eq:op} defined for a finite Lie algebra.

Recall also the positive subalgebra $\null^L\n_+$ of $\Lang$. We define a Lie group $\null^L N_+$ whose Lie algebra is $\null^L \n_+$ by exponentiation and we consider the group $\Nl$ of $\null^L N_+$-valued meromorphic functions\footnote{The rigorous definition of $\Nl$ requires a treatment of infinite sums by appropriate completions. More details about this can be found in~\cite{Lacroix:2018fhf}.}. The group $\Nl$ acts on the space $\op{\Lang}$ by gauge transformations. We then define the space of $\Lang$-opers as the quotient
\begin{equation*}
\Op{\Lang} = \op{\Lang} / \Nl.
\end{equation*}
This definition is the equivalent for an affine algebra of the definition \eqref{Eq:Op} of opers of a finite algebra.

\paragraph{$\bm{\Lang}$-Miura opers.} We define the space of $\Lang$-Miura opers as
\begin{equation*}
\mop{\Lang} = \bigl\lbrace \p_z + \pch_{-1} + A(z), \, A\in \h^* \otimes \M \bigr\rbrace.
\end{equation*}
As $\h^*$ coincides with the Cartan subalgebra $\null^L\h$ of $\Lang$, this is the equivalent for the affine algebra $\Lang$ of the definition \eqref{Eq:mOp} of Miura opers associated with a finite algebra.

The Miura opers considered originally by Drinfeld and Sokolov are connections of similar forms but with $A(z)$ valued in $\text{Span}(\alpha_0,\cdots,\alpha_\ell)$ instead of the whole Cartan subalgebra $\null^L\h = \h^*$. In particular, this definition does not allow Miura opers for which the weight $A(z)$ contains a term proportional to $\rho$. As we will see, such a term will play a major role in our description of $\Lang$-opers.

\subsection{Quasi-canonical form and residual gauge transformation}

\paragraph{Quasi-canonical form.} Recall from Theorem \ref{Thm:Can} that finite opers possess a unique canonical form, expressed with the elements $p_d$ labelled by the exponents $d$ of the finite algebra. In Subsection \ref{SubSec:AffineLangExpo}, we defined the equivalent of these elements for the affine algebra $\Lang$: the $\pch_d$'s, labelled by the exponents $d\in E$ of $\Lang$. In particular, the objects $\pch_d$ for positive exponents $d\in E_+$ belong to the positive Borel subalgebra $\null^L\bo_+$. The affine equivalent of the existence of a canonical form is the following result:

\begin{theorem}\label{Thm:QuasiCan}
Let \emph{$[\nabla] \in \Op{\Lang}$} be a $\Lang$-oper. Then $[\nabla]$ possesses a representative of the form
\begin{equation}\label{Eq:QuasiCan}
\p_z + \pch_{-1} + c^0(z) \rho + \sum_{d\in E_+} c^d(z) \pch_d,
\end{equation}
where the $c^d(z)$'s ($d\in E_+ \cup \lbrace 0 \rbrace$) are meromorphic functions of $z$.
\end{theorem}

We refer to the preprint~\cite{Lacroix:2018fhf} for the proof of this theorem, which is based on an algorithm similar to the one in the finite case (see the proof of Theorem \ref{Thm:Can}). Let us comment on a few aspects related to this result. Note first that Theorem \ref{Thm:QuasiCan} does not state that the representative \eqref{Eq:QuasiCan} is unique (as we shall see later, it is indeed not unique). We shall not call \eqref{Eq:QuasiCan} a canonical form but a \textbf{quasi-canonical form} of $[\nabla]$, to stress this non-unicity.

Note also that contrarily to the finite case, the quasi-canonical form still possesses a component valued in the Cartan subalgebra $\null^L\h$, namely the term $c^0(z) \rho$. This is due to the fact that $\rho$ does not belong to the derived subalgebra of $\Lang$: it can never be obtained as a linear combination of Lie brackets in
$\Lang$. Indeed, if $\nabla$ is in $\op{\Lang}$, because of that property, a gauge transformation of $\nabla$ by any $g\in\Nl$ does not change the coefficient $c^0(z)$ of $\rho$ in $\nabla$. Thus, all representatives of the oper $[\nabla]$ share this same coefficient $c^0(z)$. In particular, one cannot find a quasi-canonical form of $[\nabla]$ where this coefficient vanishes.

\paragraph{Residual gauge transformation.} Let us prove that the quasi-canonical form $\nabla$ given by \eqref{Eq:QuasiCan} is not unique. For future convenience, we will write the coefficient $c^0$ of $\rho$ in $\nabla$ as
\begin{equation*}
c^0(z) = - \frac{\vp(z)}{h^\vee}.
\end{equation*}
For this subsection, one can consider the function $\vp(z)$ to be any rational function ; for the applications to quantum AGM, it will be the twist function \eqref{Eq:TwistAGM} of the model.

For each positive exponent $d\in E_+$, distinct from 1, let us consider a meromorphic function $f^d(z) \in \M$. The exponential
\begin{equation}\label{Eq:Residual}
g(z) = \exp \left( -\sum_{d\in E_+\setminus\lbrace 1 \rbrace} f^d(z) \pch_d \right)
\end{equation}
is an element of the group $\Nl$. Using the property $\Ad_{\exp(m)} = \exp(\ad_m)$ and the commutators of the $\pch_d$'s and $\rho$ given by Theorem \ref{Thm:Heisenberg}, one can compute the gauge transformation of $\nabla$ by $g$. One then finds
\begin{equation*}
\widetilde{\nabla} = g\nabla g^{-1} = \p_z + \pch_{-1} - \frac{\vp(z)}{h^\vee} \rho + \sum_{d\in E_+} \widetilde{c}^{\,d}(z) \pch_d,
\end{equation*}
where $\widetilde{c}^{\,1}(z)=c^1(z)$ and
\begin{equation*}
\widetilde{c}^{\,d}(z) = c^d(z) + \p_z f^d(z) - \frac{d\,\vp(z) }{h^\vee}f^d(z).
\end{equation*}
We will call $\p_z f^d(z) - \dfrac{d\,\vp(z) }{h^\vee}f^d(z)$ the \textbf{$\bm{d}$-twisted derivative} of $f^d(z)$.

The connection $\widetilde\nabla$ is of the same form than the initial connection $\nabla$. In particular, it is also a quasi-canonical representative of the oper $[\nabla]$. We will call \textbf{residual gauge transformations} the gauge transformations which preserve the form of a quasi-canonical representative (they encode the remaining gauge freedom of the oper once we find its quasi-canonical form). The gauge transformations by elements of the form \eqref{Eq:Residual} are thus residual gauge transformations. In the preprint~\cite{Lacroix:2018fhf}, we also prove that these are the unique ones. Thus, the oper $[\nabla]$ is entirely characterised by the functions $\vp(z)$ and $c^d(z)$'s, up to the transformation
\begin{equation}\label{Eq:TwistedDer}
c^d(z) \longmapsto c^d(z) + \p_z f^d(z) - \frac{d\,\vp(z) }{h^\vee}f^d(z), \;\;\;\;\; \text{ for } \; d \in E_+\setminus\lbrace 1 \rbrace.
\end{equation}

Note that the exponent $1$ has a singular behaviour compared to all other positive exponents: there is no residual gauge transformations modifying the coefficient $c^1(z)$ of $\pch_1$ in $\nabla$. Indeed, if one considers a gauge transformation of $\nabla$ by $g(z)=\exp\bigl( -f(z)\pch_1 \bigr)$, the coefficient $c^1(z)$ is shifted by the 1-twisted derivative of $f$ but the fact that
\begin{equation*}
[\pch_1, \pch_{-1} ] = \delta
\end{equation*}
creates a term $f(z)\delta$ in $g\nabla g^{-1}$. Thus, the representative $g\nabla g^{-1}$ of the oper is not in quasi-canonical form. We will come back on this fact later.

\subsection[Hypergometric functions on $\Lang$-opers]{Hypergometric functions on $\bm{\Lang}$-opers}

Recall that the FFR approach for finite $\g$-Gaudin models involves functions on $\Lang$-opers. To generalise this approach to quantum AGM, we will need to understand what are the functions on $\Lang$-opers, for $\g$ affine. As explained above, the coefficients $c^d(z)$ of a quasi-canonical form \eqref{Eq:QuasiCan} are not gauge-invariant. Thus, they do not define a function on the space of opers. To find such functions, one needs to extract from $c^d(z)$ a quantity invariant under the residual gauge transformations \eqref{Eq:TwistedDer}.

As these residual gauge transformations involve the function $\vp(z)$ appearing in the coefficient of $\rho$ (which then depends on the considered oper), one expects gauge-invariant quantities to be defined in a way depending on $\vp(z)$. Thus, such a gauge-invariant quantity would only define a function on the space $\Opp{\Lang}$ of $\Lang$-opers sharing the same function $\vp(z)$ (as this function appears in the coefficient of $\rho$, which is independent of the choice of representative, the definition of this space makes sense).

\paragraph{The case $\bm{\vp=0}$.} To gain some intuition, let us first consider the case where $\vp(z)=0$ (as explained in Subsection \ref{SubSec:AffineOpers}, this is the case considered initially by Drinfeld and Sokolov in~\cite{Drinfeld:1984qv}). The residual gauge transformation \eqref{Eq:TwistedDer} simply reduces to
\begin{equation*}
c^d(z) \longmapsto c^d(z) + \p_z f^d(z).
\end{equation*}
It is then clear that one can construct gauge-invariant quantities by considering integrals
\begin{equation*}
\oint_\gamma c^d(z) \, \dd z
\end{equation*}
of $c^d(z)$ over any closed contour $\gamma$ (on which $c^d(z)$ is regular).

\paragraph{The case $\bm{\vp \neq 0}$.} Let us come back to the general case $\vp\neq 0$. We introduce the function
\begin{equation*}
\Ps(z) = \exp \left( \int \vp(z) \, \dd z \right),
\end{equation*}
where $\int \vp(z) \, \dd z$ denotes any primitive of $\vp$. The function $\Ps$ satisfies
\begin{equation*}
\p_z \log \bigl( \Ps(z) \bigr) = \vp(z).
\end{equation*}
The residual gauge transformation \eqref{Eq:TwistedDer} can then be rewritten
\begin{equation*}
\Ps(z)^{-d/h^\vee} c^d(z) \longmapsto \Ps(z)^{-d/h^\vee} c^d(z) + \p_z \left( \Ps(z)^{-d/h^\vee} f^d(z) \right).
\end{equation*}
Thus, one can construct gauge-invariant quantities as integrals 
\begin{equation}\label{Eq:TwistInt}
\oint_\gamma \Ps(z)^{-d/h^\vee} c^d(z) \, \dd z
\end{equation}
over closed contours $\gamma$ on which $\Ps(z)^{-d/h^\vee} c^d(z)$ is regular. The gauge-invariance of these quantities can be seen as the fact that for any function $f(z)$, one has
\begin{equation}\label{Eq:TwistDerInt}
\oint_\gamma \Ps(z)^{-d/h^\vee} \left( \p_z f(z) - \frac{d\,\vp(z) }{h^\vee}f(z)) \right) \dd z = 0,
\end{equation}
\textit{i.e.} the integral of a $d$-twisted derivative pondered by the function $\Ps(z)^{-d/h^\vee}$ vanishes.

\paragraph{$\bm{\vp}$ with simple poles and hypergeometric integrals.} So far, we left an issue undiscussed. Indeed, as $\vp(z)$ is a rational function of $z$, the function $\Ps(z)$ is in general multi-valued and should be more rigorously defined on a multi-sheeted covering of $\CP$. We thus should restrict equation \eqref{Eq:TwistInt} to contours $\gamma$ on which there exists a single-valued branch of the function $\Ps(z)$.

We will now restrict to the case where $\vp(z)$ possesses simple poles at points $\zb=(z_1,\cdots,z_N)\in\C^N$ (it is thus of the form \eqref{Eq:TwistAGM}, as the twist function of an AGM with simple poles at $\zb$). We then consider the space
\begin{equation*}
\Oppz \subset \Opp{\Lang}
\end{equation*}
of $\Lang$-opers which are regular everywhere except at the $z_i$'s. For a function $\vp$ of the form \eqref{Eq:TwistAGM}, the function $\Ps$ is given formally by
\begin{equation*}
\Ps(z) = \prod_{i=1}^N (z-z_i)^{k_i}.
\end{equation*}
In this case, examples of contours $\gamma$ on which there exists a single-valued branch of this function are known. They are called the \textbf{Pochhammer contours}, which are closed contours winding in a particular way around the $z_i$'s. We will not enter into more details about these contours here and refer to the preprint~\cite{Lacroix:2018fhf}.\\

Let us summarise and combine the observations made in this subsection. Let $[\nabla]$ be an oper in $\Oppz$. We consider a quasi-canonical form of $[\nabla]$:
\begin{equation*}
\p_z + \pch_{-1} - \frac{\vp(z)}{h^\vee} \rho + \sum_{d\in E_+} c^d(z) \pch_d.
\end{equation*}
As $[\nabla]$ is regular on $\C\setminus\lbrace z_1,\cdots,z_N\rbrace$, the coefficients $c^d(z)$ can be chosen to be regular everywhere expect at the $z_i$'s. Let $\gamma$ be a Pochhammer contour and $d$ a positive exponent greater than one. The quantity
\begin{equation}\label{Eq:Idg}
I^d_\gamma \bigl( [\nabla] \bigr) = \oint_\gamma \Ps(z)^{-d/h^\vee} c^d(z) \, \dd z
\end{equation}
is well-defined, as $\Ps$ has a single-valued branch along $\gamma$ and $c^d$ is regular on $\gamma$. It is called an \textbf{hypergeometric integral}. Moreover, it is gauge-invariant and thus depends only on the oper $[\nabla]$.  We thus constructed a function
\begin{equation*}
I^d_\gamma : \Oppz \longrightarrow \C
\end{equation*}
on the space of opers $\Oppz$ \textit{via} an hypergeometric integral.

As the residual gauge transformation does not affect the coefficient $c^1(z)$, any quantity extracted from this coefficient (residue, integral, ...) also defines a function on $\Oppz$.

\section{Hierarchies of quantum AGM: some conjectures and a first result}
\label{Sec:AffineHam}

Based on the description of functions on $\Lang$-opers presented in the previous subsection, we present conjectures about hierarchies of quantum AGM and their spectrum. We support these conjectures by different observations and by presenting a first step towards their proofs, based on a second preprint~\cite{Lacroix:2018cag} of B. Vicedo, C.A.S. Young and myself. In this section, we then consider a quantum AGM with twist function \eqref{Eq:TwistAGM}.

\subsection{Higher-degrees Hamiltonians}

\begin{conjecture}\label{Conj:AffineHam}
There exist operators $S^d(z)$ of the Gaudin model, labelled by positive exponents $d\in E_+$ greater than one, which satisfy the following properties.
\begin{enumerate}[(i)]
\setlength\itemsep{0.1em}
\item $S^d(z)$ is of degree $d+1$ ;
\item For any $p,q\in E_+\setminus\lbrace 1 \rbrace$, we have
\begin{equation*}
\bigl[ S^p(z), S^q(w)  \bigr] = \bigl(h^\vee \p_z - p \vp(z) \bigr) A_{pq}(z,w) + \bigl(h^\vee \p_w - q \vp(w) \bigr) B_{pq}(z,w)
\end{equation*}
for some operators $A_{pq}(z,w)$ and $B_{pq}(z,w)$.
\item For any $p\in E_+\setminus\lbrace 1 \rbrace$, we have
\begin{equation*}
\bigl[ S^p(z), \Hs(w)  \bigr] = \bigl(h^\vee \p_z - p\vp(z) \bigr) A_{p1}(z,w)
\end{equation*}
for some operators $A_{p1}(z,w)$.
\end{enumerate}
\end{conjecture}

\noi If this conjecture holds, then for any Pochhammer contour $\gamma$, we define
\begin{equation}\label{Eq:HamAffine}
\Hs^d_\gamma = \oint_\gamma \Ps(z)^{-d/h^\vee} S^d(z) \, \dd z.
\end{equation}
Using the property \eqref{Eq:TwistDerInt} that the integral of a $d$-twisted derivative pondered by $\Ps(z)^{-d/h^\vee}$ vanishes, we would then prove that for any $p,q\in E_+\setminus\lbrace 1 \rbrace$ and any Pochhammer contours $\gamma$ and $\gamma'$,
\begin{equation*}
\bigl[ \Hs^p_\gamma, \Hs^q_{\gamma'} \bigr] = 0.
\end{equation*}
Similarly, we would get
\begin{equation*}
\bigl[ \Hs^p_\gamma, \Hs(z) \bigr] = 0
\end{equation*}
for all $z\in\C$. This way, we would have constructed an infinite number of commuting higher-degrees Hamiltonians, labelled by the positive exponents of $\g$ (which are the same as the ones of $\Lang$) and by Pochhammer contours. This would define a quantum hierarchy of the model.\\

In a second preprint~\cite{Lacroix:2018cag} with B. Vicedo and C.A.S. Young, we obtained a first step in the proof of Conjecture \ref{Conj:AffineHam}. Indeed, we constructed for untwisted affine algebras of type A, which possess $2$ as an exponent, the \textbf{cubic operator $\bm{S^2(z)}$}. This operator satisfies the properties (ii) and (iii) of Conjecture \ref{Conj:AffineHam}, ensuring the existence of cubic Hamiltonians $\Hs^2_\gamma$, which commute between themselves and with the quadratic Hamiltonian $\Hs(z)$. The construction of the operator $S^2(z)$ and the proof of the properties (ii) and (iii) are based on vertex algebras techniques. We will not discuss this further here and refer to the preprint~\cite{Lacroix:2018cag}.

\subsection{Spectrum of the affine Gaudin Hamiltonians}

We now consider the Hilbert space $\Hl$, as in Subsection \ref{SubSec:QuantAGM}. As the Hamiltonians $\Hs(z)$ and $\Hs^d_\gamma$ commute (assuming Conjecture \ref{Conj:AffineHam}), they can be diagonalised simultaneously on $\Hl$. We already know how to diagonalise the quadratic Hamiltonian $\Hs(z)$, using the Bethe ansatz.

The eigenvalue of $\Hs(z)$ on an on-shell Bethe vector $\Pcwo$ is given by equation \eqref{Eq:EigenAffine} in terms of the weight \eqref{Eq:LambdaAffine} associated with $\Pcwo$. From this weight, we define the $\Lang$-Miura oper:
\begin{equation*}
\nabla^{\bm c}_{\bm w} = \p_z + \pch_{-1} - \lambda^{\bm c}(z,\bm w),
\end{equation*}
in a similar way to the finite case (see Section \ref{Sec:FFR}). According to equation \eqref{Eq:LambdaTwist}, the coefficient of $\rho$ in $\nabla^{\bm c}_{\bm w}$ is
\begin{equation*}
c^0(z) = -\frac{\vp(z)}{h^\vee}.
\end{equation*}
Thus the oper $[\nabla^{\bm c}_{\bm w}]$ belongs to the space $\Opp{\Lang}$. Moreover, one shows (see~\cite{Lacroix:2018fhf}) that the Bethe equations for $\Pcwo$ implies that $[\nabla^{\bm c}_{\bm w}]$ is regular at all Bethe roots $w_j$'s (as in the finite case). The oper $[\nabla^{\bm c}_{\bm w}]$ then belongs to the space $\Oppz$. We consider a quasi-canonical form of this oper:
\begin{equation}\label{Eq:BetheQC}
[\nabla^{\bm c}_{\bm w}]_{\text{qc}} = \p_z + \pch_{-1} - \frac{\vp(z)}{h^\vee} + \sum_{d\in E_+} \Cc^d_{\bm c}(z,\bm w).
\end{equation}
One then shows (see also~\cite{Lacroix:2018fhf}) that the eigenvalue of $\Hs(z)$ on $\Pcwo$ coincides with $\Cc^1_{\bm c}(z,\bm w)$, up to a (model-independent) global factor. Thus, this eigenvalue can be read from the oper $[\nabla^{\bm c}_{\bm w}]$.\\

We now formulate another conjecture about the Bethe ansatz for higher-degree Hamiltonians and the description of their eigenvalues in terms of the oper $[\nabla^{\bm c}_{\bm w}]$.

\begin{conjecture}\label{Conj:Eigen}
Let $d$ be a positive exponent greater than $1$ and $\gamma$ be a Pochhammer contour. The on-shell Bethe state $\Pcwo$ in $\Hl$ is an eigenvector of $\Hs^d_\gamma$ and its eigenvalue is given by
\begin{equation}\label{Eq:HdGammaEigen}
I^d_\gamma \bigl( [\nabla^{\bm c}_{\bm w}] \bigr) = \oint_\gamma \Ps(z)^{-d/h^\vee} \Cc^d_{\bm c}(z,\bm w) \, \dd z.
\end{equation}
\end{conjecture}

This conjecture is an affine version of the FRR approach developed in Section \ref{Sec:FFR} for finite Gaudin models (or at least of the FFR reformulation of the Bethe ansatz). In our second preprint~\cite{Lacroix:2018cag}, we have checked this conjecture for the cubic Hamiltonians $\Hs^2_\gamma$ (see discussion at the end of previous subsection) and for a Bethe vector $\Pcwo$ with one excitation. We will not present this computation here.\\

As explained in Subsection \ref{Sec:AffineOper}, the exponent $1$ has a different behaviour than other positive exponents in the quasi-canonical form. We recover this fact in Conjectures \ref{Conj:AffineHam} and \ref{Conj:Eigen}. Indeed, there exists a quadratic Hamiltonian $\Hs(z)$ depending on the spectral parameter $z$ whereas the construction of the higher-degree Hamiltonians $\Hs^d_\gamma$ necessitates an integral \eqref{Eq:HamAffine} on the spectral parameter. The analogue of this fact in terms of opers is that the coefficient $\Cc_{\bm{c}}^1(z,\bm w)$ of the quasi-canonical form \eqref{Eq:BetheQC} is gauge-invariant whereas the coefficients $\Cc^d_{\bm c}(z,\bm w)$ ($d>1$) are not and give gauge-invariant quantities only after taking integrals \eqref{Eq:HdGammaEigen}.

\subsection{Classical limit}

We close this section by a brief discussion about the classical limit of this conjectured quantum hierarchy. To take this classical limit, one has to reintroduce the Planck constant $\hbar$ in the computation. One of the effect of this is to rescale the levels $k_i$'s of the model to $\dfrac{k_i}{\hbar}$. The expression \eqref{Eq:HamAffine} then becomes
\begin{equation}\label{Eq:ClassicalLimit}
\Hs^d_\gamma = \oint_\gamma \exp\left(-\frac{d \, \chi(z)}{\hbar} \right) S^d(z) \, \dd z,
\end{equation}
where $\chi(z)$ is a primitive of $\vp(z)/h^\vee$. The classical limit $\hbar \rightarrow 0$ of this integral is given by a saddle-point approximation (we refer to~\cite{Reshetikhin:1994qw} for the saddle-point approximation method for integrals over Pochhammer contours). In particular, the effect of this saddle point approximation is to localise the spectral parameter $z$ to extrema of $\chi(z)$. By definition, these extrema are exactly the \textbf{zeros of the twist function} $\bm\vp$. Moreover, for generic values of the sites $z_i$'s and the levels $k_i$'s there are as many zeros of $\vp$ as there are independent Pochhammer contours $\gamma$. One then expects that for any zero $x$ of $\vp$, there exists a Pochhammer contour $\gamma$ such that the classical limit of the operator \eqref{Eq:ClassicalLimit} is the evaluation $S^d(x)$.

The evaluation of the spectral parameter at zeros of the twist function is the method we used in~\cite{Lacroix:2017isl} (and Subsection \ref{SubSec:HierarchyAGM} of this thesis) to construct an infinite hierarchy for classical AGM. We thus conjecture that the evaluation $S^d(x)$ obtained above as a classical limit of the quantum Hamiltonian $\Hs^d_\gamma$ corresponds to the charge of degree $d+1$ in the classical hierarchy constructed by evaluation at the zero $x$.

As explained in the introduction of this chapter, the construction of the classical hierarchy is faced with the non-existence of invariant polynomials on $\g$ of degrees greater than 2: this difficulty is then overcome by the evaluation at zeros of the twist function. According to the discussion above, the quantum analogue of this procedure consists in taking integrals over Pochhammer contours pondered by powers of the function $\Ps(z)$. As there exists an invariant polynomial of degree two (the quadratic form $\fd$) and a corresponding quantum Casimir, one can construct a quadratic Hamiltonian $\Hs(z)$ depending on the spectral parameter, without needing to take such an integral. We recover here the difference between the exponent 1 and other positive exponents.

\cleardoublepage
\chapter{Conclusion and perspectives}
\label{Chap:Conclusion}

\section{The key role of the twist function}

The first part of this thesis concerned integrable field theories with \textbf{twist function}. It aimed to point out the fundamental role played by this function in the integrable structure of these theories. In particular, it exhibited how the general formalism based on this twist function allows the development of model-independent methods to study these field theories.

A key result in this direction is the content of Chapter \ref{Chap:LocalCharges}, which shows the importance of the \textbf{zeros} of the twist function. Indeed, under some regularity condition, one can associate with any of these zeros an infinity of conserved local charges in involution. Moreover, these charges form an integrable hierarchy, in the sense that their Hamiltonian flows generate compatible integrable systems on the same phase space.

Chapter \ref{Chap:PLie} focused on Yang-Baxter type deformations and revealed the role played by the \textbf{poles} of the twist function. Indeed, these deformations are characterised by the splitting of a double pole of the twist function into two simple poles. Studying the monodromy of the Lax matrix at these simple poles, we proved that, for all these models, one can extract conserved non-local charges forming a $q$-deformed Poisson-Hopf algebra. Moreover, we showed that these charges are associated with non-local Poisson-Lie deformed symmetries of these models.

Important examples of field theories with twist function are the integrable $\s$-models and their deformations. In particular, we were able to apply the results of Chapter \ref{Chap:PLie} to all Yang-Baxter type deformations of $\s$-models and even more generally, the results of Chapter \ref{Chap:LocalCharges} to all integrable $\s$-models, deformed or not. Doing so, we recovered some already know results for particular models and exhibited that these are actually part of more general, model-independent, constructions, which rely on the twist function and apply to a larger class of models.

Note also that in Chapter \ref{Chap:Models}, we showed that the Bi-Yang-Baxter model is a model with twist function and that it can be seen as the combination of two Yang-Baxter type deformations. Thus, this model enters in the framework of Chapters \ref{Chap:LocalCharges} and \ref{Chap:PLie}, showing that it admits an infinite local hierarchy and two deformed Poisson-Lie symmetries.

\section{Affine Gaudin models}

The second part of this thesis focused on Gaudin models and in particular \textbf{affine Gaudin models}. Indeed, the latter can classically be realised as integrable field theories with twist function. In this construction, the poles of the twist function are the sites of the Gaudin models. Moreover, as explained in Chapter \ref{Chap:GaudinClass}, all zeros of the twist function naturally satisfy the regularity condition necessary to apply the construction of local charges in involution of Chapter \ref{Chap:LocalCharges}. Thus, classical affine Gaudin models possess infinite integrable hierarchies.

One of the main subject addressed in this thesis is the quantisation of this hierarchy. More precisely, in Chapter \ref{Chap:QuantumAffine}, we conjectured that the quantum Hamiltonians in this hierarchy take the form of hypergeometric integrals (over contours determined by the poles of the twist function) and that their spectrum is described by hypergeometric functions on affine opers.  Moreover, we constructed the first higher-degree Hamiltonian in this hierarchy (the cubic one for algebras of type A) and verified that it agrees with the conjectures mentioned above.

These conjectures on quantum affine Gaudin models and their spectrum are motivated by an analogy with the situation for finite Gaudin models. Indeed, for these models, the higher-degrees Hamiltonians and their spectrum can be described in terms of finite opers by the Feigin-Frenkel-Reshetikhin approach. The above conjecture for affine Gaudin model is thus a conjecture about an affine version of this approach. In Chapter \ref{Chap:QuantumFinite}, we also lay the foundations for another generalisation of this approach, related to cyclotomic finite Gaudin models.

\section{Perspectives}
\label{Sec:Perspectives}

The work presented in this thesis can be pursued in various directions.

\paragraph{Integrable $\bm\s$-models and their deformations.} Although the Hamiltonian analysis of most deformed integrable $\s$-models has been carried out, it is still lacking for the multi-parameter deformation of the PCM introduced recently in~\cite{Delduc:2017fib}. It would be interesting to prove the Hamiltonian integrability of this model and determine its twist function. Another interesting question is the construction of deformations of $\Z_T$-coset $\s$-models for $T>2$, which are believed to exist but which have never been explicitly constructed.

\paragraph{FFR approach beyond the Bethe ansatz.} As explained in Section \ref{Sec:FFR}, the FFR approach gives a theoretical description of the entire spectrum of finite Gaudin models in terms of opers. In particular, it ensures the existence of such an oper for any eigenvector, including the ones which are not described by the Bethe ansatz (either because the Hilbert space is not a tensor product of highest-weight vectors or because the Bethe ansatz is not complete). It would be interesting to find explicit constructions of these opers, at least in particular examples.

\paragraph{Quantum hierarchy of affine Gaudin models.} A natural project in the continuity of this PhD is to prove the conjectures made in the preprint~\cite{Lacroix:2018fhf} (see Chapter \ref{Chap:QuantumAffine}) about the quantisation of the hierarchy of affine Gaudin models. More precisely, it would be interesting to construct all higher-degrees Hamiltonians of affine Gaudin models, for example using vertex algebra techniques, as done for the cubic charge in~\cite{Lacroix:2018cag} (see also Chapter \ref{Chap:QuantumAffine}). The next natural step would then be to describe the spectrum of this hierarchy, either through the Bethe ansatz or by proving the conjectured generalisation of the FFR approach for affine models.

The quantum hierarchy conjectured in Chapter \ref{Chap:QuantumAffine} concerns the simplest affine Gaudin models: complex, without cyclotomy and with only simple poles. If this hierarchy exists, it is also natural to search for its generalisation when removing these constraints, aiming for the construction of a quantum hierarchy for the most general dihedral affine Gaudin model with arbitrary multiplicities.

\paragraph{ODE/IM correspondence.} An affine FFR approach, as mentioned in the previous paragraph, would describe the spectrum of affine Gaudin models in terms of affine opers. There exists another result in the literature which relates the spectrum of some integrable field theories with affine opers, the so-called ODE/IM correspondence (initially proposed in~\cite{Dorey:1998pt} and~\cite{Bazhanov:1998wj}, see also the review~\cite{Dorey:2007zx}). More precisely, this correspondence relates the value of quantum Integrals of Motion (IM) with some Ordinary Differential Equations (ODE), encoded in the form of affine opers~\cite{Feigin:2007mr}. Examples of models for which an ODE/IM correspondence has been proposed are the quantum KdV theory~\cite{Bazhanov:1998wj,Bazhanov:2003ni}, the quantum Boussinesq theory~\cite{Bazhanov:2001xm}, affine Toda field theories~\cite{Lukyanov:2010rn,Dorey:2012bx,Adamopoulou:2014fca,Ito:2013aea,Ito:2015nla,Ito:2016qzt} and the Fateev model~\cite{Lukyanov:2013wra,Bazhanov:2013cua} (which can be seen as a deformed integrable $\s$-model~\cite{Fateev:1996ea}, more precisely the Bi-Yang-Baxter model on $SU(2)$~\cite{Hoare:2014pna}).

These models were realised as affine Gaudin models in~\cite{Feigin:2007mr} and~\cite{Vicedo:2017cge}. It was proposed in these same articles that the ODE/IM correspondence for these models originates from an affine generalisation of the FFR approach, as the latter would also involve affine opers. However, the usual approach of ODE/IM uses affine opers in a different way than the FFR approach. More precisely, it relates spectral determinants of the ODE (encoded in affine opers) with the eigenvalues of particular operators associated with integrable systems (as for example $Q$-operators). This relation is based on the observation that these spectral determinants and these eigenvalues satisfy the same functional relations, which encode the Bethe equations of the underlying quantum integrable model (see the above references and~\cite{Dorey:1999pv,Dorey:2000ma,Dorey:2006an,Masoero:2015lga,Masoero:2015rcz}).

It would be interesting to understand how these results, which use spectral determinants of affine opers, can be related with the FFR approach, which involves functions on affine opers. In the long-term, this could provide a proof of an ODE/IM correspondence for affine Gaudin models through the FFR theory. A particularly interesting observation concerning this is the fact~\cite{Bazhanov:2013cua} that the ODE/IM correspondence for the Fateev model involves hypergometric integrals over Pochhammer contours. This is in striking resemblance with the description of functions on affine opers that we presented in~\cite{Lacroix:2018fhf} (see Chapter \ref{Chap:QuantumAffine}).

It can be also interesting to note that functional relations like the ones appearing in the ODE/IM correspondence can be reinterpreted in a more abstract way as relations between representations of some algebraic objects. For example, the functional relations proved in~\cite{Masoero:2015lga,Masoero:2015rcz} have been recently related to representations of quantum affine Borel algebras in~\cite{Frenkel:2016gxg}.

\appendix

\part{Appendices}

\cleardoublepage
\chapter{Lie algebras}
\label{App:Lie}

\section{Generalities}
\label{App:LieGen}

In this section, we introduce a few basic facts and notations about Lie algebras. We will consider Lie algebras over a field $\Kb$, which we restrict to be the real numbers $\R$ or the complex numbers $\C$. A Lie algebra is then a $\Kb$-vector space $\g$, equipped with a skew-symmetric bilinear map:
\begin{equation*}
[\cdot,\cdot] : \g \times \g \longrightarrow \g,
\end{equation*} 
which satisfies the Jacobi identity
\begin{equation*}
\forall \; X,Y,Z \in \g, \;\;\;\;\; \bigl[ X, [Y,Z] \bigr] + \bigl[ Y, [Z,X] \bigr] + \bigl[ Z, [X,Y] \bigr] = 0.
\end{equation*}
We will assume that $\g$ is of finite dimension. Let us fix a basis $\lbrace I^a \rbrace$ ($a=1,\cdots,n$) of $\g$. The \textbf{structure constants} of $\g$ are defined by
\begin{equation*}
[I^a,I^b] = \fs{ab}{c}  I^c,
\end{equation*}
where a summation is implied on the repeated index $c$. The Jacobi identity is then equivalent to
\begin{equation*}
\fs{ae}{d}  \ft{bc}{e} + \fs{be}{d}  \ft{ca}{e} + \fs{ce}{d}  \ft{ab}{e} = 0.
\end{equation*}

\noi For $X$ in $\g$, we define its adjoint action as the following endomorphism of $\g$
\begin{equation*}
\begin{array}{rccc}
\ad_X : & \g & \longrightarrow & \g \\
        &  Y & \longmapsto     & [X,Y]
\end{array}.
\end{equation*}
By the Jacobi identity, the application $\ad : X \mapsto \ad_X$ is then a Lie algebra morphism from $\g$ to $\End(\g)$ (the algebra of endomorphisms of $\g$):
\begin{equation*}
\ad_{[X,Y]} = \bigl[ \ad_X, \ad_Y \bigr] = \ad_X \circ \ad_Y - \ad_Y \circ \ad_X.
\end{equation*}
We define the Killing form on $\g$ as the bilinear form
\begin{equation*}
\begin{array}{rccc}
\kappa : & \g \times \g & \longrightarrow & \Kb \\
         &  (X,Y)       & \longmapsto     & \Tr\bigl( \ad_X \circ \ad_Y \bigr)
\end{array}.
\end{equation*}
By cyclicity of the trace, it is clear that $\kappa$ is symmetric. The main property of $\kappa$ is its ad-invariance:
\begin{equation*}
\kappa \bigl( [X,Y], Z \bigr) = \kappa \bigl( X,[Y,Z] \bigr).
\end{equation*}
Writing the Killing form in the basis $\lbrace I^a \rbrace$ as
\begin{equation*}
\kappa^{ab} = \kappa\bigl(I^a,I^b\bigr),
\end{equation*}
we then have
\begin{equation*}
\kappa^{ab} = \fs{ac}{d}\ft{bd}{c}.
\end{equation*}
The ad-invariance equation then becomes
\begin{equation}\label{Eq:AdInvBasis}
\kappa^{ad}\,\ft{bc}{d}+\kappa^{db}\,\fs{ac}{d} = 0.
\end{equation}

\section{Semi-simple finite-dimensional complex Lie algebras}
\label{App:SemiSimple}

A Lie algebra $\g$ is said to be simple if it does not admit non trivial ideals and semi-simple if it is a direct sum of simple algebras. Semi-simple algebras are characterised by the Cartan criteria:

\begin{theorem}\label{Thm:CartanCriteria}
$\g$ is semi-simple if and only if its Killing form $\kappa$ is non-degenerate.
\end{theorem}

The semi-simple complex Lie algebras of finite dimension have been classified by Cartan using the notions of Cartan subalgebra and root system. Although it will be an important formalism for this thesis, we will not derive this classification here as it is a classical result (see for example~\cite{Humphreys:1980dw}). We will simply summarise its main results and introduce the notations and conventions we use in this thesis.

\subsection{Cartan-Weyl basis}
\label{App:CartanWeyl}

\paragraph{Cartan-Weyl basis.} The fundamental objects describing a semi-simple complex Lie algebra $\g$ are the following:
\begin{itemize}
\item $\h$ a Cartan subalgebra of $\g$, of dimension $\ell$ the rank of $\g$,
\item $\Delta \subset \h^*$ and $\Delta^\vee \subset \h$ the associated sets of roots and coroots,
\item $\g=\h\oplus\n_+\oplus\n_-$ the Cartan-Weyl decomposition,
\item $\lbrace E_\alpha, \alpha >0 \rbrace$ and $\lbrace E_{-\alpha}, \alpha >0 \rbrace$ the associated bases of the nilpotent subalgebras $\n_\pm$,
\item $\alpha_1,\cdots,\alpha_\ell \in \Delta$ the simple roots and $\ach_1,\cdots,\ach_\ell \in \Delta^\vee$ the associated coroots.\vspace{4pt}
\end{itemize}
We consider the normalisation of the $E_\alpha$'s such that
\begin{equation}\label{Eq:KillingE}
\kappa\left(E_\alpha,E_\beta\right)=\delta_{\alpha,-\beta},
\end{equation}
For $\alpha\in\Delta$, we define $H_\alpha \in \h$ \textit{via} the Killing form isomorphism between $\h$ and $\h^*$:
\begin{equation*}
\forall X\in\h, \; \kappa(H_\alpha,X) = \alpha(X).
\end{equation*}
In particular, for simple roots, let us denote $H_i=H_{\alpha_i}$. Then, $\lbrace H_i, \, i=1,\cdots,\ell \rbrace$ is a basis of $\h$. The element $H_\alpha$ is related to the coroot $\ach$ associated with $\alpha$ by the relation:
\begin{equation*}
\ach = \frac{2 H_\alpha}{\kappa(H_\alpha,H_\alpha)}.
\end{equation*}
We will denote by $\Delta_\pm$ the set of positive and negative roots. For $\alpha\in\Delta_+$, we will sometimes use the notation
\begin{equation*}
F_\alpha = E_{-\alpha}.
\end{equation*}

\paragraph{Commutation relations.} The element $E_\alpha$ is characterised by the commutation relation
\begin{equation*}
\forall X\in\h, \; \left[X, E_\alpha \right] = \alpha(X) E_\alpha. 
\end{equation*}
Moreover, we have
\begin{equation}\label{ComEF}
\left[E_{\alpha}, E_{-\alpha}\right]=H_\alpha \; \; \; \; \text{and} \; \; \; \; \left[E_{\alpha_i},E_{-\alpha_j}\right] = \delta_{ij}H_i.
\end{equation}
Finally, the structure of $\n_\pm$ is given by:
\begin{equation}\label{ComEE}
\left[ E_\alpha, E_\beta \right] = N_{\alpha,\beta} E_{\alpha+\beta},
\end{equation}
with $N_{\alpha,\beta}$ a real skew-symmetric normalisation constant. Moreover, one has $N_{-\alpha,-\beta}=-N_{\alpha,\beta}$. For $i\in\lbrace 1,\cdots,\ell \rbrace$, we define the so-called Chevalley generators
\begin{equation*}
E_i = E_{\alpha_i} \;\;\;\; \text{ and } \;\;\;\; F_i = F_{\alpha_i} = E_{-\alpha_i}.
\end{equation*}
Then, the $E_i$'s and $F_i$'s generate the whole algebra $\g$.

\subsection{Simple roots, coroots, weights and coweights.}
\label{App:BasesCartan}

In this appendix, we recall the main properties of some bases of the Cartan subalgebra $\h$. As it is equipped with the non-degenerate Killing form $\kappa(\cdot,\cdot)$, there exists a natural isomorphism
\begin{equation*}
\zeta : \h^* \rightarrow \h,
\end{equation*}
between $\h$ and its dual $\h^*$. It is characterised by the relation
\begin{equation*}
\forall \, \lambda \in \h^*, \; \; \forall \, X \in \h,  \; \; \kappa\bigl( \zeta(\lambda),X \bigr) = \lambda(X).
\end{equation*}
By this isomorphism, one induces a bilinear form $(\cdot,\cdot)$ on $\h^*$:
\begin{equation*}
\forall	 \, \lambda,\mu \in \h^*, \; \; ( \lambda, \mu ) = \kappa\bigl( \zeta(\lambda), \zeta(\mu) \bigr).
\end{equation*}

\paragraph{Roots and co-roots.} A basis of $\h^*$ is given by the simple roots $\alpha_1,\cdots,\alpha_\ell$. Using the notations of Subsection \ref{App:CartanWeyl}, the corresponding basis $\lbrace \zeta(\alpha_i) \rbrace$ of $\h$ is $\lbrace H_1,\cdots,H_\ell \rbrace$. We will often use
\begin{equation*}
d_i = \frac{(\alpha_i,\alpha_i)}{2} = \frac{\kappa(H_i,H_i)}{2}.
\end{equation*}
The simple co-roots are given by
\begin{equation*}
\ach_i = \frac{2H_i}{(\alpha_i,\alpha_i)}=d_i^{-1} H_i,
\end{equation*}
which form another basis of $\h$. The Cartan matrix $A=\bigl(a_{ij}\bigr)_{i,j=1,\cdots,\ell}$ is then given by
\begin{equation}\label{Eq:CartanMat}
a_{ij}=\alpha_j(\ach_i)=\frac{2(\alpha_i,\alpha_j)}{(\alpha_i,\alpha_i)}.
\end{equation}
Let us introduce the diagonal matrix $D=\diag(d_1,\cdots,d_\ell)$. Then one has $A=DB$ with $B=\bigl(b_{ij}\bigr)_{i,j=1,\cdots,\ell}$ the symmetric matrix defined by
\begin{equation*}
b_{ij} = (\alpha_i, \alpha_j) = \kappa(H_i,H_j).
\end{equation*}

\paragraph{Fundamental weights and co-weights.} We define the fundamental weights $\omega_i \in \h^*$ as the dual basis of the co-roots $\ach_i$:
\begin{equation*}
\omega_i(\ach_i)=\delta_{ij}.
\end{equation*}
By the Killing form duality, the
\begin{equation*}
P_i=\zeta(\omega_i)
\end{equation*}
form a basis of $\h$. Moreover, we have the relation
\begin{equation}\label{AlphaP}
\alpha_i(P_j) = d_i \delta_{ij}.
\end{equation} 
In the same way, one defines the fundamental co-weights $\och_i \in \h$ as the dual basis of the simple roots:
\begin{equation*}
\alpha_j(\och_i)=\delta_{ij},
\end{equation*}
which simply relates them to the weights by
\begin{equation*}
\och_i=d_i^{-1}P_i.
\end{equation*}
The scalar products between coweights are given by
\begin{equation*}
\kappa(\och_i,\och_j) = m_{ij}, \;\;\;\; \text{ where } \;\;\;\; M = \bigl(m_{ij}\bigr)_{i,j=1,\cdots,\ell} = B^{-1},
\end{equation*}
with the matrix $B$ as defined in the previous paragraph.

\subsection{Classification}

A complex semi-simple Lie algebra is uniquely determined, up to isomorphisms, by its Cartan matrix \eqref{Eq:CartanMat}. One can show that this matrix satisfies the following conditions
\begin{enumerate}[(a)]
\setlength\itemsep{0.1em}
\item $a_{ii}=2$ for all $i\in\lbrace 1,\cdots,\ell\rbrace$ ;
\item $a_{ij} \in - \mathbb{N}$ for $i\neq j$ ;
\item if $a_{ij}=0$, then $a_{ji}=0$ ;
\item $A=DB$ with $D$ diagonal and $B$ symmetric definite positive.
\end{enumerate}

Conversely, given any matrix $A$ satisfying these conditions, there exists a semi-simple complex Lie algebra with Cartan matrix $A$. The classification of semi-simple complex Lie algebras is then reduced to the classification of such matrices.

This classification is a classic result (see for example~\cite{Humphreys:1980dw}), that we shall not describe in detail here. Semi-simple complex Lie algebras are divided in seven different types, from A to G. Elements of type $X$ are denoted $X_\ell$, where $\ell$ is the size of the corresponding Cartan matrix (or equivalently the rank of the corresponding Lie algebra). The types A, B, C and D are infinite families corresponding to classical Lie algebras:\vspace{-2pt}
\begin{itemize}
\setlength\itemsep{0.5em}
\item $A_\ell$ is the special linear algebra $\sl(\ell+1,\C) = \bigl\lbrace M\in M_{\ell+1}(\C) \; | \; \Tr(M)=0 \bigr\rbrace$,
\item $B_\ell$ is the orthogonal algebra $\so(2\ell+1,\C) = \bigl\lbrace M\in M_{2\ell+1}(\C) \; | \; \null^t M + M = 0 \bigr\rbrace$,
\item $C_\ell$ is the symplectic algebra $\mathfrak{sp}(2\ell,\C) = \bigl\lbrace M\in M_{2\ell}(\C) \; | \; \null^t M J + J M = 0 \bigr\rbrace$ with $J = \begin{pmatrix}
0 & I_n \\
-I_n & 0
\end{pmatrix}$,
\item $D_\ell$ is the orthogonal algebra $\so(2\ell,\C) = \bigl\lbrace M\in M_{2\ell}(\C) \; | \; \null^t M + M = 0 \bigr\rbrace$.
\end{itemize}
At the contrary, the so-called exceptional types E, F and G are finite families. More precisely, there are five exceptional Lie algebras: $E_6$, $E_7$, $E_8$, $F_4$ and $G_2$.

\subsection{Split quadratic Casimir}
\label{App:Casimir}

Let us fix a basis $\lbrace I^a \rbrace$ of $\g$. Recall the Killing form $\kappa^{ab}$ expressed in this basis. As $\g$ is semi-simple, it is non-degenerate (see Theorem \ref{Thm:CartanCriteria}). We can then define the inverse $\kappa_{ab}$ of the Killing form, satisfying $\kappa^{ac}\kappa_{cb} = \delta^a_{\;\, b}$. In this paragraph, we use the tensorial notations $\bm{\underline{i}}$ defined in Section \ref{Sec:Ham}. We define the split quadratic Casimir of $\g$ as
\begin{equation*}
C\ti{12} = \kappa_{ab} \; I^a \otimes I^b.
\end{equation*}
One checks that it does not depend on the choice of basis. The ad-invariance equation \eqref{Eq:AdInvBasis} translates to
\begin{equation}\label{Eq:CasIdentity}
\left[ C\ti{12}, X\ti{1} + X\ti{2} \right] = 0, \;\;\;\; \forall \; X \in \g.
\end{equation}
Moreover, using the fact that $\kappa_{ab}$ is the inverse of $\kappa^{ab}$, we get the following completeness relation:
\begin{equation}\label{Eq:CasComp}
\kappa\ti{2} \left( C\ti{12}, X\ti{2} \right) = X, \;\;\;\; \forall \; X \in \g.
\end{equation}
In the Cartan-Weyl basis described in Subsection \ref{App:CartanWeyl}, we get
\begin{equation*}
C\ti{12} = H\ti{12} + \sum_{\alpha\in\Delta} E_\alpha \otimes E_{-\alpha} = H\ti{12} + \sum_{\alpha\in\Delta_+} \bigl( E_\alpha \otimes F_\alpha + F_\alpha \otimes E_\alpha \bigr),
\end{equation*}
where $H\ti{12}$ belongs to $\h\otimes\h$. More precisely, $H\ti{12}$ can be expressed in various way in terms of the bases of $\h$ described in Subsection \ref{App:BasesCartan}. In particular, we have
\begin{equation*}
H\ti{12} = \sum_{i,j=1}^\ell m_{ij} H_i \otimes H_j = \sum_{i=1}^\ell P_i \otimes \ach_i.
\end{equation*}

\section{Real forms}
\label{App:RealForms}

\subsection{Generalities}

Let us consider a complex Lie algebra $\g$ of dimension $n$, with basis $\lbrace I^a, \, a=1,\cdots,n \rbrace$. It can be seen as a real Lie algebra of dimension $2n$, with basis $\lbrace I^a, \, a=1,\cdots,n \rbrace \sqcup \lbrace i \, I^a, \, a=1,\cdots,n \rbrace$. We will call this real algebra the realification $\g_\R$ of $\g$. A \textbf{real form} of $\g$ is then a (real) subalgebra of $\g_\R$ of dimension $n$.

In other words, a real form amounts to the choice of a basis $\lbrace I^a, \, a=1,\cdots,n \rbrace$ of $\g$ whose structure constants are real. The real form is then $\g_0 = \Span_\R \bigl( I^a,\, a=1,\cdots,\ell \bigr)$. In this case, we recover $\g$ as the complexification of $\g_0$:
\begin{equation*}
\g = \g_0^\C = \g_0 \otimes_\R \C = \g_0 \oplus i\,\g_0.
\end{equation*}
Let us then define
\begin{equation*}
\begin{array}{rccc}
\tau : & \g = \g_0 \otimes_\R \C & \longrightarrow & \g  = \g_0 \otimes_\R \C\\
       &    X \otimes z       & \longmapsto     & X \otimes \bar{z}
\end{array}.
\end{equation*}
One easily checks the following properties of $\tau$:
\begin{enumerate}[(a)]
\setlength\itemsep{0.1em}
\item $\tau$ is antilinear, \textit{i.e.} $\tau(\lambda X + \mu Y) = \bar{\lambda}\, \tau(X) + \bar{\mu}\, \tau(Y)$ for all $X,Y\in\g$ and $\lambda,\mu\in\C$ ;
\item $\tau$ is an involution, \textit{i.e.} $\tau^2 = \Id$ ;
\item $\tau$ is a Lie algebra automorphism, \textit{i.e.} $\tau\bigl([X,Y]\bigr) = \bigl[ \tau(X),\tau(Y) \bigr]$ ;
\item the subalgebra of fixed points of $\tau$ is the real form $\g_0 = \g^\tau$.
\end{enumerate}
Conversely, such a antilinear involutive automorphism of $\g$ defines a real form of $\g$:

\begin{theorem}\label{Thm:RealForms}
The real forms of $\g$ are in one-to-one correspondence with the antilinear involutive automorphisms of $\g$ (as their subalgebra of fixed points).
\end{theorem}

A natural question at this point is whether two real forms of $\g$ can be isomorphic and whether we can classify the isomorphism equivalence classes of real forms of $\g$. The answer is yes:

\begin{proposition}\label{Prop:RealFormsIso}
Let $\tau$ and $\theta$ be two antilinear involutive automorphisms of $\g$. Then the two real forms $\g^\tau$ and $\g^\theta$ are isomorphic if and only if there exists an automorphism $\s\in\Aut(\g)$ of $\g$ such that $\tau=\s\circ\theta\circ\s^{-1}$.
\end{proposition}

\subsection{Real semi-simple algebras}
\label{App:RealSS}

\begin{proposition}
A real Lie algebra is semi-simple if and only if it is the real form of a semi-simple complex Lie algebra.
\end{proposition}

Combining this proposition with Theorem \ref{Thm:RealForms} and Proposition \ref{Prop:RealFormsIso}, we see that the classification of real semi-simple Lie algebras reduces to the classification of complex ones $\g$ and of their antilinear involutive automorphisms, up to conjugacy in $\Aut(\g)$. The classification of complex semi-simple Lie algebras was described in Appendix \ref{App:SemiSimple}, using the notions of root system and Cartan-Weyl basis. Their antilinear involutive automorphisms, up to conjugacy in $\Aut(\g)$, have also been entirely classified~\cite{Frappat:2000}.

We will not describe here this whole classification. More precisely, we will restrict ourselves to the so-called split and non-split real forms, which exist for all complex Lie algebra $\g$. We consider the Cartan-Weyl basis $\lbrace H_i, E_\alpha \rbrace$ of $\g$ (cf. appendix \ref{App:SemiSimple}). As the $E_{\pm\alpha_i}$'s form generators of $\g$, an automorphism $\tau$ of $\g$ is completely described by its action on it.

\paragraph{Split real form.} Let us consider the semi-linear involutive automorphism $\tau$ given by:
\begin{equation*}
\tau \left(E_{\pm\alpha_i} \right) = E_{\pm\alpha_i}.
\end{equation*}
As the normalisation constants $N_{\alpha,\beta}$ in equation \eqref{ComEE} are real, one gets for any root $\alpha$:
\begin{equation}\label{Eq:SplitE}
\tau \left( E_\alpha \right) = E_{\alpha}.
\end{equation}
Hence, one also has:
\begin{equation}\label{Eq:SplitH}
\tau \left( H_\alpha \right) = H_\alpha.
\end{equation}
As a consequence, a basis of the real subalgebra fixed by $\tau$ is given by the Cartan-Weyl basis $\lbrace H_i, E_\alpha \rbrace$ of $\g$ itself. This way, we obtain the so-called split real form:
\begin{equation}\label{BasisSplit}
\g_0 = \left( \bigoplus_{i=1}^\ell \R \, H_i \right) \oplus \left( \bigoplus_{\alpha\in\Delta} \R \, E_\alpha \right).
\end{equation}
For example, for the simple algebra of type A, $\g=\sl(n,\C)$, the split real form is simply $\g_0=\sl(n,\R)$.

\paragraph{Non-split real forms.} Another possibility for defining $\tau$ is:
\begin{equation*}
\tau \left(E_{\pm\alpha_i} \right) = -\lambda_i E_{\mp\alpha_i},
\end{equation*}
where $\lambda_i=\pm 1$. Using equation \eqref{ComEE}, one obtains for any positive root $\alpha=p_1\alpha_1+\cdots+p_\ell\alpha_\ell$:
\begin{equation}\label{Eq:NonSplitE}
\tau \left(E_\alpha \right) = -\lambda_{\alpha} E_{-\alpha},
\end{equation}
with $\lambda_\alpha= \lambda_1^{p_1}\cdots\lambda_\ell^{p_\ell} \, \in \lbrace +1,-1 \rbrace$. In the same way, using \eqref{ComEF}, one has:
\begin{equation}\label{Eq:NonSplitH}
\tau(H_\alpha)=- H_\alpha.
\end{equation}
The real subalgebra $\g_0$ of elements fixed by $\tau$ is called a non-split real form of $\g$. A basis of $\g_0$ is given by:
\begin{equation}\label{BasisNonSplit}
T_i=iH_i, \; \; \; B_\alpha = \frac{i}{\sqrt{2}} \left(E_\alpha+\lambda_\alpha E_{-\alpha} \right), \; \; \; C_\alpha = \frac{1}{\sqrt{2}} \left(E_\alpha-\lambda_\alpha E_{-\alpha} \right).
\end{equation}
If one chooses $\lambda_1=\cdots=\lambda_\ell=1$, then $\lambda_\alpha=1$ for any root $\alpha$ and we get the so-called compact real form of $\g$.

Let us discuss a few examples of non-split real forms. For a simple algebra of type A, $\g=\sl(n,\C)$, the non-split real forms are the $\su(p,q,\R)$ (unitary algebra for a metric of signature $(p,q)$, with $n=p+q$). In particular, the compact real form is the unitary algebra $\su(n,\R)$. In the same way, for simple algebras of type B and D, $\g=\so(n,\C)$, the non-split real forms are the $\so(p,q,\R)$ (orthogonal algebra for a metric of signature $(p,q)$, with $n=p+q$). The compact real form is then $\so(n,\R)$.

\section{Finite order automorphisms of Lie algebras}
\label{App:Torsion}

\paragraph{Generalities.} Let us consider a complex Lie algebra $\g$ and $\s$ an automorphism of $\g$ of finite order $T\in\Z_{\geq 1}$. We define
\begin{equation*}
\omega=\exp\left(\frac{2i\pi}{T}\right).
\end{equation*}
We define the eigenspaces of $\s$:
\begin{equation*}
\g^{(p)} = \left\lbrace x\in\g \; | \; \s(x)=\omega^p x \right\rbrace, \;\;\;\; p=0,\cdots,T-1.
\end{equation*}
These eigenspaces form a $\Z_T$-gradation of $\g$ (where $\Z_T=\Z/T\Z$ is the cyclic group of order $T$):
\begin{equation}\label{Eq:TGradation}
\g = \bigoplus_{p=0}^{T-1} \g^{(p)}, \;\;\;\;\; \text{ with } \;\;\;\;\; [\g^{(p)},\g^{(q)}]=\g^{(p+q \; \text{mod} \; T)}.
\end{equation}
Conversely, a $\Z_T$-gradation \eqref{Eq:TGradation} defines a unique automorphism $\s$ of order $T$ (we define $\s$ as acting as the multiplication by $\omega^p$ on $\g^{(p)}$). In particular, $\g^{(0)}$ is a subalgebra of $\g$ and, for any $p\in \lbrace 0,\cdots,T-1 \rbrace$, $\g^{(p)}$ is a $\g^{(0)}$-module.\\

We define $\pi^{(p)}$ to be the projection on $\g^{(p)}$ in the decomposition \eqref{Eq:TGradation}. These projectors then satisfy $\pi^{(p)}\pi^{(q)} =  \delta_{p,q} \pi^{(p)}$. They can be expressed in terms of $\s$ as
\begin{equation}\label{Eq:ProjectorP}
\pi^{(p)} = \frac{1}{T} \sum_{k=0}^{T-1} \omega^{-kp} \s^k.
\end{equation}

\paragraph{Inner automorphisms.} Let $G$ be a connected Lie group with Lie algebra $\g$. Any element $g\in G$ acts on $\g$ by the adjoint (conjugacy) action:
\begin{equation*}
\begin{array}{rccc}
\Ad_g : & \g & \longrightarrow & \g \\
        & X  & \longmapsto     & gXg^{-1}
\end{array}.
\end{equation*}
This is an automorphism of the Lie algebra $\g$, called an inner automorphism. We will denote by $\Inn(\g)$ the set of such automorphism: it is a subgroup of $\Aut(\g)$ and moreover, the map
\begin{equation*}
\begin{array}{rccc}
\Ad : & G & \longrightarrow & \Inn(\g) \\
      & g & \longmapsto     & \Ad_g
\end{array}
\end{equation*}
is a Lie group morphism. It induces an automorphism of Lie algebras
\begin{equation*}
\begin{array}{rccc}
\ad : & \g & \longrightarrow & \ad(\g) \\
      & X & \longmapsto     & \ad_X
\end{array},
\end{equation*}
where
\begin{equation*}
\begin{array}{rccc}
\ad_X : & \g & \longrightarrow & \g \\
        & Y  & \longmapsto     & [X,Y]
\end{array}.
\end{equation*}
We recall in particular that for all $X\in\g$, we have
\begin{equation*}
\Ad_{\exp(X)} = \exp \left( \ad_X \right).
\end{equation*}

Let us now suppose that $\g$ is semi-simple. We will use the notations of Appendix \ref{App:CartanWeyl}. If $X$ is an element of $\h$ and $z$ is a complex number, we define the group element
\begin{equation*}
z^X = \exp \left( \log(z) X \right),
\end{equation*}
where $\log$ is a determination of the logarithm on $\C$. We can then define $\Ad_{z^X} \in \Inn(\g)$, which acts on the basis $\lbrace H_i, E_\alpha \rbrace$ of $\g$ as
\begin{equation*}
\Ad_{z^X} (H_i) = H_i \;\;\;\; \text{ and } \;\;\;\; \Ad_{z^X} (E_\alpha) = z^{\alpha(X)} E_\alpha.
\end{equation*} 
We will be particularly interested in the case where $z=\omega=\exp\left( \frac{2i\pi}{T} \right)$ and $X$ is such that $\alpha(X)\in\Z$ for all $\alpha\in\Delta$, as the automorphism $\Ad_{\omega^X}$ is then of order $T$. Recall the coweights $\lwb \och_i, i=1,\cdots,\ell \rwb$ defined in Subsection \ref{App:BasesCartan}. The space of $X\in\h$ such as above is then the lattice of these coweights:
\begin{equation}\label{Eq:WeightLattice}
\Lambda(\h) = \lwb X \in \h \; \bigl| \; \alpha(X)\in\Z, \, \forall \alpha\in\Delta \rwb = \lwb \sum_{i=1}^\ell m_i \och_i, \; m_i \in \Z \rwb.
\end{equation}

\paragraph{Diagram automorphisms of semi-simple Lie algebras.} We still suppose that $\g$ is a finite dimensional semi-simple Lie algebra. Recall the Cartan matrix $A$ of $\g$, defined in \eqref{Eq:CartanMat}. We say that a permutation $\mu$ of $\lbrace 1,\cdots,\ell \rbrace$ is a diagram automorphism of $A$ if
\begin{equation*}
a_{\mu(i) \mu(j)} = a_{ij}, \;\;\; \forall \; i,j \in \lbrace 1,\cdots,\ell \rbrace.
\end{equation*}
Recall also the Chevalley generators $C=(E_i,F_i)_{i=1,\cdots,\ell}$ of $\g$. As they generate the algebra $\g$, an automorphism of $\g$ is entirely defined by its action on the $E_i$'s and $F_i$'s. One can show that there exists a unique automorphism $\mu_C$ of $\g$ such that
\begin{equation*}
\mu_C(E_i) = E_{\mu(i)} \;\;\;\; \text{ and } \;\;\;\; \mu_C(F_i) = F_{\mu(i)}.
\end{equation*}
$\mu_C$ is then called an outer automorphism of $\g$. It has the same order than the permutation $\mu$ of $\lbrace 1,\cdots,\ell \rbrace$: in particular, it is of finite order. Note here that the definition of $\mu_C$  from the permutation $\mu$ depends on the choice of the Chevalley generators $E_i$'s and $F_i$'s. Two choices $C=(E_i,F_i)_{i=1,\cdots,\ell}$ and $C'=(E'_i,F'_i)_{i=1,\cdots,\ell}$ of such generators are always related by an inner automorphism $\gamma\in\Inn(\g)$. The corresponding automorphisms $\mu_C$ and $\mu_{C'}$ of $\g$ are then related by conjugacy by $\gamma$, \textit{i.e.} $\mu'=\gamma\circ\mu\circ\gamma^{-1}$.

A choice of Chevalley generators $C=(E_i,F_i)_{i=1,\cdots,\ell}$ also determines a Cartan subalgebra $\h$ by specifying the element $H_i$'s (see Appendix \ref{App:CartanWeyl}), as one have $[E_i,F_i]=H_i$. The action of $\mu_C$ on the $H_i$'s is simply
\begin{equation*}
\mu_C(H_i) = H_{\mu(i)}.
\end{equation*}

\paragraph{Classification of finite order automorphisms of semi-simple Lie algebras.} The automorphisms of finite order of $\g$ are classified by the following theorem~\cite{Kac:1990gs}.

\begin{theorem}
Let $\s$ be an automorphism of order $T$ of $\g$. Then there exist a diagram automorphism $\mu$, a choice of Chevalley generators $C=(E_i,F_i)_{i=1,\cdots,\ell}$ (hence also a choice of Cartan subalgebra $\h$) and an element $X\in\Lambda(\h)$ (see previous paragraph), such that
\begin{equation*}
\s = \mu_C \circ \Ad_{\omega^X} = \Ad_{\omega^X} \circ \mu_C.
\end{equation*}
Moreover, one has $\mu_C(X)=X$. If we write $X = \sum_{i=1}^\ell m_i \och_i$ (see equation \eqref{Eq:WeightLattice}), this is equivalent to $m_i=m_{\mu(i)}$ for all $i\in\lwb 1,\cdots,\ell \rwb$. The action of $\s$ on the Chevalley generators is then given by
\begin{equation*}
\s(E_i) = \omega^{m_i} E_{\mu(i)} \;\;\;\; \text{ and } \;\;\;\; \s(F_i) = \omega^{-m_i} F_{\mu(i)}.
\end{equation*}
\end{theorem}

\paragraph{$\bm{\Z_T}$-gradings of real Lie algebras.} In the first paragraph of this section, we have seen that for a complex Lie algebra $\g$, there is a one-to-one correspondence between $\Z_T$-gradings of $\g$ and automorphisms of $\g$ of finite order $T$. We will now discuss the $\Z_T$-gradings of real Lie algebras. The following lemma, whose demonstration is straightforward, relates them to the $\Z_T$-gradings of complex Lie algebras (recall the complexification of an algebra defined in Appendix \ref{App:RealForms}).

\begin{lemma}\label{Lem:RealGradings}
Let $\g_0$ be a real Lie algebra and $\g=\g_0 \otimes \C$ be its complexification. If $\g_0$ is equipped with a $\Z_T$-grading $\g_0=\bigoplus_{p=0}^{T-1} \g_0^{(p)}$, then
\begin{equation*}
\g = \bigoplus_{p=0}^{T-1} \g^{(p)}, \;\;\;\;\; \g^{(p)} = \g^{(p)}_0 \otimes \C
\end{equation*}
is a $\Z_T$-grading of $\g$ and $\g^{(p)}_0 = \g^{(p)} \cap \g_0$.
Conversely, if $\g$ is equipped with a $\Z_T$-grading $\g=\bigoplus_{p=0}^{T-1} \g^{(p)}$ such that $\g_0$ is a graded subalgebra of $\g$, \textit{i.e.}
\begin{equation*}
\g_0 = \bigoplus_{p=0}^{T-1} \g^{(p)}\cap \g_0,
\end{equation*}
then the $\g^{(p)}_0 = \g^{(p)}\cap \g_0$'s define a $\Z_T$-grading of $\g$.
\end{lemma}

According to the lemma, classifying all $\Z_T$ gradings of $\g_0$ is equivalent to classifying the $\Z_T$-gradings of the complexification $\g$ such that $\g_0$ is a graded subalgebra of $\g$. Recall that $\g_0$ can be seen as the subalgbera $\g^\tau$ of fixed point of an antilinear involutive automorphism $\tau$ of $\g$ and that the $\Z_T$-gradings of $\g$ are in one-to-one correspondence with automorphisms of order $T$ of $\g$ (see first paragraph of this section).

\begin{theorem}\label{Thm:Dihedrality}
Let $\g$ be a complex Lie algebra. We suppose that it possesses an antilinear involutive automorphism $\tau$ and a $\Z_T$-grading $\g=\bigoplus_{p=0}^{T-1} \g^{(p)}$, associated with an automorphism $\s$. Then the following points are equivalent:
\begin{enumerate}[(i)]
\setlength\itemsep{0.1em}
\item $\g^\tau$ is a graded subalgebra of $\g$,
\item $\tau\circ\s=\s^{-1}\circ\tau$ (dihedrality condition),
\item $\tau$ stabilises the eigenspaces $\g^{(p)}$ of $\s$,
\item $\tau$ commutes with the projections $\pi^{(p)}$ on $\g^{(p)}$,
\item there exists a common basis of eigenvectors of $\s$ and $\tau$.\vspace{0pt}
\end{enumerate}
\end{theorem}

\begin{corollary}\label{Cor:RealGradings}
Let $\g_0$ be a real Lie algebra. We denote by $\g$ its complexificaton and by $\tau$ the antilinear involutive automorphism of $\g$ such that $\g_0=\g^\tau$. Then the $\Z_T$-gradings of $\g_0$ are in one-to-one correspondence with the automorphisms $\s$ of order $T$ satisfying the dihedrality condition $\tau \circ \s = \s^{-1} \circ \tau$.
\end{corollary}

\begin{proof}
The corollary is straightforward from the theorem and lemma \ref{Lem:RealGradings}. Let us then prove the theorem. We will show the following cyclic sequence of implications: (i)$\Rightarrow$(v)$\Rightarrow$(ii)$\Rightarrow$(iv)$\Rightarrow$(iii)$\Rightarrow$(i).\\

\noi \underline{(i)$\Rightarrow$(v):} Let us suppose that $\g^\tau$ is graded, \textit{i.e.} that
\begin{equation*}
\g^\tau = \bigoplus_{p=0}^{T-1} \g^\tau \cap \g^{(p)}.
\end{equation*}
For $p\in\lbrace 0,\cdots,T-1 \rbrace$, let $X^{(p)}_1, \cdots, X^{(p)}_{m_p}$ be a basis of $\g^\tau \cap \g^{(p)}$, with $m_p=\dim \left( \g^\tau \cap \g^{(p)} \right)$. As the $X^{(p)}_k$'s are in $\g^{(p)}$, they are eigenvectors of $\s$ (of eigenvalue $\omega^p$) and as they are in $\g^\tau$, they are eigenvectors of $\tau$ (of eigenvalue $1$). In the same way, $i X^{(p)}_k$ is also an eigenvector of $\s$ (of eigenvalue $\omega^p$) and of $\tau$ (of eigenvalue $-1$). the family $B=\bigsqcup_{p=0}^{T-1}\lbrace X^{(p)}_k, \, k=1,\cdots,m_p \rbrace$ is a basis of $\g^\tau$ and $B'=\bigsqcup_{p=0}^{T-1}\lbrace i X^{(p)}_k, \, k=1,\cdots,m_p \rbrace$ is a basis of $i \g^\tau$. Thus, $B \sqcup B'$ is a basis of $\g$ composed of eigenvectors of $\s$ and $\tau$.\\

\noi \underline{(v)$\Rightarrow$(ii):} Let us suppose that there exists a common basis of eigenvectors of $\s$ and $\tau$. To prove the identity $\tau\circ\s=\s^{-1}\circ\tau$, it is enough to prove it on each element of this basis. Let then $X$ be in this basis: there exist  $\epsilon\in\lbrace +1,-1 \rbrace$ and $p\in\lbrace 0,\cdots,T-1 \rbrace$ such that $\tau(X)=\epsilon X$ and $\s(X) = \omega^p X$ (and thus $\s^{-1}(X)=\omega^{-p}X$). As $\s$ is linear and $\tau$ is antilinear, we have
\begin{equation*}
\tau\circ\s (X) = \tau\bigl( \omega^{p} X \bigr) = \omega^{-p} \tau(X) = \omega^{-p} \epsilon X = \epsilon \s^{-1}(X) = \s^{-1} (\epsilon X) = \s^{-1} \circ \tau (X).
\end{equation*}
This then proves (ii).\\

\noi \underline{(ii)$\Rightarrow$(iv):} We suppose that $\tau\circ\s=\s^{-1}\circ\tau$. By recursion on $k\in\Z$, it implies $\tau\circ\s^k = \s^{-k} \circ \tau$. Recall the expression \eqref{Eq:ProjectorP} of the projector $\pi^{(p)}$. For any $X\in\g$, we then have
\begin{eqnarray*}
\tau \circ \pi^{(p)} (X) &=& \tau \left( \frac{1}{T} \sum_{k=0}^{T-1} \omega^{-kp}\, \s^k(X) \right) = \frac{1}{T} \sum_{k=0}^{T-1} \omega^{kp}\, \tau \circ \s^k (X) \\
&=& \frac{1}{T} \sum_{k=0}^{T-1} \omega^{kp}\, \s^{-k} \circ \tau(X) = \frac{1}{T} \sum_{j=1}^T \omega^{Tp-jp}\, \s^{j-T} \circ \tau(X) \\
&=& \frac{1}{T} \sum_{j=0}^{T-1} \omega^{-jp}\, \s^j \circ \tau (X) = \pi^{(p)} \circ \tau (X),
\end{eqnarray*}
where we used the antilinearity of $\tau$ for the second equality, the identity $\tau\circ\s^k = \s^{-k} \circ \tau$ for the third, the map $k \mapsto j=T-k$ for the fourth and the facts that $\omega^T=1$ and $\s^T=\Id$ for the fifth. This then proves (iv).\\

\noi \underline{(iv)$\Rightarrow$(iii):} Let us suppose that $\tau$ commutes with the projectors $\pi^{(p)}$'s. Let then $X$ be in $\g^{(p)}$. As $X=\pi^{(p)}(X)$, we have
\begin{equation*}
\tau (X) = \tau \circ \pi^{(p)} (X) = \pi^{(p)} \circ \tau(X),
\end{equation*}
hence $\tau(X)$ belongs to the image of $\pi^{(p)}$, \textit{i.e.} to $\g^{(p)}$. This proves (iii).\\

\noi \underline{(iii)$\Rightarrow$(i):} As the $\g^{(p)}$'s form a direct sum of $\g$, it is clear that the subspaces $\g^{(p)}\cap\g^\tau$ of $\g^\tau$ have a trivial intersection. Let us now suppose that $\tau$ stabilises the $\g^{(p)}$'s. Let $X$ be in $\g^\tau$. As an element of $\g$, it can be decomposed as
\begin{equation*}
X = \sum_{p=0}^{T-1} X^{(p)}, \;\;\;\; \text{ with } \;\;\;\; X^{(p)} = \pi^{(p)} (X) \in \g^{(p)}.
\end{equation*}
Moreover, as $X\in\g^\tau$, we also have
\begin{equation*}
X = \tau(X) = \sum_{p=0}^{T-1} \tau\bigl( X^{(p)} \bigr).
\end{equation*}
Yet, by hypothesis, $\tau\bigl( X^{(p)} \bigr)$ belongs to $\g^{(p)}$. By unicity of the decomposition along the direct sum of $\g^{(p)}$'s, we then have $\tau\bigl( X^{(p)} \bigr)=X^{(p)}$, \textit{i.e.} $X^{(p)} \in \g^\tau$. Thus, we have
\begin{equation*}
\g^\tau = \bigoplus_{p=0}^{T-1} \g^{(p)}\cap\g^\tau.
\end{equation*}
This proves (i).
\end{proof}

\section{Path-ordered exponentials}
\label{App:Pexp}

In this section, we recall some properties of path-ordered exponentials. Let $\g$ be a Lie algebra and $G$ be a connected Lie group with Lie algebra $\g$. Consider a $\g$-valued field $\Lc(x)$ and the path-ordered exponential
\begin{equation*}
T(x,y)= \Pexp \left( -\int_{y}^x \dd z \, \Lc(z) \right).
\end{equation*}
$T(x,y)$ is a $G$-valued field verifying the differential equations
\begin{subequations}\label{Eq:DerPexp}
\begin{align}
\bigl( \p_x T(x,y) \bigr) T(x,y)^{-1} &= -\Lc(x), \\
T(x,y)^{-1} \bigl( \p_{y} T(x,y) \bigr) &= \Lc(y),
\end{align}
\end{subequations}
and the initial condition $T(x,x)=\Id$. Under an infinitesimal transformation $\delta \Lc$ of $\Lc$, the path-ordered exponential is transformed by
\begin{equation}\label{dPExp}
\delta T(x,y) = - \int_{y}^x \dd z \; T(x,z) \delta\Lc(z) T(z,y).
\end{equation}
This formula allows one to compute the variations of $T(x,y)$ or Poisson brackets of $T(x,y)$ with other observables.\\

\noi In particular, if $\Lc$ is the spatial component of a zero curvature equation
\begin{equation*}
\p_t \Lc - \p_x \mathcal{M} + [\mathcal{M},\mathcal{L}] = 0,
\end{equation*}
one finds from equation \eqref{dPExp} that
\begin{equation*}
\p_t T(x,y) = T(x,y)\mathcal{M}(y)-\mathcal{M}(x)T(x,y).
\end{equation*}

Suppose now that we are given a Lie homomorphism $\s: G \rightarrow F$, from $G$ to another Lie group $F$. It induces a Lie algebra homomorphism $\s_\g : \g \rightarrow \f$. The image of the path-ordered exponential $T$ by the homomorphism $\s$ is simply
\begin{equation}\label{AutoPExp}
\s \bigl( T(x,y) \bigr) = \Pexp_F \left( -\int_{y}^x \dd z \, \s_\g \bigl( \Lc(z) \bigr) \right),
\end{equation}
where $\Pexp_F$ designates the path-ordered exponential in the group $F$.

\cleardoublepage
\chapter{Poisson and symplectic geometries}
\label{App:Poisson}

In this appendix, we describe the basic notions of Poisson geometry and symplectic geometry. We will restrict by simplicity to the case of manifolds of finite dimension. The notation $\F[M]$ then stands for the smooth functions on $M$, valued in $\R$. The notions described here generalise to infinite dimensional manifolds, when taking appropriate care of the good definition(s) of infinite dimensional differential manifolds and of the subtleties which come with it. This is useful for field theories (where $M$ is then the manifold of the fields configurations and $\F[M]$ the space of ``smooth'' functionals on it). However, we shall not enter into these technical considerations here to keep the appendix to a minimum and refer to~\cite{Abraham:1978} for a more complete presentation.

\section{Poisson and symplectic manifolds}

\begin{definition} A Poisson manifold $M$ is a differential manifold such that the space of functions $\F[M]$ on $M$ is equipped with a Poisson bracket
\begin{equation*}
\begin{array}{rccl}
\lwb \cdot, \cdot \rwb : & \F[M] \times \F[M] & \longrightarrow & \F[M] \\
 & (f, g) & \rightarrow & \lwb f, g \rwb
\end{array},
\end{equation*}
which is a skew-symmetric bilinear derivation satisfying the Jacobi identity:
\begin{equation}\label{Eq:JacobiPB}
\forall f,g,h \in \F[M], \;\;\; \lwb f, \lwb g, h \rwb \rwb + \lwb g, \lwb h, f \rwb \rwb + \lwb h, \lwb f, g \rwb \rwb = 0.
\end{equation}
\end{definition}

Let $\lbrace x^i \rbrace$ be (local) coordinates on $M$. Then, as the Poisson bracket is a derivation (on the left and on the right), one can rewrite it in these coordinates as
\begin{equation}\label{Eq:PBLocal}
\lwb f, g \rwb = \sum_{i,j} P^{ij} \frac{\p f}{\p x^i} \frac{\p g}{\p x^j},
\end{equation}
where $P^{ij}=-P^{ji}$ are functions in $\F[M]$. The Jacobi identity is then equivalent to
\begin{equation}\label{Eq:JacobiLocal}
P^{il} \frac{\p P^{jk}}{\p x^l} + P^{jl} \frac{\p P^{ki}}{\p x^l} + P^{kl} \frac{\p P^{ij}}{\p x^l} = 0,
\end{equation}
where a summation on the repeated index $l$ is implied. The Poisson bracket is said to be non-degenerate if the matrix $P^{ij}$ is invertible (in any coordinate chart).

\begin{definition}
A symplectic manifold is a differential manifold $M$ which possesses a closed non-degenerate 2-form $\omega$.
\end{definition}

\begin{proposition}\label{Prop:Symp2Poiss}
The symplectic manifolds are exactly the Poisson manifolds with non-degenerate Poisson brackets.
\end{proposition}
\begin{proof}
Suppose that we have a non-degenerate Poisson bracket $\PB$, given locally by some functional $P^{ij}$'s. The matrix $P^{ij}$ admits an inverse $P_{ij}$, satisfying $P^{ik}P_{kj}=\delta^i_{\;j}$. Let us then define the 2-form:
\begin{equation}\label{Eq:SymplecPB}
\omega = P_{ij}\, \dd x^i \wedge \dd x^j.
\end{equation}
It is non-degenerate, as $P_{ij}$ is invertible. Moreover, one checks that the Jacoby identity \eqref{Eq:JacobiLocal} translates in the closeness of the 2-form $\omega$. Conversely, if one as a symplectic form $\omega$ on $M$, writing it in a local coordinates as \eqref{Eq:SymplecPB} and inverting the matrix $P_{ij}$, one gets a non-degenerate Poisson bracket on $M$. 
\end{proof}

\section{The Kirillov-Kostant bracket}
\label{App:KK}

Let $\g$ be a real Lie algebra (see Appendix \ref{App:Lie} for definitions and basics results on Lie algebras). We consider the dual $\g^*$ of $\g$, defined as the vector space of linear forms on $\g$. As $\g^*$ is a vector space, its tangent space at a point $\xi\in\g^*$ is identified with $\g^*$ itself. If $f$ is a real-valued smooth function on $\g^*$ (\textit{i.e.} an element of $\F[\g^*]$), the differential $\dd_\xi f$ of $f$ at a point $\xi\in\g^*$ is then a linear map from $\g^*$ to $\R$, hence an element of $(\g^*)^* = \g$. We then define the bracket of two functions $f,g$ in $\F[\g^*]$ as the function
\begin{equation}\label{Eq:KKB}
\lwb f, g \rwb_* (\xi) = \xi \bigl( \left[\dd_\xi f, \dd_\xi g \right] \bigr).
\end{equation}
As the Lie bracket $[\cdot,\cdot]$ is linear and skew-symmetric, it is clear that the bracket $\PB_*$ is linear and skew-symmetric. Moreover, by the Leibniz rule $\dd(fg) = f \dd g + g \dd f$ for the differential, the bracket $\PB_*$ is a derivation. Finally, the Jacobi identity on the Lie bracket $[\cdot,\cdot]$ implies the one for the bracket $\PB_*$. The dual space $\g^*$ is thus naturally equipped with a Poisson bracket $\PB_*$, called the \textbf{Kirillov-Kostant} bracket.\\

The definition \eqref{Eq:KKB} is an abstract definition of $\PB_*$ for all functions $f$ and $g$ in $\F[\g^*]$. Let us now give a more explicit formula for the bracket $\PB_*$. As $\g^*$ is a vector space, it is easy to find smooth functions on it: the linear forms in $(\g^*)^*$. These can be canonically identified with elements of $\g$: for any $X\in\g$, the map
\begin{equation*}
\begin{array}{rccc}
X_\circ : & \g^* & \longrightarrow & \R \\
          &  \xi & \longmapsto     & \xi(X)
\end{array}
\end{equation*}
is a (linear) function on $\g^*$ and thus an element of $\F[\g^*]$. On these linear functions, the Kirillov-Kostant bracket then coincides with the Lie bracket:
\begin{equation}\label{Eq:KKXY}
\forall \; X,Y \in \g, \;\; \lwb X_\circ, Y_\circ \rwb_* = [X,Y]_\circ.
\end{equation}
In particular, a choice of basis $\lwb I^a \rwb$ of $\g$ defines a choice of coordinate functions $X^a=I^a_\circ$ on $\g^*$. The ``fundamental'' brackets of these coordinate functions is then
\begin{equation*}
\lbrace X^a, X^b \rbrace_* = \fs{ab}{c} X^c,
\end{equation*}
where the $\ft{ab}{c}$ are the structure constants of the Lie algebra $\g$ (see Appendix \ref{App:LieGen}).\\

Suppose that $\g$ is semi-simple. Then its Killing form $\kappa^{ab}$ is non-degenerate and possesses an inverse $\kappa_{ab}$. One can encode all coordinates functions $X^a$ in the $\g$-valued function
\begin{equation*}
X = \kappa_{ab} X^a I^b \in \F[\g^*] \otimes \g.
\end{equation*}
As $X$ is a $\g$-valued function, we can describe the Poisson bracket of its components using the tensorial notations $\ti{i}$ introduced in Subsection \ref{Sec:Ham}. We then find that the Kirillov-Kostant bracket can be written in a compact way as
\begin{equation*}
\lbrace X\ti{1}, X\ti{2} \rbrace_* = \left[ C\ti{12}, X\ti{1} \right] = - \left[ C\ti{12}, X\ti{2} \right],
\end{equation*}
where $C\ti{12}$ is the split quadratic Casimir of $\g$ (see Appendix \ref{App:Casimir}).

\section{Poisson maps and canonical transformations}

\begin{definition}
Let $M$ and $N$ be two Poisson manifolds, with brackets $\PB_M$ and $\PB_N$. A differentiable map $\vp : M \rightarrow N$ is said to be a \textbf{Poisson map} if it preserves the Poisson bracket:
\begin{equation*}
\forall\; f, g \in \F[N], \;\;\;\; \lbrace f \circ \vp, g \circ \vp \rbrace_M = \lbrace f,g \rbrace_N \circ \vp.
\end{equation*}
\end{definition}

\begin{definition}
Let $M$ be a symplectic manifold with symplectic form $\omega$. A diffeomorphism $\vp : M \mapsto M$ is said to be a \textbf{canonical transformation} if it preserves the symplectic form:
\begin{equation*}
\vp^* \omega = \omega,
\end{equation*}
where $\vp^* \omega$ is the pullback of $\omega$ by the diffeomorphism $\vp$.
\end{definition}

\begin{proposition}
Let $M$ be a symplectic manifold and thus a Poisson manifold (proposition \ref{Prop:Symp2Poiss}). Then the canonical transformations of $M$ are exactly the Poisson diffeomorphisms from $M$ to itself.
\end{proposition}

Let us consider a one-parameter Lie group of diffeomorphisms $\lbrace \vp_\alpha,\, \alpha\in\R \rbrace$ of $M$ (\textit{i.e.} satisfying $\vp_\alpha \circ \vp_\beta = \vp_{\alpha+\beta}$ and $\vp_0 = \Id_M$). Then the infinitesimal transformation around $\alpha=0$ defines a vector field $X$ on $M$:
\begin{equation*}
\forall \; p \in M, \;\;\;\; X(p) = \left. \frac{\p \vp_\alpha(p)}{\p \alpha} \right|_{\alpha=0} \in T_p M.
\end{equation*}
Reciprocally, given a vector field $X$ on $M$, its flow defines a one-parameter Lie group of diffeomorphisms of $M$. Recall that vector fields $\X[M]$ act on the space of smooth functions $\F[M]$ as derivations: we will denote this action by a dot.

\begin{proposition}
If the diffeomorphisms $\lbrace \vp_\alpha,\, \alpha\in\R \rbrace$ are Poisson maps, the vector field $X$ satisfies:
\begin{equation}\label{Eq:CanVec}
\forall \; f,g\in\F[M], \;\;\;\; \lbrace X.f, g \rbrace + \lbrace f, X.g \rbrace = X. \lbrace f, g \rbrace.
\end{equation}
\end{proposition}

A vector field $X$ satisfying \eqref{Eq:CanVec} is said to be Poisson. We will denote by $\Pc[M]$ the Poisson vector fields of $M$. If $M$ is a symplectic manifold, it is also a Poisson manifold by Proposition \ref{Prop:Symp2Poiss}. Then, the Poisson vector fields are the ones preserving the symplectic form:

\begin{proposition}\label{Prop:SympVect}
Let $M$ be a symplectic manifold with symplectic form $\omega$ and $X\in\X[M]$ be a vector field on $M$. Then $X$ is Poisson if and only if $\Lc_X \omega=0$, where $\Lc_X$ denotes the Lie derivative along $X$.
\end{proposition}

Important examples of Poisson vector fields are given by the so-called Hamiltonian vector fields, which are associated with functions on the Poisson manifold.

\begin{definition}
Let $M$ be a Poisson manifold and $f\in\F[M]$ be a function from $M$ to $\R$. Then, the Hamiltonian vector field associated with $f$ is the unique $V_f \in \X[M]$ such that $V_f.g = \lbrace f,g \rbrace$ for all $g\in\F[M]$ (this vector field exists as $\lbrace f, \cdot \rbrace$ acts as a derivation on $\F[M]$).
\end{definition}

\begin{proposition}
Let $M$ be a Poisson manifold. The map
\begin{equation*}
\begin{array}{rccc}
\mathcal{V} : & \F[M] & \longrightarrow & \X[M] \\
              &   f   & \longmapsto     & V_f 
\end{array}
\end{equation*}
is a derivation from the algebra $\F[M]$ to the $\F[M]$-module $\X[M]$. Moreover, it is a Lie homomorphism if $\F[M]$ is equipped with the Poisson bracket $\PB$ and $\X[M]$ is equipped with the Lie bracket $[\cdot,\cdot]$ of vector fields, \textit{i.e.}
\begin{equation}\label{Eq:HamVectLie}
\forall \; f,g\in\F[M], \;\;\; V_{\lbrace f,g \rbrace} = \left[ V_f, V_g \right].
\end{equation}
Finally, for all $f\in\F[M]$, the Hamiltonian vector field $V_f$ is Poisson. Thus $\mathcal{V}$ is $\Pc[M]$-valued.
\end{proposition}

\begin{proof}
It is clear that $\mathcal{V}$ is a derivation as $\PB$ is a derivation on the left. Moreover, for all $f,g,h\in\F[M]$, we have
\begin{eqnarray*}
\left[ V_f, V_g \right].h &=& V_f.\bigl(V_g.h\bigr) - V_g.\bigl(V_f.h\bigr) \\
 &=& \bigl\lbrace f, \lbrace g, h \rbrace \bigr\rbrace - \bigl\lbrace g, \lbrace f, h \rbrace \bigr\rbrace \\
 &=& \bigl\lbrace \lbrace f,  g \rbrace, h  \bigr\rbrace \hspace{95pt} \text{ by the Jacobi identity \eqref{Eq:JacobiPB}} \\
 &=& V_{\lbrace f,g \rbrace}.h,
\end{eqnarray*}
hence equation \eqref{Eq:HamVectLie}. In the same way, using the Jacoby identity, we have
\begin{equation*}
\lbrace V_f.g, h \rbrace + \lbrace g, V_f.h \rbrace = \bigl\lbrace \lbrace f, g \rbrace, h \bigr\rbrace + \bigl\lbrace g, \lbrace f, h \rbrace \bigr\rbrace = \bigl\lbrace f, \lbrace g, h \rbrace \bigr\rbrace = V_f.\lbrace g,h \rbrace.
\end{equation*}
This proves that $V_f$ is a Poisson vector field.
\end{proof}

It is natural to ask whether every Poisson vector field can be written as a Hamiltonian vector field. The answer is yes when $M$ is symplectic and simply connected.

\begin{theorem}\label{Thm:SympHam}
Let $M$ be a symplectic manifold with symplectic form $\omega$. Then, the Hamiltonian vector field associated with $f\in\F[M]$ is the unique vector field $V_f$ such that
\begin{equation*}
\dd f = \iota_{V_f} \omega,
\end{equation*}
where $\iota_{V_f}$ denotes the interior derivative with respect to $V_f$. Moreover, if $M$ is simply connected, the map $\mathcal{V}:\F[M] \mapsto \Pc[M]$ is surjective.
\end{theorem}
\begin{proof}
Let us consider local coordinates $x^i$ on $M$ and the associated expressions \eqref{Eq:PBLocal} and \eqref{Eq:SymplecPB} of the Poisson bracket and the symplectic form on $M$. For $f\in\F[M]$, we have $V_f.x^i = \lbrace f, x^i \rbrace = P^{ki}\frac{\p f}{\p x^k}$. Thus
\begin{equation*}
\iota_{V_f} \omega = \omega(V_f,\cdot) = P_{ij} \, \dd x^i (V_F) \, \dd x^j = P_{ij} \bigl(V_f.x^i\bigr) \, \dd x^j = \underbrace{P_{ij} P^{ki}}_{\delta^k_{\;j}} \frac{\p f}{\p x^k} \,\dd x^j = \frac{\p f}{\p x^k} \dd x^k = \dd f,
\end{equation*}
which proves the first part of the theorem (the uniqueness of $V_f$ defined this way comes from the non-degeneracy of $\omega$).

Let us now consider a Poisson vector field $X$ and let us define the 1-form $\alpha = \iota_X \omega$. By the Cartan identity, we have
\begin{equation*}
\dd \alpha = \dd \bigl(\iota_X \omega \bigr) = - \iota_X \bigl( \dd \omega \bigr) + \Lc_X \omega = 0,
\end{equation*}
as $\dd \omega = 0$ ($\omega$ is closed) and $\Lc_X \omega=0$ ($X$ is a Poisson vector field, see Proposition \ref{Prop:SympVect}). Thus, the 1-form $\alpha$ is closed. If $M$ is simply-connected, $X$ is also exact by the Poincar\'e lemma, \textit{i.e.} $\alpha=\dd f$ for some $f$ in $\F[M]$. We then conclude that $X=V_f$ by the first part of the theorem.
\end{proof}

\section{Hamiltonian action of Lie groups}
\label{App:HamAction}

Let $G$ be a Lie group, with Lie algebra $\g$. We suppose that it acts smoothly on a differentiable manifold $M$, through the action
\begin{equation*}
\begin{array}{ccc}
 G \times M & \longrightarrow & M \\
 (g,p)      & \longmapsto     & \rho(g,p)
\end{array}.
\end{equation*}
This action can be seen as a group homomorphism $\rho : g \mapsto \rho(g,\cdot)$ from $G$ to the group $\Diff{M}$ of diffeomorphisms of $M$. Taking the differential of this homomorphism at the identity, we get a linear map
\begin{equation*}
\begin{array}{rcccc}
\delta : & \g & \longrightarrow & \X[M] \\
         & \epsilon  & \longmapsto     & \delta_\epsilon
\end{array}
\end{equation*}
from the Lie algebra $\g=T_eG$ to the space of vector fields $\X[M]=T_\Id \Diff{M}$. As $\rho$ is a group homomorphism, the map $\delta$ is an homomorphism of Lie algebra from $\g$ to $\X[M]$, equipped with the Lie bracket of vector fields. In other words, we have for any $\epsilon,\epsilon' \in \g$:
\begin{equation}\label{Eq:DeltaLie}
\delta_{[\epsilon,\epsilon']} = \left[ \delta_\epsilon, \delta_{\epsilon'} \right].
\end{equation}

Let us now suppose that the action of $G$ conserves the Poisson bracket, \textit{i.e.} that for any fixed $g\in G$, the map $\rho(g,\cdot):M\rightarrow M$ is Poisson. Then, the map $\delta$ is valued in $\Pc[M]$. We shall further suppose that the infinitesimal actions $\delta_\epsilon$ are Hamiltonian for any $\epsilon\in\g$ (this is directly the case if $M$ is symplectic and simply connected, by Theorem \ref{Thm:SympHam}). Then there exists a linear map
\begin{equation*}
\mu: \g \longrightarrow \F[M]
\end{equation*}
such that for any $\epsilon \in \g$, 
\begin{equation*}
\delta_\epsilon = V_{\mu(\epsilon)} = \lbrace \mu(\epsilon),  \cdot \rbrace.
\end{equation*}
We call $\mu$ the \textbf{moment map}. It is easy to see that
\begin{equation}\label{Eq:DeltaVm}
\delta = \mathcal{V} \circ \mu,
\end{equation}
as for any $\epsilon\in\g$, $\mathcal{V}\circ \mu(\epsilon) =\mathcal{V}\bigl(\mu(\epsilon)\bigr)=V_{\mu(\epsilon)}=\delta_\epsilon$. As a linear map from $\g$ to $\F[M]$, it can be seen as an element of $\g^* \otimes \F[M]$, \textit{i.e.} as a $\g^*$-valued function on $M$:
\begin{equation*}
Q: M \longrightarrow \g^*.
\end{equation*}
In terms of this function, we have, for all $\epsilon\in\g$ and $f\in\F[M]$,
\begin{equation}\label{Eq:DeltaMomentMap}
\delta_\epsilon f =\langle \epsilon, \lbrace Q, f \rbrace \rangle,
\end{equation}
where $\langle \cdot, \cdot \rangle$ represents the pairing on $\g\times\g^*$.\\

Let us now suppose that $\g$ is semi-simple. Then there is a natural isomorphism between $\g^*$ and $\g$ \textit{via} the Killing form $\kappa$ on $\g$, which is non-degenerate (see Appendix \ref{App:SemiSimple}). We can then consider the image $m$ of $Q$ under this isomorphism, which is a $\g$-valued function on $M$. The infinitesimal action $\delta_\epsilon$ of $\epsilon\in\g$ can then be written
\begin{equation}\label{Eq:MomentMapSS}
\delta_\epsilon  = V_{\kappa(m,X)} = \kappa\bigl(\epsilon,\lwb m, \cdot \rwb \bigr). 
\end{equation}
If $f$ is a $\g$-valued function on $M$ whose Poisson bracket vanishes with all other functions in $\F[M]$ (such a function is called a Casimir of $\bigl(\F[M],\PB\bigr)$), then note that one can replace $m$ by $m+f$ in the equation \eqref{Eq:MomentMapSS}. It is a classical result~\cite{Abraham:1978} that using this freedom, one can always choose $m$ to satisfy the Kirillov-Kostant bracket defined in subsection \ref{App:KK}:
\begin{equation}\label{Eq:KKm}
\lwb m\ti{1}, m\ti{2} \rwb = \left[ C\ti{12}, m\ti{1} \right].
\end{equation}
This is due to the fact that $\delta$ is a Lie algebra homomorphism (and that a semi-simple Lie algebra has no non trivial central extension). Conversely, if one has a $\g$-valued map $m$ satisfying the Kostant-Kirillov bracket \eqref{Eq:KKm}, one can define a Lie algebra homomorphism $\delta$ from $\g$ to $\X[M]$ by equation \eqref{Eq:MomentMapSS}, \textit{i.e.} an infinitesimal action of $\g$ on $M$.

The present subsection concerned left actions of the group $G$ and thus of the Lie algebra $\g$. One can also develop a similar formalism for right actions. In this case, there is still a moment map $m$ such that equation \eqref{Eq:MomentMapSS} holds. However, the Poisson bracket \eqref{Eq:KKm} changes into minus itself.

\cleardoublepage
\chapter{$\Rc$-matrices and classical Yang-Baxter equations}
\label{App:RMat}

\section{Classical Yang-Baxter Equations}
\label{App:CYBE}

Let $\g$ be a real Lie algebra, with Lie bracket $\LB$ (see Appendix \ref{App:Lie}). In this appendix, we describe the (modified) Classical Yang-Baxter Equation, denoted (m)CYBE, on $\g$. We will present both the ``operator'' form and the ``matricial'' form of this equation and explain the link between these two forms. We refer to \cite{SemenovTianShansky:1983ik,Babebook,Vicedo:2010qd,Drinfeld:1983ky} for more details.

\paragraph{mCYBE in ``operator'' form.} Let $R:\g\rightarrow \g$ be a linear operator on $\g$. We say that $R$ is a solution of the \textbf{operator mCYBE on $\bm{\g}$} if it satisfies
\begin{equation*}
\forall \, X,Y \in \g, \; \; [RX,RY]-R\bigl([RX,Y]+[X,RY]\bigr) = - \alpha [X,Y],
\end{equation*}
where $\alpha$ is a real number. We note that rescaling the operator $R$ into $\lambda R$ sends solutions of the mCYBE for $\alpha$ to solutions for $\lambda^2 \alpha$. Thus, by an appropriate rescaling, we reduce the study of the mCYBE to three cases: $\alpha=0$, $\alpha=1$ and $\alpha=-1$. We will then consider the equation
\begin{equation}\label{Eq:AppmCYBE}
\forall \, X,Y \in \g, \; \; [RX,RY]-R\bigl([RX,Y]+[X,RY]\bigr) = - c^2 [X,Y],
\end{equation}
with $c=0$ (homogeneous case), $c=1$ (split case) and $c=i$ (non-split case). When in the homogeneous case $c=0$, we then talk of the CYBE instead of the modified CYBE. Note that $R=\pm c \,\Id$ is a solution of the mCYBE. Solutions of the (resp. homogeneous, split, non-split) mCYBE \eqref{Eq:AppmCYBE} are called (resp. homogeneous, split, non-split) $R$-matrices. For $R$ such a matrix, we define the $R$-bracket on $\g$ as
\begin{equation*}
[X,Y]_R = [ RX, Y ] + [ X,RY ], \;\;\;\; \forall \, X,Y\in\g.
\end{equation*}
This bracket possesses many interesting properties, including being a Lie bracket. As it is one of the main tool for the Chapter \ref{Chap:PLie} of this thesis, it is described in detail in this chapter and we will not study it further in this appendix. We will just rewrite the mCYBE in a more compact way using the $R$-bracket:
\begin{equation}\label{Eq:mCYBEbracketR}
\forall \, X,Y \in \g, \; \; [RX,RY] + c^2 [X,Y] = R\bigl([X,Y]_R\bigr).
\end{equation}

\paragraph{mCYBE in ``matricial'' form.} In this section, we will use the tensorial notation $\ti{i}$ introduced in Section \ref{Sec:Ham}. Let us suppose that $\g$ admits a non-degenerate invariant form $\kappa$, as for example the Killing form for semi-simple Lie algebras. Then, one can define the split quadratic Casimir $C\ti{12}$ in $\g \otimes \g$ (see Appendix \ref{App:Casimir}).

Let $R\ti{12}$ be a matrix in $\g \otimes \g$. We say that $R$ is a solution of the \textbf{matricial mCYBE on $\bm{\g}$} if it satisfies
\begin{equation}\label{Eq:AppmCYBEMat}
\lsb R\ti{12}, R\ti{13} \rsb + \lsb R\ti{12}, R\ti{23} \rsb + \lsb R\ti{32}, R\ti{13} \rsb = -c^2\lsb C\ti{12}, C\ti{13} \rsb.
\end{equation}
As for the operator case, a rescaling of $R$ allows to consider $c\in\lbrace 0,1,i \rbrace$. Using equation \eqref{Eq:QuadCas}, one checks that $\pm c \, C\ti{12}$ is a solution of the mCYBE.

It is often useful to restrict to skew-symmetric matrices $R\ti{12}=-R\ti{21}$. In this case, one can rewrite the mCYBE on $R$ in (the more often used) form
\begin{equation*}
R\ti{12} = - R\ti{21} \;\;\;\; \text{ and } \;\;\;\; \lsb R\ti{12}, R\ti{13} \rsb + \lsb R\ti{12}, R\ti{23} \rsb + \lsb R\ti{13}, R\ti{23}  \rsb = -c^2\lsb C\ti{12}, C\ti{13} \rsb.
\end{equation*}

We end this subsection by a remark on the right-hand side of the mCYBE \eqref{Eq:AppmCYBEMat}. Using the Jacoby identity and the fundamental property \eqref{Eq:CasIdentity} of the split Casimir, one finds that
\begin{equation*}
\Bigl[ \lsb C\ti{12}, C\ti{13} \rsb, X\ti{1}+X\ti{2}+X\ti{3} \Bigr] = 0, \;\;\;\; \forall \, X\in\g.
\end{equation*}

\paragraph{Kernels of operators.} Let us consider a matrix $R\ti{12} \in \g \otimes \g$. We associate with this matrix a linear operator $R$ on $\g$ by
\begin{equation*}
R(X) = \kappa\ti{2} \left( R\ti{12}, X\ti{2} \right), \;\;\;\; \forall \; X\in\g.
\end{equation*}
We then say that $R\ti{12}$ is the kernel of $R$. The following lemma collects some basic facts about kernels.

\begin{lemma}\label{Lem:Kernels} We denote by $\null^t$ the transpose with respect to the bilinear form $\kappa$.\vspace{-5pt}
\begin{enumerate}[(i)]
\setlength\itemsep{0.1em}
\item The kernel of a linear operator $R$ on $\g$ always exists and is unique.
\item The kernel of the identity is $C\ti{12}$. More generally the kernel of $R$ is $R\ti{1}C\ti{12}$.
\item The kernel of $\null^t R$ is $R\ti{21}$. In particular $R$ is skew-symmetric with respect to $\kappa$, \textit{i.e.} $R=-\null^t R$, if and only if $R\ti{12}=-R\ti{21}$.
\end{enumerate}
\end{lemma}
\begin{proof}
The point (i) is a consequence of the non-degeneracy of $\kappa$. The point (ii) is a rewriting of the completeness equation \eqref{Eq:CasComp}. Finally, let $X,Y$ be any elements of $\g$. We then have
\begin{equation*}
\kappa \bigl(\null^t R (X), Y) = \kappa \bigl( X, R(Y) \bigr) = \kappa \ti{1}\left( X, \kappa\ti{2}\bigl( R\ti{12}, Y\ti{2} \bigr) \right) = \kappa\ti{2} \left( \kappa\ti{1}\bigl( X\ti{1},R\ti{12} \bigr), Y\ti{2} \right) = \kappa\ti{1} \left( \kappa\ti{2}\bigl( R\ti{21}, X\ti{2} \bigr), Y\ti{1} \right).
\end{equation*}
As this is true for all $Y\in\g$, we get, by non-degeneracy of $\kappa$,
\begin{equation*}
\null^t R (X) = \kappa\ti{2}\bigl( R\ti{21}, X\ti{2} \bigr),
\end{equation*}
which proves point (iii).
\end{proof}

\noi The following proposition relates the operator and matricial forms of the mCYBE:

\begin{proposition}\label{Prop:Rop2Rmat}
Let $R$ be a linear operator on $\g$, with kernel $R\ti{12}$. Then $R$ is solution of the operator mCYBE \eqref{Eq:AppmCYBE} if and only if $R\ti{12}$ is a solution of \eqref{Eq:AppmCYBEMat}.
\end{proposition}
\begin{proof}
We denote
\begin{equation*}
\Yc\ti{123} = \lsb R\ti{12}, R\ti{13} \rsb + \lsb R\ti{12}, R\ti{23} \rsb + \lsb R\ti{32}, R\ti{13} \rsb
\end{equation*}
the left-hand side of $\eqref{Eq:AppmCYBEMat}$. For $X,Y$ in $\g$, we have
\begin{eqnarray*}
\kappa\ti{23} \bigl( \Yc\ti{123}, X\ti{2}Y\ti{3} \bigr)
 &=& \left[ \kappa\ti{2}\bigl(R\ti{12},X\ti{2}\bigr), \kappa\ti{3}\bigl(R\ti{13},Y\ti{3}\bigr) \right] + \kappa\ti{2} \left( \left[ R\ti{12}, \kappa\ti{3}\bigl( R\ti{23}, Y\ti{3} \bigr) \right], X\ti{2} \right) \\
 & & \hspace{80pt} + \, \kappa\ti{3} \left( \left[ \kappa\ti{2}\bigl( R\ti{32}, X\ti{2} \bigr), R\ti{13} \right], Y\ti{3} \right) \\[4pt]
 &=& [ RX, RY ] + \kappa\ti{2} \left( \bigl[R\ti{12},RY\ti{2}\bigr], X\ti{2} \right) + \kappa\ti{3} \left( \bigl[RX\ti{3},R\ti{13}\bigr],Y\ti{3} \right) \\[4pt]
 &=& [ RX, RY ] - \kappa\ti{2}\bigl( R\ti{12}, [R,RY]\ti{2}) - \kappa\ti{3}\bigl(R\ti{13},[RX,Y]\ti{3}\bigr)  \\[4pt]
 &=& [RX,RY] - R\bigl([RX,Y]+[X,RY]\bigr).
\end{eqnarray*}
In the same way, we have
\begin{equation*}
\kappa\ti{23} \bigl( \lsb C\ti{12}, C\ti{13} \rsb, X\ti{2}Y\ti{3} \bigr) = \left[ \kappa\ti{2}\bigl(C\ti{12},X\ti{2}\bigr), \kappa\ti{3}\bigl(C\ti{13},Y\ti{3}\bigr) \right] = [X,Y].
\end{equation*}
Thus, one has
\begin{equation*}
\kappa\ti{23} \bigl( \Yc\ti{123} + c^2\lsb C\ti{12}, C\ti{13} \rsb, X\ti{2}Y\ti{3} \bigr) = [RX,RY] - R\bigl([RX,Y]+[X,RY]\bigr) + c^2 [X,Y],
\end{equation*}
hence the proposition.
\end{proof}

\section{Adler-Kostant-Symes scheme}
\label{App:AKS}

\paragraph{AKS construction.} In this appendix, we present a general scheme to construct $R$-matrices on a Lie algebra $\g$, called the Adler-Kostant-Symes (AKS) scheme~\cite{Adler:1979ib,Kostant:1979qu,Symes:1981}. At least for the beginning of this section, we shall consider $\g$ to be a complex Lie algebra (we will explain later how to apply this construction to real Lie algberas). We suppose that $\g$ can be decomposed as
\begin{equation}\label{Eq:gABC}
\g = A \oplus B \oplus C,
\end{equation}
where $A$, $B$ and $C$ are three subalgebras of $\g$ and $\oplus$ denotes a direct sum as a vector space. Note that we do not require $A$, $B$ and $C$ to be in direct sum as Lie algebras, \textit{i.e.} to satisfy $[A,B]=[A,C]=[B,C]=0$. However, we will impose the following conditions on $C$:
\begin{equation}\label{Eq:AKScondC}
[C,C] = 0, \;\;\;\; [C,A] \subset A \;\;\;\; \text{and} \;\;\;\ [C,B] \subset B.
\end{equation}
In particular, $C$ must be an abelian subalgebra of $\g$. Note that if $\g=A\oplus B$ with $A$ and $B$ subalgebras, we are in the situation described above by defining $C=\lbrace 0 \rbrace$ (without any assumptions on $A$ and $B$).

We will denote by $\pi_S$ ($S=A,B,C$) the projectors along the decomposition \eqref{Eq:gABC}. Moreover, for $X$ in $\g$, we will denote by simplicity $X_S = \pi_S X$. Let $\phi_C$ be a linear operator from $C$ to itself: we extend it to an operator of the whole algebra $\g$ by letting $\phi_C(A)=\phi_C(B)=0$. Let us then define the AKS operator $R$ on $\g$ as:
\begin{equation}\label{Eq:RAKS}
R = c\bigl(\pi_A - \pi_B) + \phi_C.
\end{equation}

\begin{theorem}\label{Thm:AKS}
$R$ is a solution of the operator mCYBE \eqref{Eq:AppmCYBE}.
\end{theorem}
\begin{proof}
Let us first compute the $R$-bracket associated with the operator $R$. As  $\g=A\oplus B\oplus C$, we have $X=X_A+X_B+X_C$ for all $X\in\g$. One then finds
\begin{eqnarray*}
[X,Y]_R
 &=& [RX,Y] + [X,RY] \\
 &=& [c(X_A-X_B)+\phi_C X,Y_A+Y_B+Y_C] + [X_A+X_B+X_C,c(Y_A-Y_B)+\phi_C Y] \\
 &=& c \left( [X_A,Y_A]+\Ccancel[red]{[X_A,Y_B]}-\Ccancel[blue]{[X_B,Y_A]}-[X_B,Y_B] \right)\\
 & & \hspace{40pt} +c[X_A,Y_C]-c[X_B,Y_C]+[\phi_CX,Y_A]+[\phi_CX,Y_B] \\
 & & +c \left( [X_A,Y_A]-\Ccancel[red]{[X_A,Y_B]}+\Ccancel[blue]{[X_B,Y_A]}-[X_B,Y_B] \right)\\
 & & \hspace{40pt} +c[X_C,Y_A]-c[X_C,Y_B]+[X_A,\phi_CY]+[X_B,\phi_CY],
\end{eqnarray*}
where we used the fact that $[C,C]=0$. Thus, one has
\begin{eqnarray}
[X,Y]_R &=& \underbrace{2c [X_A,Y_A] + c[X_A,Y_C] + c[X_C,Y_A] + [\phi_CX,Y_A] + [X_A,\phi_CX]}_{\in A} \\
& & \hspace{20pt} -\underbrace{\Bigl( 2c [X_B,Y_B] + c[X_B,Y_C] + c[X_C,Y_B] - [\phi_CX,Y_B] - [X_B,\phi_CX] \Bigr)}_{\in B}, \notag
\end{eqnarray}
using the fact that $[A,C] \subset A$ and $[B,C] \subset C$. As $[X,Y]_R$ does not contain any element of $C$, one has
\begin{equation*}
R\bigl([X,Y]_R\bigr) = c(\pi_A-\pi_B) [X,Y]_R.
\end{equation*}
We then find
\begin{eqnarray*}
R\bigl([X,Y]_R\bigr) &=& c^2\Bigl(2 [X_A,Y_A] + 2 [X_B,Y_B] + [X_A,Y_C] + [X_C,Y_A] + c[X_B,Y_C] + c[X_C,Y_B] \Bigr) \\
 & & \hspace{90pt} +c \Bigl( [\phi_CX,Y_A-Y_B] + [X_A-X_B,\phi_CX] \Bigr).
\end{eqnarray*}
In the same way, we have
\begin{equation*}
[RX,RY]+c^2[X,Y] = [c(X_A-X_B)+\phi_C X, c(Y_A-Y_B)+\phi_C Y ] + c^2 [X_A+X_B+X_C,Y_A+Y_B+Y_C].
\end{equation*}
Developping this expression and using $[C,C]=0$, one finds that
\begin{equation*}
R\bigl([X,Y]_R\bigr) = [RX,RY]+c^2[X,Y],
\end{equation*}
\textit{i.e.} $R$ satisfies the operator mCYBE \eqref{Eq:AppmCYBE}.
\end{proof}

\paragraph{Standard AKS operator.} Let us consider the $R$-matrix \eqref{Eq:RAKS} constructed from the AKS construction. We will say that $R$ is standard if we chose $\phi_C=0$ and $c\neq 0$, so that
\begin{equation}\label{Eq:StandAKS}
R=c(\pi_A-\pi_B)
\end{equation}
is either a split or a non-split $R$-matrix (not a homogeneous one).

\begin{proposition}
Let $R$ be a solution of the operator mCYBE \eqref{Eq:AppmCYBE}. Then $R$ is a standard AKS operator if and only if $R^3=c^2 R$.
\end{proposition}
\begin{proof}
It is clear that $R=c(\pi_A-\pi_B)$ satisfies $R^3=c^2 R$. Conversely let us suppose that we have a solution $R$ of \eqref{Eq:AppmCYBE} satisfying $R^3=c^2 R$. Then $R$ is diagonalisable and have eigenvalues in $\lbrace 0, c, -c \rbrace$. We then define $A=\Ker(R-c)$, $B=\Ker(R+c)$ and $C=\Ker(R)$. As $R$ is diagonalisable, one has the vector space decomposition $\g = A \oplus B \oplus C$ and $R$ can be re-expressed as $R=c(\pi_A-\pi_B)$.

To prove that $R$ is an AKS operator, we now need to show that $A$, $B$ and $C$ are subalgebras of $\g$ and that $C$ satisfies the condition \eqref{Eq:AKScondC}. Recall the expression \eqref{Eq:mCYBEbracketR} of the mCYBE. We will distinguish several cases.

For example let us suppose that $X$ and $Y$ are in $A$. We then have $RX=cX$ and $RY=cY$. Thus, one has
\begin{equation*}
[X,Y]_R = [RX,Y]+[X,RY] = 2c[X,Y] \;\;\;\; \text{and} \;\;\;\; [RX,RY]+c^2[X,Y] = 2c^2[X,Y].
\end{equation*}
We then get by \eqref{Eq:mCYBEbracketR} that
\begin{equation*}
(R-c) \bigl( [X,Y] \bigr) = 0.
\end{equation*}
Thus, $[X,Y]\in A = \Ker(R-c)$. We then deduce that $A$ is a subalgebra of $\g$.

In the same way, choosing appropriately $X$ and $Y$ either in $A$, $B$ or $C$, one shows from \eqref{Eq:mCYBEbracketR} that $B$ is also a subalgebra of $\g$, that $C$ is abelian and that $C$ satisfies \eqref{Eq:AKScondC}. This ends the demonstration of the proposition.
\end{proof}
\noi One can note that a standard AKS operator $R$ satisfies $R^2=c^2 \Id$ if and only if $C=0$.\\

Let us now suppose that $\g$ is equipped with an invariant non-degenerate bilinear form $\kappa$ (for example the Killing form if $\g$ is semi-simple). We can then associate a kernel $R\ti{12}\in\g\otimes\g$ with the standard AKS operator $R$, which satisfies the matricial mCYBE \eqref{Eq:AppmCYBEMat} (see Section \ref{App:CYBE}). Recall from Lemma \ref{Lem:Kernels} that $R\ti{12}$ is skew-symmetric if and only if $R$ is skew-symmetric with respect to $\kappa$. The following proposition gives a necessary and sufficient condition to happen. 

\begin{proposition}\label{Prop:SkewAKS}
The standard AKS operator $R$ is skew-symmetric if and only if $A$ and $B$ are $\kappa$-isotropic and $C$ is $\kappa$-orthogonal to $A$ and $B$.

\noi In this case, $\kappa$ pairs non-degenerately $A$ and $B$ and restricts to a non-degenerate form on $C$. If $\lbrace I^a \rbrace$ is a basis of $A$, there exists a corresponding dual basis $\lbrace I_a \rbrace$ in $B$. Then, the kernel $R\ti{12}$ of $R$ is
\begin{equation}\label{Eq:RSkew}
R\ti{12} = c\bigl(I^a \otimes I_a - I_a \otimes I^a\bigr).
\end{equation}
\end{proposition}
\begin{proof}
For $E$ a subspace of $\g$, we denote by $E^\perp$ the orthogonal of $E$ with respect to $\kappa$:
\begin{equation*}
E^\perp = \lbrace X \in \g \; | \; \kappa(X,Y)=0, \, \forall \, Y\in E \rbrace.
\end{equation*}
It is a standard result on quadratic forms that
\begin{equation*}
\null^t \pi_A = \pi_{A'}, \;\;\;\; \null^t \pi_B = \pi_{B'} \;\;\;\; \text{and} \;\;\;\; \null^t \pi_C = \pi_{C'},
\end{equation*}
with $\pi_{A'}$, $\pi_{B'}$ and $\pi_{C'}$ the projections along the decomposition
\begin{equation*}
\g = A' \oplus B' \oplus C', \;\;\;\; \text{ with } \;\;\;\; A'=(B\oplus C)^\perp, \; B'=(A\oplus C)^\perp, \; C'=(A\oplus B)^\perp  .
\end{equation*}
Thus, one has
\begin{equation*}
\null^t R = c(\pi_{A'}-\pi_{B'}),
\end{equation*}
hence $R=-\null^t R$ if and only if $A=B'=(A\oplus C)^\perp$ and $B=A'=(B\oplus C)^\perp$.

If this the case, then one has $\kappa(A,A)=\kappa(B,B)=\kappa(A,C)=\kappa(B,C)=0$, \textit{i.e.} $A$ and $B$ are $\kappa$-isotropic and $C$ is $\kappa$-orthogonal to $A$ and $B$. Conversely, if we suppose the latter, we have $A \subseteq (A \oplus C)^\perp=B'$ and  $B\subseteq(B\oplus C)^\perp=A'$: we then conclude that these inclusions are equalities from dimension considerations. This proves the first part of the proposition.

Let us now suppose that we are in the case described above. As $\kappa$ is non-degenerate and $A$ does not pair with $A \oplus C$, $\kappa$ should pair non-degenerately $A$ and $B$. In the same way, $C$ does not pair with $A \oplus B$ so it should pair non-degenerately with itself. The expression \eqref{Eq:RSkew} then follows from the fact that the kernel of $\pi_A$ and $\pi_B$ are respectively $I^a \otimes I_a$ and $I_a \otimes I^a$, which is straightforward to prove.
\end{proof}

Let us suppose that we are in the case described by Proposition \ref{Prop:SkewAKS}. Let us also consider a basis $\lbrace J^b \rbrace$ of $C$ and the corresponding dual basis $\lbrace J_b \rbrace$, which is then also in $C$. The Casimir can then be written as
\begin{equation*}
C\ti{12} = I^a \otimes I_a + I_a \otimes I^a + J^b \otimes J_b.
\end{equation*}
We will use the two matrices
\begin{subequations}\label{Eq:KernelRpm}
\begin{eqnarray}
R^+\ti{12} = R\ti{12} + c\,C\ti{12} &=& c \bigl(2 \underbrace{I^a \otimes I_a}_{\in A \otimes B} + \underbrace{J^b \otimes J_b}_{\in C\otimes C} \bigr), \\
R^-\ti{12} = R\ti{12} - c\,C\ti{12} &=& -c \bigl(2 \underbrace{I_a \otimes I^a}_{\in B \otimes A} + \underbrace{J^b \otimes J_b}_{\in C\otimes C} \bigr),
\end{eqnarray}
\end{subequations}
which are the kernels of $R^\pm=R \pm c \, \Id$.

\paragraph{AKS matrices for real algebras.} Let us consider a real Lie algebra $\g_0$. We denote by $\g$ its complexification and by $\tau$ the antilinear involutive automorphism of $\g$ such that $\g_0$ is the real form $\g_0=\g^\tau$ of $\g$ (see Appendix \ref{App:RealForms}). Let us suppose that $\g$ admits a $R$-matrix $R$.

\begin{proposition}
The operator $R$ induces a $R$-matrix of $\g_0$ if and only if $R \circ \tau = \tau \circ R$.
\end{proposition}
\begin{proof}
If $R$ stabilises $\g_0$, the induced operator of $\g_0$ also satisfies the mCYBE. Thus one needs to find a \textit{sine qua non} condition for $R$ to stabilise $\g_0=\g^\tau=\left\lbrace X\in\g \; | \; \tau(X)=X \right\rbrace$. It is a classical fact from linear algebra that $R$ and $\tau$ commute if and only if $R$ stabilises all eigenspaces of $\tau$ (as $\tau$ is diagonalisable). Thus $R \circ \tau = \tau \circ R$ implies that $R$ stabilises $\g_0$. Conversely, if $R$ stabilises $\g_0$, it also stabilises the other eigenspace $i \g_0$ of $\tau$, as $R$ is $\C$-linear: then $R$ and $\tau$ commute.
\end{proof}

Conversely, if one has a $R$-matrix on $\g_0$, it can be extended to a matrix $R$ on $\g$, also satisfying the mCYBE equation. In this case, this matrix $R$ commutes with $\tau$.\\

Let us now suppose that $\g$ admits an AKS decomposition \eqref{Eq:gABC}, as described in the first paragraph of this section. We can then apply the AKS scheme to find $R$ matrices on $\g$.

\begin{proposition}\label{Prop:AKSreal}
Let $R$ be the standard AKS operator \eqref{Eq:StandAKS} on $\g$. Then $R$ induces a $R$-matrix on $\g_0$ if and only if:\vspace{-5pt}
\begin{itemize}
\setlength\itemsep{0.1em}
\item $\tau(A)=A$, $\tau(B)=B$ and $\tau(C)=C$, in the split case $c=1$,
\item $\tau(A)=B$ (hence also $\tau(B)=A$) and $\tau(C)=C$, in the non-split case $c=i$.
\end{itemize}
\end{proposition}
\begin{proof}
For $S=A,B,C$, we define $\overline{S}=\tau(S)$ and $\overline{\pi}_S = \tau \circ \pi_S \circ \tau^{-1}$. We have
\begin{equation*}
\overline{\pi}_S \circ \overline{\pi}_{S'} = \tau \circ \bigl( \pi_S \circ \pi_{S'} \bigr) \circ \tau^{-1} = \delta_{SS'}\, \tau \circ \pi_S \circ \tau^{-1} = \delta_{SS'} \,\overline{\pi}_S
\end{equation*}
and
\begin{equation*}
\overline{\pi}_A+\overline{\pi}_B+\overline{\pi}_C=\tau\circ\bigl(\pi_A+\pi_B+\pi_C\bigr)\circ\tau^{-1} = \tau\circ\Id\circ\tau^{-1}=\Id.
\end{equation*}
Thus, the $\overline{\pi}_S$'s form a complete set of projectors. As $\text{Im}(\overline{\pi}_S) = \tau\bigl( \text{Im} (\pi_S) \bigr) = \overline{S}$, they are the projectors associated with the decomposition $\g = \overline{A} \oplus \overline{B} \oplus \overline{C}$. We thus have
\begin{equation*}
\tau \circ \pi_S \circ \tau^{-1} = \pi_{\overline{S}}.
\end{equation*}
In particular, we have
\begin{equation*}
\tau \circ R \circ \tau^{-1} = \overline{c} \bigl( \pi_{\overline{A}} - \pi_{\overline{B}} \bigr). 
\end{equation*}
We will then distinguish the split and non split cases.

If $c=1$ (split case), we see that $\tau$ commutes with $R$ if and only if $\pi_A - \pi_B = \pi_{\overline{A}} - \pi_{\overline{B}}$. As these two operators are diagonalisable with eigenvalues $\lbrace 0,1,-1\rbrace$, they coincide if and only if their eigenspaces are the same, thus if and only if $A=\overline{A}$, $B=\overline{B}$ and $C=\overline{C}$.

If $c=i$ (non-split case), we then have that $\tau$ commutes with $R$ if and only if $\pi_A - \pi_B = \pi_{\overline{B}} - \pi_{\overline{A}}$. Using similar arguments as above, this is equivalent to $A=\overline{B}$ and $B=\overline{A}$ and $C=\overline{C}$.
\end{proof}

Note that in the split case, the subalgebras $S=A,B,C$ are stabilised under the action of $\tau$. Thus one can define $S_0=S^\tau$, which are subalgebras of $\g_0$. We then get a decomposition $\g_0 = A_0 \oplus B_0 \oplus C_0$, which satisfies the conditions of the AKS construction. The standard $R$-matrix on $\g_0$ that one would obtain from this construction then coincides with the restriction of $R$ on $\g_0$.

At the contrary, in the non-split case, the subalgebras $S=A,B,C$ are not stabilised by $\tau$. Thus, they do not possess real forms in $\g_0$ and we cannot interpret the $R$-matrix on $\g_0$ as an AKS operator constructed from real subalgebras of $\g_0$.

\section[Standard $R$-matrices on finite semi-simple Lie algebras.]{Standard $\bm{R}$-matrices on finite semi-simple Lie algebras.}
\label{App:StandardFinite}

Let us consider a finite dimensional complex semi-simple Lie algebra $\g$. It is then described by the formalism developed in Section \ref{App:SemiSimple}. In particular, it possesses a Cartan-Weyl decomposition:
\begin{equation*}
\g = \n_+ \oplus \n_- \oplus \h.
\end{equation*}
The subalgebras $A=\n_+$, $B=\n_-$ and $C=\h$ satisfy the conditions required for the AKS scheme described in Subsection \ref{App:AKS}. Let us then consider the corresponding standard AKS operator
\begin{equation}\label{Eq:AKSss}
R = c(\pi_+-\pi_-),
\end{equation}
where we abbreviated the projections on $\n_\pm$ as $\pi_\pm$.\\

The algebra $\g$ is equipped with a non-degenerate invariant form, the Killing form $\kappa$. The nilpotent subalgebras $\n_\pm$ are isotropic with respect to $\kappa$ and are orthogonal to the Cartan subalgbera $\h$ (see Subsection \ref{App:CartanWeyl}). According to Proposition \ref{Prop:SkewAKS}, the matrix $R$ is thus skew-symmetric.

A basis of $\n_\pm$ is given by the $E_\alpha$'s with $\alpha$ in the positive root system $\Delta_+$. According to equation \eqref{Eq:KillingE}, the corresponding dual basis in $\n_-$ is $\lbrace F_\alpha = E_{-\alpha} \rbrace_{\alpha\in\Delta_+}$. The kernel of $R$ is then given by Proposition \ref{Prop:SkewAKS} as
\begin{equation}\label{Eq:KernelStandardSS}
R\ti{12} = c \sum_{\alpha\in\Delta_+} \bigl( E_\alpha \otimes F_\alpha - F_\alpha \otimes E_\alpha \bigr).
\end{equation}
The matrices $R^\pm\ti{12}$ introduced in equation \eqref{Eq:KernelRpm} are then given by
\begin{equation*}
R^+\ti{12} = 2c \sum_{\alpha\in\Delta_+} E_\alpha \otimes F_\alpha + c \, H\ti{12} \;\;\;\; \text{ and } \;\;\;\; R^-\ti{12} = -2c \sum_{\alpha\in\Delta_+} F_\alpha \otimes E_\alpha - c\,H\ti{12},
\end{equation*}
with $H\ti{12}\in\h\otimes\h$ as introduced in Subsection \ref{App:Casimir}.\\

So far, we constructed a standard AKS operator $R$ on the complex Lie algebra $\g$. In Subsection \ref{App:RealSS}, we discussed real forms of $\g$ and in particularly the split and non-split real forms.

Let us consider the split real form. By equations \eqref{Eq:SplitE} and \eqref{Eq:SplitH}, we see that in this case, the subalgebras $\n_+$, $\n_-$ and $\h$ are stabilised by $\tau$. According to Proposition \ref{Prop:AKSreal}, the split AKS matrix \eqref{Eq:AKSss} for $c=1$ then induces a $R$-matrix on the split real form of $\g$.

Let us consider now the non-split real form. By equations \eqref{Eq:NonSplitE} and \eqref{Eq:NonSplitH}, we see that $\tau$ stabilises $\h$ and exchange $\n_+$ and $\n_-$. By Proposition \ref{Prop:AKSreal}, the non-split AKS matrix \eqref{Eq:AKSss} for $c=i$ then restricts to the non-split real form of $\g$.

\section[Standard $\Rc$-matrices on loop algebras.]{Standard $\bm{\Rc}$-matrices on loop algebras.}
\label{App:StandardLoop}

\paragraph{Loop algebras.} Let us consider a complex Lie algebra $\g$, equipped with an invariant non-degenerate bilinear form $\kappa$. We introduce the loop algebra
\begin{equation*}
\Lc(\g) = \g \pl = \g \otimes \C\pl,
\end{equation*}
where $\C\pl$ denotes the complex Laurent series in a formal variable $\lambda$ (formal power series in $\lambda$ with a finite number of negative powers). We define a Lie bracket on $\Lc(\g)$ by extending the one on $\g$:
\begin{equation*}
[X \otimes f, Y \otimes g ] = [X,Y] \otimes fg, \;\;\;\; \forall \; X, Y \in \g \; \text{ and } \; \forall \; f,g \in \C\pl.
\end{equation*}
Moreover, we extend the bilinear form $\kappa$ on $\Lc(\g)$ as
\begin{equation*}
\langle X \otimes f, Y \otimes g \rangle = \kappa(X,Y) \; \res_{\lambda=0} f(\lambda)g(\lambda)\dd \lambda, \;\;\;\; \forall \; X, Y \in \g \; \text{ and } \; \forall \; f,g \in \C\pl. 
\end{equation*}
One can see elements of $\Lc(\g)$ as $\g$-valued Laurent series in the variable $\lambda$. In this case, the bracket defined above is the point-wise bracket
\begin{equation*}
\forall \, M,N\in\Lc(\g), \;\;\;\; [M,N](\lambda) = \bigl[ M(\lambda),N(\lambda) \bigr].
\end{equation*}
In the same way, the bilinear form above is
\begin{equation}\label{Eq:LoopForm}
\forall \, M,N\in\Lc(\g), \;\;\;\; \langle M,N \rangle = \res_{\lambda=0} \kappa \bigl( M(\lambda), N(\lambda) \bigr) \dd \lambda = \oint \kappa\bigl( M(\lambda), N(\lambda) \bigr) \dd\lambda,
\end{equation}
where the integral is taken on a closed contour around $0$. This bilinear form on $\Lc(\g)$ is invariant and non-degenerate, as $\kappa$ is invariant and non-degenerate.

\paragraph{Standard AKS operator on $\bm{\Lc(\g)}$.} The loop algebra $\Lc(\g)$ admits a natural vector space decomposition
\begin{equation}\label{Eq:LoopDecomposition}
\Lc(\g) = \g\pl = \g[[\lambda]] \oplus \lambda^{-1} \g[\lambda^{-1}]
\end{equation}
into series of positive and strictly negative powers of the loop parameter $\lambda$, respectively. The subspaces $\g[[\lambda]]$ and $\lambda^{-1} \g[\lambda^{-1}]$ are subalgebras of $\Lc(\g)$. Thus, the decomposition \eqref{Eq:LoopDecomposition} satisfies the condition of the AKS scheme described in Section \ref{App:AKS}, with $A=\g[[\lambda]]$, $B=\lambda^{-1} \g[\lambda^{-1}]$ and $C=\lbrace 0 \rbrace$. We denote by $\pi_+$ and $\pi_-$ the projectors along this decomposition. The standard AKS operator
\begin{equation}\label{Eq:LoopAKS}
\Rc = \pi_+ -\pi_-
\end{equation}
is thus a solution of the (split) operator mCYBE \eqref{Eq:AppmCYBE} on $\Lc(\g)$.\\

The algebra $\Lc(\g)$ is equipped with the invariant non-degenerate bilinear form \eqref{Eq:LoopForm}. If $M,N$ belong to $\g[[\lambda]]$, the power series $\kappa\bigl(M(\lambda),N(\lambda)\bigr)$ is composed of positive powers of $\lambda$ and thus has no residue, hence $\langle M,N \rangle=0$. Thus, the subalgebra $\g[[\lambda]]$ is isotropic with respect to $\langle\cdot,\cdot\rangle$. In the same way, if $M,N$ belong to $\lambda^{-1} \g[\lambda^{-1}]$, the power series $\kappa\bigl(M(\lambda),N(\lambda)\bigr)$ is composed of powers of $\lambda$ stritcly inferior to one and thus has no residue. Thus, the subalgebra $\lambda^{-1} \g[\lambda^{-1}]$ is also isotropic. According to Proposition \ref{Prop:SkewAKS}, the operator $\Rc$ is thus skew-symmetric.

\paragraph{Kernels and standard $\bm{\Rc}$-matrix on $\bm{\Lc(\g)}$.} As $\Lc(\g)$ is equipped with a non-degenerate bilinear form, one should be able to consider kernels of operators on $\Lc(\g)$, as described in section \ref{App:CYBE}. However, due to the infinite dimensional nature of $\Lc(\g)$, the definition and manipulation of these kernels are quite subtle. Indeed, the kernel of an operator of $\Lc(\g)$ is not technically an element of the tensor product $\Lc(\g) \otimes \Lc(\g)$ but an element of a completion of this space. We will not enter into details here as this is a technical and subtle matter: we will only present the main ideas needed for this thesis and refer to~\cite{Vicedo:2010qd} for details and rigorous proofs.

The completion of $\Lc(\g) \otimes \Lc(\g)$ mentioned above is composed of infinite series in two variables $\lambda$ and $\mu$, valued in $\g\otimes\g$ (where the variables $\lambda$ and $\mu$ to be the loop variables of respectively the left and right tensor factor in $\Lc(\g) \otimes \Lc(\g)$). These can be seen as distributions in the two variables $\lambda$ and $\mu$. For example, the quadratic Casimir $\widetilde{C}\ti{12}$ of $\Lc(\g)$ is related to the Casimir $C\ti{12}$ of $\g$ by
\begin{equation}\label{Eq:CasLoop}
\widetilde{C}\ti{12}(\lambda,\mu) = C\ti{12} \, \delta(\lambda-\mu),
\end{equation}
where $\delta$ is the Dirac $\delta$-distribution. Without developping the rigorous mathematical formalism behind this, let us motivate this expression for $\Ct$. Let $M\in\Lc(\g)$, which can then be seen as a $\g$-valued Laurent series in $\lambda$. By the second equality in \eqref{Eq:LoopForm}, we then have
\begin{equation*}
\langle \widetilde{C}\ti{12}, M\ti{2} \rangle\ti{2} (\lambda) = \oint \kappa\ti{2} \bigl( \widetilde{C}\ti{12}(\lambda,\mu), M\ti{2}(\mu) \bigr) \dd\mu,
\end{equation*}
where the integral is taken on $\mu$ as it is the loop variable associated with the second tensor factor $\Lc(\g)$. Using the expression \eqref{Eq:CasLoop} of $\Ct$, one then gets
\begin{equation*}
\langle \widetilde{C}\ti{12}, M\ti{2} \rangle\ti{2} (\lambda) = \oint \kappa\ti{2}\bigl( C\ti{12}, M\ti{2}(\mu) \bigr) \delta(\lambda-\mu) \dd \mu = \kappa\ti{2}\bigl(C\ti{12},M\ti{2}(\lambda)\bigr) = M(\lambda),
\end{equation*} 
where we used the definition of the $\delta$-distribution for the second equality and the completeness relation for the finite Casimir $C\ti{12}$ for the last one. Thus, we get
\begin{equation*}
\langle \widetilde{C}\ti{12}, M\ti{2} \rangle\ti{2} = M,
\end{equation*}
as expected for the Casimir of $\Lc(\g)$, as it is the kernel of the identity operator.\\

Let $\Rct\ti{12}$ denote the kernel of the AKS operator $\Rc$ defined in \eqref{Eq:LoopAKS}. It is valued in the completion of $\Lc(\g) \otimes \Lc(\g)$ mentioned above and thus is a $\g\otimes\g$-valued distribution in the two variables $\lambda$ and $\mu$. Moreover, it is a solution of the matricial mCYBE \eqref{Eq:AppmCYBEMat} on $\Lc(\g)$. According to the expression \eqref{Eq:CasLoop} of the Casimir of $\Lc(\g)$, this equation then reads:
\begin{eqnarray}\label{Eq:mCYBEloop}
&&\bigl[ \Rct\ti{12}(\lambda_1,\lambda_2), \Rct\ti{13}(\lambda_1,\lambda_3) \bigr] + \bigl[ \Rct\ti{12}(\lambda_1,\lambda_2), \Rct\ti{23}(\lambda_2,\lambda_3) \bigr] + \bigl[ \Rct\ti{32}(\lambda_3,\lambda_2), \Rct\ti{13}(\lambda_1,\lambda_3) \bigr] \\
&& \hspace{280pt} = \lsb C\ti{12}, C\ti{13} \rsb \delta(\lambda_1-\lambda_2) \delta(\lambda_1-\lambda_3). \notag
\end{eqnarray}
The right-hand side of this equation is called a contact term. It is non-zero only if the loop variables $\lambda_i$'s coincide. Computing the kernel $\Rct\ti{12}$, one finds
\begin{equation*}
\Rct\ti{12}(\lambda,\mu) = \text{p.v.} \; \frac{C\ti{12}}{\mu-\lambda},
\end{equation*}
where $\text{p.v.}$ denotes the principal value (see~\cite{Vicedo:2010qd} for the precise definition). $\Rct\ti{12}(\lambda,\mu)$ is thus a distribution on the two variables $\lambda$ and $\mu$, satisfying the mCYBE \eqref{Eq:mCYBEloop}. Let us consider the same kernel but without the principal value: we then get the \textbf{standard $\bm{\Rc}$-matrix on $\bm{\Lc(\g)}$}, 
\begin{equation*}
\Rc^0\ti{12}(\lambda,\mu) = \frac{C\ti{12}}{\mu-\lambda}
\end{equation*}
as introduced in \eqref{Eq:R0NonTwisted}. Contrarily to $\Rct\ti{12}(\lambda,\mu)$, it is a meromorphic function of $\lambda$ and $\mu$, valued in $\g\otimes\g$, and not a distribution. For different values of the loop variables $\lambda$ and $\mu$, the fraction $(\mu-\lambda)^{-1}$ coincide with its principal value, hence the kernels $\Rc^0\ti{12}$ and $\Rct\ti{12}$ coincide for $\lambda \neq \mu$. Computing the left-hand side of \eqref{Eq:mCYBEloop} for $\Rc^0$ instead of $\Rct$, one can show that considering the meromorphic function instead of its principal value as for effect to discard the distribution-valued contact term, which is not zero only when the three loop variables are equal. Thus, one finds that $\Rc^0$ is a solution of the non-modified CYBE on $\Lc(\g)$
\begin{equation*}
\lsb \Rc^0\ti{12}(\lambda_1,\lambda_2), \Rc^0\ti{13}(\lambda_1,\lambda_3) \rsb + \lsb \Rc^0\ti{12}(\lambda_1,\lambda_2), \Rc^0\ti{23}(\lambda_2,\lambda_3) \rsb + \lsb \Rc^0\ti{32}(\lambda_3,\lambda_2), \Rc^0\ti{13}(\lambda_1,\lambda_3) \rsb = 0.
\end{equation*}

\paragraph{Twist function and kernel.} In the previous paragraph, we constructed the standard $\Rc$-matrix on $\Lc(\g)$ by considering the kernel of the operator \eqref{Eq:LoopAKS} with respect to the bilinear form \eqref{Eq:LoopForm}. Let us slightly modify the bilinear form:
\begin{equation}\label{Eq:LoopFormTwist}
\forall \, M,N\in\Lc(\g), \;\;\;\; \langle M,N \rangle_\vp = \res_{\lambda=0} \kappa \bigl( M(\lambda), N(\lambda) \bigr) \vp(\lambda) \dd \lambda = \oint \kappa\bigl( M(\lambda), N(\lambda) \bigr) \vp(\lambda) \dd\lambda,
\end{equation}
where $\vp$ is a rational functional of $\lambda$.

This bilinear form is also invariant and non-degenerate on $\Lc(\g)$. One can then define the corresponding kernel of the operator $\Rc$, which will then be a solution of the mCYBE \eqref{Eq:mCYBEloop}. As in the previous paragraph, considering a meromorphic function instead of its principal value, one obtains a matrix $\Rc\ti{12}(\lambda,\mu)$ solution of the non-modified CYBE. This matrix reads
\begin{equation}\label{Eq:RPhiApp}
\Rc\ti{12}(\lambda,\mu) = \frac{C\ti{12}}{\mu-\lambda} \vp(\mu)^{-1} = \Rc^0\ti{12}(\lambda,\mu) \vp(\mu)^{-1}.
\end{equation}
We recognize here the $\Rc$-matrix \eqref{Eq:DefR} introduced in Chapter \ref{Chap:Lax}. The function $\vp$ is then what we called the \textbf{twist function} in this chapter.

Note that for $\vp$ different from a constant function, the subalgebras $\g[[\lambda]]$ and $\lambda^{-1} \g[\lambda^{-1}]$ of $\Lc(\g)$ are not isotropic with respect to the bilinear form $\langle\cdot,\cdot\rangle_\vp$. Thus, we get a non skew-symmetric kernel $\Rc\ti{12}(\lambda,\mu)$, as we can directly observe on equation \eqref{Eq:RPhiApp}.

\paragraph{Twisted $\bm{\Rc}$-matrices.} We end this section by discussing the loop interpretation of the twisted standard $\Rc$-matrices \eqref{Eq:RCyc} on $\Lc(\g)$. Suppose that we are given an automorphism $\s$ of $\g$, of finite order $T$. Let $\omega$ be a $T^{\rm th}$-root of unity. We define an endomorphism $\widehat{\s}$ on $\Lc(\g)$ by
\begin{equation*}
\forall \; M\in\Lc(\g), \;\;\;\; \widehat\s (M) (\lambda) = \s \bigl( M(\omega^{-1}\lambda) \bigr).
\end{equation*}
This is an automorphism of $\Lc(\g)$, also of order $T$. We define the twisted loop algebra by $\s$ as the subalgebra of fixed points of $\widehat{\s}$:
\begin{equation*}
\Lc(\g,\s) = \Lc(\g)^{\widehat{\s}}.
\end{equation*}
The elements of $\Lc(\g,\s)$ are then equivariant $\g$-valued Laurent series, in the sense that:
\begin{equation*}
\forall \; M\in\Lc(\g), \;\;\;\; M\in\Lc(\g,\s) \Longleftrightarrow \s\bigl( M(\lambda) \bigr) = M(\omega\lambda).
\end{equation*}
The subalgebras $\g[[\lambda]]$ and $\lambda^{-1} \g[\lambda^{-1}]$ of $\Lc(\g)$ are stabilised by $\widehat{\s}$. Thus, the decomposition \eqref{Eq:LoopDecomposition} of $\Lc(\g)$ induces the following decomposition of $\Lc(\g,\s)$;
\begin{equation*}
\Lc(\g,\s) = \g[[\lambda]]^{\widehat{\s}} \oplus \lambda^{-1} \g[\lambda^{-1}]^{\widehat{\s}}.
\end{equation*}
This defines an AKS decomposition on $\Lc(\g,\s)$. We denote by $\pi_{+}^\s$ and $\pi_-^\s$ the projectors along this decomposition. Applying the AKS scheme, we get an operator
\begin{equation*}
\Rc^\s = \pi_{+}^\s - \pi_{-}^\s
\end{equation*}
of $\Lc(\g,\s)$, solution of the operator mCYBE.

Recall the invariant non-degenerate bilinear form \eqref{Eq:LoopFormTwist} on $\Lc(\g)$, twisted by the function $\vp(\lambda)$. This form reduces to an invariant non-degenerate form on $\Lc(\g,\s)=\Lc(\g)^{\widehat{\s}}$ if it is invariant under the action of $\widehat{\s}$, \textit{i.e.} if
\begin{equation*}
\langle \widehat{\s}(M), \widehat{\s}(N) \rangle_\vp = \langle M,N \rangle_\vp, \;\;\;\ \forall \, M,N \in \Lc(\g).
\end{equation*}
This is true if and only if the twist function $\vp$ satisfies the equivariance condition
\begin{equation*}
\vp(\omega\lambda) = \omega^{-1} \vp(\lambda).
\end{equation*}
We recognize here the equivariance condition \eqref{Eq:TwistEqui} discussed in Chapter \ref{Chap:Lax}. We will now suppose that this condition is verified, so that $\langle\cdot,\cdot\rangle_\vp$ defines an invariant non-degenerate bilinear form on $\Lc(\g,\s)$. We can then construct the kernel of the operator $\Rc^\s$. As in the previous subsections, considering meromorphic functions instead of principal values, we get from this kernel a matrix solution of the non-modified CYBE:
\begin{equation*}
\Rc\ti{12}(\lambda,\mu) = \Rc^0\ti{12}(\lambda,\mu) \vp(\mu)^{-1},
\end{equation*}
where
\begin{equation*}
\Rc^0\ti{12}(\lambda,\mu) = \frac{1}{T}\sum_{k=0}^{T-1} \frac{\s\ti{1}^k C\ti{12}}{\mu-\omega^{-k}\lambda}.
\end{equation*}
We recognize here the standard $\Rc$-matrix twisted by $\s$, as introduced in \eqref{Eq:RCyc}.

\part{Bibliography}

\bibliography{fr,fr2,fr3,gaudin,neumann,ater-sylvain,Sylvain}

\providecommand{\href}[2]{#2}\begingroup\raggedright\begin{thebibliography}{100}

\bibitem{Delduc:2015xdm}
F.~Delduc, S.~Lacroix, M.~Magro and B.~Vicedo, \emph{{On the Hamiltonian
  integrability of the bi-Yang-Baxter $\sigma$-model}},
  \href{https://doi.org/10.1007/JHEP03(2016)104}{\emph{JHEP} {\bfseries 03}
  (2016) 104} [\href{https://arxiv.org/abs/1512.02462}{{\ttfamily
  1512.02462}}].

\bibitem{Delduc:2016ihq}
F.~Delduc, S.~Lacroix, M.~Magro and B.~Vicedo, \emph{{On q-deformed symmetries
  as Poisson-Lie symmetries and application to Yang-Baxter type models}},
  \href{https://doi.org/10.1088/1751-8113/49/41/415402}{\emph{J. Phys.}
  {\bfseries A49} (2016) 415402}
  [\href{https://arxiv.org/abs/1606.01712}{{\ttfamily 1606.01712}}].

\bibitem{Lacroix:2017isl}
S.~Lacroix, M.~Magro and B.~Vicedo, \emph{{Local charges in involution and
  hierarchies in integrable sigma-models}},
  \href{https://doi.org/10.1007/JHEP09(2017)117}{\emph{JHEP} {\bfseries 09}
  (2017) 117} [\href{https://arxiv.org/abs/1703.01951}{{\ttfamily
  1703.01951}}].

\bibitem{Lacroix:2016mpg}
S.~Lacroix and B.~Vicedo, \emph{{Cyclotomic Gaudin models, Miura opers and flag
  varieties}}, \href{https://doi.org/10.1007/s00023-017-0616-8}{\emph{Annales
  Henri Poincar\'e} {\bfseries 19} (2018) 71}
  [\href{https://arxiv.org/abs/1607.07397}{{\ttfamily 1607.07397}}].

\bibitem{Lacroix:2018fhf}
S.~Lacroix, B.~Vicedo and C.~A.~S. Young, \emph{{Affine Gaudin models and
  hypergeometric functions on affine opers}},
  \href{https://arxiv.org/abs/1804.01480}{{\ttfamily 1804.01480}}.

\bibitem{Lacroix:2018cag}
S.~Lacroix, B.~Vicedo and C.~A.~S. Young, \emph{{Cubic hypergeometric integrals
  of motion in affine Gaudin models}},
  \href{https://arxiv.org/abs/1804.06751}{{\ttfamily 1804.06751}}.

\bibitem{Maldacena:1997re}
J.~M. Maldacena, \emph{{The Large N limit of superconformal field theories and
  supergravity}}, \href{https://doi.org/10.1023/A:1026654312961}{\emph{Adv.
  Theor. Math. Phys.} {\bfseries 2} (1998) 231}
  [\href{https://arxiv.org/abs/hep-th/9711200}{{\ttfamily hep-th/9711200}}].

\bibitem{Gubser:1998bc}
S.~Gubser, I.~R. Klebanov and A.~M. Polyakov, \emph{{Gauge theory correlators
  from noncritical string theory}},
  \href{https://doi.org/10.1016/S0370-2693(98)00377-3}{\emph{Phys. Lett.}
  {\bfseries B428} (1998) 105}
  [\href{https://arxiv.org/abs/hep-th/9802109}{{\ttfamily hep-th/9802109}}].

\bibitem{Witten:1998qj}
E.~Witten, \emph{{Anti-de Sitter space and holography}},
  \href{https://doi.org/10.4310/ATMP.1998.v2.n2.a2}{\emph{Adv. Theor. Math.
  Phys.} {\bfseries 2} (1998) 253}
  [\href{https://arxiv.org/abs/hep-th/9802150}{{\ttfamily hep-th/9802150}}].

\bibitem{Metsaev:1998it}
R.~Metsaev and A.~A. Tseytlin, \emph{{Type IIB superstring action in $AdS_5
  \times S^5$ background}},
  \href{https://doi.org/10.1016/S0550-3213(98)00570-7}{\emph{Nucl. Phys.}
  {\bfseries B533} (1998) 109}
  [\href{https://arxiv.org/abs/hep-th/9805028}{{\ttfamily hep-th/9805028}}].

\bibitem{Bena:2003wd}
I.~Bena, J.~Polchinski and R.~Roiban, \emph{{Hidden symmetries of the AdS$_5$
  $\times$ S$^5$ superstring}},
  \href{https://doi.org/10.1103/PhysRevD.69.046002}{\emph{Phys. Rev.}
  {\bfseries D69} (2004) 046002}
  [\href{https://arxiv.org/abs/hep-th/0305116}{{\ttfamily hep-th/0305116}}].

\bibitem{Magro:2008dv}
M.~Magro, \emph{{The classical exchange algebra of $AdS_5 \times S^5$ string
  theory}}, \href{https://doi.org/10.1088/1126-6708/2009/01/021}{\emph{JHEP}
  {\bfseries 0901} (2009) 021}
  [\href{https://arxiv.org/abs/0810.4136}{{\ttfamily 0810.4136}}].

\bibitem{Vicedo:2009sn}
B.~Vicedo, \emph{{Hamiltonian dynamics and the hidden symmetries of the $AdS_5
  \times S^5$ superstring}},
  \href{https://doi.org/10.1007/JHEP01(2010)102}{\emph{JHEP} {\bfseries 1001}
  (2010) 102} [\href{https://arxiv.org/abs/0910.0221}{{\ttfamily 0910.0221}}].

\bibitem{Beisert:2010jr}
N.~{Beisert}, C.~{Ahn}, L.~F. {Alday}, Z.~{Bajnok}, J.~M. {Drummond},
  L.~{Freyhult} et~al., \emph{{Review of AdS/CFT Integrability: An Overview}},
  \href{https://doi.org/10.1007/s11005-011-0529-2}{\emph{Lett. Math. Phys.}
  {\bfseries 99} (2012) 3} [\href{https://arxiv.org/abs/1012.3982}{{\ttfamily
  1012.3982}}].

\bibitem{Lax:1968fm}
P.~D. Lax, \emph{{Integrals of Nonlinear Equations of Evolution and Solitary
  Waves}}, \href{https://doi.org/10.1002/cpa.3160210503}{\emph{Commun. Pure
  Appl. Math.} {\bfseries 21} (1968) 467}.

\bibitem{Gelfand:1951}
I.~Gelfand and B.~Levitan, \emph{{On The Determination of a Differential
  Equation by its Spectral Function}}, {\emph{Izvestiya Akad. Nauk SSR, Ser.
  Mat.} {\bfseries 15} (1951) 309}.

\bibitem{Marchenko:1955}
V.~A. Marchenko, \emph{{On the reconstruction of the potential energy from
  phases of the scattered waves.}}, {\emph{Dokl. Akad. Nauk SSSR} {\bfseries
  104} (1955) 695}.

\bibitem{Gardner:1967}
C.~S. Gardner, J.~M. Greene, M.~D. Kruskal and R.~M. Miura, \emph{{Method for
  Solving the Korteweg-deVries Equation}},
  \href{https://doi.org/10.1103/PhysRevLett.19.1095}{\emph{Physical Review
  Letters} {\bfseries 19} (1967) 1095}.

\bibitem{Zakharov:1970tma}
V.~E. Zakharov and A.~B. Shabat, \emph{{Exact theory of two-dimensional
  self-focusing and one-dimensional self-modulation of waves in nonlinear
  media}}, {\emph{Sov. Phys. JETP} {\bfseries 34} (1972) 62}.

\bibitem{Ablowitz:1973}
M.~J. Ablowitz, D.~J. Kaup, A.~C. Newell and H.~Segur, \emph{{Method for
  Solving the Sine-Gordon Equation}},
  \href{https://doi.org/10.1103/PhysRevLett.30.1262}{\emph{Phys. Rev. Lett.}
  {\bfseries 30} (1973) 1262}.

\bibitem{Babebook}
O.~Babelon, D.~Bernard and M.~Talon, \emph{{Introduction to Classical
  Integrable Models}}. Cambridge University Press, 2003.

\bibitem{fadtakbook87}
L.~Faddeev and L.~L.A.~Takhtajan, \emph{{Hamiltonian methods in the theory of
  solitons}}. Springer, 1987.

\bibitem{Torrielli:2016ufi}
A.~Torrielli, \emph{{Lectures on Classical Integrability}},
  \href{https://doi.org/10.1088/1751-8113/49/32/323001}{\emph{J. Phys.}
  {\bfseries A49} (2016) 323001}
  [\href{https://arxiv.org/abs/1606.02946}{{\ttfamily 1606.02946}}].

\bibitem{Sklyanin:1982tf}
E.~Sklyanin, \emph{{Some algebraic structures connected with the Yang-Baxter
  equation}}, {\emph{Funct. Anal. Appl.} {\bfseries 16} (1982) 263}.

\bibitem{Maillet:1985fn}
J.~M. Maillet, \emph{{Kac-Moody algebra and extended Yang-Baxter relations in
  the $O(N)$ non-linear sigma model}},
  \href{https://doi.org/10.1016/0370-2693(85)91075-5}{\emph{Phys. Lett.}
  {\bfseries B162} (1985) 137}.

\bibitem{Maillet:1985ek}
J.~M. Maillet, \emph{{New integrable canonical structures in two-dimensional
  models}}, \href{https://doi.org/10.1016/0550-3213(86)90365-2}{\emph{Nucl.
  Phys.} {\bfseries B269} (1986) 54}.

\bibitem{HALDANE1983464}
F.~Haldane, \emph{{Continuum dynamics of the 1-D Heisenberg antiferromagnet:
  Identification with the O(3) nonlinear sigma model}},
  \href{https://doi.org/http://dx.doi.org/10.1016/0375-9601(83)90631-X}{\emph{Physics
  Letters A} {\bfseries 93} (1983) 464 }.

\bibitem{Cherednik:1981df}
I.~Cherednik, \emph{{Relativistically invariant quasiclassical limits of
  integrable two-dimensional quantum models}},
  \href{https://doi.org/10.1007/BF01086395}{\emph{Theor. Math. Phys.}
  {\bfseries 47} (1981) 422}.

\bibitem{Balog:1993es}
J.~Balog, P.~Forg\'acs, Z.~Horv\'ath and L.~Palla, \emph{{A New family of SU(2)
  symmetric integrable sigma models}},
  \href{https://doi.org/10.1016/0370-2693(94)90213-5}{\emph{Phys. Lett.}
  {\bfseries B324} (1994) 403}
  [\href{https://arxiv.org/abs/hep-th/9307030}{{\ttfamily hep-th/9307030}}].

\bibitem{Rajeev:1988hq}
S.~Rajeev, \emph{{Nonabelian bosonization without Wess-Zumino terms. I. New
  current algebra}},
  \href{https://doi.org/10.1016/0370-2693(89)91528-1}{\emph{Phys. Lett.}
  {\bfseries B217} (1989) 123}.

\bibitem{Fateev:1996ea}
V.~Fateev, \emph{{The sigma model (dual) representation for a two-parameter
  family of integrable quantum field theories}},
  \href{https://doi.org/10.1016/0550-3213(96)00256-8}{\emph{Nucl. Phys.}
  {\bfseries B473} (1996) 509}.

\bibitem{Fateev:1992tk}
V.~Fateev, E.~Onofri and A.~B. Zamolodchikov, \emph{{Integrable deformations of
  the O(3) sigma model. The sausage model}},
  \href{https://doi.org/10.1016/0550-3213(93)90001-6}{\emph{Nucl. Phys.}
  {\bfseries B406} (1993) 521}.

\bibitem{Klimcik:2002zj}
C.~Klim\v{c}ik, \emph{{Yang-Baxter sigma models and dS/AdS T duality}},
  {\emph{JHEP} {\bfseries 0212} (2002) 051}
  [\href{https://arxiv.org/abs/hep-th/0210095}{{\ttfamily hep-th/0210095}}].

\bibitem{Klimcik:2008eq}
C.~Klim\v{c}ik, \emph{{On integrability of the Yang-Baxter $\sigma$-model}},
  \href{https://doi.org/10.1063/1.3116242}{\emph{J. Math. Phys.} {\bfseries 50}
  (2009) 043508} [\href{https://arxiv.org/abs/0802.3518}{{\ttfamily
  0802.3518}}].

\bibitem{Delduc:2013fga}
F.~Delduc, M.~Magro and B.~Vicedo, \emph{{On classical $q$-deformations of
  integrable $\sigma$-models}},
  \href{https://doi.org/10.1007/JHEP11(2013)192}{\emph{JHEP} {\bfseries 1311}
  (2013) 192} [\href{https://arxiv.org/abs/1308.3581}{{\ttfamily 1308.3581}}].

\bibitem{Hoare:2014pna}
B.~Hoare, R.~Roiban and A.~Tseytlin, \emph{{On deformations of $AdS_n \times
  S^n$ supercosets}},
  \href{https://doi.org/10.1007/JHEP06(2014)002}{\emph{JHEP} {\bfseries 1406}
  (2014) 002} [\href{https://arxiv.org/abs/1403.5517}{{\ttfamily 1403.5517}}].

\bibitem{Delduc:2013qra}
F.~Delduc, M.~Magro and B.~Vicedo, \emph{{Integrable deformation of the $AdS_5
  \times S^5$ superstring action}},
  \href{https://doi.org/10.1103/PhysRevLett.112.051601}{\emph{Phys. Rev. Lett.}
  {\bfseries 112} (2014) 051601}
  [\href{https://arxiv.org/abs/1309.5850}{{\ttfamily 1309.5850}}].

\bibitem{Delduc:2014kha}
F.~Delduc, M.~Magro and B.~Vicedo, \emph{{Derivation of the action and
  symmetries of the $q$-deformed $AdS_5 \times S^5$ superstring}},
  \href{https://doi.org/10.1007/JHEP10(2014)132}{\emph{JHEP} {\bfseries 1410}
  (2014) 132} [\href{https://arxiv.org/abs/1406.6286}{{\ttfamily 1406.6286}}].

\bibitem{Klimcik:2014bta}
C.~Klim\v{c}ik, \emph{{Integrability of the Bi-Yang-Baxter $\sigma$-model}},
  \href{https://doi.org/10.1007/s11005-014-0709-y}{\emph{Lett. Math. Phys.}
  {\bfseries 104} (2014) 1095}
  [\href{https://arxiv.org/abs/1402.2105}{{\ttfamily 1402.2105}}].

\bibitem{Kawaguchi:2011pf}
I.~Kawaguchi and K.~Yoshida, \emph{{Hybrid classical integrability in squashed
  sigma models}},
  \href{https://doi.org/10.1016/j.physletb.2011.09.117}{\emph{Phys. Lett.}
  {\bfseries B705} (2011) 251}
  [\href{https://arxiv.org/abs/1107.3662}{{\ttfamily 1107.3662}}].

\bibitem{Kawaguchi:2012gp}
I.~Kawaguchi, T.~Matsumoto and K.~Yoshida, \emph{{On the classical equivalence
  of monodromy matrices in squashed sigma model}},
  \href{https://doi.org/10.1007/JHEP06(2012)082}{\emph{JHEP} {\bfseries 1206}
  (2012) 082} [\href{https://arxiv.org/abs/1203.3400}{{\ttfamily 1203.3400}}].

\bibitem{Delduc:2017brb}
F.~Delduc, T.~Kameyama, M.~Magro and B.~Vicedo, \emph{{Affine $q$-deformed
  symmetry and the classical Yang-Baxter $\sigma$-model}},
  \href{https://doi.org/10.1007/JHEP03(2017)126}{\emph{JHEP} {\bfseries 03}
  (2017) 126} [\href{https://arxiv.org/abs/1701.03691}{{\ttfamily
  1701.03691}}].

\bibitem{Kawaguchi:2012ve}
I.~Kawaguchi, T.~Matsumoto and K.~Yoshida, \emph{{The classical origin of
  quantum affine algebra in squashed sigma models}},
  \href{https://doi.org/10.1007/JHEP04(2012)115}{\emph{JHEP} {\bfseries 1204}
  (2012) 115} [\href{https://arxiv.org/abs/1201.3058}{{\ttfamily 1201.3058}}].

\bibitem{Kawaguchi:2014qwa}
I.~Kawaguchi, T.~Matsumoto and K.~Yoshida, \emph{{Jordanian deformations of the
  $AdS_5 \times S^5$ superstring}},
  \href{https://doi.org/10.1007/JHEP04(2014)153}{\emph{JHEP} {\bfseries 1404}
  (2014) 153} [\href{https://arxiv.org/abs/1401.4855}{{\ttfamily 1401.4855}}].

\bibitem{Sfetsos:2013wia}
K.~Sfetsos, \emph{{Integrable interpolations: From exact CFTs to non-Abelian
  T-duals}}, \href{https://doi.org/10.1016/j.nuclphysb.2014.01.004}{\emph{Nucl.
  Phys.} {\bfseries B880} (2014) 225}
  [\href{https://arxiv.org/abs/1312.4560}{{\ttfamily 1312.4560}}].

\bibitem{Hollowood:2014rla}
T.~J. Hollowood, J.~L. Miramontes and D.~M. Schmidtt, \emph{{Integrable
  deformations of strings on symmetric spaces}},
  \href{https://doi.org/10.1007/JHEP11(2014)009}{\emph{JHEP} {\bfseries 1411}
  (2014) 009} [\href{https://arxiv.org/abs/1407.2840}{{\ttfamily 1407.2840}}].

\bibitem{Hollowood:2014qma}
T.~J. Hollowood, J.~L. Miramontes and D.~M. Schmidtt, \emph{{An integrable
  deformation of the $AdS_5 \times S^5$ superstring}},
  \href{https://doi.org/10.1088/1751-8113/47/49/495402}{\emph{J. Phys.}
  {\bfseries A47} (2014) 495402}
  [\href{https://arxiv.org/abs/1409.1538}{{\ttfamily 1409.1538}}].

\bibitem{Wess:1971yu}
J.~Wess and B.~Zumino, \emph{{Consequences of anomalous Ward identities}},
  \href{https://doi.org/10.1016/0370-2693(71)90582-X}{\emph{Phys. Lett.}
  {\bfseries 37B} (1971) 95}.

\bibitem{Novikov:1982ei}
S.~P. Novikov, \emph{{The Hamiltonian formalism and a many valued analog of
  Morse theory}},
  \href{https://doi.org/10.1070/RM1982v037n05ABEH004020}{\emph{Usp. Mat. Nauk}
  {\bfseries 37N5} (1982) 3}.

\bibitem{Witten:1983ar}
E.~Witten, \emph{{Nonabelian Bosonization in Two Dimensions}},
  \href{https://doi.org/10.1007/BF01215276}{\emph{Commun.Math.Phys.} {\bfseries
  92} (1984) 455}.

\bibitem{Abdalla:1982yd}
E.~Abdalla, M.~Forger and M.~Gomes, \emph{{On the origin of anomalies in the
  quantum nonlocal charge for the generalized nonlinear sigma models}},
  \href{https://doi.org/10.1016/0550-3213(82)90238-3}{\emph{Nucl. Phys.}
  {\bfseries B210} (1982) 181}.

\bibitem{Delduc:2014uaa}
F.~Delduc, M.~Magro and B.~Vicedo, \emph{{Integrable double deformation of the
  principal chiral model}},
  \href{https://doi.org/10.1016/j.nuclphysb.2014.12.018}{\emph{Nucl. Phys.}
  {\bfseries B891} (2015) 312}
  [\href{https://arxiv.org/abs/1410.8066}{{\ttfamily 1410.8066}}].

\bibitem{Delduc:2017fib}
F.~Delduc, B.~Hoare, T.~Kameyama and M.~Magro, \emph{{Combining the
  bi-Yang-Baxter deformation, the Wess-Zumino term and TsT transformations in
  one integrable $\sigma$-model}},
  \href{https://doi.org/10.1007/JHEP10(2017)212}{\emph{JHEP} {\bfseries 10}
  (2017) 212} [\href{https://arxiv.org/abs/1707.08371}{{\ttfamily
  1707.08371}}].

\bibitem{Lukyanov:2012zt}
S.~L. Lukyanov, \emph{{The integrable harmonic map problem versus Ricci flow}},
  \href{https://doi.org/10.1016/j.nuclphysb.2012.08.002}{\emph{Nucl. Phys.}
  {\bfseries B865} (2012) 308}
  [\href{https://arxiv.org/abs/1205.3201}{{\ttfamily 1205.3201}}].

\bibitem{Arutyunov:2013ega}
G.~Arutyunov, R.~Borsato and S.~Frolov, \emph{{S-matrix for strings on
  $\eta$-deformed $AdS_{5} \times S^5$}},
  \href{https://doi.org/10.1007/JHEP04(2014)002}{\emph{JHEP} {\bfseries 1404}
  (2014) 002} [\href{https://arxiv.org/abs/1312.3542}{{\ttfamily 1312.3542}}].

\bibitem{Arutyunov:2015qva}
G.~Arutyunov, R.~Borsato and S.~Frolov, \emph{{Puzzles of $\eta$-deformed
  AdS$_5 \times$ S$^5$}},
  \href{https://doi.org/10.1007/JHEP12(2015)049}{\emph{JHEP} {\bfseries 12}
  (2015) 049} [\href{https://arxiv.org/abs/1507.04239}{{\ttfamily
  1507.04239}}].

\bibitem{Borsato:2016hud}
R.~Borsato, \emph{{Integrable strings for AdS/CFT}}, Ph.D. thesis, 2015.
\newblock \href{https://arxiv.org/abs/1605.03173}{{\ttfamily 1605.03173}}.

\bibitem{Beisert:2008tw}
N.~Beisert and P.~Koroteev, \emph{{Quantum Deformations of the One-Dimensional
  Hubbard Model}},
  \href{https://doi.org/10.1088/1751-8113/41/25/255204}{\emph{J.Phys.A}
  {\bfseries A41} (2008) 255204}
  [\href{https://arxiv.org/abs/0802.0777}{{\ttfamily 0802.0777}}].

\bibitem{Beisert:2010kk}
N.~Beisert, \emph{{The Classical Trigonometric r-Matrix for the
  Quantum-Deformed Hubbard Chain}},
  \href{https://doi.org/10.1088/1751-8113/44/26/265202}{\emph{J. Phys. A}
  {\bfseries A44} (2011) 265202}
  [\href{https://arxiv.org/abs/1002.1097}{{\ttfamily 1002.1097}}].

\bibitem{Hoare:2011wr}
B.~Hoare, T.~J. Hollowood and J.~L. Miramontes, \emph{{q-Deformation of the
  $AdS_5 \times S^5$ Superstring S-matrix and its Relativistic Limit}},
  \href{https://doi.org/10.1007/JHEP03(2012)015}{\emph{JHEP} {\bfseries 1203}
  (2012) 015} [\href{https://arxiv.org/abs/1112.4485}{{\ttfamily 1112.4485}}].

\bibitem{Staudacher:2004tk}
M.~Staudacher, \emph{{The factorized S-matrix of CFT/AdS}},
  \href{https://doi.org/10.1088/1126-6708/2005/05/054}{\emph{JHEP} {\bfseries
  0505} (2005) 054} [\href{https://arxiv.org/abs/hep-th/0412188}{{\ttfamily
  hep-th/0412188}}].

\bibitem{Beisert:2005tm}
N.~Beisert, \emph{{The su(2$/$2) dynamic S-matrix}},
  \href{https://doi.org/10.4310/ATMP.2008.v12.n5.a1}{\emph{Adv.Theor.Math.Phys.}
  {\bfseries 12} (2008) 945}
  [\href{https://arxiv.org/abs/hep-th/0511082}{{\ttfamily hep-th/0511082}}].

\bibitem{Beisert:2006qh}
N.~Beisert, \emph{{The Analytic Bethe Ansatz for a Chain with Centrally
  Extended su(2$/$2) Symmetry}},
  \href{https://doi.org/10.1088/1742-5468/2007/01/P01017}{\emph{J.Stat.Mech.}
  {\bfseries 0701} (2007) P01017}
  [\href{https://arxiv.org/abs/nlin/0610017}{{\ttfamily nlin/0610017}}].

\bibitem{Janik:2006dc}
R.~A. Janik, \emph{{The AdS$_5$ $\times$ S$^5$ superstring worldsheet S-matrix
  and crossing symmetry}},
  \href{https://doi.org/10.1103/PhysRevD.73.086006}{\emph{Phys. Rev.}
  {\bfseries D73} (2006) 086006}
  [\href{https://arxiv.org/abs/hep-th/0603038}{{\ttfamily hep-th/0603038}}].

\bibitem{Arutyunov:2006iu}
G.~Arutyunov and S.~Frolov, \emph{{On AdS$_5$ $\times$ S$^5$ string S-matrix}},
  \href{https://doi.org/10.1016/j.physletb.2006.06.064}{\emph{Phys. Lett.}
  {\bfseries B639} (2006) 378}
  [\href{https://arxiv.org/abs/hep-th/0604043}{{\ttfamily hep-th/0604043}}].

\bibitem{Beisert:2006ib}
N.~Beisert, R.~Hernandez and E.~Lopez, \emph{{A crossing-symmetric phase for
  AdS$_5$ $\times$ S$^5$ strings}},
  \href{https://doi.org/10.1088/1126-6708/2006/11/070}{\emph{JHEP} {\bfseries
  0611} (2006) 070} [\href{https://arxiv.org/abs/hep-th/0609044}{{\ttfamily
  hep-th/0609044}}].

\bibitem{Beisert:2006ez}
N.~Beisert, B.~Eden and M.~Staudacher, \emph{{Transcendentality and crossing}},
  \href{https://doi.org/10.1088/1742-5468/2007/01/P01021}{\emph{J.Stat.Mech.}
  {\bfseries 0701} (2007) P01021}
  [\href{https://arxiv.org/abs/hep-th/0610251}{{\ttfamily hep-th/0610251}}].

\bibitem{Arutyunov:2006yd}
G.~Arutyunov, S.~Frolov and M.~Zamaklar, \emph{{The Zamolodchikov-Faddeev
  algebra for AdS$_5$ $\times$ S$^5$ superstring}},
  \href{https://doi.org/10.1088/1126-6708/2007/04/002}{\emph{JHEP} {\bfseries
  0704} (2007) 002} [\href{https://arxiv.org/abs/hep-th/0612229}{{\ttfamily
  hep-th/0612229}}].

\bibitem{Arutyunov:2009ga}
G.~Arutyunov and S.~Frolov, \emph{{Foundations of the AdS$_5$ $\times$ S$^5$
  Superstring. Part I}},
  \href{https://doi.org/10.1088/1751-8113/42/25/254003}{\emph{J. Phys.}
  {\bfseries A42} (2009) 254003}
  [\href{https://arxiv.org/abs/0901.4937}{{\ttfamily 0901.4937}}].

\bibitem{Ahn:2010ka}
C.~Ahn and R.~I. Nepomechie, \emph{{Review of AdS/CFT Integrability, Chapter
  III.2: Exact World-Sheet S-Matrix}},
  \href{https://doi.org/10.1007/s11005-011-0478-9}{\emph{Lett.Math.Phys.}
  {\bfseries 99} (2012) 209} [\href{https://arxiv.org/abs/1012.3991}{{\ttfamily
  1012.3991}}].

\bibitem{Borsato:2016ose}
R.~Borsato and L.~Wulff, \emph{{Target space supergeometry of $\eta$ and
  $\lambda$-deformed strings}},
  \href{https://doi.org/10.1007/JHEP10(2016)045}{\emph{JHEP} {\bfseries 10}
  (2016) 045} [\href{https://arxiv.org/abs/1608.03570}{{\ttfamily
  1608.03570}}].

\bibitem{Borsato:2016zcf}
R.~Borsato, A.~A. Tseytlin and L.~Wulff, \emph{{Supergravity background of
  $\lambda$-deformed model for AdS$_2 \times$ S$^2$ supercoset}},
  \href{https://doi.org/10.1016/j.nuclphysb.2016.02.018}{\emph{Nucl. Phys.}
  {\bfseries B905} (2016) 264}
  [\href{https://arxiv.org/abs/1601.08192}{{\ttfamily 1601.08192}}].

\bibitem{Chervonyi:2016ajp}
Y.~Chervonyi and O.~Lunin, \emph{{Supergravity background of the
  $\lambda$-deformed AdS$_3 \times$ S$^3$ supercoset}},
  \href{https://doi.org/10.1016/j.nuclphysb.2016.07.023}{\emph{Nucl. Phys.}
  {\bfseries B910} (2016) 685}
  [\href{https://arxiv.org/abs/1606.00394}{{\ttfamily 1606.00394}}].

\bibitem{Chervonyi:2016bfl}
Y.~Chervonyi and O.~Lunin, \emph{{Generalized $\lambda$-deformations of AdS$_p
  \times$ S$^p$}},
  \href{https://doi.org/10.1016/j.nuclphysb.2016.10.014}{\emph{Nucl. Phys.}
  {\bfseries B913} (2016) 912}
  [\href{https://arxiv.org/abs/1608.06641}{{\ttfamily 1608.06641}}].

\bibitem{Sfetsos:2014cea}
K.~Sfetsos and D.~C. Thompson, \emph{{Spacetimes for $\lambda$-deformations}},
  \href{https://doi.org/10.1007/JHEP12(2014)164}{\emph{JHEP} {\bfseries 1412}
  (2014) 164} [\href{https://arxiv.org/abs/1410.1886}{{\ttfamily 1410.1886}}].

\bibitem{Demulder:2015lva}
S.~Demulder, K.~Sfetsos and D.~C. Thompson, \emph{{Integrable
  $\lambda$-deformations: Squashing Coset CFTs and $AdS_5\times S^5$}},
  \href{https://doi.org/10.1007/JHEP07(2015)019}{\emph{JHEP} {\bfseries 07}
  (2015) 019} [\href{https://arxiv.org/abs/1504.02781}{{\ttfamily
  1504.02781}}].

\bibitem{Arutyunov:2015mqj}
G.~Arutyunov, S.~Frolov, B.~Hoare, R.~Roiban and A.~A. Tseytlin, \emph{{Scale
  invariance of the $\eta$-deformed $AdS_5 \times S^5$ superstring, T-duality
  and modified type II equations}},
  \href{https://doi.org/10.1016/j.nuclphysb.2015.12.012}{\emph{Nucl. Phys.}
  {\bfseries B903} (2016) 262}
  [\href{https://arxiv.org/abs/1511.05795}{{\ttfamily 1511.05795}}].

\bibitem{Wulff:2016tju}
A.~A. Tseytlin and L.~Wulff, \emph{{Kappa-symmetry of superstring sigma model
  and generalized 10d supergravity equations}},
  \href{https://doi.org/10.1007/JHEP06(2016)174}{\emph{JHEP} {\bfseries 06}
  (2016) 174} [\href{https://arxiv.org/abs/1605.04884}{{\ttfamily
  1605.04884}}].

\bibitem{Hoare:2015wia}
B.~Hoare and A.~A. Tseytlin, \emph{{Type IIB supergravity solution for the
  T-dual of the $\eta$-deformed AdS$_{5} \times$ S$^{5}$ superstring}},
  \href{https://doi.org/10.1007/JHEP10(2015)060}{\emph{JHEP} {\bfseries 10}
  (2015) 060} [\href{https://arxiv.org/abs/1508.01150}{{\ttfamily
  1508.01150}}].

\bibitem{Kameyama:2016yuv}
T.~Kameyama and K.~Yoshida, \emph{{Generalized quark-antiquark potentials from
  a $q$-deformed AdS$_5 \times $S$^5$ background}},
  \href{https://doi.org/10.1093/ptep/ptw059}{\emph{PTEP} {\bfseries 2016}
  (2016) 063B01} [\href{https://arxiv.org/abs/1602.06786}{{\ttfamily
  1602.06786}}].

\bibitem{vanTongeren:2015uha}
S.~J. van Tongeren, \emph{{Yang-Baxter deformations, AdS/CFT, and
  twist-noncommutative gauge theory}},
  \href{https://doi.org/10.1016/j.nuclphysb.2016.01.012}{\emph{Nucl. Phys.}
  {\bfseries B904} (2016) 148}
  [\href{https://arxiv.org/abs/1506.01023}{{\ttfamily 1506.01023}}].

\bibitem{vanTongeren:2016eeb}
S.~J. van Tongeren, \emph{{Almost abelian twists and AdS/CFT}},
  \href{https://doi.org/10.1016/j.physletb.2016.12.002}{\emph{Phys. Lett.}
  {\bfseries B765} (2017) 344}
  [\href{https://arxiv.org/abs/1610.05677}{{\ttfamily 1610.05677}}].

\bibitem{Araujo:2017jap}
T.~Araujo, I.~Bakhmatov, E.~O. Colgáin, J.-i. Sakamoto, M.~M. Sheikh-Jabbari
  and K.~Yoshida, \emph{{Conformal Twists, Yang-Baxter $\sigma$-models and
  Holographic Noncommutativity}},
  \href{https://arxiv.org/abs/1705.02063}{{\ttfamily 1705.02063}}.

\bibitem{Vicedo:2017cge}
B.~Vicedo, \emph{{On integrable field theories as dihedral affine Gaudin
  models}},  \href{https://arxiv.org/abs/1701.04856}{{\ttfamily 1701.04856}}.

\bibitem{Evans:1999mj}
J.~Evans, M.~Hassan, N.~MacKay and A.~Mountain, \emph{{Local conserved charges
  in principal chiral models}},
  \href{https://doi.org/10.1016/S0550-3213(99)00489-7}{\emph{Nucl. Phys.}
  {\bfseries B561} (1999) 385}
  [\href{https://arxiv.org/abs/hep-th/9902008}{{\ttfamily hep-th/9902008}}].

\bibitem{Evans:2000qx}
J.~Evans and A.~Mountain, \emph{{Commuting charges and symmetric spaces}},
  \href{https://doi.org/10.1016/S0370-2693(00)00566-9}{\emph{Phys. Lett.}
  {\bfseries B483} (2000) 290}
  [\href{https://arxiv.org/abs/hep-th/0003264}{{\ttfamily hep-th/0003264}}].

\bibitem{Zamolodchikov:1978xm}
A.~B. Zamolodchikov and A.~B. Zamolodchikov, \emph{{Factorized S Matrices in
  Two-Dimensions as the Exact Solutions of Certain Relativistic Quantum Field
  Models}}, \href{https://doi.org/10.1016/0003-4916(79)90391-9}{\emph{Annals of
  Physics} {\bfseries 120} (1979) 253}.

\bibitem{Faddeev:1979gh}
L.~Faddeev, E.~Sklyanin and L.~Takhtajan, \emph{{The Quantum Inverse Problem
  Method. 1}}, {\emph{Theor. Math. Phys.} {\bfseries 40} (1980) 688}.

\bibitem{Faddeev:1982}
L.~Faddeev, \emph{{Integrable models in 1+1 dimensional quantum field theory}},
   in \emph{{Les Houches Summer School in Theoretical Physics: Recent Advances
  in Field Theory and Statistical Mechanics Les Houches, France, August
  2-September 10, 1982}}, 1982.

\bibitem{Korepin:1997}
V.~Korepin, N.~Bogoliubov and A.~Izergin, \emph{Quantum Inverse Scattering
  Method and Correlation Functions}, Cambridge Monographs on Mathematical
  Physics. Cambridge University Press, 1997.

\bibitem{Freidel:1991jx}
L.~Freidel and J.~M. Maillet, \emph{{Quadratic algebras and integrable
  systems}}, \href{https://doi.org/10.1016/0370-2693(91)91566-E}{\emph{Phys.
  Lett.} {\bfseries B262} (1991) 278}.

\bibitem{Freidel:1991jv}
L.~Freidel and J.~M. Maillet, \emph{{On classical and quantum integrable field
  theories associated to Kac-Moody current algebras}},
  \href{https://doi.org/10.1016/0370-2693(91)90479-A}{\emph{Phys. Lett.}
  {\bfseries B263} (1991) 403}.

\bibitem{Luscher:1977uq}
M.~L\"uscher, \emph{{Quantum Nonlocal Charges and Absence of Particle
  Production in the Two-Dimensional Nonlinear Sigma Model}},
  \href{https://doi.org/10.1016/0550-3213(78)90211-0}{\emph{Nucl. Phys.}
  {\bfseries B135} (1978) 1}.

\bibitem{Gromov:2013pga}
N.~Gromov, V.~Kazakov, S.~Leurent and D.~Volin, \emph{{Quantum Spectral Curve
  for Planar $\mathcal{N} =$ Super-Yang-Mills Theory}},
  \href{https://doi.org/10.1103/PhysRevLett.112.011602}{\emph{Phys. Rev. Lett.}
  {\bfseries 112} (2014) 011602}
  [\href{https://arxiv.org/abs/1305.1939}{{\ttfamily 1305.1939}}].

\bibitem{Gaudin_76a}
M.~Gaudin, \emph{{Diagonalisation d'une classe d'hamiltoniens de spin}},
  \href{https://doi.org/10.1051/jphys:0197600370100108700}{\emph{Journal de
  Physique} {\bfseries 37 (10)} (1976) 1087}.

\bibitem{Gaudin_book83}
M.~Gaudin, \emph{{La fonction d'onde de Bethe}}. Masson, 1983.

\bibitem{Jurco:1989mg}
B.~Jur\v{c}o, \emph{{Classical Yang-Baxter equations and quantum integrable
  systems}}, \href{https://doi.org/10.1063/1.528305}{\emph{J. Math. Phys.}
  {\bfseries 30} (1989) 1289}.

\bibitem{Babujian:1993ts}
H.~M. Babujian and R.~Flume, \emph{{Off-shell Bethe Ansatz equation for Gaudin
  magnets and solutions of Knizhnik-Zamolodchikov equations}},
  \href{https://doi.org/10.1142/S0217732394001891}{\emph{Mod. Phys. Lett.}
  {\bfseries A9} (1994) 2029}
  [\href{https://arxiv.org/abs/hep-th/9310110}{{\ttfamily hep-th/9310110}}].

\bibitem{Varchenko:1991}
V.~Schechtman and A.~Varchenko, \emph{{Arrangements of Hyperplanes and
  Lie-algebra Homology}},
  \href{https://doi.org/10.1007/BF01243909}{\emph{Inventiones Mathematicae}
  {\bfseries 106} (1991) 139}.

\bibitem{Reshetikhin:1994qw}
N.~Reshetikhin and A.~Varchenko, \emph{{Quasiclassical asymptotics of solutions
  to the KZ equations}},
  \href{https://arxiv.org/abs/hep-th/9402126}{{\ttfamily hep-th/9402126}}.

\bibitem{Feigin:1994in}
B.~Feigin, E.~Frenkel and N.~Reshetikhin, \emph{{Gaudin model, Bethe ansatz and
  correlation functions at the critical level}},
  \href{https://doi.org/10.1007/BF02099300}{\emph{Commun. Math. Phys.}
  {\bfseries 166} (1994) 27}
  [\href{https://arxiv.org/abs/hep-th/9402022}{{\ttfamily hep-th/9402022}}].

\bibitem{mukhin_2009a}
E.~Mukhin, V.~Tarasov and A.~Varchenko, \emph{{Bethe algebra of gaudin model,
  calogero-moser space and cherednik algebra}},
  \href{https://arxiv.org/abs/arXiv:0906.5185 [math.QA]}{{\ttfamily
  arXiv:0906.5185 [math.QA]}}.

\bibitem{Talalaev:2004qi}
D.~Talalaev, \emph{{Quantization of the Gaudin system}},
  \href{https://arxiv.org/abs/hep-th/0404153}{{\ttfamily hep-th/0404153}}.

\bibitem{Chervov:2006xk}
A.~Chervov and D.~Talalaev, \emph{{Quantum spectral curves, quantum integrable
  systems and the geometric Langlands correspondence}},
  \href{https://arxiv.org/abs/hep-th/0604128}{{\ttfamily hep-th/0604128}}.

\bibitem{Tarasov:2006}
E.~Mukhin, V.~Tarasov and A.~Varchenko, \emph{{Bethe eigenvectors of higher
  transfer matrices}},
  \href{https://doi.org/10.1088/1742-5468/2006/08/P08002}{\emph{Journal of
  Statistical Mechanics: Theory and Experiment} {\bfseries 8} (2006) 08002}
  [\href{https://arxiv.org/abs/math/0605015}{{\ttfamily math/0605015}}].

\bibitem{Molev:2013}
A.~I. Molev, \emph{{Feigin-Frenkel center in types B, C and D}},
  \href{https://doi.org/10.1007/s00222-012-0390-7}{\emph{Inventiones
  Mathematicae} {\bfseries 191} (2013) 1}.

\bibitem{Rybnikov:2008}
L.~Rybnikov, \emph{{Uniqueness of higher Gaudin hamiltonians}},
  \href{https://doi.org/10.1016/S0034-4877(08)80013-4}{\emph{Reports on
  Mathematical Physics} {\bfseries 61} (2008) 247}
  [\href{https://arxiv.org/abs/math/0608588}{{\ttfamily math/0608588}}].

\bibitem{Frenkel:1995zp}
E.~Frenkel, \emph{{Affine algebras, Langlands duality and Bethe ansatz}},  in
  \emph{{Mathematical physics. Proceedings, 11th International Congress, Paris,
  France, July 18-22, 1994}}, 1995,
  \href{https://arxiv.org/abs/q-alg/9506003}{{\ttfamily q-alg/9506003}}.

\bibitem{Frenkel:2003qx}
E.~Frenkel, \emph{{Opers on the projective line, flag manifolds and bethe
  ansatz}}, \href{https://doi.org/10.1.1.233.8419}{\emph{Moscow Mathematical
  Journal} {\bfseries 4} (2004) 655}
  [\href{https://arxiv.org/abs/math/0308269}{{\ttfamily math/0308269}}].

\bibitem{Mukhin:2005}
E.~Mukhin and A.~Varchenko, \emph{{Miura Opers and Critical Points of Master
  Functions}}, \href{https://doi.org/10.2478/BF02479193}{\emph{Central European
  Journal of Mathematics} {\bfseries 3} (2005) 155}
  [\href{https://arxiv.org/abs/math/0312406}{{\ttfamily math/0312406}}].

\bibitem{Frenkel:2004qy}
E.~Frenkel, \emph{{Gaudin model and opers}},
  \href{https://arxiv.org/abs/math/0407524}{{\ttfamily math/0407524}}.

\bibitem{Chervov:2004ty}
A.~Chervov and D.~Talalaev, \emph{{Universal G-oper and Gaudin eigenproblem}},
  \href{https://arxiv.org/abs/hep-th/0409007}{{\ttfamily hep-th/0409007}}.

\bibitem{Chervov:2009ck}
A.~Chervov and D.~Talalaev, \emph{{KZ equation, G-opers and quantum
  Drinfeld-Sokolov reduction}},
  \href{https://doi.org/10.1007/s10958-009-9415-1}{\emph{J.Math.Sci.}
  {\bfseries 158} (2009) 904}
  [\href{https://arxiv.org/abs/hep-th/0607250}{{\ttfamily hep-th/0607250}}].

\bibitem{Rybnikov:2016}
L.~Rybnikov, \emph{{A proof of the Gaudin Bethe Ansatz conjecture}},
  {\emph{ArXiv e-prints} (2016) }
  [\href{https://arxiv.org/abs/1608.04625}{{\ttfamily 1608.04625}}].

\bibitem{Feigin:2006xs}
B.~Feigin, E.~Frenkel and V.~Toledano~Laredo, \emph{{Gaudin models with
  irregular singularities}},
  \href{https://doi.org/10.1016/j.aim.2009.09.007}{\emph{Adv. Math.} {\bfseries
  223} (2010) 873} [\href{https://arxiv.org/abs/math/0612798}{{\ttfamily
  math/0612798}}].

\bibitem{Sklyanin:1996pr}
E.~Sklyanin and T.~Takebe, \emph{{Algebraic Bethe ansatz for XYZ Gaudin
  model}}, \href{https://doi.org/10.1016/0375-9601(96)00448-3}{\emph{Phys.
  Lett.} {\bfseries A219} (1996) 217}
  [\href{https://arxiv.org/abs/q-alg/9601028}{{\ttfamily q-alg/9601028}}].

\bibitem{Skrypnyk:2006}
T.~Skrypnyk, \emph{{Integrable quantum spin chains, non-skew symmetric
  r-matrices and quasigraded Lie algebras}},
  \href{https://doi.org/10.1016/j.geomphys.2006.02.002}{\emph{Journal of
  Geometry and Physics} {\bfseries 57} (2006) 53}.

\bibitem{Crampe:2007cj}
N.~Crampe, L.~Frappat and E.~Ragoucy, \emph{{Thermodynamical limit of general
  gl(N) spin chains II: Excited states and energies}},
  \href{https://doi.org/10.1088/1742-5468/2008/01/P01015}{\emph{J. Stat. Mech.}
  {\bfseries 0801} (2008) P015}
  [\href{https://arxiv.org/abs/0710.5904}{{\ttfamily 0710.5904}}].

\bibitem{Vicedo:2014zza}
B.~Vicedo and C.~A.~S. Young, \emph{{Cyclotomic Gaudin models: construction and
  Bethe ansatz}},
  \href{https://doi.org/10.1007/s00220-016-2601-3}{\emph{Commun. Math. Phys.}
  {\bfseries 343} (2016) 971}
  [\href{https://arxiv.org/abs/1409.6937}{{\ttfamily 1409.6937}}].

\bibitem{Skrypnyk:2013usa}
T.~Skrypnyk, \emph{{'$Z_{2}$-graded' Gaudin models and analytical Bethe
  ansatz}}, \href{https://doi.org/10.1016/j.nuclphysb.2013.01.013}{\emph{Nucl.
  Phys.} {\bfseries B870} (2013) 495}.

\bibitem{Vicedo_161109059}
B.~Vicedo and C.~A.~S. Young, \emph{{Cyclotomic Gaudin models with irregular
  singularities}},
  \href{https://doi.org/10.1016/j.geomphys.2017.07.013}{\emph{Journal of
  Geometry and Physics} {\bfseries 121} (2017) 247}
  [\href{https://arxiv.org/abs/1611.09059}{{\ttfamily 1611.09059}}].

\bibitem{Feigin:2007mr}
B.~Feigin and E.~Frenkel, \emph{{Quantization of soliton systems and Langlands
  duality}},  in \emph{{Exploring new structures and natural constructions in
  mathematical physics}}, vol.~61 of \emph{Advanced Studies in Pure
  Mathematics}, p.~185, 2011, \href{https://arxiv.org/abs/0705.2486}{{\ttfamily
  0705.2486}}.

\bibitem{Bazhanov:1998wj}
V.~V. Bazhanov, S.~L. Lukyanov and A.~B. Zamolodchikov, \emph{{Spectral
  determinants for Schrodinger equation and Q operators of conformal field
  theory}}, \href{https://doi.org/10.1023/A:1004838616921}{\emph{J. Statist.
  Phys.} {\bfseries 102} (2001) 567}
  [\href{https://arxiv.org/abs/hep-th/9812247}{{\ttfamily hep-th/9812247}}].

\bibitem{Bazhanov:2003ni}
V.~V. Bazhanov, S.~L. Lukyanov and A.~B. Zamolodchikov, \emph{{Higher level
  eigenvalues of Q operators and Schroedinger equation}},
  \href{https://doi.org/10.4310/ATMP.2003.v7.n4.a4}{\emph{Adv. Theor. Math.
  Phys.} {\bfseries 7} (2003) 711}
  [\href{https://arxiv.org/abs/hep-th/0307108}{{\ttfamily hep-th/0307108}}].

\bibitem{Dorey:1998pt}
P.~Dorey and R.~Tateo, \emph{{Anharmonic oscillators, the thermodynamic Bethe
  ansatz, and nonlinear integral equations}},
  \href{https://doi.org/10.1088/0305-4470/32/38/102}{\emph{J. Phys.} {\bfseries
  A32} (1999) L419} [\href{https://arxiv.org/abs/hep-th/9812211}{{\ttfamily
  hep-th/9812211}}].

\bibitem{SemenovTianShansky:1983ik}
M.~Semenov-Tian-Shansky, \emph{{What is a classical $r$-matrix?}},
  {\emph{Funct. Anal. Appl.} {\bfseries 17} (1983) 259}.

\bibitem{Maillet:1985ec}
J.~M. Maillet, \emph{{Hamiltonian structures for integrable classical theories
  from graded Kac-Moody algebras}},
  \href{https://doi.org/10.1016/0370-2693(86)91289-X}{\emph{Phys. Lett.}
  {\bfseries B167} (1986) 401}.

\bibitem{Sevostyanov:1995hd}
A.~Sevostyanov, \emph{{The Classical $R$ matrix method for nonlinear sigma
  model}}, \href{https://doi.org/10.1142/S0217751X96001978}{\emph{Int. J. Mod.
  Phys.} {\bfseries A11} (1996) 4241}
  [\href{https://arxiv.org/abs/hep-th/9509030}{{\ttfamily hep-th/9509030}}].

\bibitem{Vicedo:2010qd}
B.~Vicedo, \emph{{The classical R-matrix of AdS/CFT and its Lie dialgebra
  structure}}, \href{https://doi.org/10.1007/s11005-010-0446-9}{\emph{Lett.
  Math. Phys.} {\bfseries 95} (2011) 249}
  [\href{https://arxiv.org/abs/1003.1192}{{\ttfamily 1003.1192}}].

\bibitem{Zakharov:1973pp}
V.~Zakharov and A.~Mikhailov, \emph{{Relativistically invariant two-dimensional
  models in field theory integrable by the inverse problem technique}},
  {\emph{Sov. Phys. JETP} {\bfseries 47} (1978) 1017}.

\bibitem{Pohlmeyer:1975nb}
K.~Pohlmeyer, \emph{{Integrable hamiltonian systems and interactions through
  quadratic constraints}},
  \href{https://doi.org/10.1007/BF01609119}{\emph{Commun. Math. Phys.}
  {\bfseries 46} (1976) 207}.

\bibitem{Eichenherr:1979ci}
H.~Eichenherr and M.~Forger, \emph{{On the Dual Symmetry of the Nonlinear Sigma
  Models}}, \href{https://doi.org/10.1016/0550-3213(79)90276-1}{\emph{Nucl.
  Phys.} {\bfseries B155} (1979) 381}.

\bibitem{dirac1964lectures}
P.~A.~M. Dirac, \emph{Lectures on Quantum Mechanics}. Belfer Graduate School of
  Science, Yeshiva University, 1964.

\bibitem{Henneaux:1992ig}
M.~Henneaux and C.~Teitelboim, \emph{{Quantization of gauge systems}}.
  Princeton University Press, 1992.

\bibitem{Delduc:2012qb}
F.~Delduc, M.~Magro and B.~Vicedo, \emph{{Alleviating the non-ultralocality of
  coset sigma models through a generalized Faddeev-Reshetikhin procedure}},
  \href{https://doi.org/10.1007/JHEP08(2012)019}{\emph{JHEP} {\bfseries 1208}
  (2012) 019} [\href{https://arxiv.org/abs/1204.0766}{{\ttfamily 1204.0766}}].

\bibitem{Young:2005jv}
C.~A.~S. Young, \emph{{Non-local charges, Z(m) gradings and coset space
  actions}}, \href{https://doi.org/10.1016/j.physletb.2005.10.090}{\emph{Phys.
  Lett.} {\bfseries B632} (2006) 559}
  [\href{https://arxiv.org/abs/hep-th/0503008}{{\ttfamily hep-th/0503008}}].

\bibitem{Beisert:2012ue}
N.~Beisert and F.~Luecker, \emph{{Construction of Lax Connections by
  exponentiation}}, \href{https://doi.org/10.1063/1.4769824}{\emph{J. Math.
  Phys.} {\bfseries 53} (2012) 122304}
  [\href{https://arxiv.org/abs/1207.3325}{{\ttfamily 1207.3325}}].

\bibitem{Bykov:2017vsm}
D.~V. Bykov, \emph{{Cyclic gradings of Lie algebras and Lax pairs for
  $\sigma$-models}},
  \href{https://doi.org/10.1134/S0040577916120060}{\emph{Theor. Math. Phys.}
  {\bfseries 189} (2016) 1734}.

\bibitem{Ke:2011zzb}
S.-M. Ke, X.-Y. Li, C.~Wang and R.-H. Yue, \emph{{Classical exchange algebra of
  the nonlinear sigma model on a supercoset target with Z(2n) grading}},
  \href{https://doi.org/10.1088/0256-307X/28/10/101101}{\emph{Chin. Phys.
  Lett.} {\bfseries 28} (2011) 101101}.

\bibitem{Berkovits:1999zq}
N.~Berkovits, M.~Bershadsky, T.~Hauer, S.~Zhukov and B.~Zwiebach,
  \emph{{Superstring theory on $AdS_2 \times S^2$ as a coset supermanifold}},
  \href{https://doi.org/10.1016/S0550-3213(99)00683-5}{\emph{Nucl. Phys.}
  {\bfseries B567} (2000) 61}
  [\href{https://arxiv.org/abs/hep-th/9907200}{{\ttfamily hep-th/9907200}}].

\bibitem{deVega:1979zy}
H.~J. de~Vega, \emph{{Field Theories With an Infinite Number of Conservation
  Laws and Backlund Transformations in Two-dimensions}},
  \href{https://doi.org/10.1016/0370-2693(79)90971-7}{\emph{Phys. Lett.}
  {\bfseries 87B} (1979) 233}.

\bibitem{Kawaguchi:2011mz}
I.~Kawaguchi, D.~Orlando and K.~Yoshida, \emph{{Yangian symmetry in deformed
  WZNW models on squashed spheres}},
  \href{https://doi.org/10.1016/j.physletb.2011.06.007}{\emph{Phys. Lett.}
  {\bfseries B701} (2011) 475}
  [\href{https://arxiv.org/abs/1104.0738}{{\ttfamily 1104.0738}}].

\bibitem{Kawaguchi:2013gma}
I.~Kawaguchi and K.~Yoshida, \emph{{A deformation of quantum affine algebra in
  squashed Wess-Zumino-Novikov-Witten models}},
  \href{https://doi.org/10.1063/1.4880341}{\emph{J. Math. Phys.} {\bfseries 55}
  (2014) 062302} [\href{https://arxiv.org/abs/1311.4696}{{\ttfamily
  1311.4696}}].

\bibitem{Vicedo:2015pna}
B.~Vicedo, \emph{{Deformed integrable $\sigma$-models, classical $R$-matrices
  and classical exchange algebra on Drinfel'd doubles}},
  \href{https://doi.org/10.1088/1751-8113/48/35/355203}{\emph{J. Phys.}
  {\bfseries A48} (2015) 355203}
  [\href{https://arxiv.org/abs/1504.06303}{{\ttfamily 1504.06303}}].

\bibitem{Hoare:2015gda}
B.~Hoare and A.~Tseytlin, \emph{{On integrable deformations of superstring
  sigma models related to $AdS_n \times S^n$ supercosets}},
  \href{https://doi.org/10.1016/j.nuclphysb.2015.06.001}{\emph{Nucl.Phys.}
  {\bfseries B897} (2015) 448}
  [\href{https://arxiv.org/abs/1504.07213}{{\ttfamily 1504.07213}}].

\bibitem{Klimcik:2015gba}
C.~Klim\v{c}ik, \emph{{$\eta$ and $\lambda$ deformations as ${\cal E}$-models}},
  \href{https://doi.org/10.1016/j.nuclphysb.2015.09.011}{\emph{Nucl. Phys.}
  {\bfseries B900} (2015) 259}
  [\href{https://arxiv.org/abs/1508.05832}{{\ttfamily 1508.05832}}].

\bibitem{Klimcik:1995jn}
C.~Klim\v{c}ik, \emph{{Poisson-Lie T duality}},
  \href{https://doi.org/10.1016/0920-5632(96)00013-8}{\emph{Nucl.Phys.Proc.Suppl.}
  {\bfseries 46} (1996) 116}
  [\href{https://arxiv.org/abs/hep-th/9509095}{{\ttfamily hep-th/9509095}}].

\bibitem{Klimcik:1995dy}
C.~Klim\v{c}ik and P.~Severa, \emph{{Poisson-Lie T duality and loop groups of
  Drinfeld doubles}},
  \href{https://doi.org/10.1016/0370-2693(96)00025-1}{\emph{Phys. Lett.}
  {\bfseries B372} (1996) 65}
  [\href{https://arxiv.org/abs/hep-th/9512040}{{\ttfamily hep-th/9512040}}].

\bibitem{Klimcik:1996br}
C.~Klim\v{c}ik and P.~Severa, \emph{{T duality and the moment map}},  in
  \emph{Cargese 1996, Quantum fields and quantum space time}, pp.~323--329,
  1996, \href{https://arxiv.org/abs/hep-th/9610198}{{\ttfamily
  hep-th/9610198}}.

\bibitem{Hoare:2014oua}
B.~Hoare, \emph{{Towards a two-parameter q-deformation of $AdS_3 \times S^3
  \times M^4$ superstrings}},
  \href{https://doi.org/10.1016/j.nuclphysb.2014.12.012}{\emph{Nucl. Phys.}
  {\bfseries B891} (2015) 259}
  [\href{https://arxiv.org/abs/1411.1266}{{\ttfamily 1411.1266}}].

\bibitem{Evans:2000hx}
J.~M. Evans, M.~Hassan, N.~J. MacKay and A.~J. Mountain, \emph{{Conserved
  charges and supersymmetry in principal chiral and WZW models}},
  \href{https://doi.org/10.1016/S0550-3213(00)00257-1}{\emph{Nucl. Phys.}
  {\bfseries B580} (2000) 605}
  [\href{https://arxiv.org/abs/hep-th/0001222}{{\ttfamily hep-th/0001222}}].

\bibitem{Evans:2005zd}
J.~M. Evans and C.~A.~S. Young, \emph{{Higher-spin conserved currents in
  supersymmetric sigma models on symmetric spaces}},
  \href{https://doi.org/10.1016/j.nuclphysb.2005.04.022}{\emph{Nucl. Phys.}
  {\bfseries B717} (2005) 327}
  [\href{https://arxiv.org/abs/hep-th/0501090}{{\ttfamily hep-th/0501090}}].

\bibitem{Drinfeld:1986in}
V.~Drinfeld, \emph{{Quantum groups}},
  \href{https://doi.org/10.1007/BF01247086}{\emph{J. Sov. Math.} {\bfseries 41}
  (1988) 898}.

\bibitem{Drinfeld:1983ky}
V.~G. Drinfeld, \emph{{Hamiltonian structures of lie groups, lie bialgebras and
  the geometric meaning of the classical Yang-Baxter equations}}, {\emph{Sov.
  Math. Dokl.} {\bfseries 27} (1983) 68}.

\bibitem{SemenovTianShansky:1985my}
M.~A. Semenov-Tian-Shansky, \emph{{Dressing transformations and Poisson group
  actions}}, \href{https://doi.org/10.2977/prims/1195178514}{\emph{Publ. Res.
  Inst. Math. Sci. Kyoto} {\bfseries 21} (1985) 1237}.

\bibitem{Semenov-nonv}
M.~A. Semenov-Tyan-Shanskii, \emph{{Poisson-Lie groups. The quantum duality
  principle and the twisted quantum double}},
  \href{https://doi.org/10.1007/BF01083527}{\emph{Theoretical and Mathematical
  Physics} {\bfseries 93} (1992) 1292}.

\bibitem{luweinstein1990_a}
J.-H. Lu and A.~Weinstein, \emph{{Poisson-Lie groups, Dressing transformations,
  and Bruhat decompositions}}, {\emph{Journal of Differential Geometry}
  {\bfseries 31} (1990) 501}.

\bibitem{lu_1990_phd}
J.-H. Lu, \emph{{Multiplicative and Affine Poisson structures on Lie groups}},
  Ph.D. thesis, Berkeley, 1990.

\bibitem{Babelon:1991ah}
O.~Babelon and D.~Bernard, \emph{{Dressing symmetries}},
  \href{https://doi.org/10.1007/BF02097626}{\emph{Commun. Math. Phys.}
  {\bfseries 149} (1992) 279}
  [\href{https://arxiv.org/abs/hep-th/9111036}{{\ttfamily hep-th/9111036}}].

\bibitem{Falceto:1992bf}
F.~Falceto and K.~Gawedzki, \emph{{Lattice Wess-Zumino-Witten model and quantum
  groups}}, \href{https://doi.org/10.1016/0393-0440(93)90056-K}{\emph{J. Geom.
  Phys.} {\bfseries 11} (1993) 251}
  [\href{https://arxiv.org/abs/hep-th/9209076}{{\ttfamily hep-th/9209076}}].

\bibitem{Feher:2002fx}
L.~Feher, \emph{{Dynamical r matrices and Poisson-Lie symmetries in the chiral
  wznw model}}, {\emph{PoS} (2002) 012}
  [\href{https://arxiv.org/abs/hep-th/0212006}{{\ttfamily hep-th/0212006}}].

\bibitem{Chari_Pressley_1994}
V.~Chari and A.~Pressley, \emph{{Quantum Groups}}. Cambridge University Press,
  1994.

\bibitem{Kosmann-Schwarzbach2004}
Y.~Kosmann-Schwarzbach, \emph{Integrability of Nonlinear Systems}, ch.~{Lie
  Bialgebras, Poisson Lie Groups, and Dressing Transformations}, pp.~107--173.
\newblock Springer Berlin Heidelberg, Berlin, Heidelberg, 2004.

\bibitem{Ballesteros_2009}
A.~Ballesteros, E.~Celeghini and M.~A. del Olmo, \emph{{Poisson-Hopf limit of
  quantum algebras}}, {\emph{J. Phys. A: Math. Theor.} {\bfseries 42} (2009)
  275202} [\href{https://arxiv.org/abs/0903.2178}{{\ttfamily 0903.2178}}].

\bibitem{Ruegg:1993eq}
H.~Ruegg, \emph{{Integrable Systems, Quantum Groups, and Quantum Field
  Theories, NATO ASI Series Volume 409}}, ch.~{$q$-deformation of semisimple
  and non-semisimple Lie algebras}, pp.~45--81.
\newblock Springer, 1993.

\bibitem{MacKay:1992he}
N.~J. MacKay, \emph{{On the classical origins of Yangian symmetry in integrable
  field theory}}, \href{https://doi.org/10.1016/0370-2693(93)91310-J,
  10.1016/0370-2693(92)90280-H}{\emph{Phys. Lett.} {\bfseries B281} (1992) 90}.

\bibitem{Bernard:1992ya}
D.~Bernard, \emph{{An Introduction to Yangian Symmetries}},
  \href{https://doi.org/10.1142/S0217979293003371}{\emph{Int.J.Mod.Phys.}
  {\bfseries B7} (1993) 3517}
  [\href{https://arxiv.org/abs/hep-th/9211133}{{\ttfamily hep-th/9211133}}].

\bibitem{Kac:1990gs}
V.~G. Kac, \emph{{Infinite dimensional Lie algebras}}. Cambridge University
  Press, 1990.

\bibitem{Dixmier:1977}
J.~Dixmier, \emph{Enveloping Algebras}, North-Holland mathematical library.
  Akademie-Verlag, 1977.

\bibitem{Neumann1859}
C.~Neumann, \emph{{De problemate quodam mechanico, quod ad primam integralium
  ultraellipticorum classem revocatur}}, {\emph{Journal f\"ur die reine und
  angewandte Mathematik} {\bfseries 56} (1859) 46}.

\bibitem{Uhlenbeck:1982}
K.~Uhlenbeck, \emph{{Equivariant Harmonic-maps Into Spheres}}, {\emph{Lecture
  Notes In Mathematics} {\bfseries 949} (1982) 146}.

\bibitem{Avan:1991ib}
J.~Avan and M.~Talon, \emph{{Alternative Lax structures for the classical and
  quantum Neumann model}},
  \href{https://doi.org/10.1016/0370-2693(91)90805-Z}{\emph{Phys. Lett.}
  {\bfseries B268} (1991) 209}.

\bibitem{Ragnisco1997}
O.~Ragnisco and Y.~B. Suris, \emph{{On the r-Matrix Structure of the Neumann
  System and its Discretizations}}, pp.~285--300.
\newblock Birkh\"auser Boston, Boston, MA, 1997.

\bibitem{Arutyunov:2014cda}
G.~Arutyunov and D.~Medina-Rincon, \emph{{Deformed Neumann model from spinning
  strings on $(AdS_5 \times S^5)_\eta$}},
  \href{https://doi.org/10.1007/JHEP10(2014)050}{\emph{JHEP} {\bfseries 1410}
  (2014) 50} [\href{https://arxiv.org/abs/1406.2536}{{\ttfamily 1406.2536}}].

\bibitem{Kuznetsov:1992}
V.~B. Kuznetsov, \emph{{Isomorphism of the N-dimensional Neumann System and the
  N-site Gaudin Magnet}},
  \href{https://doi.org/10.1007/BF01075058}{\emph{Functional Analysis and its
  Applications} {\bfseries 26} (1992) 302}.

\bibitem{Takiff:1971}
S.~Takiff, \emph{{Rings Of Invariant Polynomials For A Class Of Lie Algebras}},
  \href{https://doi.org/{10.2307/1995803}}{\emph{Transactions of the American
  Mathematical Society} {\bfseries 160} (1971) 249}.

\bibitem{Skrypnyk:2005}
T.~Skrypnyk, \emph{{New integrable Gaudin-type systems, classical r-matrices
  and quasigraded Lie algebras}},
  \href{https://doi.org/10.1016/j.physleta.2004.11.041}{\emph{Physics Letters
  A} {\bfseries 334} (2005) 390}.

\bibitem{Etingof:1998}
P.~Etingof, I.~Frenkel and A.~Kirillov, \emph{Lectures on Representation Theory
  and Knizhnik-Zamolodchikov Equations}, Mathematical surveys and monographs.
  American Mathematical Society, 1998.

\bibitem{Babichenko:2012uq}
A.~Babichenko and D.~Ridout, \emph{{Takiff superalgebras and Conformal Field
  Theory}}, \href{https://doi.org/10.1088/1751-8113/46/12/125204}{\emph{J.
  Phys.} {\bfseries A46} (2013) 125204}
  [\href{https://arxiv.org/abs/1210.7094}{{\ttfamily 1210.7094}}].

\bibitem{Mukhin:2007}
E.~Mukhin and A.~Varchenko, \emph{{Multiple orthogonal polynomials and a
  counterexample to the gaudin bethe ansatz conjecture}},
  \href{https://doi.org/10.1090/S0002-9947-07-04217-1}{\emph{Transactions of
  the American Mathematical Society} {\bfseries 359} (2007) 5383}.

\bibitem{Humphreys:1980dw}
J.~E. Humphreys, \emph{{Introduction to Lie Algebras and Representation
  Theory}}. Springer, 1980.

\bibitem{Humphreys:1972}
J.~Humphreys, \emph{Representations of Semisimple Lie Algebras in the BGG
  Category O}. American Mathematical Soc., 1972.

\bibitem{Hartshorne:1977}
R.~Hartshorne, \emph{Algebraic Geometry}, Encyclopaedia of mathematical
  sciences. Springer, 1977.

\bibitem{Kostant:1959}
B.~Kostant, \emph{{The Principal 3-dimensional Subgroup and the Betti Numbers
  of a Complex Simple Lie Group}},
  \href{https://doi.org/10.2307/2372999}{\emph{American Journal of Mathematics}
  {\bfseries 81} (1959) 973}.

\bibitem{Kostant:1963}
B.~Kostant, \emph{{Lie Group Representations on Polynomial Rings}},
  \href{https://doi.org/10.2307/2373130}{\emph{American Journal of Mathematics}
  {\bfseries 85} (1963) }.

\bibitem{Kostant:1978}
B.~Kostant, \emph{{Whittaker Vectors and Representation Theory}},
  \href{https://doi.org/10.1007/BF01390249}{\emph{Inventiones mathematicae}
  {\bfseries 48} (1978) 101}.

\bibitem{Drinfeld:1984qv}
V.~G. Drinfeld and V.~V. Sokolov, \emph{{Lie algebras and equations of
  Korteweg-de Vries type}}, \href{https://doi.org/10.1007/BF02105860}{\emph{J.
  Sov. Math.} {\bfseries 30} (1984) 1975}.

\bibitem{Frenkel:book}
E.~Frenkel, \emph{Langlands Correspondence for Loop Groups}, Cambridge Studies
  in Advanced Mathematics. Cambridge University Press, 2007.

\bibitem{Beilinson:1995}
A.~Beilinson and V.~Drinfeld, \emph{{Quantisation of Hitchin's integrable
  systems and Hecke eigensheaves}}, {\emph{Preprint} (1995) }.

\bibitem{Mukhin_2002a}
E.~{Mukhin} and A.~{Varchenko}, \emph{{Critical points of master functions and
  flag varieties}},
  \href{https://doi.org/10.1142/S0219199704001288}{\emph{Communications in
  Contemporary Mathematics} {\bfseries 6} (2004) 111}
  [\href{https://arxiv.org/abs/math/0209017}{{\ttfamily math/0209017}}].

\bibitem{Varchenko:2015}
A.~Varchenko and C.~A.~S. Young, \emph{{Populations of Solutions to Cyclotomic
  Bethe Equations}},
  \href{https://doi.org/10.3842/SIGMA.2015.091}{\emph{Symmetry, Integrability
  and Geometry: Methods and Applications} {\bfseries 11} (2015) }.

\bibitem{Chari:1984}
V.~Chari and S.~Ilangovan, \emph{{On the Harishchandra Homomorphism for
  Infinite-dimensional Lie-algebras}},
  \href{https://doi.org/10.1016/0021-8693(84)90185-6}{\emph{Journal of Algebra}
  {\bfseries 90} (1984) 476}.

\bibitem{Bazhanov:2013cua}
V.~V. Bazhanov and S.~L. Lukyanov, \emph{{Integrable structure of Quantum Field
  Theory: Classical flat connections versus quantum stationary states}},
  \href{https://doi.org/10.1007/JHEP09(2014)147}{\emph{JHEP} {\bfseries 09}
  (2014) 147} [\href{https://arxiv.org/abs/1310.4390}{{\ttfamily 1310.4390}}].

\bibitem{Dorey:2007zx}
P.~Dorey, C.~Dunning and R.~Tateo, \emph{{The ODE/IM Correspondence}},
  \href{https://doi.org/10.1088/1751-8113/40/32/R01}{\emph{J. Phys.} {\bfseries
  A40} (2007) R205} [\href{https://arxiv.org/abs/hep-th/0703066}{{\ttfamily
  hep-th/0703066}}].

\bibitem{Bazhanov:2001xm}
V.~V. Bazhanov, A.~N. Hibberd and S.~M. Khoroshkin, \emph{{Integrable structure
  of W(3) conformal field theory, quantum Boussinesq theory and boundary affine
  Toda theory}},
  \href{https://doi.org/10.1016/S0550-3213(01)00595-8}{\emph{Nucl. Phys.}
  {\bfseries B622} (2002) 475}
  [\href{https://arxiv.org/abs/hep-th/0105177}{{\ttfamily hep-th/0105177}}].

\bibitem{Lukyanov:2010rn}
S.~L. Lukyanov and A.~B. Zamolodchikov, \emph{{Quantum Sine(h)-Gordon Model and
  Classical Integrable Equations}},
  \href{https://doi.org/10.1007/JHEP07(2010)008}{\emph{JHEP} {\bfseries 07}
  (2010) 008} [\href{https://arxiv.org/abs/1003.5333}{{\ttfamily 1003.5333}}].

\bibitem{Dorey:2012bx}
P.~Dorey, S.~Faldella, S.~Negro and R.~Tateo, \emph{{The Bethe Ansatz and the
  Tzitzeica-Bullough-Dodd equation}},
  \href{https://doi.org/10.1098/rsta.2012.0052}{\emph{Phil. Trans. Roy. Soc.
  Lond.} {\bfseries A371} (2013) 20120052}
  [\href{https://arxiv.org/abs/1209.5517}{{\ttfamily 1209.5517}}].

\bibitem{Adamopoulou:2014fca}
P.~Adamopoulou and C.~Dunning, \emph{{Bethe Ansatz equations for the classical
  $A_n^{(1)}$ affine Toda field theories}},
  \href{https://doi.org/10.1088/1751-8113/47/20/205205}{\emph{J. Phys.}
  {\bfseries A47} (2014) 205205}
  [\href{https://arxiv.org/abs/1401.1187}{{\ttfamily 1401.1187}}].

\bibitem{Ito:2013aea}
K.~Ito and C.~Locke, \emph{{ODE/IM correspondence and modified affine Toda
  field equations}},
  \href{https://doi.org/10.1016/j.nuclphysb.2014.06.007}{\emph{Nucl. Phys.}
  {\bfseries B885} (2014) 600}
  [\href{https://arxiv.org/abs/1312.6759}{{\ttfamily 1312.6759}}].

\bibitem{Ito:2015nla}
K.~Ito and C.~Locke, \emph{{ODE/IM correspondence and Bethe ansatz for affine
  Toda field equations}},
  \href{https://doi.org/10.1016/j.nuclphysb.2015.05.016}{\emph{Nucl. Phys.}
  {\bfseries B896} (2015) 763}
  [\href{https://arxiv.org/abs/1502.00906}{{\ttfamily 1502.00906}}].

\bibitem{Ito:2016qzt}
K.~Ito and H.~Shu, \emph{{ODE/IM correspondence for modified $B_2^{(1)}$ affine
  Toda field equation}},
  \href{https://doi.org/10.1016/j.nuclphysb.2017.01.009}{\emph{Nucl. Phys.}
  {\bfseries B916} (2017) 414}
  [\href{https://arxiv.org/abs/1605.04668}{{\ttfamily 1605.04668}}].

\bibitem{Lukyanov:2013wra}
S.~L. Lukyanov, \emph{{ODE/IM correspondence for the Fateev model}},
  \href{https://doi.org/10.1007/JHEP12(2013)012}{\emph{JHEP} {\bfseries 12}
  (2013) 012} [\href{https://arxiv.org/abs/1303.2566}{{\ttfamily 1303.2566}}].

\bibitem{Dorey:1999pv}
P.~Dorey and R.~Tateo, \emph{{Differential equations and integrable models: The
  SU(3) case}},
  \href{https://doi.org/10.1016/S0550-3213(99)00791-9}{\emph{Nucl. Phys.}
  {\bfseries B571} (2000) 583}
  [\href{https://arxiv.org/abs/hep-th/9910102}{{\ttfamily hep-th/9910102}}].

\bibitem{Dorey:2000ma}
P.~Dorey, C.~Dunning and R.~Tateo, \emph{{Differential equations for general
  SU(n) Bethe ansatz systems}},
  \href{https://doi.org/10.1088/0305-4470/33/47/308}{\emph{J. Phys.} {\bfseries
  A33} (2000) 8427} [\href{https://arxiv.org/abs/hep-th/0008039}{{\ttfamily
  hep-th/0008039}}].

\bibitem{Dorey:2006an}
P.~Dorey, C.~Dunning, D.~Masoero, J.~Suzuki and R.~Tateo,
  \emph{{Pseudo-differential equations, and the Bethe Ansatz for the classical
  Lie algebras}},
  \href{https://doi.org/10.1016/j.nuclphysb.2007.02.029}{\emph{Nucl. Phys.}
  {\bfseries B772} (2007) 249}
  [\href{https://arxiv.org/abs/hep-th/0612298}{{\ttfamily hep-th/0612298}}].

\bibitem{Masoero:2015lga}
D.~Masoero, A.~Raimondo and D.~Valeri, \emph{{Bethe Ansatz and the Spectral
  Theory of Affine Lie Algebra-Valued Connections I. The simply-laced Case}},
  \href{https://doi.org/10.1007/s00220-016-2643-6}{\emph{Commun. Math. Phys.}
  {\bfseries 344} (2016) 719}
  [\href{https://arxiv.org/abs/1501.07421}{{\ttfamily 1501.07421}}].

\bibitem{Masoero:2015rcz}
D.~Masoero, A.~Raimondo and D.~Valeri, \emph{{Bethe Ansatz and the Spectral
  Theory of Affine Lie algebra-Valued Connections II. The Non Simply-laced
  Case}}, \href{https://doi.org/10.1007/s00220-016-2744-2}{\emph{Commun. Math.
  Phys.} {\bfseries 349} (2017) 1063}
  [\href{https://arxiv.org/abs/1511.00895}{{\ttfamily 1511.00895}}].

\bibitem{Frenkel:2016gxg}
E.~Frenkel and D.~Hernandez, \emph{{Spectra of quantum KdV Hamiltonians,
  Langlands duality, and affine opers}},
  \href{https://arxiv.org/abs/1606.05301}{{\ttfamily 1606.05301}}.

\bibitem{Frappat:2000}
L.~Frappat, P.~Sorba and A.~Sciarrino, \emph{{Dictionary on Lie algebras and
  superalgebras}}. {Academic Press (London)}, 2000.

\bibitem{Abraham:1978}
R.~Abraham and J.~Marsden, \emph{Foundations of Mechanics}, AMS Chelsea
  publishing. American Mathematical Society, 1978.

\bibitem{Adler:1979ib}
M.~Adler, \emph{{On a Trace functional for formal pseudo differential operators
  and the symplectic structure of the Korteweg-de Vries equation}},
  \href{https://doi.org/10.1007/BF01410079}{\emph{Invent. Math.} {\bfseries 50}
  (1979) 219}.

\bibitem{Kostant:1979qu}
B.~Kostant, \emph{{The Solution to a generalized Toda lattice and
  representation theory}},
  \href{https://doi.org/10.1016/0001-8708(79)90057-4}{\emph{Adv. Math.}
  {\bfseries 34} (1979) 195}.

\bibitem{Symes:1981}
W.~W. Symes, \emph{{Systems of Toda type, inverse spectral problems, and
  representation theory}},
  \href{https://doi.org/10.1007/BF01389068}{\emph{Inventiones mathematicae}
  {\bfseries 63} (1981) 519}.

\end{thebibliography}\endgroup
\bibliographystyle{JHEP}

\end{document}